\documentclass[11pt]{article}
\usepackage{graphicx}
\usepackage{latexsym}
\usepackage{amssymb}
\usepackage{color}
\usepackage{amsfonts}
\usepackage{amsmath}
\usepackage{mathtools}
\usepackage{booktabs}
\usepackage{comment}
\usepackage{longtable}
\usepackage[center]{subfigure}
\usepackage{url}
\newcommand{\BE}{\begin{equation}}
\newcommand{\EE}{\end{equation}}
\newcommand{\BEQ}{\begin{eqnarray}}
\newcommand{\EEQ}{\end{eqnarray}}
\newcommand{\BEQA}{\begin{eqnarray*}}
\newcommand{\EEQA}{\end{eqnarray*}}
\newcommand{\BA}{\begin{array}}
\newcommand{\EA}{\end{array}}

\newtheorem{assumption}{Assumption}

\usepackage[numbers,square,sort]{natbib} 
\bibpunct{(}{)}{,}{a}{,}{,} 

\begin{document}

\title{Improving the accuracy of bubble date estimators under time-varying volatility\footnote{\footnotesize{Anton Skrobotov' research for this paper was supported by Basic Research Program of HSE University (HSE-BR-2026-003).
Eiji Kurozumi's research was supported by JSPS KAKENHI Grant Number 22K01422 and 23H00804.
Address correspondence to 

}}}
\author{
Eiji Kurozumi$^{a}$, Anton Skrobotov$^{b,c}$ \\
{\small {$^{a}$ Hitotsubashi University}}\\
{\small {$^{b}$ HSE University}}\\
}

\date{\today}
\maketitle


\begin{abstract}
In this study, we consider a four-regime bubble model under the assumption of time-varying volatility and propose an algorithm of estimating the break dates with volatility correction. First, we estimate the emerging date of the explosive bubble, its collapsing date, and its recovering date to the normal market under the assumption of homoskedasticity. Second, we collect the residuals and then employ the weighted-least-squares-based estimation of the bubble dates. Using Monte Carlo simulations, we demonstrate that the accuracy of the break date estimators improves significantly via this two-step procedure in some cases compablack to those based on the ordinary least squares {\color{black} method and to the sign-based dating procedure of Harvey, Leybourne and Zu (2020).}
\medskip

\indent \emph{Keywords}: rational bubble; change points; explosive autoregression; time-varying volatility; right-tailed unit root testing; mildly explosive; mildly integrated.

\medskip

\indent \emph{JEL Codes}: C12, C22

\end{abstract}


\newpage
\renewcommand{\baselinestretch}{1.4}

\section{Introduction}
\baselineskip= 6mm

Non-stationary volatility is sometimes observed in time series (in particular, financial data), but discussions on break date estimators under non-stationary volatility have received limited attention in the literature. One of the exceptions is \citet{harris2020level}, wherein the estimation of level shift is improved by correcting the original time series via non-parametrically estimated time-varying variance. While the explosive bubble model has been proposed by \citet{PWY2011} and extended by \citet{phillips2015a,phillips2015b} and \citet{harvey2017improving}, in which the time series is generated by a unit root process followed by an explosive regime that is again followed by a unit root regime (or with a possible stationary correction market in a recovery regime), the importance of non-stationary volatility accommodation in bubble detection methods has been discussed by \citet{HLST2016}, \citet{harvey2018testing,harvey2020sign}, \citet{phillips2020real}, \citet{astill2023cusum}, and \citet{kurozumi2022time}, among others. \citet{phillips2020real} proposed a modification of the wild bootstrap recursive algorithm (based on the expanding sample) of \citet{HLST2016} for obtaining the dates of the bubble(s) and also addressed the multiplicity testing problem. On the contrary, \citet{harvey2020sign} consideblack the maximization of {\color{black} a} sign-based {\color{black} statistic, whose null distribution does not depend on the volatility function,} for obtaining the dates of the bubble but did not provide any finite sample {\color{black} performance of the resulting date estimators; we fill this gap by including their procedure as the main competitor in our Monte Carlo experiments.} At the same time, as discussed in  \citet{harvey2017improving} and \citet{pang2021estimating} (PDC hereafter), the break date estimators based on the minimization of the sum of squablack residuals (SSR) are more accurate than the recursive method of \citet{phillips2015a,phillips2015b} under the assumption of homoskedasticity. Nevertheless, as far as we know, there are no studies that accommodate the non-stationary volatility behavior into the estimation of the bubble dates based on the minimization of the SSR.

{\color{black} Among the volatility-robust methods listed above, \citet{kurozumi2022time} propose a test based on a time-transformed statistic. Because the statistic is constructed on a transformed time scale rather than on calendar time, the resulting sequence of statistics is not indexed by the original dates and cannot be used directly to date the beginning and the end of the explosive regime. For this reason, we do not consider it as a benchmark for the dating problem. Instead, as a volatility-robust competitor that delivers estimates of the calendar dates through the maximization of a statistic, we use the sign-based procedure of \citet{harvey2020sign}.}

Recently, PDC and \citet{kurozumi2022asymptotic} investigated the asymptotic behavior of bubble date estimators. In particular, they obtained the consistency of the collapsing date estimator by minimizing the SSR using the two-regime model (even though the true model has four regimes), allowing non-stationary volatility. Owing to the consistency, one could split the whole sample at the estimated break date and consider the estimation of the date of the origination of the bubble using the sample before the estimated collapsing date and the date of the market recovery using the sample after the estimated collapsing date. This sample splitting approach closely resembles that of \citet{harvey2017improving} by minimizing the full SSR based on the four-regime model, but it is computationally less involved and, as PDC demonstrated, performs better in terms of the estimation accuracy of the break dates. On the contrary to the collapsing date of the bubble, the consistency of the dates of the origination of the bubble and the market recovery depend on the extent of the explosive and collapsing regimes. In other words, if the explosive speed is not sufficiently fast, then PDC and \citet{kurozumi2022asymptotic} obtain only the consistency of the estimators of the break fractions, not the break date.

In this study, we propose a two-step algorithm for estimating the emerging, collapsing, and recovery dates of a bubble under non-stationary volatility. First, due to the consistency of the break date (fraction) estimators regardless of heteroskedasticity, we estimate these break dates as proposed by PDC and \citet{kurozumi2022asymptotic} and collect the residuals of the fitted four-regime model. Second, we non-parametrically estimate the time-varying error variance from these residuals and perform the generalized-least-squares (GLS)-based sample splitting approach, which minimizes the weighted SSR. Monte Carlo simulations demonstrate the performance of our correction method for a model with a one-time break in volatility, especially when this break occurs at the beginning or end of the sample. The empirical application consists of different time series of cryptocurrencies for which the two methods of identifying the bubble dates are performed: one without volatility correction and another with volatility correction.

{\color{black} Our paper is also related to the recent study of \citet{kurozumi2025confidence}, who construct confidence sets for the emergence, collapse, and recovery dates of a bubble by inverting tests for the location of a break date. However, their analysis does not allow for time-varying volatility, because the shocks are assumed to have a constant variance. In contrast, we focus on point estimation of the dates under nonstationary volatility and show that its accuracy can be improved by the volatility correction. The two papers are therefore complementary, and it is of interest to investigate whether the volatility correction proposed here can also improve the accuracy of confidence sets, which we leave for future research.}

The remainder of this paper is organized as follows: Section 2 formulates the model and assumptions. In Section 3, we define the main GLS-based procedure under a general type of weights. Section 4 discusses the choice of the specific weights and proposes the new two-step algorithm. Section 5 demonstrates the finite sample performance of the estimated break dates and Section 6 provides an empirical example. Section 7 concludes the paper.

\section{Model}

Let us consider the following bubble's emerging and collapsing model for $t=1,2,\ldots,T$:
\BE
y_t=\left\{\BA{lcl}
c_0T^{-\eta_0}+y_{t-1}+\varepsilon_t & : & 1 \leq t \leq k_e,\\
\phi_a y_{t-1}+\varepsilon_t & : & k_e+1\leq t \leq k_c, \\
\phi_b y_{t-1}+\varepsilon_t & : & k_c+1\leq t \leq k_r,\\
c_1T^{-\eta_1}+y_{t-1}+\varepsilon_t & : & k_r+1\leq t\leq T,
\EA\right.
\label{model:0}
\EE
where $y_0=o_p(T^{1/2})$, $c_0\geq 0$, $\eta_0 > 1/2$, $\phi_a > 1$, $\phi_b < 1$, $c_1\geq 0$, and $\eta_1 > 1/2$. We assume that the market is normal in the first and last regimes in the sense that the time series $y_t$ is a unit root process (a random walk) with a possibly positive drift shrinking to 0. The process starts exploding at $t=k_e+1$ at a rate of $\phi_a$, which is typically only slightly greater than one and thus sometimes characterized as a mildly explosive specification. The explosive behavior stops at $t=k_c$ and $y_t$ is collapsing at a rate of $\phi_b < 1$ in the next regime, followed by the normal market regime. This model can be regarded as a structural change model with the break points being given by $k_e$, $k_c$, and $k_r$. The corresponding break fractions are defined as $\tau_e\coloneqq k_e/T$, $\tau_c\coloneqq k_c/T$, and $\tau_r\coloneqq k_r/T$, respectively. We would like to estimate these break dates as accurately as possible.

For model \eqref{model:0}, we make the following assumptions:

\begin{assumption}
\label{assumption:tau}
$0 < \tau_e < \tau_c < \tau_r < 1$.
\end{assumption}

\begin{assumption}
\label{assumption:base}
$\varepsilon_t\coloneqq\sigma_t e_t$, where $\{e_t\}\sim i.i.d.(0,1)$ with $E[e_t^4]<\infty$ and $\sigma_t\coloneqq \omega(t/T)$, where $\omega(\cdot)$ is a nonstochastic and strictly positive function on $[0,1]$, satisfying $\underline{\omega} < \omega(\cdot) <\overline{\omega} < \infty$.
\end{assumption}

By Assumption \ref{assumption:tau}, the break fractions are distinct and not too close each other. Assumption \ref{assumption:base} allows for various kinds of nonstationary unconditional volatility in the shocks, such as a volatility shift (possibly multiple times) and linear and non-linear transitions. Under Assumption \ref{assumption:base}, it is well known that the functional central limit theorem (FCLT) holds for the partial sum process of $\{\varepsilon_t\}$ normalized by $\sqrt{T}$, which weakly converges to a variance-transformed Brownian motion, as shown by \citet{CavaliereTaylor2007a,CavaliereTaylor2007b}.

\section{Individual Estimation of Break Dates}

Following PDC and \citet{kurozumi2022asymptotic}, we estimate the break dates one at a time. As model \eqref{model:0} can be expressed as
\BE
y_t
=
\left\{\BA{l}
\phi_1y_{t-1}+u_t \\
\phi_ay_{t-1}+u_t \\
\phi_by_{t-1}+u_t \\
\phi_1y_{t-1}+u_t 
\EA\right.
\quad \mbox{where $\phi_1=1$}\quad\mbox{and}\quad
u_t
\coloneqq
\left\{\BA{l}
c_0/T^{\eta_0}+\varepsilon_t \\
\varepsilon_t \\
\varepsilon_t \\
c_1/T^{\eta_1}+\varepsilon_t,
\EA\right.
\label{model:est1}
\EE
PDC and \citet{kurozumi2022asymptotic} proposed to fit the one-time structural change model without a constant and to estimate the break point by minimizing the SSR. It is shown that the estimated break date, $\hat{k}_c$, is consistent for $k_c$. We then split the whole sample into the two subsamples, and from the fist subsample before $\hat{k}_c$, the emerging date of the explosive behavior is estimated by fitting a one-time structural change model again, while $k_r$ is estimated from the second subsample after $\hat{k}_c$. These estimated break fractions, $\hat{\tau}_e\coloneqq\hat{k}_e/T$ and $\hat{\tau}_r\coloneqq\hat{k}_r/T$, are shown to be consistent, and furthermore, $\hat{k}_e$ ($\hat{k}_r$) is consistent for $k_e$ ($k_r$) if, roughly speaking, $\phi_a$ deviates from 1 sufficiently ($\phi_a-1 > 1-\phi_b$). See PDC and \citet{kurozumi2022asymptotic} for details.

Although the aforementioned estimated break dates (fractions) are consistent under nonstationary volatility in Assumption \ref{assumption:base}, the efficiency gain would be expected by estimating the break dates based on the weighted SSR following \citet{xuphillips2008adaptive}. To be more precise, let $w_t$ be a generic series of weights $w_t$, and then, the weighted SSR based on a one-time structural change model is given by
\BE
SSR(k,w_t,\phi_a,\phi_b)\coloneqq \sum_{1}^{k}{w}_t^{-2}\left(y_t-{\phi}_a y_{t-1}\right)^2+\sum_{k+1}^{T}{w}_t^{-2}\left(y_t-{\phi}_by_{t-1}\right)^2,
\label{ssr:k}
\EE
where $\sum_{t=\ell}^m$ is abbreviated just as $\sum_{\ell}^m$. As $SSR(k,w_t,\phi_a,\phi_b)$ is minimized at
\[
\hat{\phi}_a(k,w_t)\coloneqq\frac{\sum_1^k y_{t-1}y_t{w}_t^{-2}}{\sum_1^k y_{t-1}^2{w}_t^{-2}}
\quad\mbox{and}\quad
\hat{\phi}_b(k,w_t)\coloneqq\frac{\sum_{k+1}^T y_{t-1}y_t{w}_t^{-2}}{\sum_{k+1}^T y_{t-1}^2{w}_t^{-2}}
\]
for given $k$ and $w_t$, the estimator of $k_c$ is given by
\[
\hat{k}_c(w_t)\coloneqq \arg\min_{\underline{\tau}_c \leq k/T \leq \overline{\tau}_c} SSR(k,w_t),
\]
where $0<\underline{\tau}_c<\tau_c<\overline{\tau}_c < 1$ and $SSR(k,w_t)\coloneqq SSR(k,w_t,\hat{\phi}_a (k,w_t),\hat{\phi}_b(k,w_t))$. The corresponding break fraction estimator is defined as $\hat{\tau}_c(w_t)\coloneqq \hat{k}_c(w_t)/T$.

Once we obtain the estimator of $k_c$, we can move on to the estimation of $k_e$ and $k_r$. For $k_e$, on the one hand, the estimation is based on the minimization of the weighted SSR using the first sub-sample, and the estimator is defined as
\[
\hat{k}_e(w_t)\coloneqq \arg\min_{\underline{\tau}_e\leq k/T\leq \overline{\tau}_e}  SSR_1(k,w_t)
\]
where $0 <\underline{\tau}_e < \tau_e < \overline{\tau}_e <{\hat{\tau}}_c$ and
\[
SSR_1(k,w_t) \coloneqq \sum_{1}^k{w}_t^{-2}\left(y_t-\hat{\phi}_c(k,w_t)y_{t-1}\right)^2+\sum_{k+1}^{\hat{k}_c(w_t)}{w}_t^{-2}\left(y_t-\hat{\phi}_d(k,w_t)y_{t-1}\right)^2
\]
\[
\mbox{with}\quad \hat{\phi}_c(k,w_t)\coloneqq\frac{\sum_{1}^k y_{t-1}y_t{w}_t^{-2}}{\sum_{1}^k y_{t-1}^2{w}_t^{-2}}\quad\mbox{and}\quad
\hat{\phi}_d(k,w_t)\coloneqq\frac{\sum_{k+1}^{\hat{k}_c(w_t)} y_{t-1}y_t{w}_t^{-2}}{\sum_{k+1}^{\hat{k}_c(w_t)} y_{t-1}^2{w}_t^{-2}}.
\]
The corresponding break fraction estimator is defined as $\hat{\tau}_e(w_t)\coloneqq \hat{k}_e(w_t)/T$. For notational convenience, we suppressed the dependence of $\hat{k}_e(w_t)$, $\hat{\tau}_e(w_t)$, and $SSR_1(k,w_t)$ on $\hat{k}_c(w_t)$.

On the other hand, for the estimation of $k_r$, we minimize the weighted SSR using the second sub-sample, and the estimator is defined as
\[
\hat{k}_r(w_t)\coloneqq \arg\min_{\underline{\tau}_r\leq k/T\leq \overline{\tau}_r}  SSR_2(k,w_t)
\]
where $\hat{\tau}_c <\underline{\tau}_r <\tau_r < \overline{\tau}_r <1$ and
\[
SSR_2(k,w_t) \coloneqq \sum_{\hat{k}_c(w_t)+1}^k{w}_t^{-2}\left(y_t-\hat{\phi}_e(k,w_t)y_{t-1}\right)^2+\sum_{k+1}^T{w}_t^{-2}\left(y_t-\hat{\phi}_f(k,w_t)y_{t-1}\right)^2
\]
\[
\mbox{with}\quad \hat{\phi}_e(k,w_t)=\frac{\sum_{\hat{k}_c(w_t)+1}^k y_{t-1}y_t{w}_t^{-2}}{\sum_{\hat{k}_c(w_t)+1}^k y_{t-1}^2{w}_t^{-2}}\quad\mbox{and}\quad
\hat{\phi}_f(k,w_t)=\frac{\sum_{k+1}^T y_{t-1}y_t{w}_t^{-2}}{\sum_{k+1}^T y_{t-1}^2{w}_t^{-2}}.
\]
The corresponding break fraction estimator is defined by $\hat{\tau}_r(w_t)\coloneqq \hat{k}_r(w_t)/T$.

We call the above-stated method the sample splitting approach based on the weighted least squares (WLS) method. Note that the special case where $w_t=1$ for all $t$ corresponds to the estimation method employed by PDC, with the sample ranging from 1 to $k_r$, and that by \citet{kurozumi2022asymptotic}.

\section{Adaptive Estimation}

To implement the sample splitting approach based on the WLS method in practice, we need to choose an appropriate weight function $w_t$. In our model, it is natural to choose the volatility function $\sigma_t$ as the weight $w_t$ to obtain the efficiency gain, but such a WLS estimation is infeasible because the volatility function is unknown. In this article, we follow \citet{xuphillips2008adaptive} and estimate $\sigma_t$ via a kernel-based method. More precisely, as the first step, we estimate $\tau_e$, $\tau_c$, and $\tau_r$ using the sample splitting approach based on the ordinary least squares (OLS) method ($w_t=1$ for all $t$), as proposed by PDC and \citet{kurozumi2022asymptotic}. Furthermore, using the estimated break dates denoted as $\hat{\tau}_e(1)$, $\hat{\tau}_c(1)$, and $\hat{\tau}_r(1)$, we estimate
\begin{equation}
    \Delta y_t={\mu}_1D_t(\hat{\tau}_e(1),\hat{\tau}_c(1))+{\mu}_2D_t(\hat{\tau}_c(1),\hat{\tau}_r(1))+{\phi}_aD_t(\hat{\tau}_e(1),\hat{\tau}_c(1))y_{t-1}+{\phi}_bD_t(\hat{\tau}_c(1),\hat{\tau}_r(1))y_{t-1}+e_t,
\end{equation}
via the OLS method and obtain the residuals $\hat{e}_t$, where $D_t(a,b)=\mathbb I(\lfloor aT\rfloor<t\leq\lfloor bT\rfloor)$, with $\mathbb{I}(\cdot)$ being the indicator function. Using these first-step residuals, the second step entails the calculation of the weight function, which is used for the WLS estimator, that is, we calculate $\hat{\sigma}_t^2$ as
\begin{equation}
    \hat{\sigma}_t^2=\sum_{t=1}^T\left(\sum_{i=1}^T{K_{it}}\right)^{-1}K_{it}\hat{e}_t^2,
\quad
where
\quad
    K_{it} = \left\{\BA{lc}
K\left(\frac{t-i}{Tb}\right) & \text{if $t\neq i$}\\
0 &  \text{if $t=i$}
\EA\right.,
\end{equation}
$K(\cdot)$ is a bounded nonnegative continuous kernel function defined on the real line with $\int_{-\infty}^{\infty}K(s)ds=1$, and $b$ is a bandwidth parameter. By plugging $\hat{\sigma}_t^2$ into $w_t^2$ in the sample splitting approach, we obtain the estimators $\hat{\tau}_e(\hat{\sigma}_t)$, $\hat{\tau}_c(\hat{\sigma}_t)$, and $\hat{\tau}_r(\hat{\sigma}_t)$. \citet{xuphillips2008adaptive} showed that the estimation accuracy of the coefficient in a stable autoregressive model improves by the adaptive (WLS) estimation, and we investigate if their observation holds for the estimation of the bubble's dates in the next section.

\section{Monte Carlo Simulations}

In this section, we examine the performance of the estimates of the bubble regime dates in finite samples when the error variance is subject to changes in volatility. 

Monte-Carlo simulations reported in this section are based on the series generated by model \eqref{model:0} with $y_0=1500$ and $\{\varepsilon_t\}\sim IIDN(0,1)$. Data are generated from this DGP for samples of $T=(400,800)$ with {\color{black} $10,000$} replications.\footnote{All simulations were programmed in R with rnorm random number generator.} We set the drift terms in the first and fourth regimes to $c_0T^{-\eta_0}=1/800$ and $c_1T^{-\eta_0}=1/800$, respectively, following PDC. In this experiment, we focus on the local to  unit root behavior characterized by $\phi_a=1+c_a/T$ and $\phi_b=1-c_b/T$, where $c_a$ takes values among $\{4,5,6\}$ whereas $c_b$ is fixed at $6$. 

For the dates of bubble regimes, we set $(\tau_e,\tau_c,\tau_r)$ to be equal to (0.4,0.6,0.7). This setting seems to be empirically relevant considering Japanese stock price, the US house price index, and cryptocurrencies. We consider the case in which there is a one-time break in volatility at date $\tau $, so that the volatility function $\sigma_t$ has the following form:
\[\sigma_t^2=s_0^2+\delta (s_1^2-s_0^2)\mathbb I(t>\lfloor \tau T\rfloor)\]
where $s_1/s_0$ takes values among $\{1/5, 5\}$ and $\tau$ takes values among $\{0.2, 0.8\}$.

{\color{black} To examine the sensitivity to the form of the volatility change, we also consider three other designs: a temporary change in volatility that coincides with the explosive regime (double shift in volatility), $\sigma_t=s_0+(s_1-s_0)\,\mathbb I(\lfloor 0.4T\rfloor<t\leq\lfloor 0.6T\rfloor)$; a smooth logistic transition in the middle of the sample, $\sigma_t=s_0+(s_1-s_0)/(1+\exp\{-50(t/T-0.5)\})$; and a trending volatility, $\sigma_t=s_0+(s_1-s_0)\,t/T$. In all the designs, we report the results for the ratio $s_0/s_1\in\{1/5,5\}$, and for the homoskedastic case $s_0/s_1=1$. For the double shift, $s_0$ is the volatility outside and $s_1$ inside the explosive regime.}

As in \cite{kurozumi2022asymptotic}, in the minimization of $SSR(k/T)$, $SSR_1(k/T)$, and $SSR_2(k/T)$, we exclude the first and last 5\% observations from the permissible break date $k$. For example, when estimating $k_r$ based on $SSR_2(k/T)$, the permissible break date $k$ ranges from $\hat{k}_c+0.05T+1$ to $0.95T$. If the break date estimate $\hat{k}_c$ exceeds $0.95T$, then we cannot estimate $k_r$; we do not include such a case in any bins of the histogram and thus the sum of the heights of the bins is not necessarily equal to one for $\hat{k}_r$ in some cases. Similarly, we cannot estimate $\hat{k}_e$ when $\hat{k}_c < 0.05T$. To save space, we pick up several selected cases in the following, and the other cases are provided in the online appendix.

{\color{black} In addition to the SSR-based estimators without and with the volatility correction (OLS and WLS hereafter), we consider the sign-based dating procedure of \citet{harvey2020sign} (sADF hereafter) as the main competitor. Like our WLS method, it is designed to be valid under nonstationary volatility, but it does not require estimating the volatility function. The procedure is based on the cumulated signs of the first differences, $C_t\coloneqq\sum_{i=2}^{t}\mathrm{sign}(\Delta y_i)$, where $\mathrm{sign}(x)\coloneqq 1-2\,\mathbb I(x\leq 0)$. For a window $(a,b]$ with $0\leq a<b\leq T$, let $\hat{\rho}(a,b)$ be the OLS estimator from the regression of $\Delta C_t$ on $C_{t-1}$ without an intercept for $t=a+1,\ldots,b$, and let $\hat{s}^2(a,b)$ denote the corresponding residual variance. The dating statistic is
\BE
sDF^*(a,b)\coloneqq\frac{\hat{\rho}(a,b)}{\sqrt{\left(\tilde{s}^2(a,b)\right)^{\epsilon}\Big/\sum_{t=a+1}^{b}C_{t-1}^2}},\qquad
\tilde{s}^2(a,b)\coloneqq\frac{b\,\hat{s}^2(0,b)-a\,\hat{s}^2(0,a)}{b-a-1},
\label{eq:sdf}
\EE
and the emerging and collapsing dates are estimated by $(\hat{k}_e,\hat{k}_c)\coloneqq\arg\max_{a,b} sDF^*(a,b)$ over the windows with $b-a\geq\pi T$. The modified variance estimator $\tilde{s}^2$ together with the power $\epsilon\in(0,1]$ is introduced in \citet{harvey2020sign} to obtain a consistent estimator of the emerging date. We set $\epsilon=0.01$ and $\pi=0.1$, which are the values used in the empirical application of \citet{harvey2020sign}, and we exclude the first and last 5\% observations from the permissible dates, as for the SSR-based estimators. The sADF procedure does not provide an estimator of the recovering date, which is a limitation of this approach relative to the sample-splitting estimators, which estimate all three dates; therefore, it is compablack with the OLS and WLS methods only for $k_e$ and $k_c$. We do not include the dating based on the time-transformed statistic of \citet{kurozumi2022time}, because it is computed on a transformed time scale and its path does not provide estimates of calendar dates.}

{\color{black} In addition to the histograms, we report the following measures of accuracy for each date $k\in\{k_e,k_c,k_r\}$ and each estimator: the bias, the RMSE, the MAE, and the probabilities $P_5$ and $P_{10}$ that the estimate is within $5$ and $10$ observations of the true date $\lfloor\tau_\cdot T\rfloor$. The bias, RMSE and MAE are computed over the replications in which the date is estimated, whereas $P_5$ and $P_{10}$ are computed over all the replications, so that the cases in which the date cannot be estimated are counted as misses; the latter are therefore free from the selection effect and comparable across the estimators. Because the tolerance is expressed in the number of observations, $P_5$ and $P_{10}$ for $T=800$ are lower than for $T=400$, and the comparison should be made for the same $T$. Table \ref{tab:acc_summary} summarizes $P_5$, and the complete tables are reported in the online appendix.}

Figure 1 presents the histograms of $\hat{k}_c$ when $\tau=0.8$, $s_0/s_1=1/5$, and $T=400$. The left-hand column shows the results based on the OLS method, while the right-hand column corresponds to the WLS-based {\color{black} method; in each panel, the histogram of the sADF estimator is overlaid in orange.} In this case, the process becomes more volatile at the end of the sample, and thus, it would be difficult to distinguish between the explosive and collapsing behavior (from $\tau=0.4$ to 0.7) and a random walk with high volatility (from $\tau=0.8$ to 1). As expected, the OLS method incorrectly chooses the end of the sample as the collapsing date when $c_a=4$, as shown in Figure 1(a), although the local peak is observed at around the true break fraction ($\tau_c=0.6$). As the size of the bubble ($c_a$) gets larger, the local peak becomes higher, as is observed in Figures 1(c) and (e) (note that the vertical axis is different depending on the value of $c_a$). On the contrary, we can observe from Figures 1(b), (d), and (f) that the WLS method can estimate the collapsing date more accurately than the OLS method; the finite sample distribution has a mode at the true break fraction and {\color{black} the probability $P_5$ that $\hat{k}_c$ is within five observations of the true date is $0.40$, $0.49$ and $0.58$ for WLS versus $0.06$, $0.11$ and $0.17$ for OLS when $c_a=4$, $5$ and $6$, respectively.} {\color{black} The sADF estimator does not select the end of the sample as the collapsing date either, and its distribution has a mode at the true break fraction. However, its peak is lower than that of the WLS-based estimator and its distribution is more dispersed over earlier dates; thus, sADF is more accurate than OLS but less accurate than WLS in this case.} {\color{black} ($P_5$ of sADF is $0.23$, $0.32$ and $0.44$ for $c_a=4$, $5$ and $6$.)} 

Figure 2 shows the histograms of $\hat{k}_c$ when $\tau=0.2$, $s_0/s_1=5$, and $T=400$. In this case, there exists a unit root regime with high volatility at the beginning of the sample, and thus, it is expected that the histograms would have positive frequencies before $\tau=0.2$. In fact, this is the case as observed in Figure 2, although the accuracy is much better than the case in Figure 1. Overall, the WLS-based method can detect the true collapsing date more often than the OLS-based method. {\color{black} For example, when $c_a=4$ and $c_b=6$, $P_5$ rises from $0.30$ to $0.70$ by introducing the adaptive procedure (from $0.41$ to $0.77$ for $c_a=5$ and from $0.52$ to $0.82$ for $c_a=6$).} We can also observe that the WLS method incorrectly detects the collapsing date at the beginning of the sample less frequently than the OLS method. {\color{black} The sADF estimator also avoids false detections at the beginning of the sample, and its peak at the true date is higher than that of OLS;} {\color{black} however, it is lower than that of WLS for all values of $c_a$ ($P_5$ is $0.53$, $0.66$ and $0.75$ for sADF versus $0.70$, $0.77$ and $0.82$ for WLS).} 

    Figure 3 presents the histograms of $\hat{k}_e$ when $\tau=0.2$, $s_0/s_1=5$, and $T=400$, which is the same case as in Figure 2. Overall, when the size of the bubble is small with $c_a=4$, it is difficult to estimate the emerging date ($\tau_e=0.4$) accurately, but for large values of $c_a$, the accuracy of $\hat{k}_e$ improves and the histograms have a peak at $0.4$, as in Figures 3(c)--(f). Again, in this case, the performance of the estimator based on the WLS method is better than that based on the OLS method. {\color{black} The sADF estimator does not produce the spurious frequencies at the beginning of the sample observed for OLS, but its distribution around $\tau_e=0.4$ is more dispersed than that of WLS, whose peak at the true date is the highest and the sharpest.} {\color{black} For $c_a=4$, $5$ and $6$, $P_5$ of $\hat{k}_e$ is $0.02$, $0.08$ and $0.19$ for OLS, $0.30$, $0.38$ and $0.45$ for WLS, and $0.23$, $0.27$ and $0.31$ for sADF.}

The other results are briefly summarized in the online appendix. Overall, Monte Carlo simulations demonstrate that the accuracy of the estimators of the break dates improves significantly in some cases, while in other cases, we cannot find any difference between the distribution of the estimator based on the OLS method and that based on the WLS method. {\color{black} The accuracy measures in Table \ref{tab:acc_summary} (and the complete tables in the online appendix) show that the relative performance of the estimators depends on where the high-volatility regime is located with respect to the bubble. Under homoskedasticity, OLS and WLS are practically indistinguishable. When the periods of high volatility lie outside the bubble episode, WLS is much more accurate than OLS: for example, $P_5$ of $\hat{k}_c$ increases from $0.10$ to $0.44$ in the shift at $\tau=0.8$ with $s_0/s_1=1/5$ and from $0.40$ to $0.74$ in the shift at $\tau=0.2$ with $s_0/s_1=5$, and gains of a similar nature, though smaller, are observed in the double-shift, logistic and trend designs with $s_0/s_1=5$. By contrast, when the high-volatility regime coincides with the bubble (the double shift with $s_0/s_1=1/5$, the shift at $\tau=0.8$ with $s_0/s_1=5$, and the logistic design with $s_0/s_1=1/5$), WLS is less accurate than OLS; for the double shift with $s_0/s_1=1/5$, $P_5$ of $\hat{k}_c$ falls from $0.25$ to $0.10$. A possible explanation is that the weights down-weight the observations from the explosive regime, which carry the most information about the dates. Therefore, the volatility correction cannot be recommended unconditionally.

The sign-based estimator sADF is invariant to the scale of the shocks and does not require the volatility correction, but it discards the information on the magnitudes of the increments. For $\hat{k}_c$, it is less accurate than OLS in the homoskedastic case and less accurate than WLS in the shift designs, but it has the highest $P_5$ in the double-shift and logistic designs with $s_0/s_1=5$. For $\hat{k}_e$, all the estimators are poorly concentrated around the true date in most of the designs, and sADF has the highest $P_5$ in most of them; the exception is the shift at $\tau=0.2$ with $s_0/s_1=5$, where WLS is clearly the most accurate. Moreover, in terms of RMSE and MAE, sADF is the most accurate of the three estimators for $\hat{k}_e$ and $\hat{k}_c$ in most of the designs, because the WLS estimates, which are more often concentrated around the true date, are also more prone to large errors. Thus, neither approach dominates the other. Because the WLS-based method is as accurate as OLS under homoskedasticity and substantially more accurate when the volatility shifts occur outside the bubble episode, but may be less accurate when the volatility is high during the bubble, we recommend reporting the WLS-based estimates together with the OLS-based ones, and, if the two differ substantially, checking the robustness of the dates with the sign-based procedure. 
Note also that sADF does not allow one to estimate $k_r$, so the sign-based approach cannot be used for the full dating of the bubble episode.}

\begin{table}[ht]\centering\scriptsize
\caption{{\color{black}Average probability $P_5=P(|\hat k-k|\leq 5)$ over $T\in\{400,800\}$ and $c_a\in\{4,5,6\}$ ($c_b=6$; 10,000 replications; $\hat k$ not estimated counts as a miss). The highest value among the estimators of each date is in bold (ties after rounding are all in bold).}}
\label{tab:acc_summary}
{\color{black}
\begin{tabular}{l ccc ccc cc}
\toprule
 & \multicolumn{3}{c}{$\hat k_e$} & \multicolumn{3}{c}{$\hat k_c$} & \multicolumn{2}{c}{$\hat k_r$}\\
\cmidrule(lr){2-4}\cmidrule(lr){5-7}\cmidrule(lr){8-9}
Volatility design & OLS & WLS & sADF & OLS & WLS & sADF & OLS & WLS\\
\midrule
homoskedastic & 0.07 & 0.07 & \textbf{0.08} & \textbf{0.42} & \textbf{0.42} & 0.28 & 0.17 & \textbf{0.18} \\
\midrule
shift, $\tau=0.8$, $s_0/s_1=1/5$ & 0.02 & 0.07 & \textbf{0.08} & 0.10 & \textbf{0.44} & 0.28 & 0.01 & \textbf{0.14} \\
shift, $\tau=0.8$, $s_0/s_1=5$ & 0.06 & 0.06 & \textbf{0.08} & \textbf{0.41} & 0.36 & 0.26 & \textbf{0.19} & 0.16 \\
\midrule
shift, $\tau=0.2$, $s_0/s_1=1/5$ & 0.05 & 0.05 & \textbf{0.06} & \textbf{0.32} & \textbf{0.32} & 0.20 & \textbf{0.11} & \textbf{0.11} \\
shift, $\tau=0.2$, $s_0/s_1=5$ & 0.08 & \textbf{0.32} & 0.22 & 0.40 & \textbf{0.74} & 0.60 & 0.34 & \textbf{0.51} \\
\midrule
double shift, $s_0/s_1=1/5$ & \textbf{0.09} & 0.02 & 0.04 & \textbf{0.25} & 0.10 & 0.10 & \textbf{0.28} & 0.13 \\
double shift, $s_0/s_1=5$ & 0.07 & 0.16 & \textbf{0.27} & 0.42 & 0.56 & \textbf{0.68} & 0.14 & \textbf{0.15} \\
\midrule
logistic, $s_0/s_1=1/5$ & 0.01 & 0.02 & \textbf{0.06} & \textbf{0.13} & 0.10 & 0.08 & \textbf{0.03} & \textbf{0.03} \\
logistic, $s_0/s_1=5$ & 0.06 & \textbf{0.07} & \textbf{0.07} & 0.48 & 0.67 & \textbf{0.70} & 0.45 & \textbf{0.57} \\
\midrule
linear trend, $s_0/s_1=1/5$ & 0.04 & 0.04 & \textbf{0.06} & 0.23 & \textbf{0.25} & 0.16 & 0.05 & \textbf{0.07} \\
linear trend, $s_0/s_1=5$ & 0.05 & 0.08 & \textbf{0.10} & 0.45 & \textbf{0.51} & 0.38 & 0.28 & \textbf{0.31} \\
\bottomrule
\end{tabular}
}
\end{table}

\section{Empirical Application}

{\color{black} In this section, we apply the OLS-based, WLS-based and sADF-based dating procedures to the daily closing prices (in US dollars, 00:00 GMT) of the twelve largest cryptocurrencies by capitalization (btc, eth, xrp, xlm, bch, ltc, eos, bnb, ada, xtz, etc, xmr). Following \citet{kurozumi2022time}, who documented explosive behavior and non-stationary volatility in these series, we treat each calendar year from 2014 to 2019 as a separate sample (365 daily observations, 366 in 2016) and use every coin for which the data are available for the whole year. This gives 45 coin-year series (3, 5, 6, 7, 12 and 12 series in 2014-2019, respectively). In contrast to a selection of cases in which the methods disagree, we report all of them: the estimated dates for every series are given in Table \ref{tab:emp_all}. The OLS and WLS estimators are implemented as in Section 4 (Gaussian kernel, bandwidth chosen by leave-one-out cross-validation), the sADF estimator is implemented as in Section 5 ($\epsilon=0.01$, $\pi=0.1$), and the first and last 5\% of the observations are excluded from the permissible dates in all cases.}

{\color{black} Because the true dates are unknown, and because a visual judgement of whether the dates are ``correct'' is subjective, we assess the estimates by three objective admissibility criteria, which were fixed before the results were computed. Given the estimated dates, we estimate the four-regime model by OLS with regime-specific intercepts and slopes and require that: (C1) the autoregressive coefficient is greater than one in the explosive regime $(\hat k_e,\hat k_c]$ and smaller than one in the collapsing regime $(\hat k_c,\hat k_r]$; (C2) the price rises over $(\hat k_e,\hat k_c]$ and falls over $(\hat k_c,\hat k_r]$; (C3) the estimated collapsing date is within $0.05T$ observations of the maximum of the price over $(\hat k_e,\hat k_r]$. A set of dates is called admissible if it satisfies (C1)-(C3). Since the sADF procedure does not estimate the recovering date, the collapsing regime for sADF is $(\hat k_c,T]$. These criteria are necessary conditions for a sensible dating rather than a test of the dates, and they are not informative about the accuracy of the estimates within an admissible regime.}

{\color{black} The results are summarized in Table \ref{tab:emp_sum}. The volatility correction matters in practice: the OLS- and WLS-based dates coincide for all three dates in only 12 of the 45 series (one series in 2015, four in 2017 and seven in 2019, and none in 2014, 2016 and 2018). At the same time, the criteria do not discriminate between the two methods. The dates are admissible in 10 series for OLS and in 11 series for WLS. Among the 33 series in which the two methods disagree, the WLS-based dates are admissible and the OLS-based dates are not in 4 series, the opposite holds in 3 series, both are admissible in 2 series and neither in 24 series; the difference between 4 and 3 is clearly not significant (exact sign test, $p$-value $=1$). We conclude that, on these data, we cannot claim that the volatility correction yields more adequate dates, and we do not draw such a conclusion. A systematic difference is observed in the location of the dates at the beginning of the sample: the OLS estimator selects the emerging date at the lower boundary of the permissible dates ($\hat k_e=0$, that is, no emerging date is found) in 13 series and the collapsing date in the first 10\% of the sample in 13 series, whereas for the WLS estimator these numbers are 4 and 4. This is consistent with the Monte Carlo evidence that OLS tends to attribute the high-volatility part of the sample to the explosive behavior.}

{\color{black} The year 2018 illustrates how the volatility correction changes the dating. The 2018 samples start after the 2017 boom and contain two collapses, in January--February and in November, so that none of them satisfies the single-bubble structure of model \eqref{model:0}. For 8 of the 12 series, the OLS estimator sets the collapsing date at 5 February 2018, that is, at the crash that follows the 2017 peak, whereas for 10 of the 12 series the WLS estimator sets it between 6 and 13 November 2018, which is the second crash. The high volatility in the first months of 2018 is down-weighted by the WLS method, which is consistent with this change. Neither choice can be called correct in this case, since the model is misspecified; the example shows that nonstationary volatility affects the dating in realistic samples and that the choice of the sample period should precede the dating. For that purpose, the procedure of \citet{phillips2015a,phillips2015b} may be useful.}

{\color{black} The sADF estimator is admissible in only one series, and the distance between its collapsing date and the maximum of the price is much larger (median 56 observations) than for the OLS and WLS estimators (13 and 17 observations). Two features explain this. First, criteria (C2) and (C3) require a decline after $\hat k_c$, whereas sADF does not estimate a recovering date and the whole remaining sample is used as the collapsing regime, which is a handicap for sADF. Second, and more importantly, the sADF statistic is built on the cumulated signs of the first differences and is therefore symmetric with respect to the direction of the movement: a persistent run of price declines also produces a large value of the statistic. In fact, the price falls over the estimated explosive window $(\hat k_e,\hat k_c]$ in 31 of the 45 series (the median price change over the window is $-33\%$). If we restrict attention to the explosive part, that is, if we require only that the autoregressive coefficient in $(\hat k_e,\hat k_c]$ is greater than one and that the price rises over this window, then the criteria are satisfied in 15, 12 and 8 series for OLS, WLS and sADF, respectively. Hence, sADF remains the least successful of the three methods, but the difference is smaller than that in Table \ref{tab:emp_sum}.}

\begin{table}[ht]
\centering
\caption{Number of series (out of 45) for which the estimated dates satisfy criteria (C1)--(C3) of Section 6, and the median distance (in observations) between $\hat k_c$ and the maximum of the price.}\label{tab:emp_sum}
\begin{tabular}{lccccc}
\toprule
 & (C1) & (C2) & (C3) & (C1)--(C3) & Median distance \\
\midrule
OLS  & 15 & 24 & 27 & 10 & 13 \\
WLS  & 12 & 26 & 24 & 11 & 17 \\
sADF & 11 &  7 &  5 &  1 & 56 \\
\bottomrule
\end{tabular}
\end{table}

\section{Conclusion}

We propose an algorithm for volatility correction in the estimation of the dates of the bubble in the four-regime model. The method consists of the following steps: estimation of the break dates without volatility correction and calculation of the residuals, and WLS-based estimation of the dates of the bubble using the weight function obtained through the non-parametric estimation of the volatility function. {\color{black} The Monte Carlo results show that the estimated break dates are as accurate as those under the homoskedasticity assumption when the volatility is constant, and substantially more accurate when the high-volatility regime lies outside the bubble episode, whereas the correction may blackuce the accuracy when the volatility is high during the bubble. The comparison with the sign-based dating procedure of \citet{harvey2020sign} shows that neither approach dominates: the WLS-based estimators are more accurate in terms of the probability of locating the dates close to the true ones in the designs with a volatility shift outside the bubble, whereas the sign-based estimators are competitive for the emerging date and in terms of RMSE and MAE, although the sign-based procedure does not estimate the recovery date. The empirical illustration using 45 coin-year series of cryptocurrencies shows that the
 volatility correction changes the estimated dates in most of the series, while objective admissibility
 criteria do not discriminate between the OLS- and WLS-based dates, which is not surprising because
 most of the samples do not satisfy the single-bubble structure of the model.}

\bibliographystyle{apalike} 
\bibliography{ref_joe}

\newpage

\begin{figure}[h!]%
\begin{center}%
\subfigure[$T=400$, $c_a=4$, $c_b=6$, $s_0/s_1=1/5$]{\includegraphics[width=0.45\linewidth]{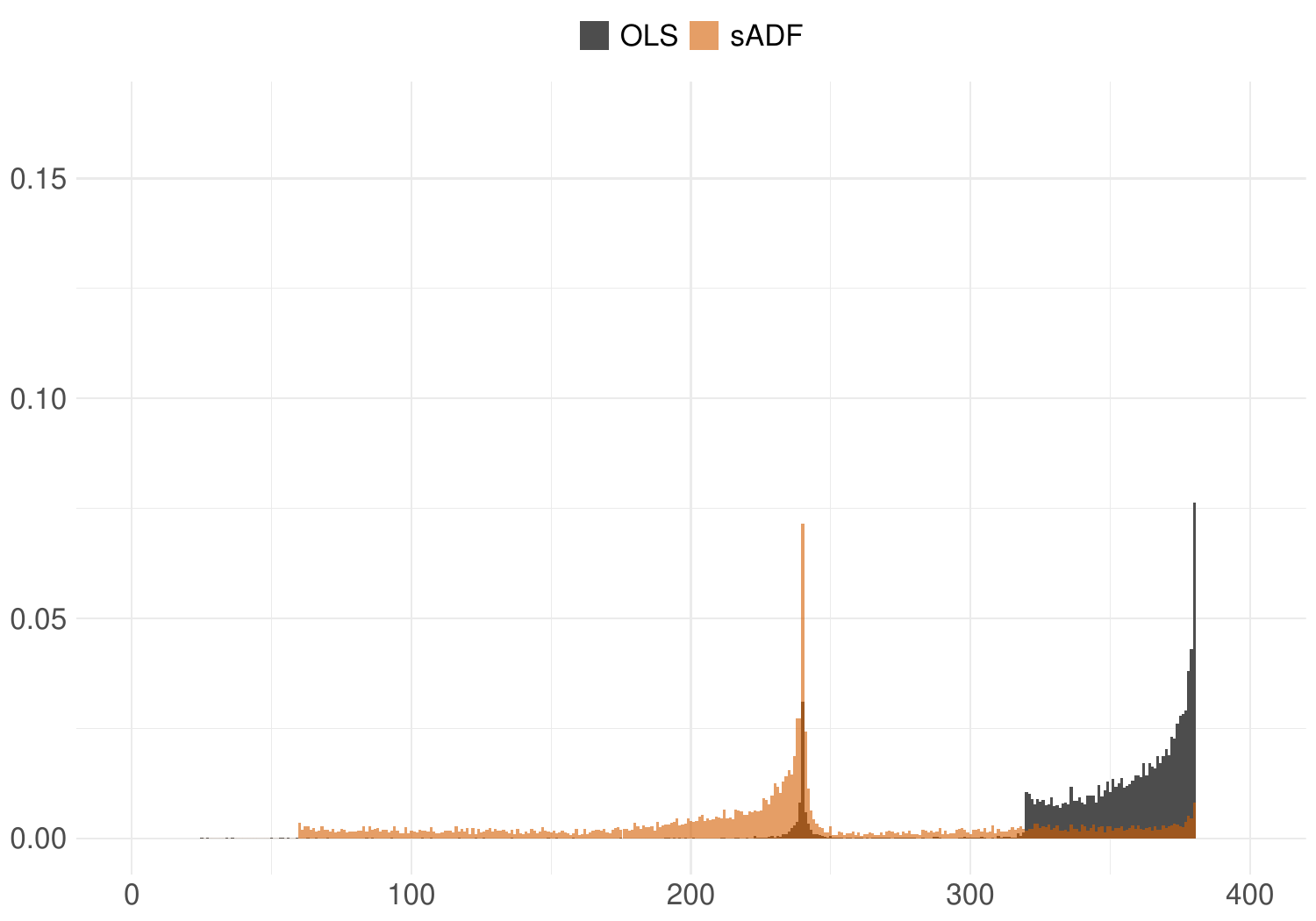}\label{fig:1:1}}
\subfigure[$T=400$, $c_a=4$, $c_b=6$, $s_0/s_1=1/5$]{\includegraphics[width=0.45\linewidth]{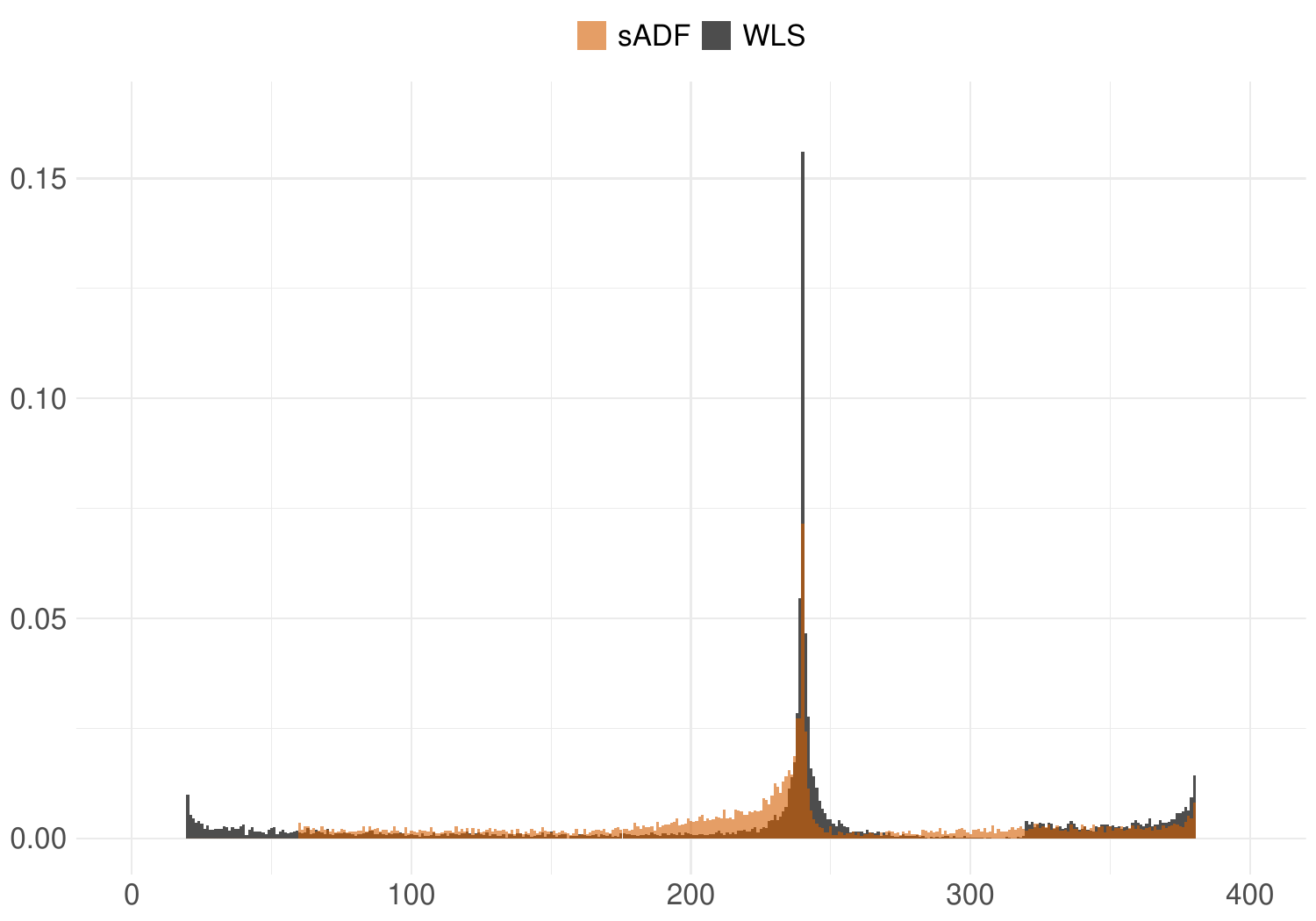}\label{fig:1:2}}\\
\subfigure[$T=400$, $c_a=5$, $c_b=6$, $s_0/s_1=1/5$]{\includegraphics[width=0.45\linewidth]{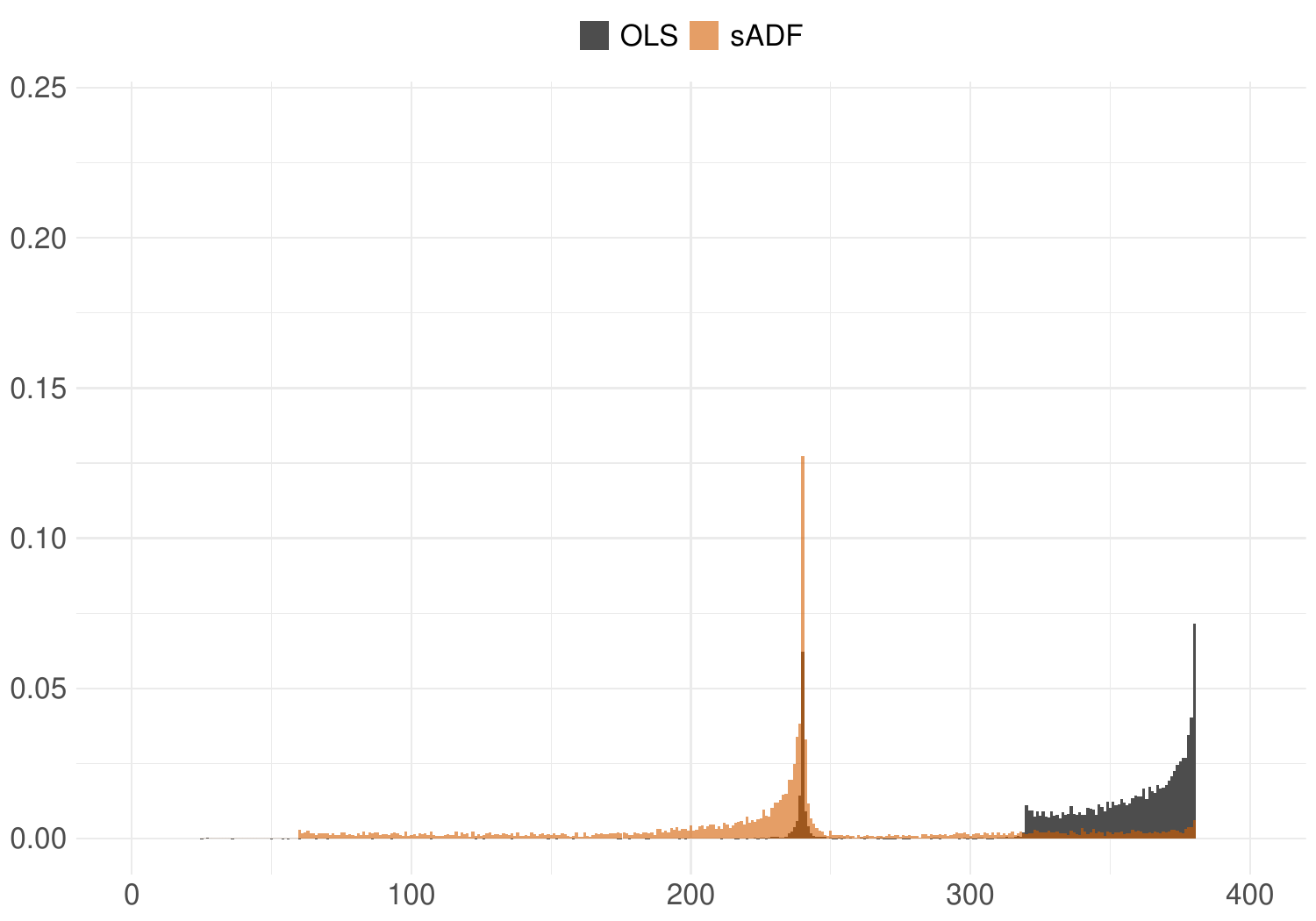}\label{fig:1:3}}
\subfigure[$T=400$, $c_a=5$, $c_b=6$, $s_0/s_1=1/5$]{\includegraphics[width=0.45\linewidth]{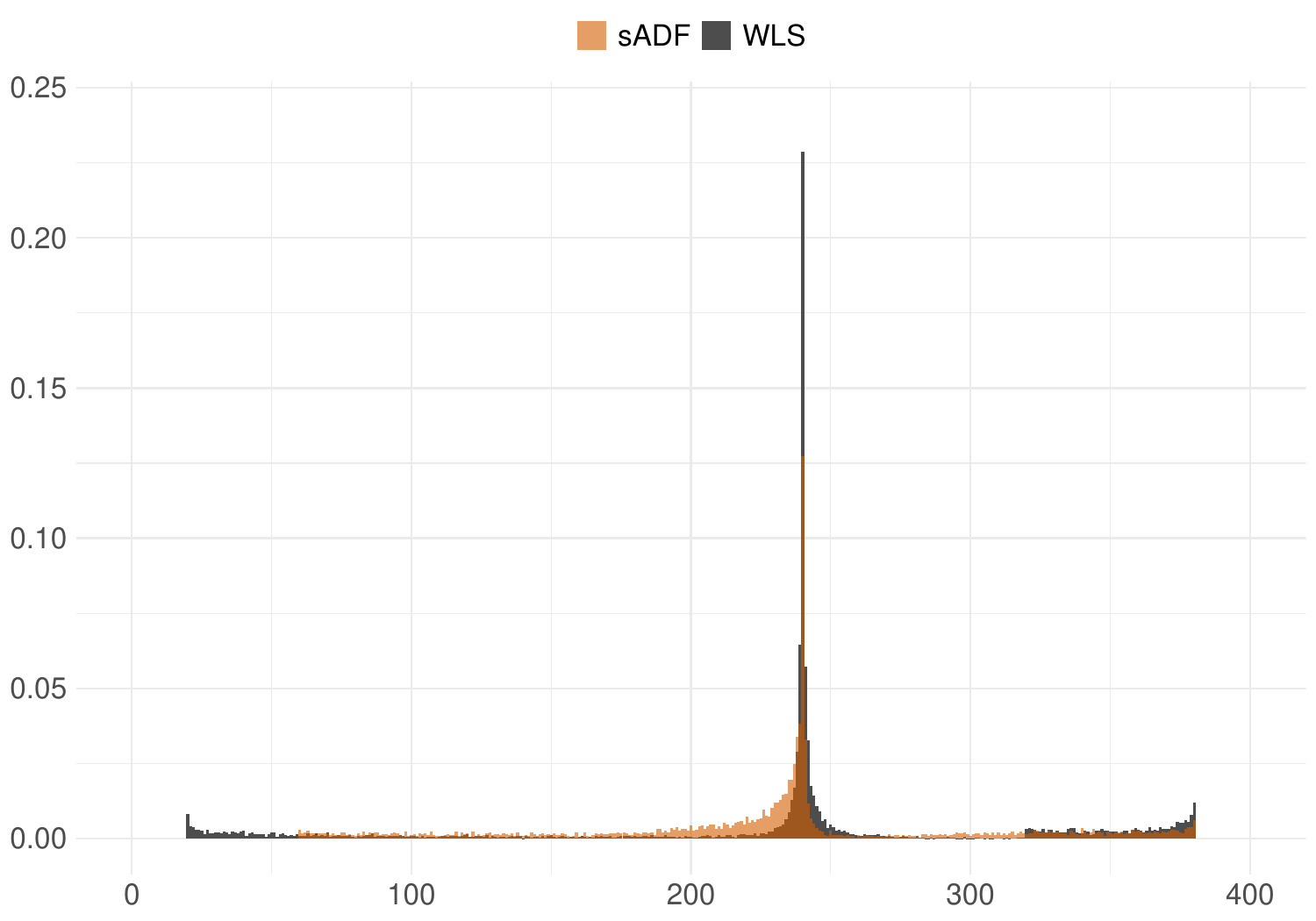}\label{fig:1:4}}\\
\subfigure[$T=400$, $c_a=6$, $c_b=6$, $s_0/s_1=1/5$]{\includegraphics[width=0.45\linewidth]{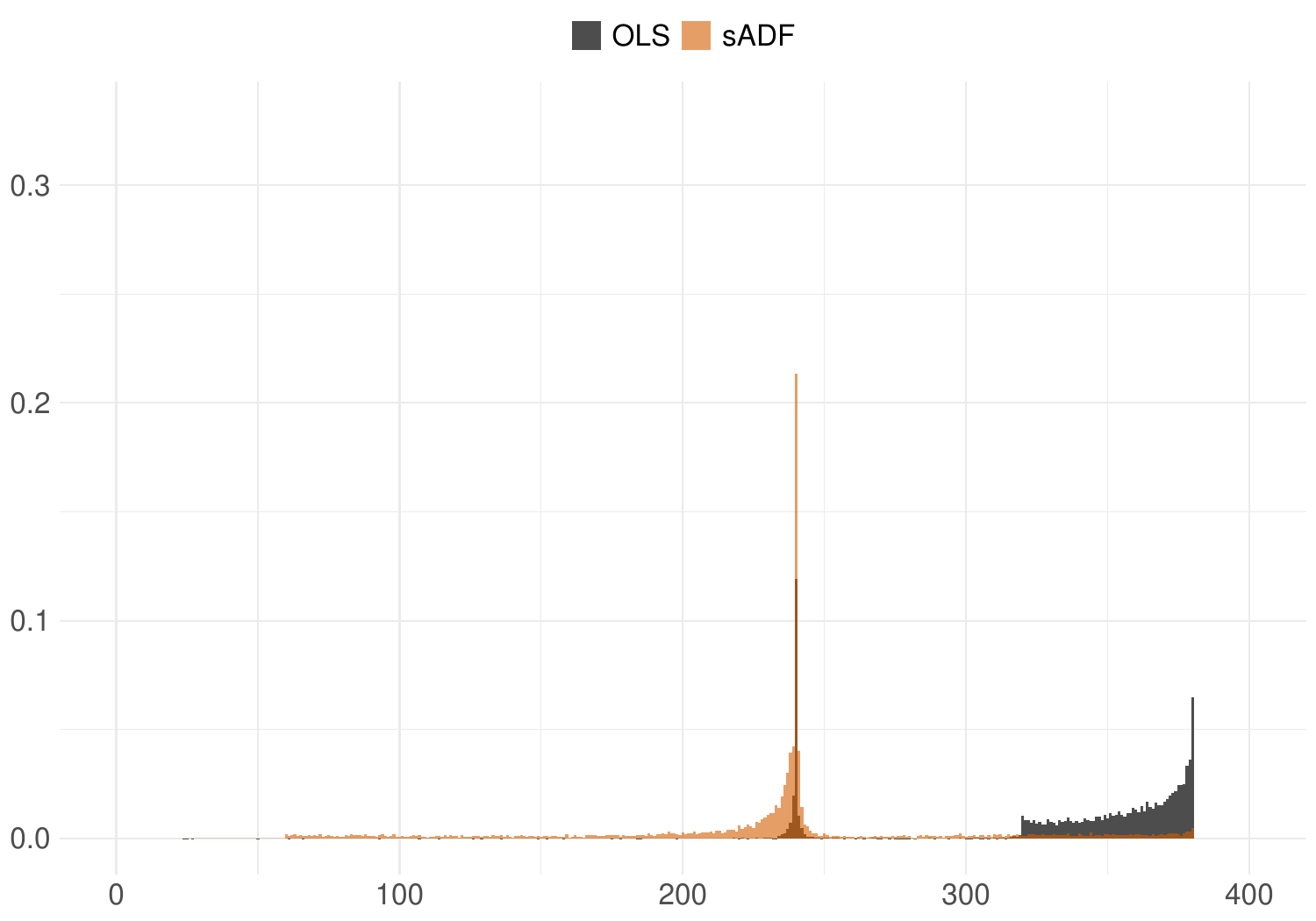}\label{fig:1:5}}
\subfigure[$T=400$, $c_a=6$, $c_b=6$, $s_0/s_1=1/5$]{\includegraphics[width=0.45\linewidth]{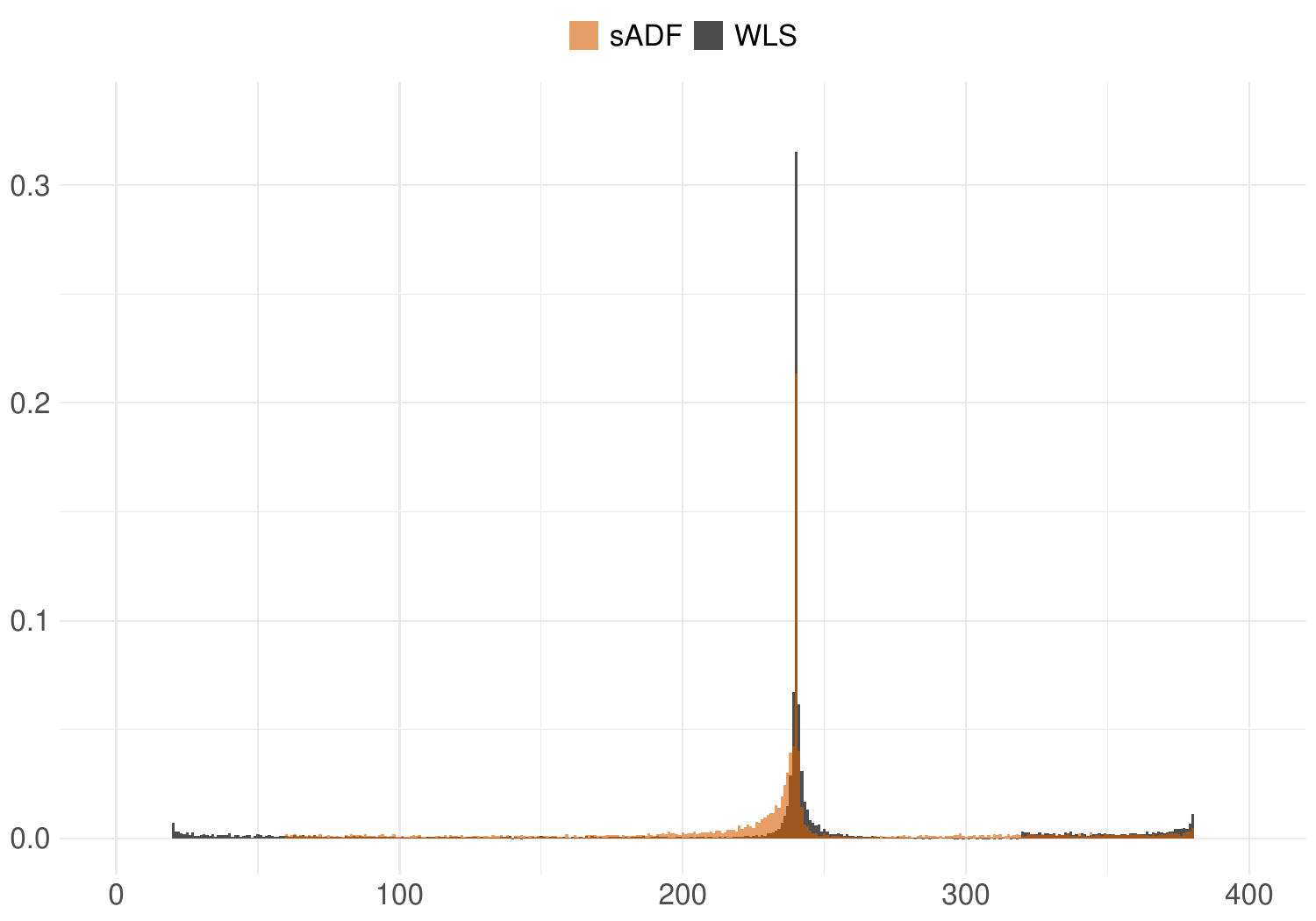}\label{fig:1:6}}\\
\end{center}%
\caption{Histograms of $\hat{k}_c$ 
for $(\tau_e,\tau_c,\tau_r)=(0.4,0.6,0.7)$,  $\tau=0.8$, $s_0/s_1=1/5$, {\color{black} $T=400$. Grey: OLS (left column) and WLS (right column); orange: sADF.}}
\label{fig1}
\end{figure}

\newpage

\begin{figure}[h!]%
\begin{center}%
\subfigure[$T=400$, $c_a=4$, $c_b=6$, $s_0/s_1=5$]{\includegraphics[width=0.45\linewidth]{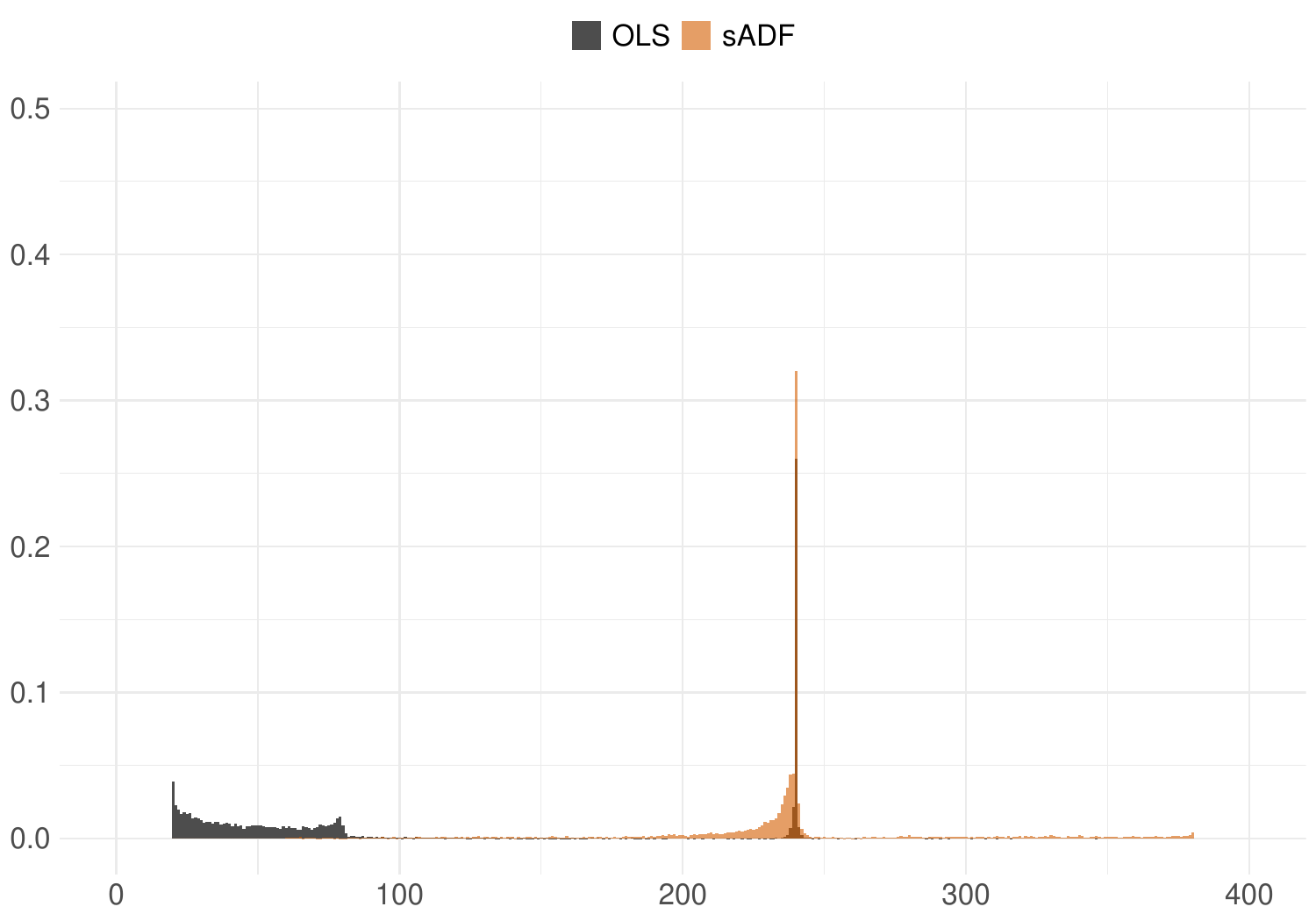}\label{fig3:5:1}}
\subfigure[$T=400$, $c_a=4$, $c_b=6$, $s_0/s_1=5$]{\includegraphics[width=0.45\linewidth]{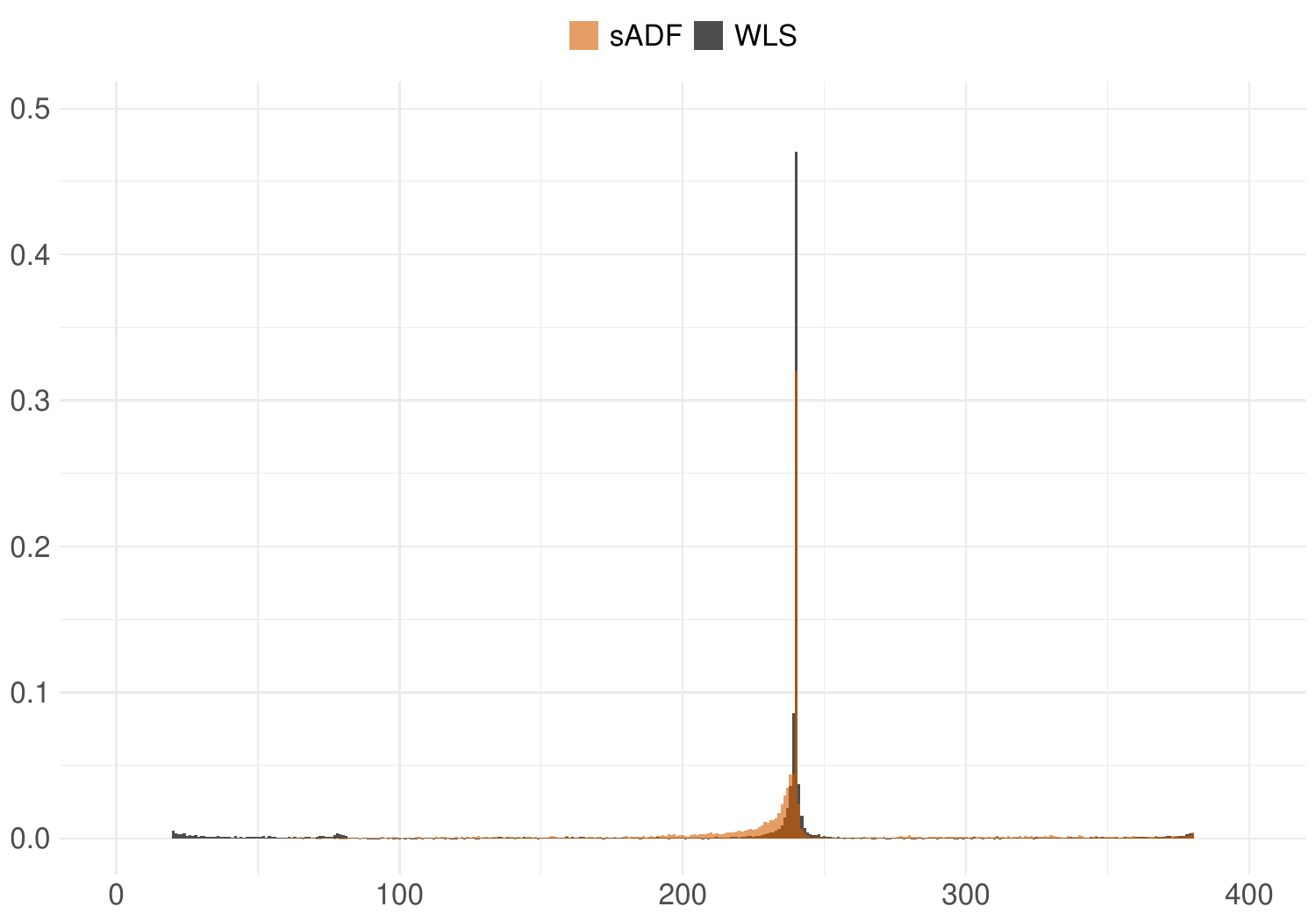}\label{fig3:5:2}}\\
\subfigure[$T=400$, $c_a=5$, $c_b=6$, $s_0/s_1=5$]{\includegraphics[width=0.45\linewidth]{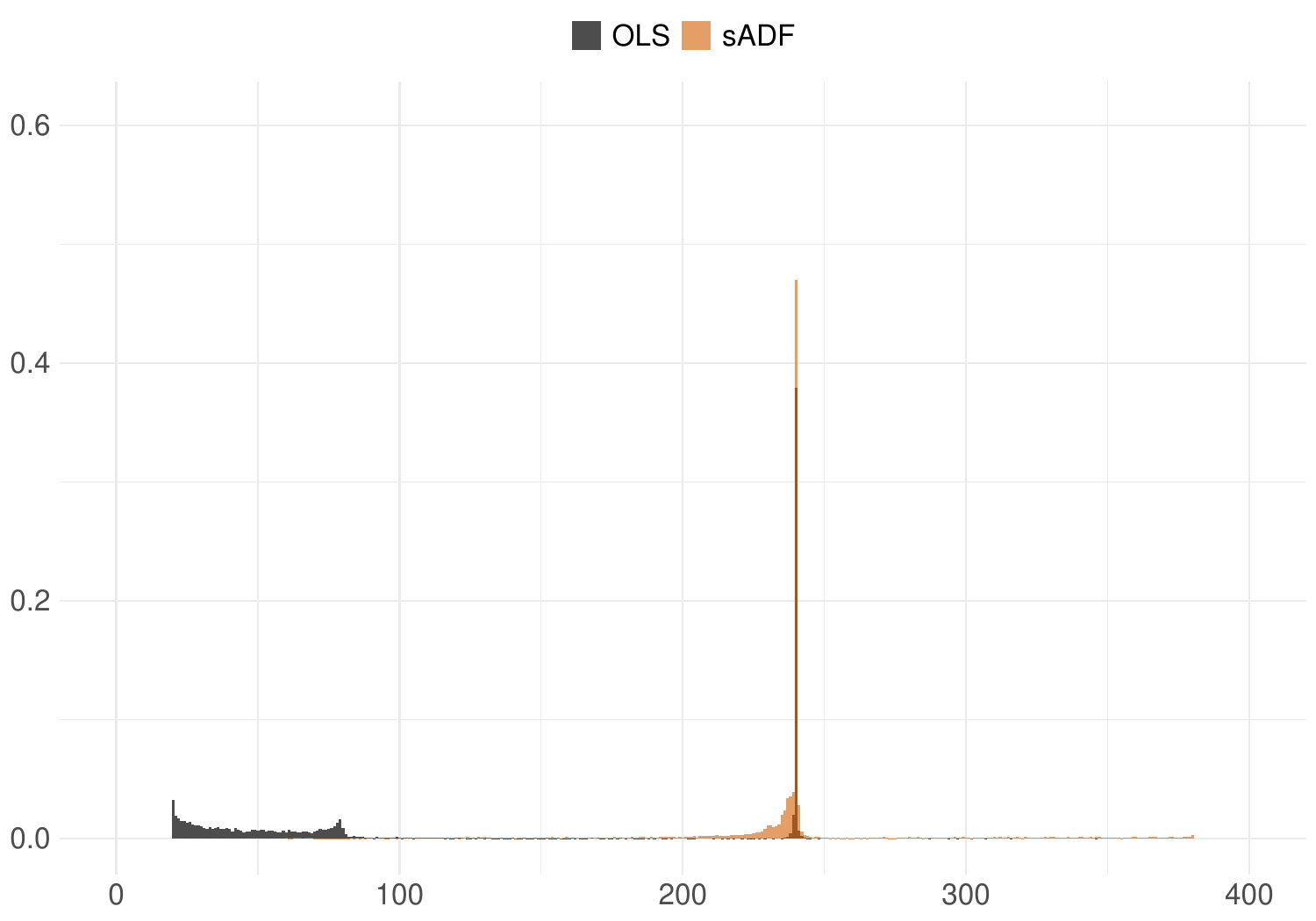}\label{fig3:5:3}}
\subfigure[$T=400$, $c_a=5$, $c_b=6$, $s_0/s_1=5$]{\includegraphics[width=0.45\linewidth]{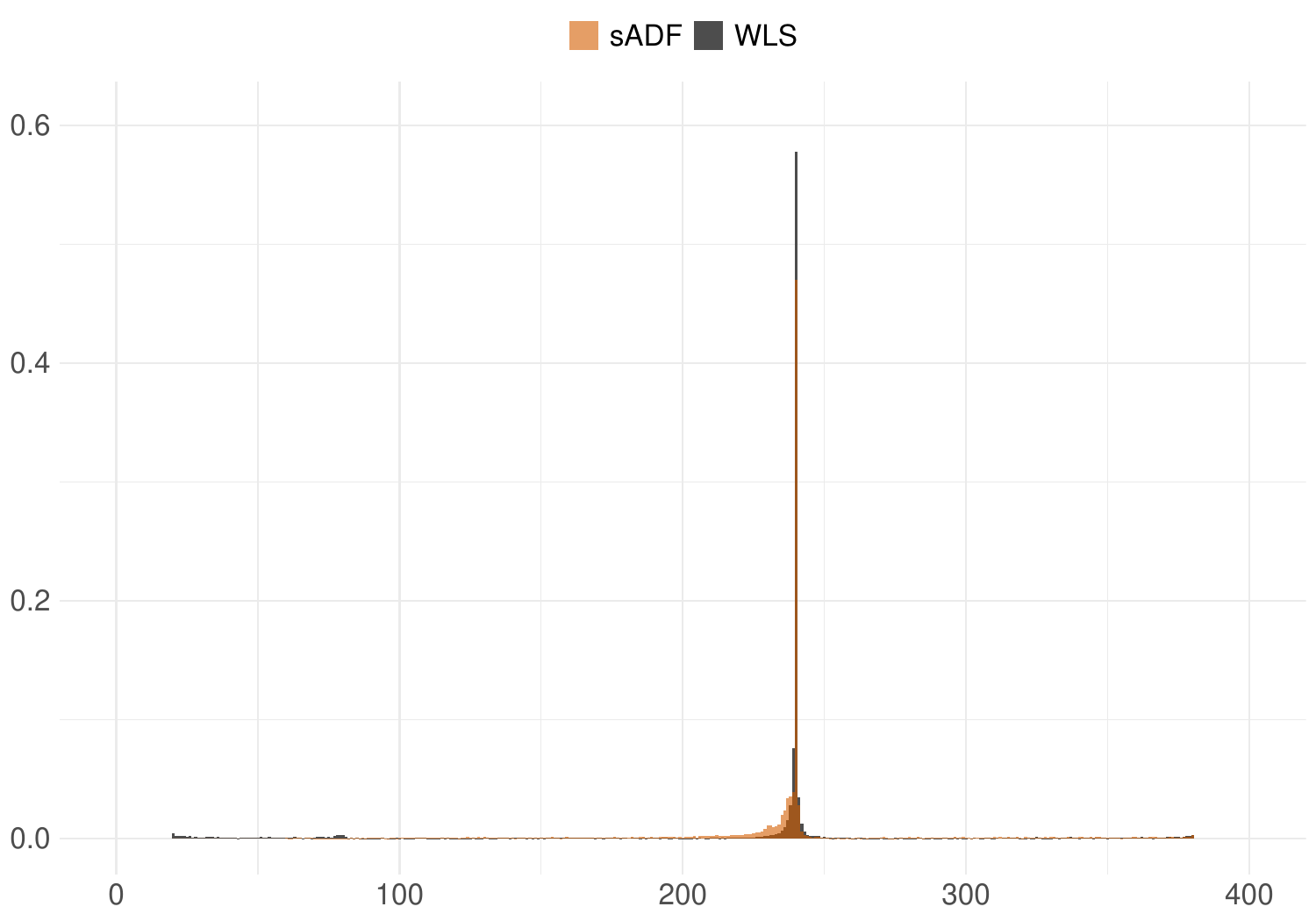}\label{fig3:5:4}}\\
\subfigure[$T=400$, $c_a=6$, $c_b=6$, $s_0/s_1=5$]{\includegraphics[width=0.45\linewidth]{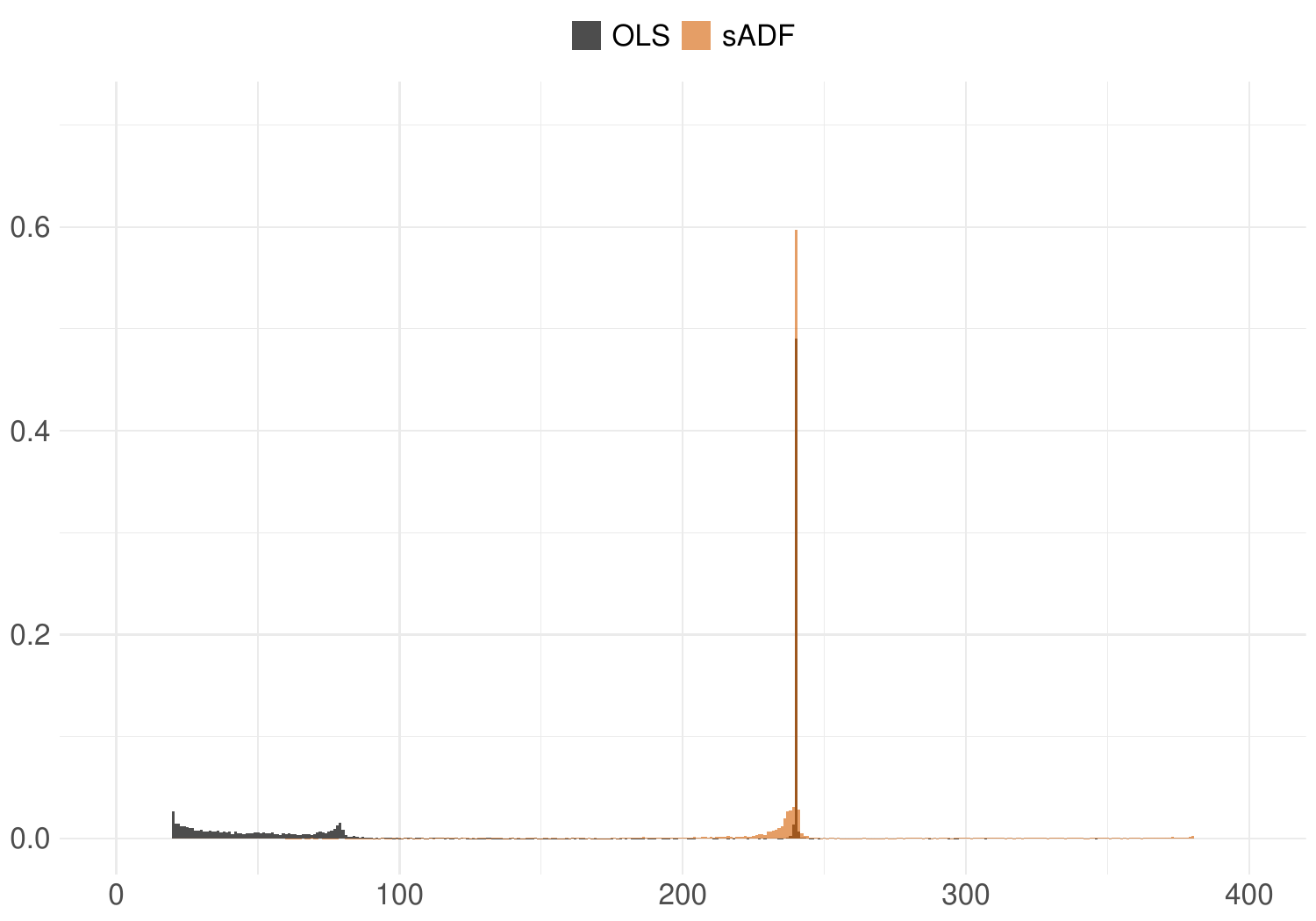}\label{fig3:5:5}}
\subfigure[$T=400$, $c_a=6$, $c_b=6$, $s_0/s_1=5$]{\includegraphics[width=0.45\linewidth]{graph_XP1/0.2_k_c_XP_T=400_5_6_Model1s0.s15.pdf}\label{fig3:5:6}}\\
\end{center}%
\caption{Histograms of $\hat{k}_c$ 
for $(\tau_e,\tau_c,\tau_r)=(0.4,0.6,0.7)$,  $\tau=0.2$, $s_0/s_1=5$, {\color{black} $T=400$. Grey: OLS (left column) and WLS (right column); orange: sADF.}}
\label{fig35}
\end{figure}

\newpage

\begin{figure}[h!]%
\begin{center}%
\subfigure[$T=400$, $c_a=4$, $c_b=6$, $s_0/s_1=5$]{\includegraphics[width=0.45\linewidth]{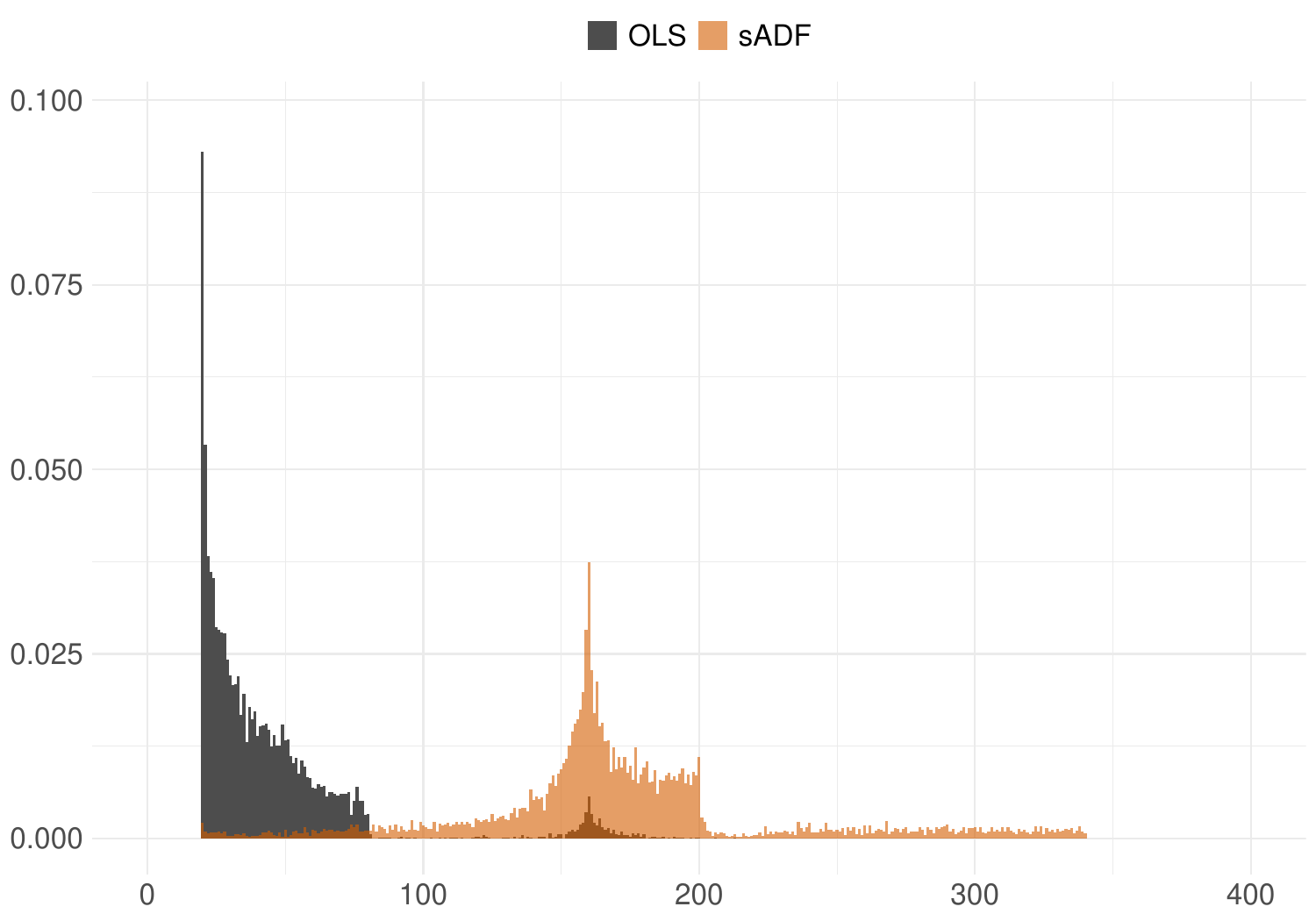}\label{fig4:5:1}}
\subfigure[$T=400$, $c_a=4$, $c_b=6$, $s_0/s_1=5$]{\includegraphics[width=0.45\linewidth]{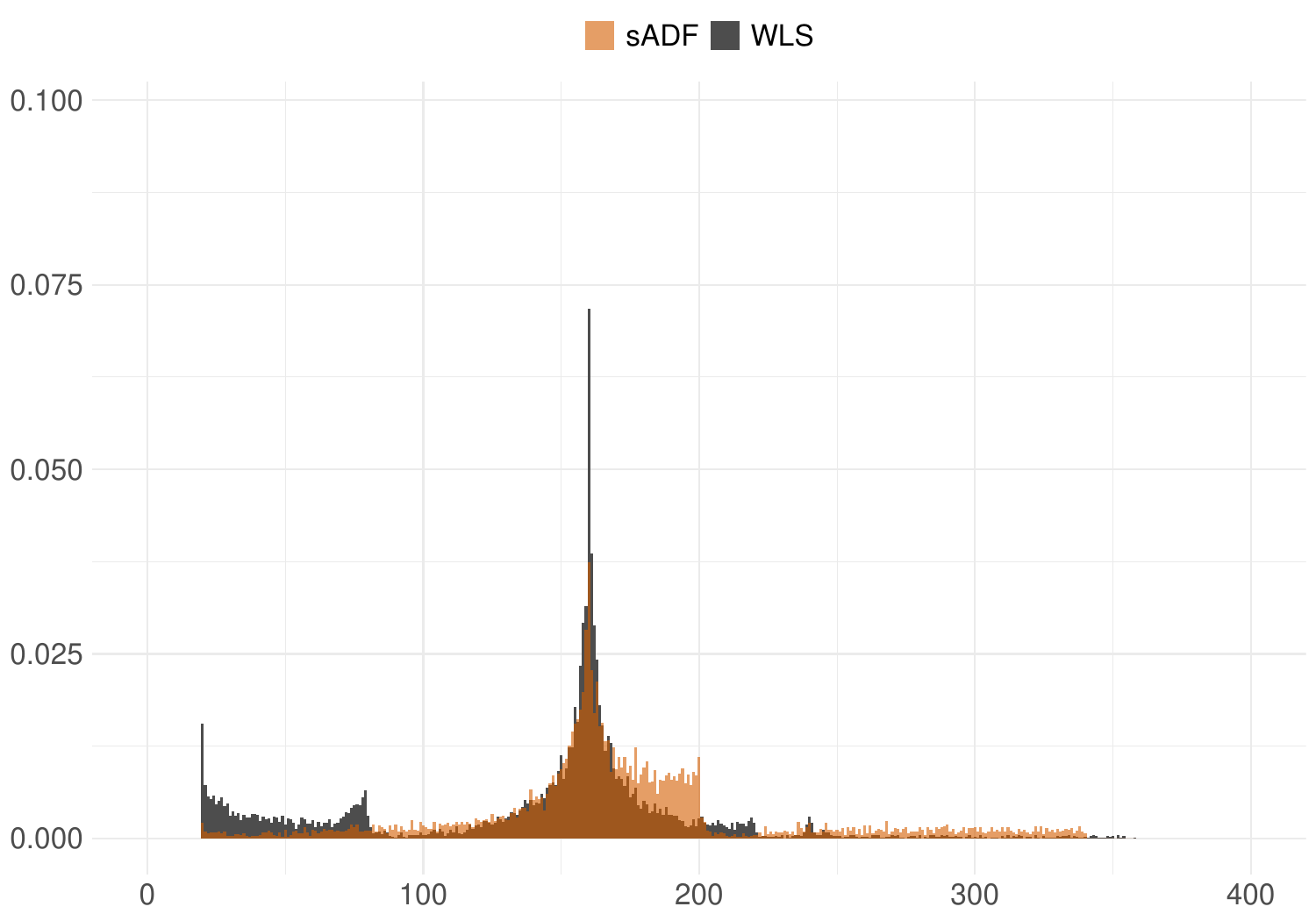}\label{fig4:5:2}}\\
\subfigure[$T=400$, $c_a=5$, $c_b=6$, $s_0/s_1=5$]{\includegraphics[width=0.45\linewidth]{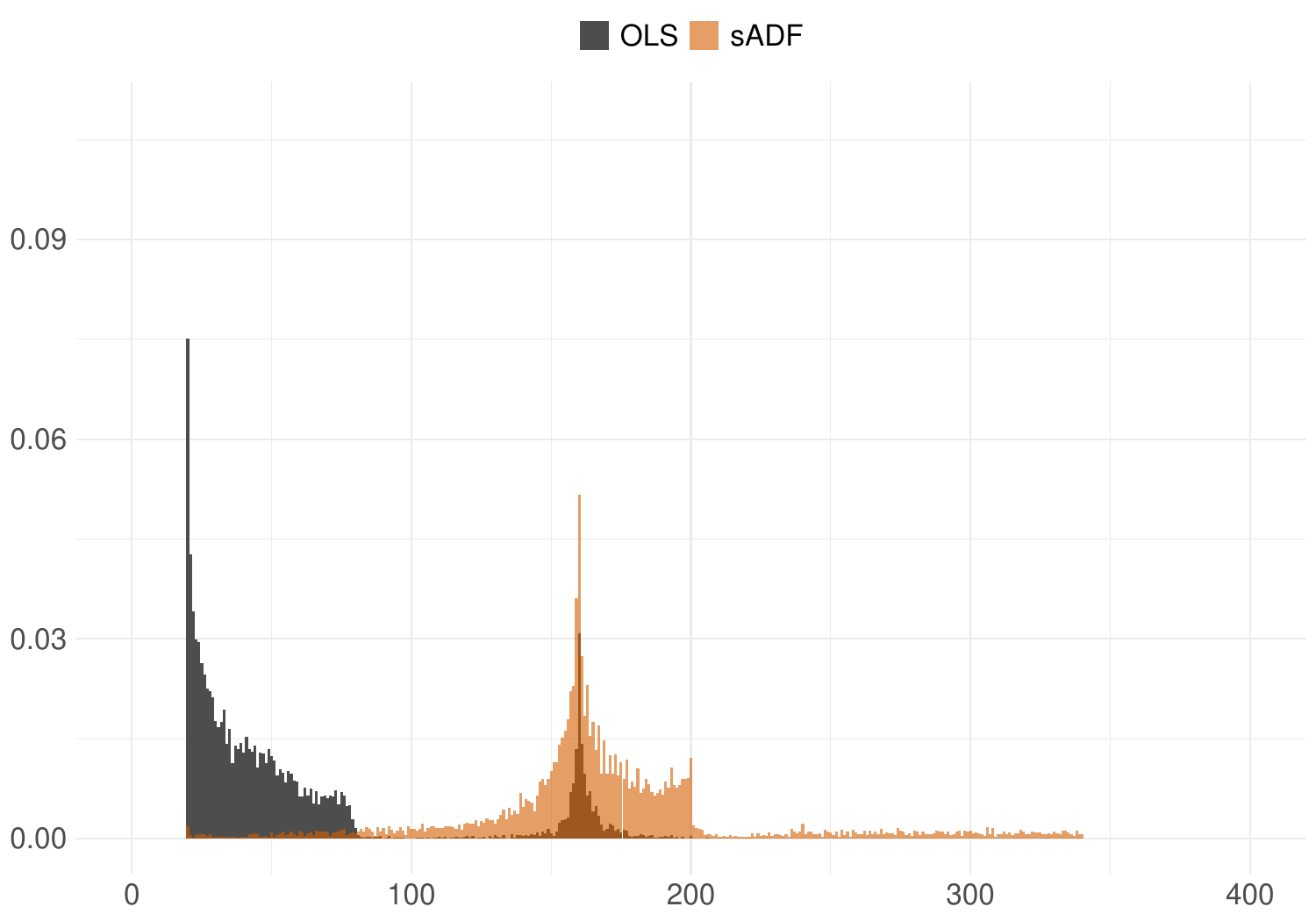}\label{fig4:5:3}}
\subfigure[$T=400$, $c_a=5$, $c_b=6$, $s_0/s_1=5$]{\includegraphics[width=0.45\linewidth]{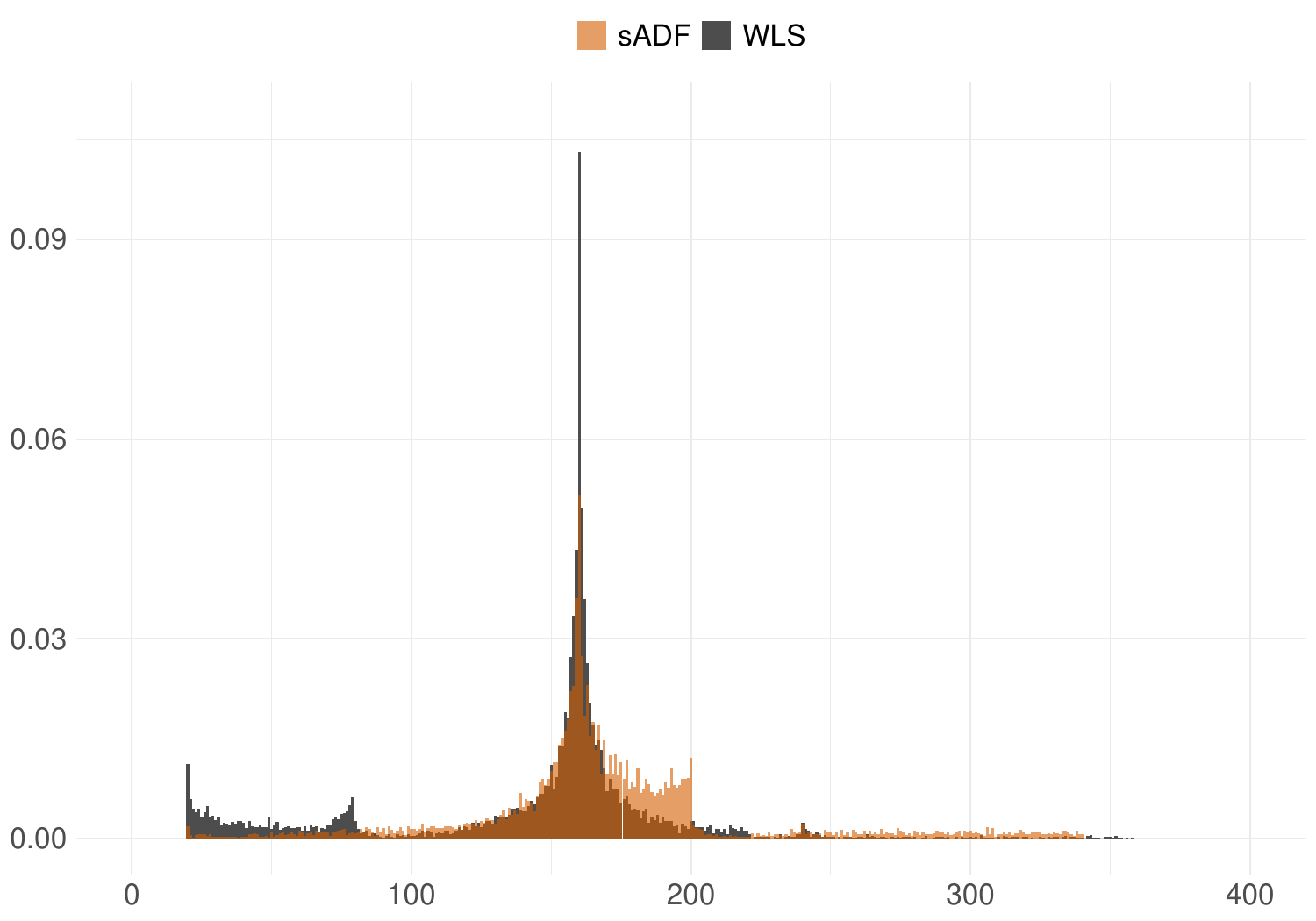}\label{fig4:5:4}}\\
\subfigure[$T=400$, $c_a=6$, $c_b=6$, $s_0/s_1=5$]{\includegraphics[width=0.45\linewidth]{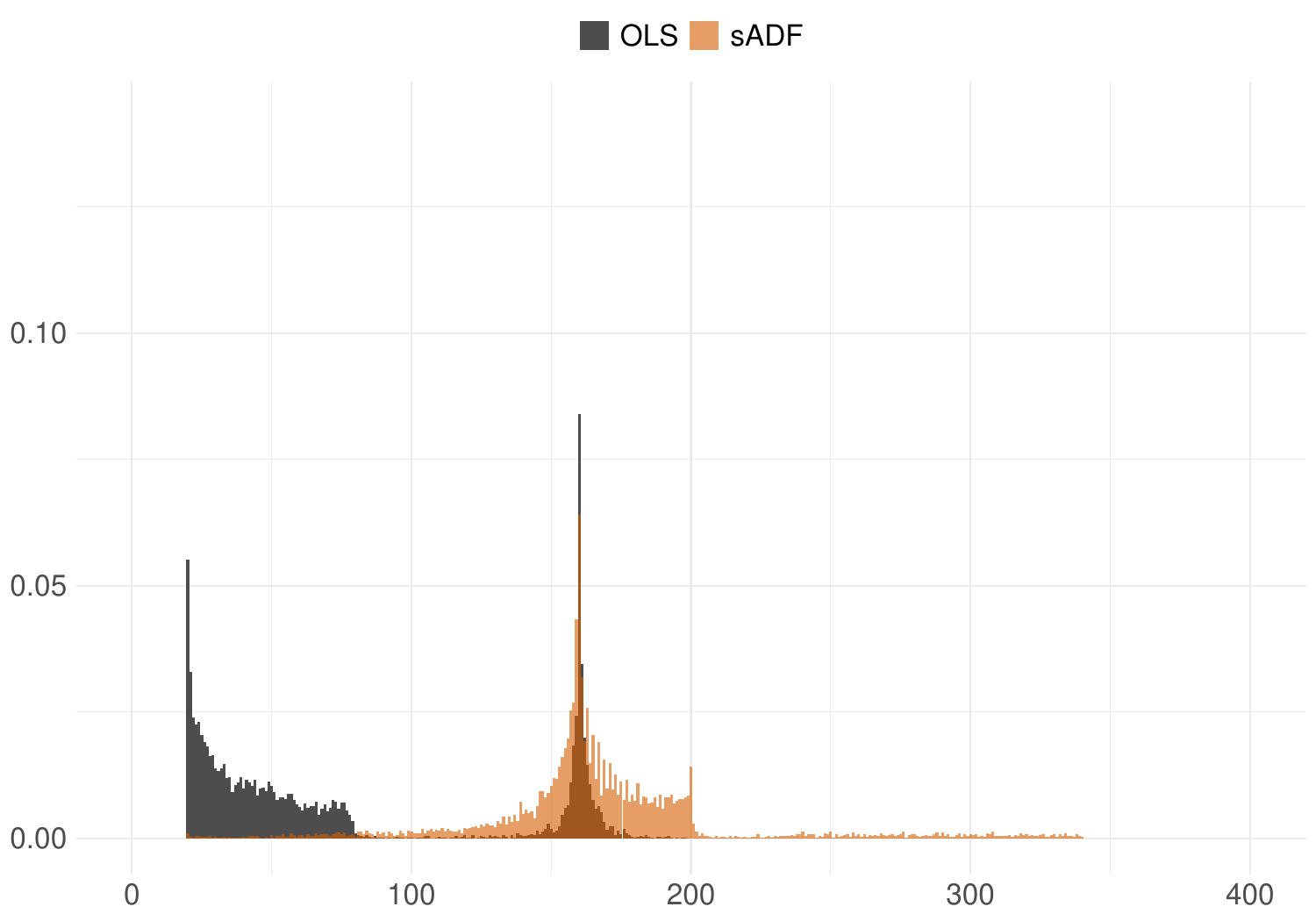}\label{fig4:5:5}}
\subfigure[$T=400$, $c_a=6$, $c_b=6$, $s_0/s_1=5$]{\includegraphics[width=0.45\linewidth]{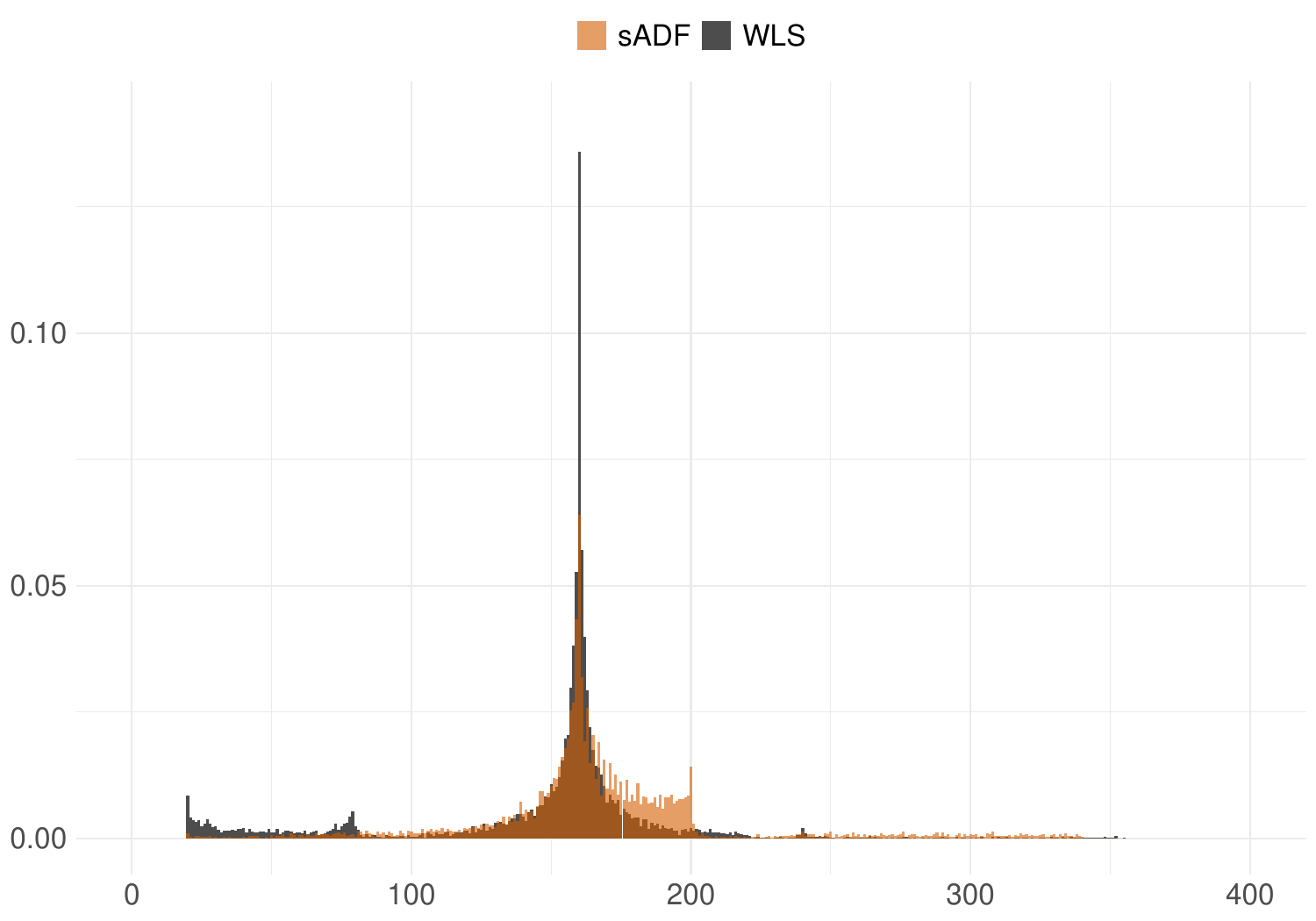}\label{fig4:5:6}}\\
\end{center}%
\caption{Histograms of $\hat{k}_e$ 
for $(\tau_e,\tau_c,\tau_r)=(0.4,0.6,0.7)$,  $\tau=0.2$, $s_0/s_1=5$, {\color{black} $T=400$. Grey: OLS (left column) and WLS (right column); orange: sADF.}}
\label{fig45}
\end{figure}

\newpage

{\scriptsize
\begin{longtable}{lr ccc ccc cc}
\caption{Estimated break dates (observation numbers within the calendar year) for all 45 coin-year series. The dates in bold are admissible according to criteria (C1)--(C3) of Section 6. sADF does not provide an estimate of the recovering date.}\label{tab:emp_all}\\
\toprule
 & & \multicolumn{3}{c}{OLS} & \multicolumn{3}{c}{WLS} & \multicolumn{2}{c}{sADF}\\
\cmidrule(lr){3-5}\cmidrule(lr){6-8}\cmidrule(lr){9-10}
Series & $N$ & $\hat k_e$ & $\hat k_c$ & $\hat k_r$ & $\hat k_e$ & $\hat k_c$ & $\hat k_r$ & $\hat k_e$ & $\hat k_c$\\
\midrule
\endfirsthead
\toprule
Series & $N$ & $\hat k_e$ & $\hat k_c$ & $\hat k_r$ & $\hat k_e$ & $\hat k_c$ & $\hat k_r$ & $\hat k_e$ & $\hat k_c$\\
\midrule
\endhead
\bottomrule
\endfoot
btc, 2014 & 364 & 0 & 25 & 50 & 99 & 221 & 277 & 198 & 277 \\
xrp, 2014 & 364 & 23 & 298 & 331 & 83 & 186 & 205 & 32 & 116 \\
ltc, 2014 & 364 & 0 & 18 & 50 & 277 & 340 & 364 & 155 & 277 \\
btc, 2015 & 364 & 0 & 19 & 235 & 0 & 19 & 263 & 270 & 307 \\
xrp, 2015 & 364 & 0 & 18 & 346 & 18 & 90 & 179 & 179 & 277 \\
xlm, 2015 & 364 & 27 & 52 & 166 & 27 & 52 & 166 & 168 & 298 \\
ltc, 2015 & 364 & \textbf{162} & \textbf{189} & \textbf{207} & 0 & 18 & 70 & 98 & 138 \\
xmr, 2015 & 364 & 41 & 90 & 117 & 0 & 19 & 90 & 51 & 90 \\
btc, 2016 & 365 & 167 & 345 & 365 & 30 & 345 & 365 & 99 & 170 \\
eth, 2016 & 365 & 47 & 72 & 167 & 47 & 72 & 263 & 294 & 339 \\
xrp, 2016 & 365 & 18 & 37 & 296 & 24 & 45 & 103 & 115 & 339 \\
xlm, 2016 & 365 & \textbf{251} & \textbf{271} & \textbf{336} & \textbf{251} & \textbf{271} & \textbf{302} & 271 & 328 \\
ltc, 2016 & 365 & 144 & 167 & 188 & 0 & 18 & 52 & 167 & 277 \\
xmr, 2016 & 365 & \textbf{230} & \textbf{248} & \textbf{306} & 248 & 347 & 365 & 42 & 79 \\
btc, 2017 & 364 & \textbf{315} & \textbf{340} & \textbf{364} & \textbf{315} & \textbf{340} & \textbf{364} & 23 & 161 \\
eth, 2017 & 364 & \textbf{136} & \textbf{162} & \textbf{196} & \textbf{136} & \textbf{162} & \textbf{196} & 28 & 75 \\
xrp, 2017 & 364 & 136 & 343 & 364 & \textbf{115} & \textbf{136} & \textbf{343} & 190 & 228 \\
xlm, 2017 & 364 & 126 & 315 & 339 & 126 & 315 & 339 & 19 & 57 \\
ltc, 2017 & 364 & \textbf{327} & \textbf{345} & \textbf{364} & \textbf{327} & \textbf{345} & \textbf{364} & 297 & 337 \\
etc, 2017 & 364 & 125 & 143 & 331 & \textbf{68} & \textbf{167} & \textbf{256} & 216 & 295 \\
xmr, 2017 & 364 & 308 & 345 & 364 & 239 & 305 & 339 & 34 & 74 \\
btc, 2018 & 364 & 0 & 35 & 61 & 35 & 316 & 340 & \textbf{167} & \textbf{204} \\
eth, 2018 & 364 & 0 & 27 & 87 & 316 & 346 & 364 & 197 & 254 \\
xrp, 2018 & 364 & 0 & 35 & 53 & 35 & 90 & 109 & 49 & 88 \\
xlm, 2018 & 364 & 0 & 35 & 118 & 283 & 314 & 340 & 120 & 167 \\
bch, 2018 & 364 & 0 & 35 & 125 & \textbf{291} & \textbf{310} & \textbf{339} & 307 & 344 \\
ltc, 2018 & 364 & 0 & 35 & 55 & 225 & 309 & 327 & 157 & 344 \\
eos, 2018 & 364 & \textbf{99} & \textbf{118} & \textbf{142} & 118 & 315 & 340 & 309 & 346 \\
bnb, 2018 & 364 & 0 & 35 & 156 & 156 & 314 & 340 & 274 & 330 \\
ada, 2018 & 364 & 27 & 75 & 121 & 75 & 315 & 340 & 292 & 340 \\
xtz, 2018 & 364 & 95 & 152 & 185 & 35 & 314 & 339 & 262 & 339 \\
etc, 2018 & 364 & 0 & 35 & 54 & 90 & 316 & 339 & 271 & 329 \\
xmr, 2018 & 364 & 0 & 35 & 63 & 225 & 316 & 339 & 307 & 344 \\
btc, 2019 & 364 & \textbf{158} & \textbf{176} & \textbf{196} & \textbf{158} & \textbf{176} & \textbf{196} & 83 & 134 \\
eth, 2019 & 364 & \textbf{158} & \textbf{176} & \textbf{196} & \textbf{28} & \textbf{176} & \textbf{196} & 302 & 344 \\
xrp, 2019 & 364 & 128 & 172 & 196 & 266 & 309 & 328 & 234 & 271 \\
xlm, 2019 & 364 & 36 & 152 & 196 & 36 & 314 & 335 & 200 & 248 \\
bch, 2019 & 364 & 28 & 176 & 194 & 28 & 176 & 194 & 306 & 344 \\
ltc, 2019 & 364 & 154 & 172 & 196 & 154 & 172 & 196 & 19 & 60 \\
eos, 2019 & 364 & 129 & 150 & 196 & 129 & 150 & 196 & 262 & 325 \\
bnb, 2019 & 364 & 144 & 171 & 196 & 144 & 171 & 196 & 128 & 171 \\
ada, 2019 & 364 & 71 & 92 & 176 & \textbf{66} & \textbf{94} & \textbf{118} & 220 & 260 \\
xtz, 2019 & 364 & 118 & 138 & 181 & 295 & 346 & 364 & 219 & 284 \\
etc, 2019 & 364 & 128 & 152 & 196 & 128 & 152 & 196 & 240 & 278 \\
xmr, 2019 & 364 & \textbf{154} & \textbf{174} & \textbf{196} & \textbf{154} & \textbf{174} & \textbf{196} & 37 & 74 \\
\end{longtable}
}

\newpage
\setcounter{page}{1}
\appendix
\renewcommand{\thesection}{\Alph{section}.}
\renewcommand{\theequation}{\Alph{section}.\arabic{equation}}
\setcounter{section}{0}
\setcounter{equation}{0}
\begin{center}
{\Large\bf Online Appendix to\\
``Improving the accuracy of bubble date estimators under time-varying volatility''\\
by Eiji Kurozumi and Anton Skrobotov}
\end{center}

This appendix contains additional figures with the histograms of the break date estimates with different combinations of $s_0$, $s_1$, $\tau$, and $T$, as described in Section 5 of the main part of the paper. Figures \ref{fig1:app}--\ref{fig26} correspond to the the case where the volatility shift occurs late in the sample at $\tau=0.8$ (Figures \ref{fig1:app}--\ref{fig6} for $\hat{k}_c$, \ref{fig1_e}--\ref{fig6_e} for $\hat{k}_e$, and \ref{fig21}--\ref{fig26} for $\hat{k}_r$), while Figures \ref{fig31}--\ref{fig26_r} are the histograms in the case where the volatility changes early in the sample at $\tau=0.2$ (Figures \ref{fig31}--\ref{fig36} for $\hat{k}_c$, \ref{fig41}--\ref{fig46} for $\hat{k}_e$, and \ref{fig21_r}--\ref{fig26_r} for $\hat{k}_r$). {\color{black} In the figures for $\hat{k}_e$ and $\hat{k}_c$, the histograms of the OLS and WLS estimators are drawn in grey and the histogram of the sADF estimator of \citet{harvey2020sign} (see Section 5) is overlaid in orange. The sADF procedure does not provide an estimator of the recovering date, so that the figures for $\hat{k}_r$ contain only the OLS and WLS estimators.The complete accuracy tables for all the designs, including the additional volatility designs described in Section 5 (the double shift, the logistic smooth transition and the linear trend), for which the histograms are not reported, are given in Appendix \ref{sec:acc_app}}

Figure \ref{fig1:app} ($\hat{k}_c$, $\tau=0.8$, $s_0/s_1 = 1/5$, $T=400$) is the same as Figure \ref{fig1} in the main text and Figure \ref{fig2} with $T=800$ is quantitatively similar to Figure \ref{fig1:app}; the sample splitting approach based on the WLS method performs better than that based on the OLS method. Figures \ref{fig3}--\ref{fig6} show that there is virtually no difference between the distributions of the estimates with and without volatility correction. {\color{black} Figures \ref{fig3} and \ref{fig4} ($s_0/s_1=1$) show virtually no difference between the distributions of the estimates with and without volatility correction, whereas in Figures \ref{fig5} and \ref{fig6} ($s_0/s_1=5$), where the high volatility covers the whole bubble episode, the WLS-based estimator is slightly less accurate than the OLS-based one ($P_5$ is lower by $0.04$--$0.06$). The sADF estimator is more accurate than OLS but less accurate than WLS in Figures \ref{fig1:app} and \ref{fig2}, whereas in Figures \ref{fig3}--\ref{fig6} its peak at the true break fraction is lower than those of both OLS and WLS.}

Figures \ref{fig1_e}--\ref{fig2_e} ($\hat{k}_e$, $\tau=0.8$, $s_0/s_1 = 1/5$) demonstrate the local peak of $\hat{\tau}_e$ at an incorrect location (at $0.6$, the location of collapse), but volatility correction blackuces this local peak and increases the peak at $0.4$, the correct location of bubble exuberance. For the cases where $s_0/s_1 = 1$ and $s_0/s_1 = 5$ in Figures \ref{fig3_e}--\ref{fig6_e}, the results are virtually the same regardless of the correction. {\color{black} In all these cases, the three estimators are poorly concentrated around the true date ($P_5$ does not exceed $0.14$), and the sADF estimator has a $P_5$ similar to or higher than those of the OLS and WLS estimators.} 

Figures \ref{fig21}--\ref{fig22} ($\hat{k}_r$, $\tau=0.8$, $s_0/s_1 = 1/5$) demonstrate the local peak at an incorrect location (at the end of the sample), but volatility correction blackuces this local peak and increases the peak at the correct location of the recovering date at $\tau_r=0.7$. Figures \ref{fig23}--\ref{fig26} show no significant difference between the two methods.

For $\tau=0.2$ and $\hat{k}_c$, we observe the difference between the with and without correction scenario only for the case of $s_0/s_1 = 5$ (Figures \ref{fig35:app} and \ref{fig36}, the former of which is the same as Figure \ref{fig35} in the main text). Figures \ref{fig31}--\ref{fig34} show similar results regardless of whether we correct for volatility or not. {\color{black} For $s_0/s_1=5$, the sADF estimator is more accurate than OLS, which has positive frequencies at the beginning of the sample, but its peak at the true date is lower than that of WLS; in Figures \ref{fig31}--\ref{fig34}, the peak of sADF is lower than those of both OLS and WLS.} 

For $\hat{k}_e$, it is difficult to interpret the performance of the estimates for $s_0/s_1 = 1/5$ (Figures \ref{fig41}--\ref{fig42}). For $s_0/s_1 = 1$, the results are virtually the same regardless of whether we correct for the volatility or not (Figures \ref{fig43}--\ref{fig44}), whereas for $s_0/s_1 = 5$, the local peak becomes higher around the true break fraction under volatility correction, as is observed in Figures \ref{fig45:app} and \ref{fig46}. {\color{black} In these figures, the sADF estimator does not produce the spurious frequencies at the beginning of the sample observed for OLS, but it is more dispersed around the true break fraction than WLS. In the other cases, the three estimators are poorly concentrated around the true date ($P_5$ does not exceed $0.14$), and the sADF estimator has a $P_5$ similar to or higher than those of the other two.}

For $\hat{k}_r$ and $s_0/s_1 = 1/5$, Figures \ref{fig21_r}--\ref{fig22_r} demonstrate the local peak at an incorrect location (at the end of the sample), but volatility correction blackuces this local peak and increases the peak at the correct location of bubble exuberance. For the case $s_0/s_1 = 1$, the results are similar regardless of whether we correct for the volatility or not (Figures \ref{fig23_r}--\ref{fig24_r}). For the case $s_0/s_1 = 5$, volatility correction blackuces the local peak at the incorrect location (corresponding to the date of collapse) and increases the peak at the correct location of resumption of the normal market (Figures \ref{fig25_r}--\ref{fig26_r}).

{\color{black} The complete accuracy tables (Appendix \ref{sec:acc_app}) show the following for the additional volatility designs. For double-shift in volatility,  with $s_0/s_1=5$ (low volatility during the explosive regime), WLS is more accurate than OLS ($P_5$ of $\hat{k}_c$ is $0.56$ versus $0.42$, and of $\hat{k}_e$ is $0.16$ versus $0.07$), but sADF is the most accurate for $\hat{k}_e$ and $\hat{k}_c$ ($0.27$ and $0.68$). With $s_0/s_1=1/5$ (high volatility during the explosive regime), OLS is the most accurate estimator for all the dates ($P_5$ of $\hat{k}_c$ is $0.25$ for OLS and $0.10$ for both WLS and sADF), and WLS cannot estimate $\hat{k}_e$ in $6$--$10\%$ of the replications. For logistic smooth transition in volatility, the results are similar to the double shift: with $s_0/s_1=5$, WLS and sADF are clearly more accurate than OLS for $\hat{k}_c$ ($0.67$ and $0.70$ versus $0.48$), whereas with $s_0/s_1=1/5$ all the estimators are inaccurate and OLS is slightly better for $\hat{k}_c$ ($0.13$ versus $0.10$ and $0.08$). For trending volatility, the gains of WLS over OLS are modest ($0.51$ versus $0.45$ for $s_0/s_1=5$ and $0.25$ versus $0.23$ for $s_0/s_1=1/5$ for $\hat{k}_c$), and sADF has the highest $P_5$ for $\hat{k}_e$.}

\newpage
\renewcommand{\thefigure}{\Alph{section}.\arabic{figure}}
\section{$\tau=0.8$, $\hat{k}_c$}
\setcounter{figure}{0}

\begin{figure}[h!]%
\begin{center}%
\subfigure[$T=400$, $c_a=4$, $c_b=6$, $s_0/s_1=1/5$]{\includegraphics[width=0.45\linewidth]{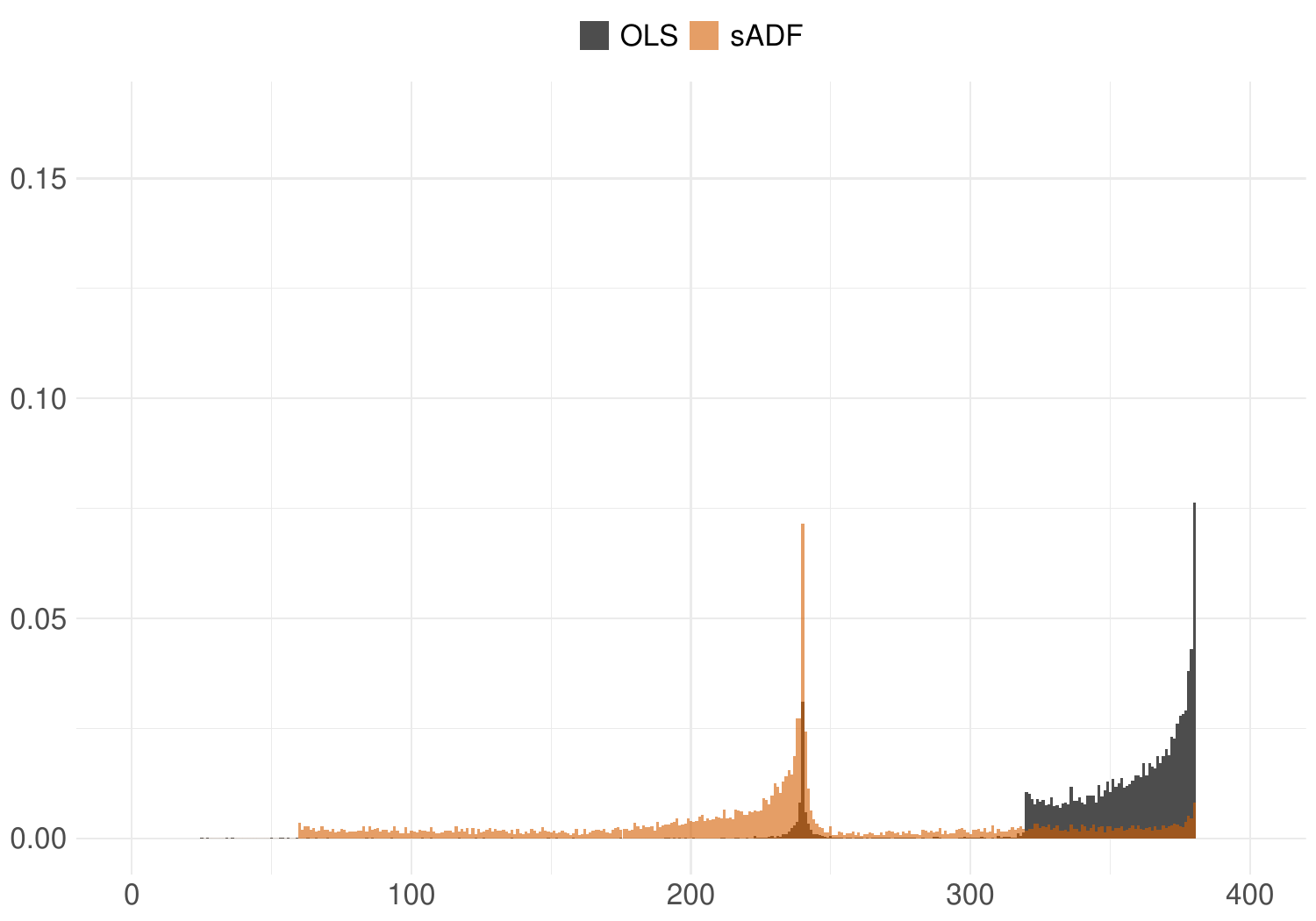}\label{fig:1:1:app}}
\subfigure[$T=400$, $c_a=4$, $c_b=6$, $s_0/s_1=1/5$]{\includegraphics[width=0.45\linewidth]{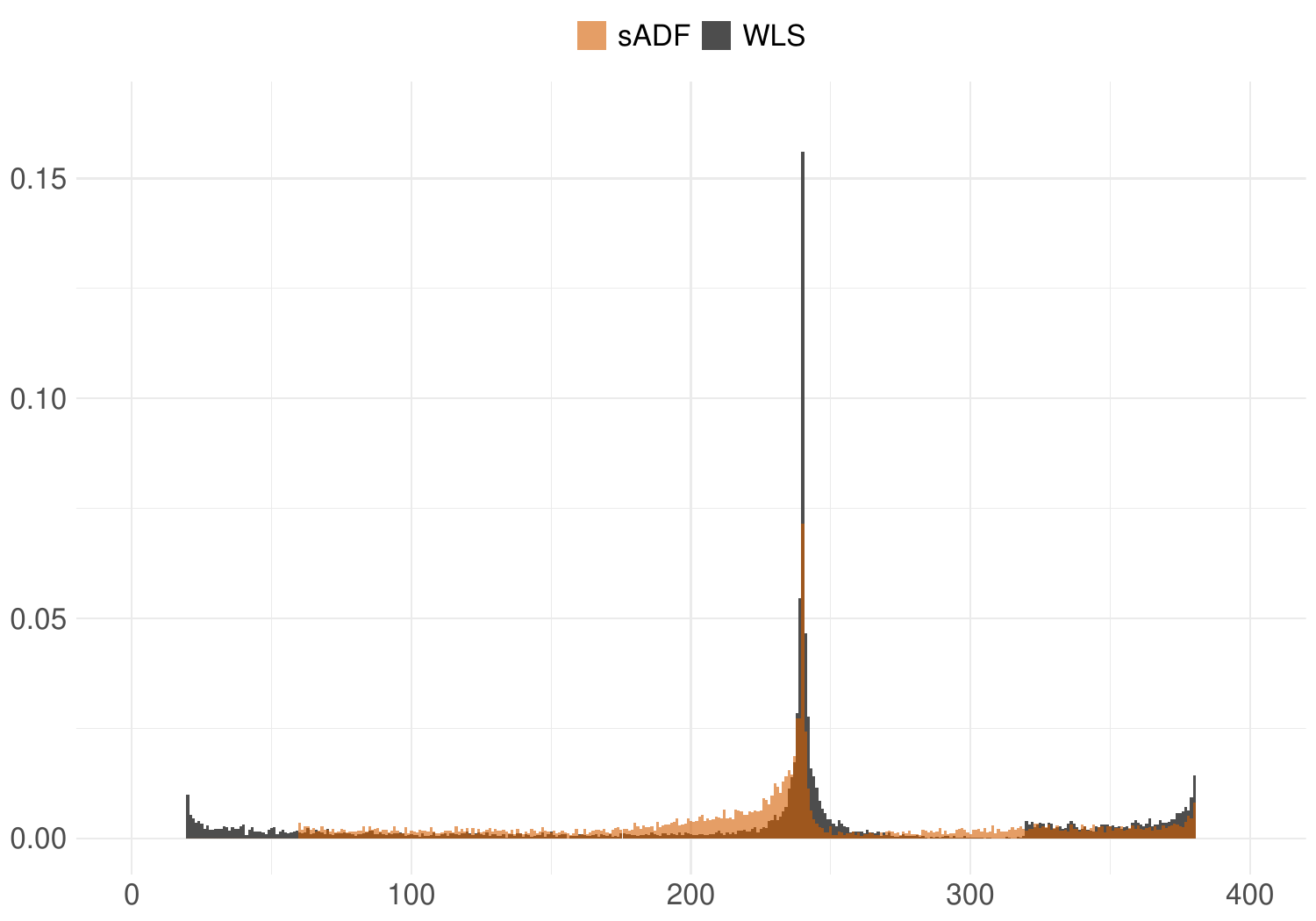}\label{fig:1:2:app}}\\
\subfigure[$T=400$, $c_a=5$, $c_b=6$, $s_0/s_1=1/5$]{\includegraphics[width=0.45\linewidth]{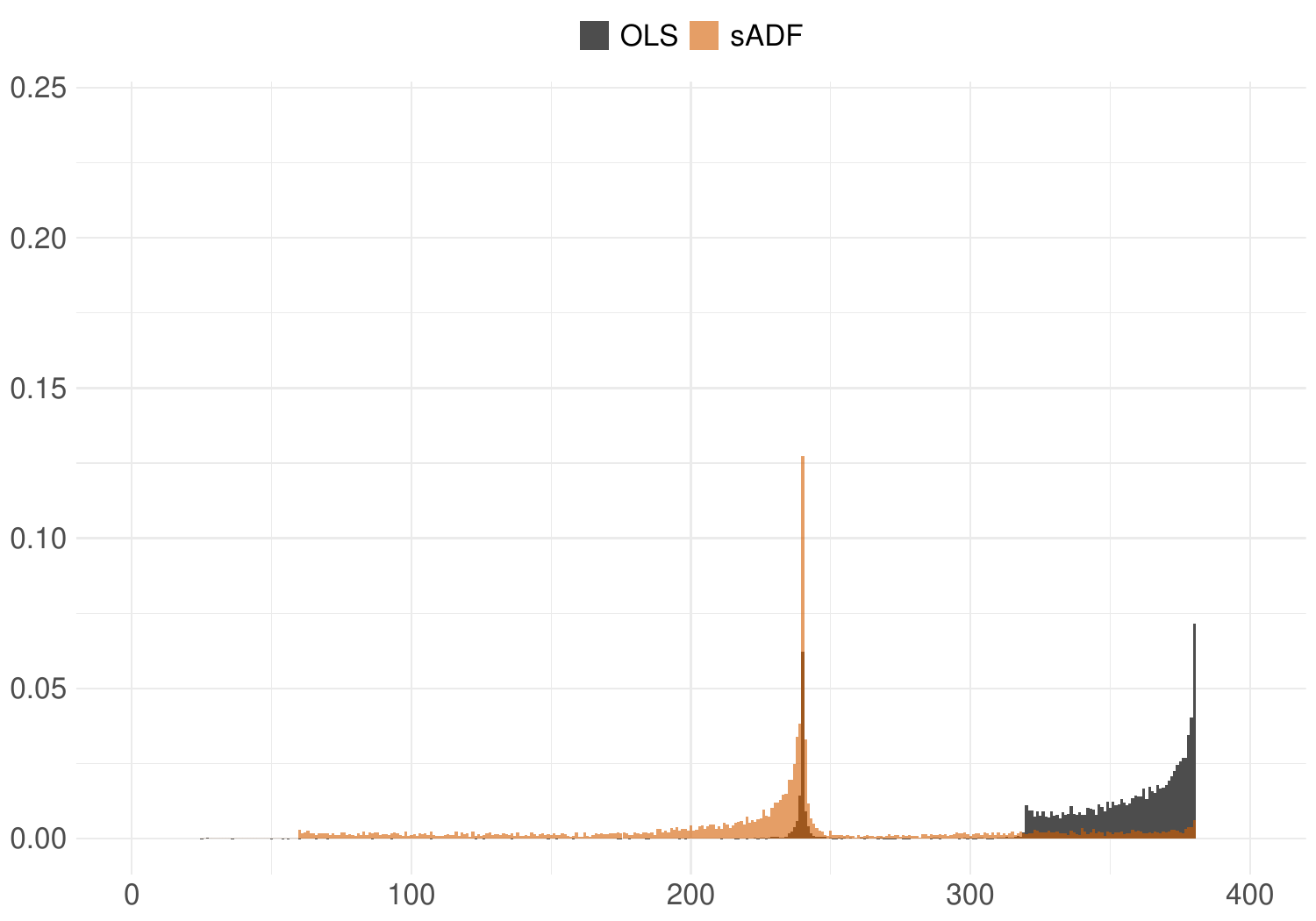}\label{fig:1:3:app}}
\subfigure[$T=400$, $c_a=5$, $c_b=6$, $s_0/s_1=1/5$]{\includegraphics[width=0.45\linewidth]{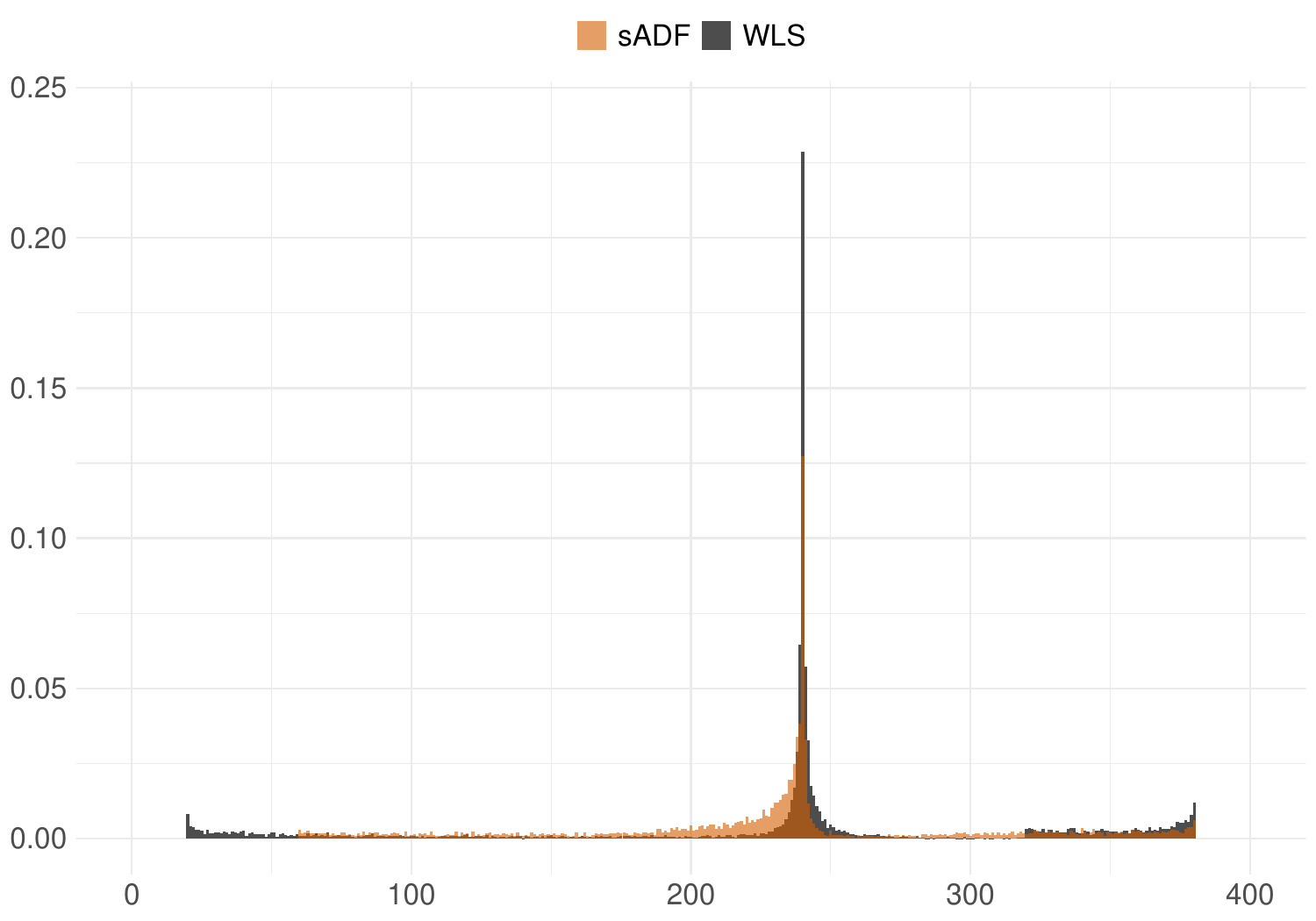}\label{fig:1:4:app}}\\
\subfigure[$T=400$, $c_a=6$, $c_b=6$, $s_0/s_1=1/5$]{\includegraphics[width=0.45\linewidth]{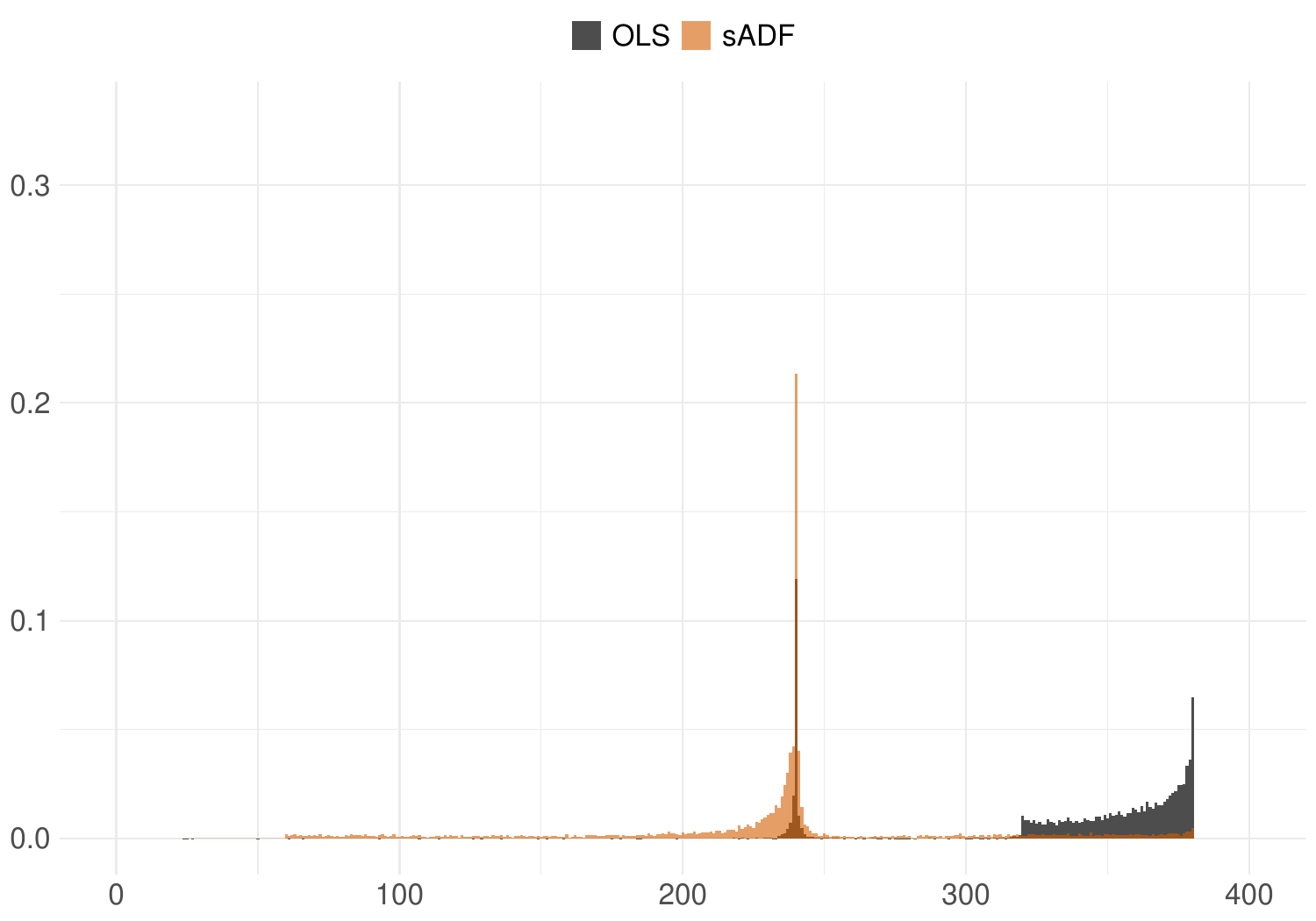}\label{fig:1:5:app}}
\subfigure[$T=400$, $c_a=6$, $c_b=6$, $s_0/s_1=1/5$]{\includegraphics[width=0.45\linewidth]{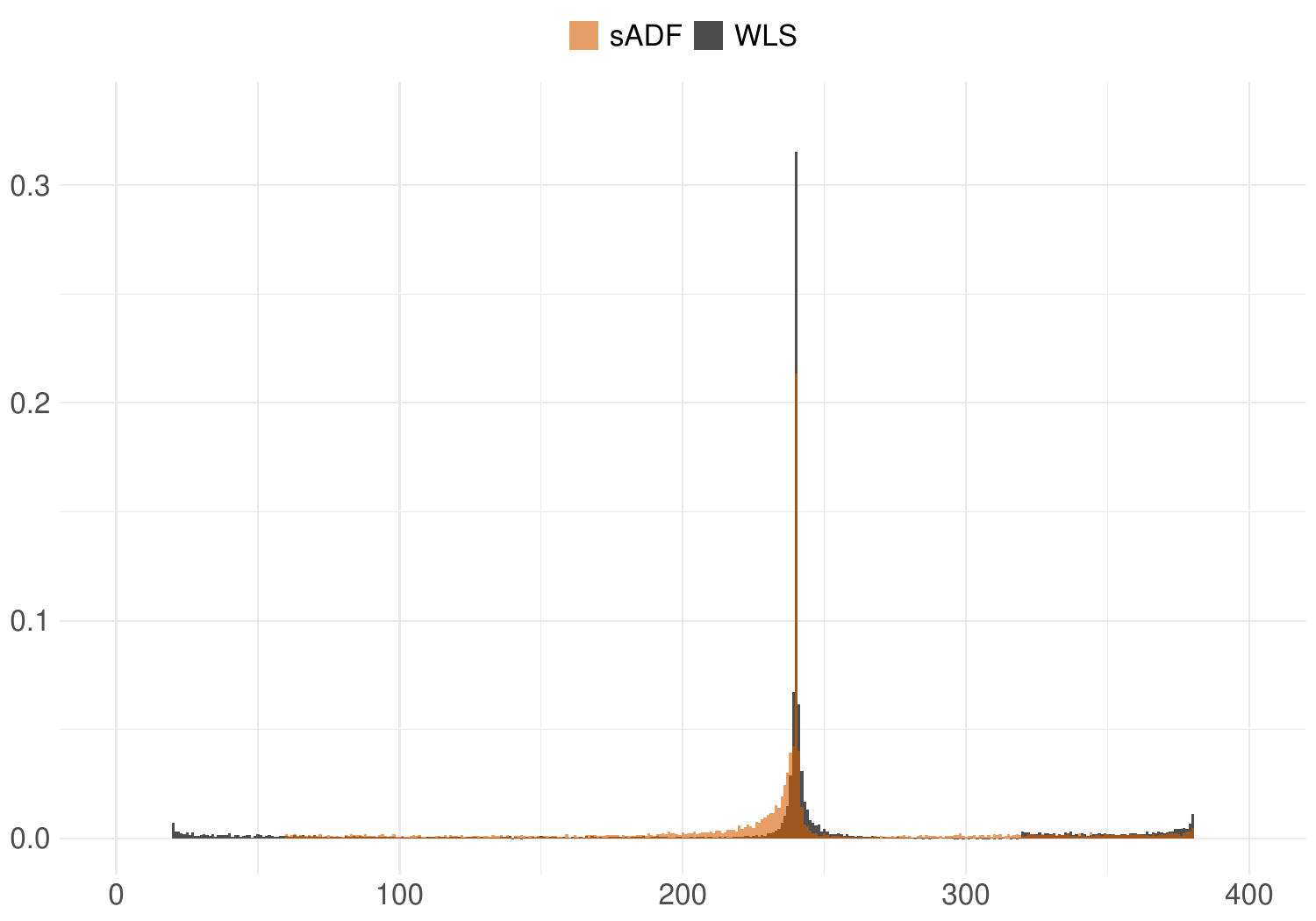}\label{fig:1:6:app}}\\
\end{center}%
\caption{Histograms of $\hat{k}_c$ 
for $(\tau_e,\tau_c,\tau_r)=(0.4,0.6,0.7)$,  $\tau=0.8$, $s_0/s_1=1/5$, $T=400$}
\label{fig1:app}
\end{figure}

\newpage

\begin{figure}[h!]%
\begin{center}%
\subfigure[$T=800$, $c_a=4$, $c_b=6$, $s_0/s_1=1/5$]{\includegraphics[width=0.45\linewidth]{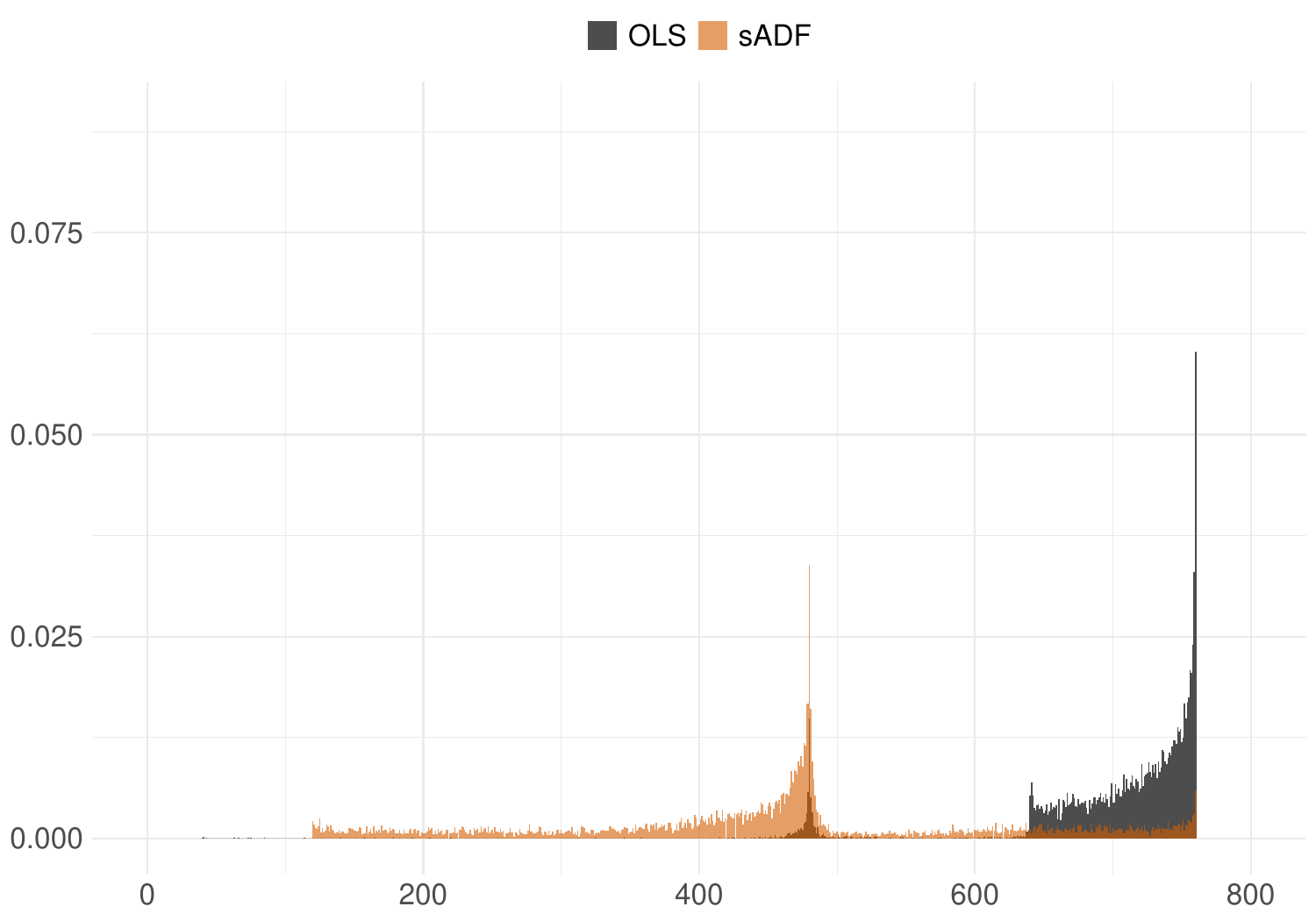}\label{fig:2:1}}
\subfigure[$T=800$, $c_a=4$, $c_b=6$, $s_0/s_1=1/5$]{\includegraphics[width=0.45\linewidth]{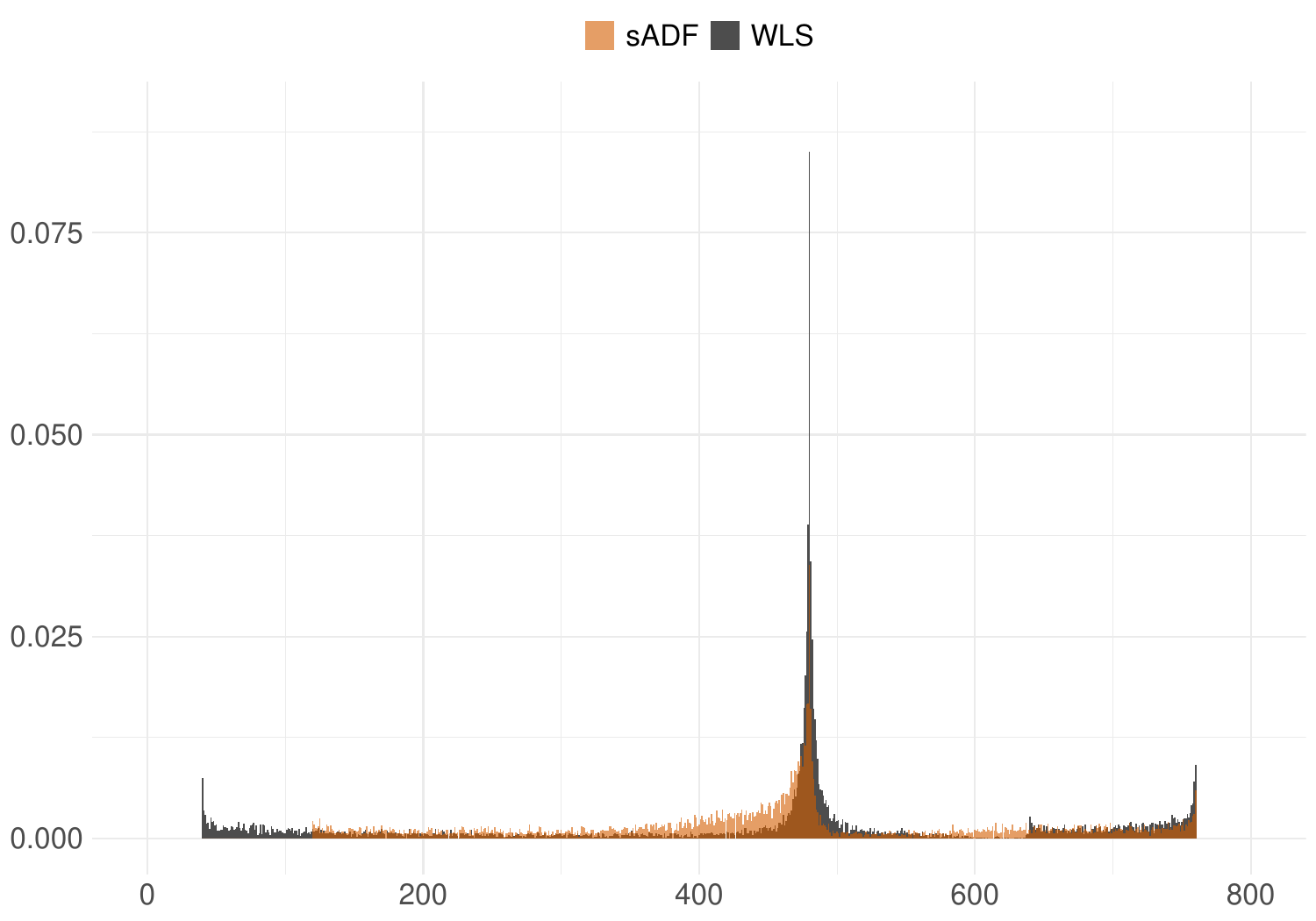}\label{fig:2:2}}\\
\subfigure[$T=800$, $c_a=5$, $c_b=6$, $s_0/s_1=1/5$]{\includegraphics[width=0.45\linewidth]{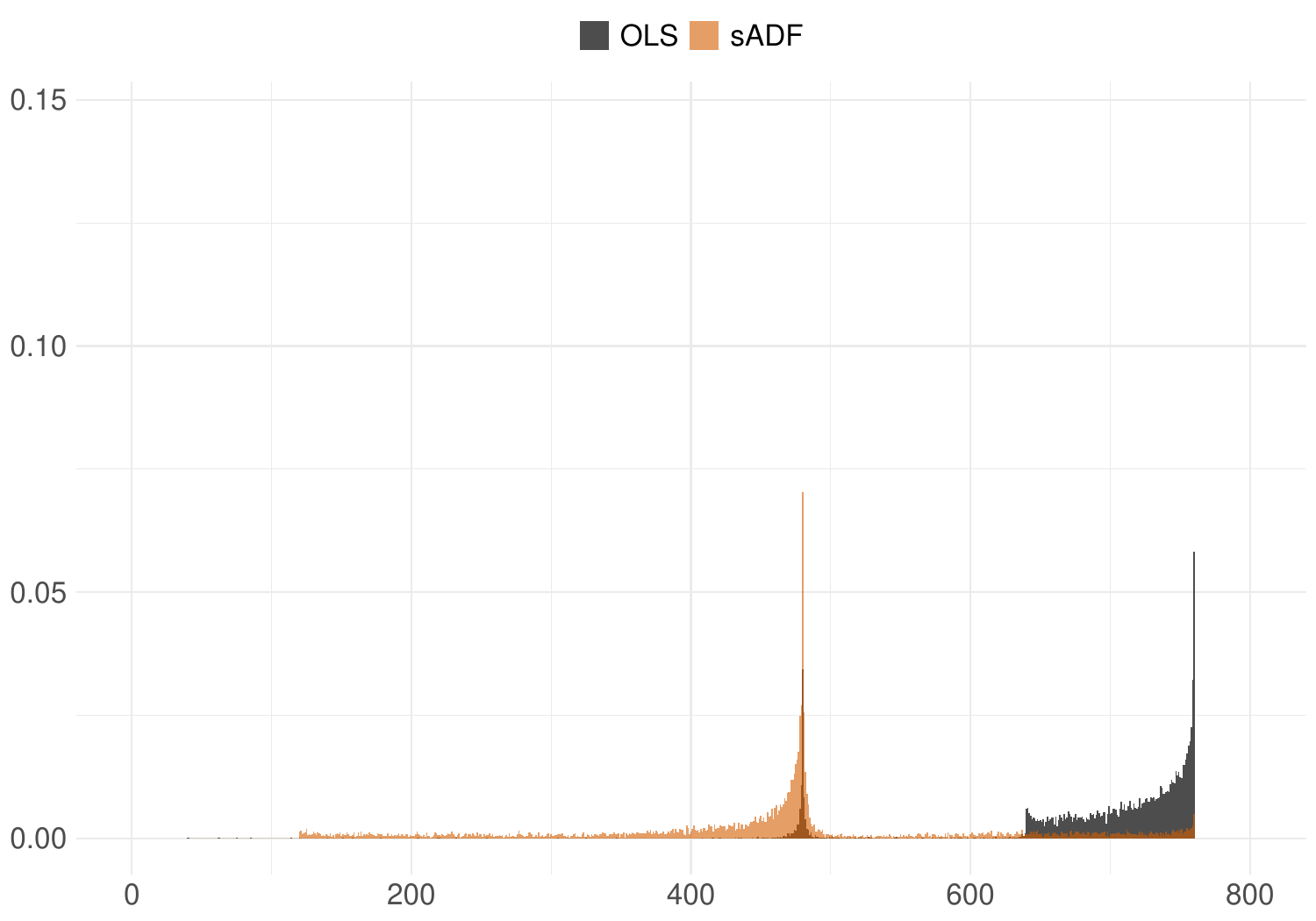}\label{fig:2:3}}
\subfigure[$T=800$, $c_a=5$, $c_b=6$, $s_0/s_1=1/5$]{\includegraphics[width=0.45\linewidth]{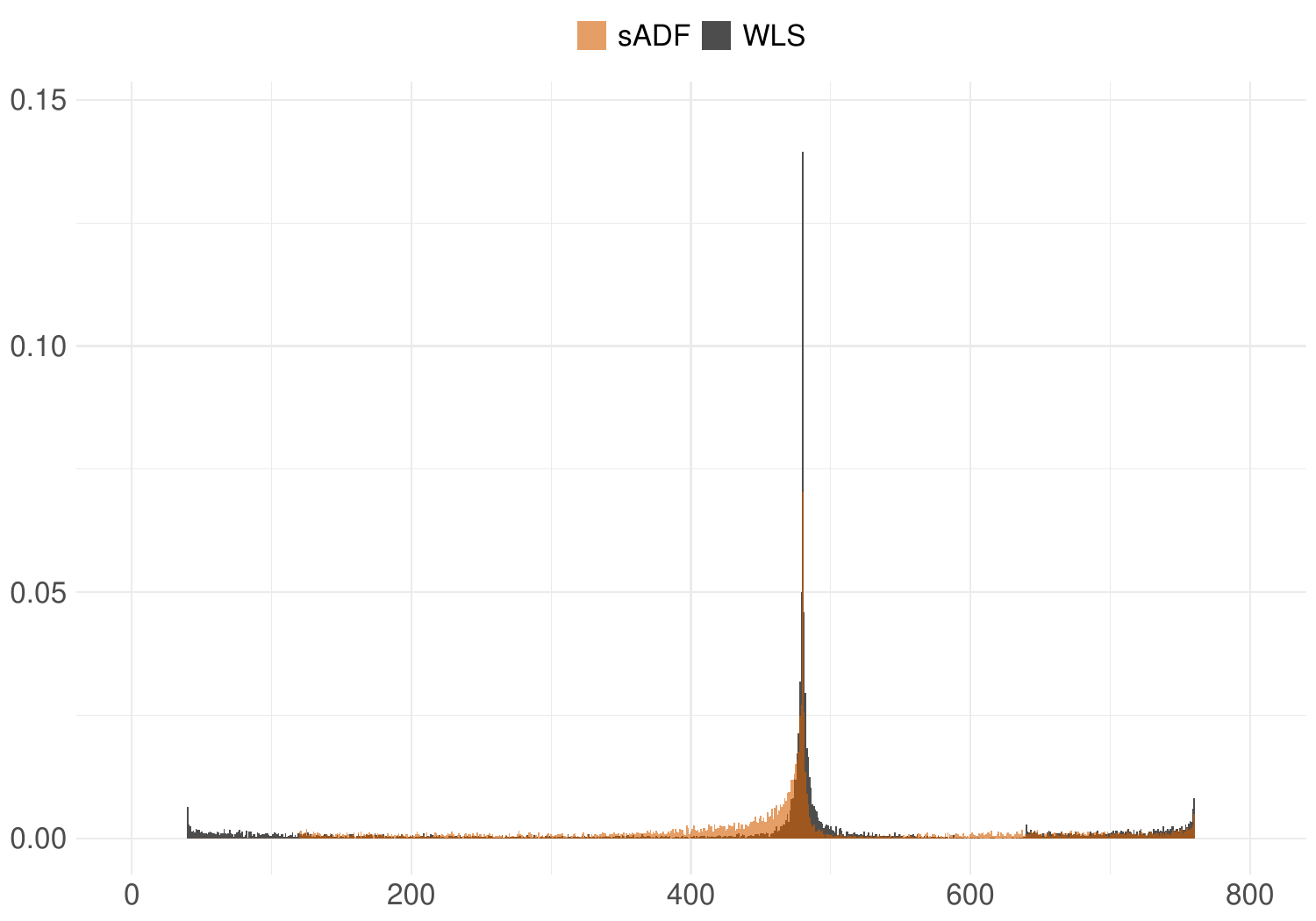}\label{fig:2:4}}\\
\subfigure[$T=800$, $c_a=6$, $c_b=6$, $s_0/s_1=1/5$]{\includegraphics[width=0.45\linewidth]{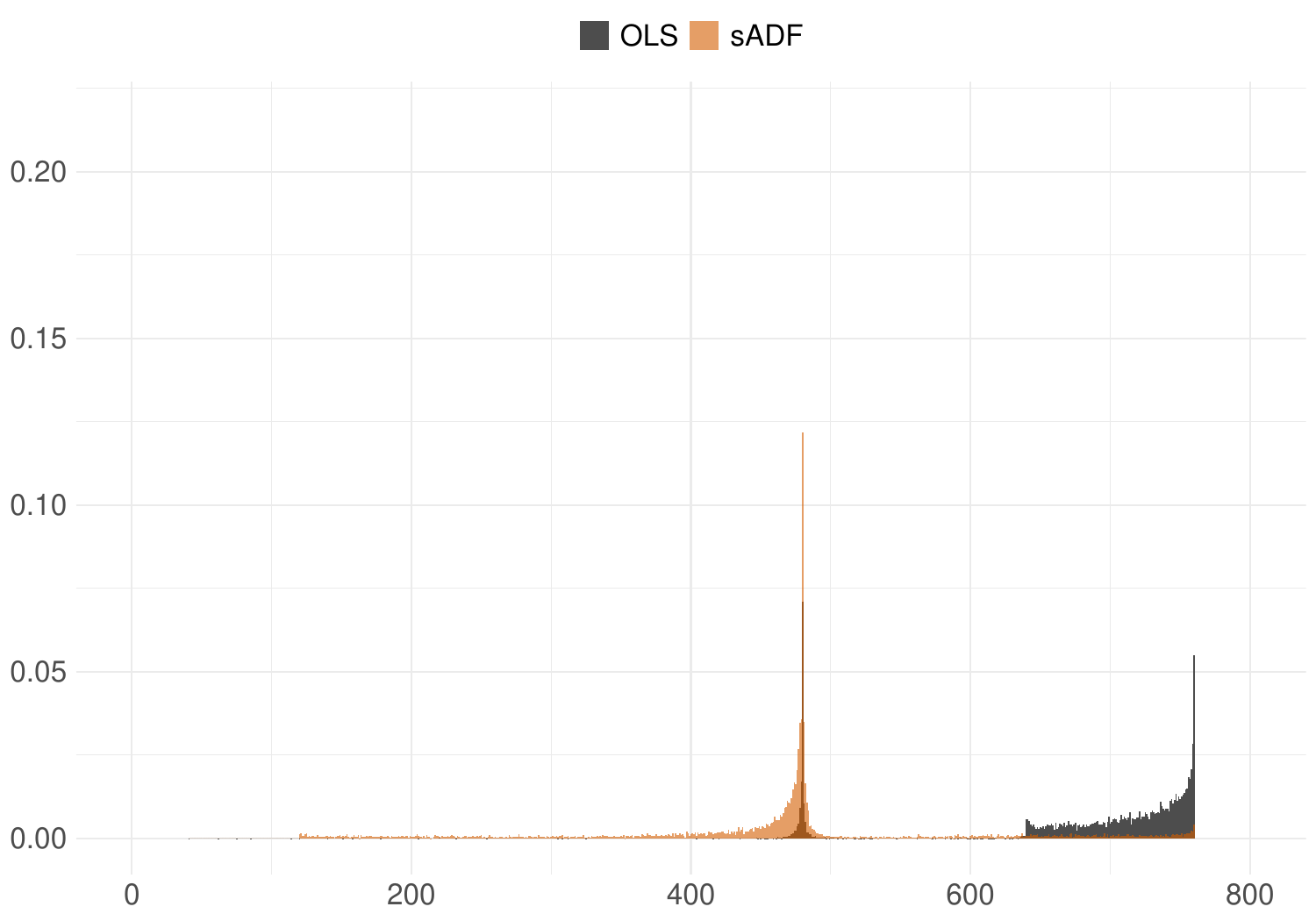}\label{fig:2:5}}
\subfigure[$T=800$, $c_a=6$, $c_b=6$, $s_0/s_1=1/5$]{\includegraphics[width=0.45\linewidth]{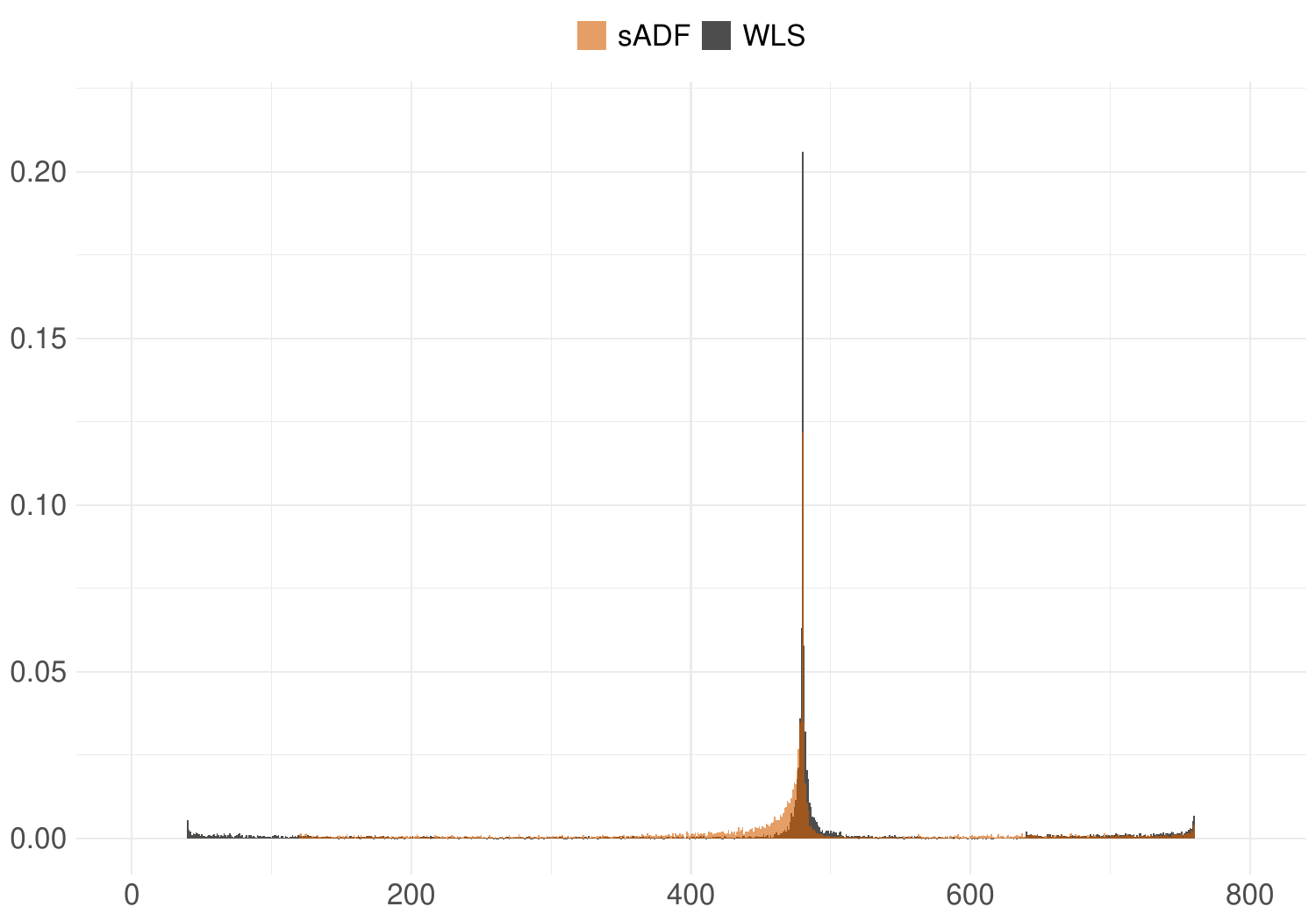}\label{fig:2:6}}\\
\end{center}%
\caption{Histograms of $\hat{k}_c$ 
for $(\tau_e,\tau_c,\tau_r)=(0.4,0.6,0.7)$,  $\tau=0.8$, $s_0/s_1=1/5$, $T=800$}
\label{fig2}
\end{figure}

\newpage

\begin{figure}[h!]%
\begin{center}%
\subfigure[$T=400$, $c_a=4$, $c_b=6$, $s_0/s_1=1$]{\includegraphics[width=0.45\linewidth]{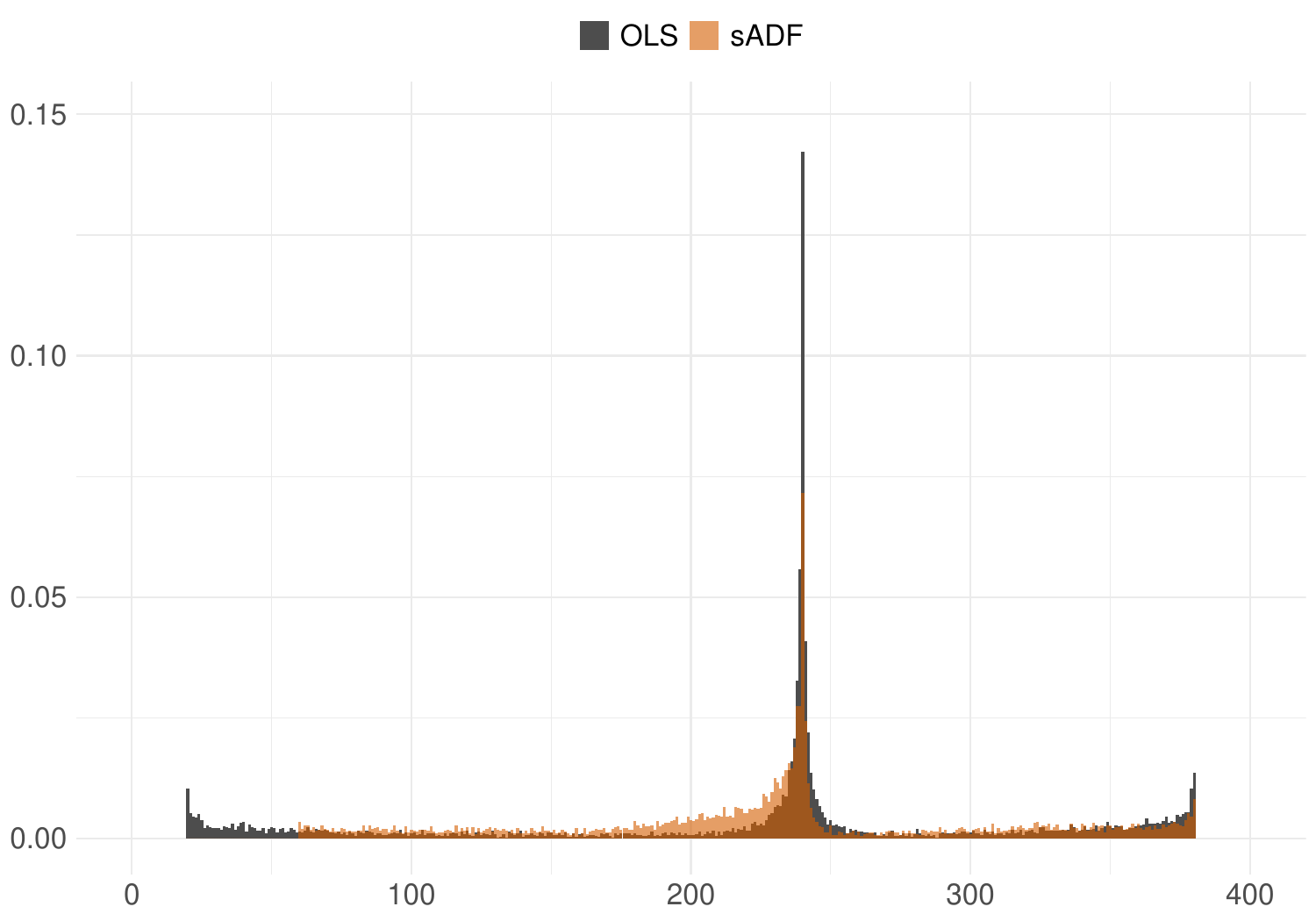}\label{fig:3:1}}
\subfigure[$T=400$, $c_a=4$, $c_b=6$, $s_0/s_1=1$]{\includegraphics[width=0.45\linewidth]{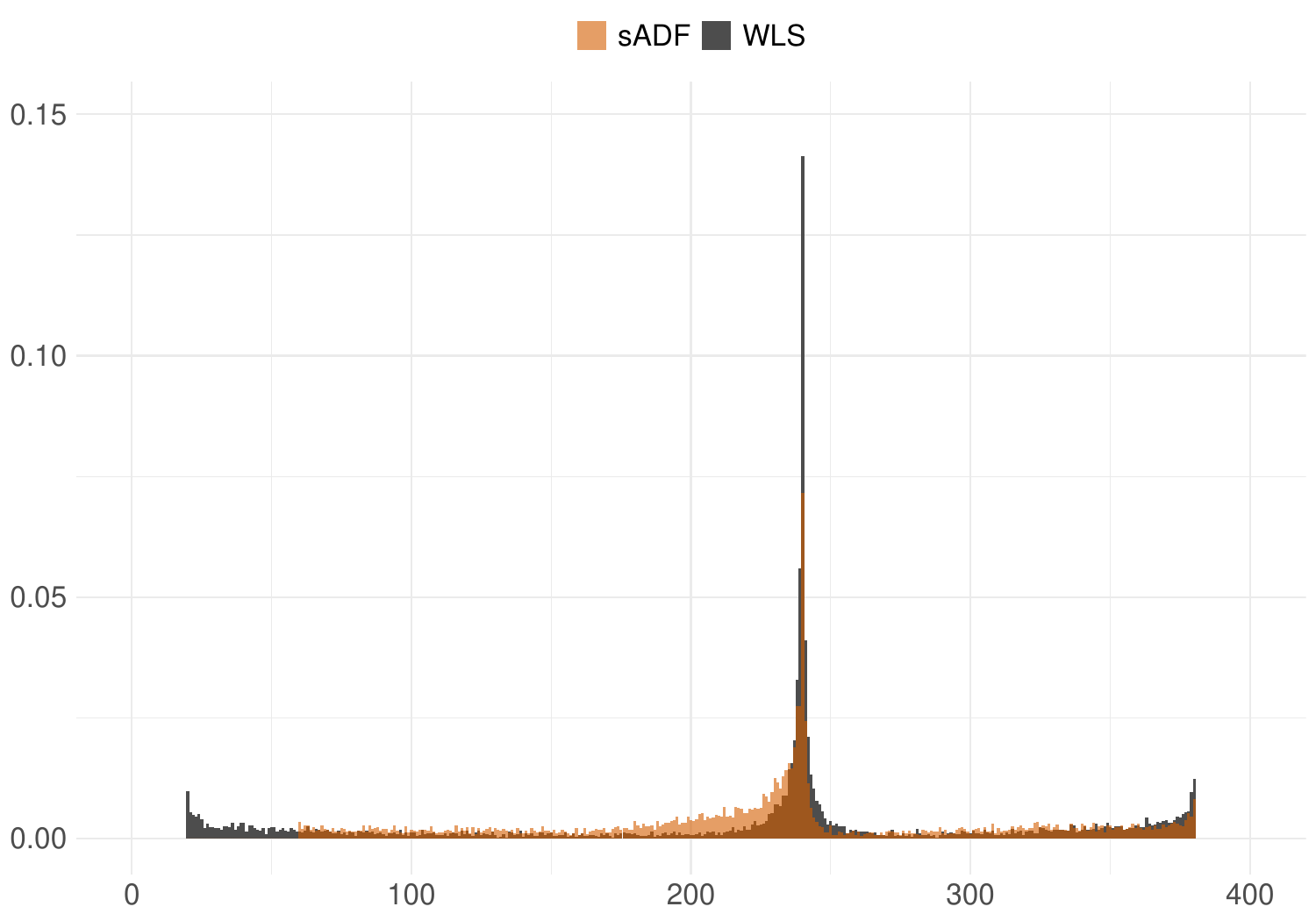}\label{fig:3:2}}\\
\subfigure[$T=400$, $c_a=5$, $c_b=6$, $s_0/s_1=1$]{\includegraphics[width=0.45\linewidth]{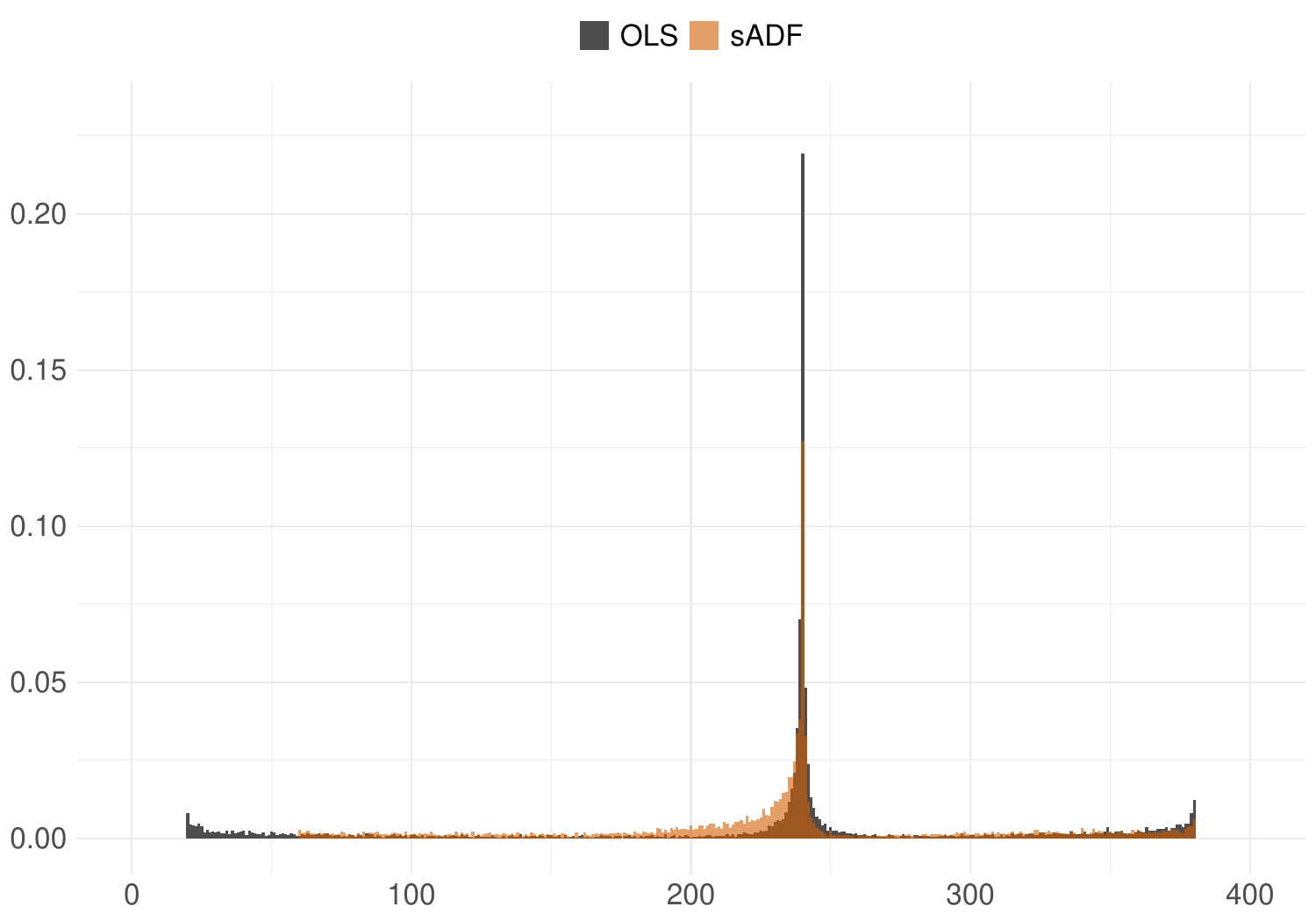}\label{fig:3:3}}
\subfigure[$T=400$, $c_a=5$, $c_b=6$, $s_0/s_1=1$]{\includegraphics[width=0.45\linewidth]{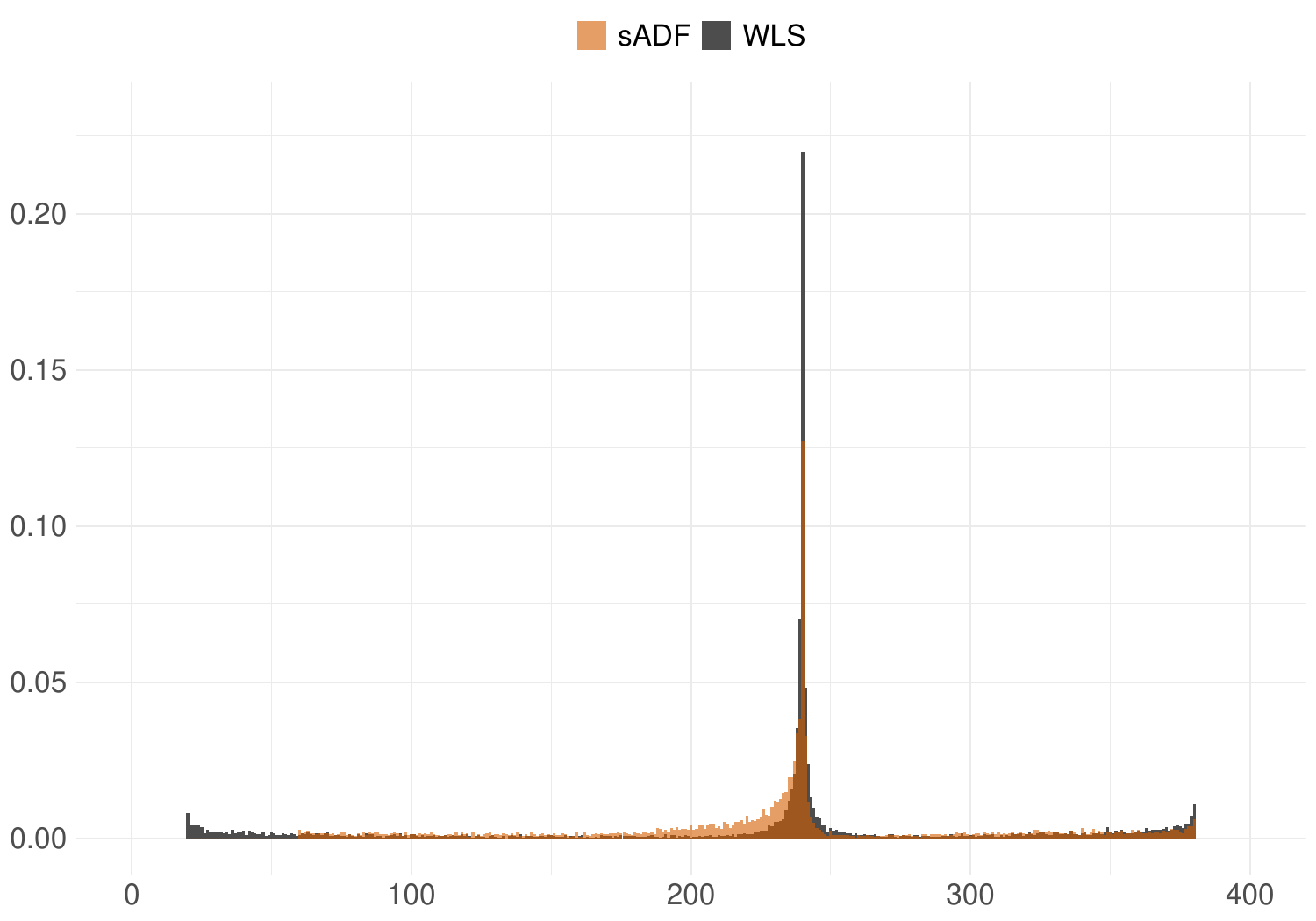}\label{fig:3:4}}\\
\subfigure[$T=400$, $c_a=6$, $c_b=6$, $s_0/s_1=1$]{\includegraphics[width=0.45\linewidth]{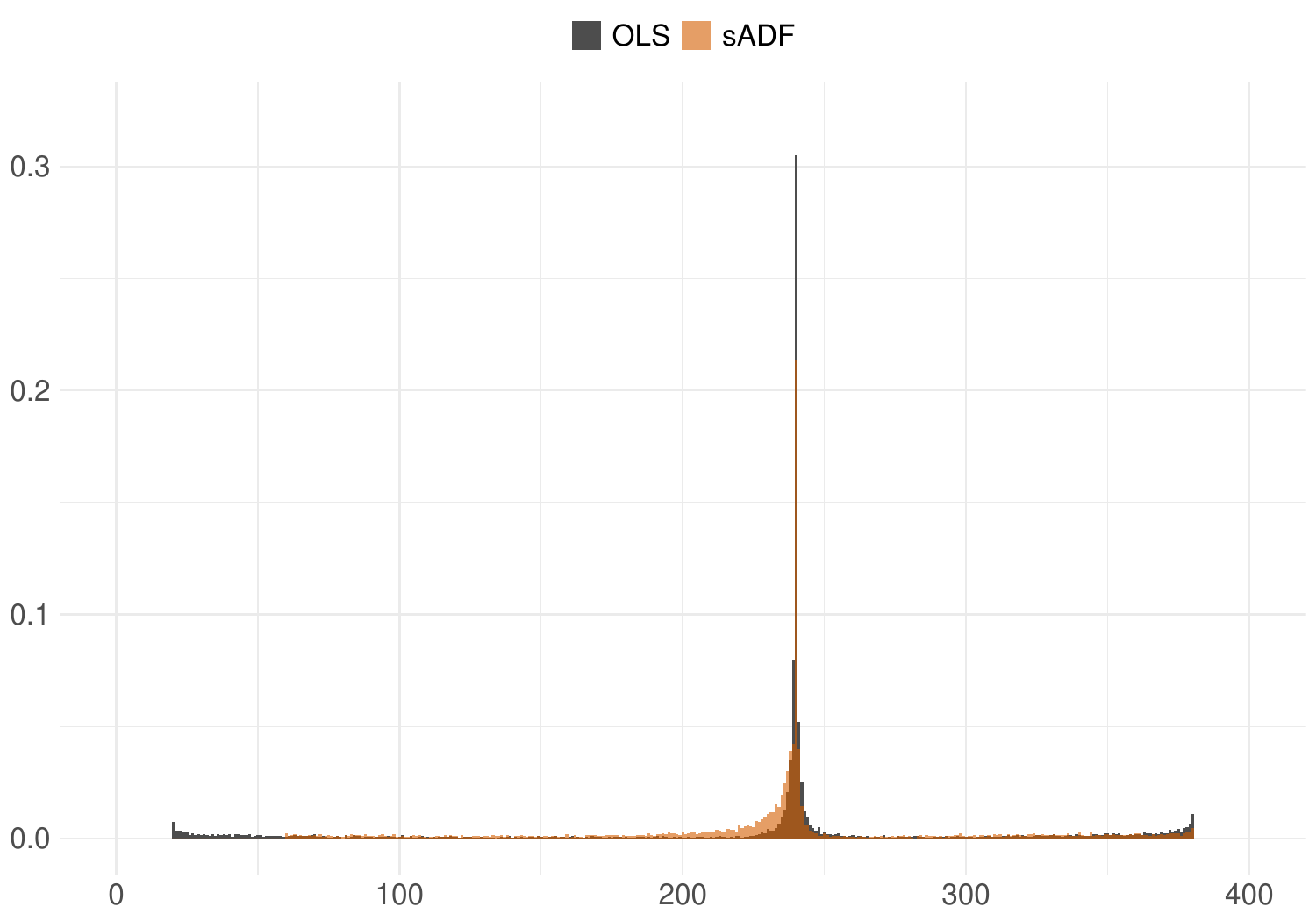}\label{fig:3:5}}
\subfigure[$T=400$, $c_a=6$, $c_b=6$, $s_0/s_1=1$]{\includegraphics[width=0.45\linewidth]{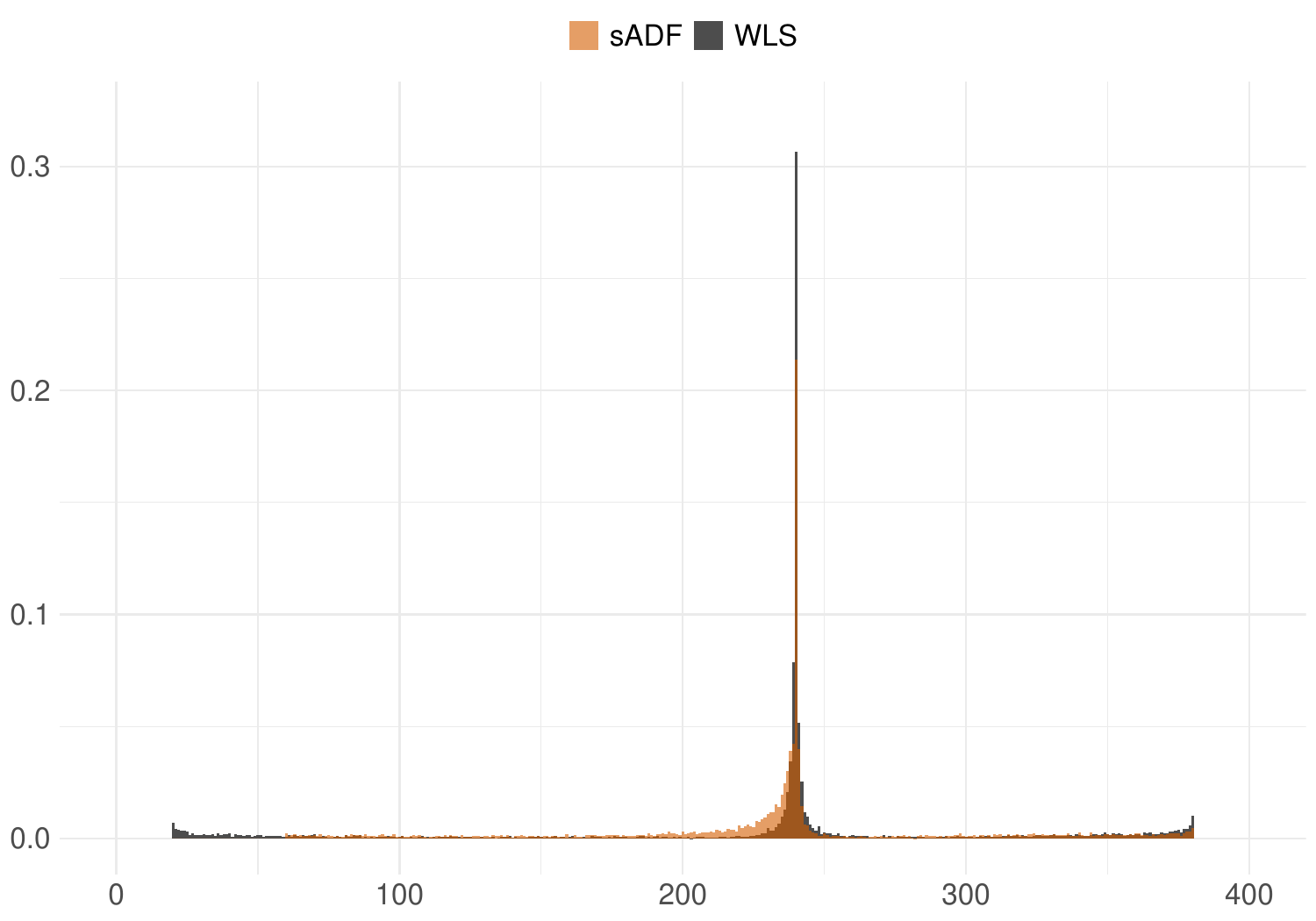}\label{fig:3:6}}\\
\end{center}%
\caption{Histograms of $\hat{k}_c$ 
for $(\tau_e,\tau_c,\tau_r)=(0.4,0.6,0.7)$,  $\tau=0.8$, $s_0/s_1=1$, $T=400$}
\label{fig3}
\end{figure}

\newpage

\begin{figure}[h!]%
\begin{center}%
\subfigure[$T=800$, $c_a=4$, $c_b=6$, $s_0/s_1=1$]{\includegraphics[width=0.45\linewidth]{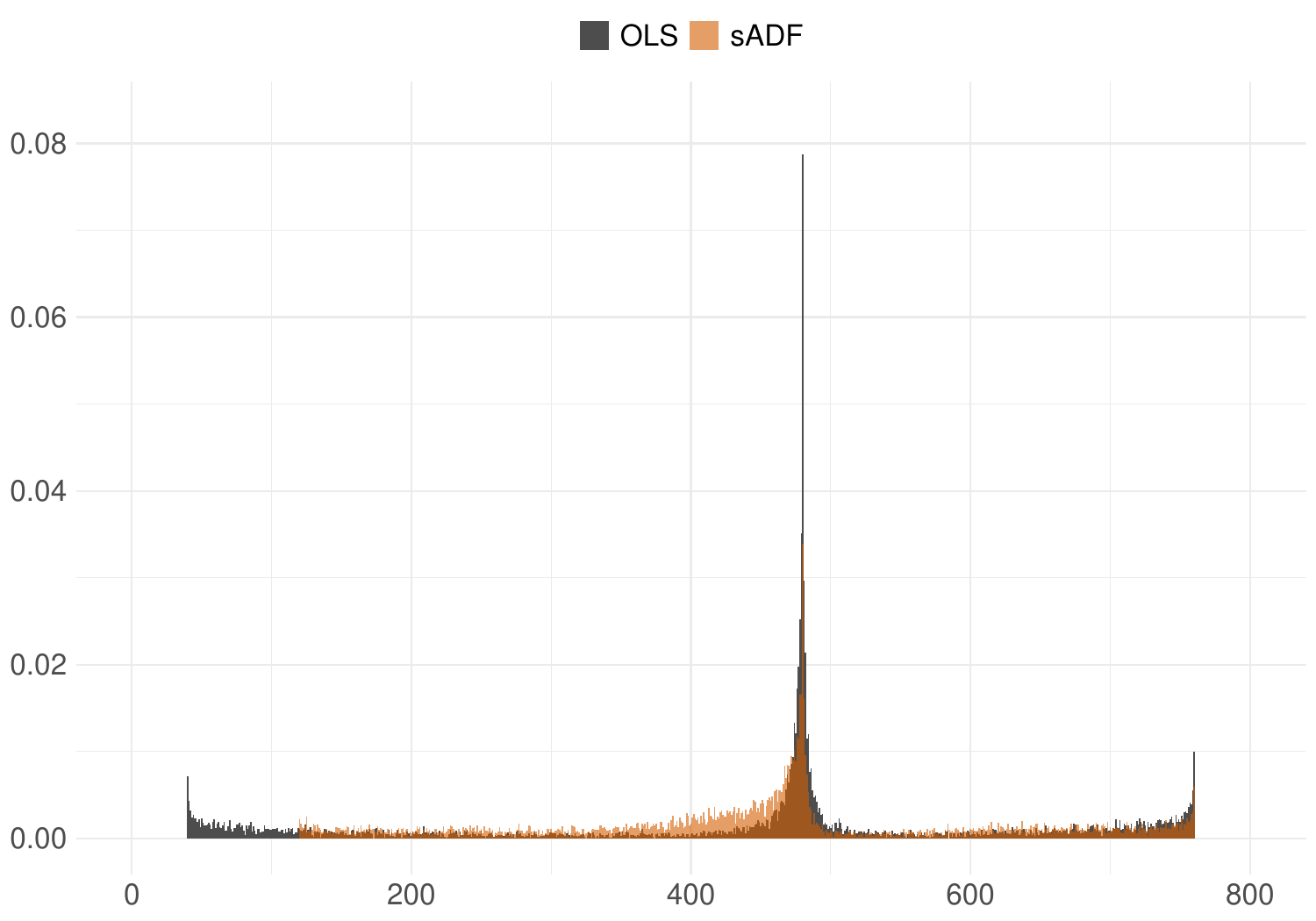}\label{fig:4:1}}
\subfigure[$T=800$, $c_a=4$, $c_b=6$, $s_0/s_1=1$]{\includegraphics[width=0.45\linewidth]{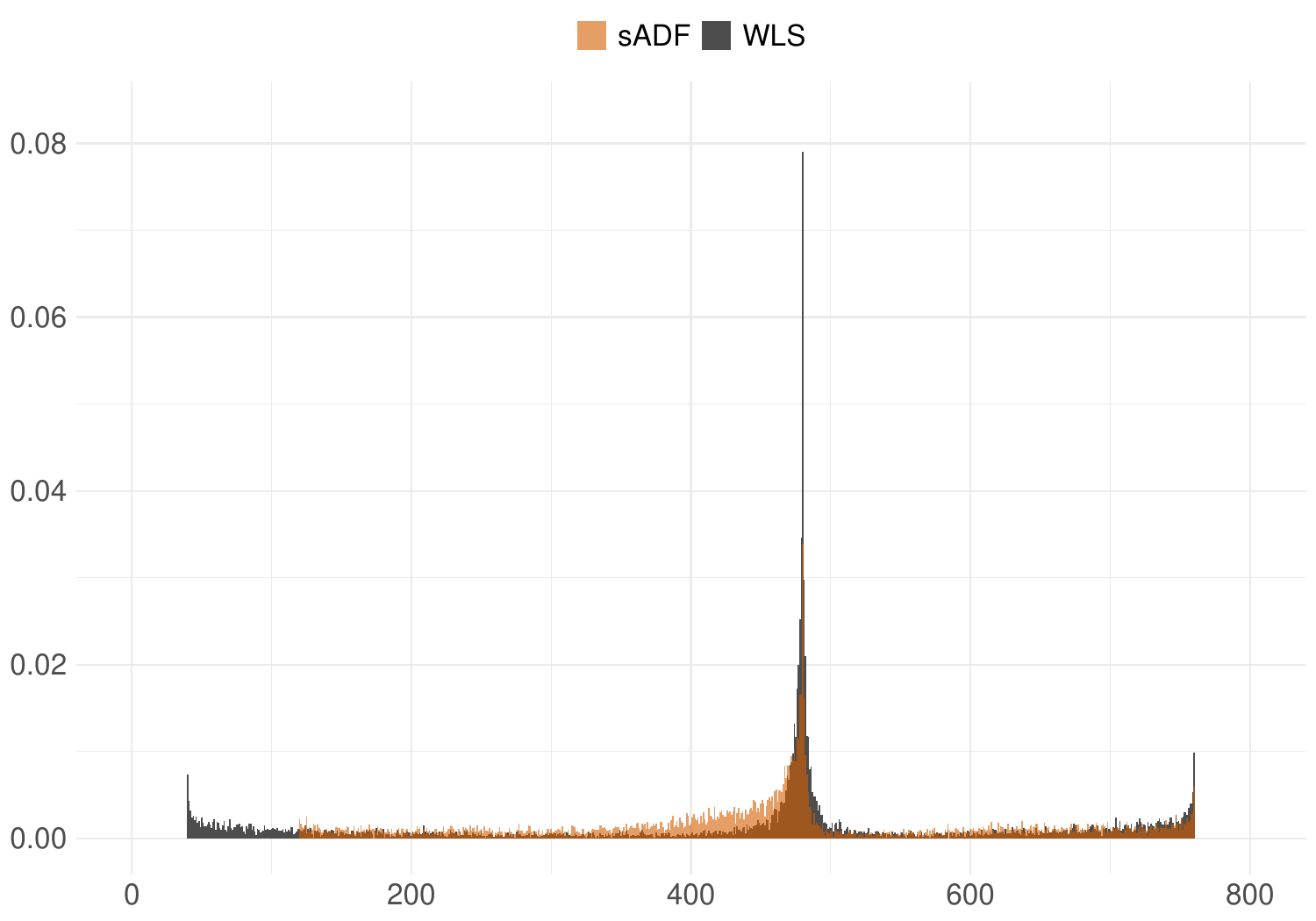}\label{fig:4:2}}\\
\subfigure[$T=800$, $c_a=5$, $c_b=6$, $s_0/s_1=1$]{\includegraphics[width=0.45\linewidth]{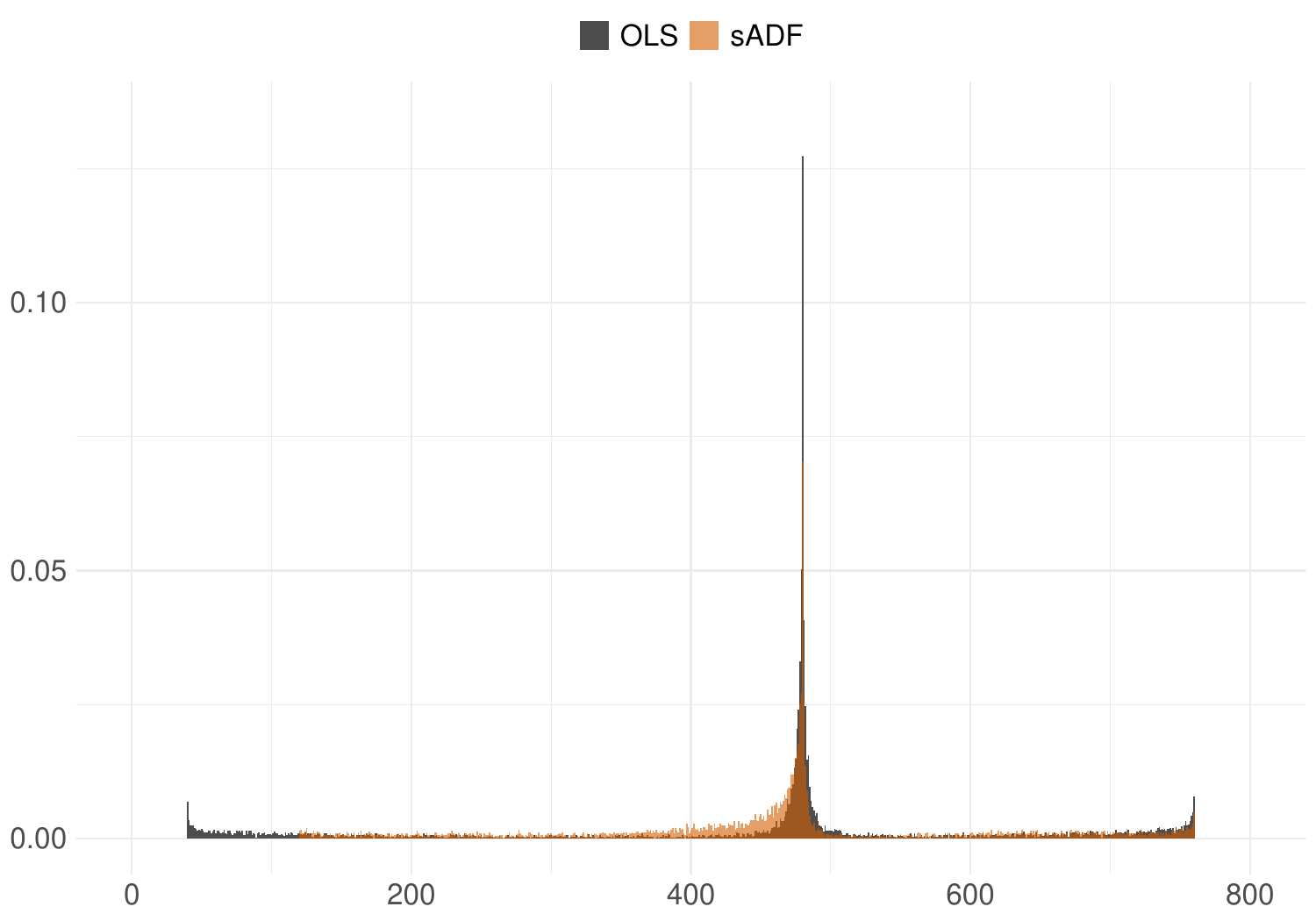}\label{fig:4:3}}
\subfigure[$T=800$, $c_a=5$, $c_b=6$, $s_0/s_1=1$]{\includegraphics[width=0.45\linewidth]{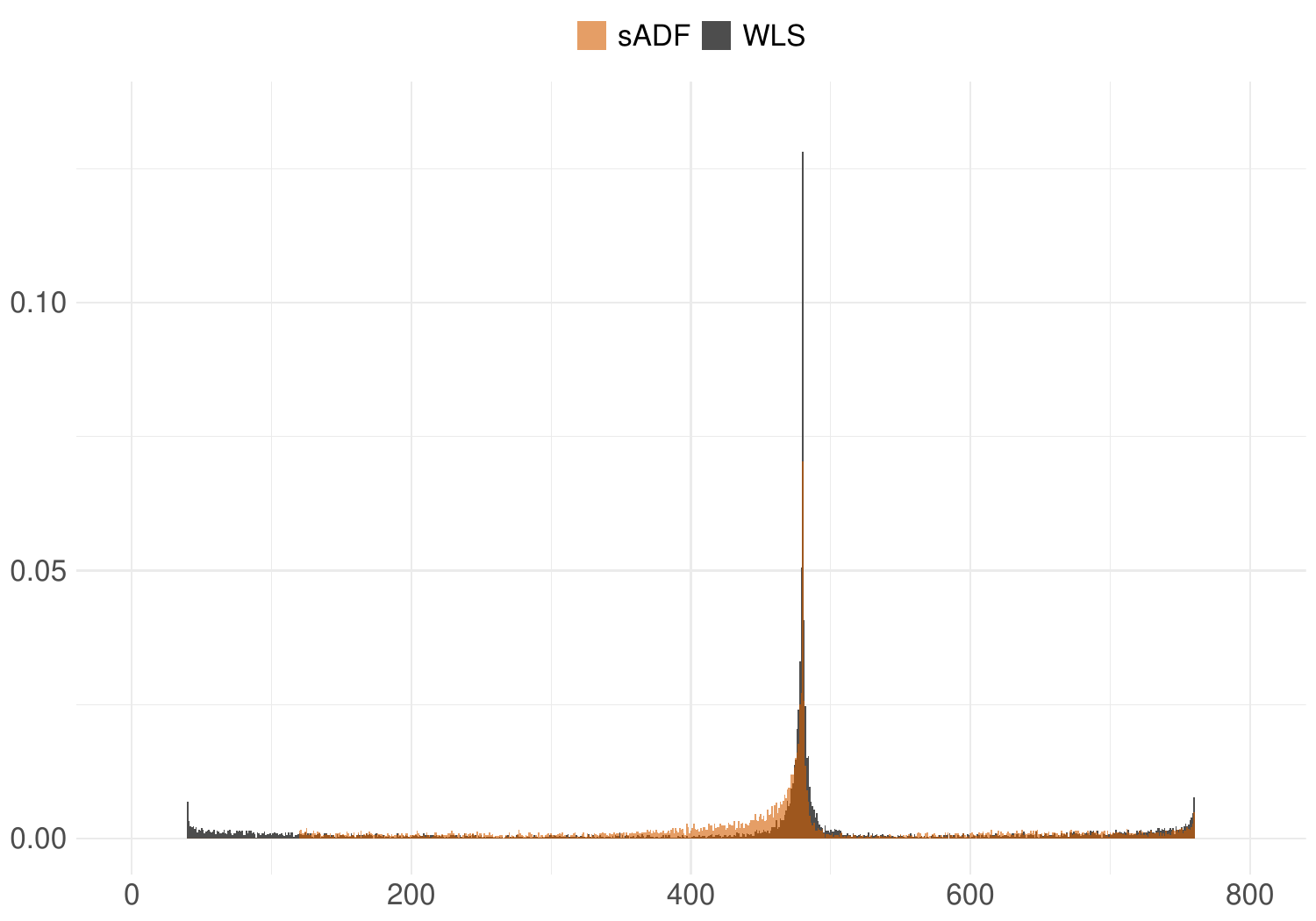}\label{fig:4:4}}\\
\subfigure[$T=800$, $c_a=6$, $c_b=6$, $s_0/s_1=1$]{\includegraphics[width=0.45\linewidth]{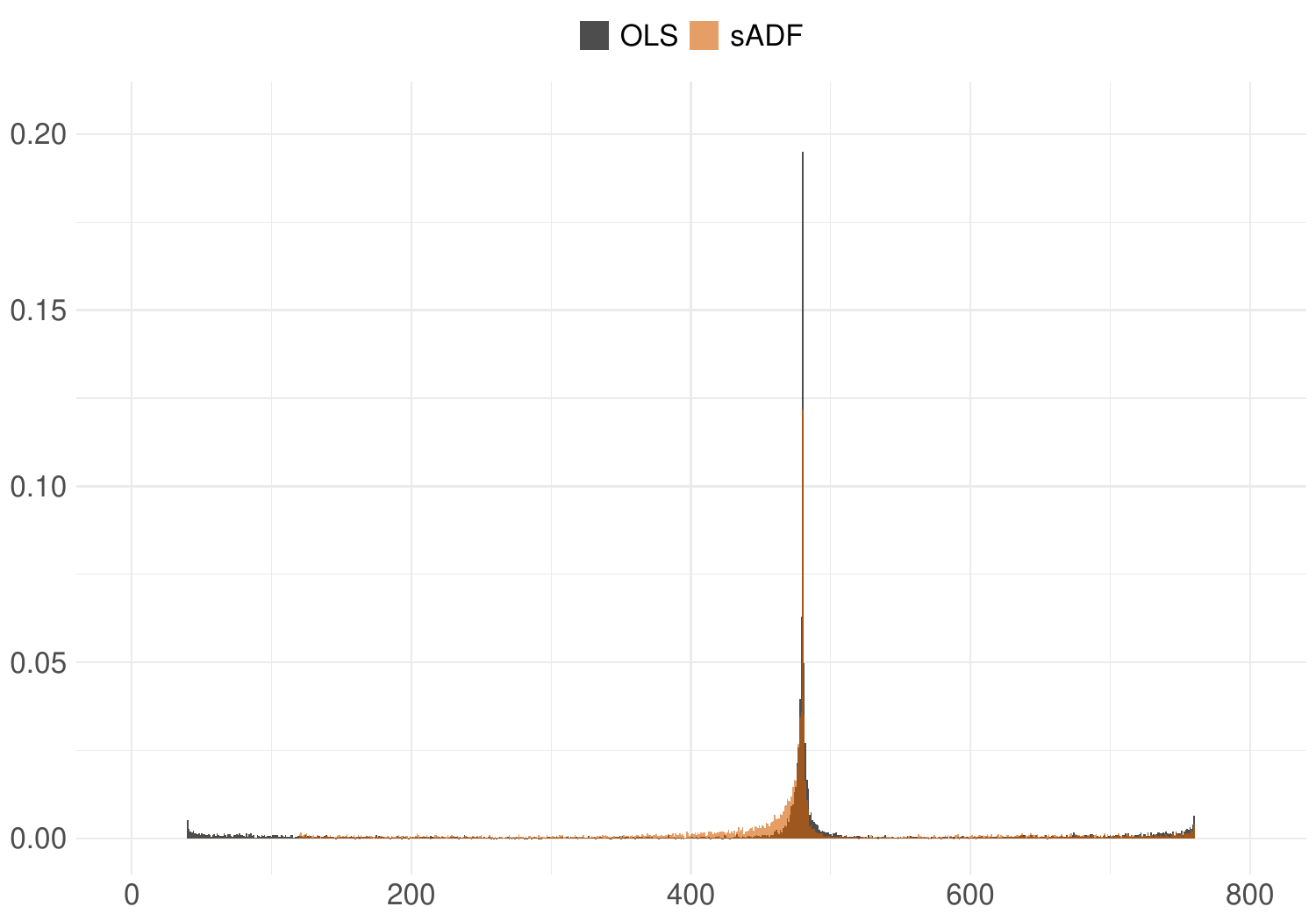}\label{fig:4:5}}
\subfigure[$T=800$, $c_a=6$, $c_b=6$, $s_0/s_1=1$]{\includegraphics[width=0.45\linewidth]{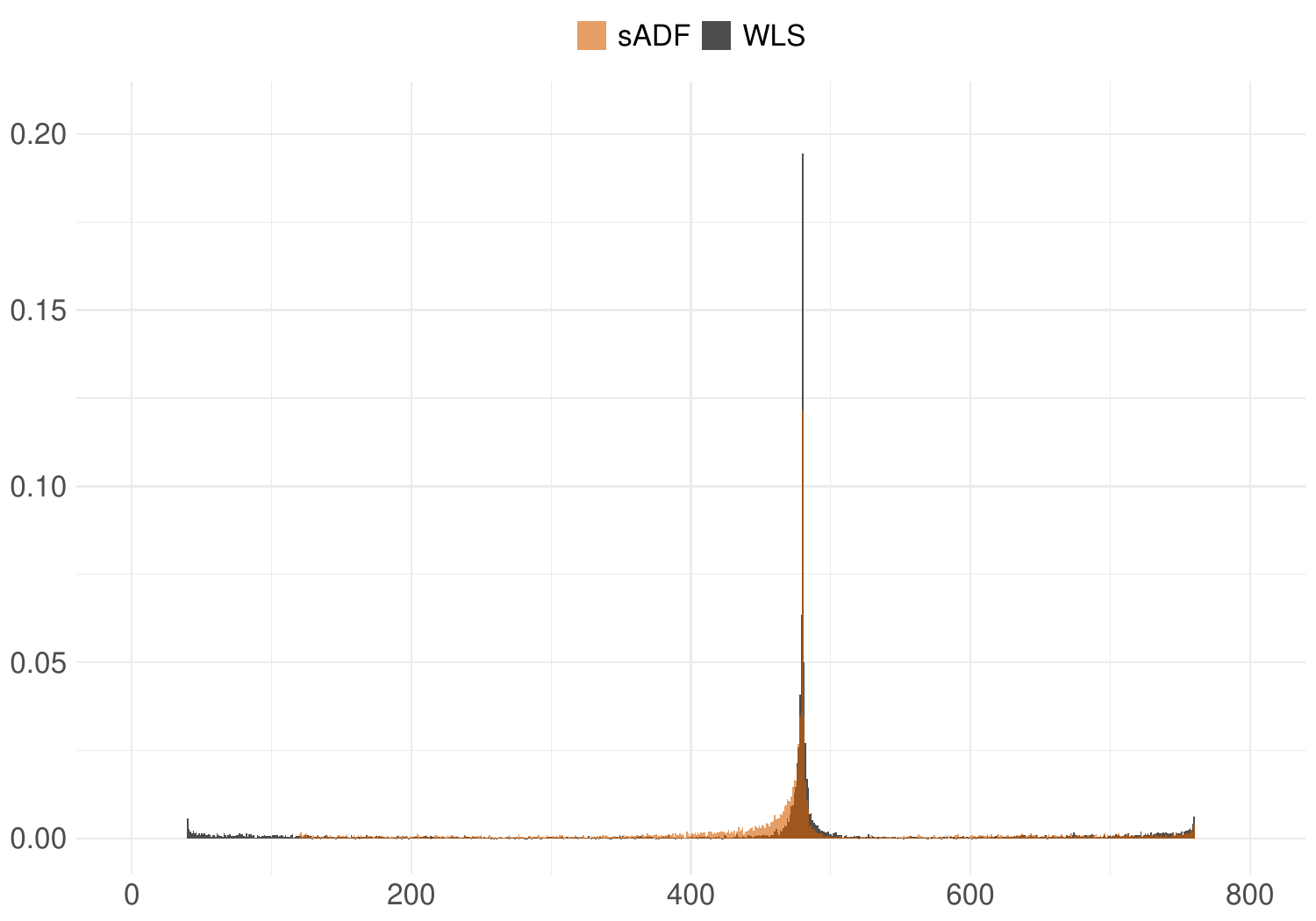}\label{fig:4:6}}\\
\end{center}%
\caption{Histograms of $\hat{k}_c$ 
for $(\tau_e,\tau_c,\tau_r)=(0.4,0.6,0.7)$,  $\tau=0.8$, $s_0/s_1=1$, $T=800$}
\label{fig4}
\end{figure}

\newpage

\begin{figure}[h!]%
\begin{center}%
\subfigure[$T=400$, $c_a=4$, $c_b=6$, $s_0/s_1=5$]{\includegraphics[width=0.45\linewidth]{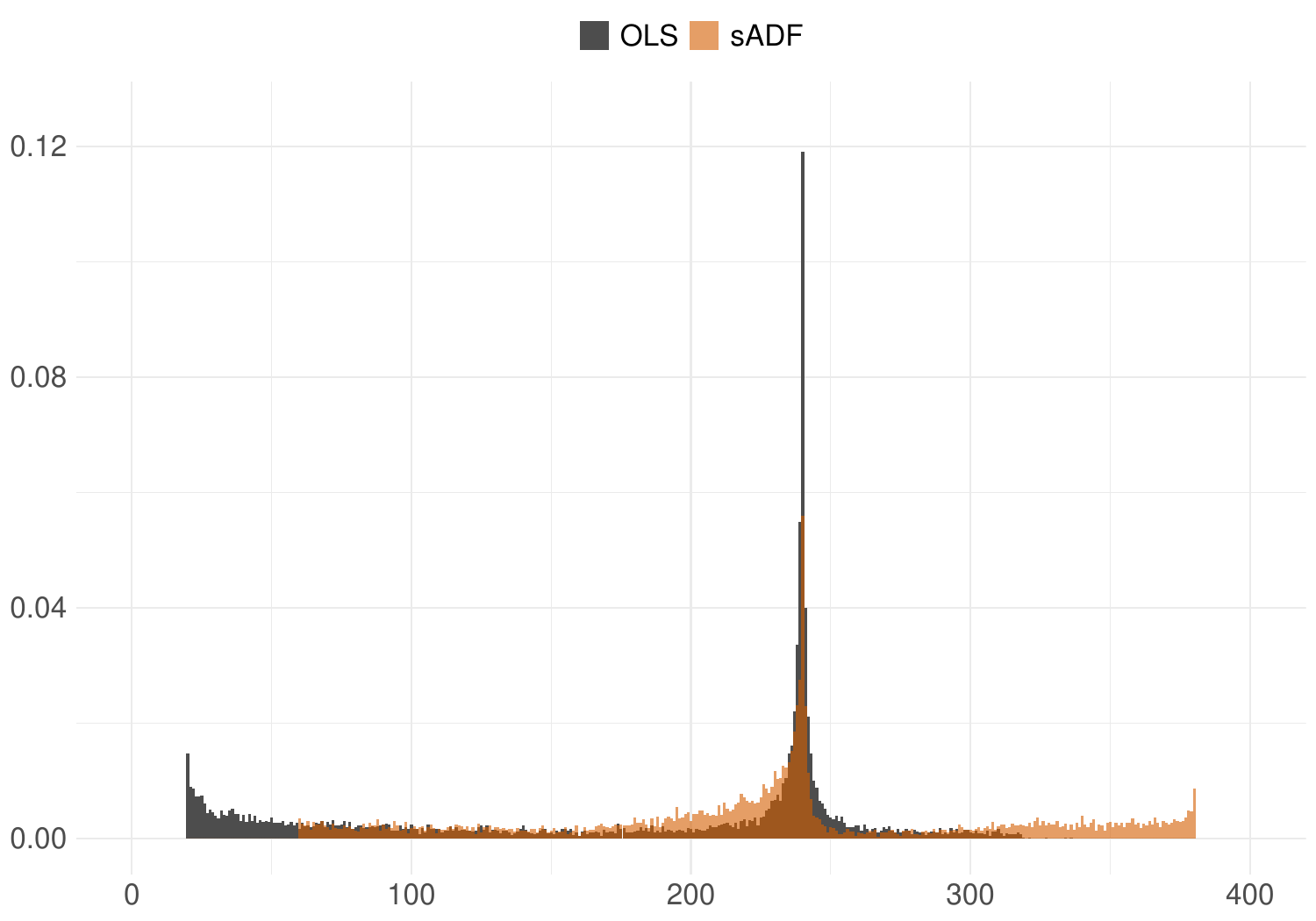}\label{fig:5:1}}
\subfigure[$T=400$, $c_a=4$, $c_b=6$, $s_0/s_1=5$]{\includegraphics[width=0.45\linewidth]{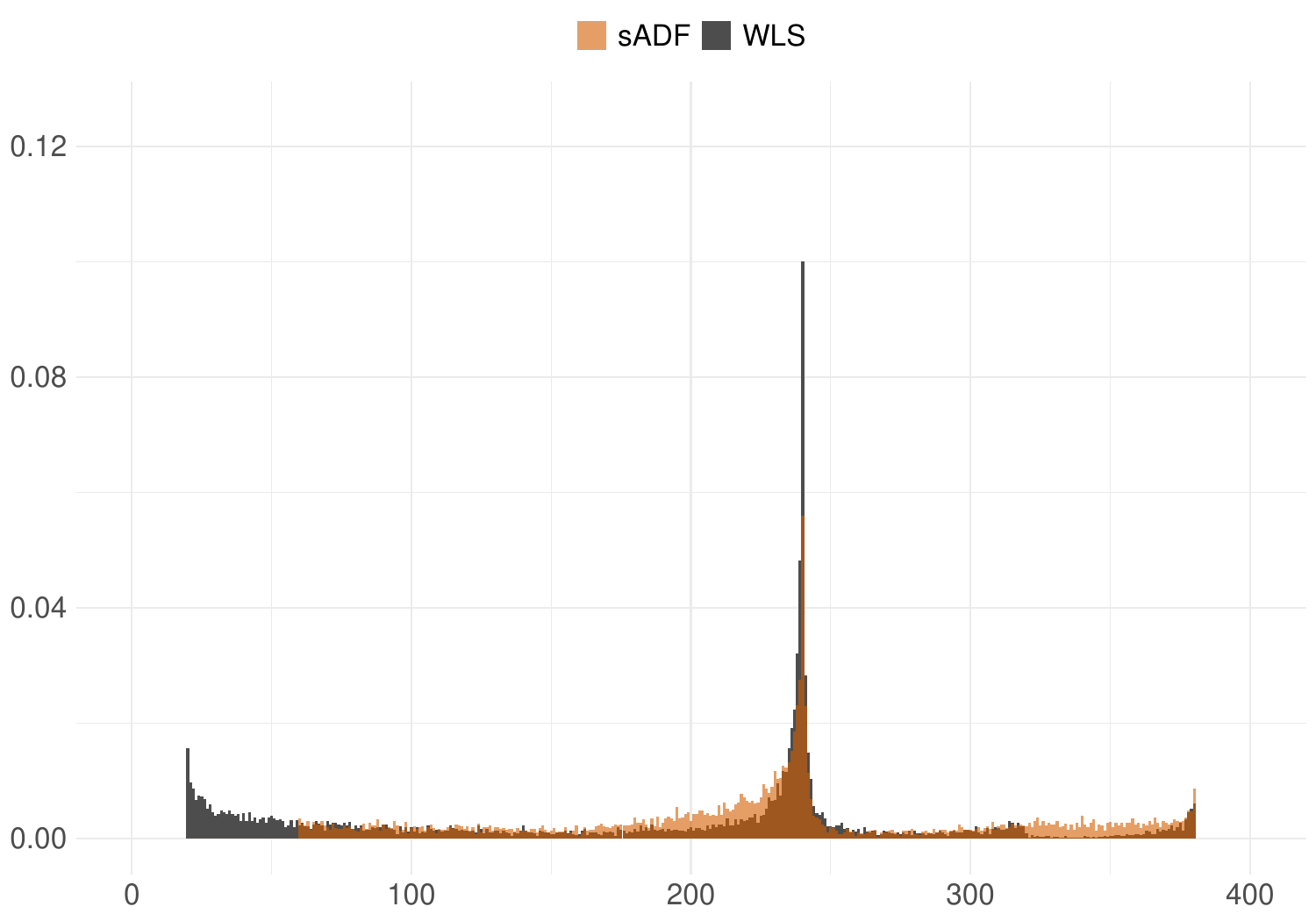}\label{fig:5:2}}\\
\subfigure[$T=400$, $c_a=5$, $c_b=6$, $s_0/s_1=5$]{\includegraphics[width=0.45\linewidth]{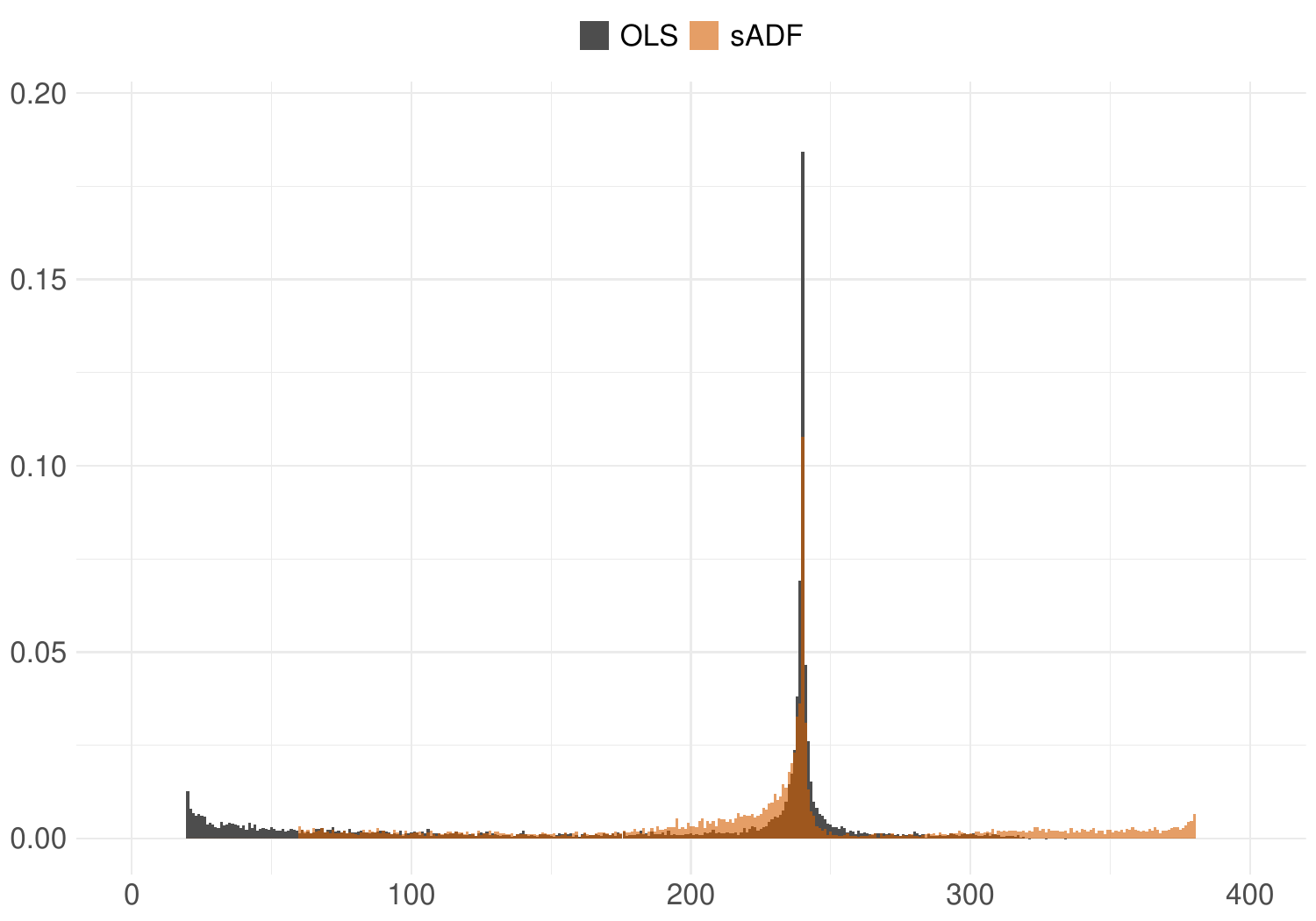}\label{fig:5:3}}
\subfigure[$T=400$, $c_a=5$, $c_b=6$, $s_0/s_1=5$]{\includegraphics[width=0.45\linewidth]{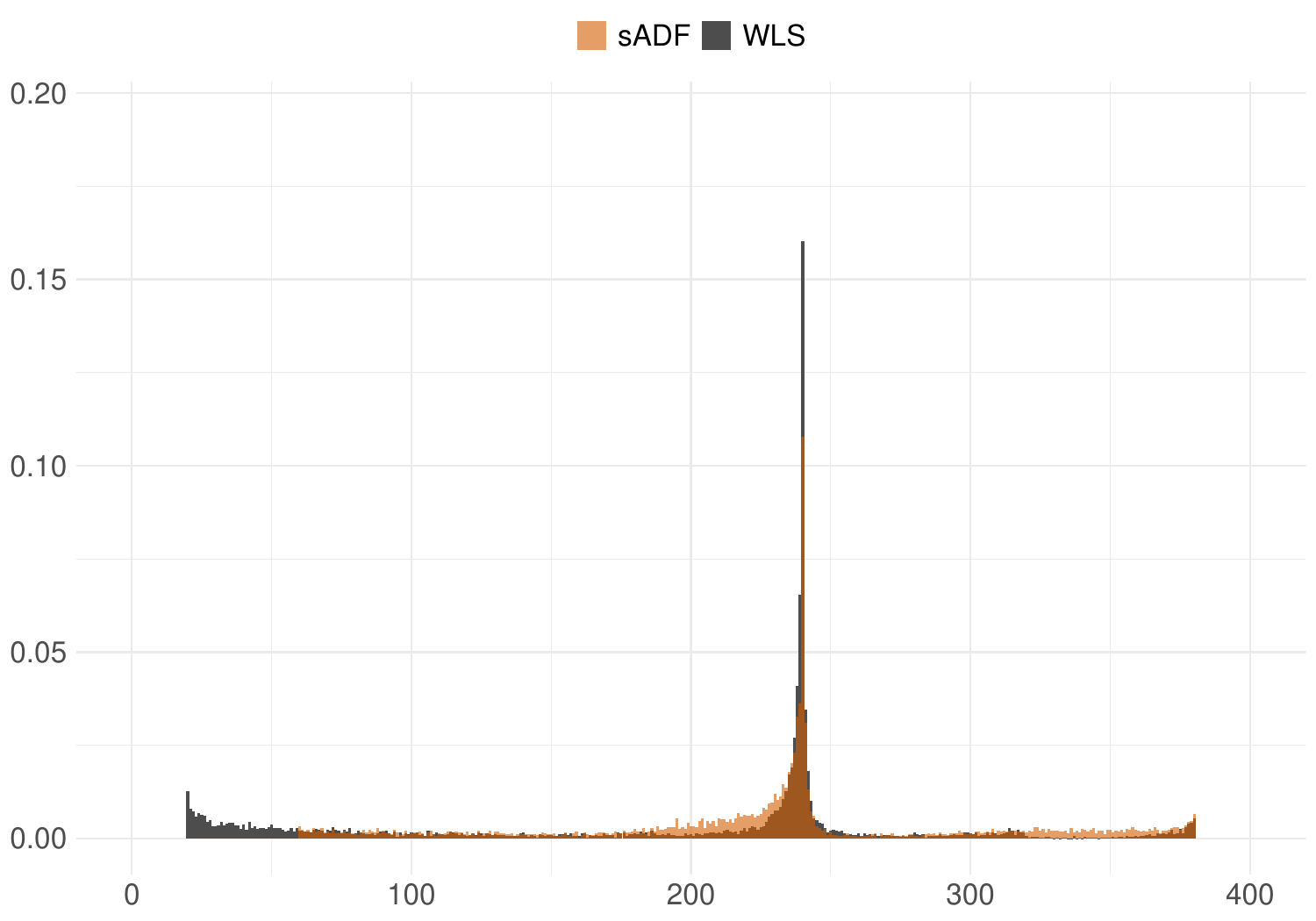}\label{fig:5:4}}\\
\subfigure[$T=400$, $c_a=6$, $c_b=6$, $s_0/s_1=5$]{\includegraphics[width=0.45\linewidth]{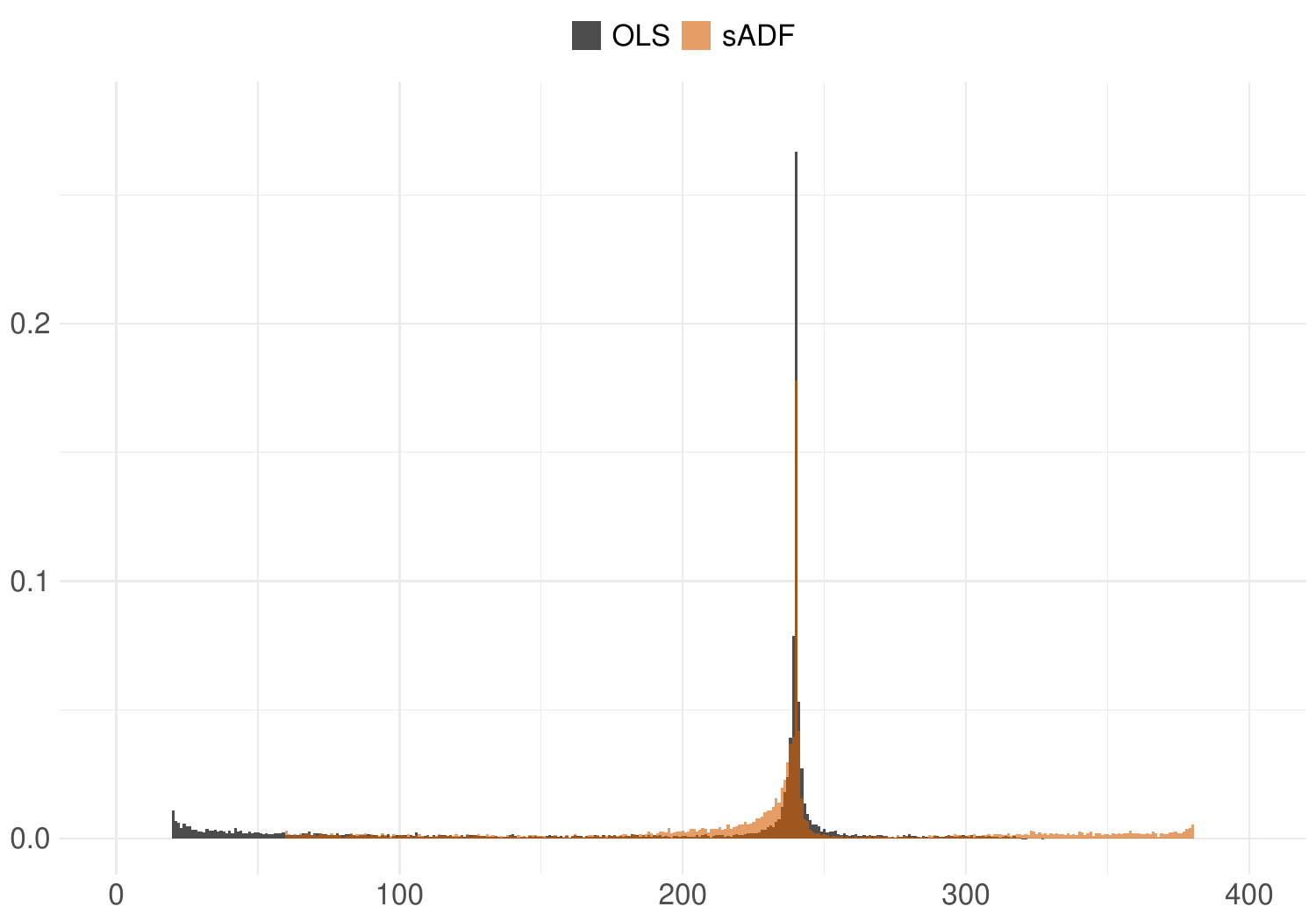}\label{fig:5:5}}
\subfigure[$T=400$, $c_a=6$, $c_b=6$, $s_0/s_1=5$]{\includegraphics[width=0.45\linewidth]{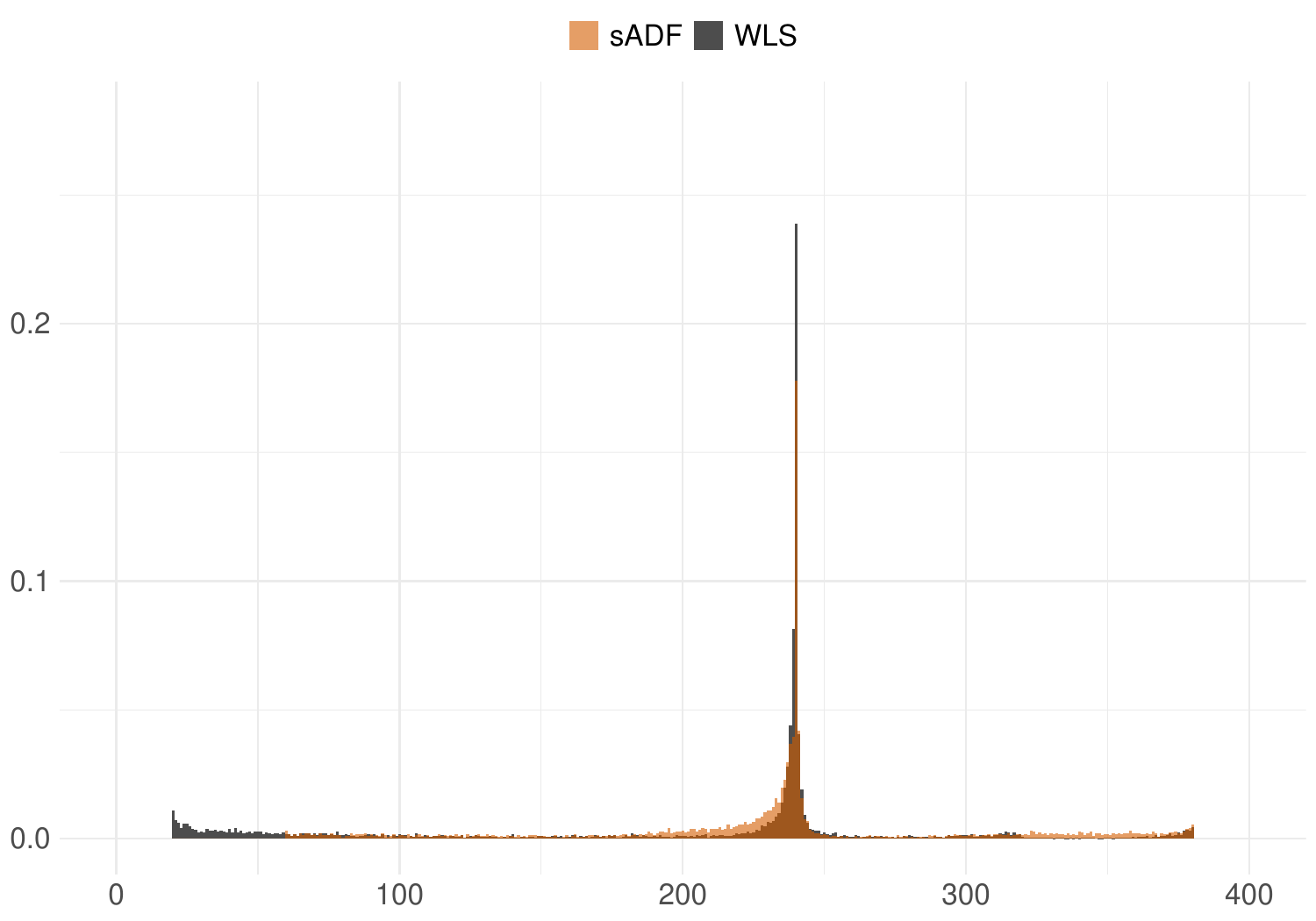}\label{fig:5:6}}\\
\end{center}%
\caption{Histograms of $\hat{k}_c$ 
for $(\tau_e,\tau_c,\tau_r)=(0.4,0.6,0.7)$,  $\tau=0.8$, $s_0/s_1=5$, $T=400$}
\label{fig5}
\end{figure}

\newpage

\begin{figure}[h!]%
\begin{center}%
\subfigure[$T=800$, $c_a=4$, $c_b=6$, $s_0/s_1=5$]{\includegraphics[width=0.45\linewidth]{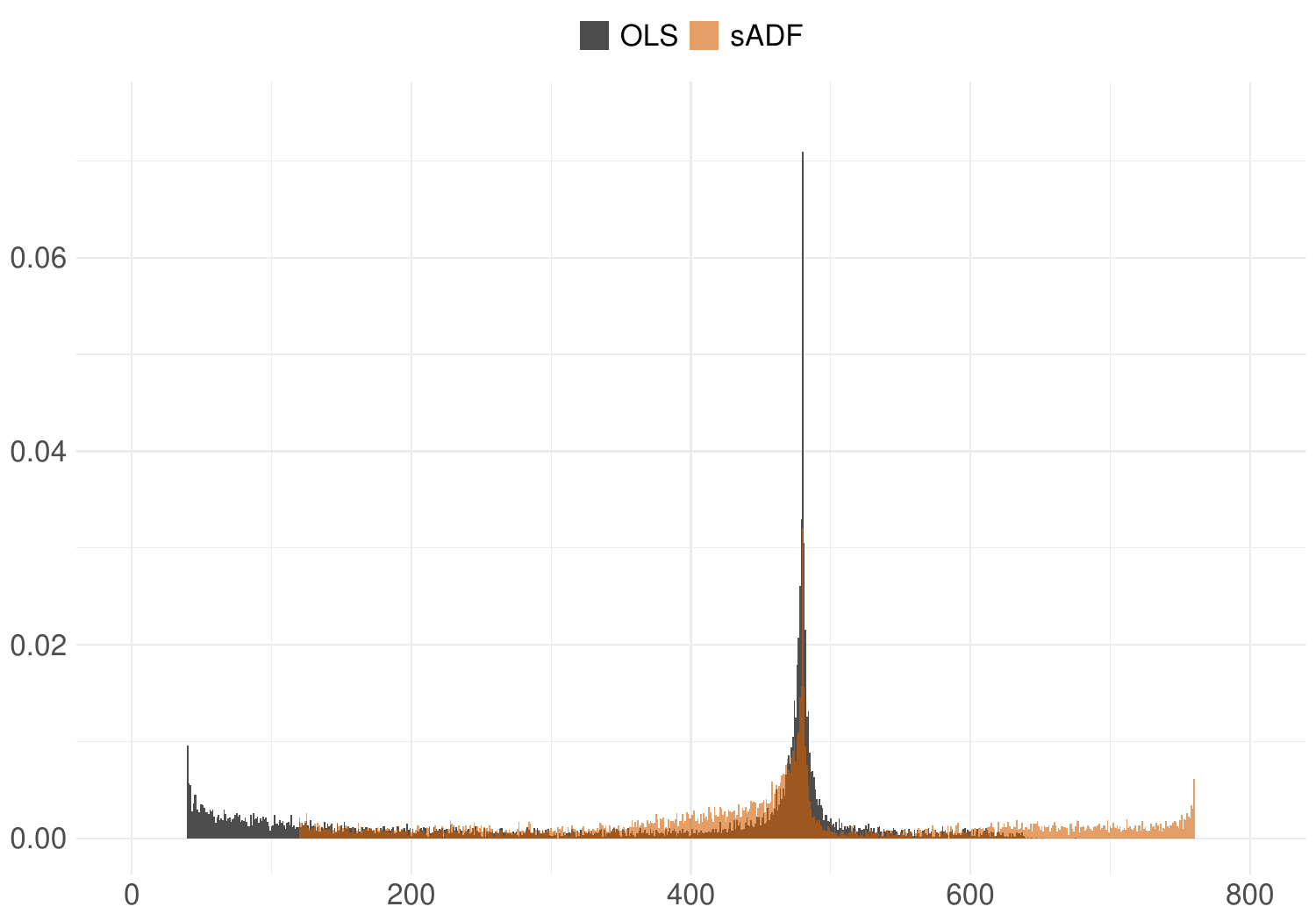}\label{fig:6:1}}
\subfigure[$T=800$, $c_a=4$, $c_b=6$, $s_0/s_1=5$]{\includegraphics[width=0.45\linewidth]{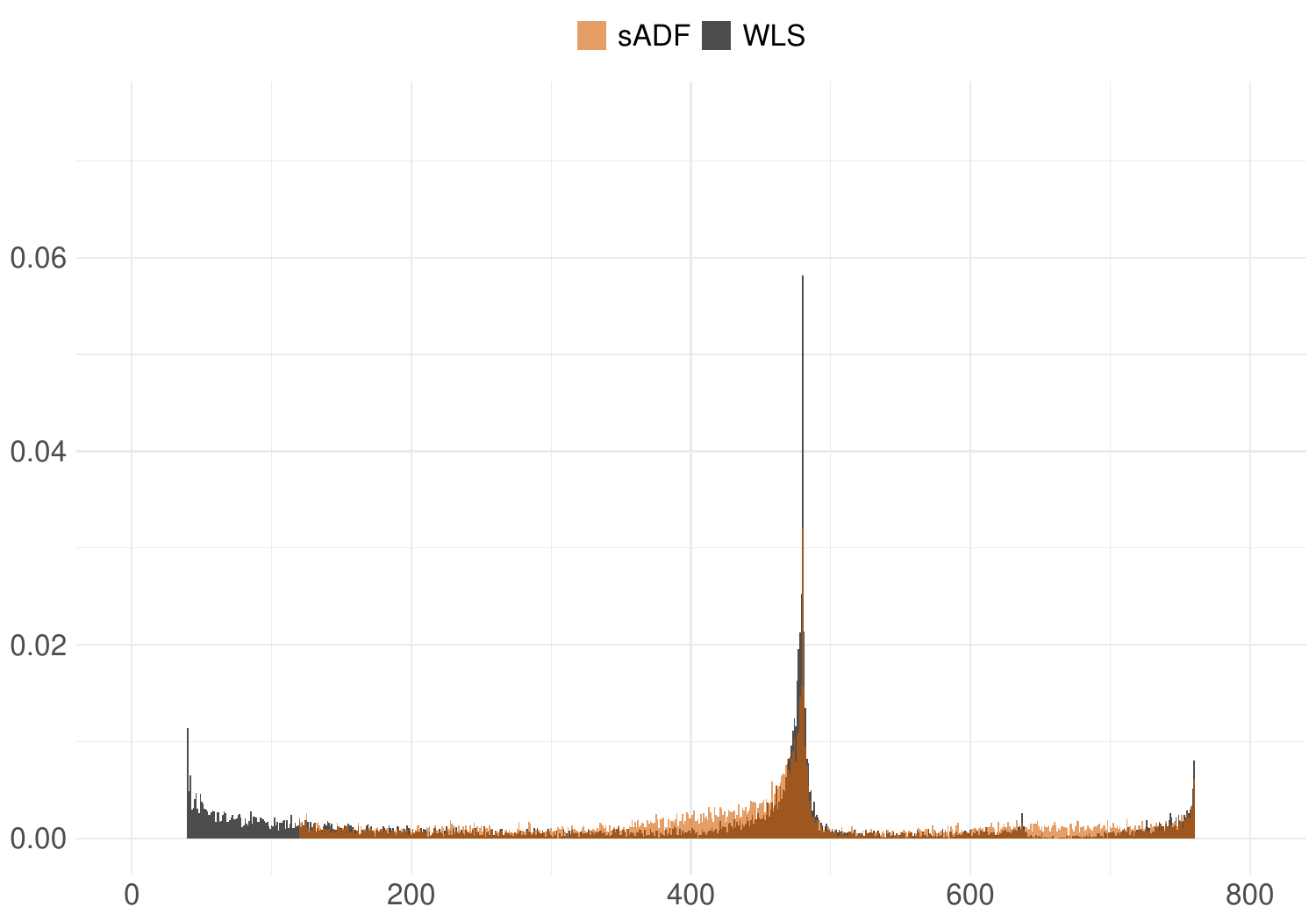}\label{fig:6:2}}\\
\subfigure[$T=800$, $c_a=5$, $c_b=6$, $s_0/s_1=5$]{\includegraphics[width=0.45\linewidth]{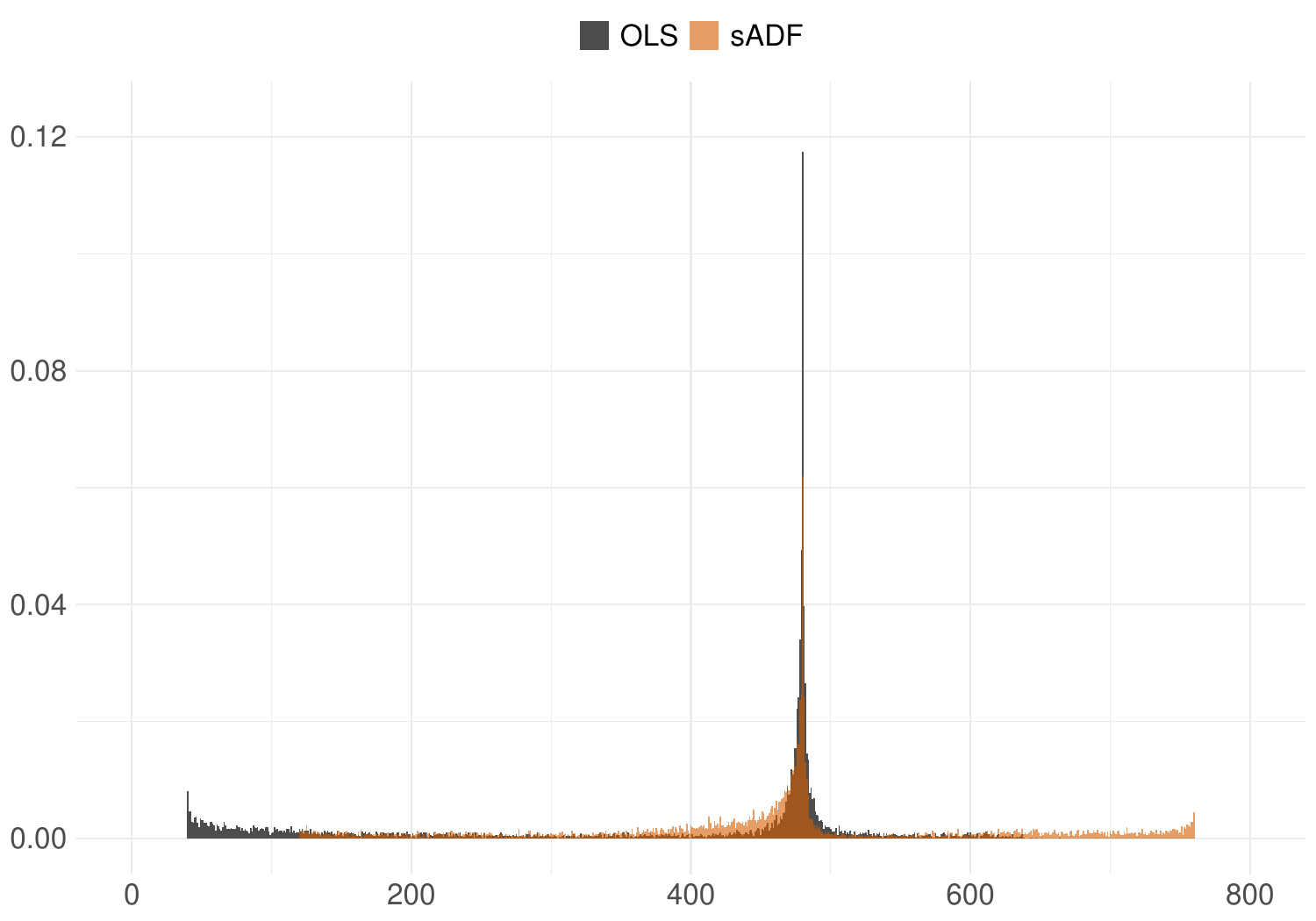}\label{fig:6:3}}
\subfigure[$T=800$, $c_a=5$, $c_b=6$, $s_0/s_1=5$]{\includegraphics[width=0.45\linewidth]{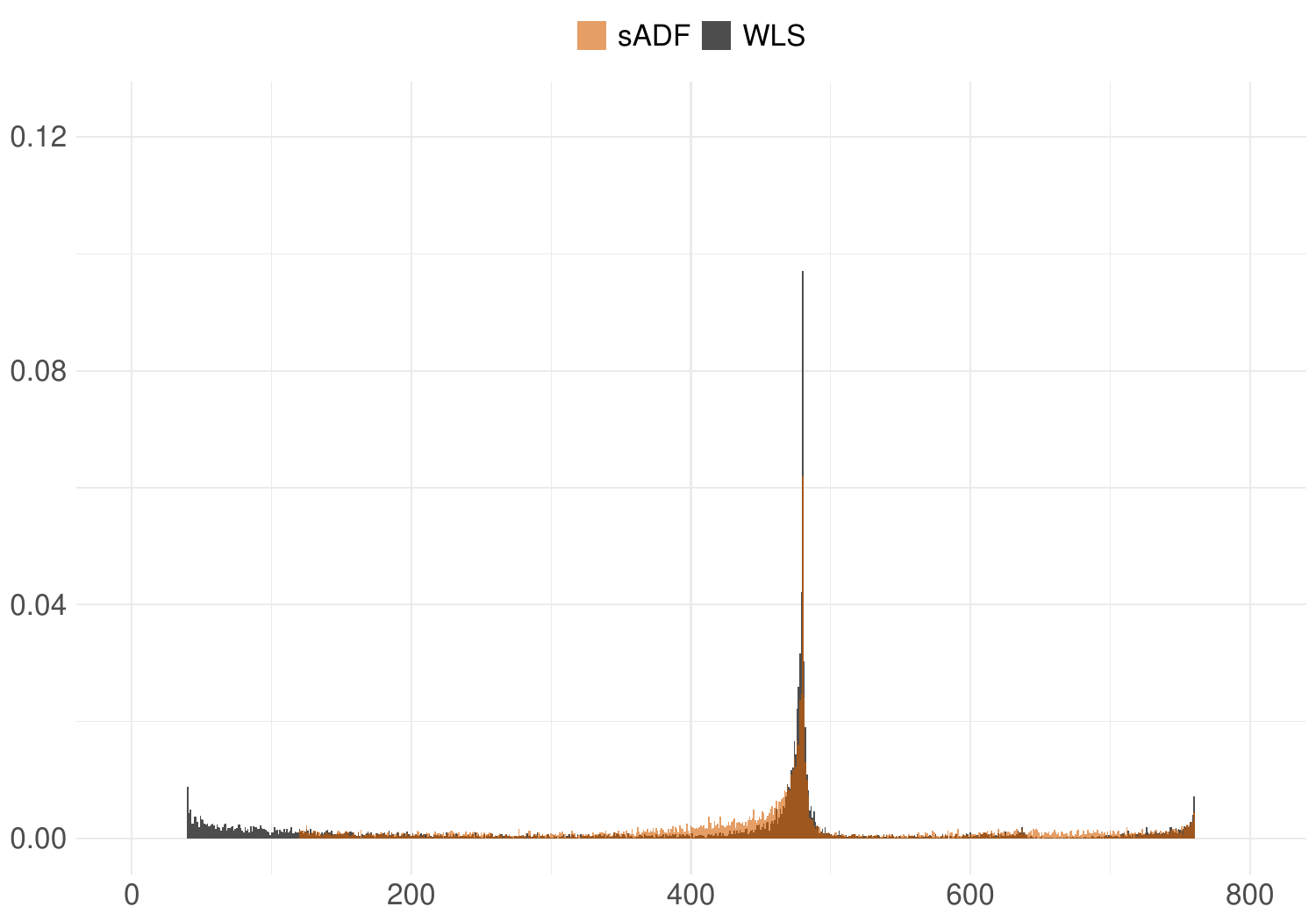}\label{fig:6:4}}\\
\subfigure[$T=800$, $c_a=6$, $c_b=6$, $s_0/s_1=5$]{\includegraphics[width=0.45\linewidth]{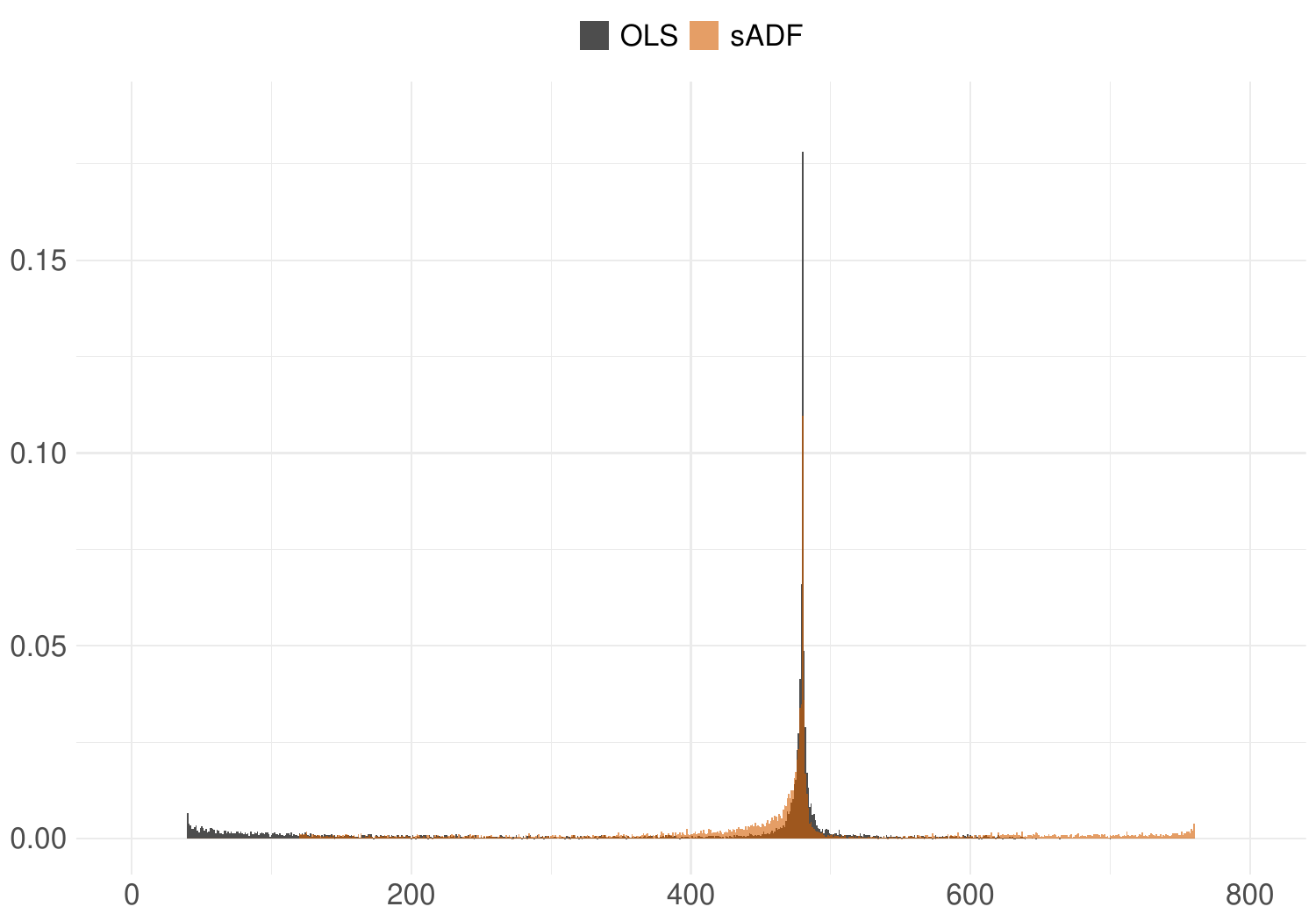}\label{fig:6:5}}
\subfigure[$T=800$, $c_a=6$, $c_b=6$, $s_0/s_1=5$]{\includegraphics[width=0.45\linewidth]{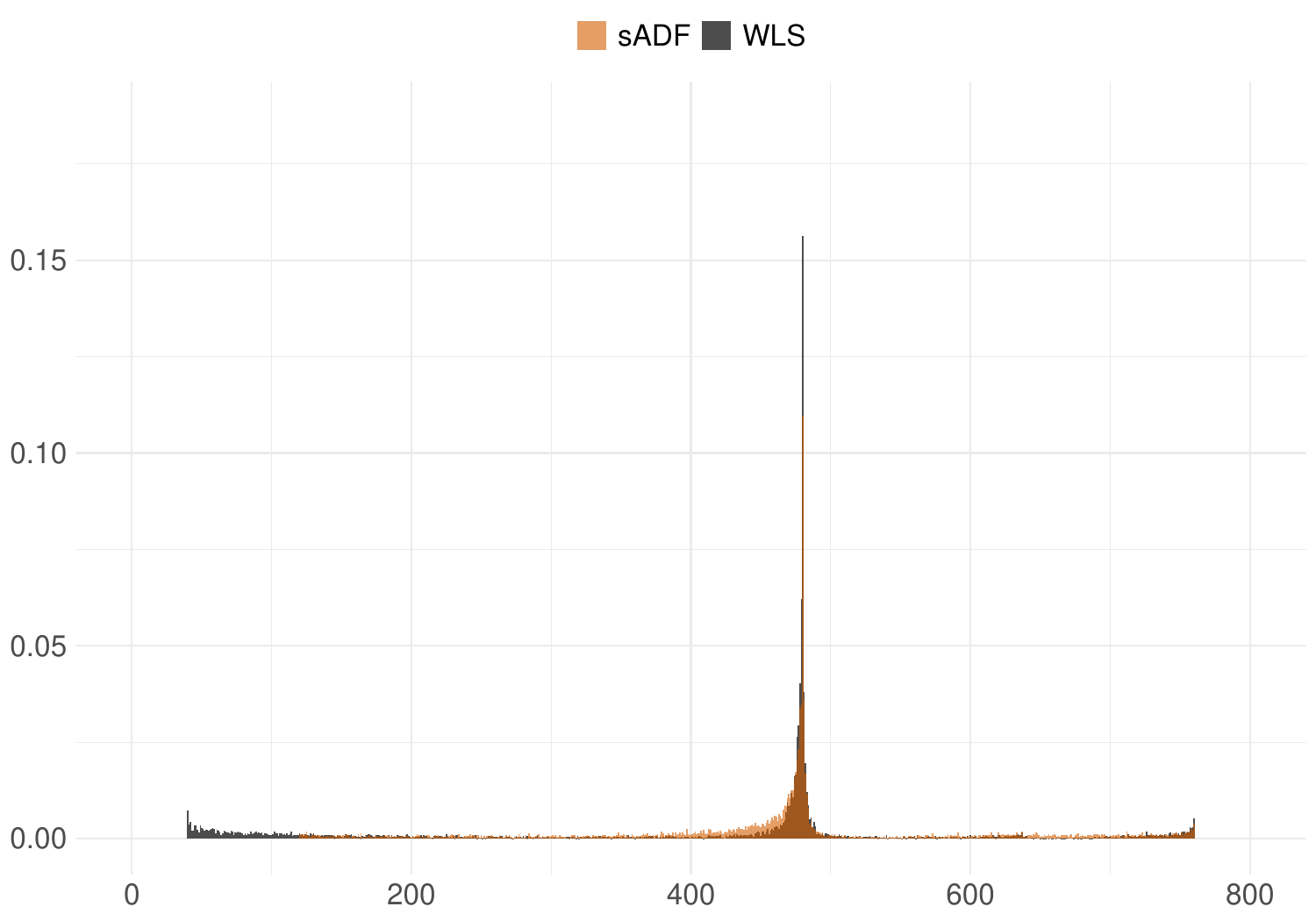}\label{fig:6:6}}\\
\end{center}%
\caption{Histograms of $\hat{k}_c$ 
for $(\tau_e,\tau_c,\tau_r)=(0.4,0.6,0.7)$,  $\tau=0.8$, $s_0/s_1=5$, $T=800$}
\label{fig6}
\end{figure}

\section{$\tau=0.8$, $\hat{k}_e$}
\setcounter{figure}{0}

\begin{figure}[h!]%
\begin{center}%
\subfigure[$T=400$, $c_a=4$, $c_b=6$, $s_0/s_1=1/5$]{\includegraphics[width=0.45\linewidth]{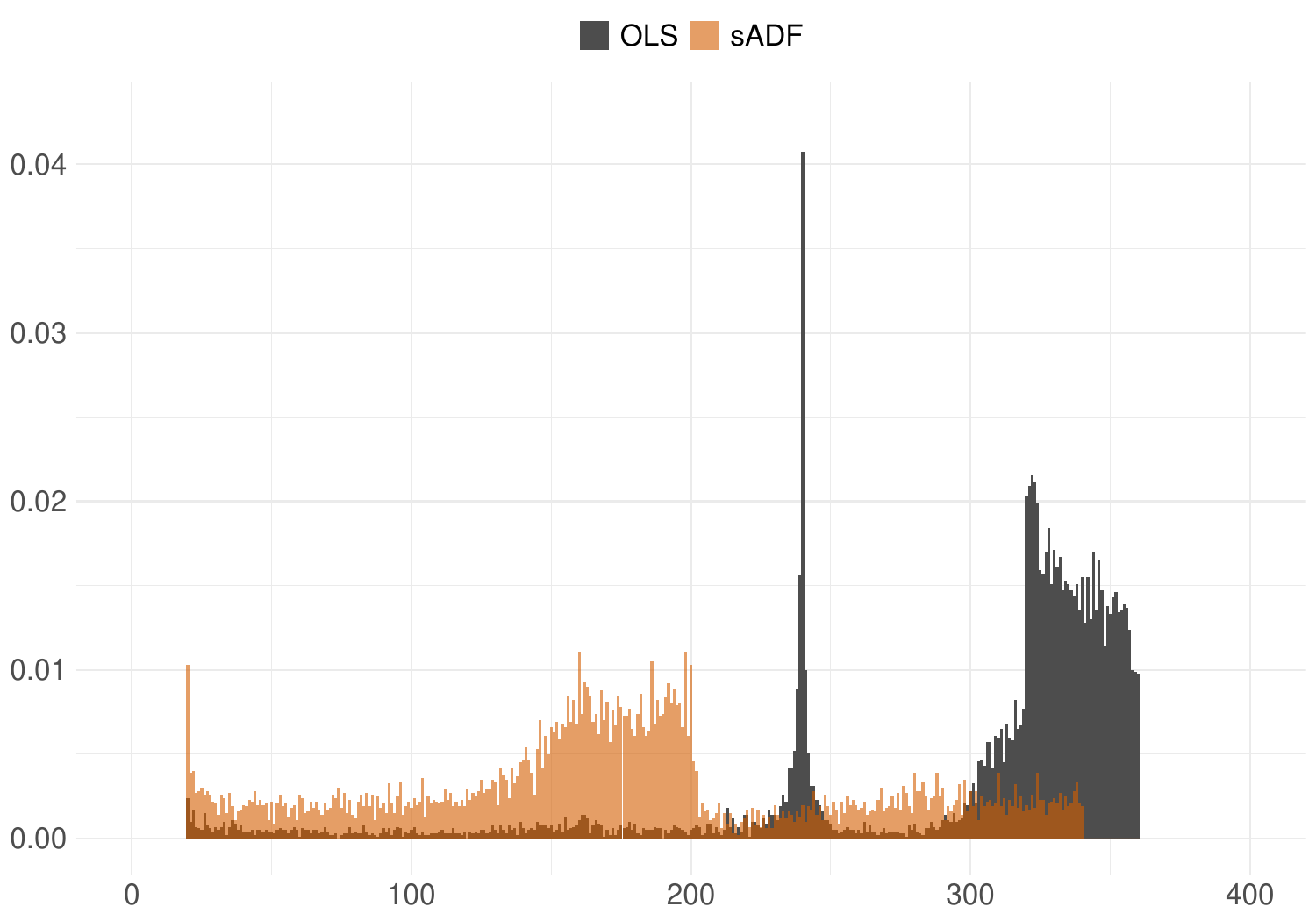}\label{fig:1e:1}}
\subfigure[$T=400$, $c_a=4$, $c_b=6$, $s_0/s_1=1/5$]{\includegraphics[width=0.45\linewidth]{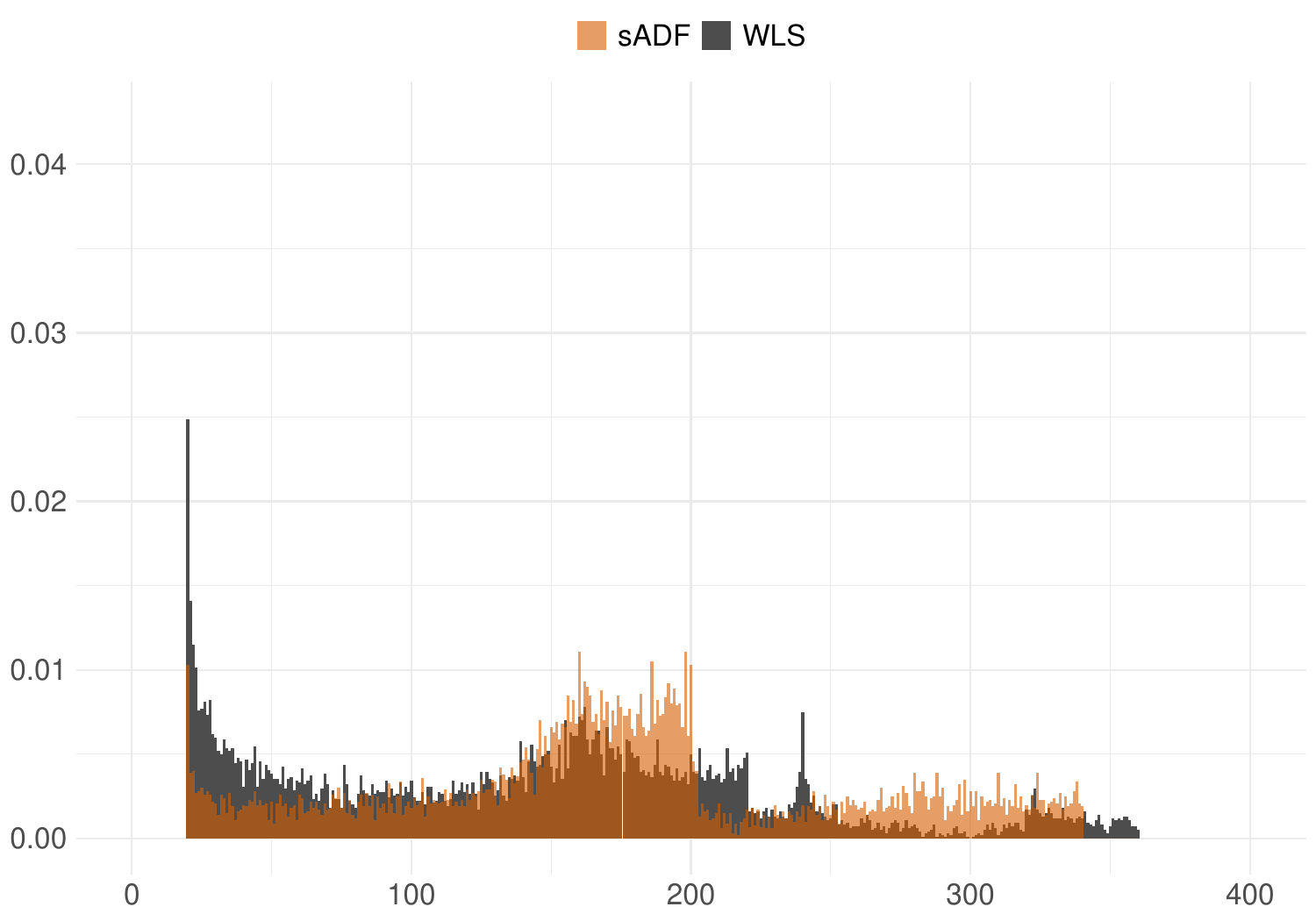}\label{fig:1e:2}}\\
\subfigure[$T=400$, $c_a=5$, $c_b=6$, $s_0/s_1=1/5$]{\includegraphics[width=0.45\linewidth]{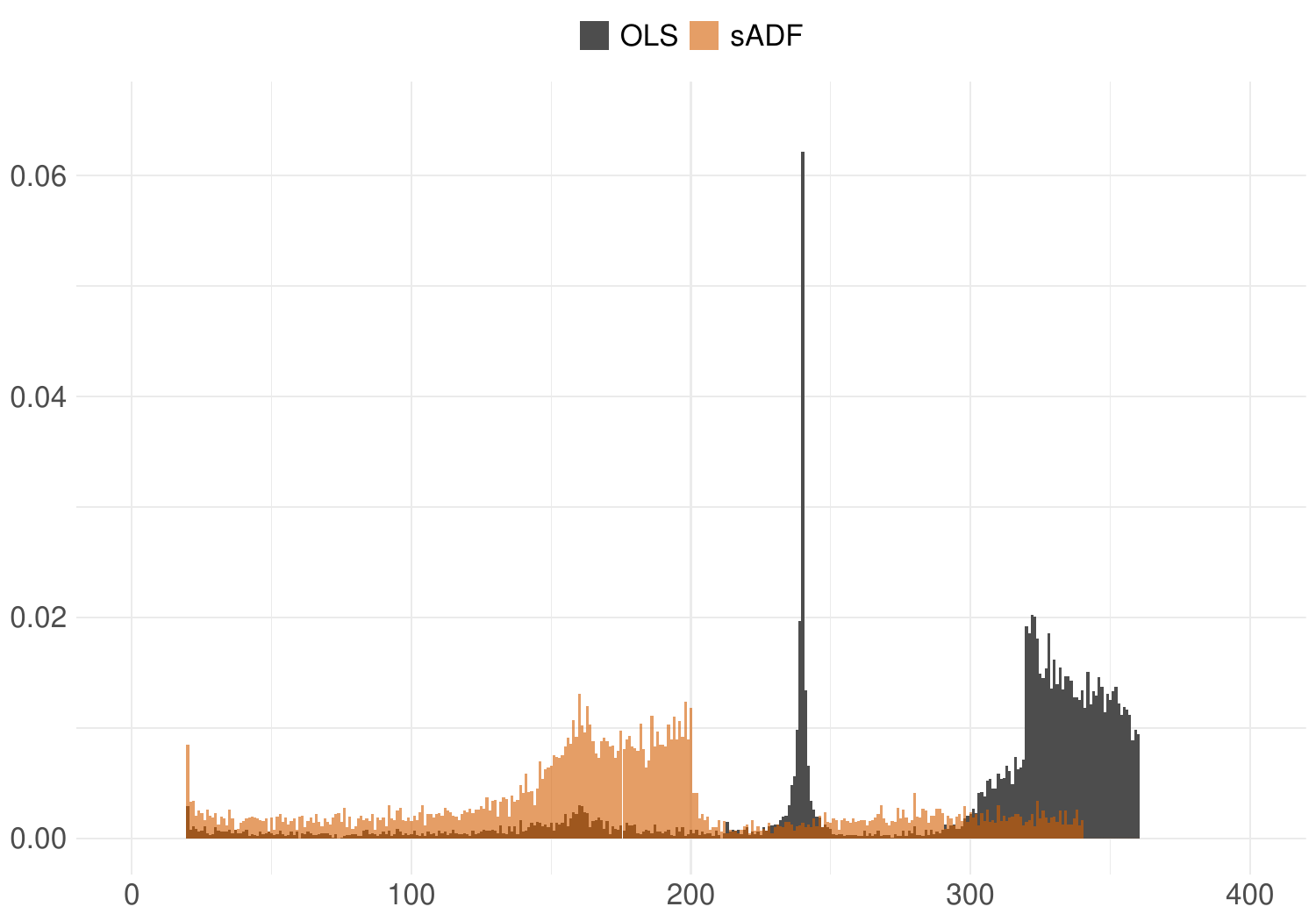}\label{fig:1e:3}}
\subfigure[$T=400$, $c_a=5$, $c_b=6$, $s_0/s_1=1/5$]{\includegraphics[width=0.45\linewidth]{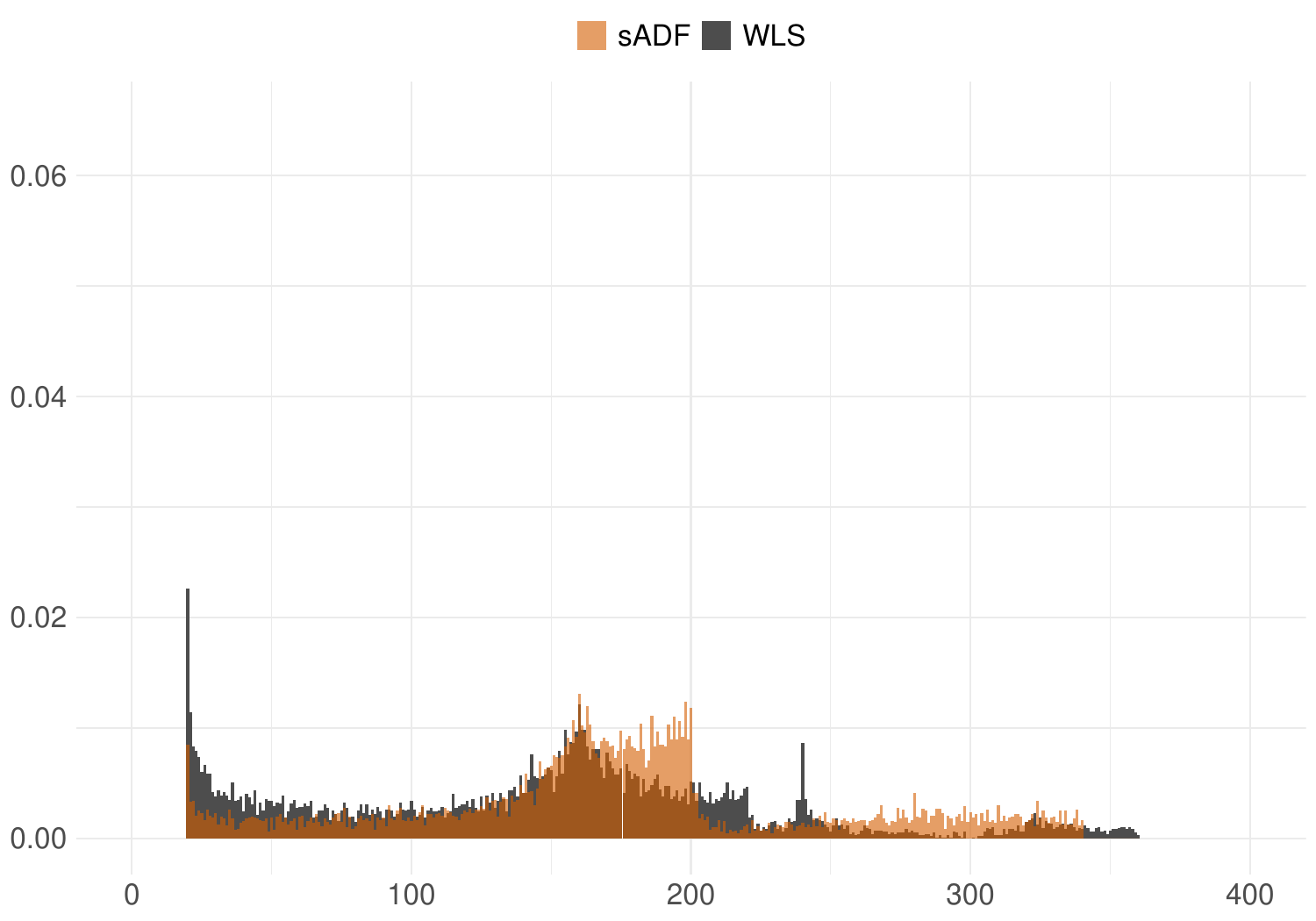}\label{fig:1e:4}}\\
\subfigure[$T=400$, $c_a=6$, $c_b=6$, $s_0/s_1=1/5$]{\includegraphics[width=0.45\linewidth]{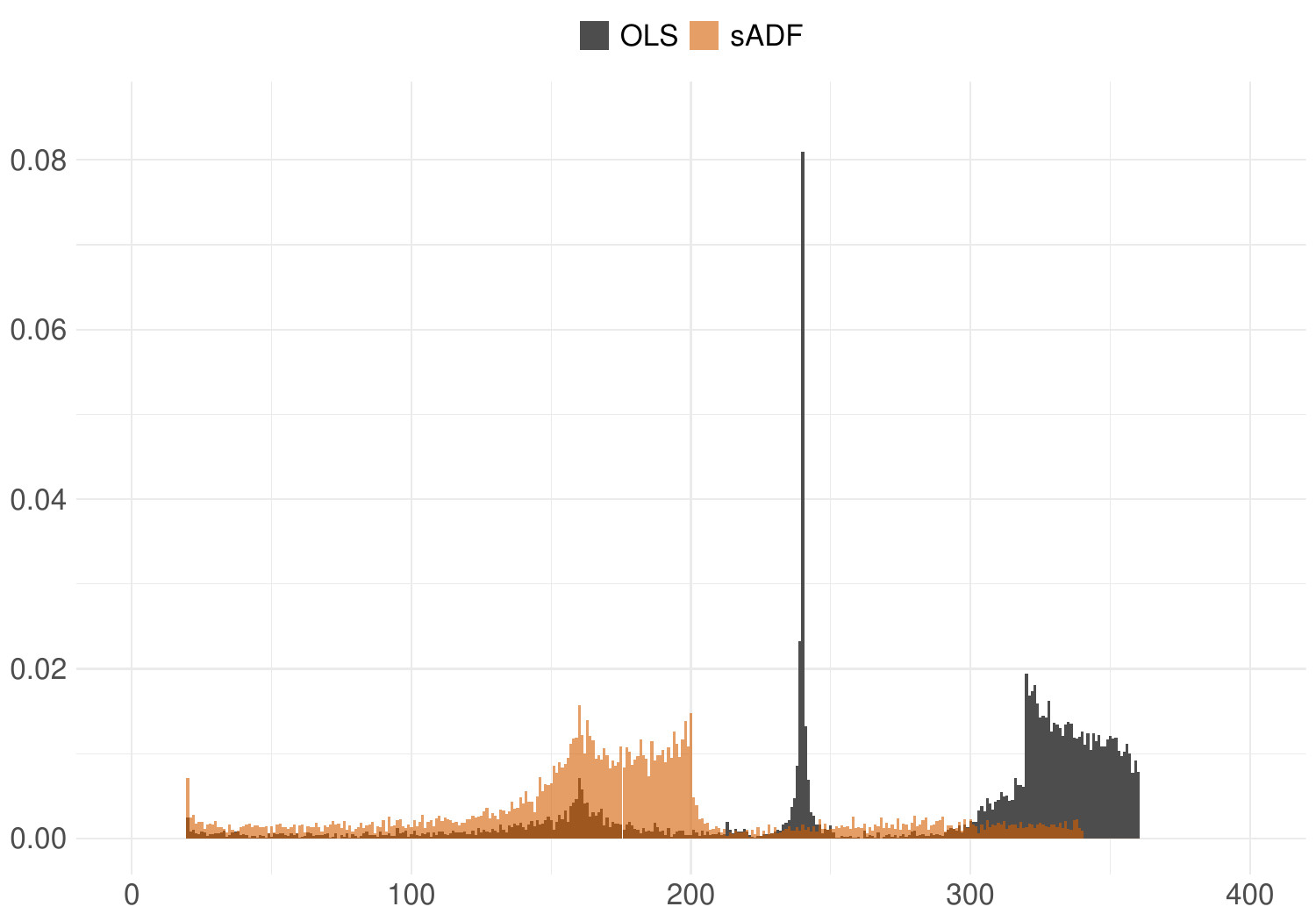}\label{fig:1e:5}}
\subfigure[$T=400$, $c_a=6$, $c_b=6$, $s_0/s_1=1/5$]{\includegraphics[width=0.45\linewidth]{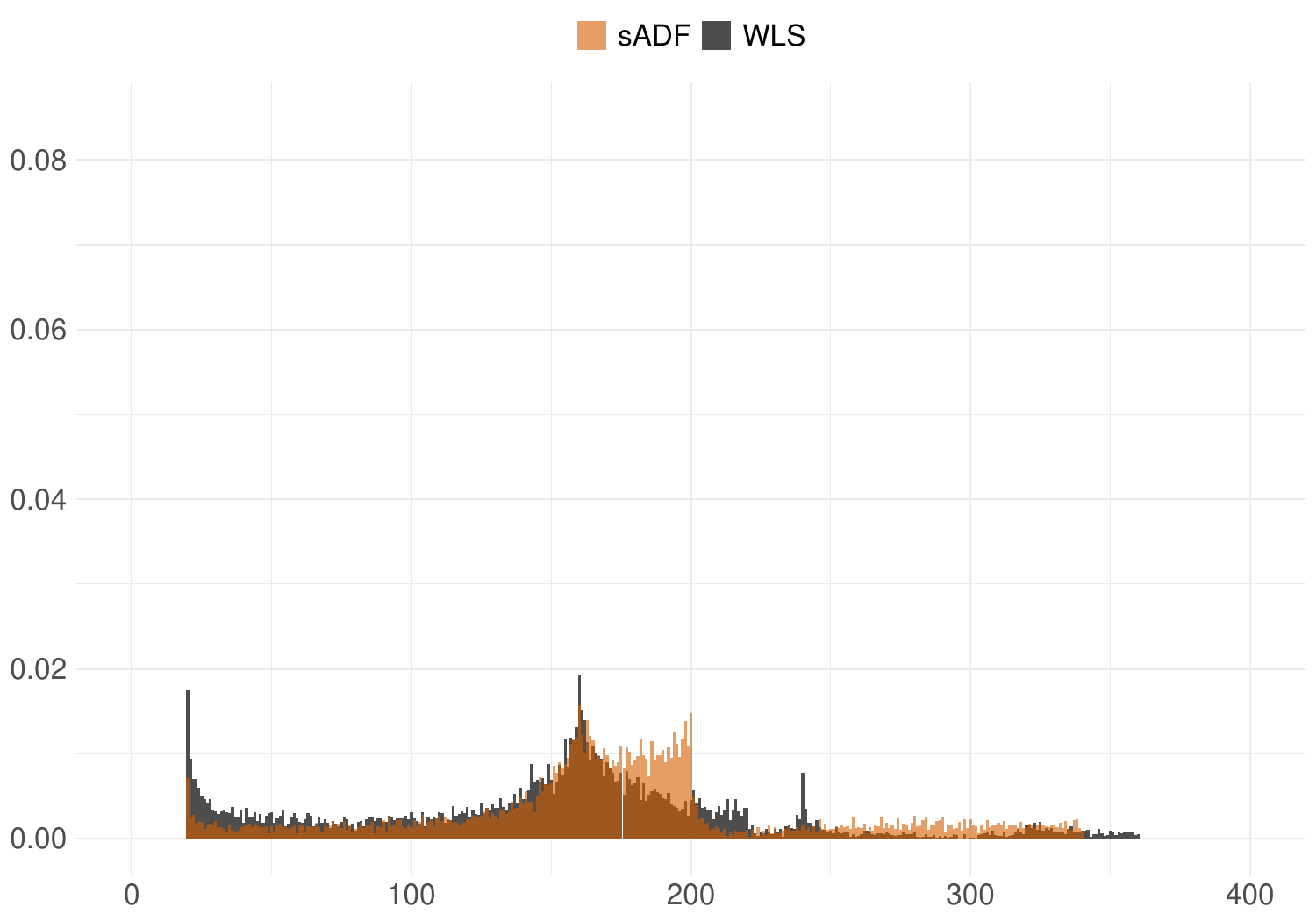}\label{fig:1e:6}}\\
\end{center}%
\caption{Histograms of $\hat{k}_e$ 
for $(\tau_e,\tau_c,\tau_r)=(0.4,0.6,0.7)$,  $\tau=0.8$, $s_0/s_1=1/5$, $T=400$}
\label{fig1_e}
\end{figure}

\newpage

\begin{figure}[h!]%
\begin{center}%
\subfigure[$T=800$, $c_a=4$, $c_b=6$, $s_0/s_1=1/5$]{\includegraphics[width=0.45\linewidth]{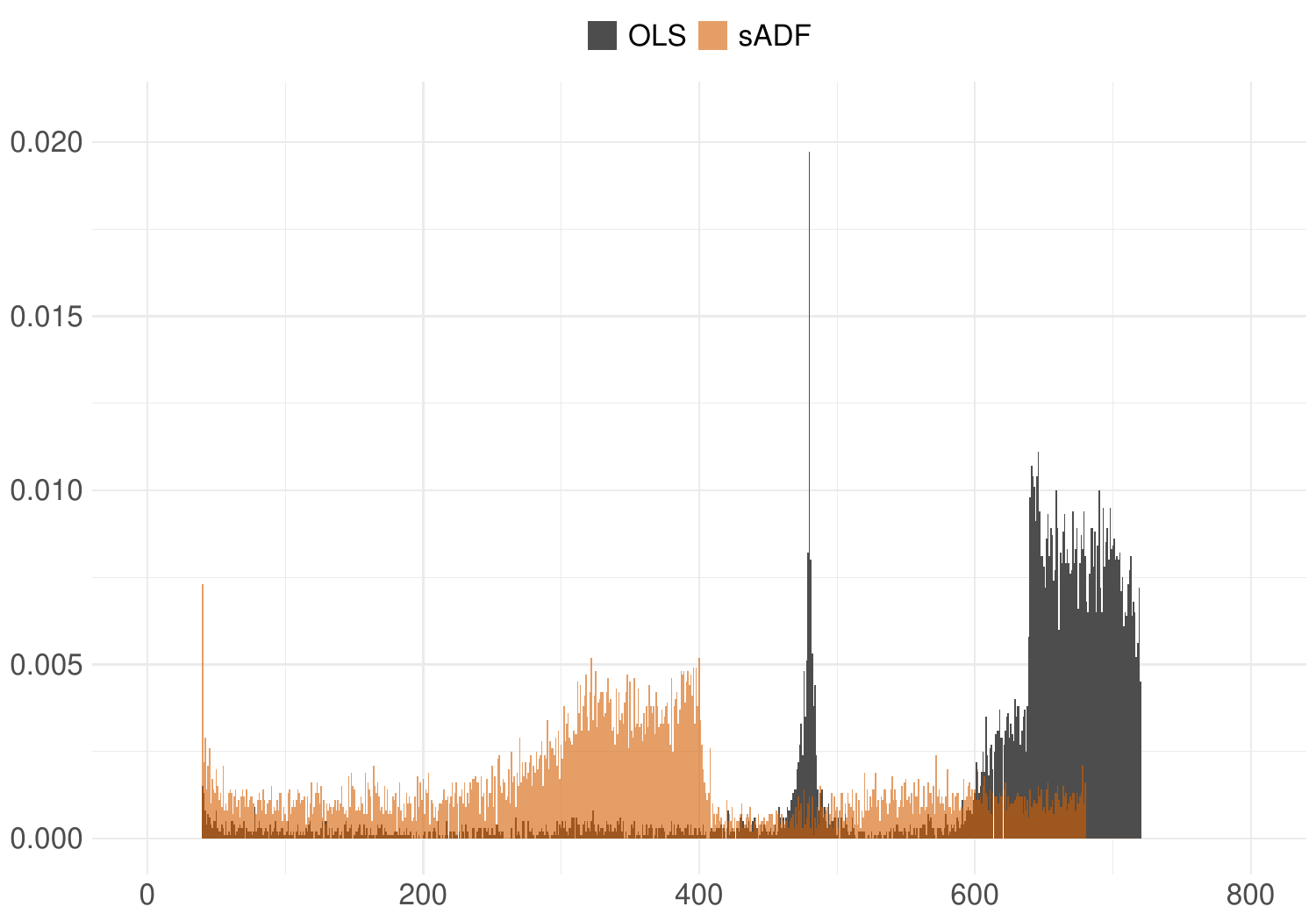}\label{fig:2e:1}}
\subfigure[$T=800$, $c_a=4$, $c_b=6$, $s_0/s_1=1/5$]{\includegraphics[width=0.45\linewidth]{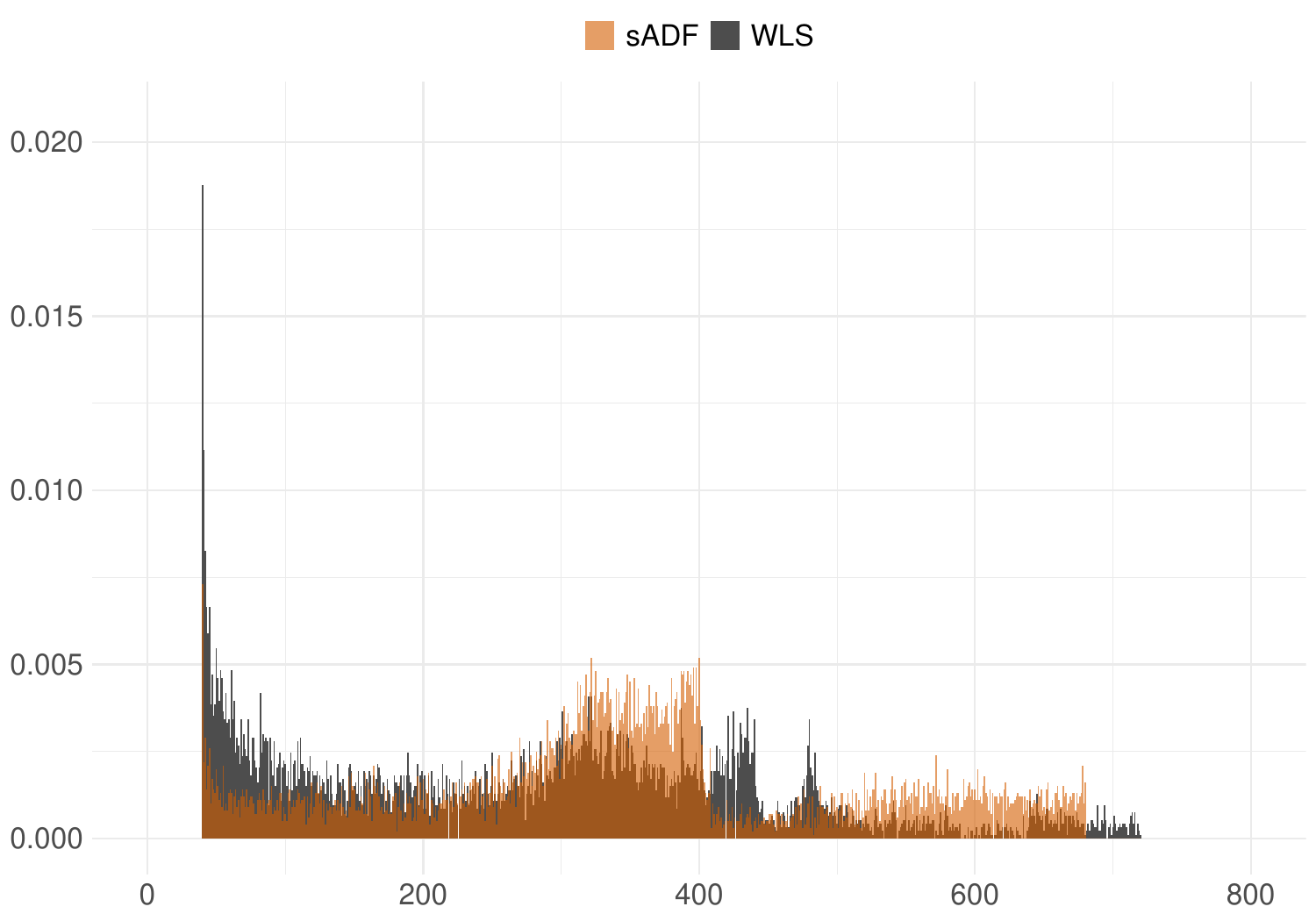}\label{fig:2e:2}}\\
\subfigure[$T=800$, $c_a=5$, $c_b=6$, $s_0/s_1=1/5$]{\includegraphics[width=0.45\linewidth]{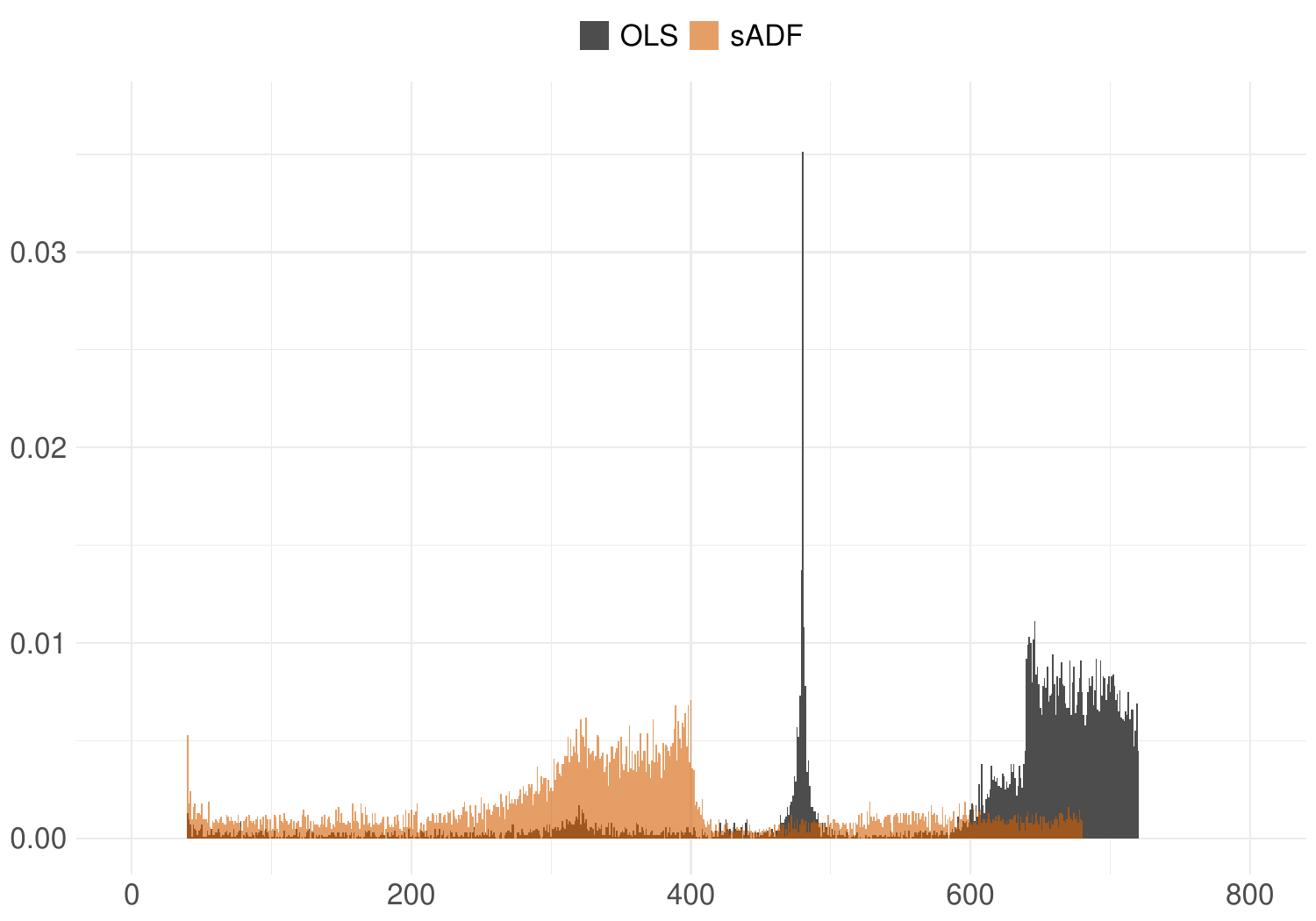}\label{fig:2e:3}}
\subfigure[$T=800$, $c_a=5$, $c_b=6$, $s_0/s_1=1/5$]{\includegraphics[width=0.45\linewidth]{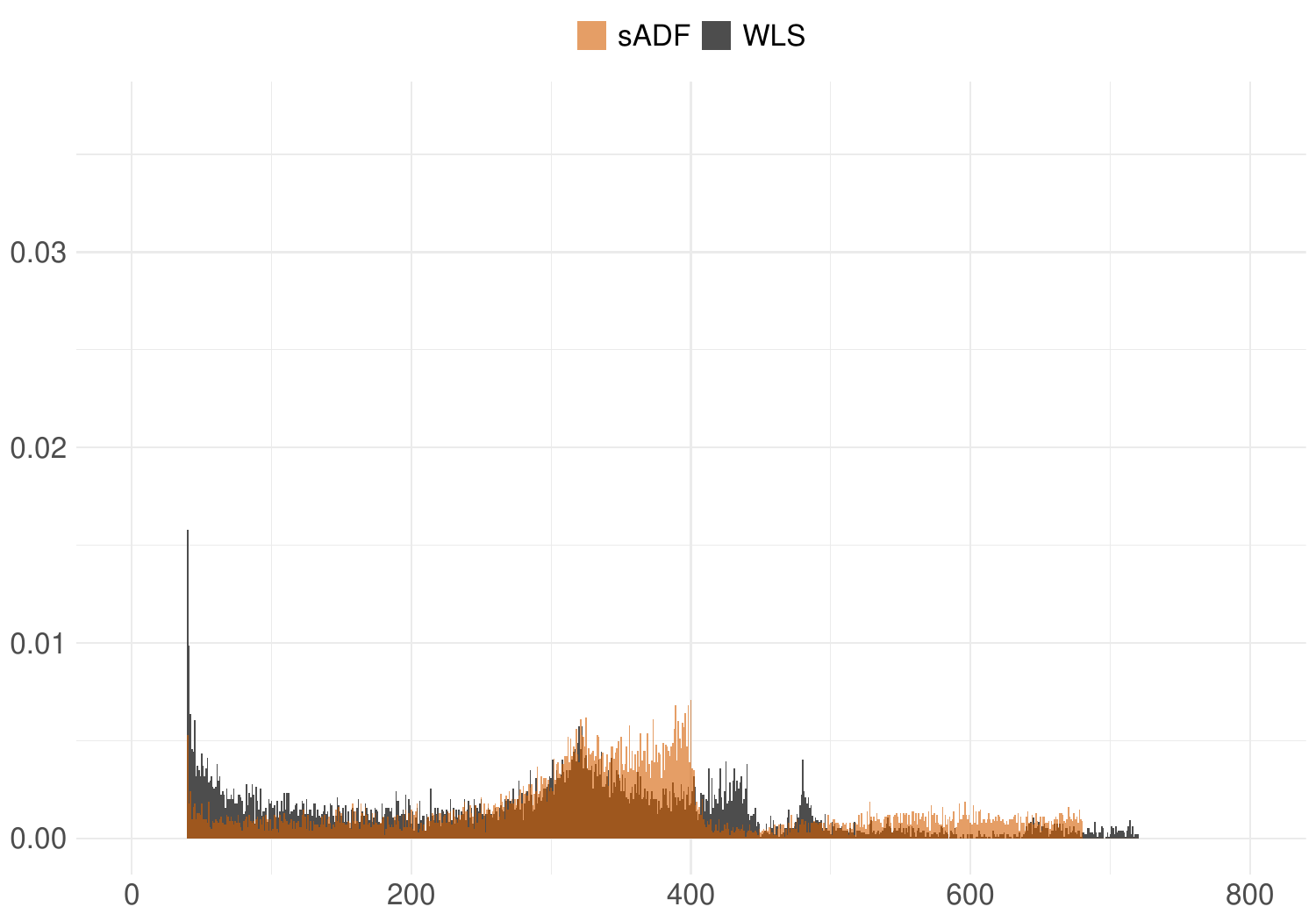}\label{fig:2e:4}}\\
\subfigure[$T=800$, $c_a=6$, $c_b=6$, $s_0/s_1=1/5$]{\includegraphics[width=0.45\linewidth]{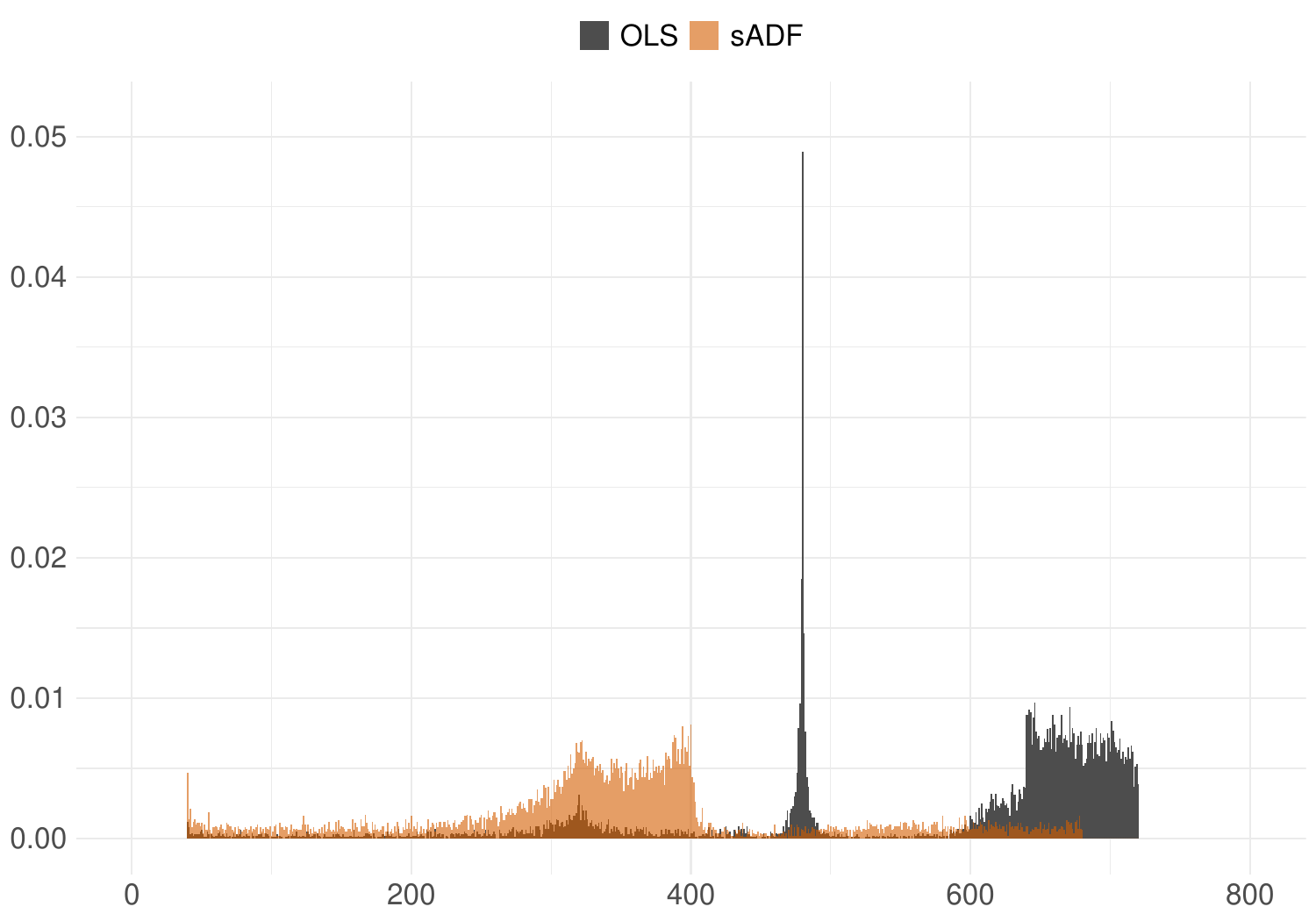}\label{fig:2e:5}}
\subfigure[$T=800$, $c_a=6$, $c_b=6$, $s_0/s_1=1/5$]{\includegraphics[width=0.45\linewidth]{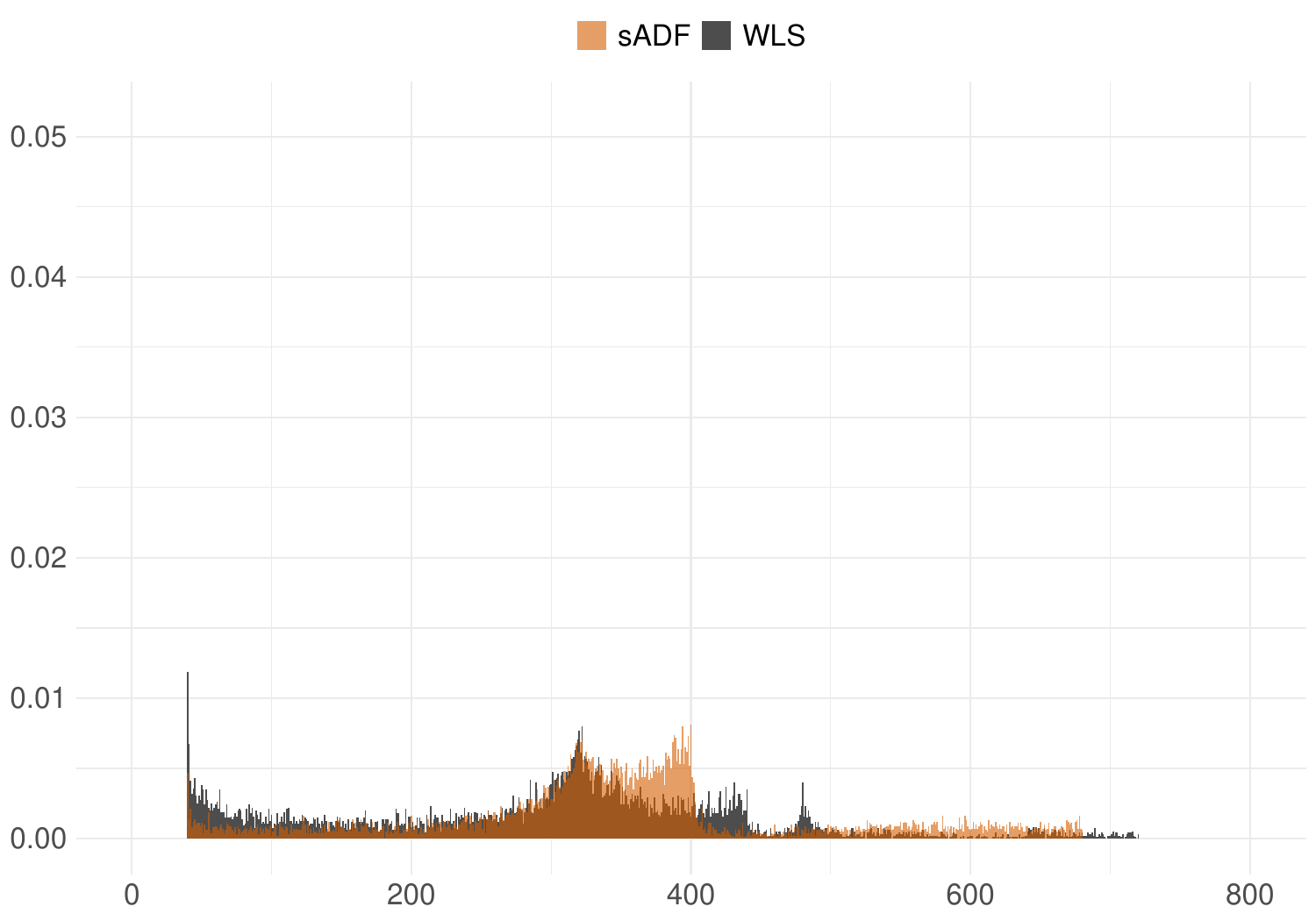}\label{fig:2e:6}}\\
\end{center}%
\caption{Histograms of $\hat{k}_e$ 
for $(\tau_e,\tau_c,\tau_r)=(0.4,0.6,0.7)$,  $\tau=0.8$, $s_0/s_1=1/5$, $T=800$}
\label{fig2_e}
\end{figure}

\newpage

\begin{figure}[h!]%
\begin{center}%
\subfigure[$T=400$, $c_a=4$, $c_b=6$, $s_0/s_1=1$]{\includegraphics[width=0.45\linewidth]{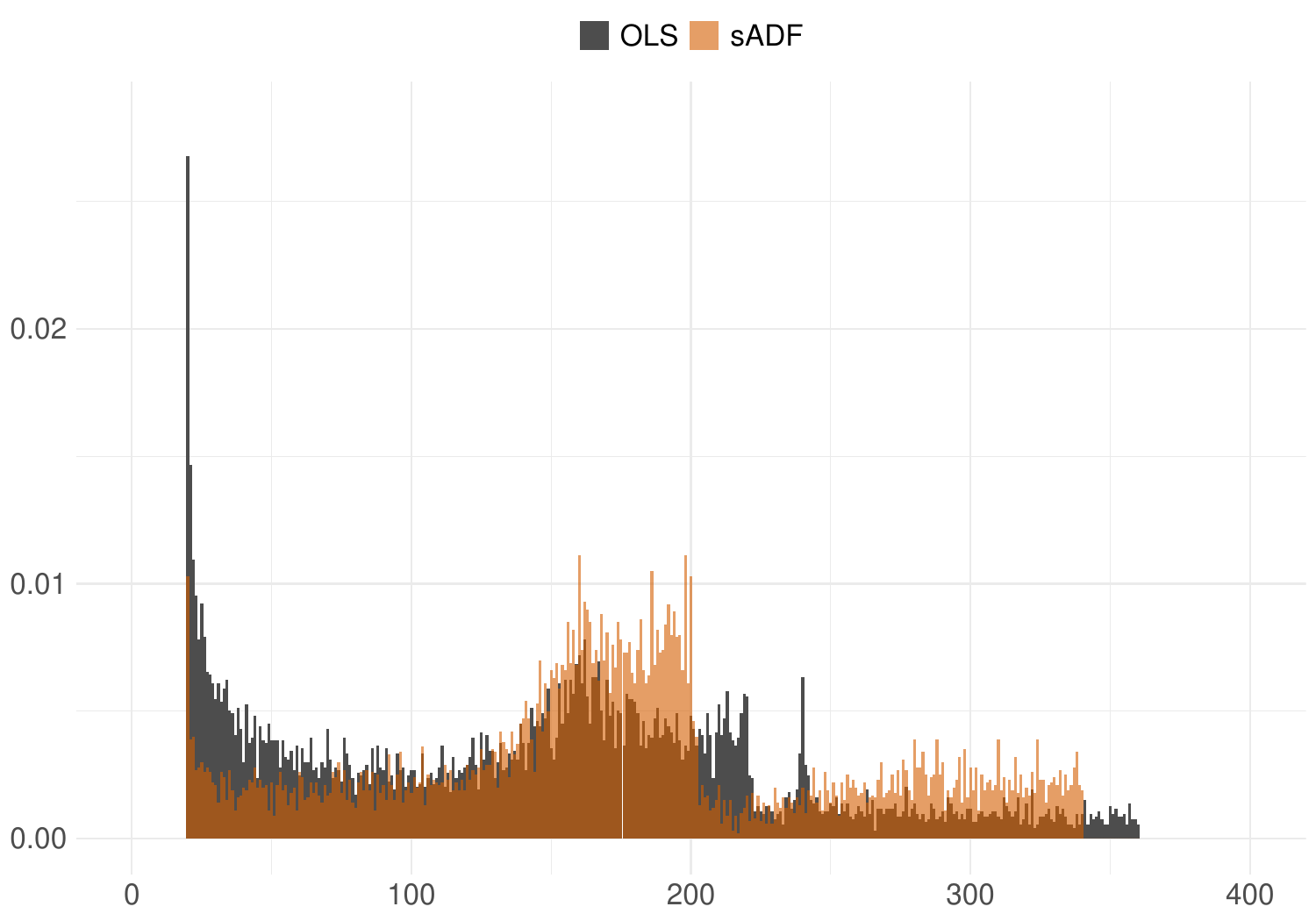}\label{fig:3e:1}}
\subfigure[$T=400$, $c_a=4$, $c_b=6$, $s_0/s_1=1$]{\includegraphics[width=0.45\linewidth]{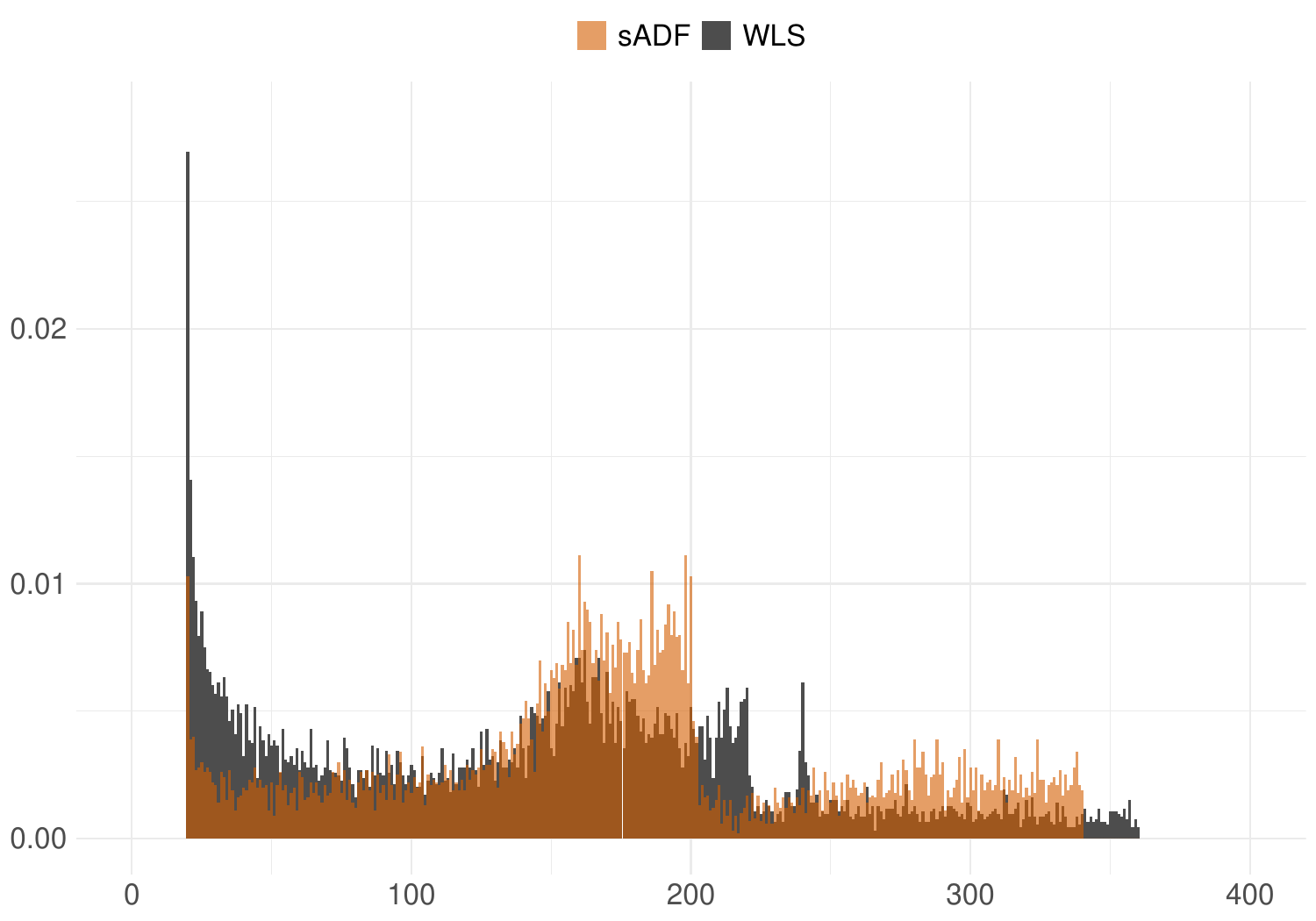}\label{fig:3e:2}}\\
\subfigure[$T=400$, $c_a=5$, $c_b=6$, $s_0/s_1=1$]{\includegraphics[width=0.45\linewidth]{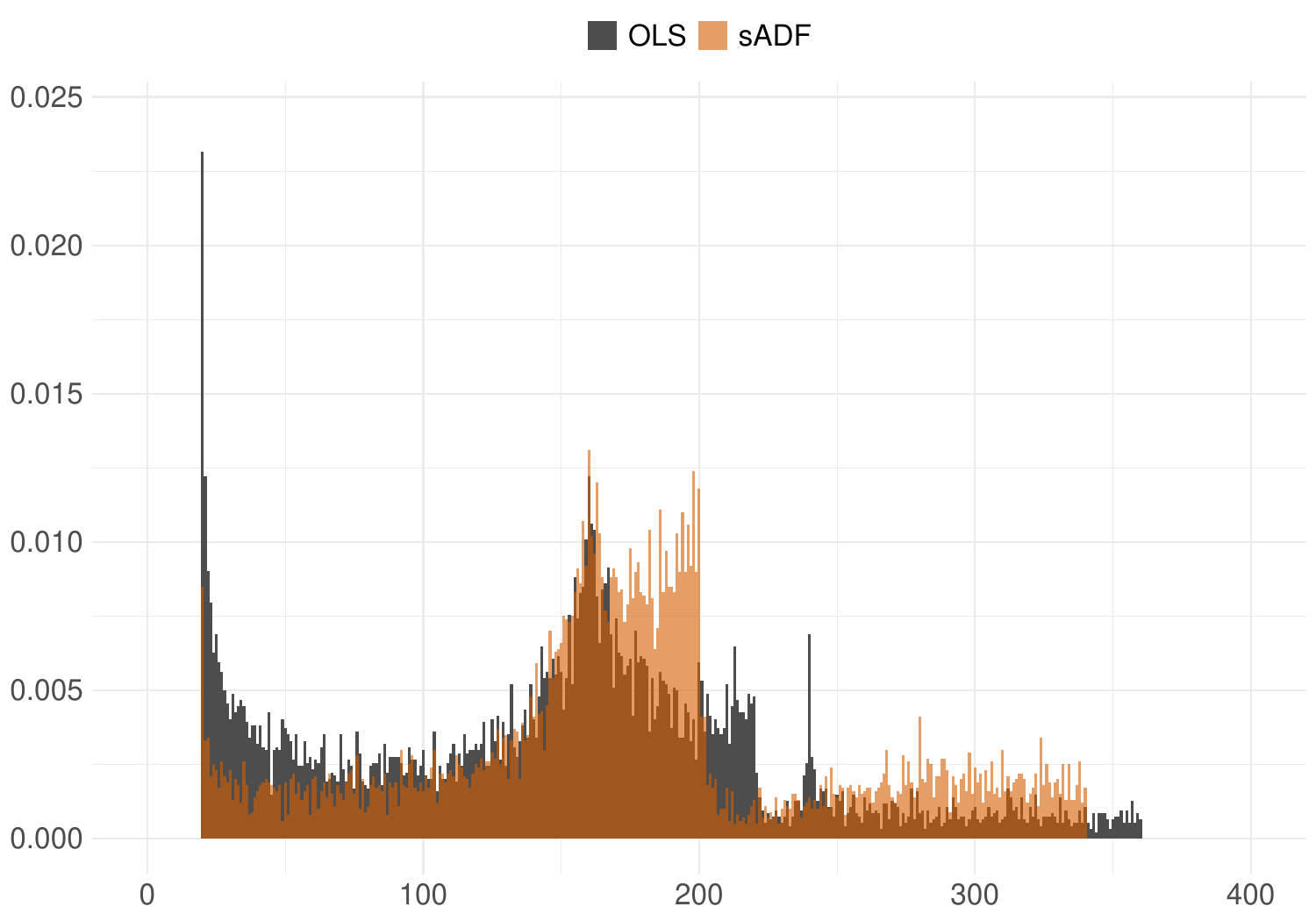}\label{fig:3e:3}}
\subfigure[$T=400$, $c_a=5$, $c_b=6$, $s_0/s_1=1$]{\includegraphics[width=0.45\linewidth]{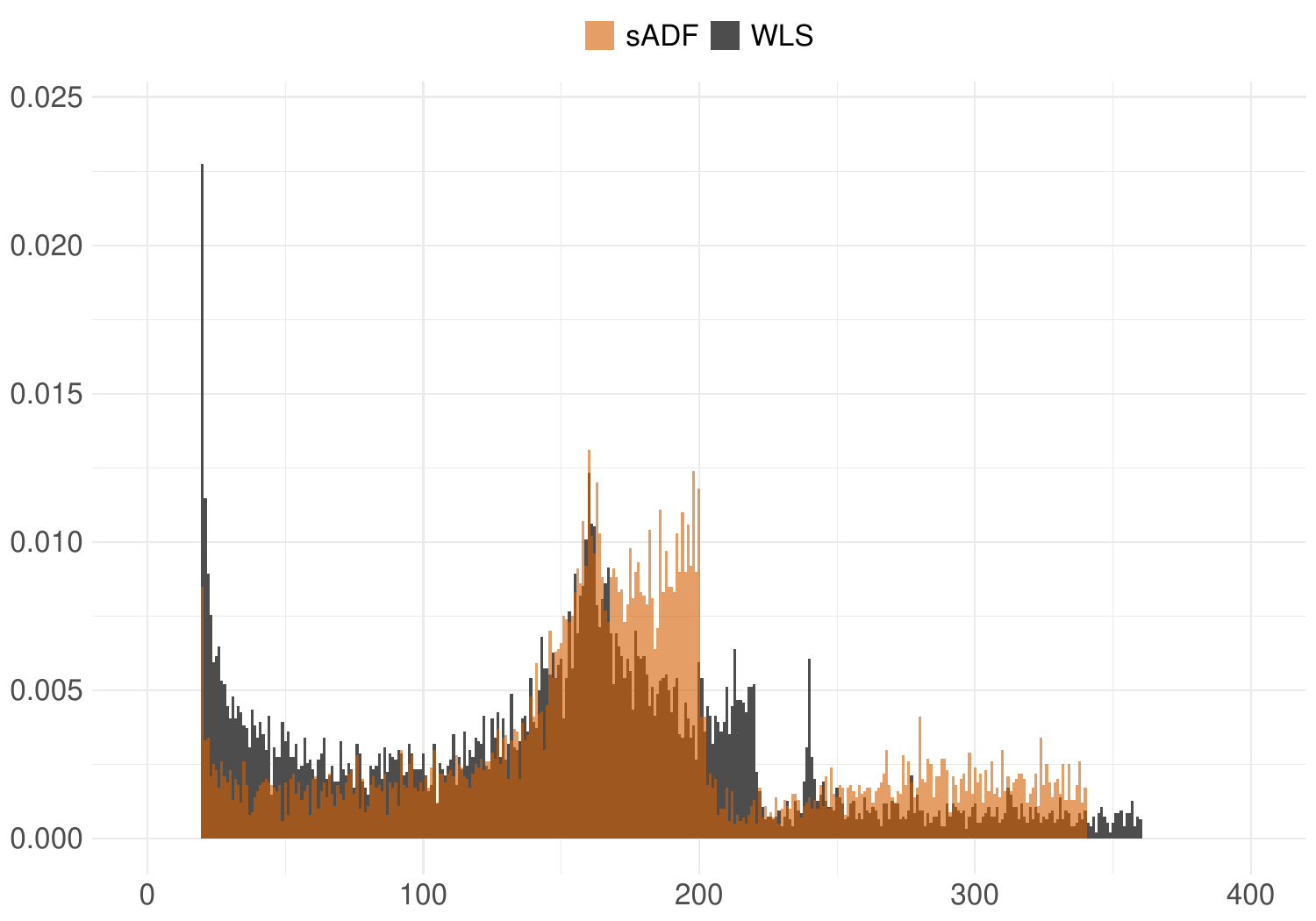}\label{fig:3e:4}}\\
\subfigure[$T=400$, $c_a=6$, $c_b=6$, $s_0/s_1=1$]{\includegraphics[width=0.45\linewidth]{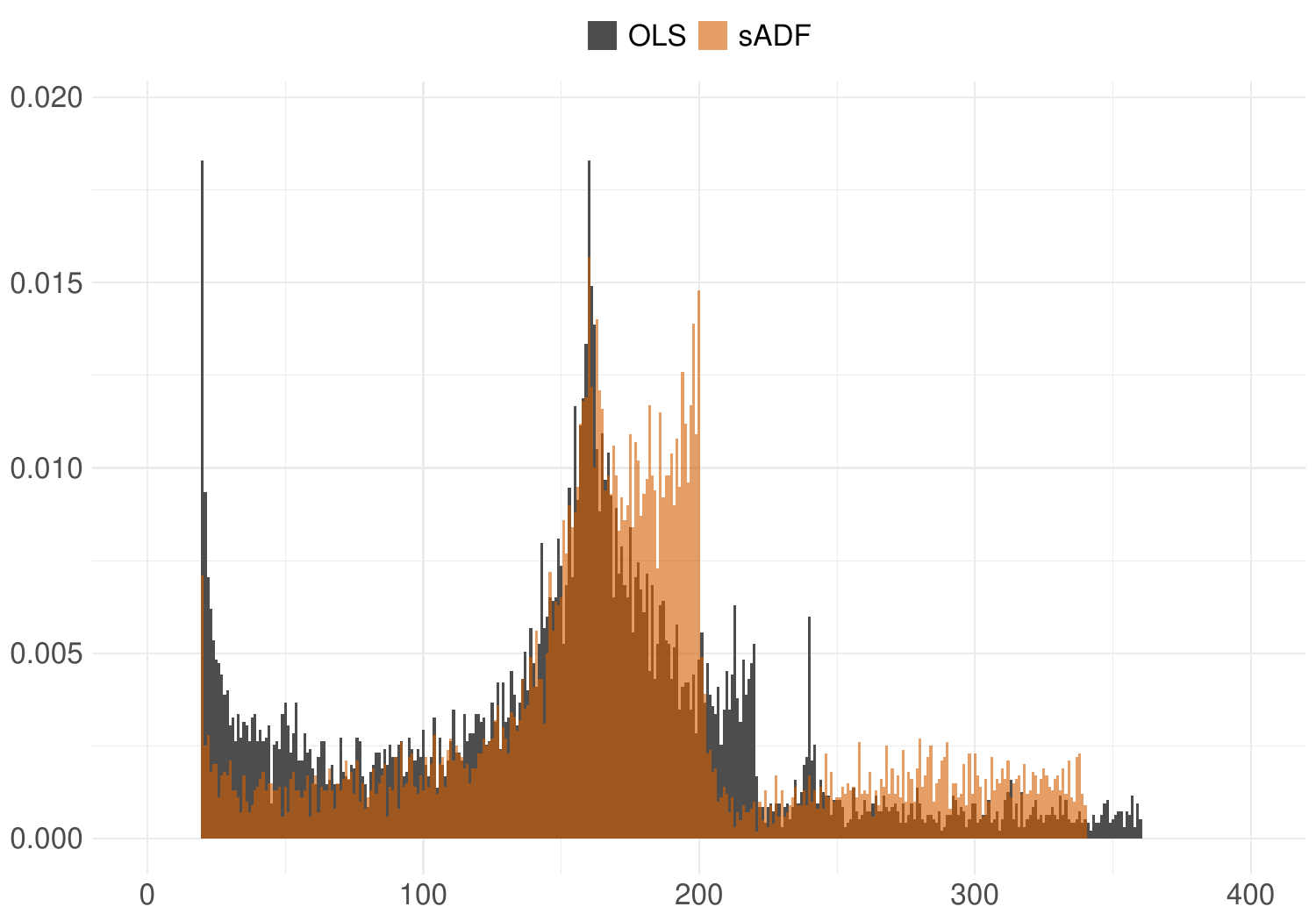}\label{fig:3e:5}}
\subfigure[$T=400$, $c_a=6$, $c_b=6$, $s_0/s_1=1$]{\includegraphics[width=0.45\linewidth]{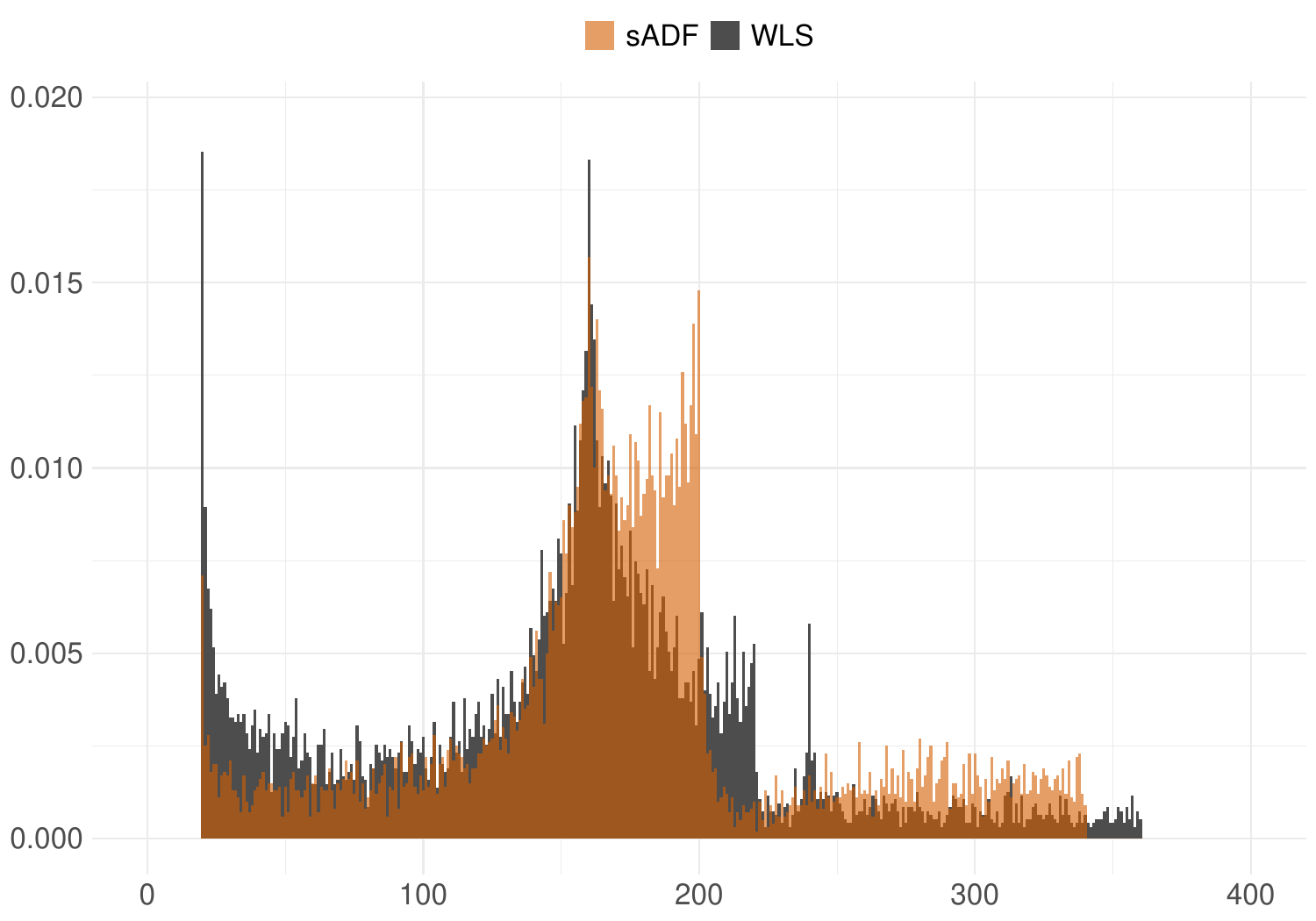}\label{fig:3e:6}}\\
\end{center}%
\caption{Histograms of $\hat{k}_e$ 
for $(\tau_e,\tau_c,\tau_r)=(0.4,0.6,0.7)$,  $\tau=0.8$, $s_0/s_1=1$, $T=400$}
\label{fig3_e}
\end{figure}

\newpage

\begin{figure}[h!]%
\begin{center}%
\subfigure[$T=800$, $c_a=4$, $c_b=6$, $s_0/s_1=1$]{\includegraphics[width=0.45\linewidth]{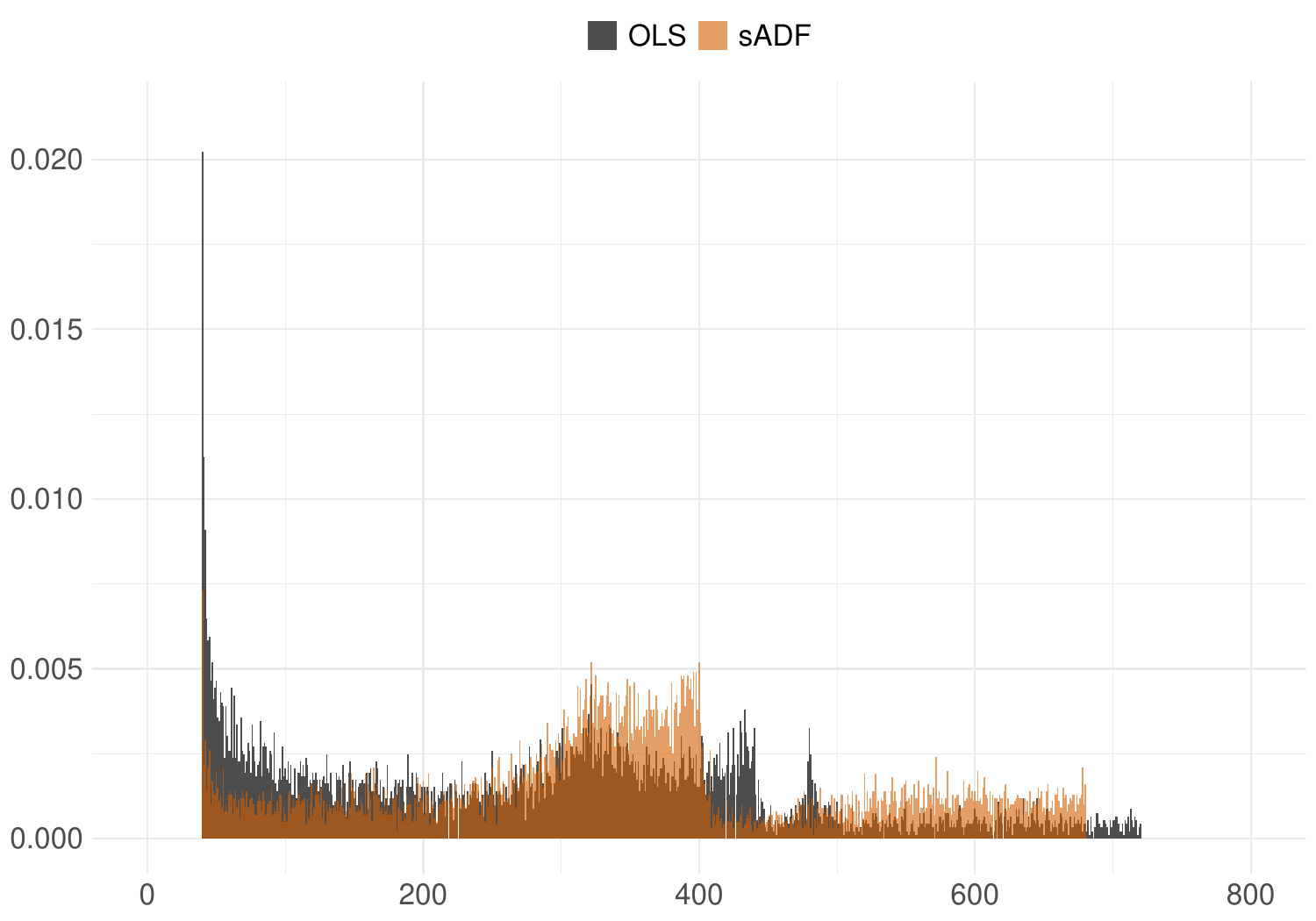}\label{fig:4e:1}}
\subfigure[$T=800$, $c_a=4$, $c_b=6$, $s_0/s_1=1$]{\includegraphics[width=0.45\linewidth]{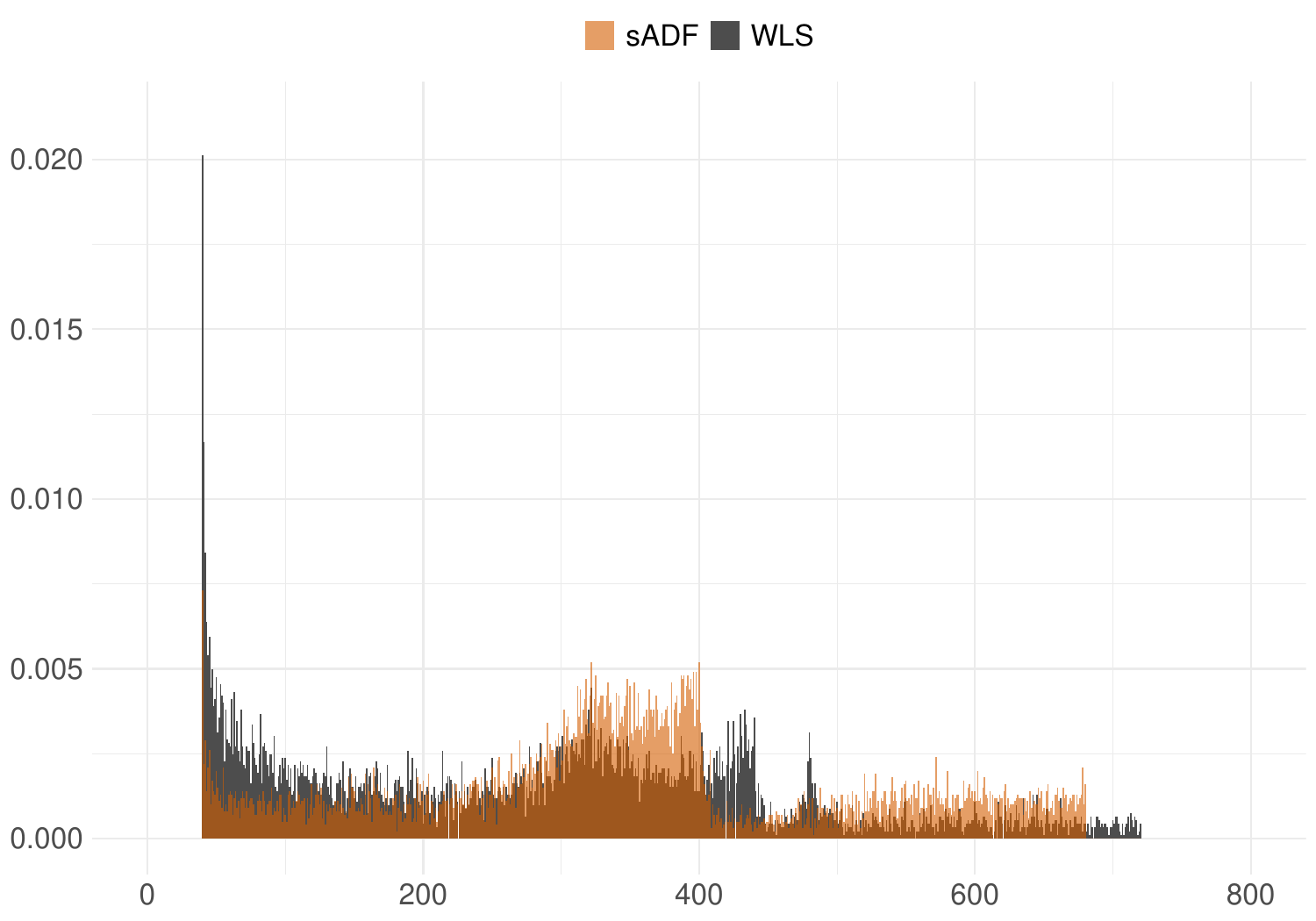}\label{fig:4e:2}}\\
\subfigure[$T=800$, $c_a=5$, $c_b=6$, $s_0/s_1=1$]{\includegraphics[width=0.45\linewidth]{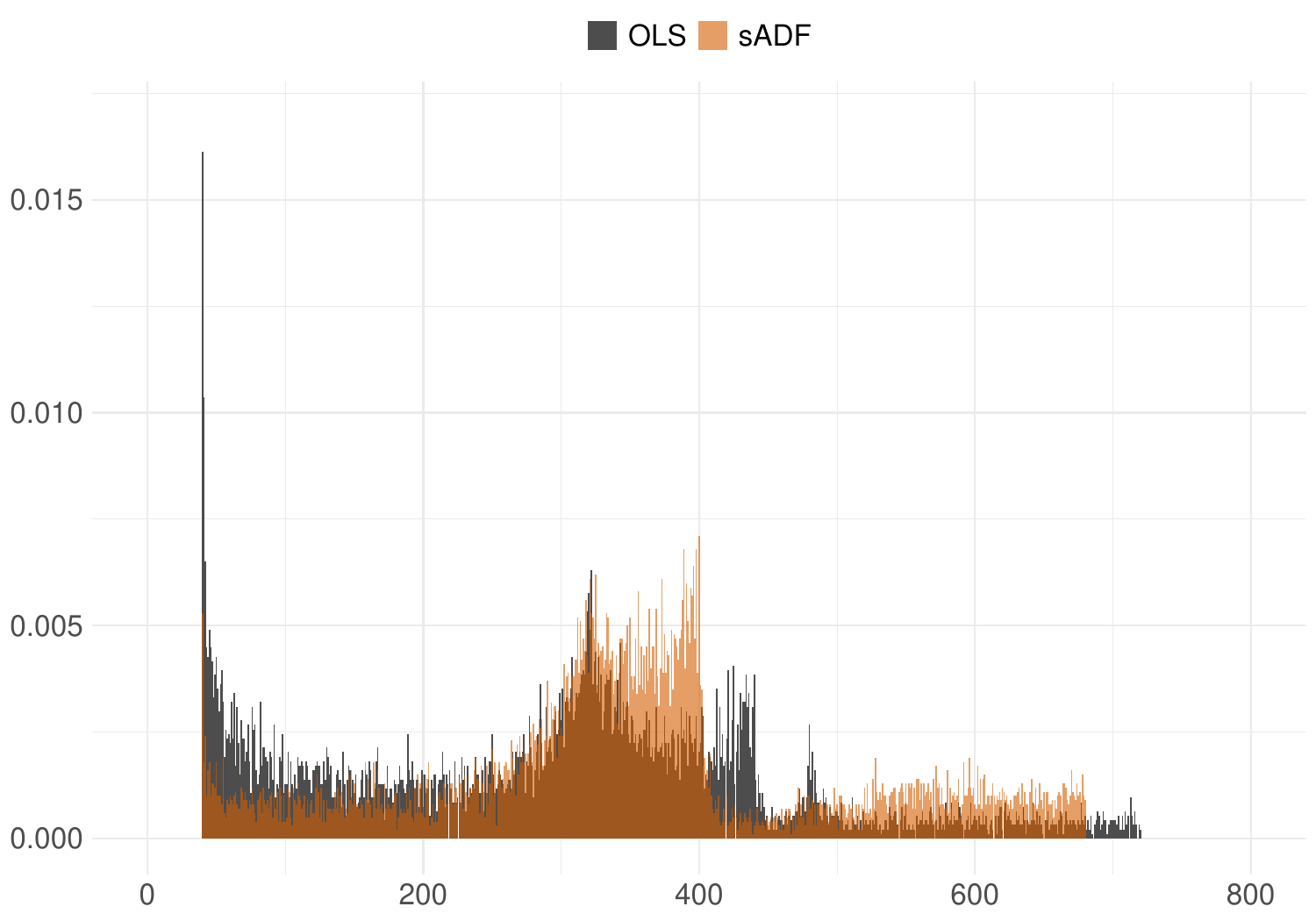}\label{fig:4e:3}}
\subfigure[$T=800$, $c_a=5$, $c_b=6$, $s_0/s_1=1$]{\includegraphics[width=0.45\linewidth]{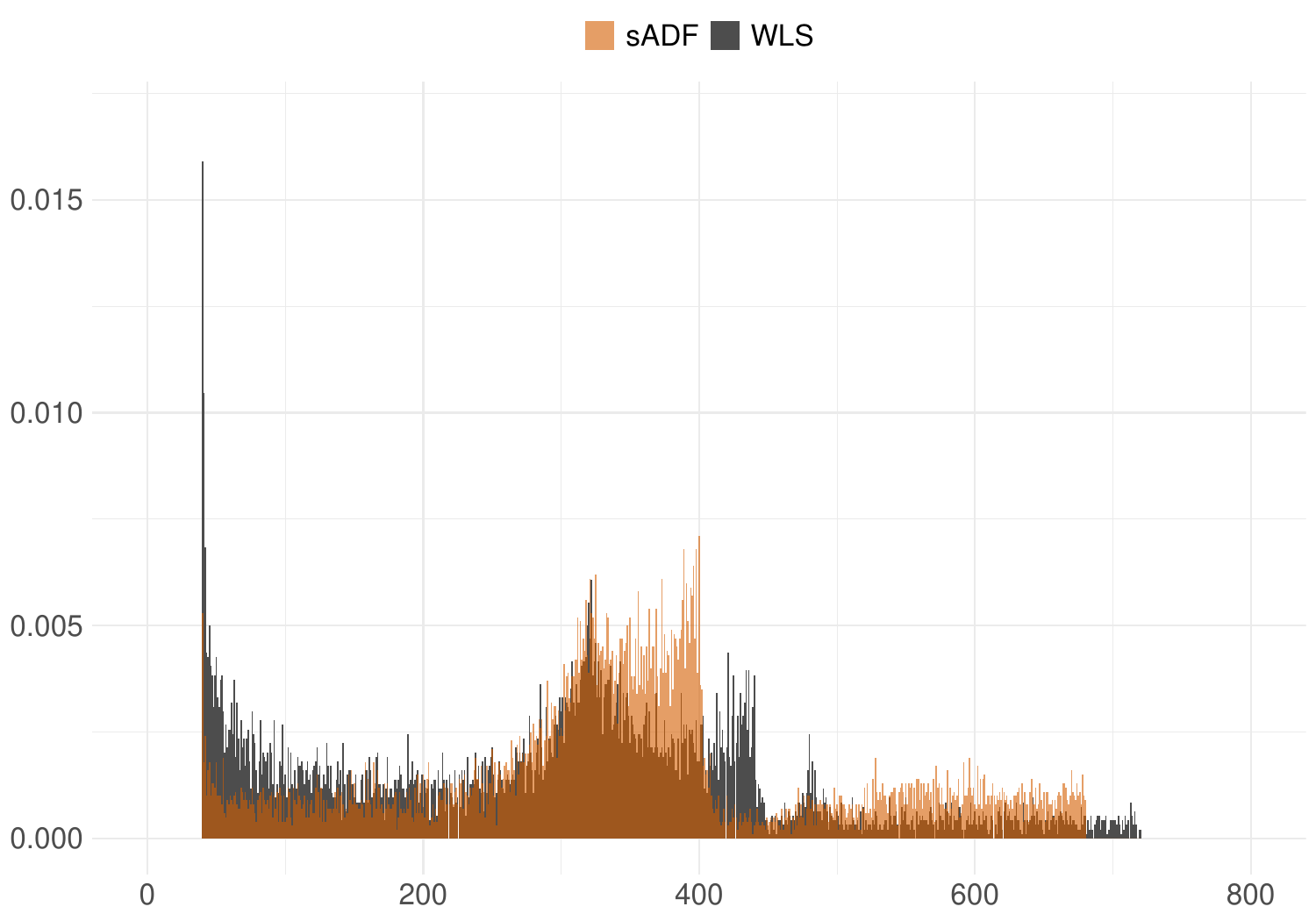}\label{fig:4e:4}}\\
\subfigure[$T=800$, $c_a=6$, $c_b=6$, $s_0/s_1=1$]{\includegraphics[width=0.45\linewidth]{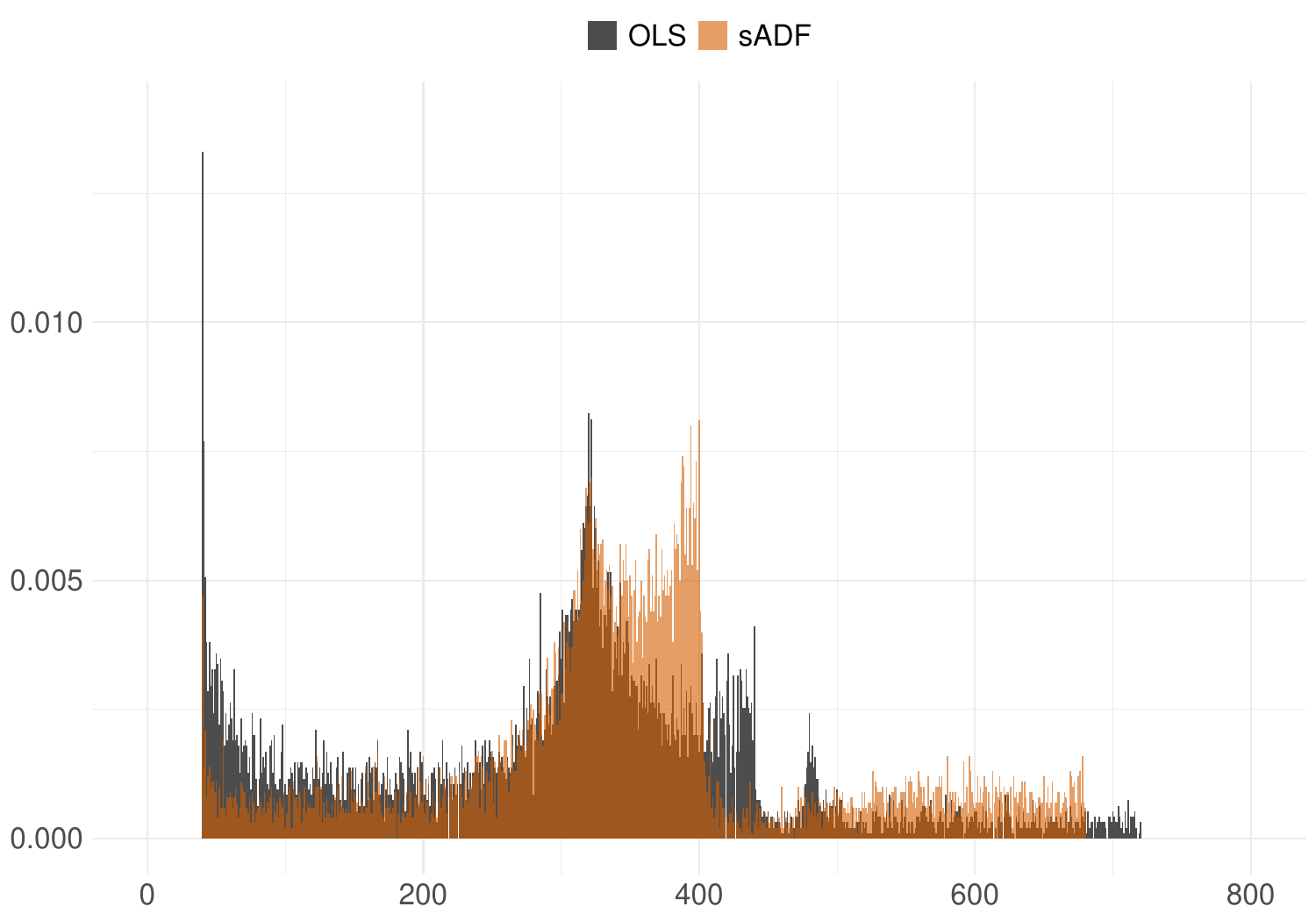}\label{fig:4e:5}}
\subfigure[$T=800$, $c_a=6$, $c_b=6$, $s_0/s_1=1$]{\includegraphics[width=0.45\linewidth]{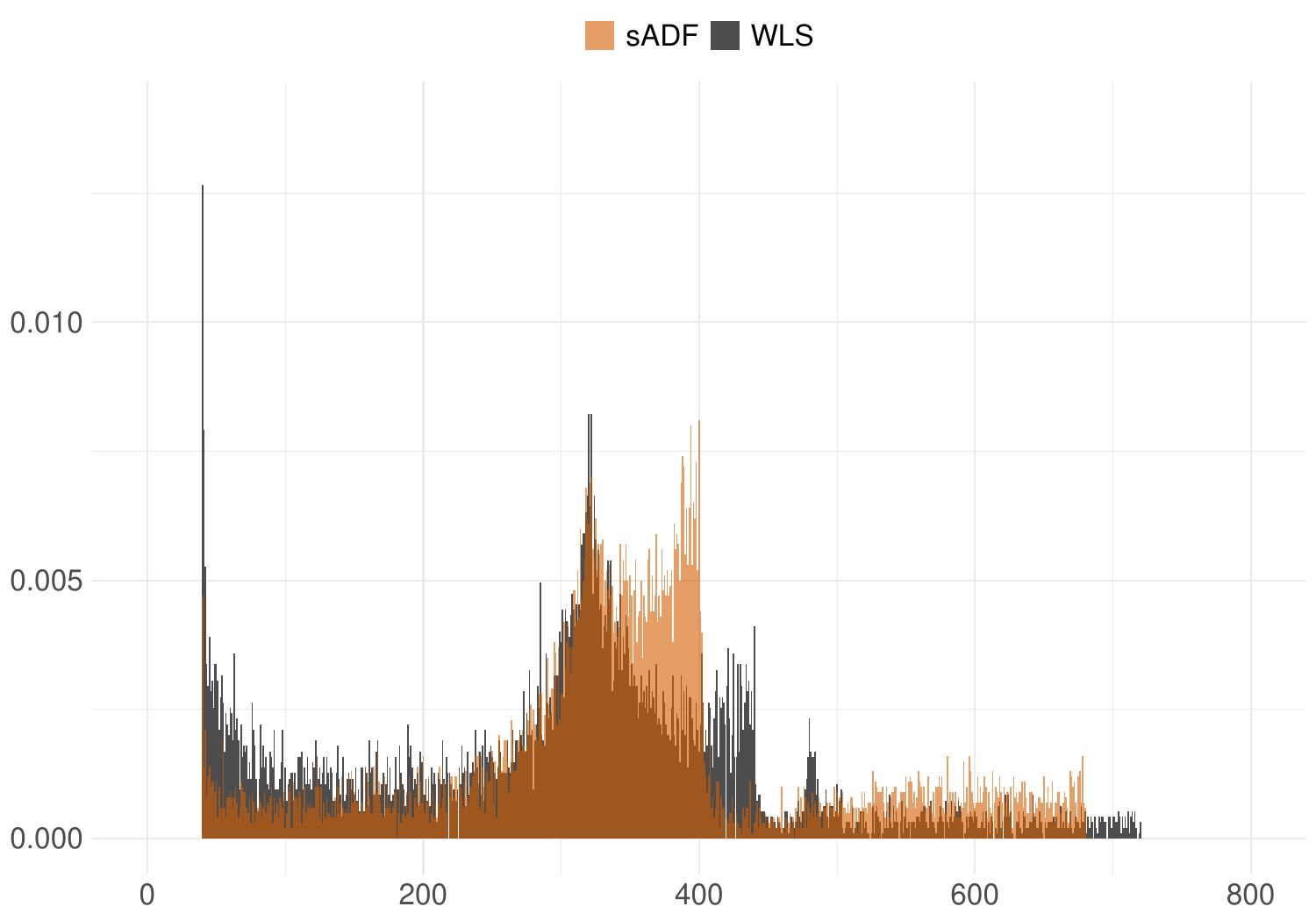}\label{fig:4e:6}}\\
\end{center}%
\caption{Histograms of $\hat{k}_e$ 
for $(\tau_e,\tau_c,\tau_r)=(0.4,0.6,0.7)$,  $\tau=0.8$, $s_0/s_1=1$, $T=800$}
\label{fig4_e}
\end{figure}

\newpage

\begin{figure}[h!]%
\begin{center}%
\subfigure[$T=400$, $c_a=4$, $c_b=6$, $s_0/s_1=5$]{\includegraphics[width=0.45\linewidth]{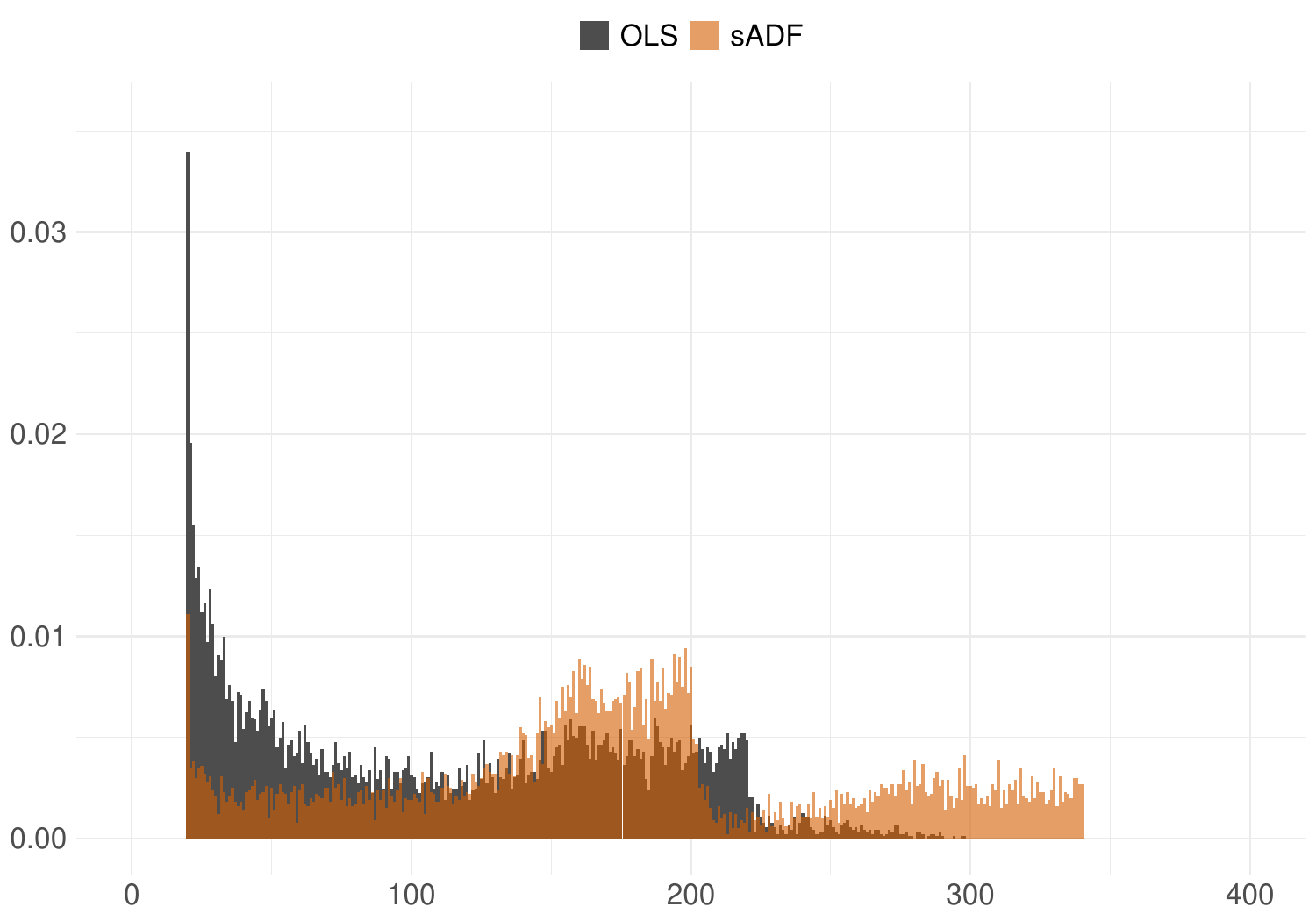}\label{fig:5e:1}}
\subfigure[$T=400$, $c_a=4$, $c_b=6$, $s_0/s_1=5$]{\includegraphics[width=0.45\linewidth]{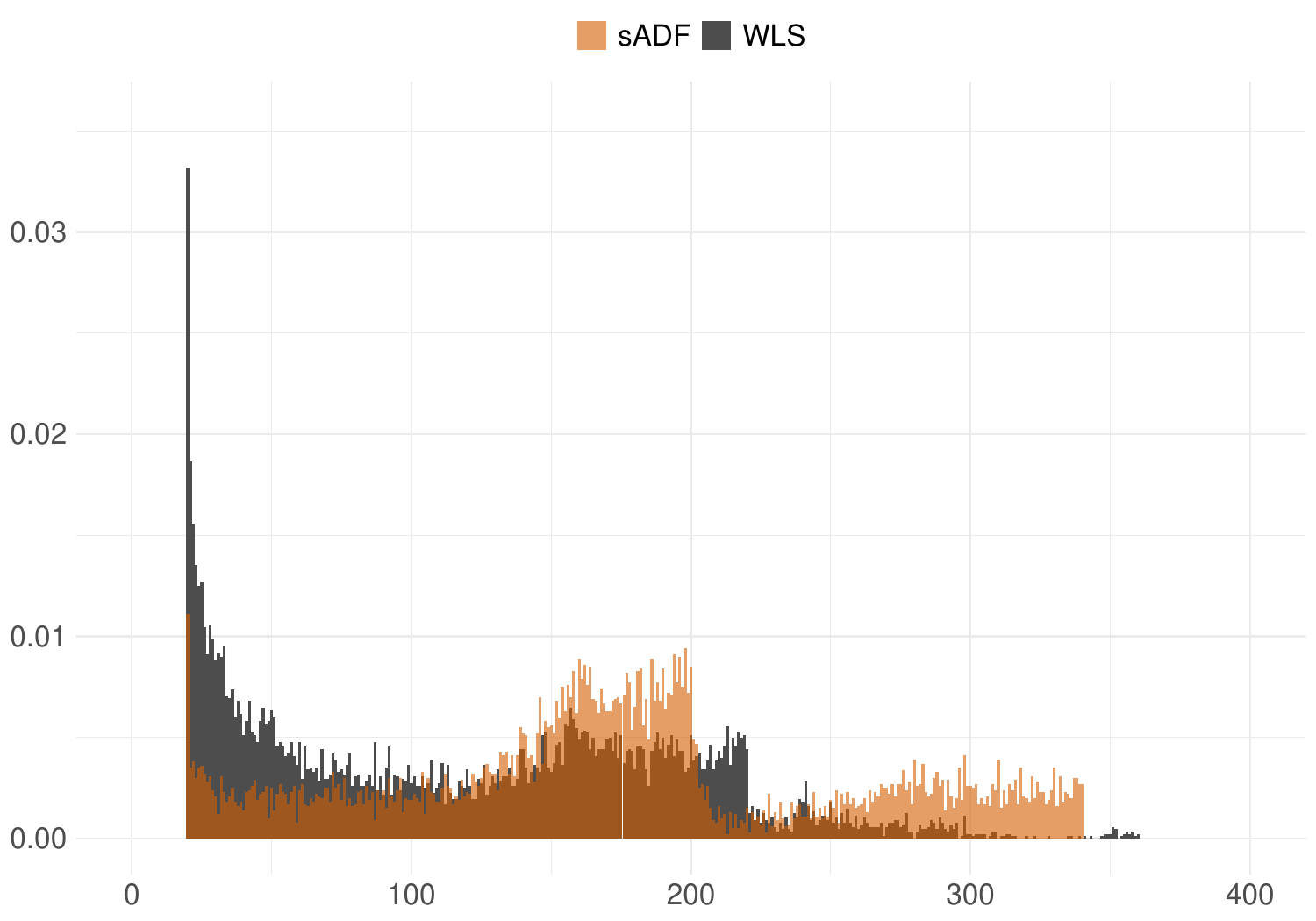}\label{fig:5e:2}}\\
\subfigure[$T=400$, $c_a=5$, $c_b=6$, $s_0/s_1=5$]{\includegraphics[width=0.45\linewidth]{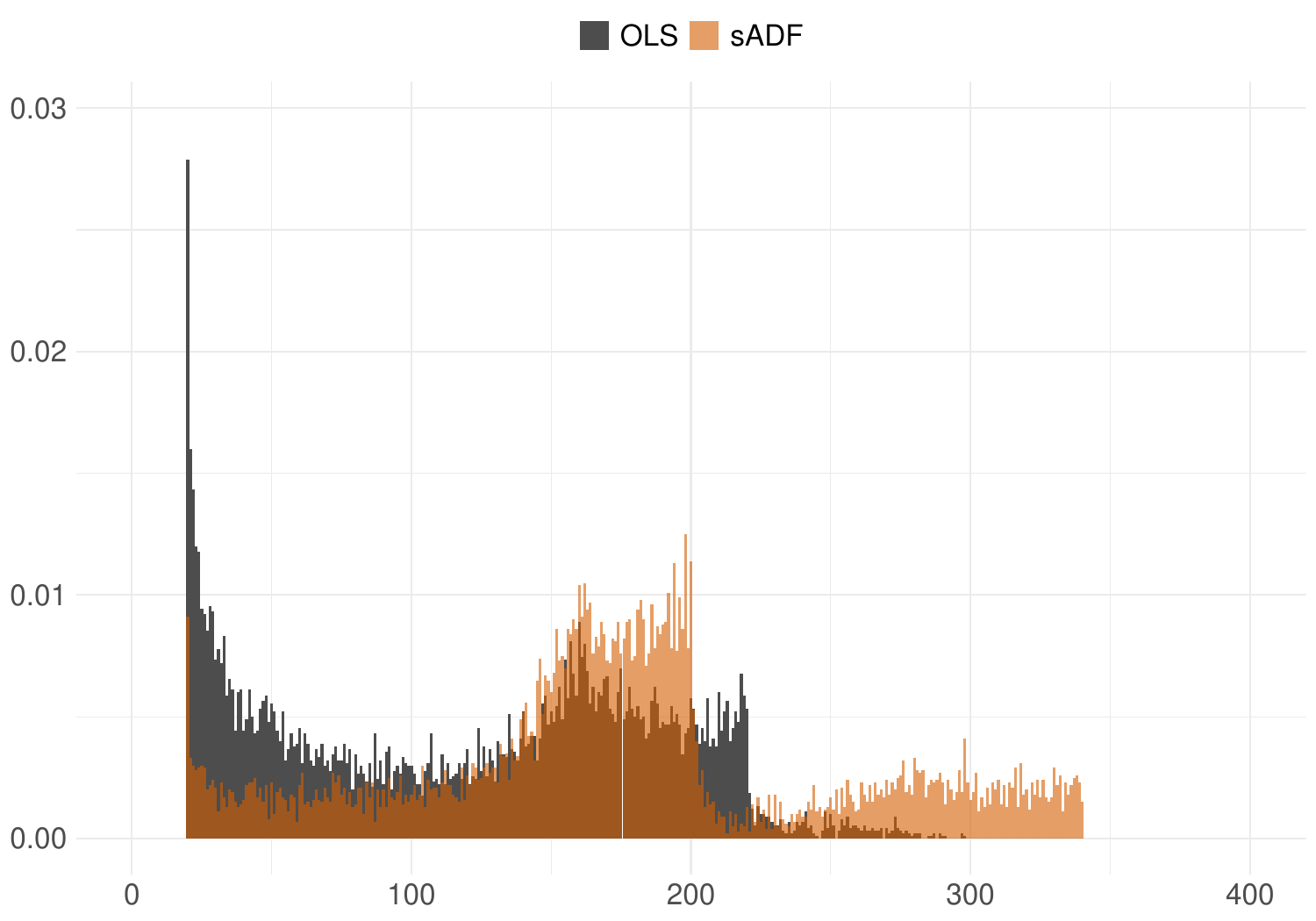}\label{fig:5e:3}}
\subfigure[$T=400$, $c_a=5$, $c_b=6$, $s_0/s_1=5$]{\includegraphics[width=0.45\linewidth]{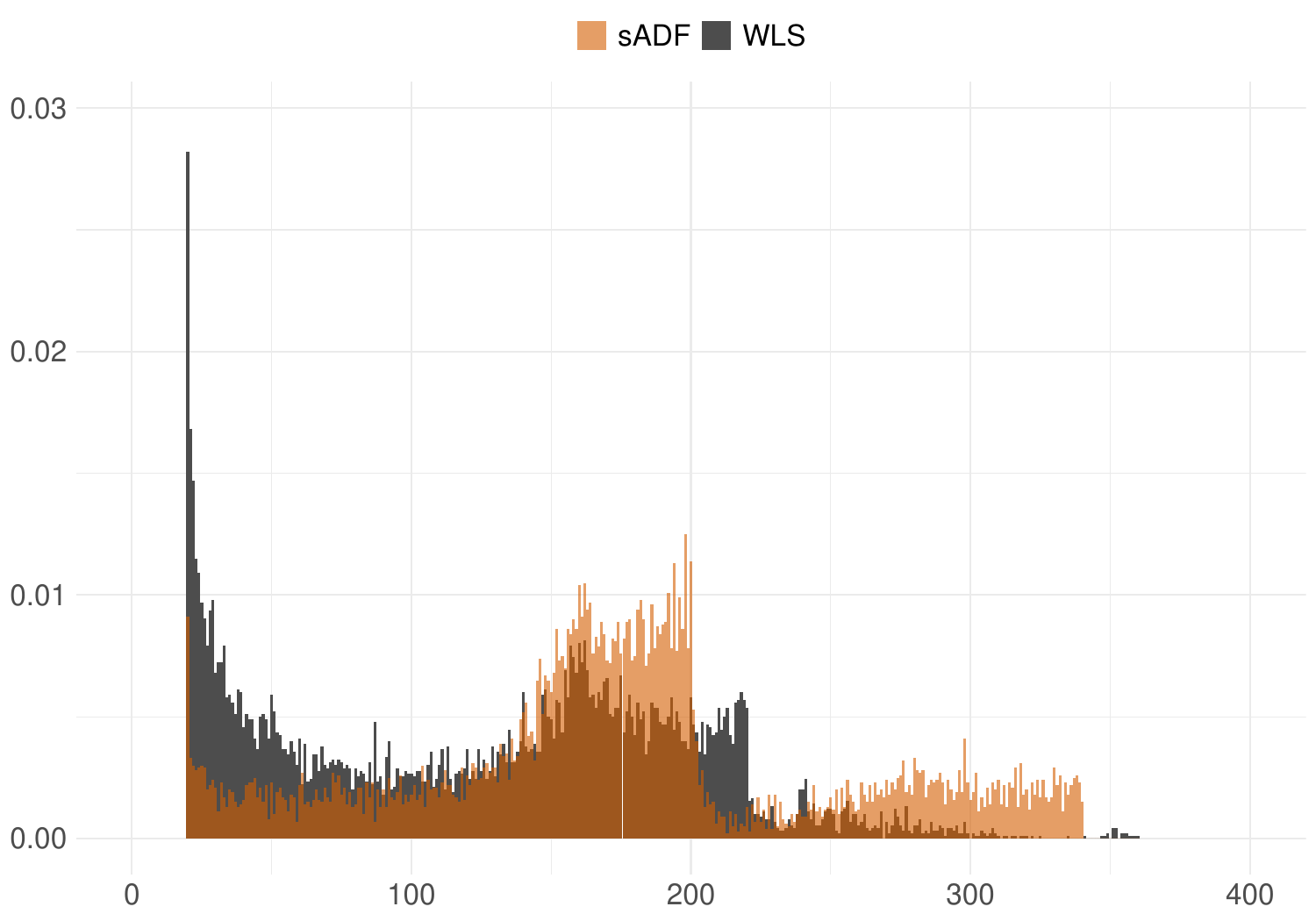}\label{fig:5e:4}}\\
\subfigure[$T=400$, $c_a=6$, $c_b=6$, $s_0/s_1=5$]{\includegraphics[width=0.45\linewidth]{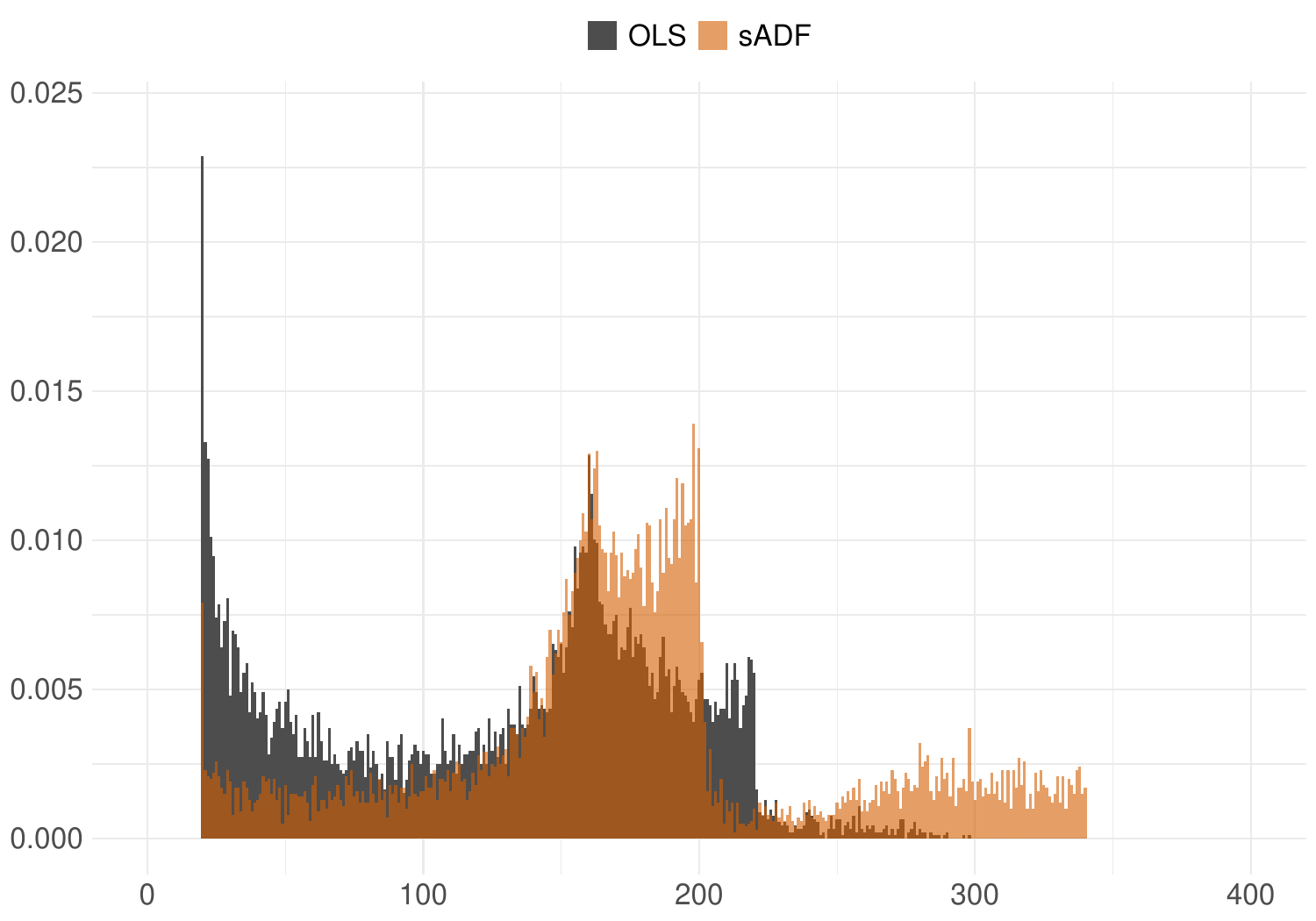}\label{fig:5e:5}}
\subfigure[$T=400$, $c_a=6$, $c_b=6$, $s_0/s_1=5$]{\includegraphics[width=0.45\linewidth]{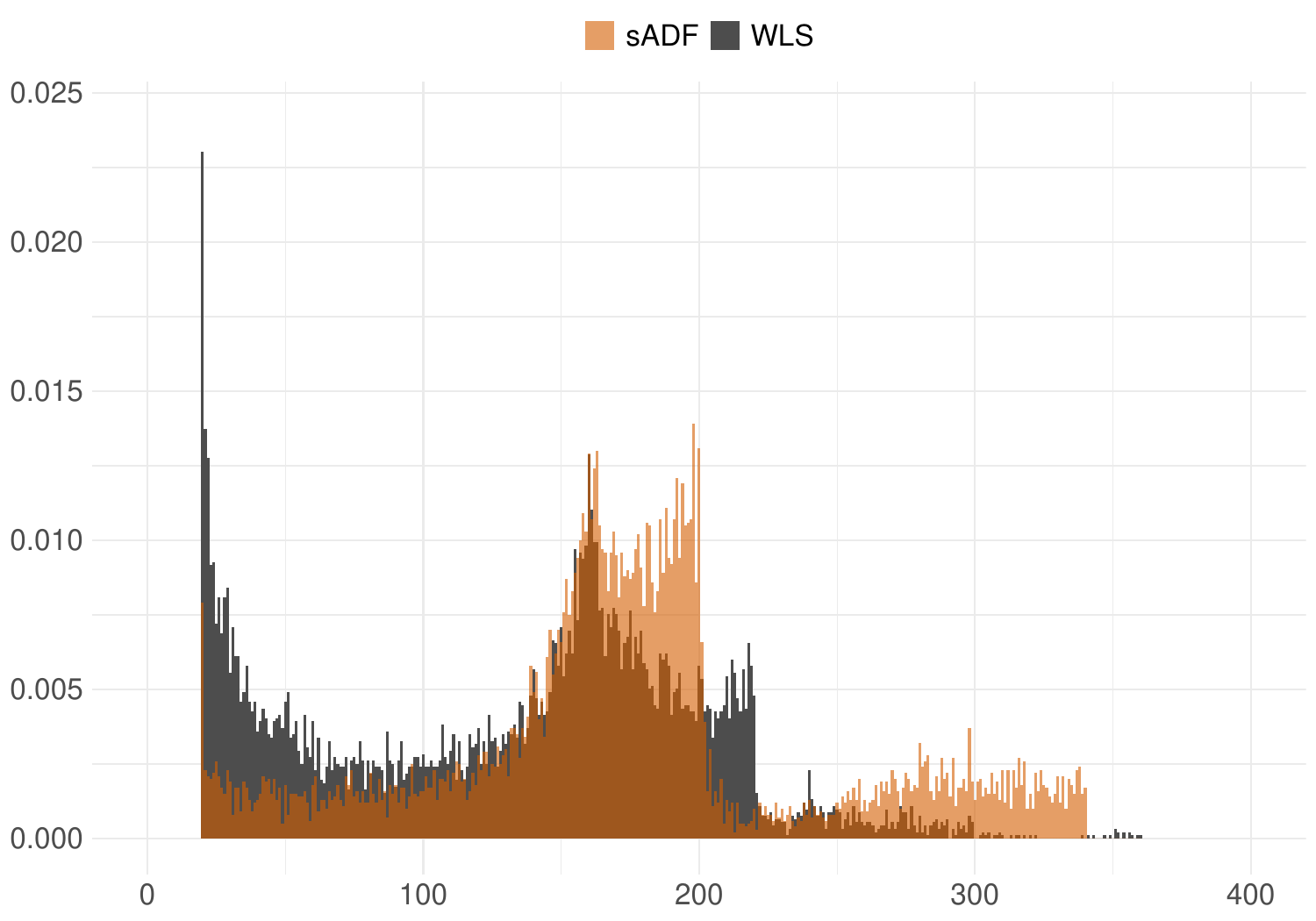}\label{fig:5e:6}}\\
\end{center}%
\caption{Histograms of $\hat{k}_e$ 
for $(\tau_e,\tau_c,\tau_r)=(0.4,0.6,0.7)$,  $\tau=0.8$, $s_0/s_1=5$, $T=400$}
\label{fig5_e}
\end{figure}

\newpage

\begin{figure}[h!]%
\begin{center}%
\subfigure[$T=800$, $c_a=4$, $c_b=6$, $s_0/s_1=5$]{\includegraphics[width=0.45\linewidth]{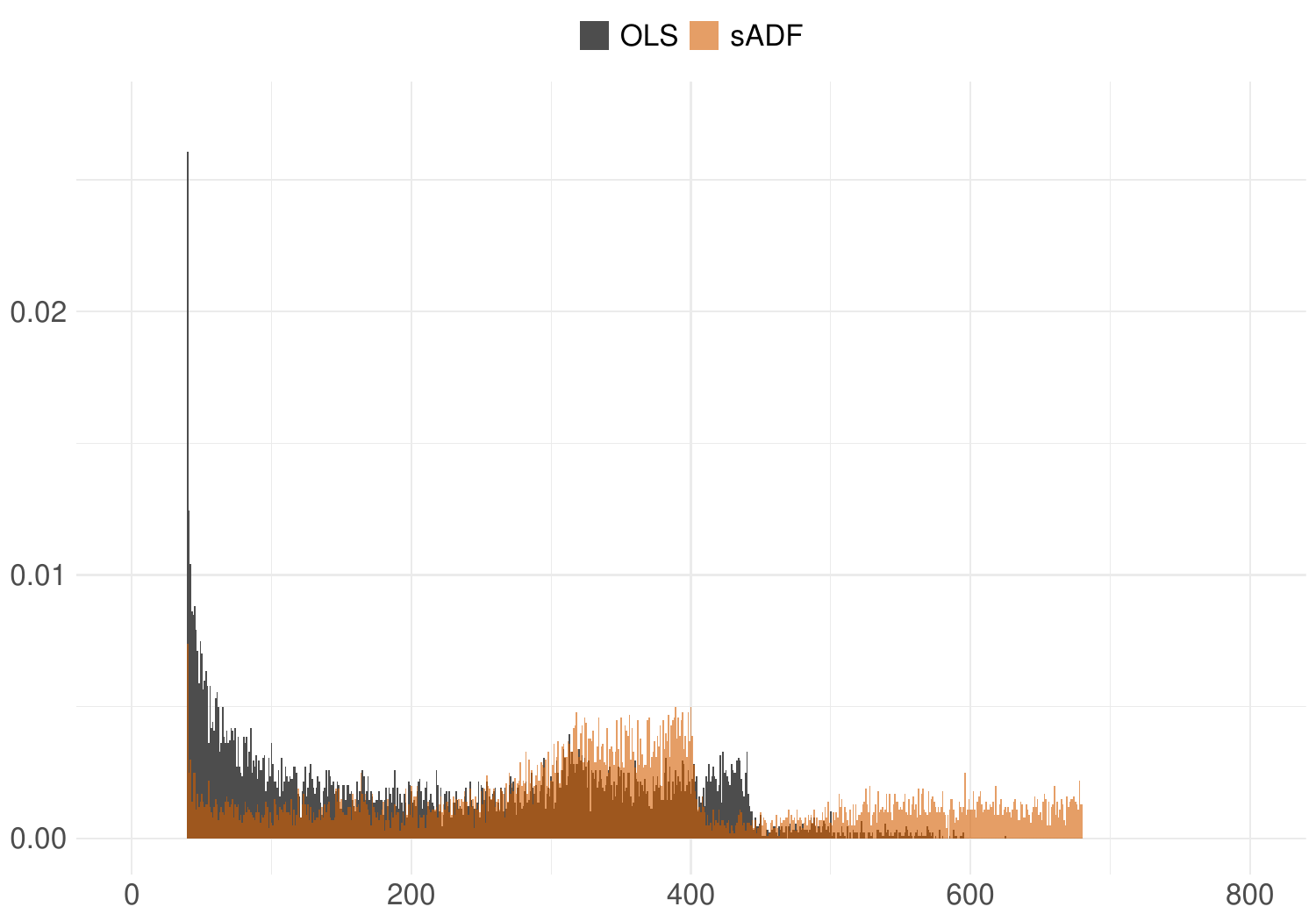}\label{fig:6e:1}}
\subfigure[$T=800$, $c_a=4$, $c_b=6$, $s_0/s_1=5$]{\includegraphics[width=0.45\linewidth]{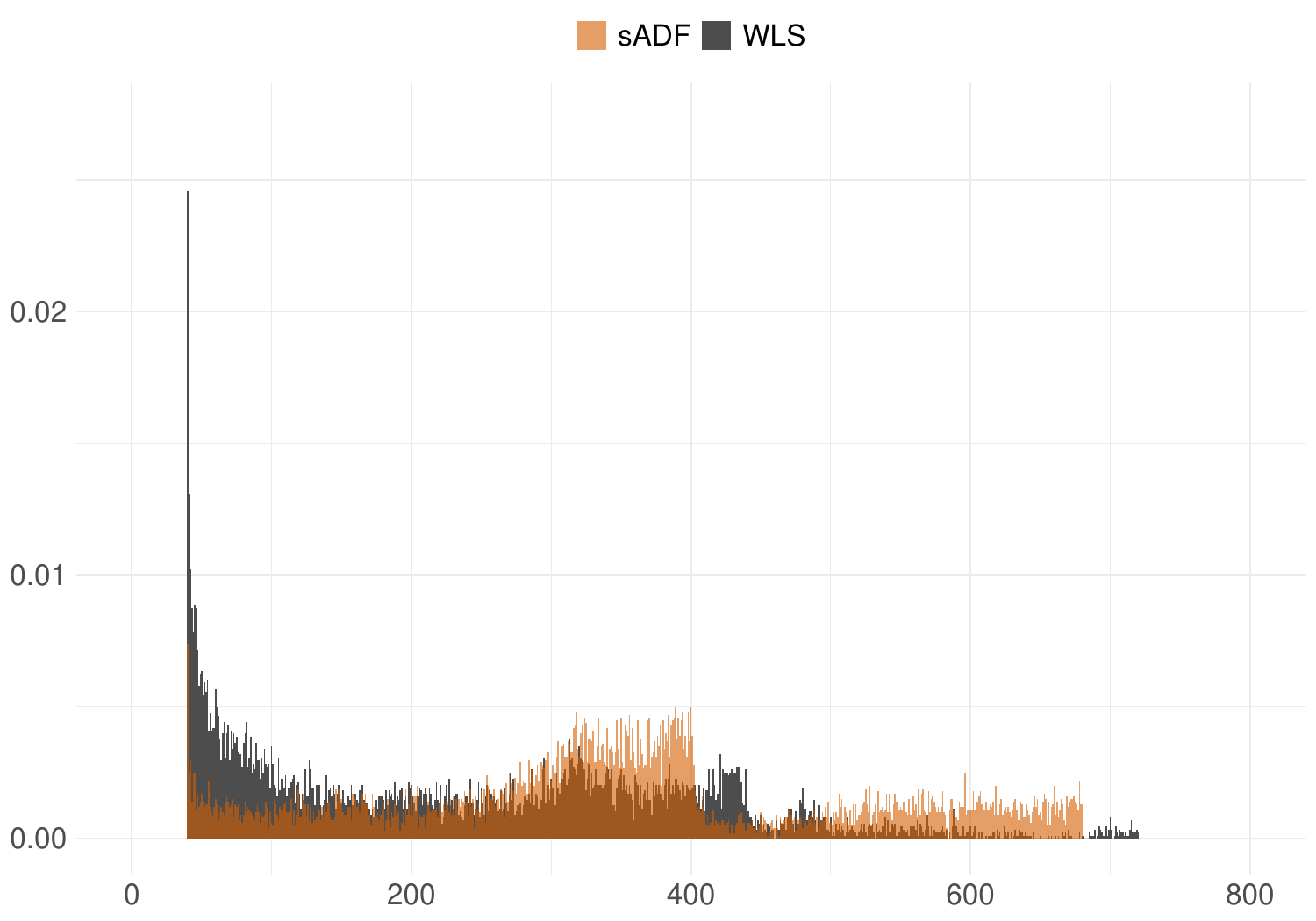}\label{fig:6e:2}}\\
\subfigure[$T=800$, $c_a=5$, $c_b=6$, $s_0/s_1=5$]{\includegraphics[width=0.45\linewidth]{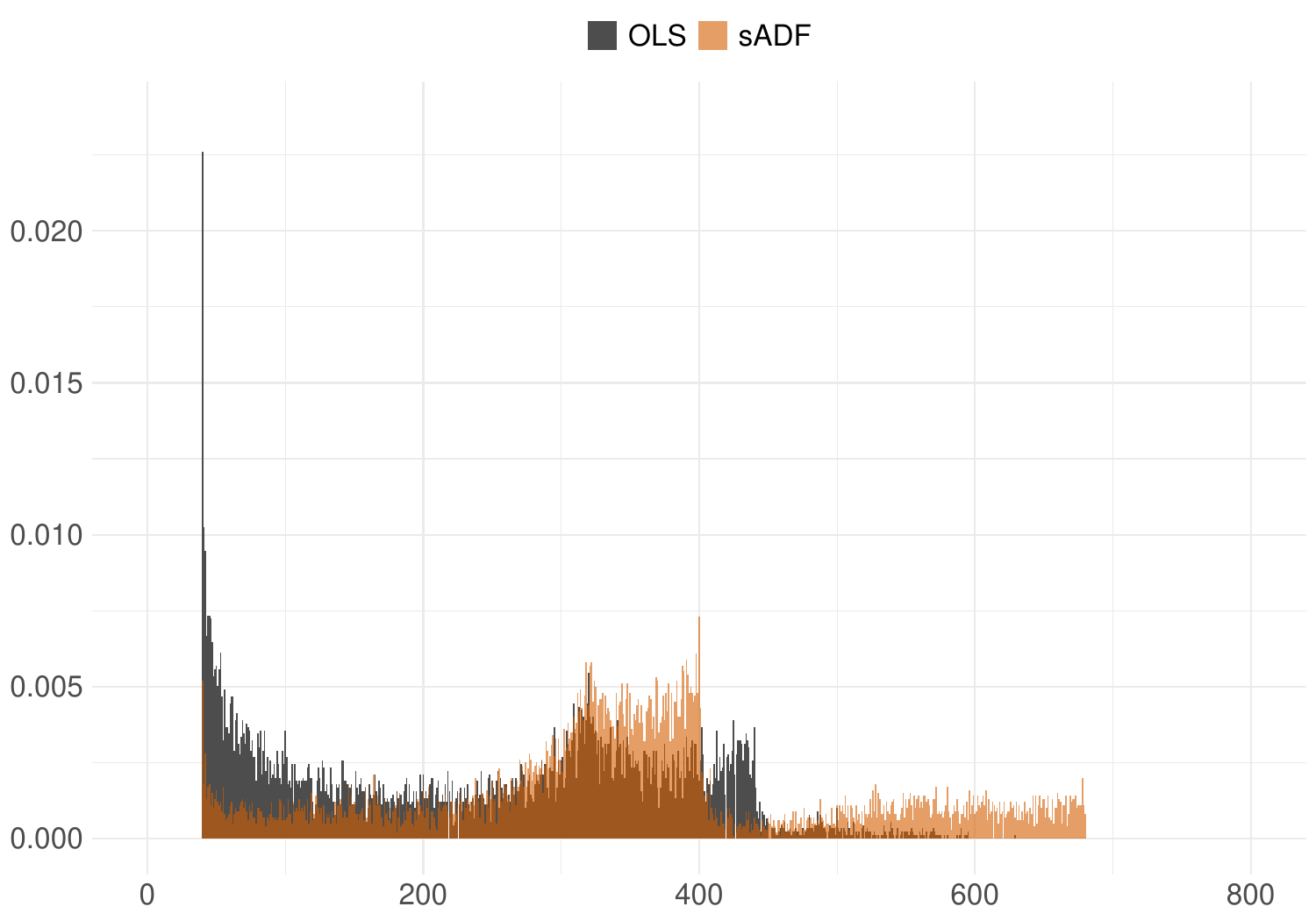}\label{fig:6e:3}}
\subfigure[$T=800$, $c_a=5$, $c_b=6$, $s_0/s_1=5$]{\includegraphics[width=0.45\linewidth]{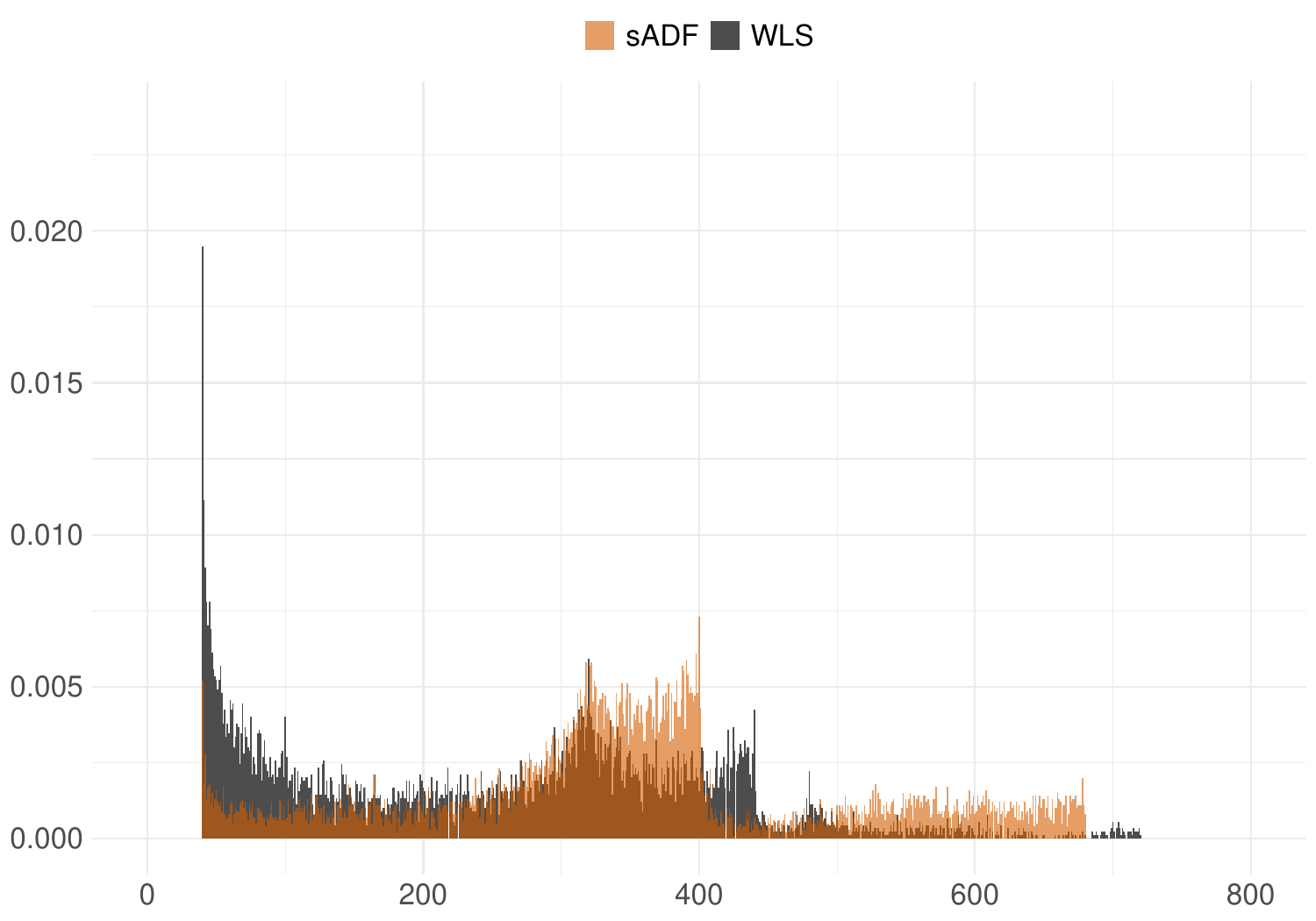}\label{fig:6e:4}}\\
\subfigure[$T=800$, $c_a=6$, $c_b=6$, $s_0/s_1=5$]{\includegraphics[width=0.45\linewidth]{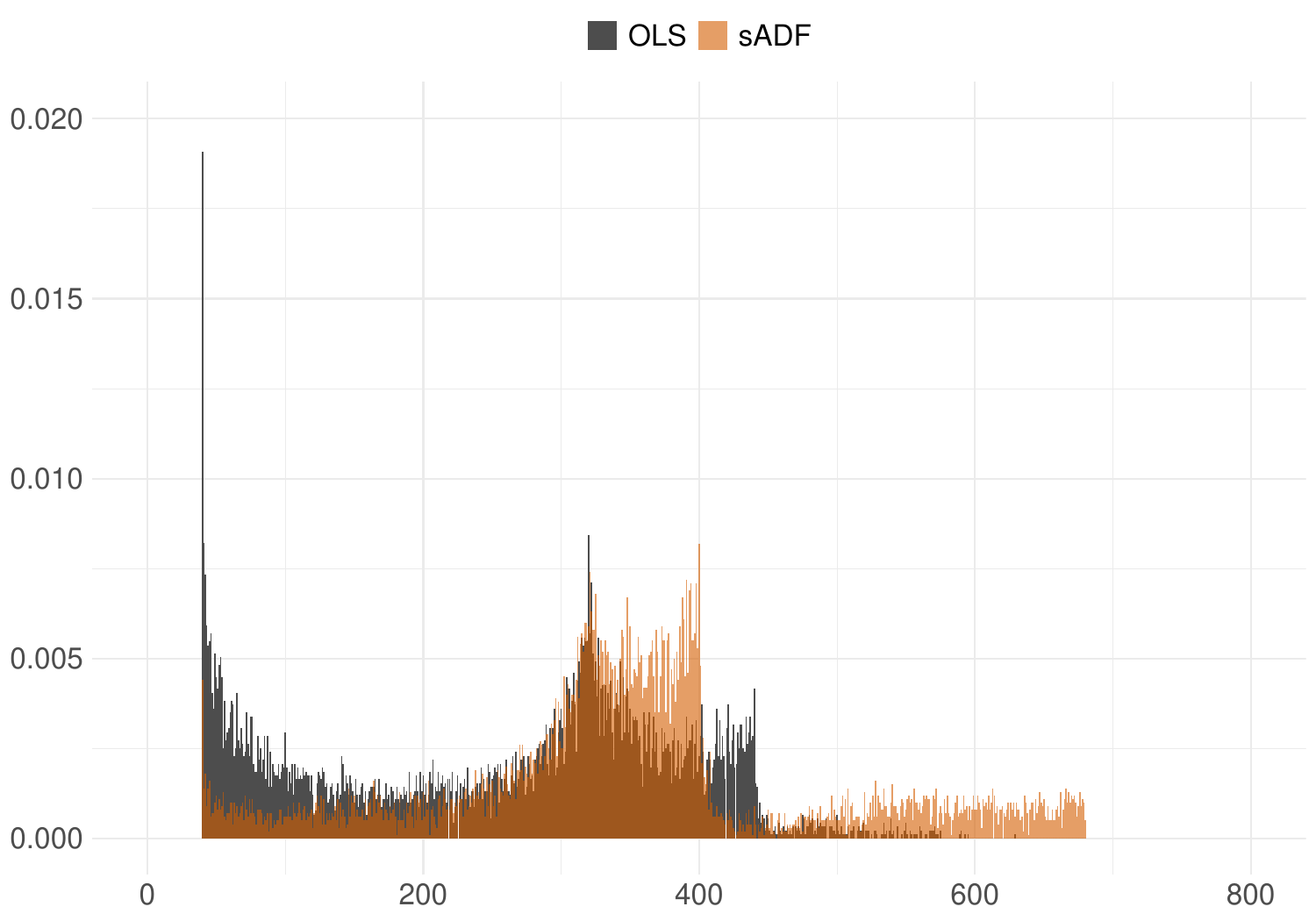}\label{fig:6e:5}}
\subfigure[$T=800$, $c_a=6$, $c_b=6$, $s_0/s_1=5$]{\includegraphics[width=0.45\linewidth]{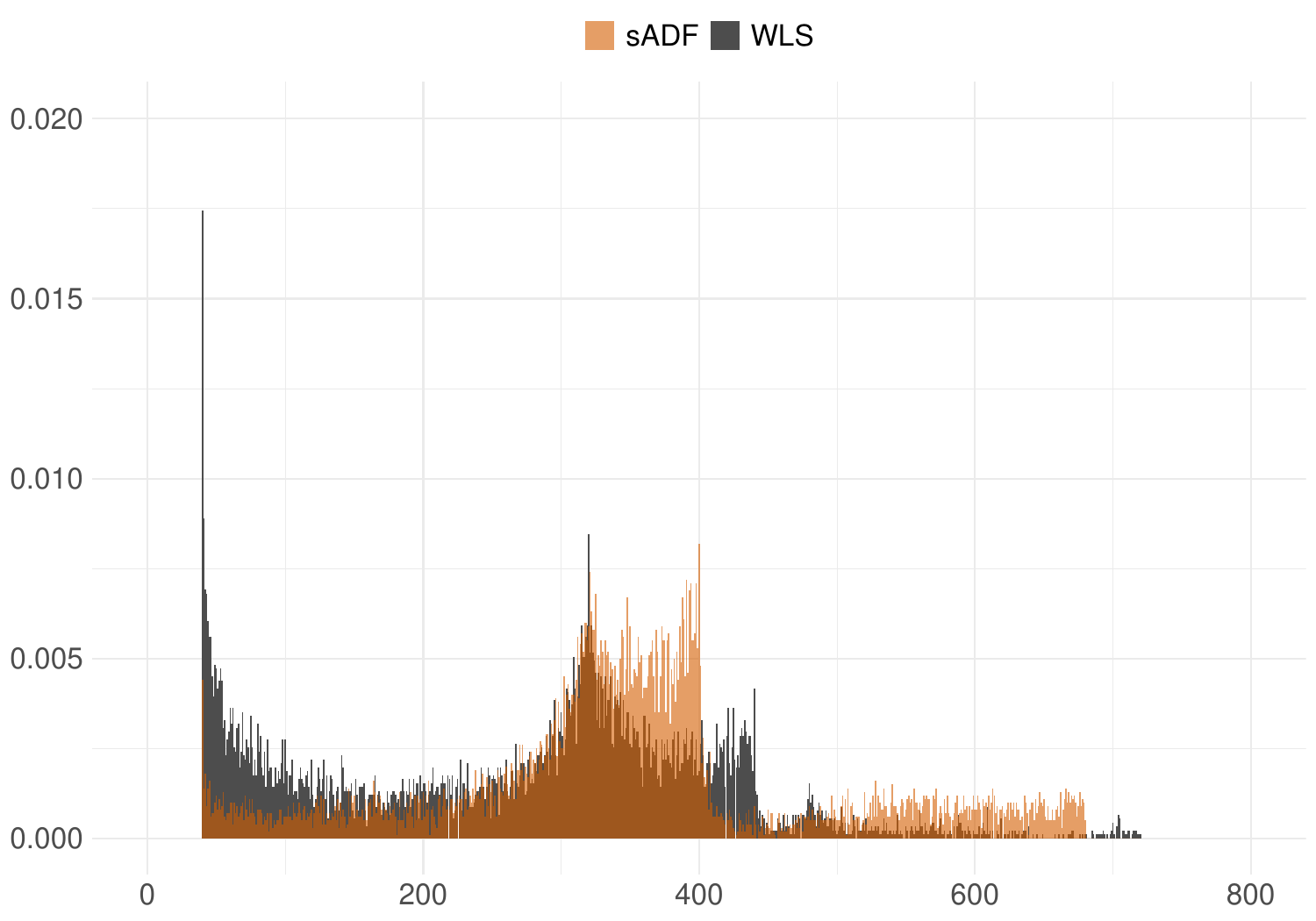}\label{fig:6e:6}}\\
\end{center}%
\caption{Histograms of $\hat{k}_e$ 
for $(\tau_e,\tau_c,\tau_r)=(0.4,0.6,0.7)$,  $\tau=0.8$, $s_0/s_1=5$, $T=800$}
\label{fig6_e}
\end{figure}

\newpage

\section{$\tau=0.8$, $\hat{k}_r$}
\setcounter{figure}{0}

\begin{figure}[h!]%
\begin{center}%
\subfigure[$T=400$, $c_a=4$, $c_b=6$, $s_0/s_1=1/5$]{\includegraphics[width=0.45\linewidth]{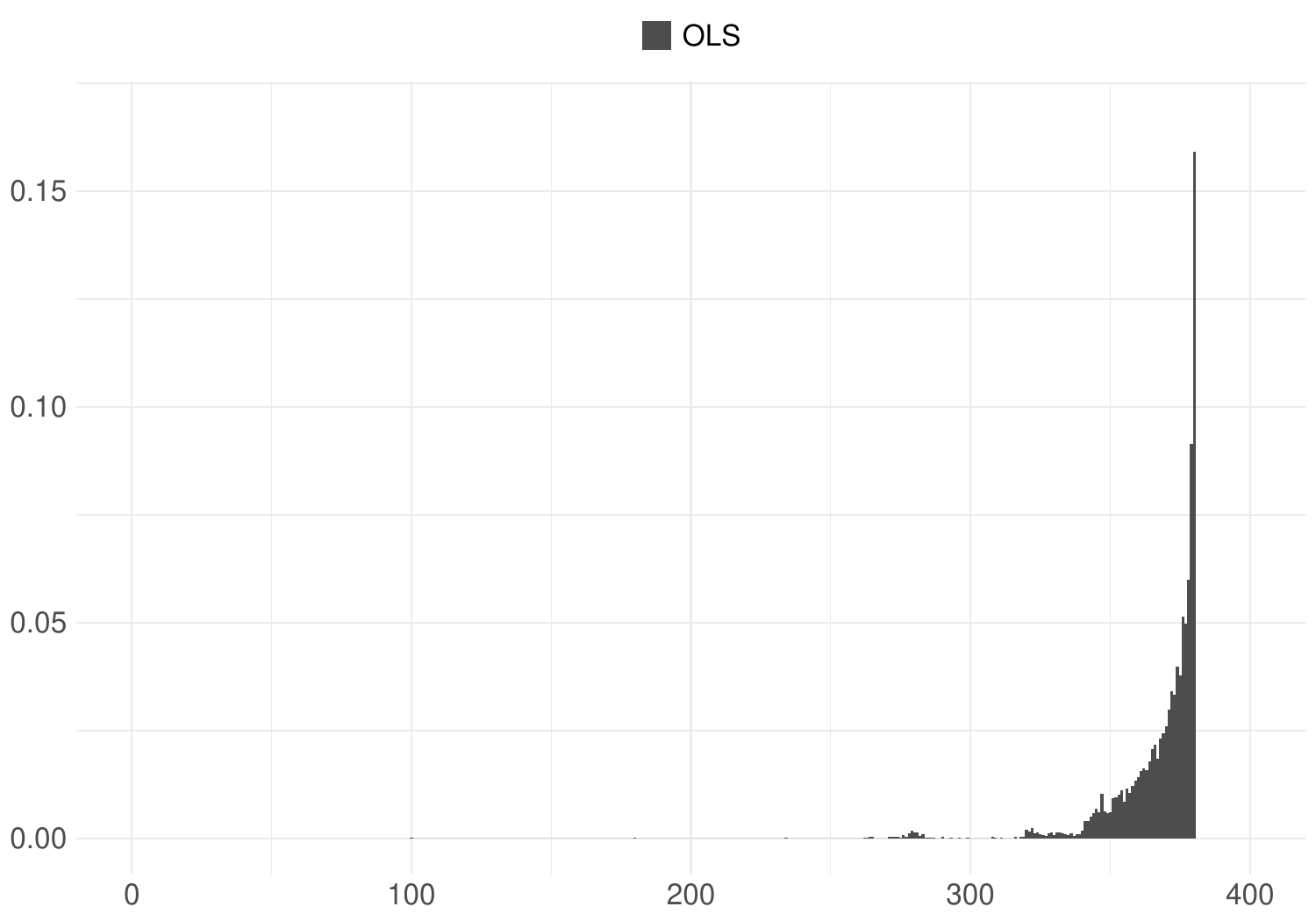}\label{fig2:1:1}}
\subfigure[$T=400$, $c_a=4$, $c_b=6$, $s_0/s_1=1/5$]{\includegraphics[width=0.45\linewidth]{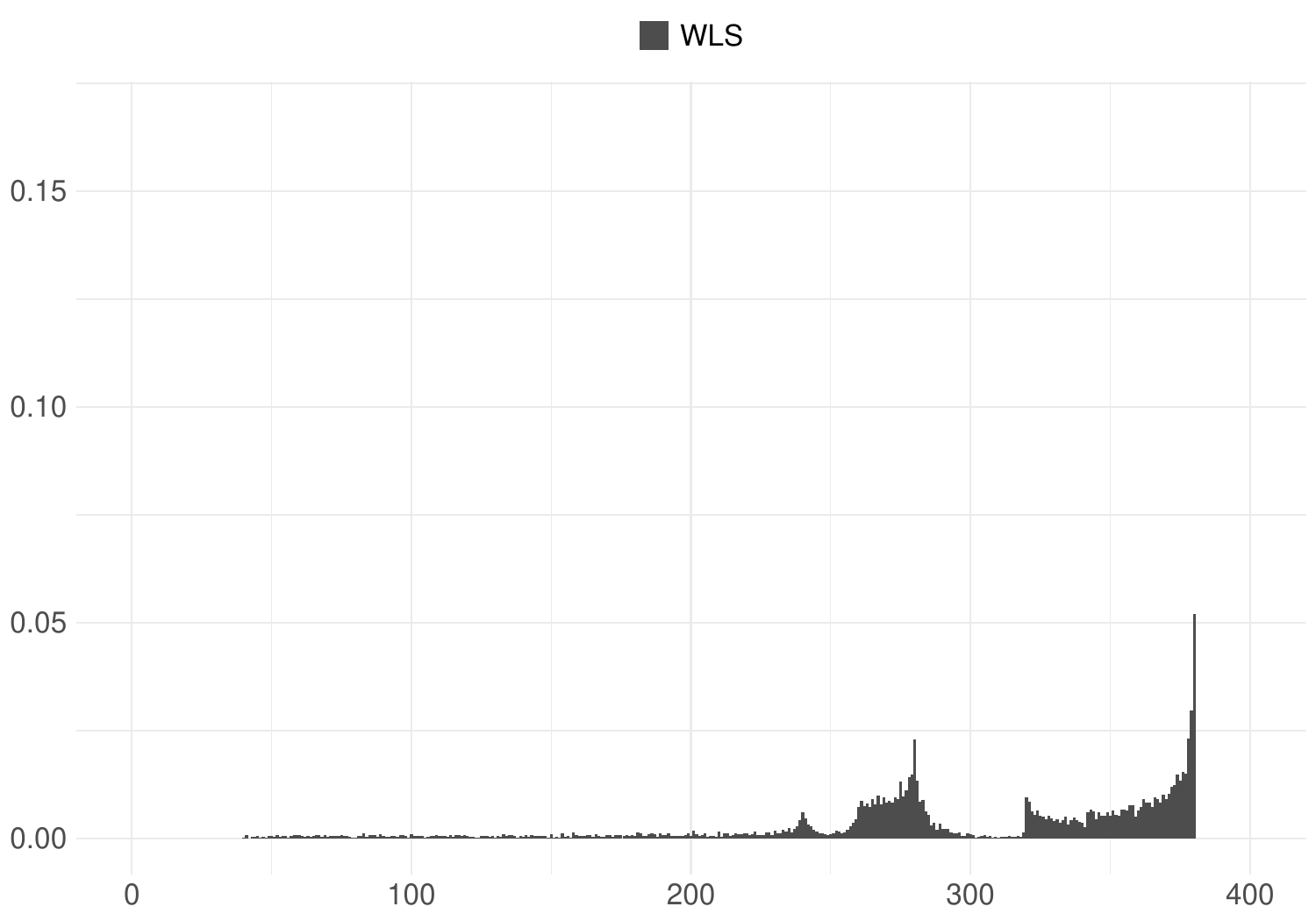}\label{fig2:1:2}}\\
\subfigure[$T=400$, $c_a=5$, $c_b=6$, $s_0/s_1=1/5$]{\includegraphics[width=0.45\linewidth]{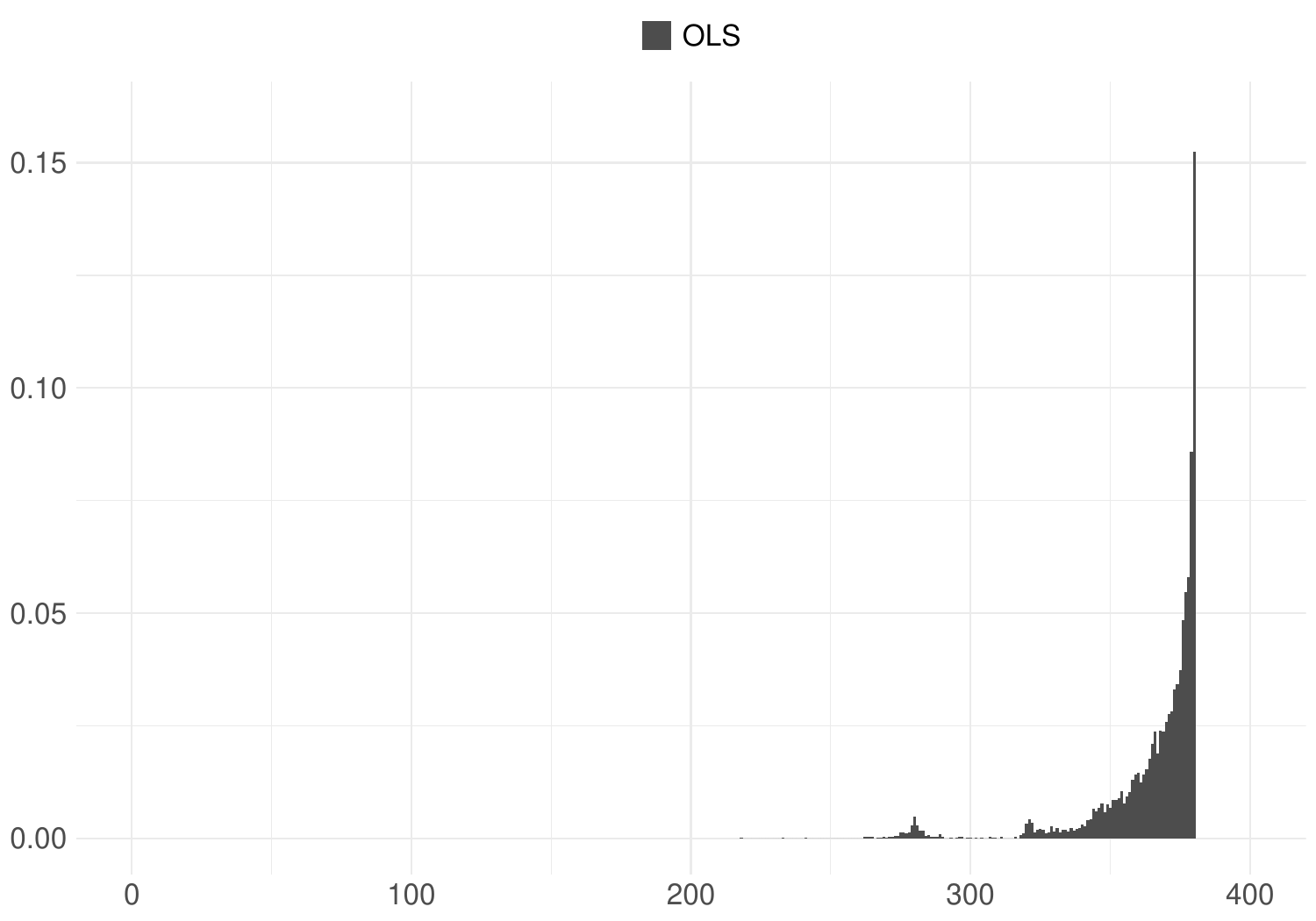}\label{fig2:1:3}}
\subfigure[$T=400$, $c_a=5$, $c_b=6$, $s_0/s_1=1/5$]{\includegraphics[width=0.45\linewidth]{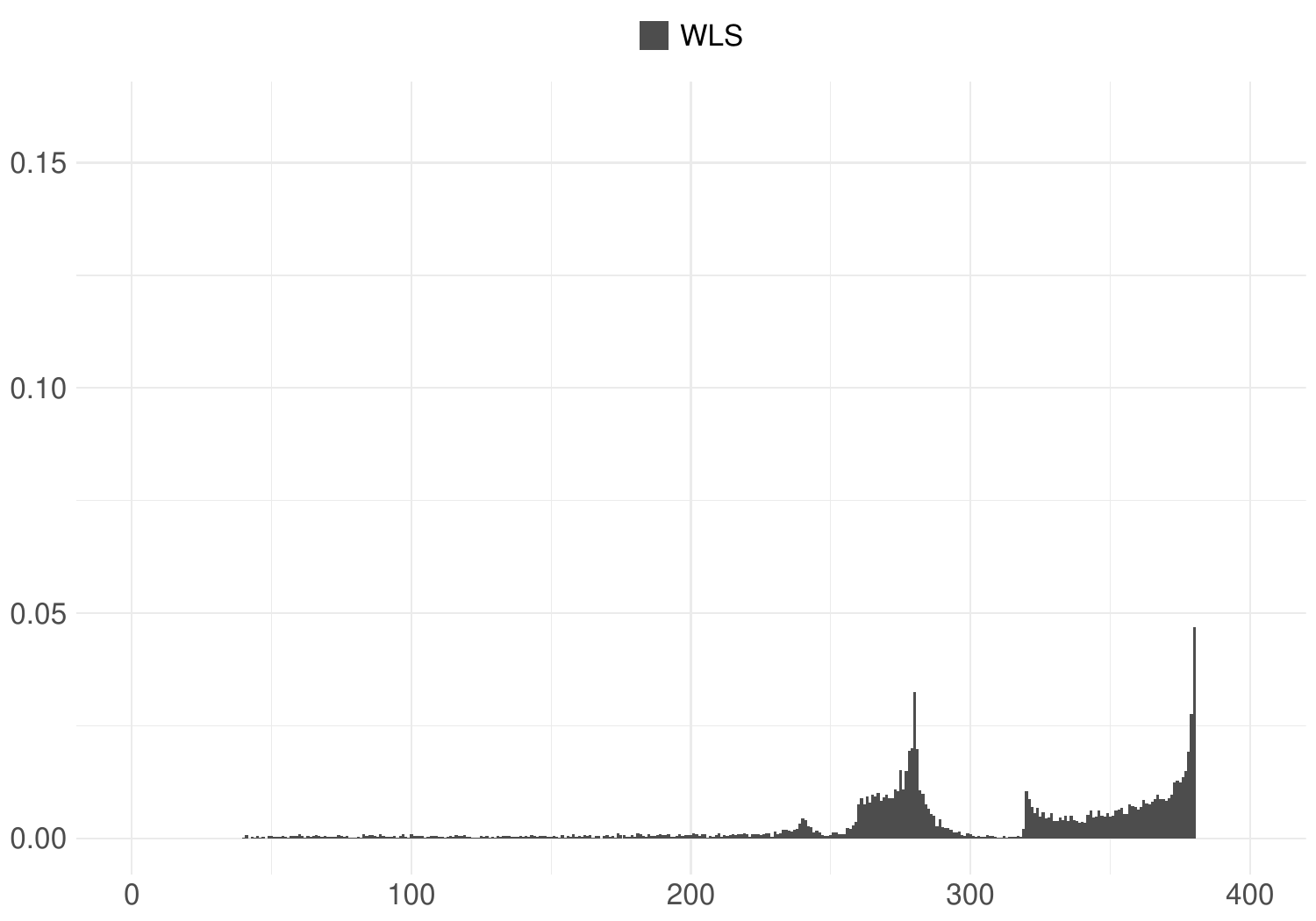}\label{fig2:1:4}}\\
\subfigure[$T=400$, $c_a=6$, $c_b=6$, $s_0/s_1=1/5$]{\includegraphics[width=0.45\linewidth]{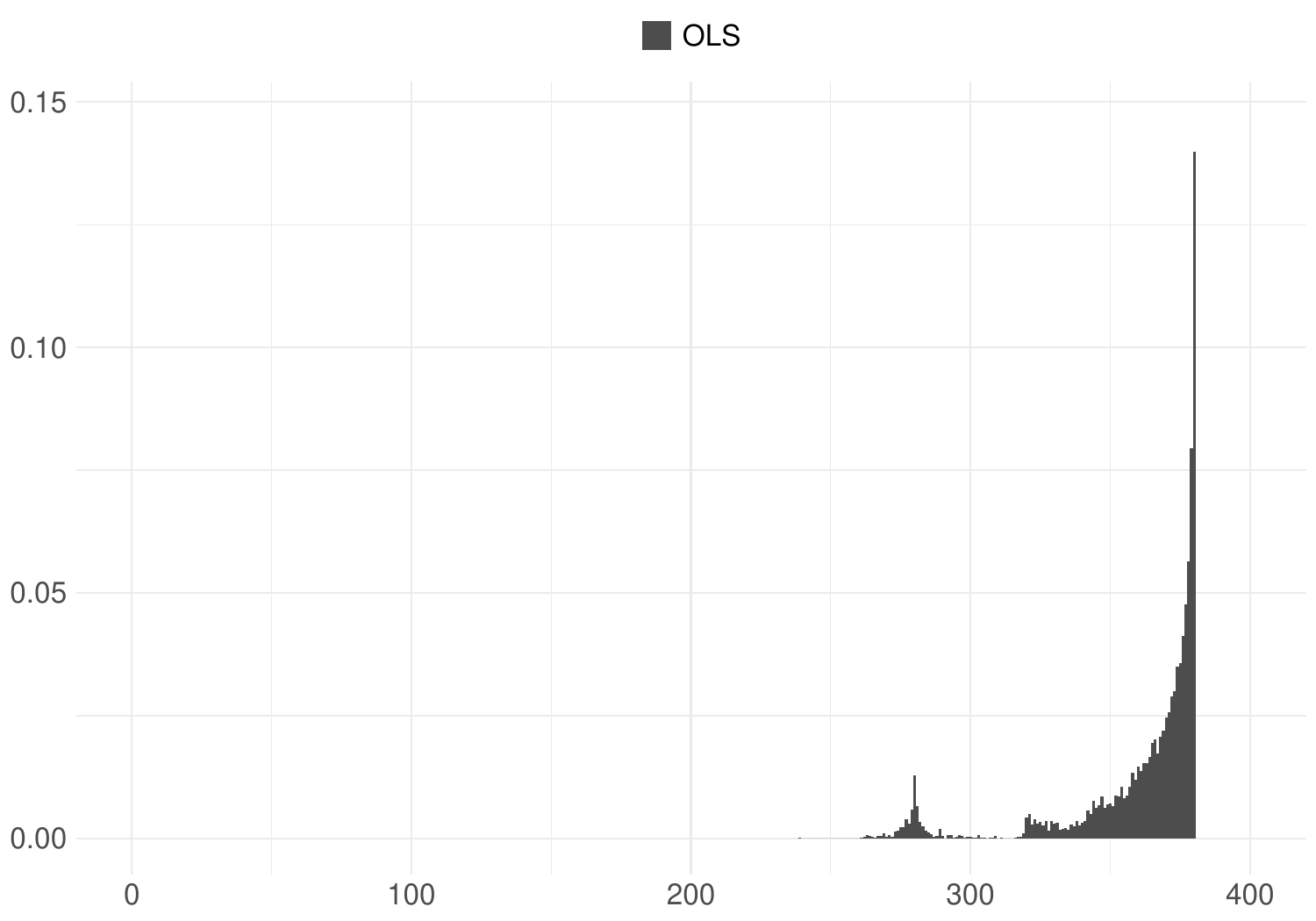}\label{fig2:1:5}}
\subfigure[$T=400$, $c_a=6$, $c_b=6$, $s_0/s_1=1/5$]{\includegraphics[width=0.45\linewidth]{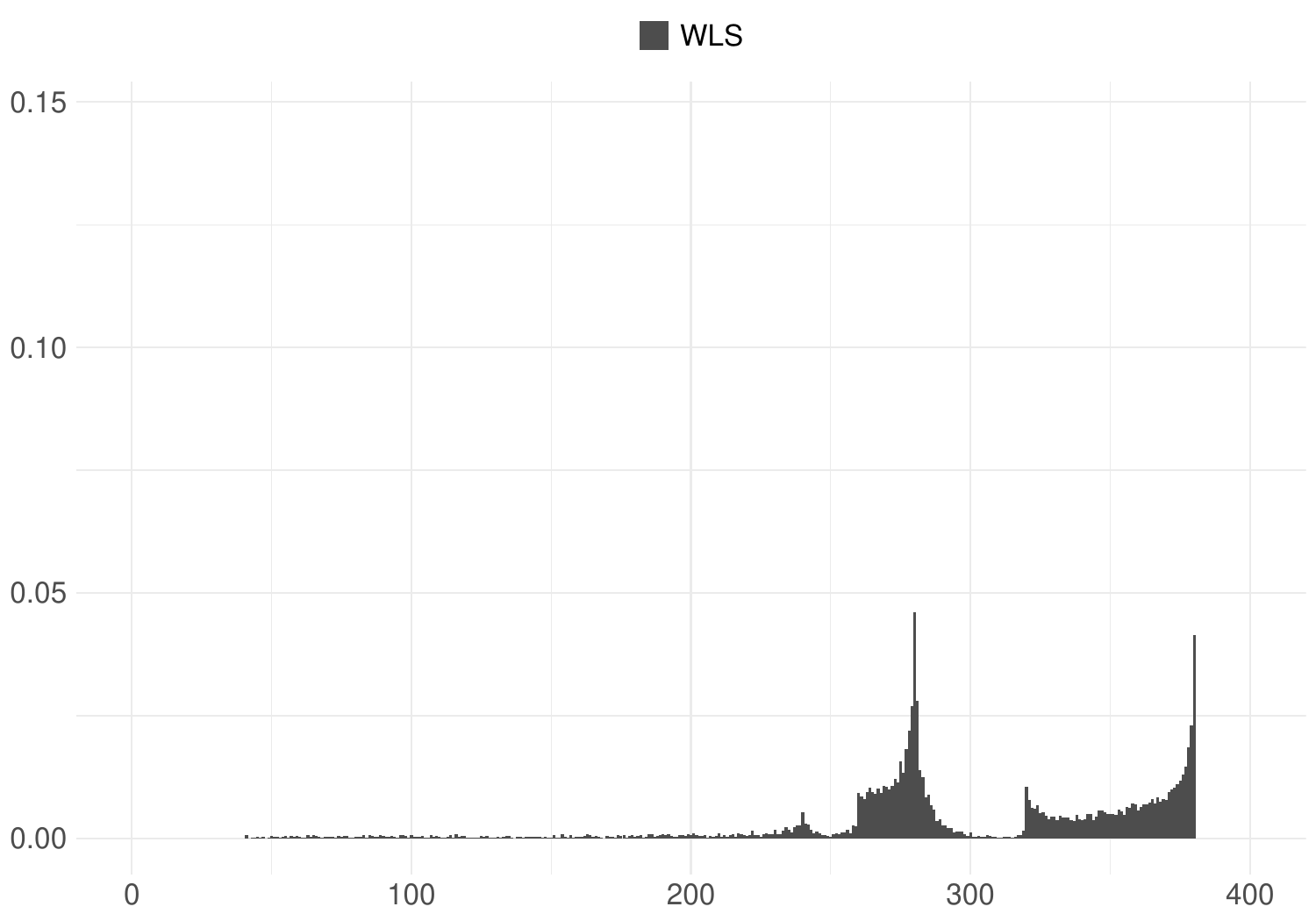}\label{fig2:1:6}}\\
\end{center}%
\caption{Histograms of $\hat{k}_r$ 
for $(\tau_e,\tau_c,\tau_r)=(0.4,0.6,0.7)$,  $\tau=0.8$, $s_0/s_1=1/5$, $T=400$}
\label{fig21}
\end{figure}

\newpage

\begin{figure}[h!]%
\begin{center}%
\subfigure[$T=800$, $c_a=4$, $c_b=6$, $s_0/s_1=1/5$]{\includegraphics[width=0.45\linewidth]{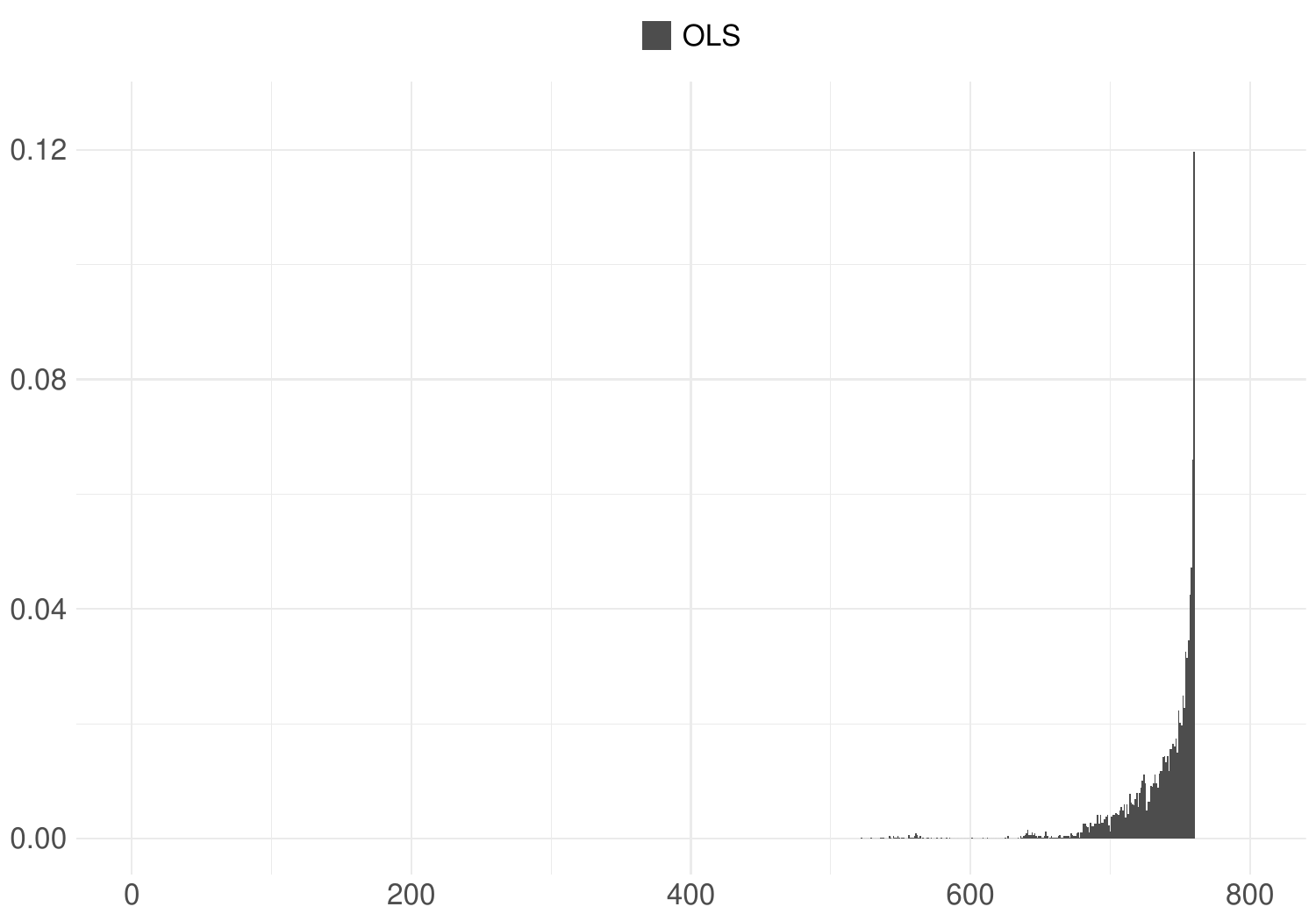}\label{fig2:2:1}}
\subfigure[$T=800$, $c_a=4$, $c_b=6$, $s_0/s_1=1/5$]{\includegraphics[width=0.45\linewidth]{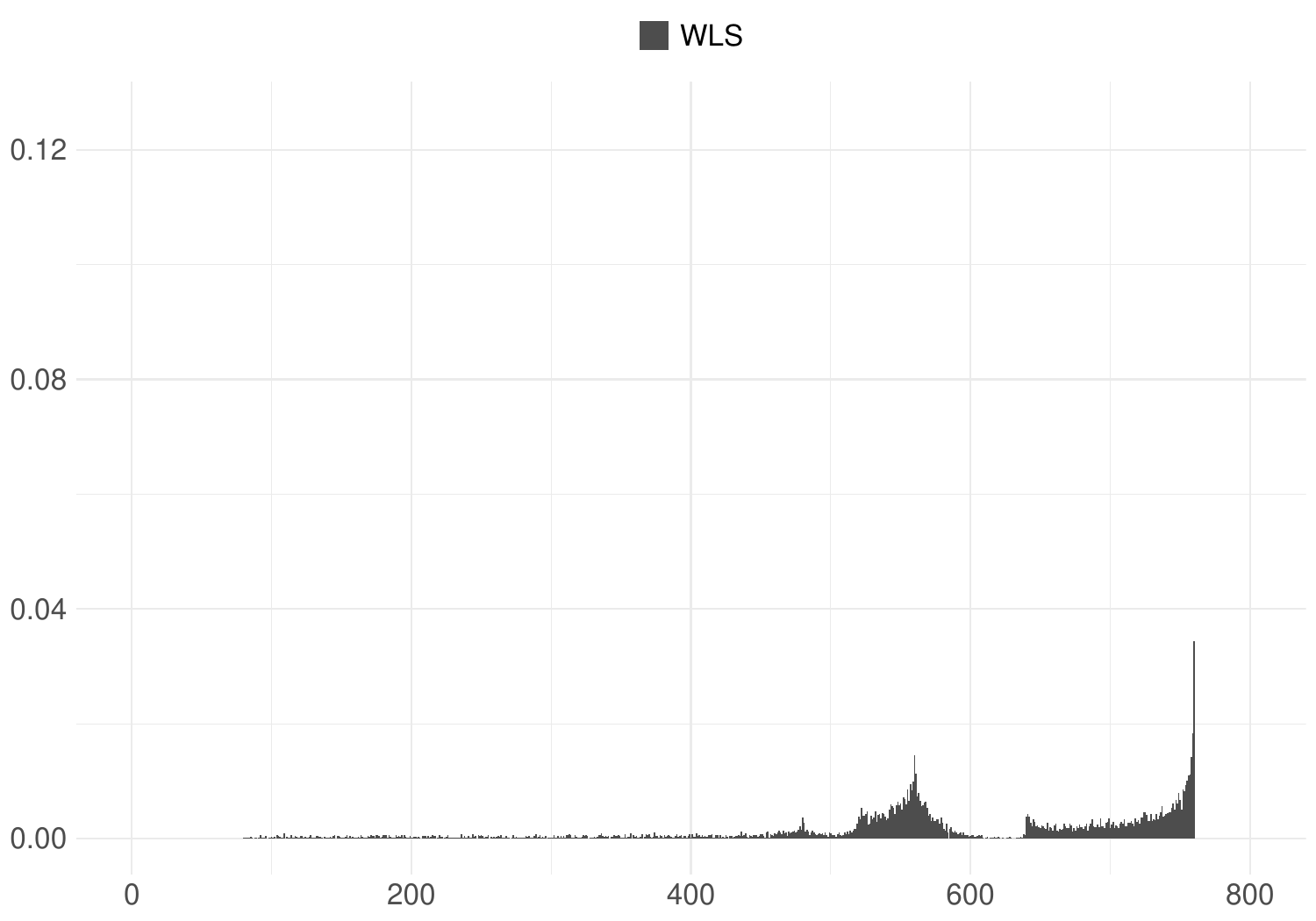}\label{fig2:2:2}}\\
\subfigure[$T=800$, $c_a=5$, $c_b=6$, $s_0/s_1=1/5$]{\includegraphics[width=0.45\linewidth]{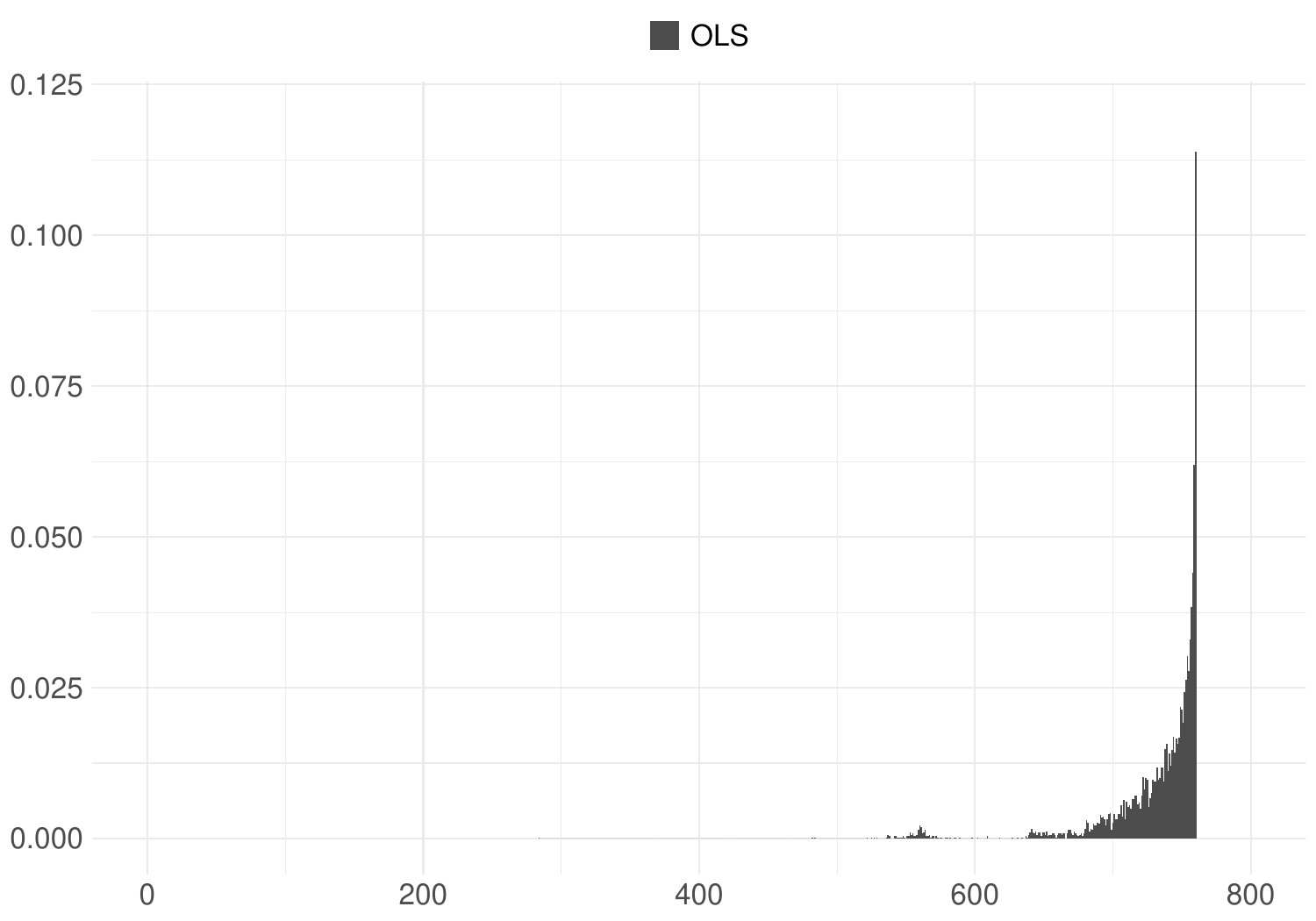}\label{fig2:2:3}}
\subfigure[$T=800$, $c_a=5$, $c_b=6$, $s_0/s_1=1/5$]{\includegraphics[width=0.45\linewidth]{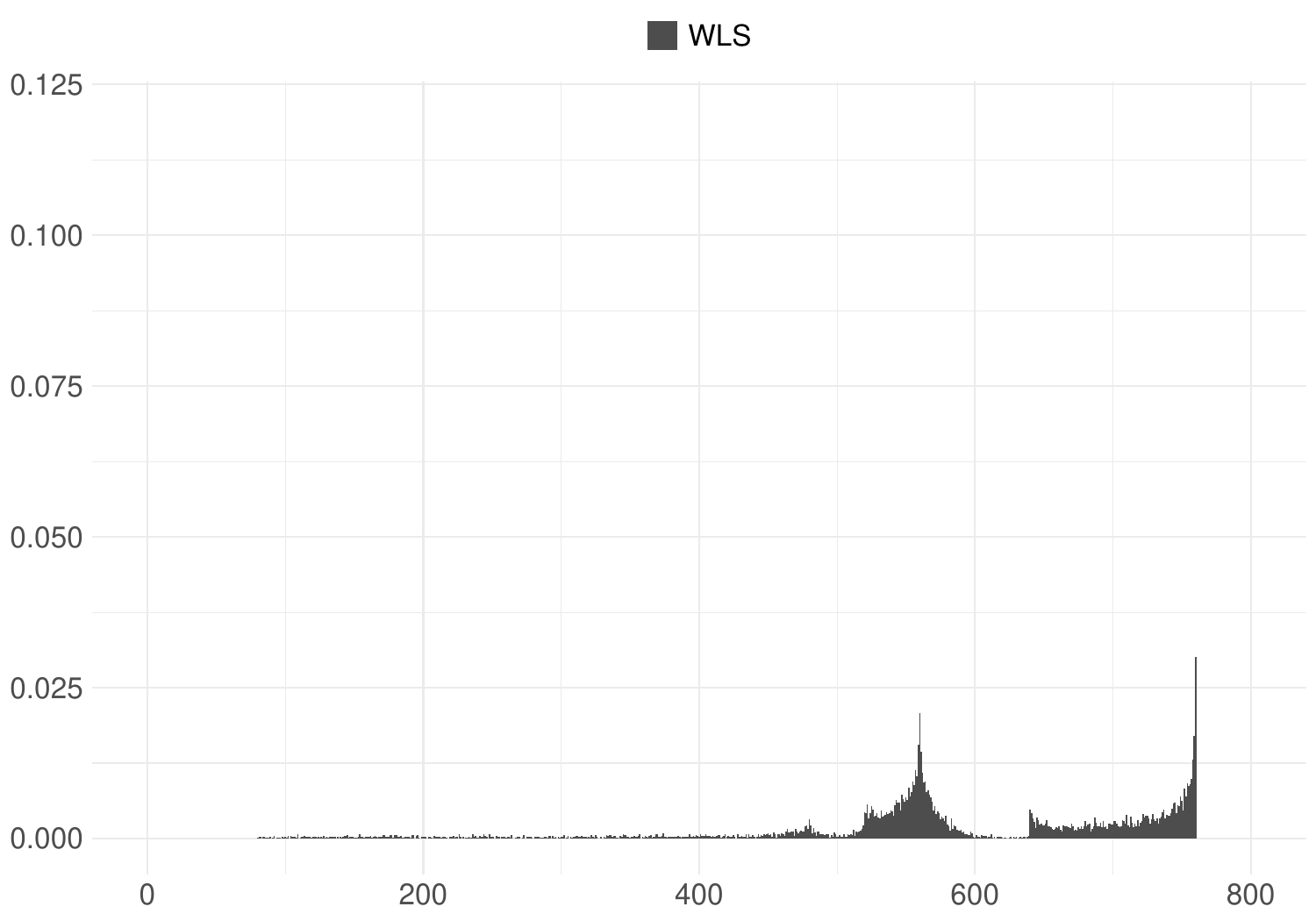}\label{fig2:2:4}}\\
\subfigure[$T=800$, $c_a=6$, $c_b=6$, $s_0/s_1=1/5$]{\includegraphics[width=0.45\linewidth]{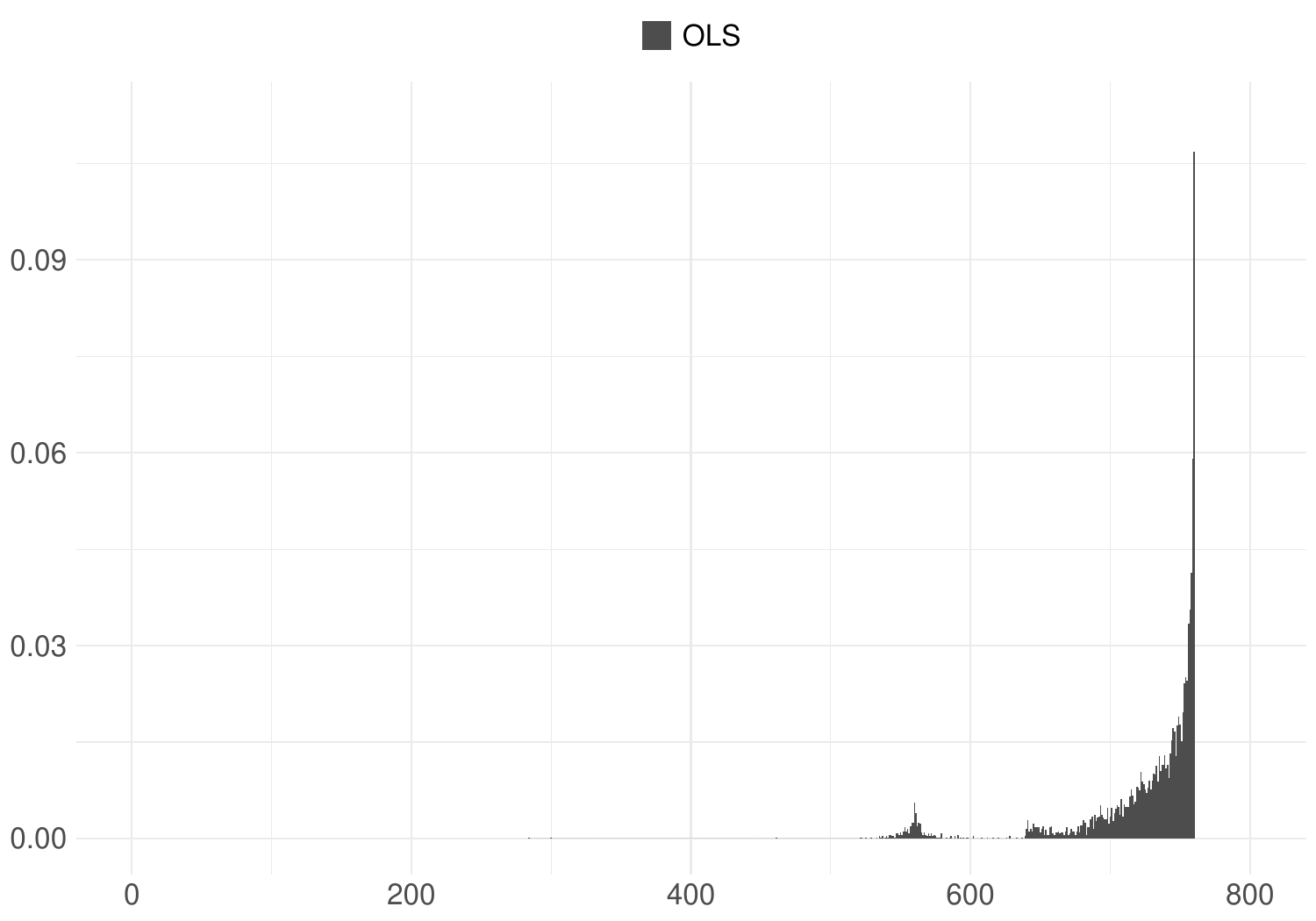}\label{fig2:2:5}}
\subfigure[$T=800$, $c_a=6$, $c_b=6$, $s_0/s_1=1/5$]{\includegraphics[width=0.45\linewidth]{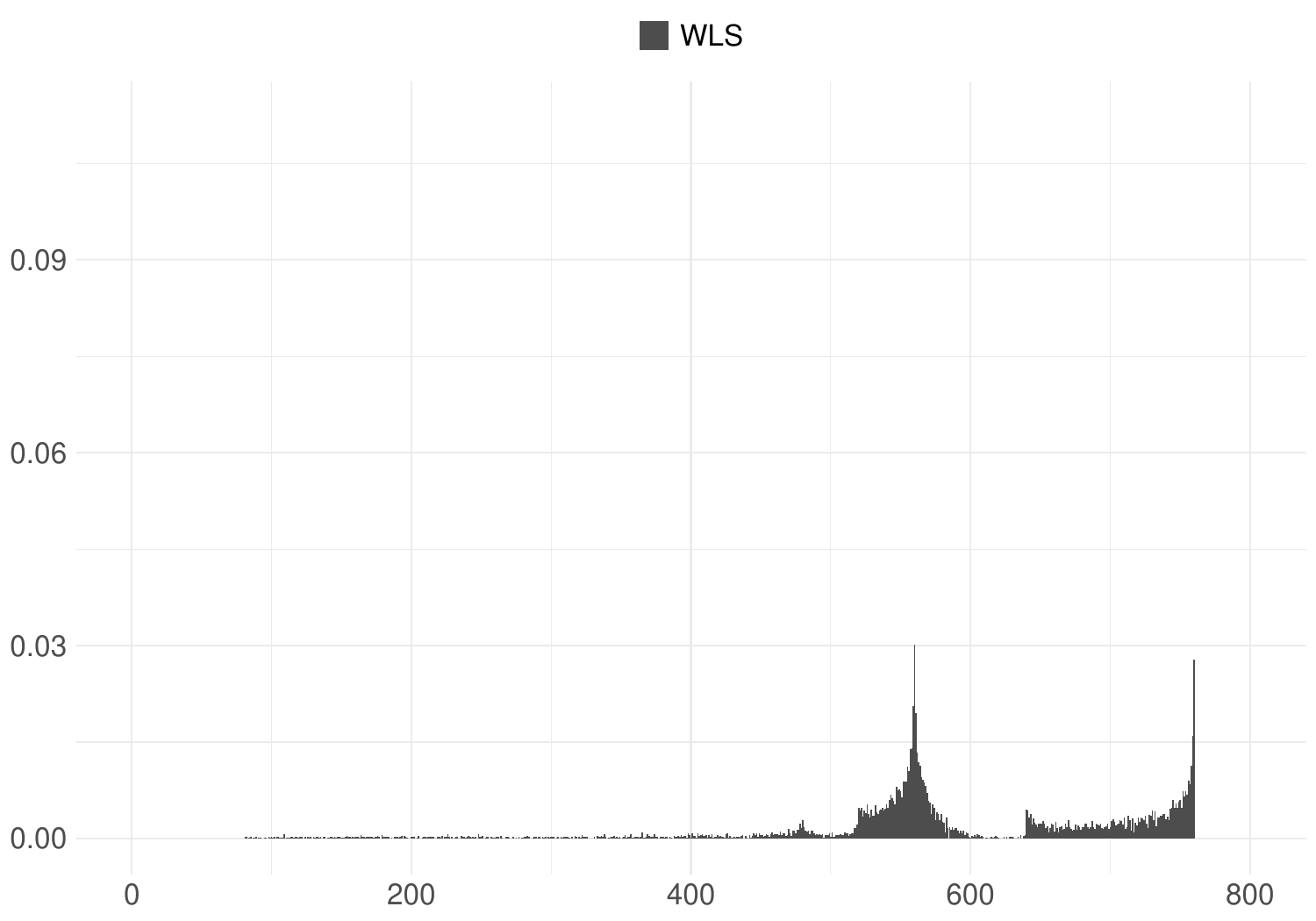}\label{fig2:2:6}}\\
\end{center}%
\caption{Histograms of $\hat{k}_r$ 
for $(\tau_e,\tau_c,\tau_r)=(0.4,0.6,0.7)$,  $\tau=0.8$, $s_0/s_1=1/5$, $T=800$}
\label{fig22}
\end{figure}

\newpage

\begin{figure}[h!]%
\begin{center}%
\subfigure[$T=400$, $c_a=4$, $c_b=6$, $s_0/s_1=1$]{\includegraphics[width=0.45\linewidth]{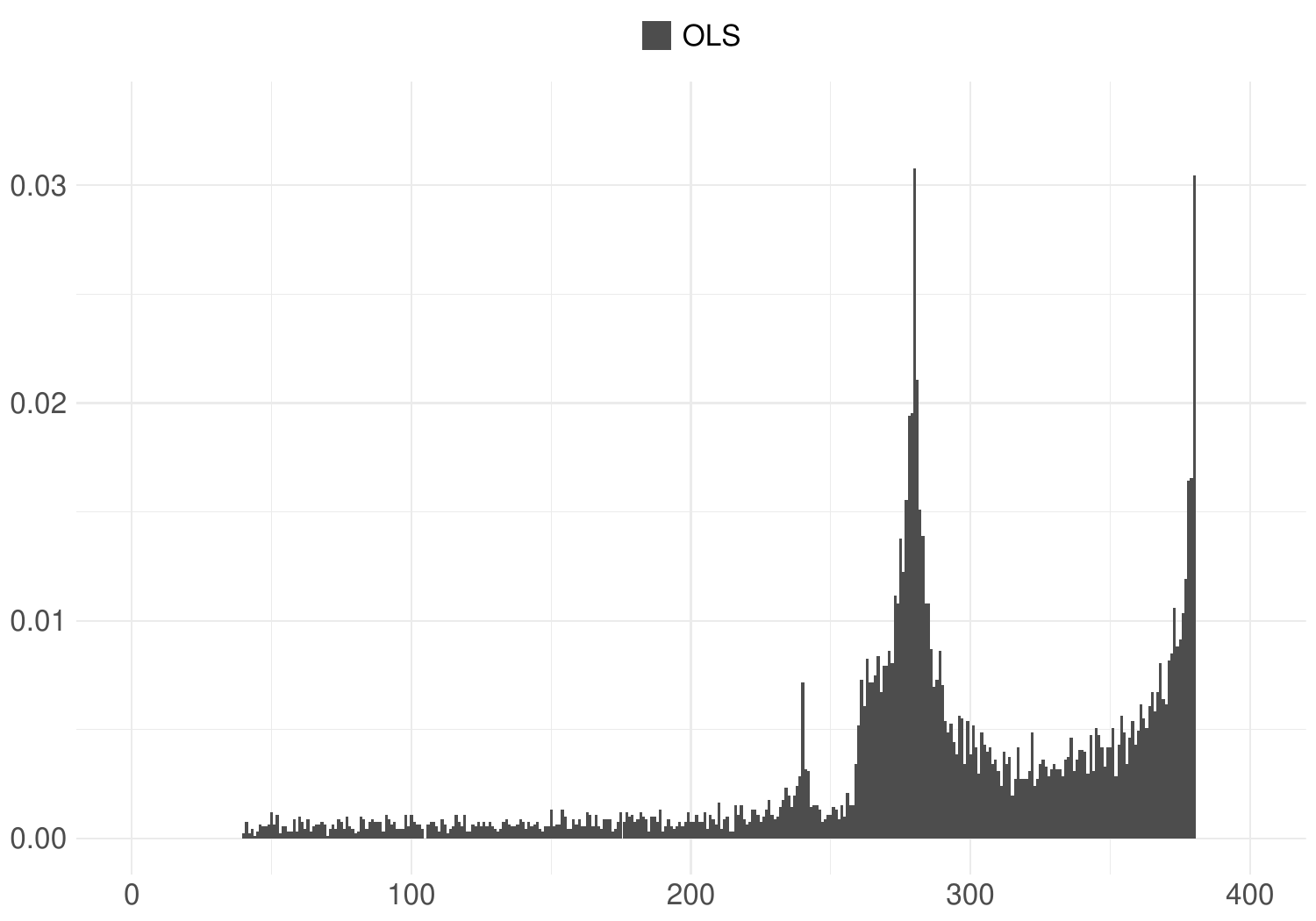}\label{fig2:3:1}}
\subfigure[$T=400$, $c_a=4$, $c_b=6$, $s_0/s_1=1$]{\includegraphics[width=0.45\linewidth]{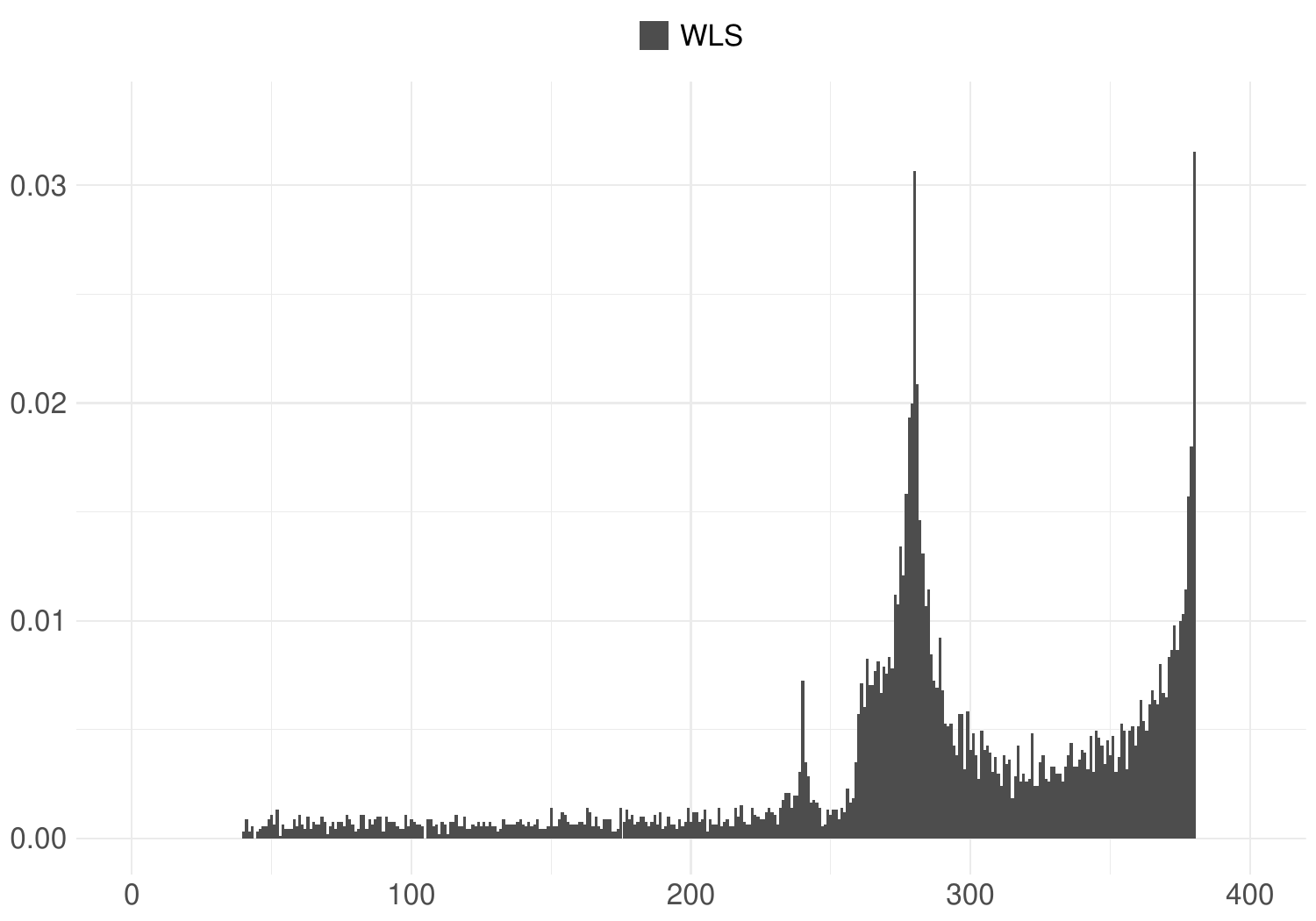}\label{fig2:3:2}}\\
\subfigure[$T=400$, $c_a=5$, $c_b=6$, $s_0/s_1=1$]{\includegraphics[width=0.45\linewidth]{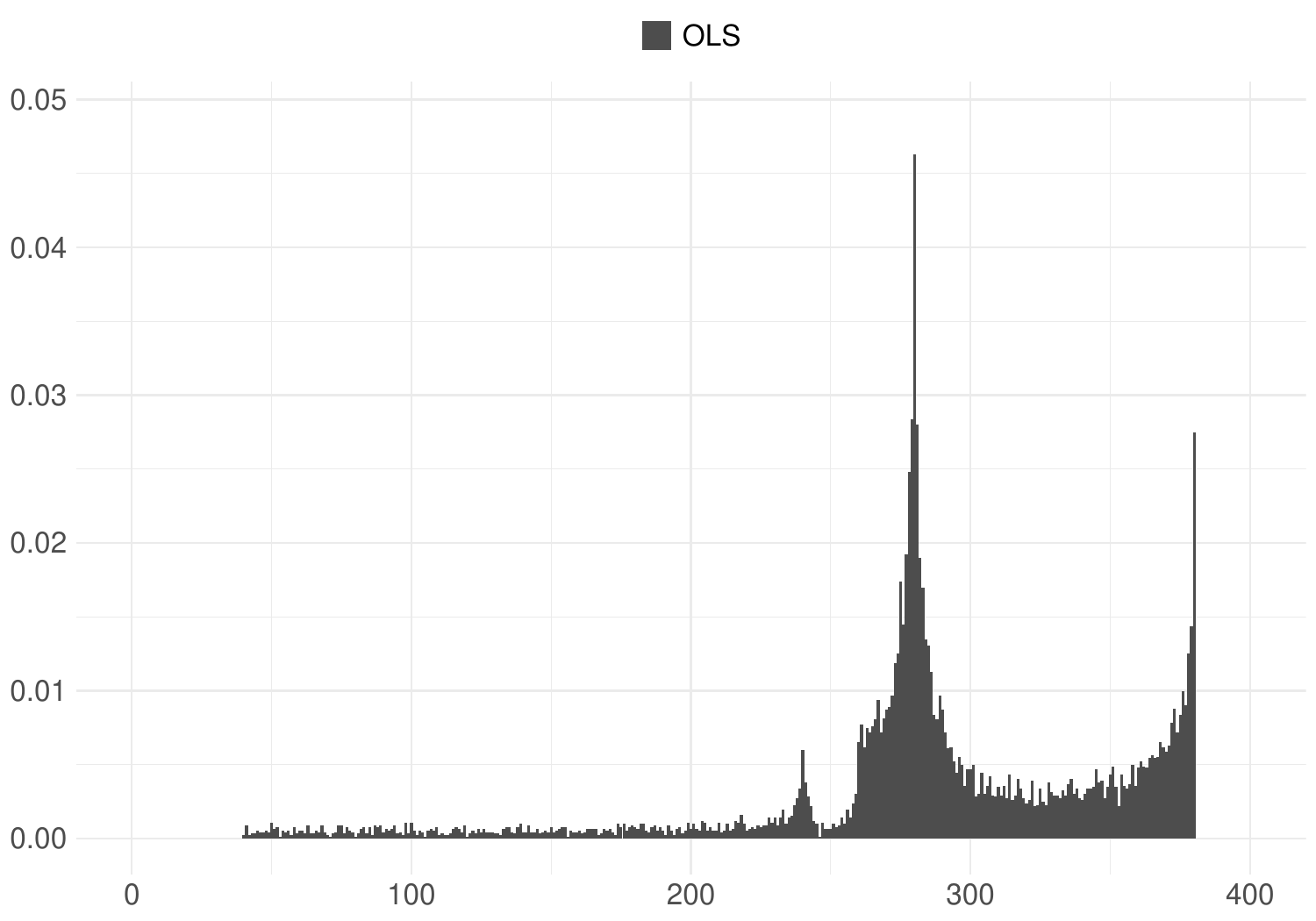}\label{fig2:3:3}}
\subfigure[$T=400$, $c_a=5$, $c_b=6$, $s_0/s_1=1$]{\includegraphics[width=0.45\linewidth]{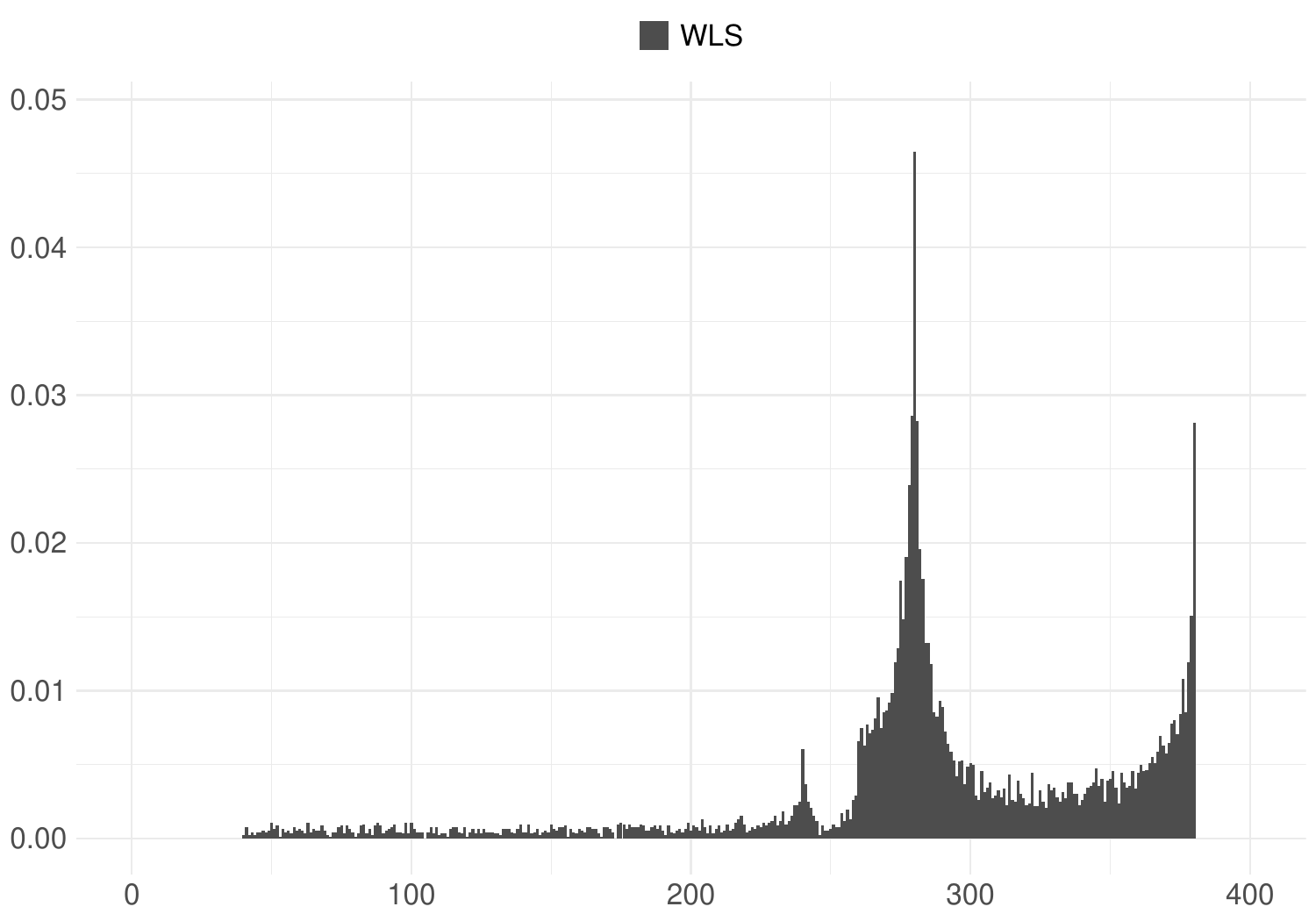}\label{fig2:3:4}}\\
\subfigure[$T=400$, $c_a=6$, $c_b=6$, $s_0/s_1=1$]{\includegraphics[width=0.45\linewidth]{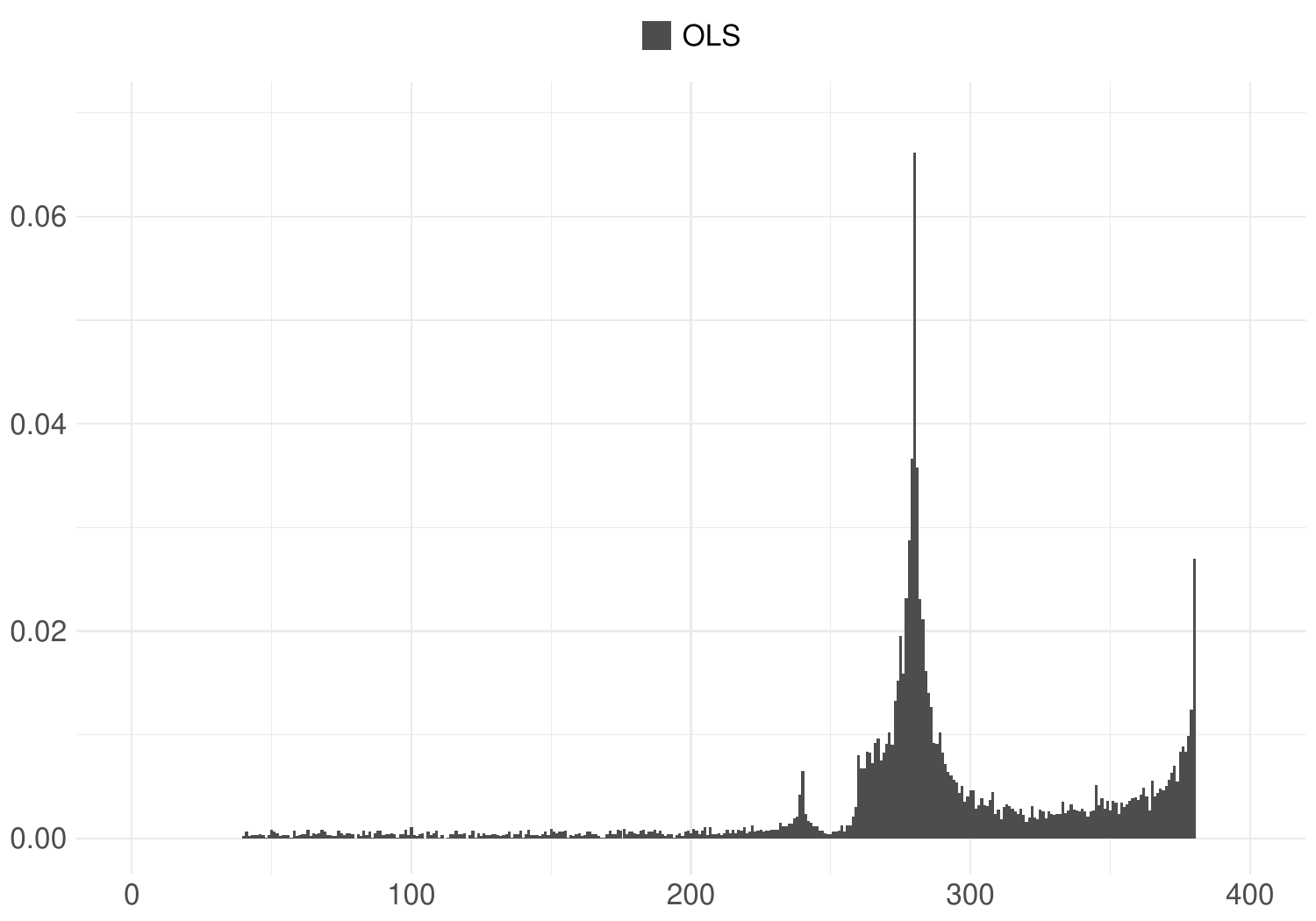}\label{fig2:3:5}}
\subfigure[$T=400$, $c_a=6$, $c_b=6$, $s_0/s_1=1$]{\includegraphics[width=0.45\linewidth]{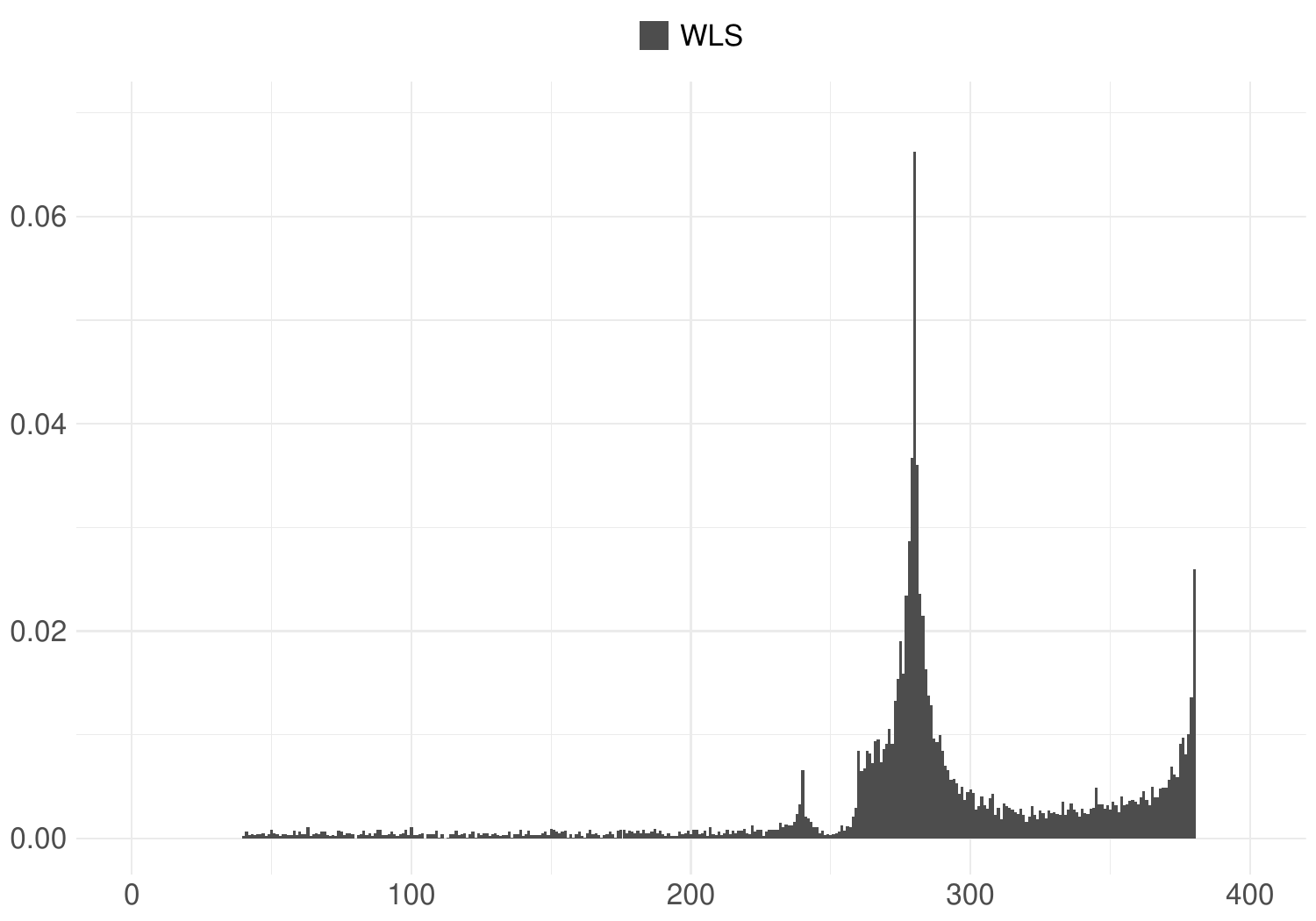}\label{fig2:3:6}}\\
\end{center}%
\caption{Histograms of $\hat{k}_r$ 
for $(\tau_e,\tau_c,\tau_r)=(0.4,0.6,0.7)$,  $\tau=0.8$, $s_0/s_1=1$, $T=400$}
\label{fig23}
\end{figure}

\newpage

\begin{figure}[h!]%
\begin{center}%
\subfigure[$T=800$, $c_a=4$, $c_b=6$, $s_0/s_1=1$]{\includegraphics[width=0.45\linewidth]{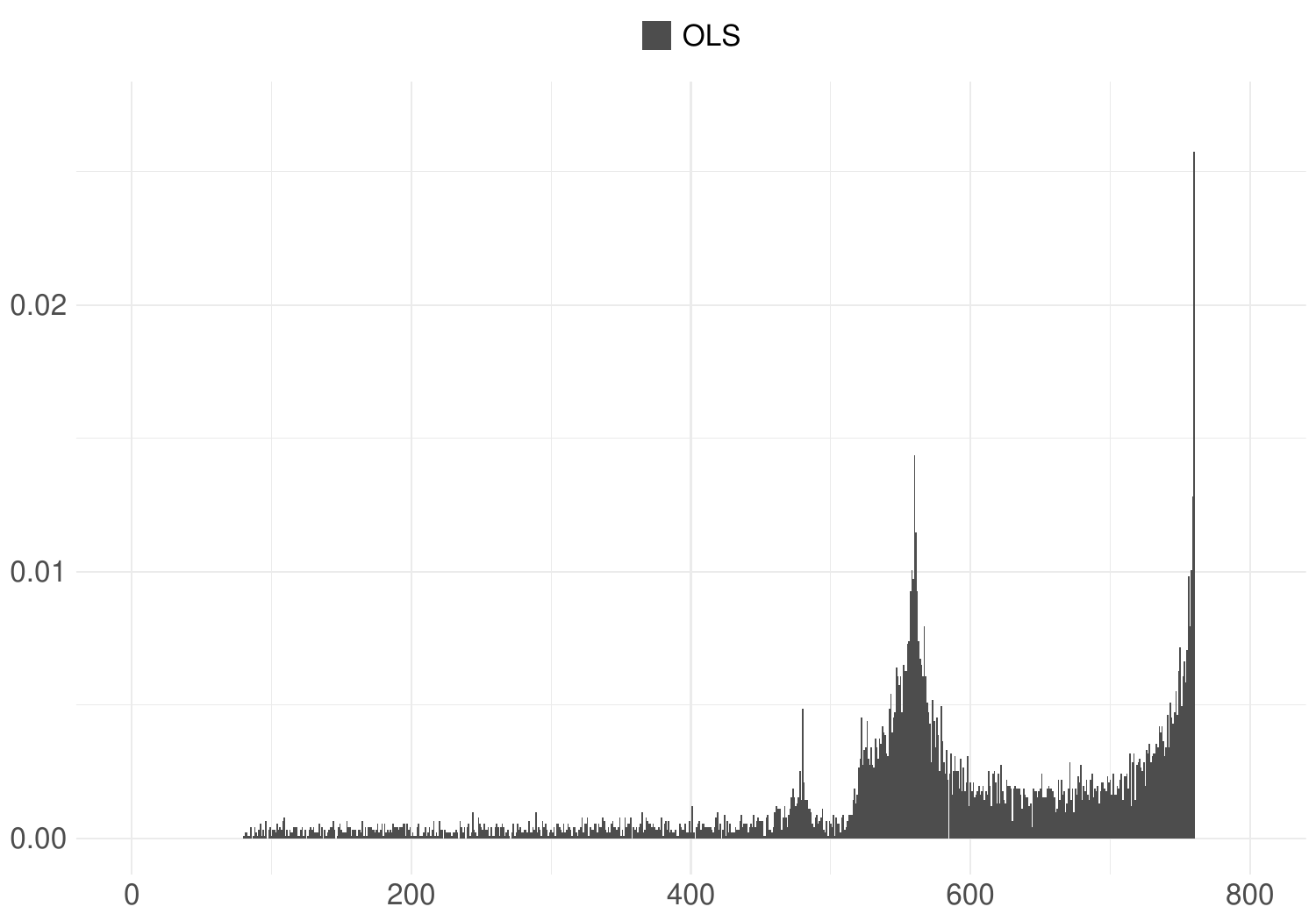}\label{fig2:4:1}}
\subfigure[$T=800$, $c_a=4$, $c_b=6$, $s_0/s_1=1$]{\includegraphics[width=0.45\linewidth]{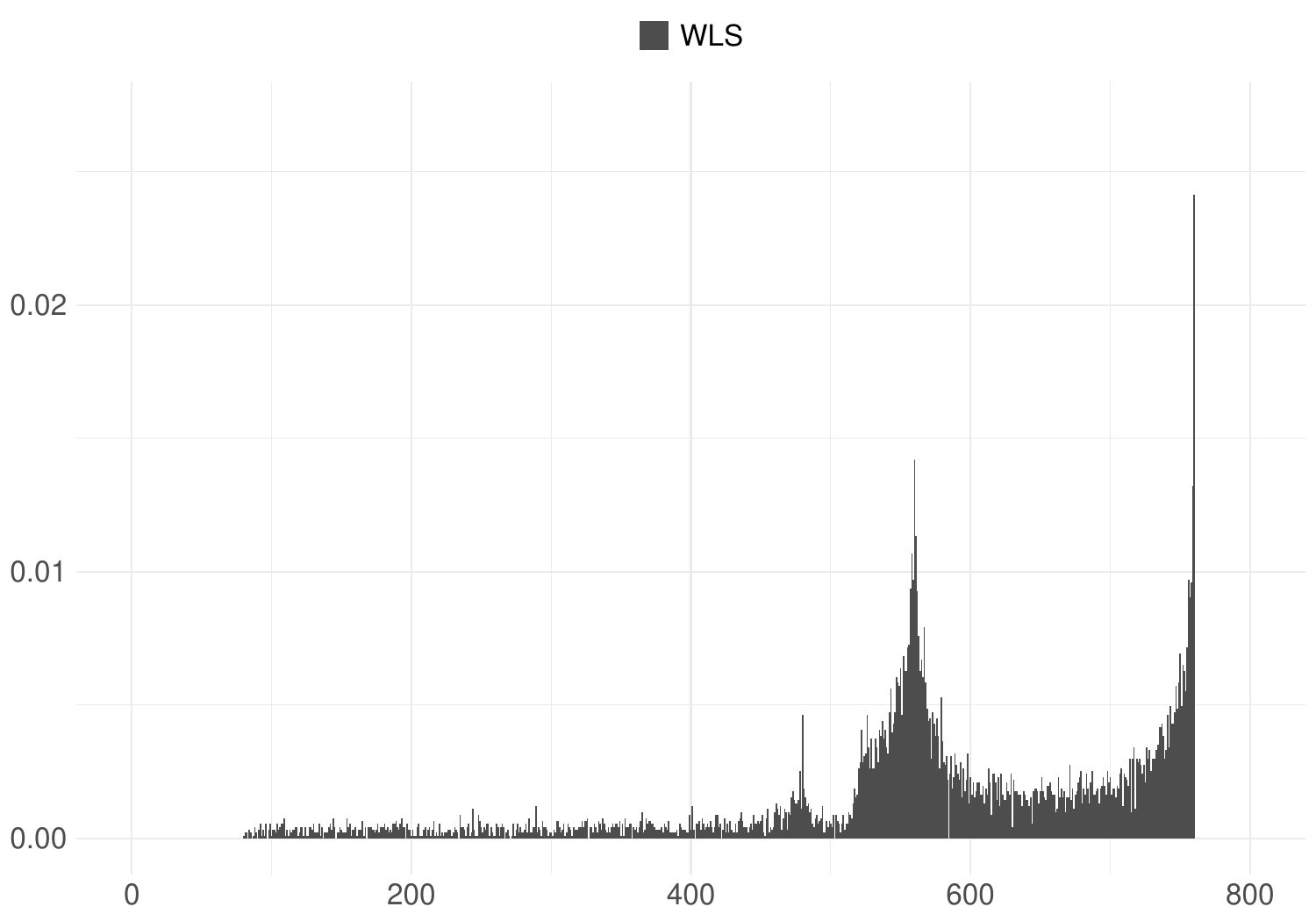}\label{fig2:4:2}}\\
\subfigure[$T=800$, $c_a=5$, $c_b=6$, $s_0/s_1=1$]{\includegraphics[width=0.45\linewidth]{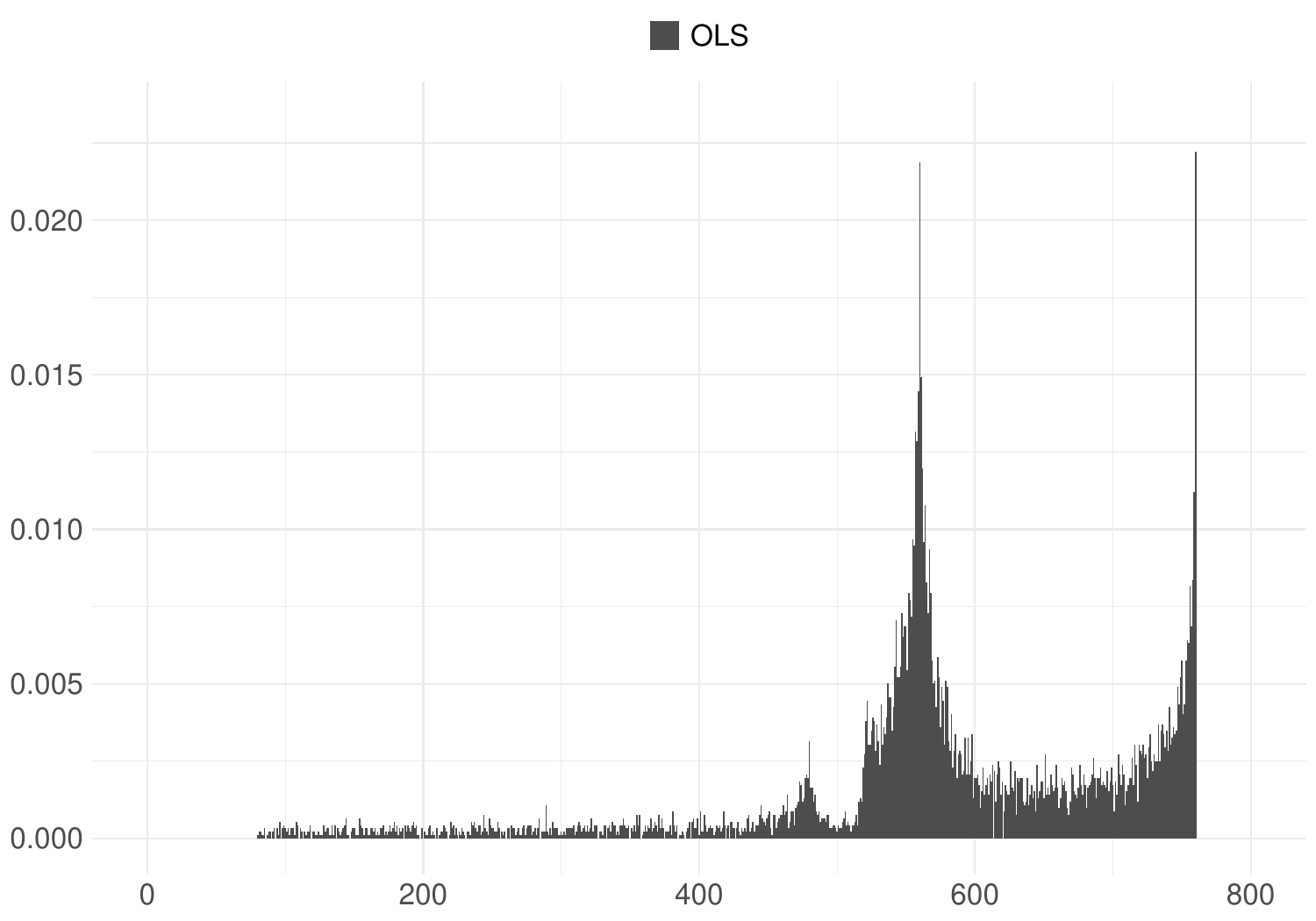}\label{fig2:4:3}}
\subfigure[$T=800$, $c_a=5$, $c_b=6$, $s_0/s_1=1$]{\includegraphics[width=0.45\linewidth]{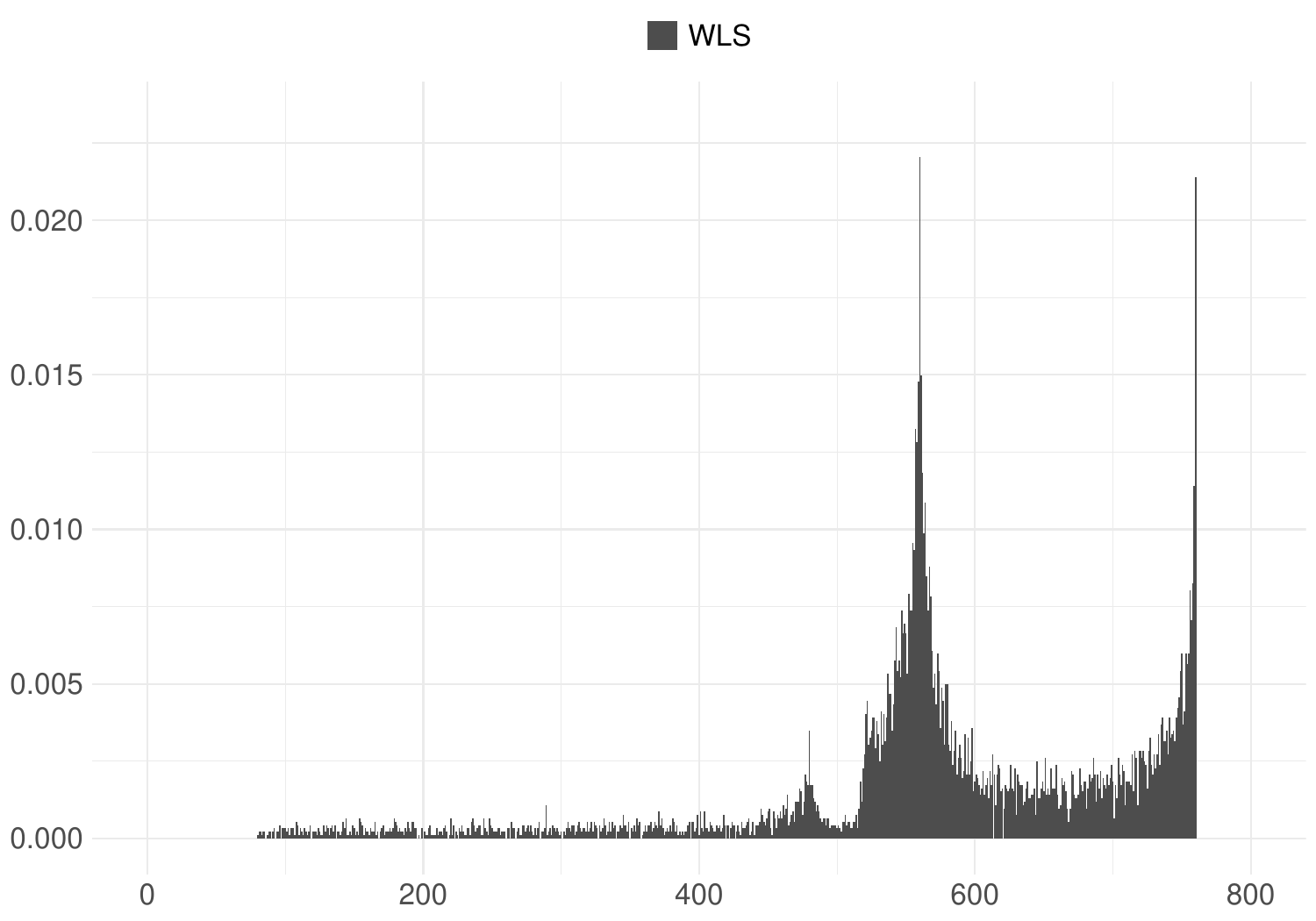}\label{fig2:4:4}}\\
\subfigure[$T=800$, $c_a=6$, $c_b=6$, $s_0/s_1=1$]{\includegraphics[width=0.45\linewidth]{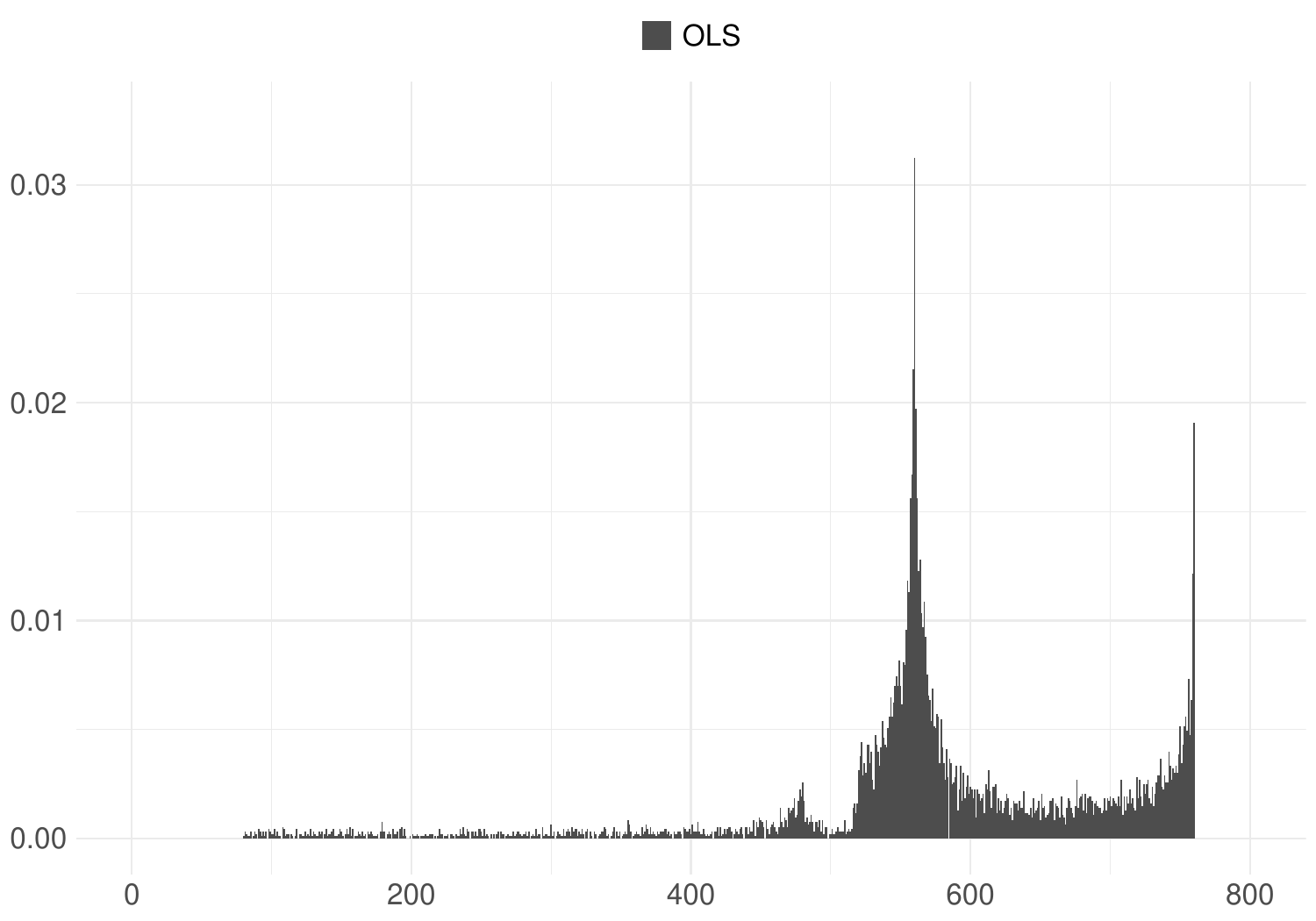}\label{fig2:4:5}}
\subfigure[$T=800$, $c_a=6$, $c_b=6$, $s_0/s_1=1$]{\includegraphics[width=0.45\linewidth]{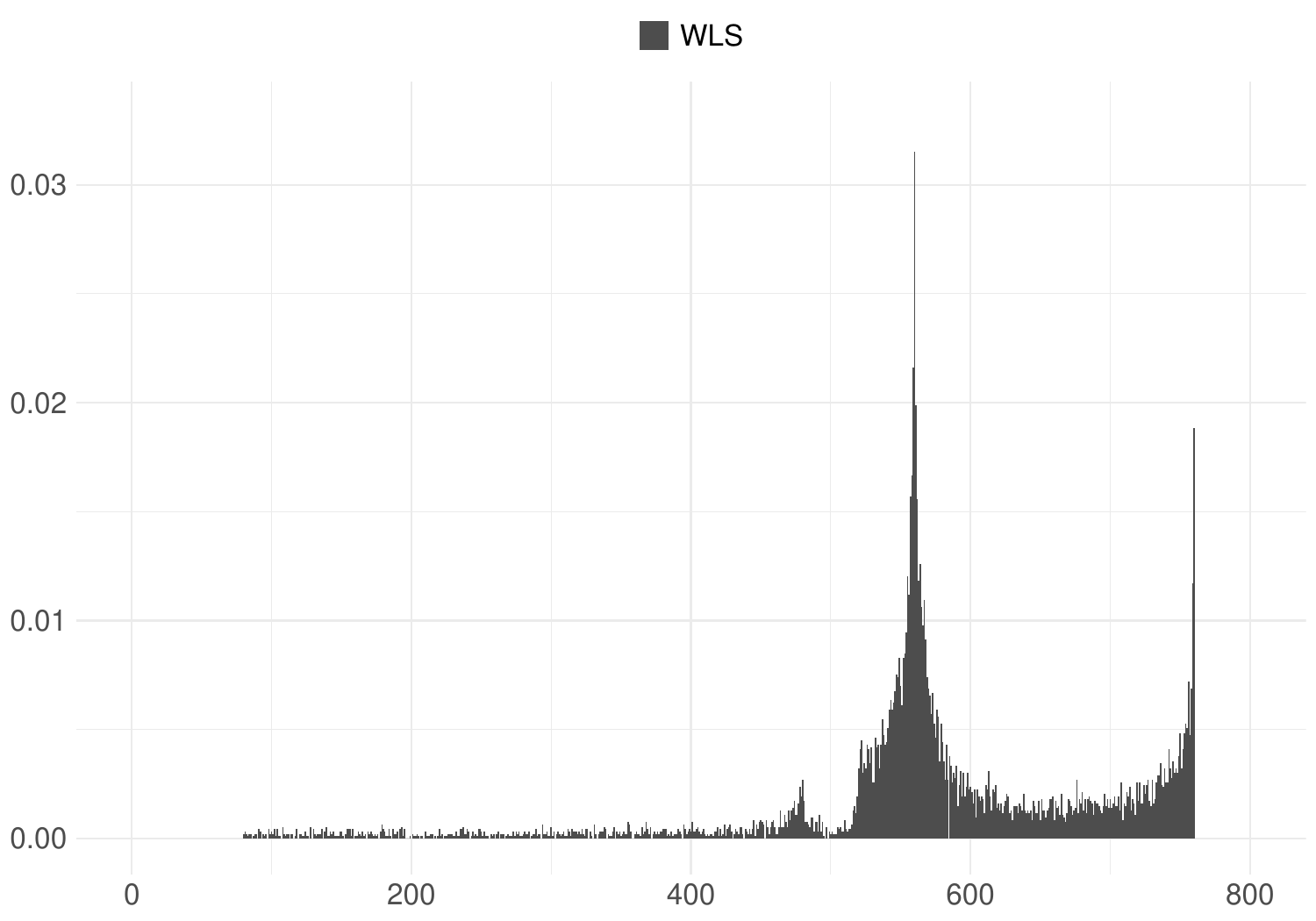}\label{fig2:4:6}}\\
\end{center}%
\caption{Histograms of $\hat{k}_r$ 
for $(\tau_e,\tau_c,\tau_r)=(0.4,0.6,0.7)$,  $\tau=0.8$, $s_0/s_1=1$, $T=800$}
\label{fig24}
\end{figure}

\newpage

\begin{figure}[h!]%
\begin{center}%
\subfigure[$T=400$, $c_a=4$, $c_b=6$, $s_0/s_1=5$]{\includegraphics[width=0.45\linewidth]{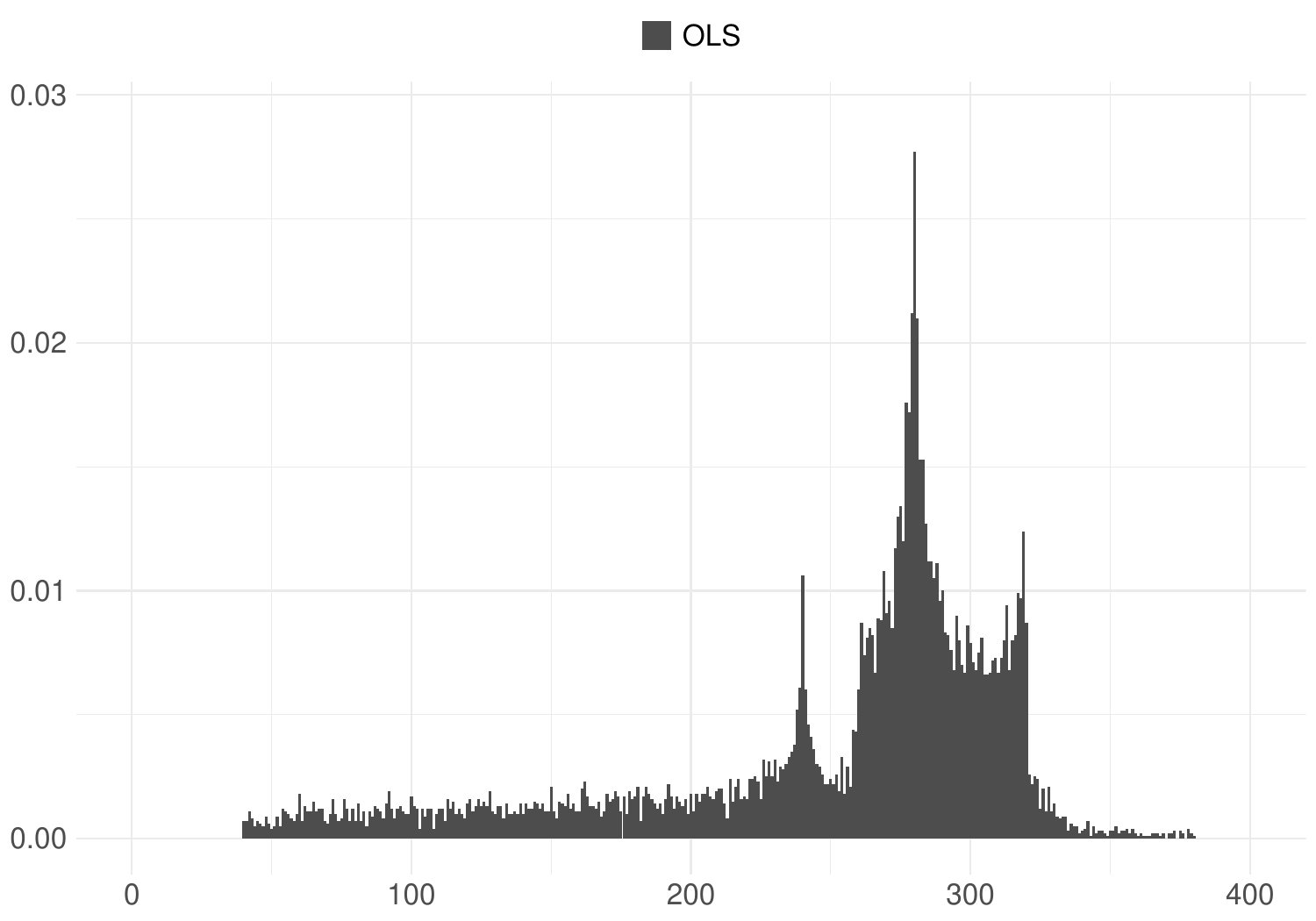}\label{fig2:5:1}}
\subfigure[$T=400$, $c_a=4$, $c_b=6$, $s_0/s_1=5$]{\includegraphics[width=0.45\linewidth]{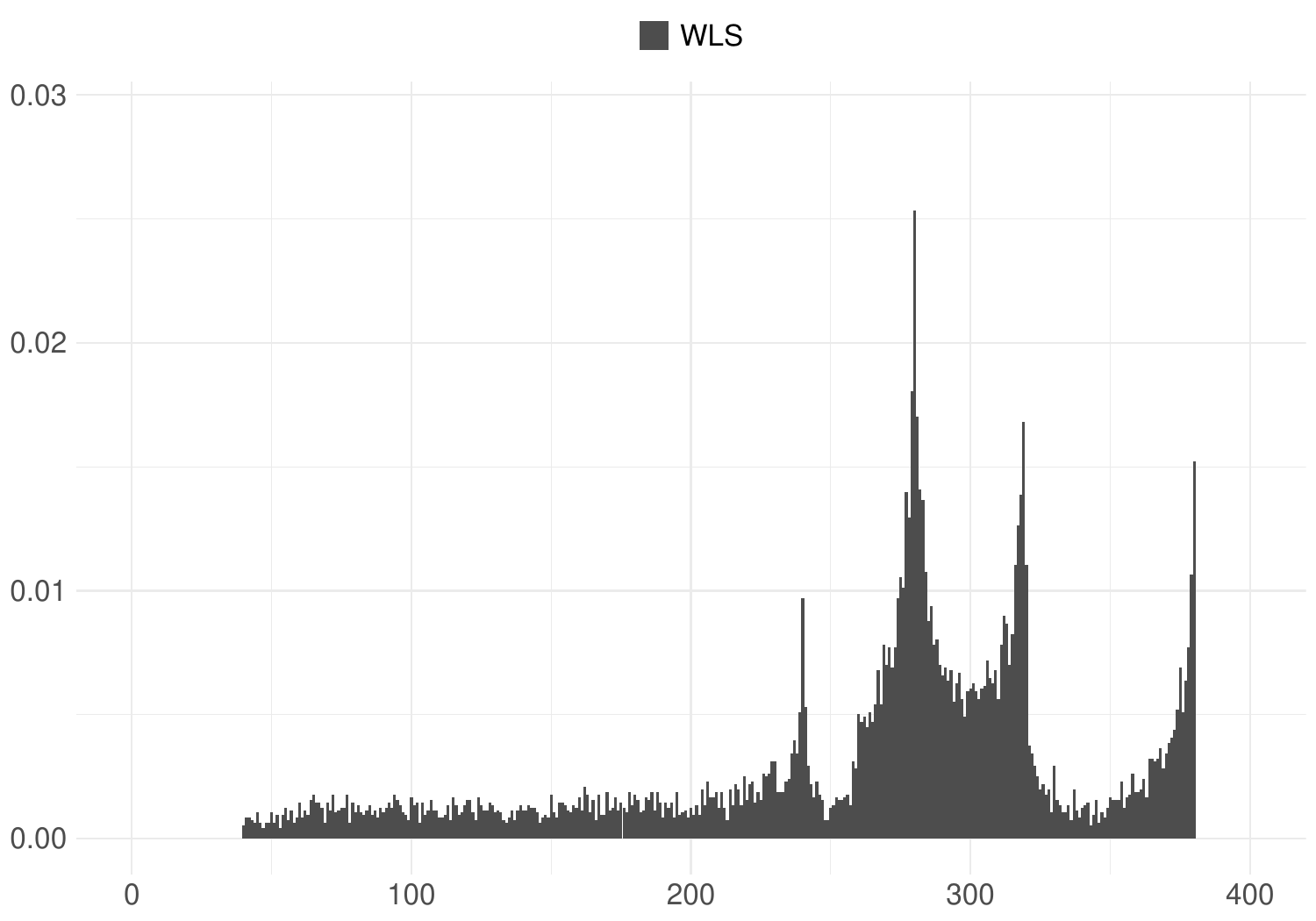}\label{fig2:5:2}}\\
\subfigure[$T=400$, $c_a=5$, $c_b=6$, $s_0/s_1=5$]{\includegraphics[width=0.45\linewidth]{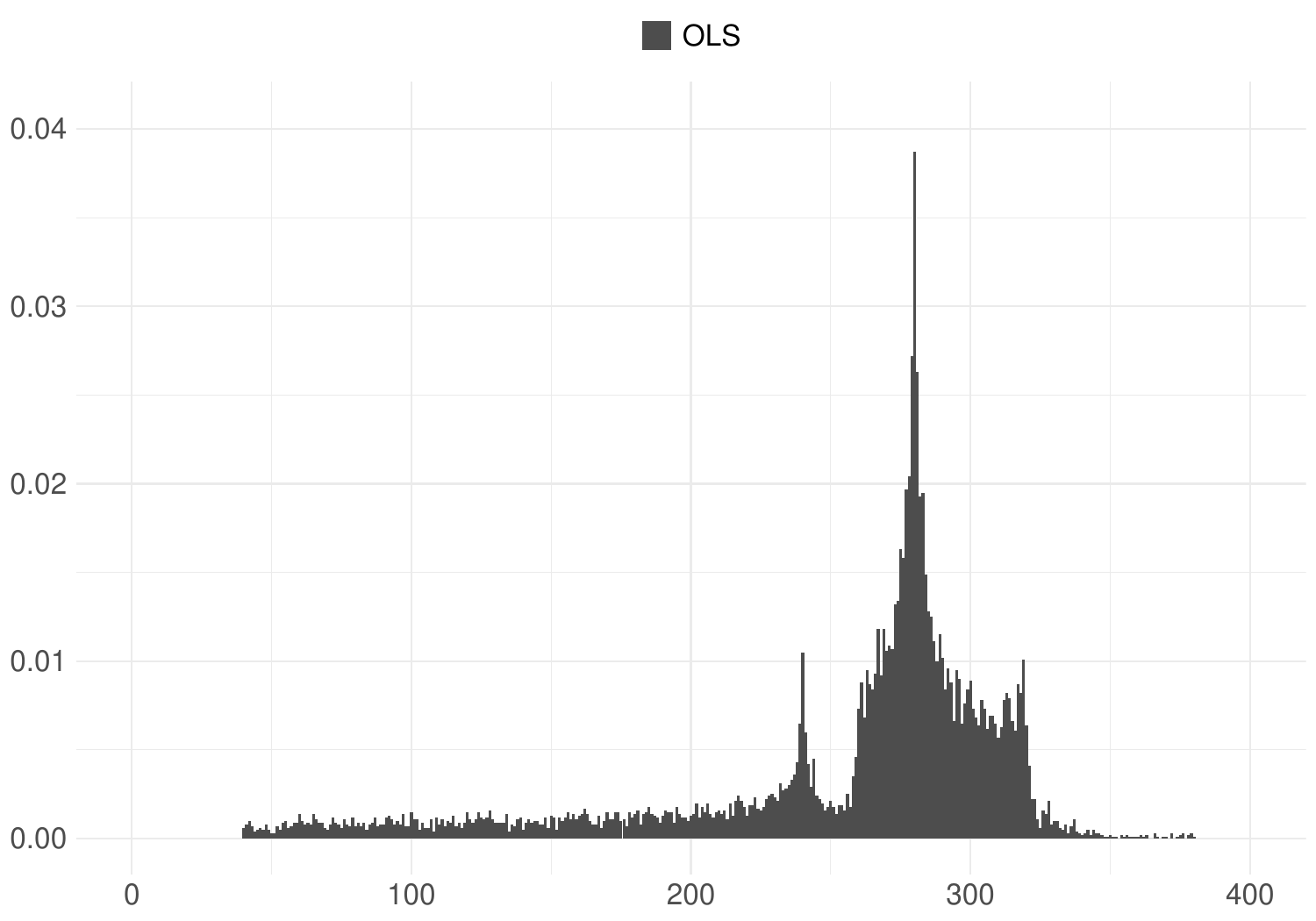}\label{fig2:5:3}}
\subfigure[$T=400$, $c_a=5$, $c_b=6$, $s_0/s_1=5$]{\includegraphics[width=0.45\linewidth]{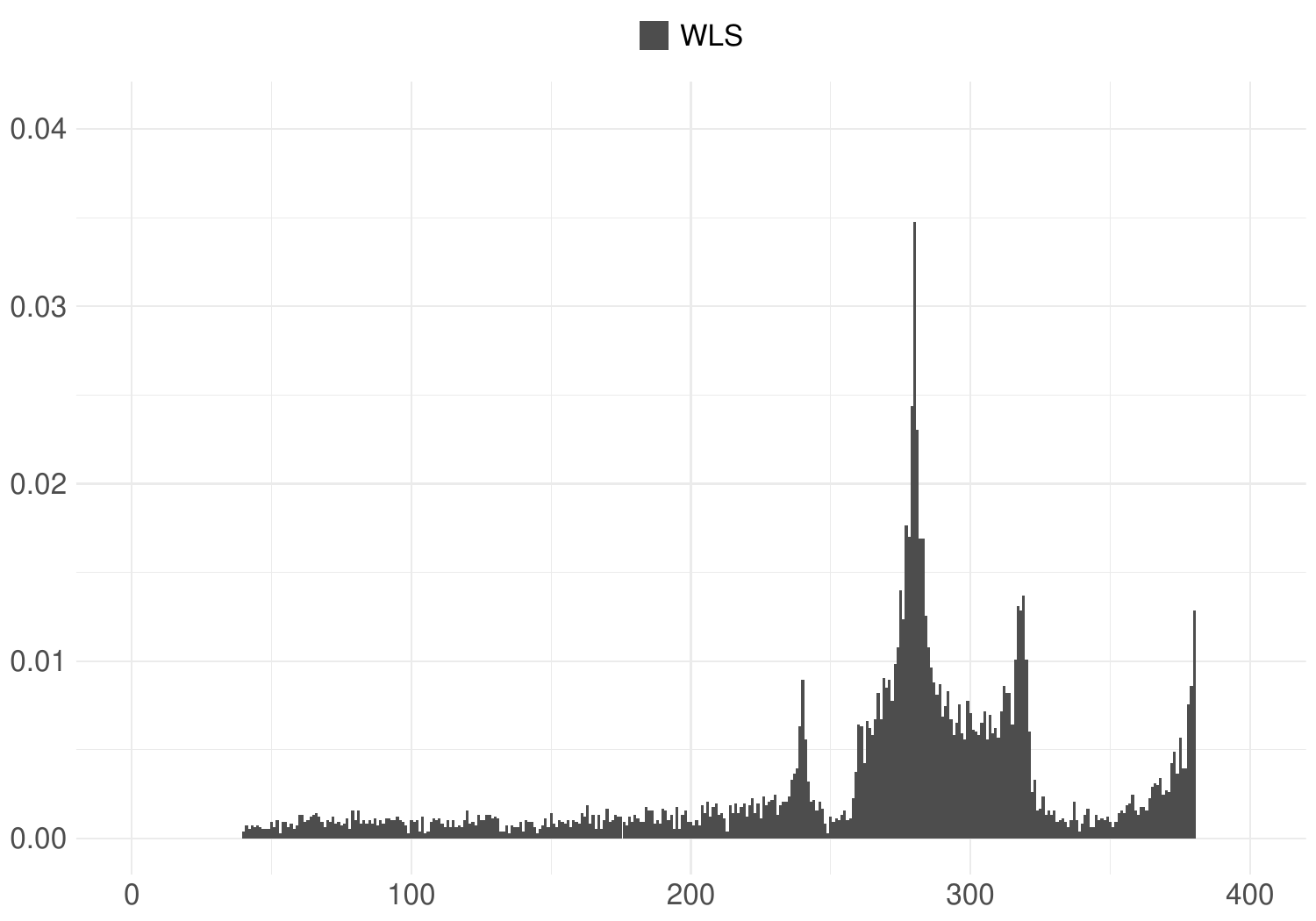}\label{fig2:5:4}}\\
\subfigure[$T=400$, $c_a=6$, $c_b=6$, $s_0/s_1=5$]{\includegraphics[width=0.45\linewidth]{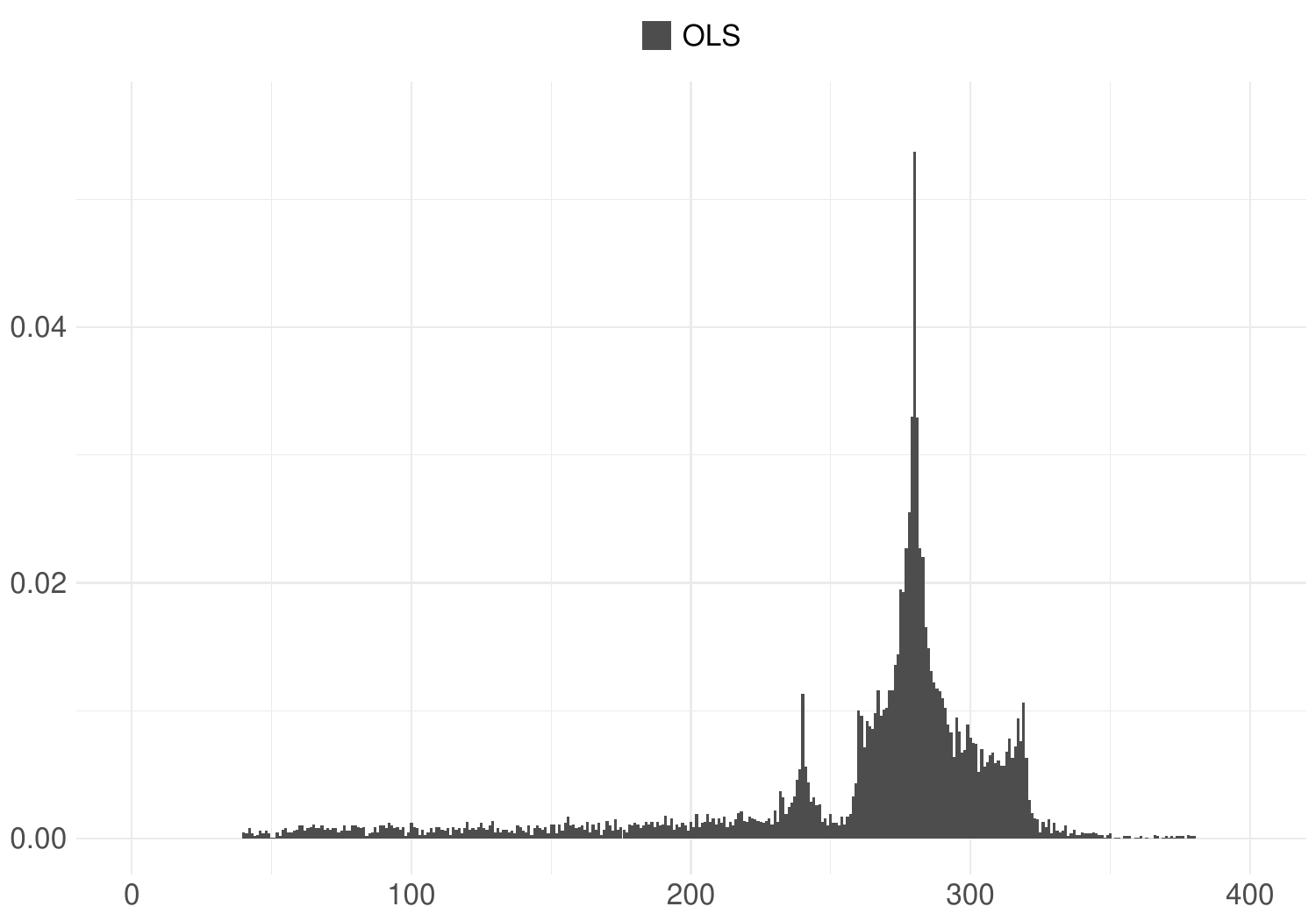}\label{fig2:5:5}}
\subfigure[$T=400$, $c_a=6$, $c_b=6$, $s_0/s_1=5$]{\includegraphics[width=0.45\linewidth]{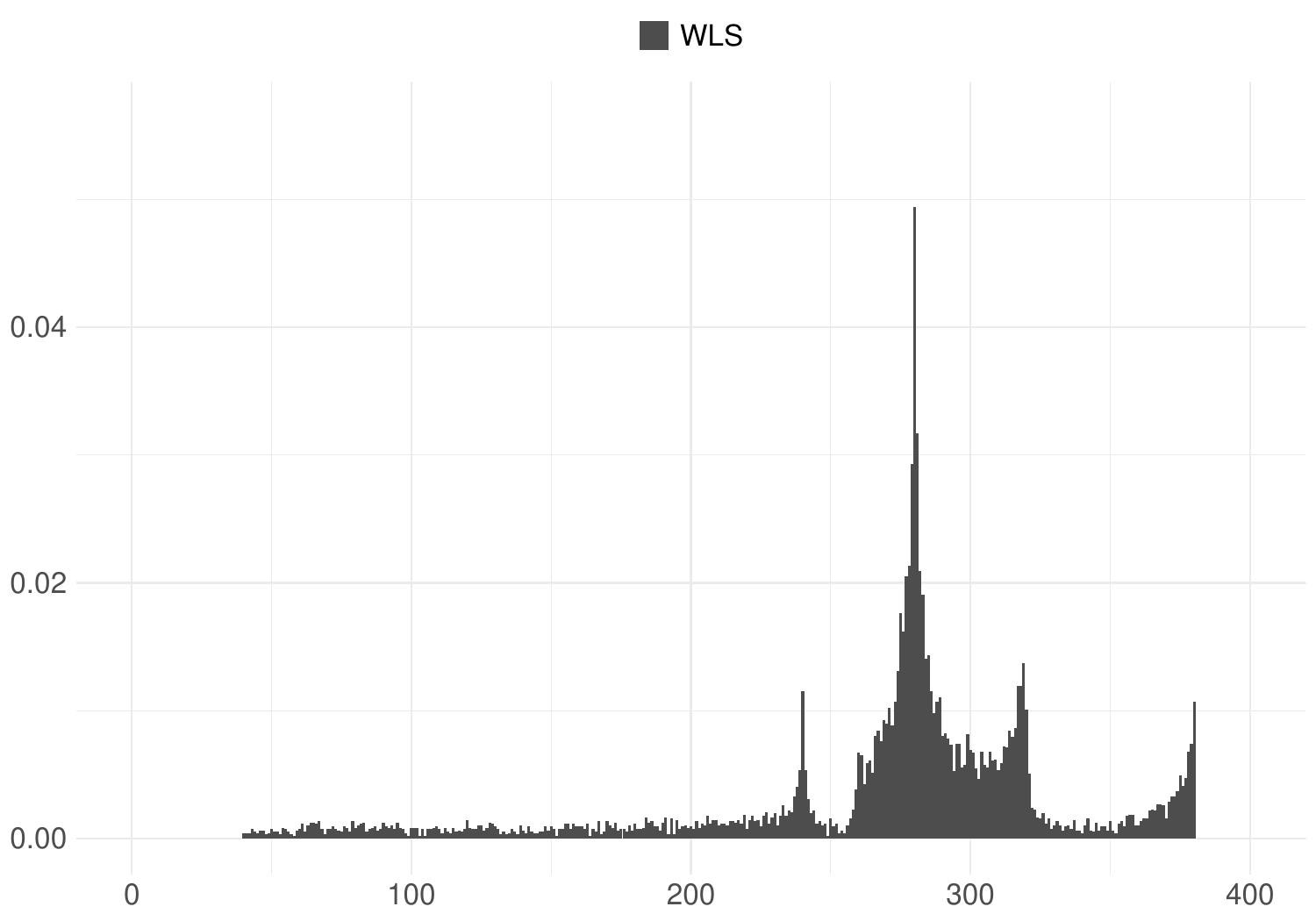}\label{fig2:5:6}}\\
\end{center}%
\caption{Histograms of $\hat{k}_r$ 
for $(\tau_e,\tau_c,\tau_r)=(0.4,0.6,0.7)$,  $\tau=0.8$, $s_0/s_1=5$, $T=400$}
\label{fig25}
\end{figure}

\newpage

\begin{figure}[h!]%
\begin{center}%
\subfigure[$T=800$, $c_a=4$, $c_b=6$, $s_0/s_1=5$]{\includegraphics[width=0.45\linewidth]{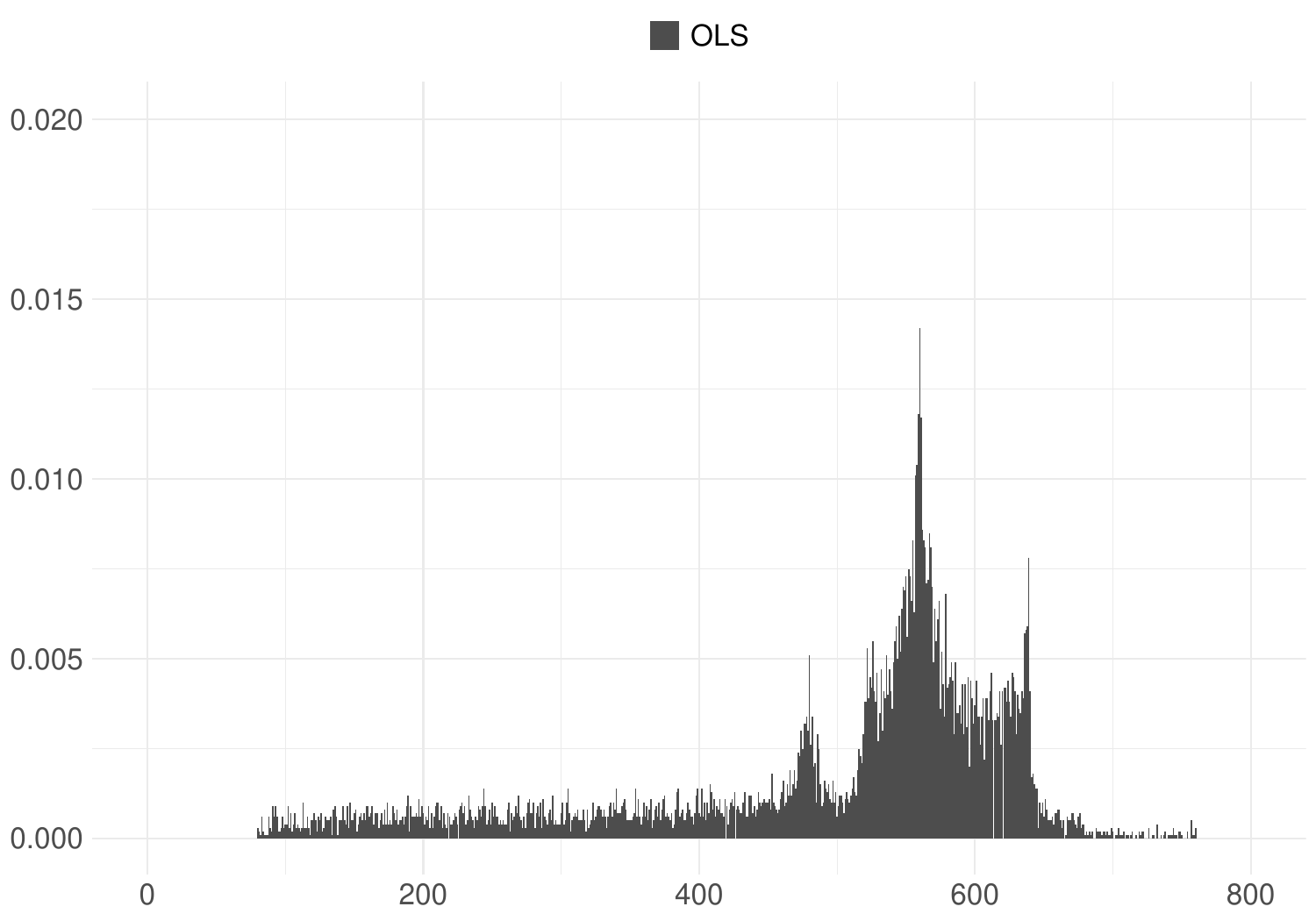}\label{fig2:6:1}}
\subfigure[$T=800$, $c_a=4$, $c_b=6$, $s_0/s_1=5$]{\includegraphics[width=0.45\linewidth]{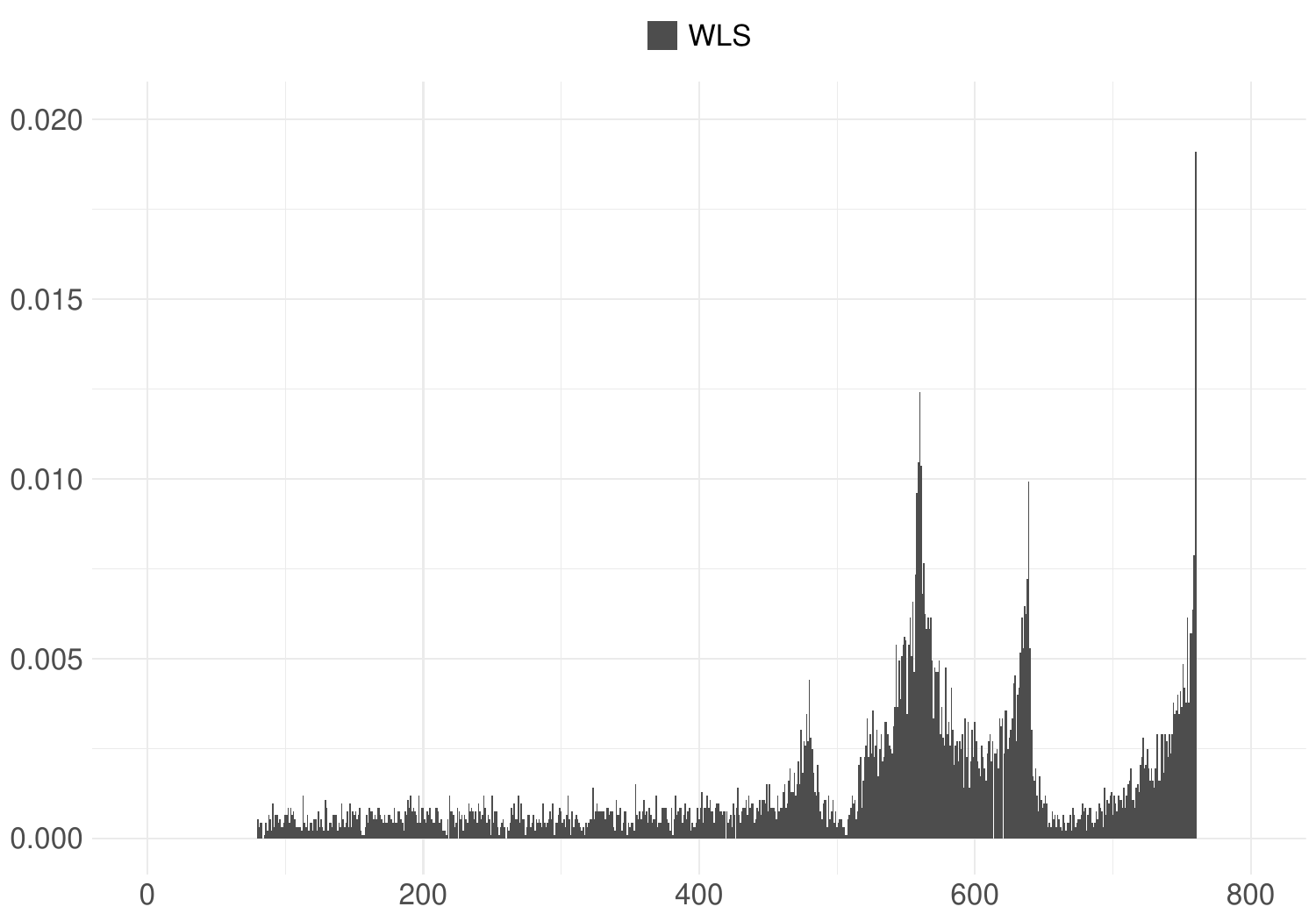}\label{fig2:6:2}}\\
\subfigure[$T=800$, $c_a=5$, $c_b=6$, $s_0/s_1=5$]{\includegraphics[width=0.45\linewidth]{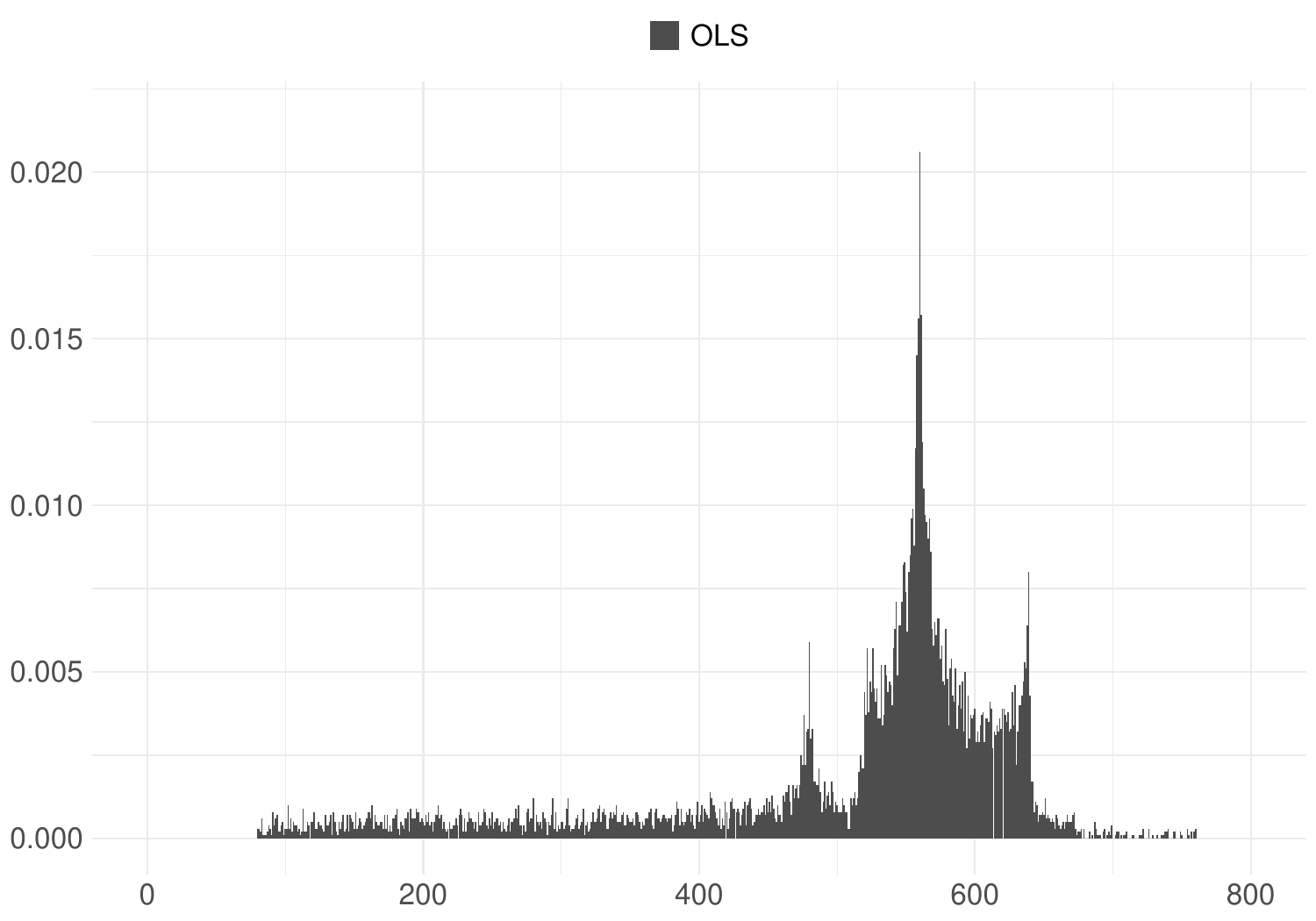}\label{fig2:6:3}}
\subfigure[$T=800$, $c_a=5$, $c_b=6$, $s_0/s_1=5$]{\includegraphics[width=0.45\linewidth]{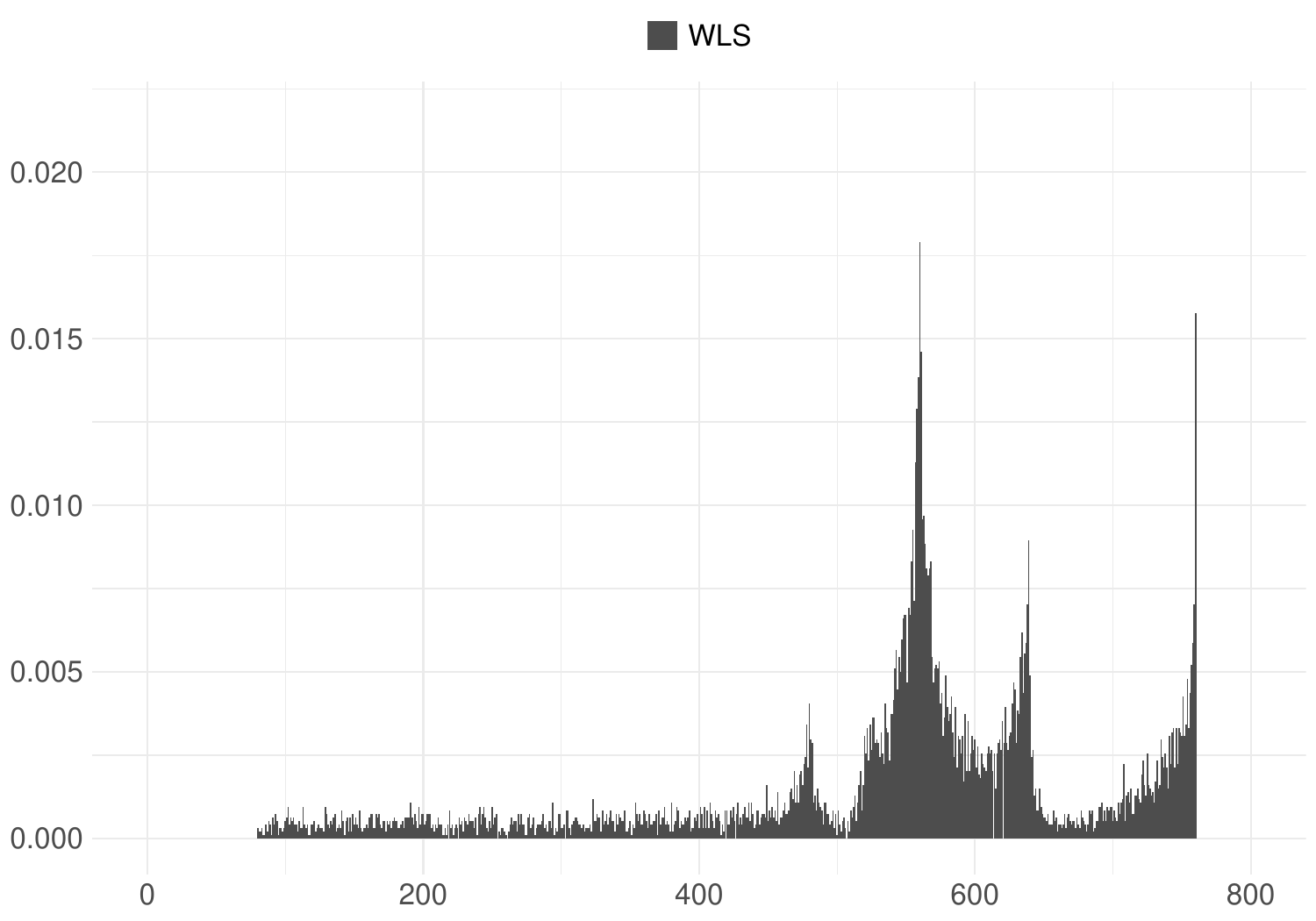}\label{fig2:6:4}}\\
\subfigure[$T=800$, $c_a=6$, $c_b=6$, $s_0/s_1=5$]{\includegraphics[width=0.45\linewidth]{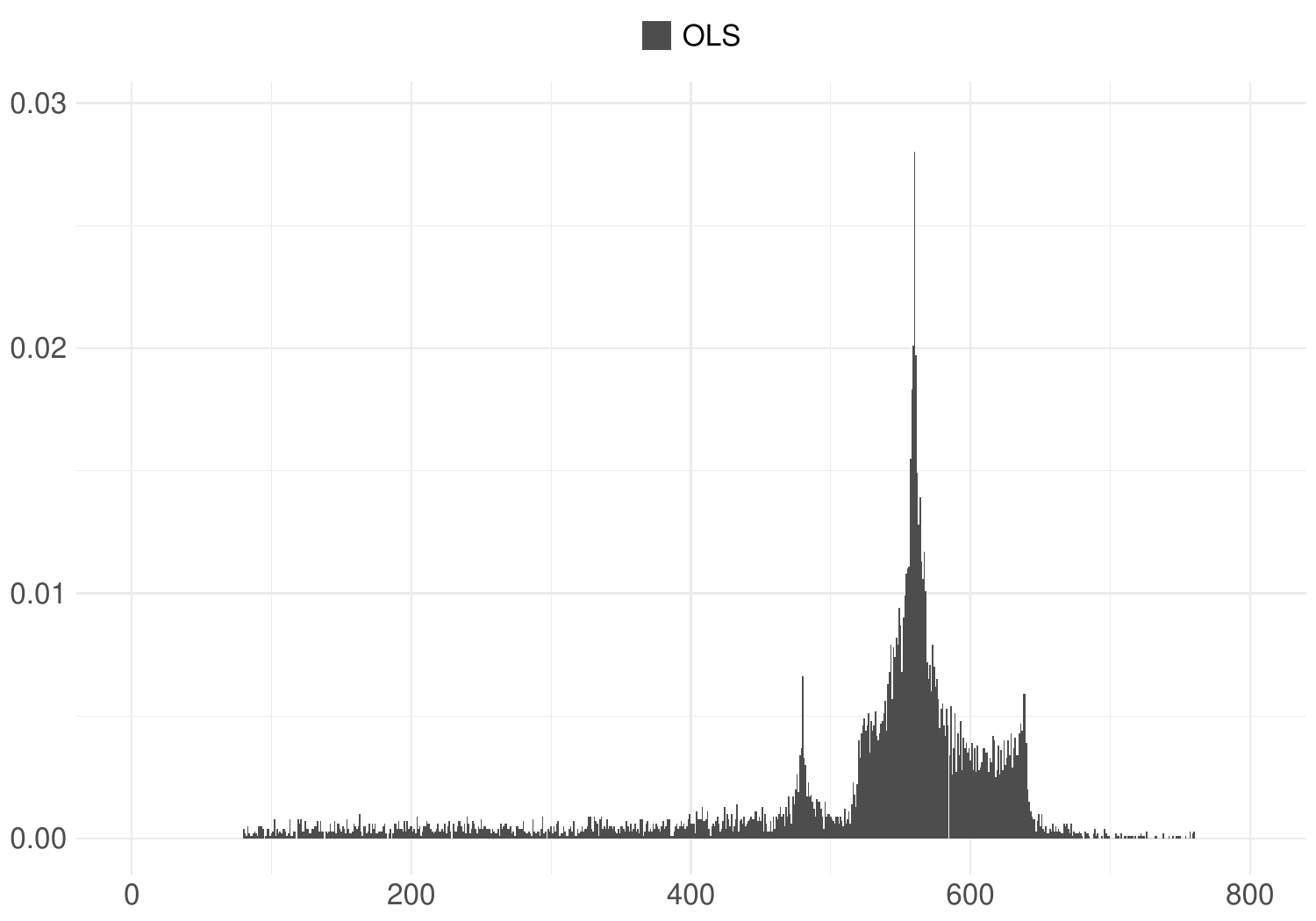}\label{fig2:6:5}}
\subfigure[$T=800$, $c_a=6$, $c_b=6$, $s_0/s_1=5$]{\includegraphics[width=0.45\linewidth]{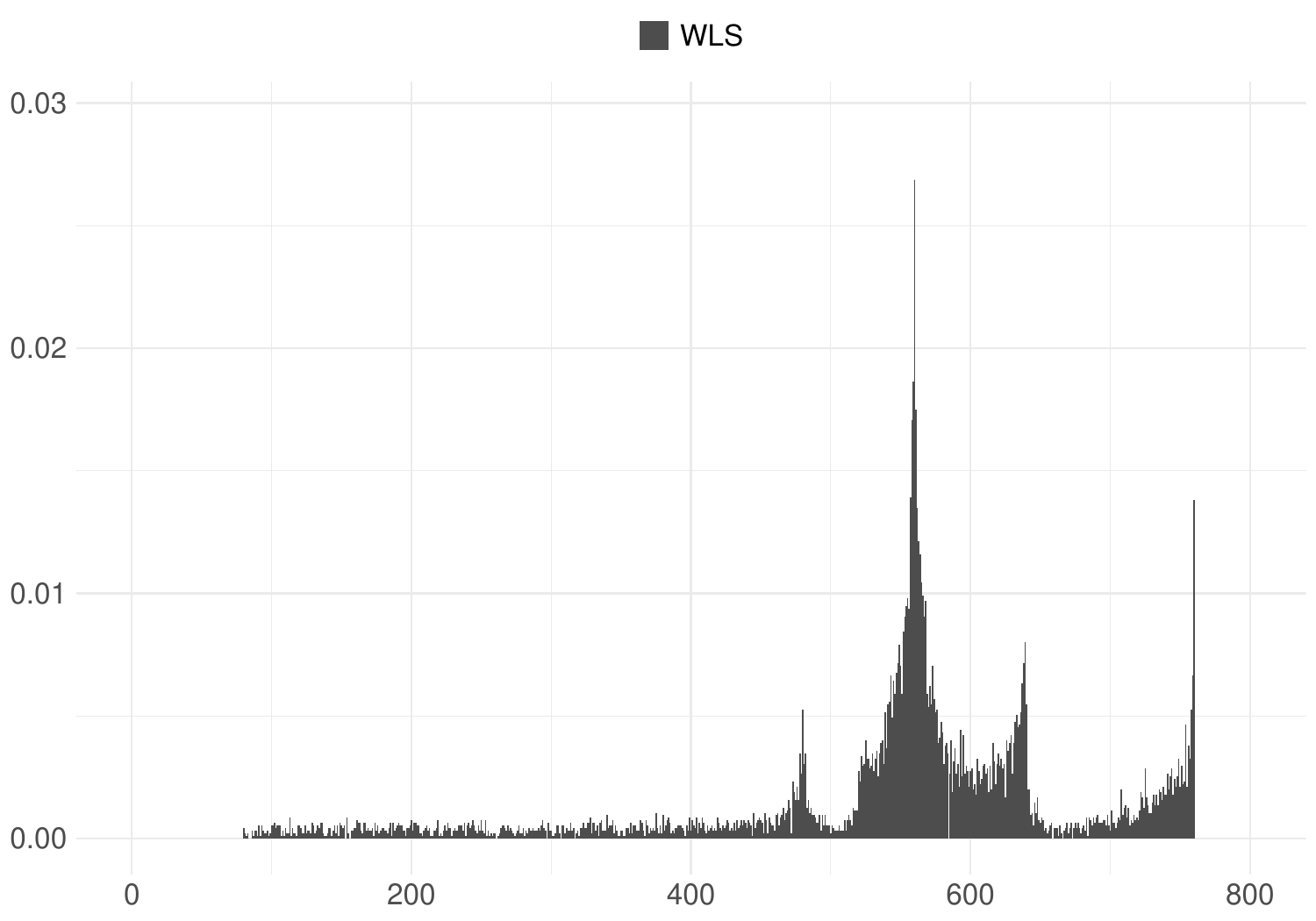}\label{fig2:6:6}}\\
\end{center}%
\caption{Histograms of $\hat{k}_r$ 
for $(\tau_e,\tau_c,\tau_r)=(0.4,0.6,0.7)$,  $\tau=0.8$, $s_0/s_1=5$, $T=800$}
\label{fig26}
\end{figure}

\newpage

\section{$\tau=0.2$, $\hat{k}_c$}
\setcounter{figure}{0}

\begin{figure}[h!]%
\begin{center}%
\subfigure[$T=400$, $c_a=4$, $c_b=6$, $s_0/s_1=1/5$]{\includegraphics[width=0.45\linewidth]{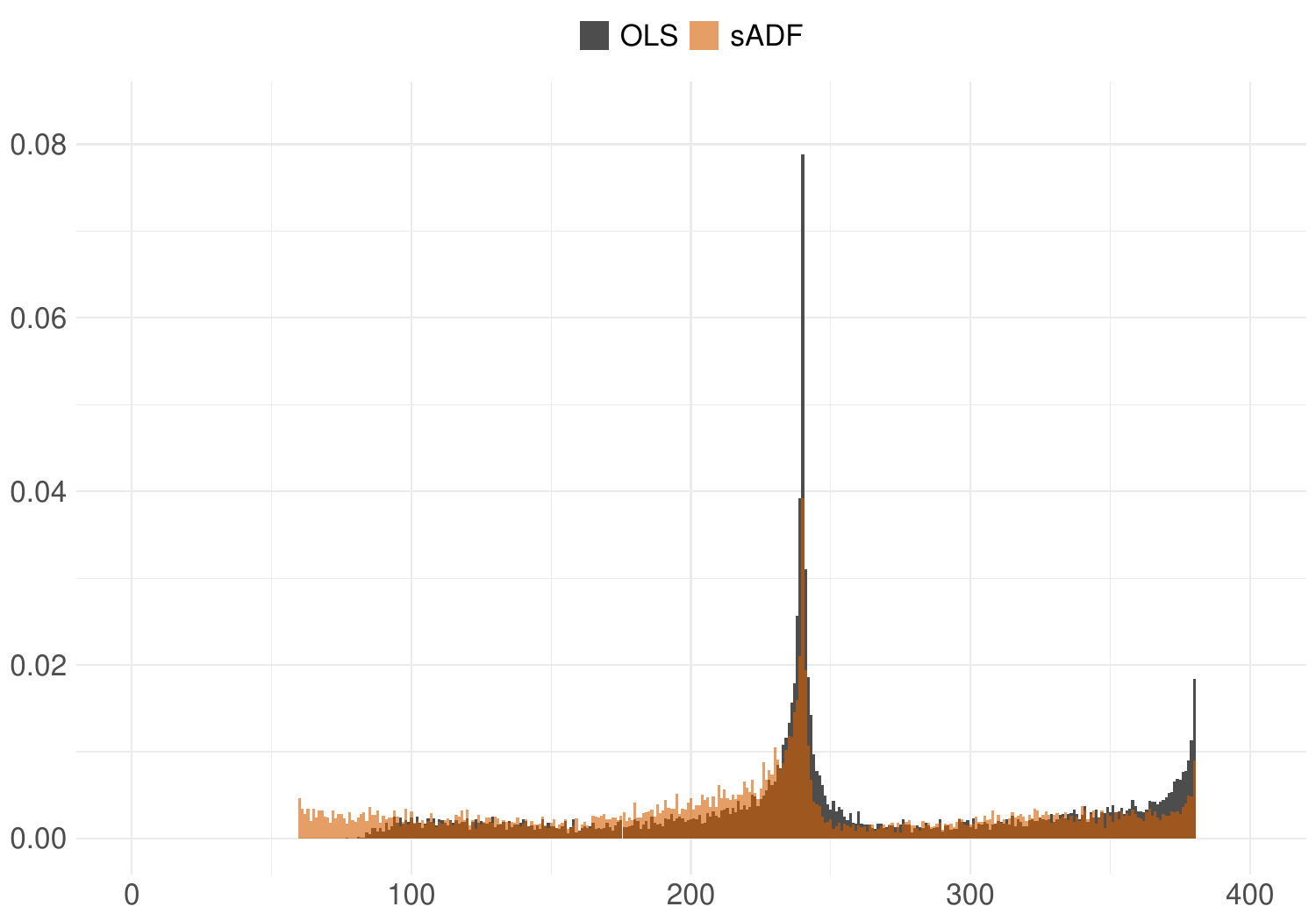}\label{fig3:1:1}}
\subfigure[$T=400$, $c_a=4$, $c_b=6$, $s_0/s_1=1/5$]{\includegraphics[width=0.45\linewidth]{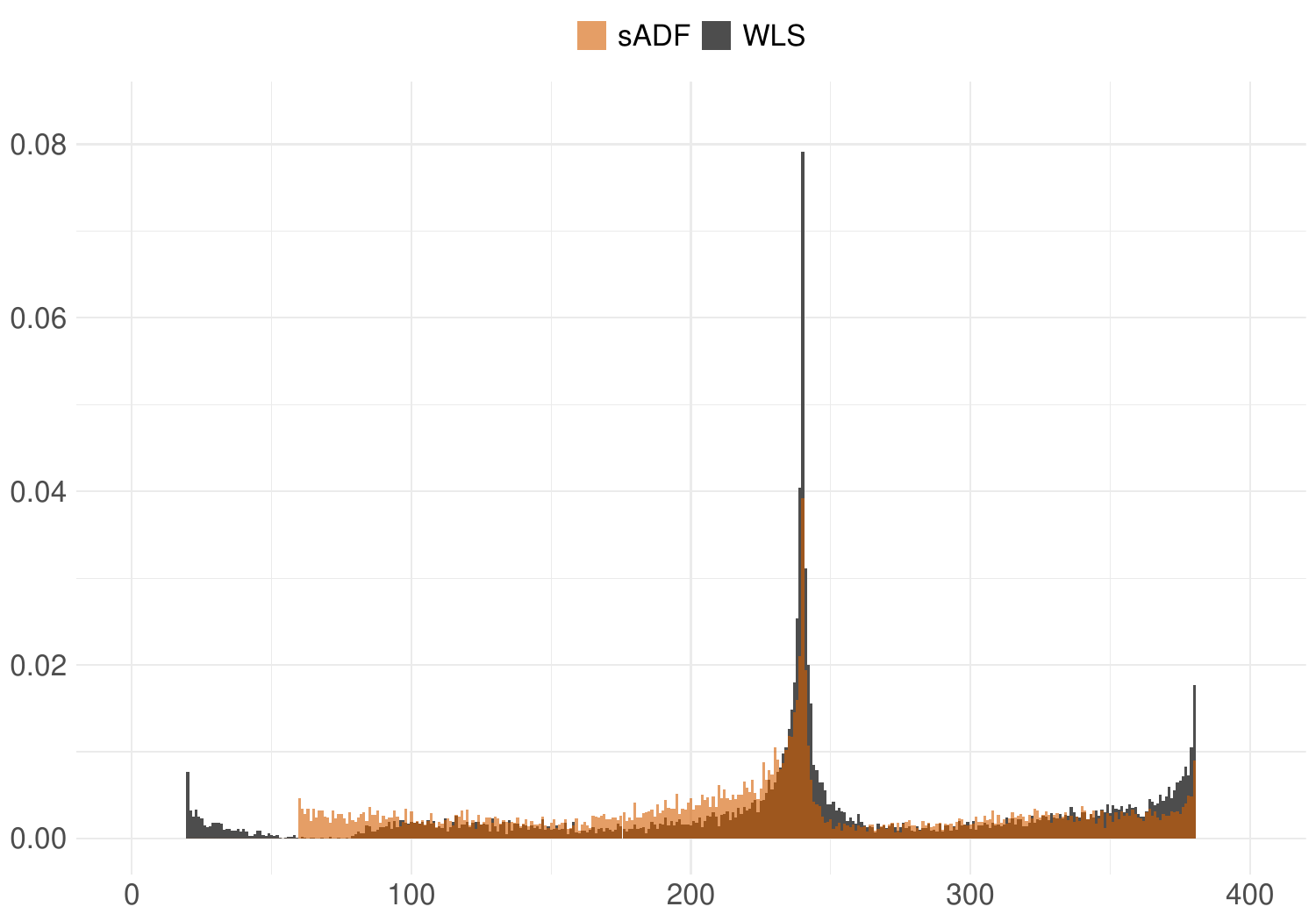}\label{fig3:1:2}}\\
\subfigure[$T=400$, $c_a=5$, $c_b=6$, $s_0/s_1=1/5$]{\includegraphics[width=0.45\linewidth]{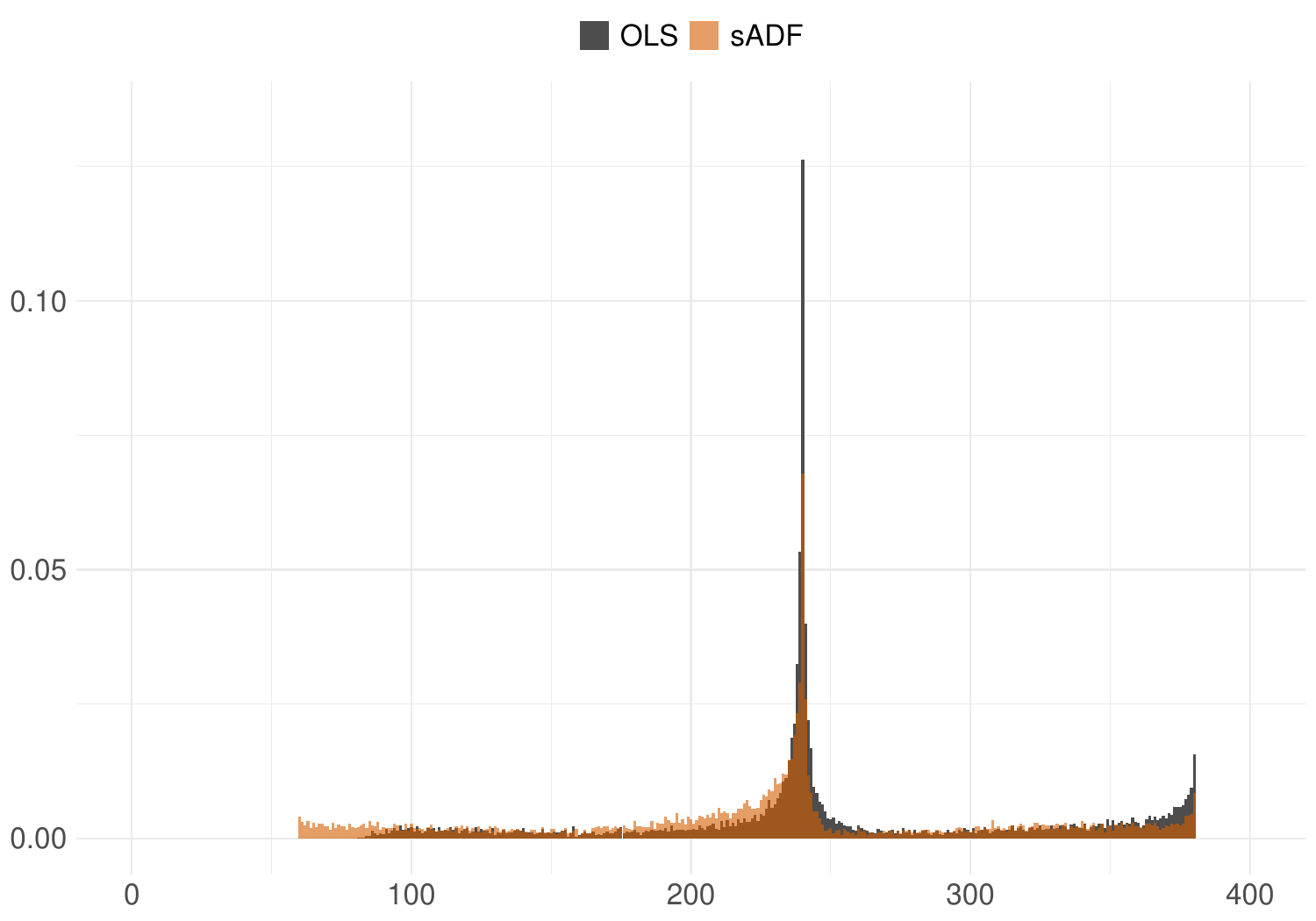}\label{fig3:1:3}}
\subfigure[$T=400$, $c_a=5$, $c_b=6$, $s_0/s_1=1/5$]{\includegraphics[width=0.45\linewidth]{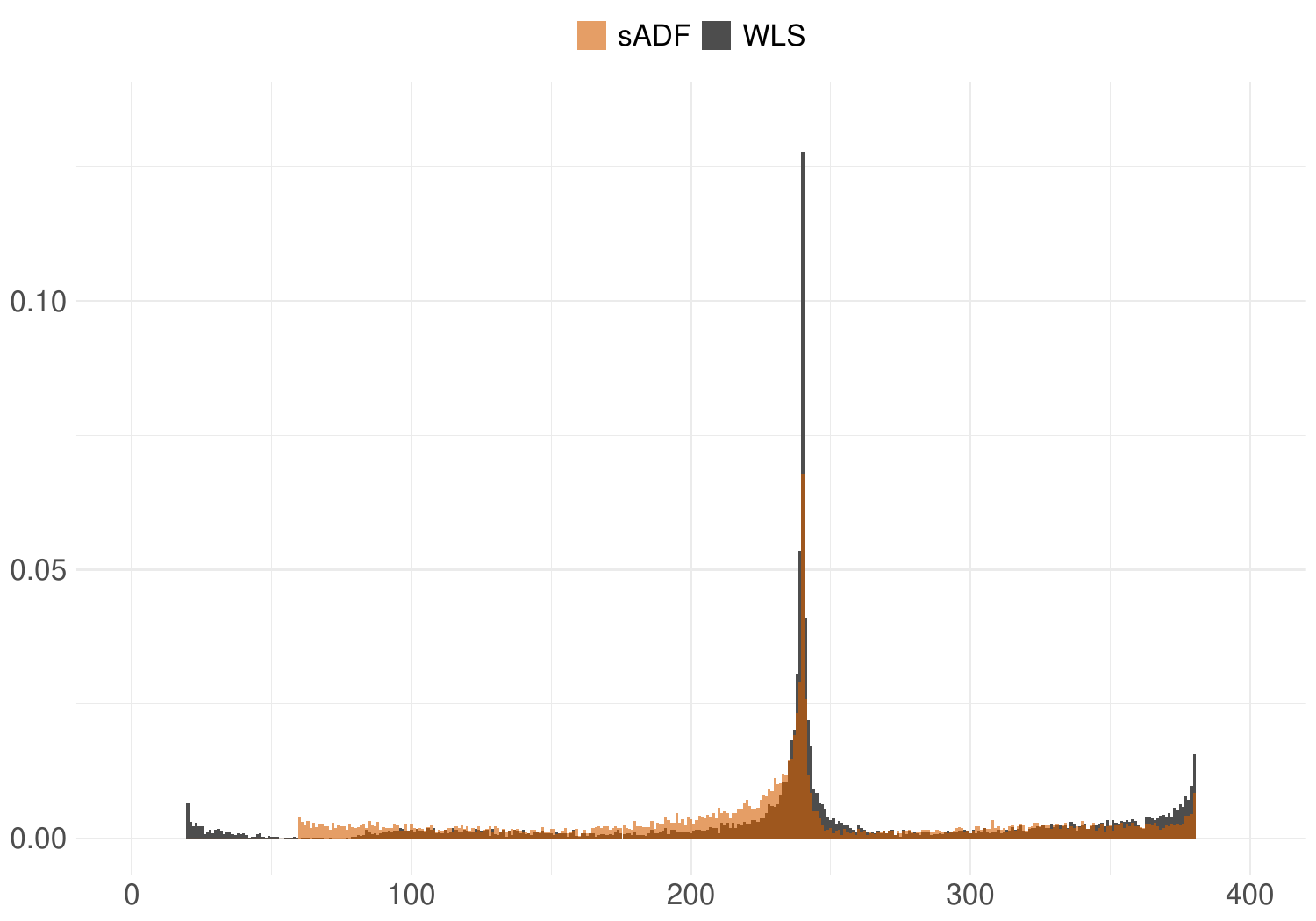}\label{fig3:1:4}}\\
\subfigure[$T=400$, $c_a=6$, $c_b=6$, $s_0/s_1=1/5$]{\includegraphics[width=0.45\linewidth]{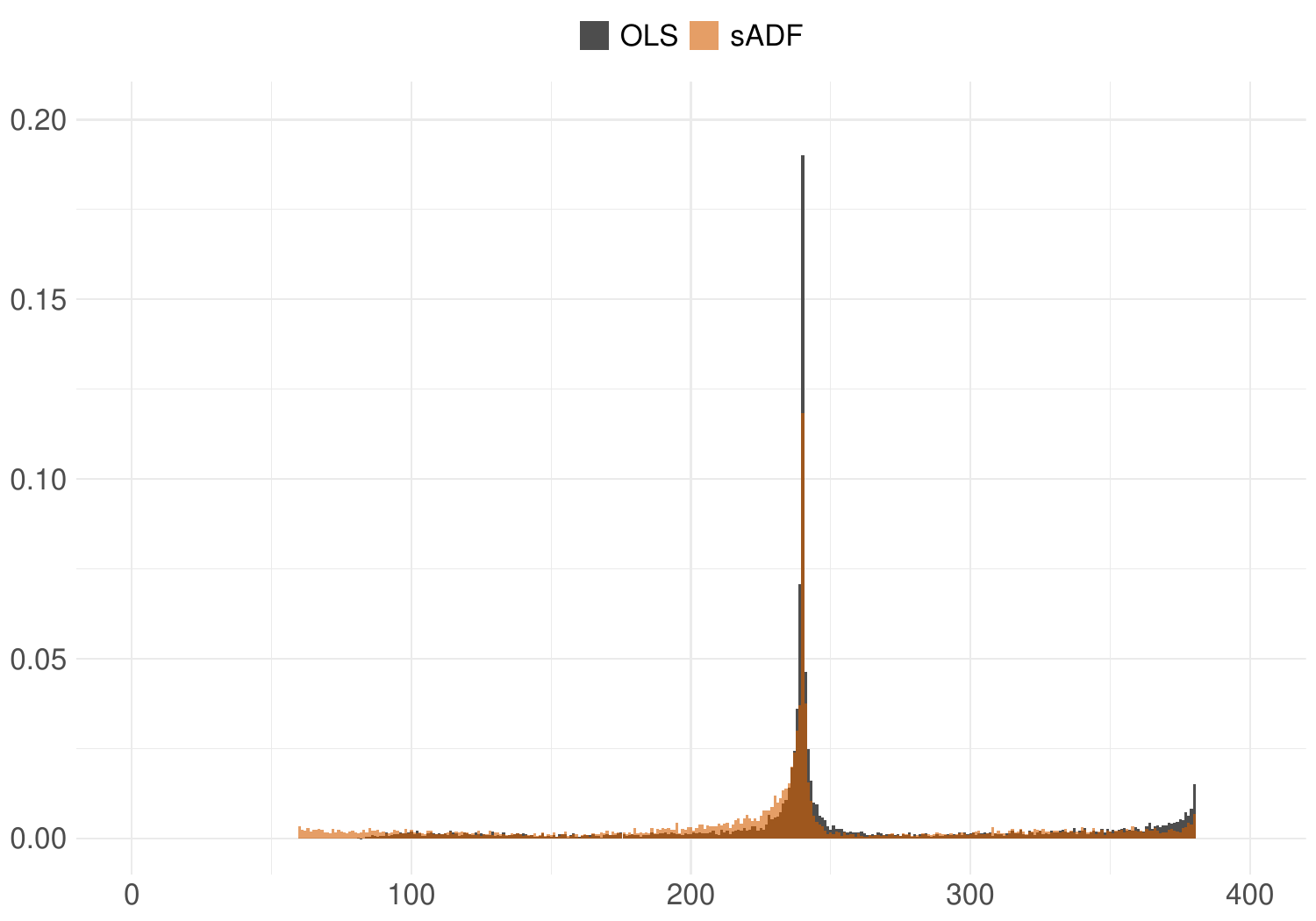}\label{fig3:1:5}}
\subfigure[$T=400$, $c_a=6$, $c_b=6$, $s_0/s_1=1/5$]{\includegraphics[width=0.45\linewidth]{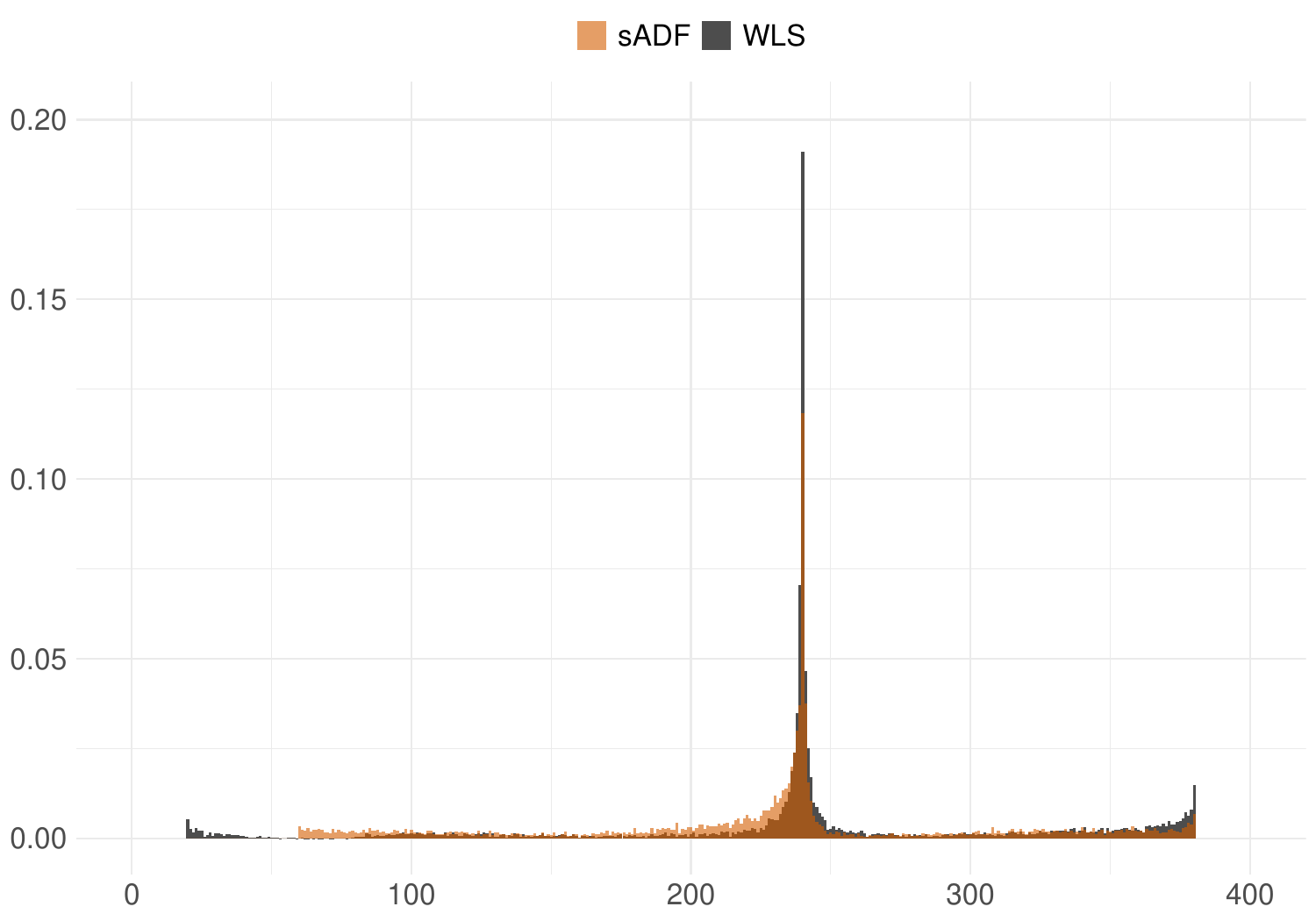}\label{fig3:1:6}}\\
\end{center}%
\caption{Histograms of $\hat{k}_c$ 
for $(\tau_e,\tau_c,\tau_r)=(0.4,0.6,0.7)$,  $\tau=0.2$, $s_0/s_1=1/5$, $T=400$}
\label{fig31}
\end{figure}

\newpage

\begin{figure}[h!]%
\begin{center}%
\subfigure[$T=800$, $c_a=4$, $c_b=6$, $s_0/s_1=1/5$]{\includegraphics[width=0.45\linewidth]{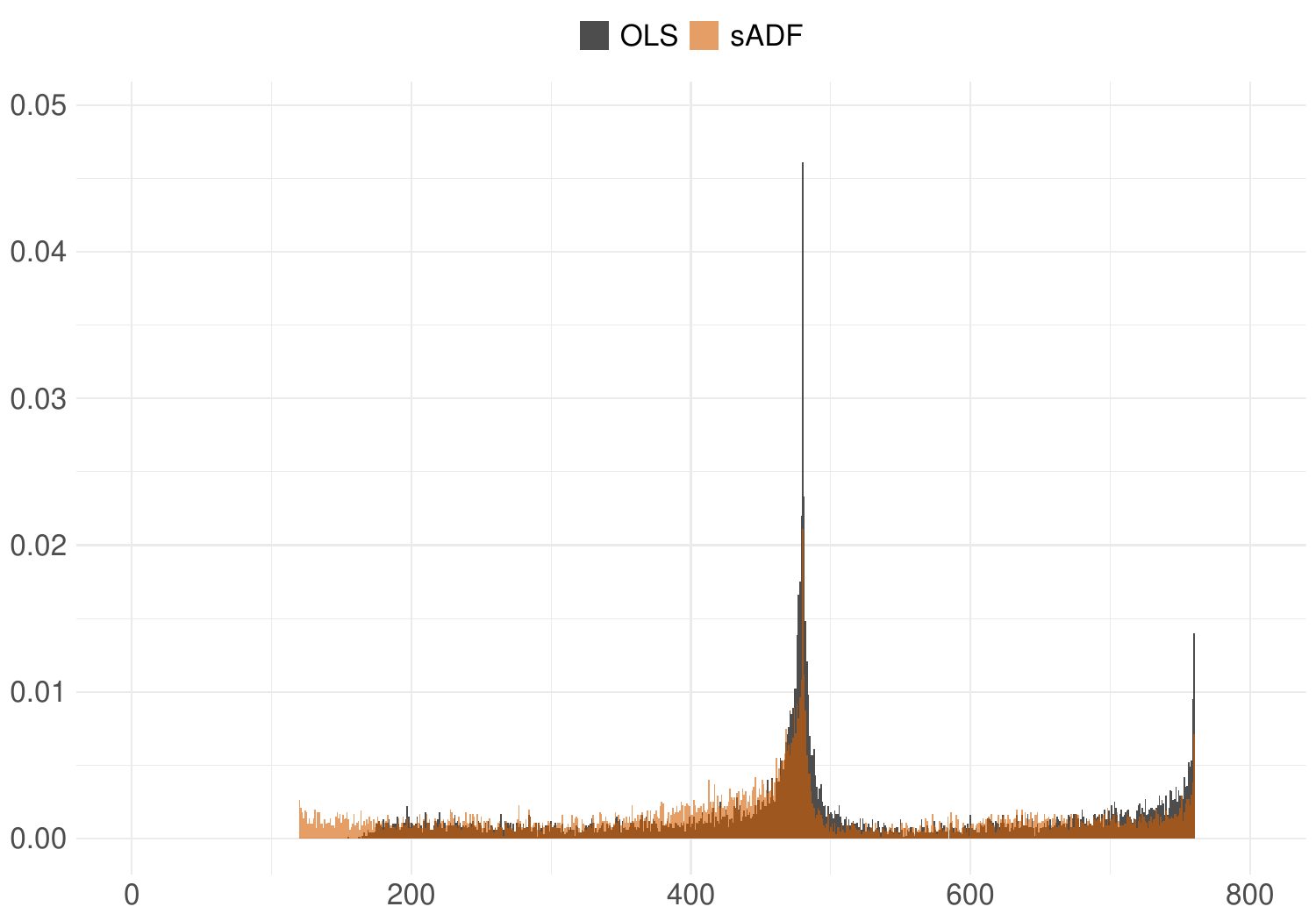}\label{fig3:2:1}}
\subfigure[$T=800$, $c_a=4$, $c_b=6$, $s_0/s_1=1/5$]{\includegraphics[width=0.45\linewidth]{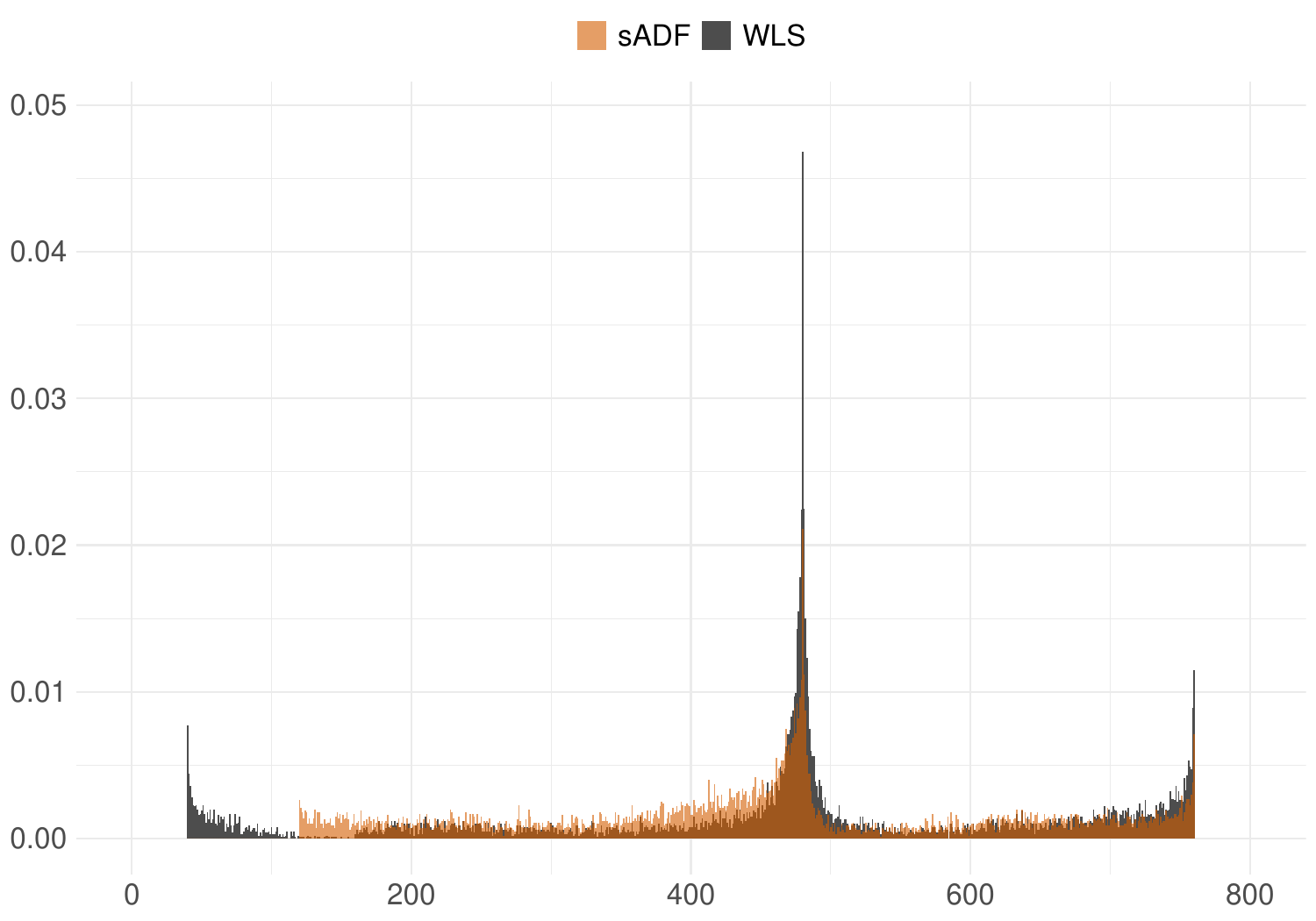}\label{fig3:2:2}}\\
\subfigure[$T=800$, $c_a=5$, $c_b=6$, $s_0/s_1=1/5$]{\includegraphics[width=0.45\linewidth]{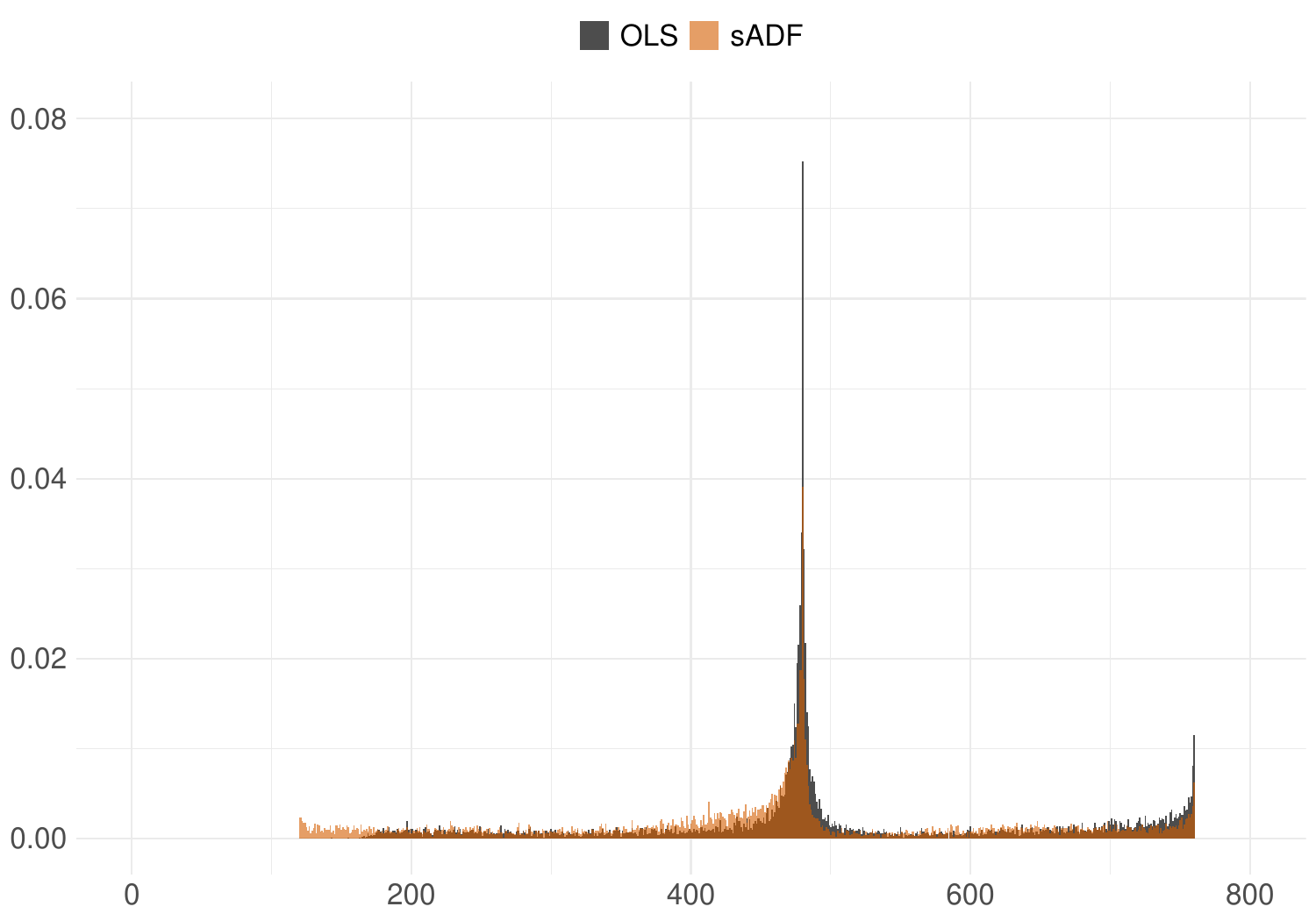}\label{fig3:2:3}}
\subfigure[$T=800$, $c_a=5$, $c_b=6$, $s_0/s_1=1/5$]{\includegraphics[width=0.45\linewidth]{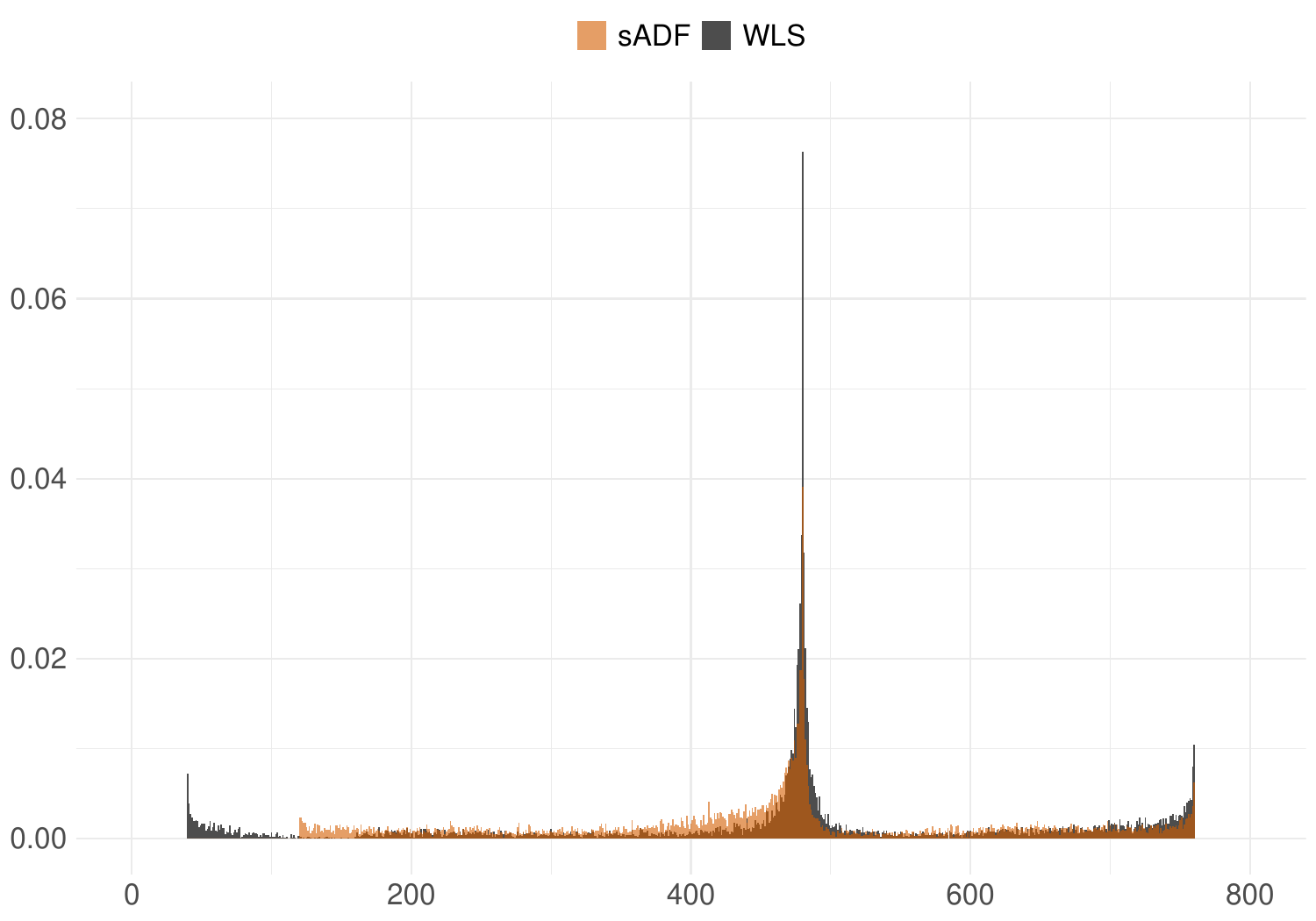}\label{fig3:2:4}}\\
\subfigure[$T=800$, $c_a=6$, $c_b=6$, $s_0/s_1=1/5$]{\includegraphics[width=0.45\linewidth]{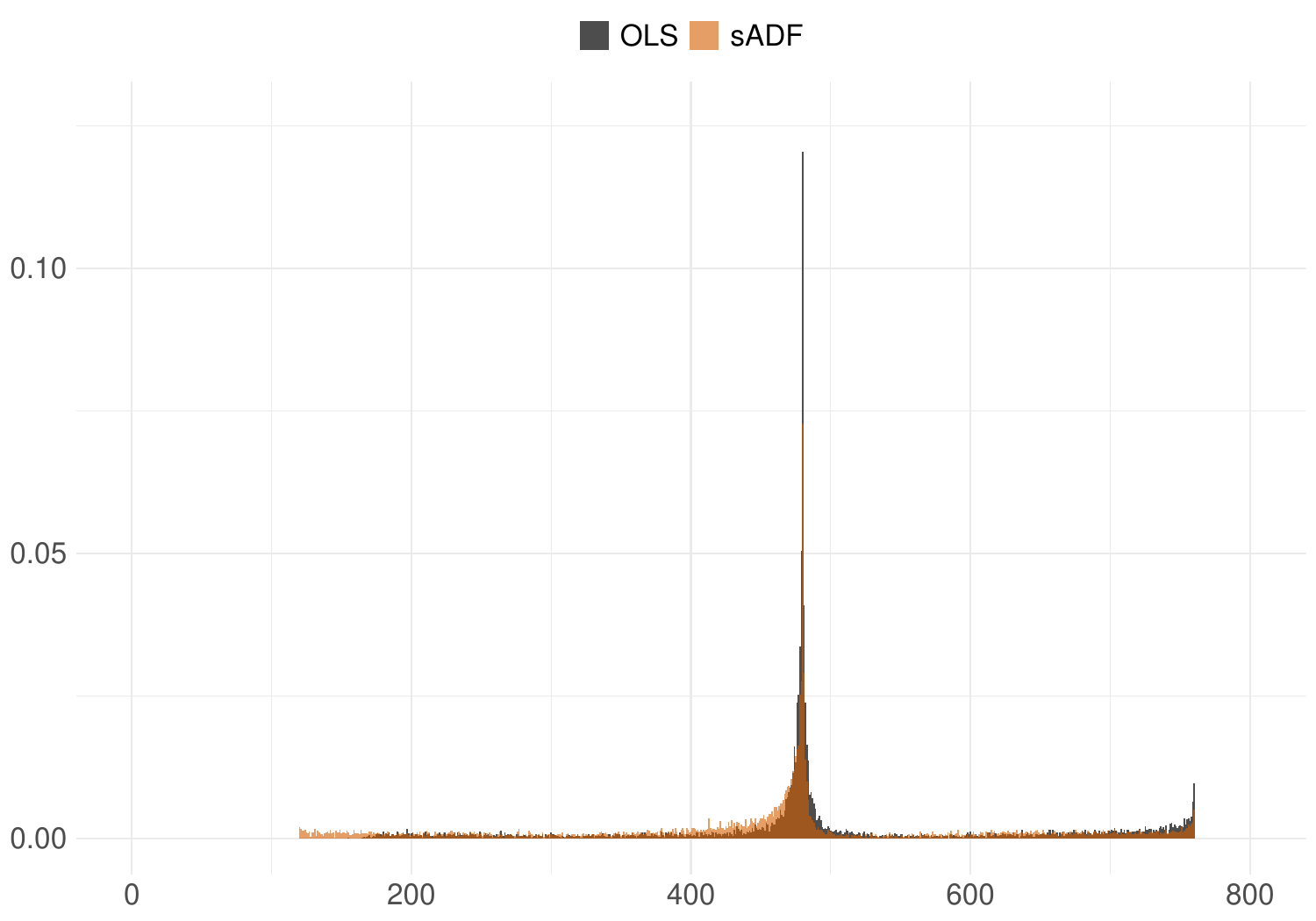}\label{fig3:2:5}}
\subfigure[$T=800$, $c_a=6$, $c_b=6$, $s_0/s_1=1/5$]{\includegraphics[width=0.45\linewidth]{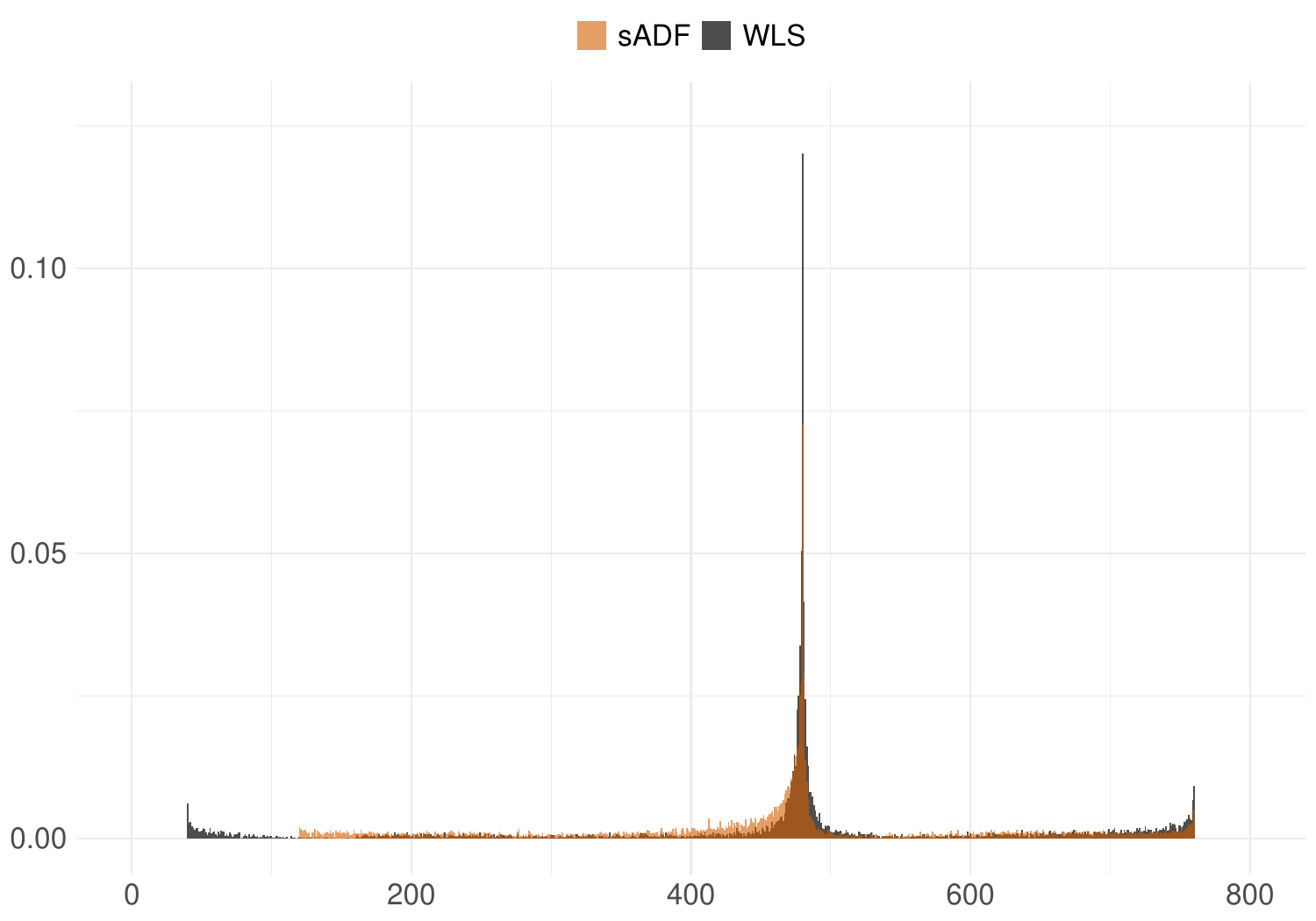}\label{fig3:2:6}}\\
\end{center}%
\caption{Histograms of $\hat{k}_c$ 
for $(\tau_e,\tau_c,\tau_r)=(0.4,0.6,0.7)$,  $\tau=0.2$, $s_0/s_1=1/5$, $T=800$}
\label{fig32}
\end{figure}

\newpage

\begin{figure}[h!]%
\begin{center}%
\subfigure[$T=400$, $c_a=4$, $c_b=6$, $s_0/s_1=1$]{\includegraphics[width=0.45\linewidth]{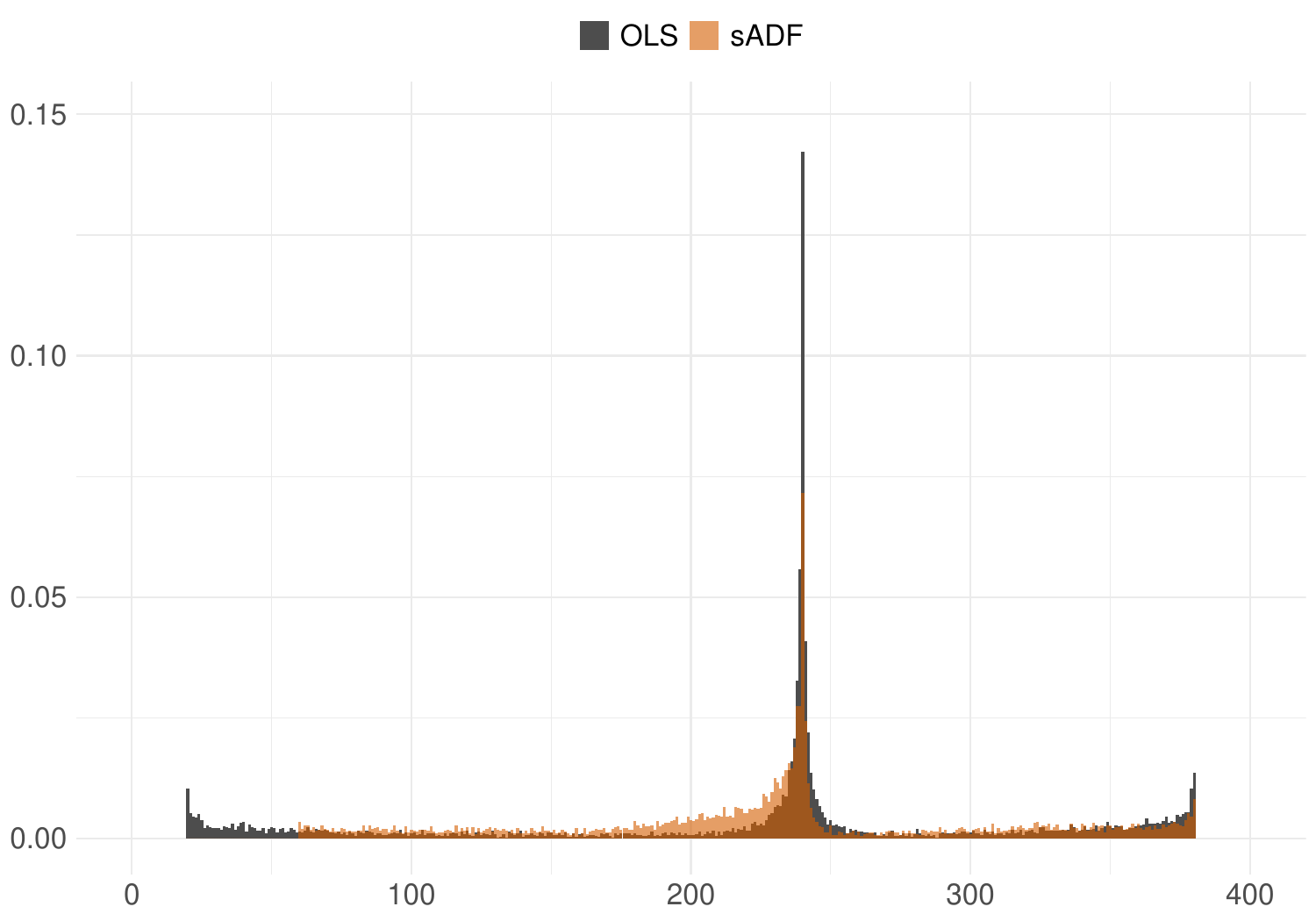}\label{fig3:3:1}}
\subfigure[$T=400$, $c_a=4$, $c_b=6$, $s_0/s_1=1$]{\includegraphics[width=0.45\linewidth]{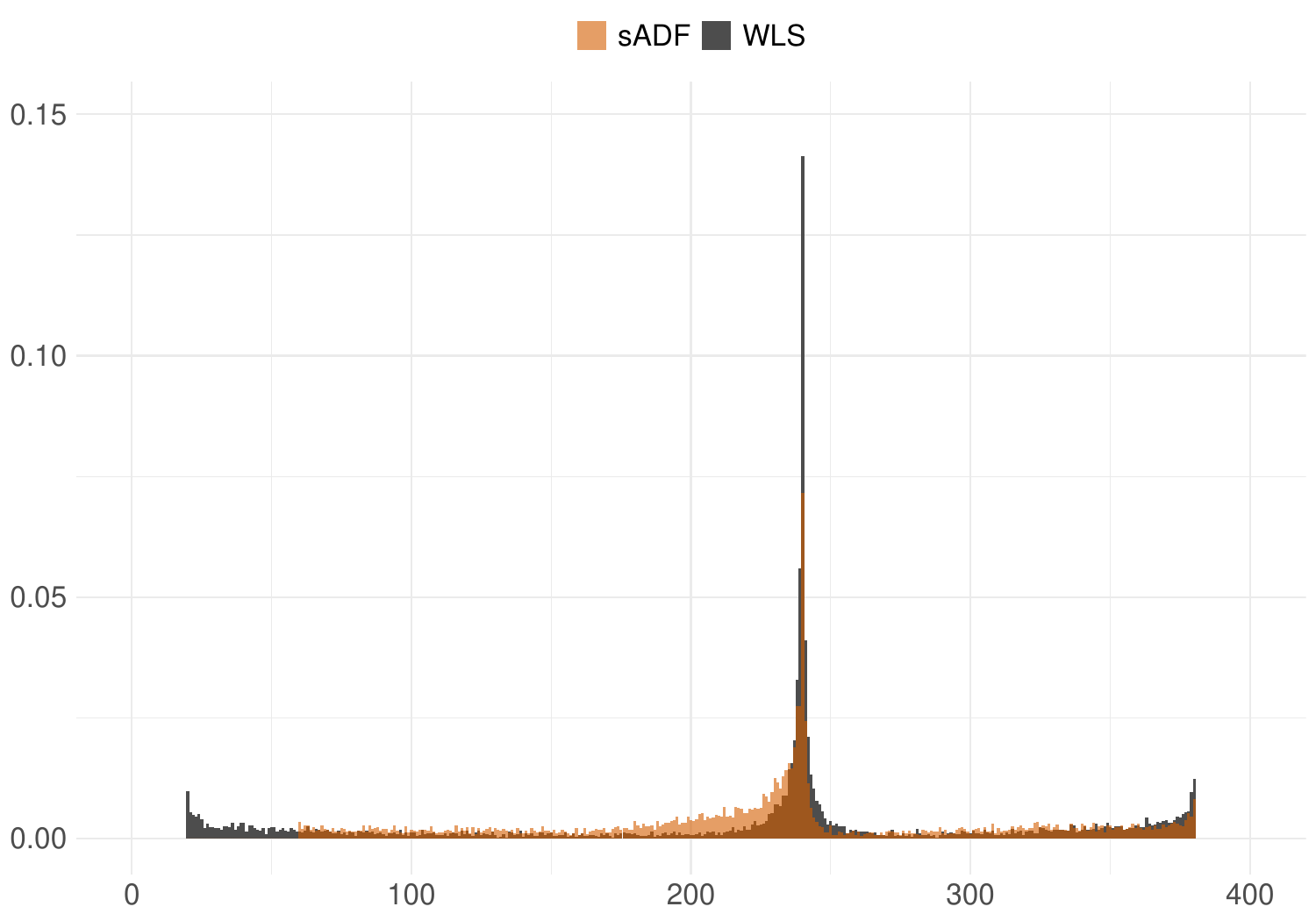}\label{fig3:3:2}}\\
\subfigure[$T=400$, $c_a=5$, $c_b=6$, $s_0/s_1=1$]{\includegraphics[width=0.45\linewidth]{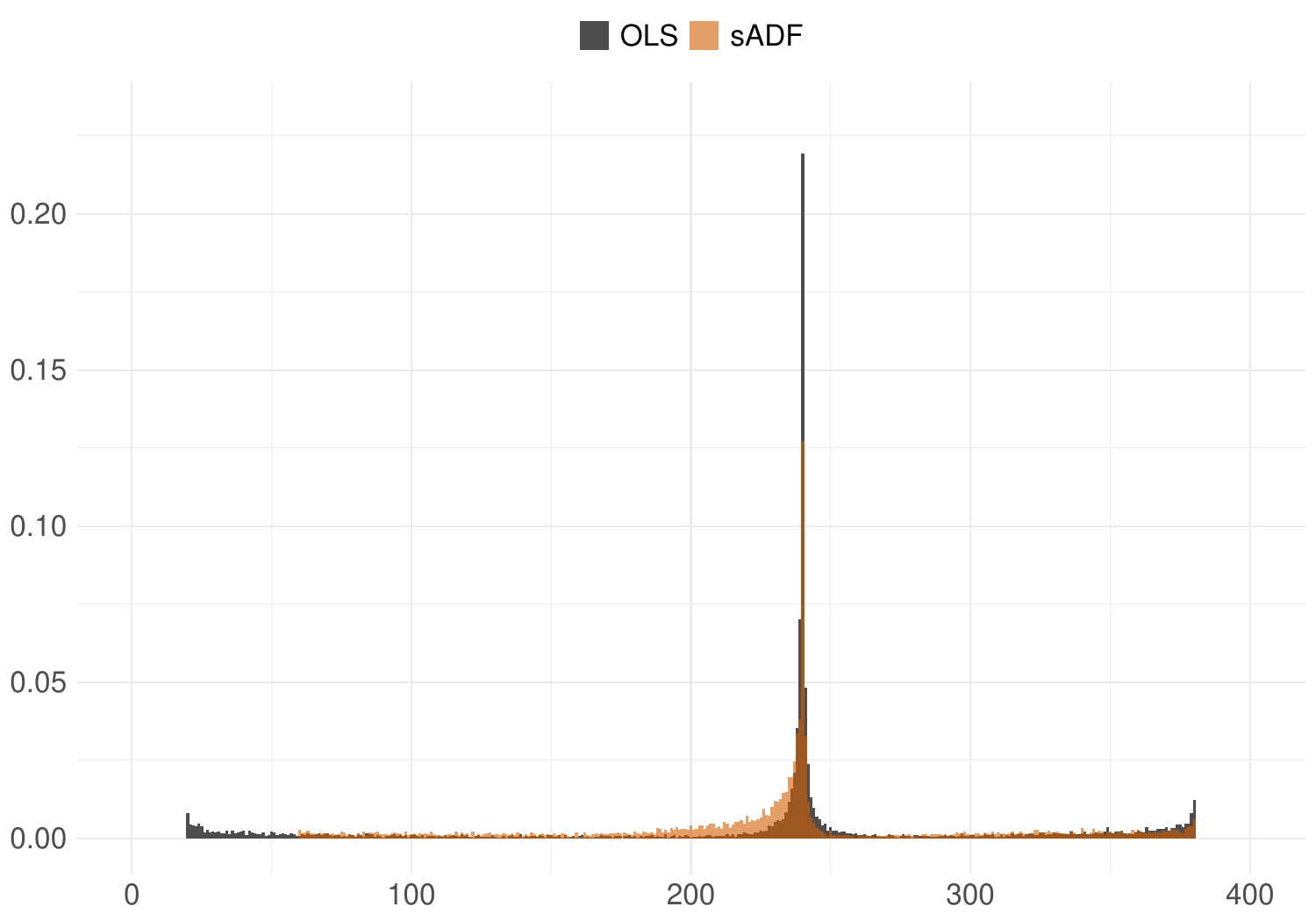}\label{fig3:3:3}}
\subfigure[$T=400$, $c_a=5$, $c_b=6$, $s_0/s_1=1$]{\includegraphics[width=0.45\linewidth]{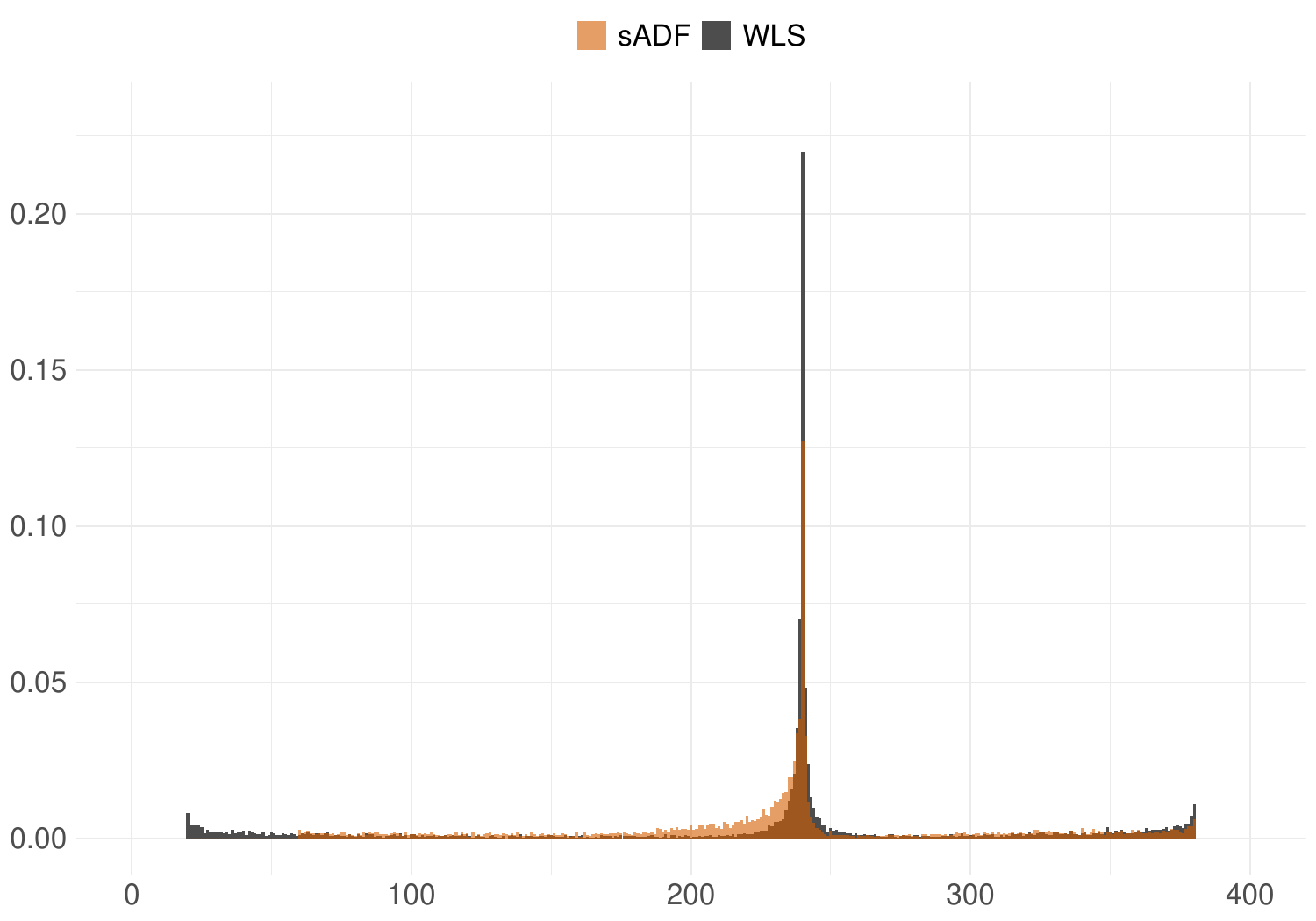}\label{fig3:3:4}}\\
\subfigure[$T=400$, $c_a=6$, $c_b=6$, $s_0/s_1=1$]{\includegraphics[width=0.45\linewidth]{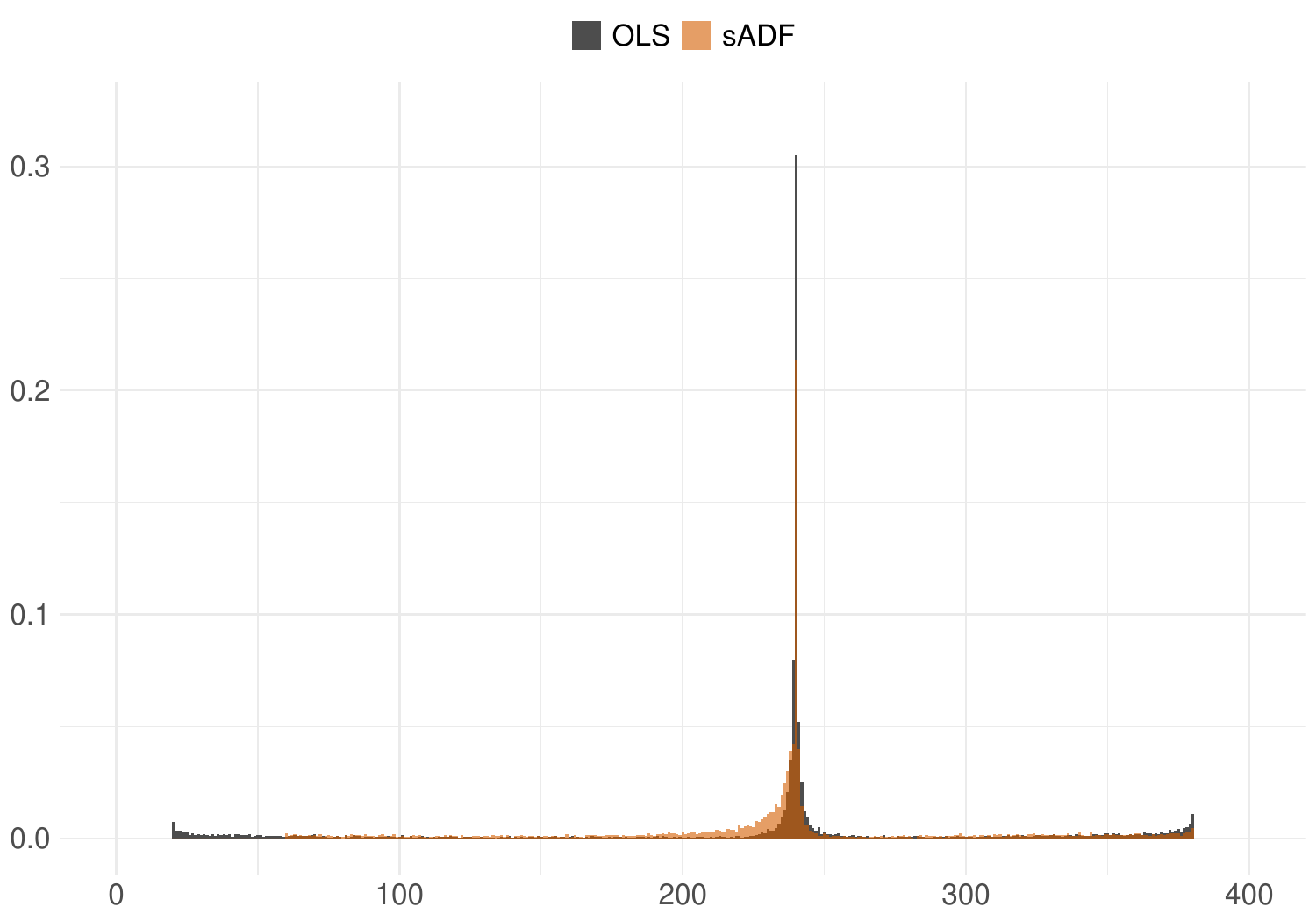}\label{fig3:3:5}}
\subfigure[$T=400$, $c_a=6$, $c_b=6$, $s_0/s_1=1$]{\includegraphics[width=0.45\linewidth]{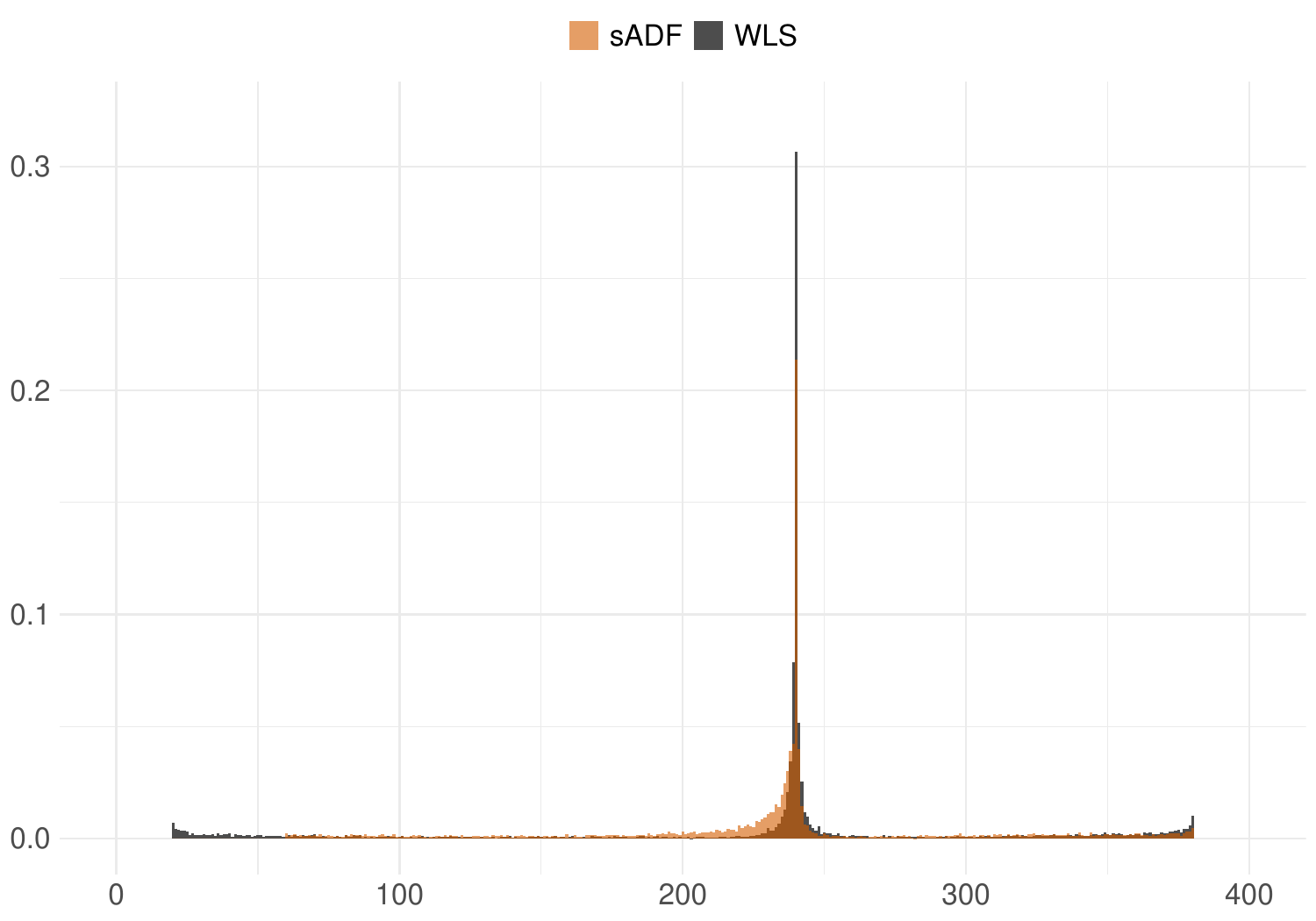}\label{fig3:3:6}}\\
\end{center}%
\caption{Histograms of $\hat{k}_c$ 
for $(\tau_e,\tau_c,\tau_r)=(0.4,0.6,0.7)$,  $\tau=0.2$, $s_0/s_1=1$, $T=400$}
\label{fig33}
\end{figure}

\newpage

\begin{figure}[h!]%
\begin{center}%
\subfigure[$T=800$, $c_a=4$, $c_b=6$, $s_0/s_1=1$]{\includegraphics[width=0.45\linewidth]{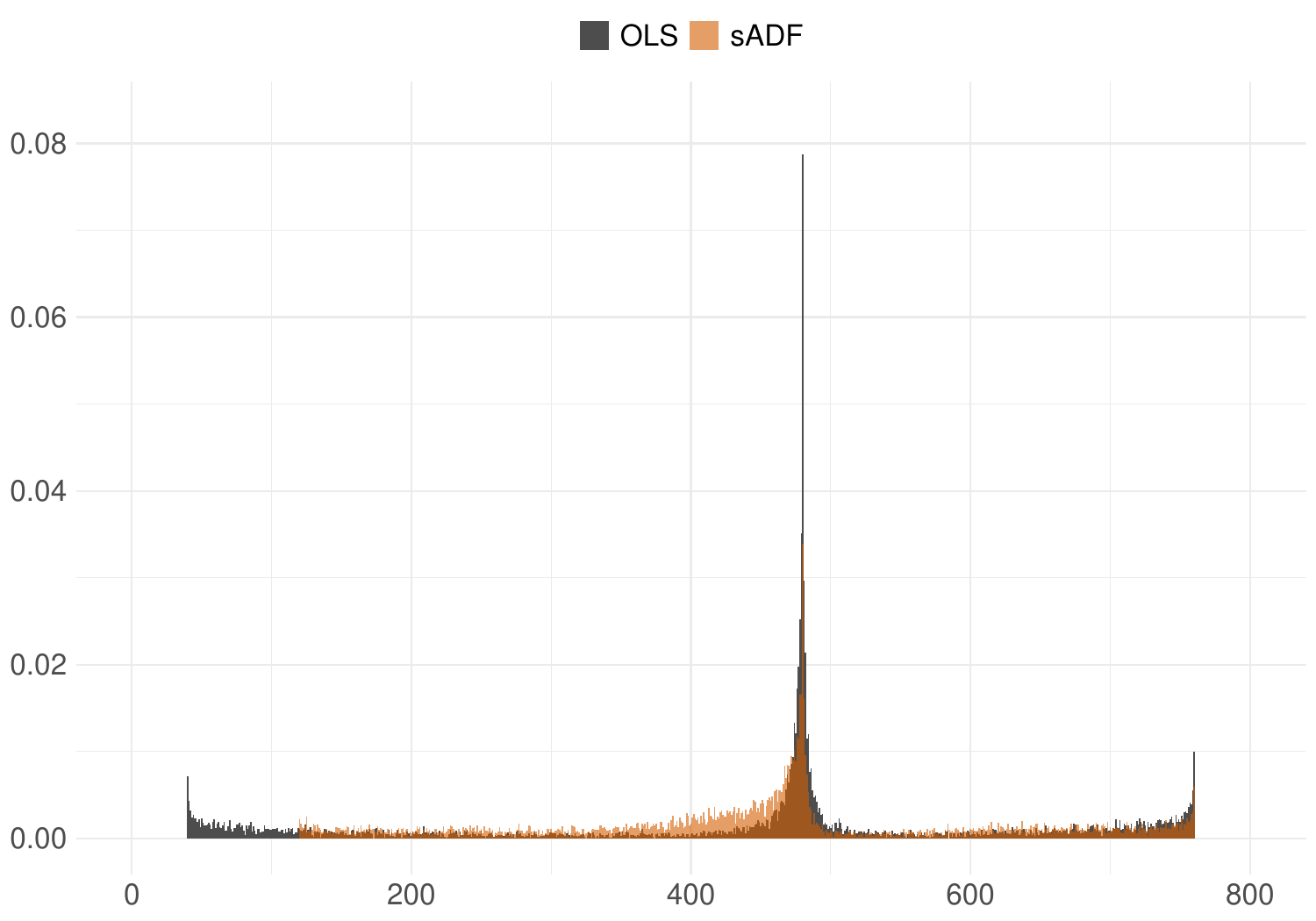}\label{fig3:4:1}}
\subfigure[$T=800$, $c_a=4$, $c_b=6$, $s_0/s_1=1$]{\includegraphics[width=0.45\linewidth]{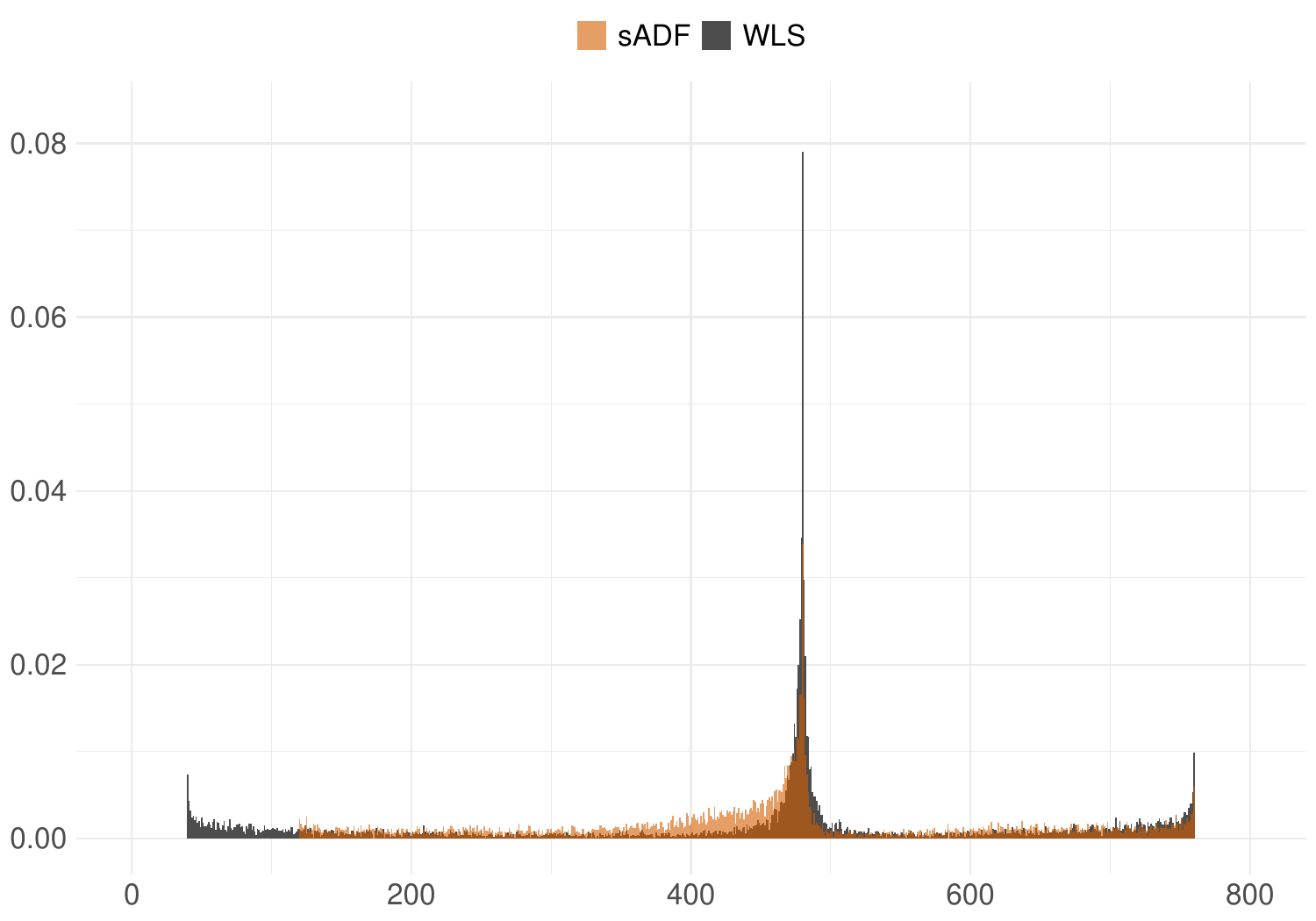}\label{fig3:4:2}}\\
\subfigure[$T=800$, $c_a=5$, $c_b=6$, $s_0/s_1=1$]{\includegraphics[width=0.45\linewidth]{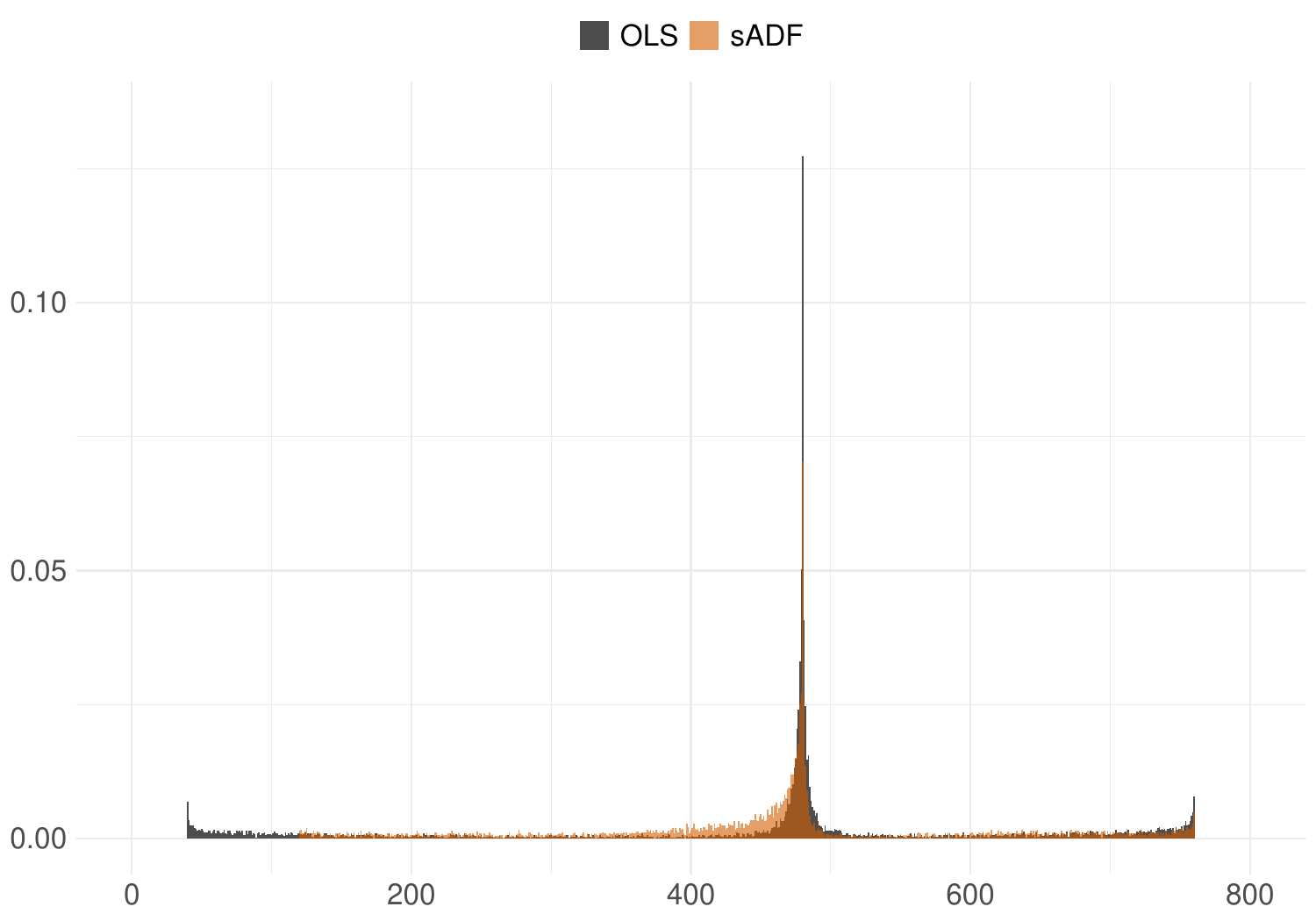}\label{fig3:4:3}}
\subfigure[$T=800$, $c_a=5$, $c_b=6$, $s_0/s_1=1$]{\includegraphics[width=0.45\linewidth]{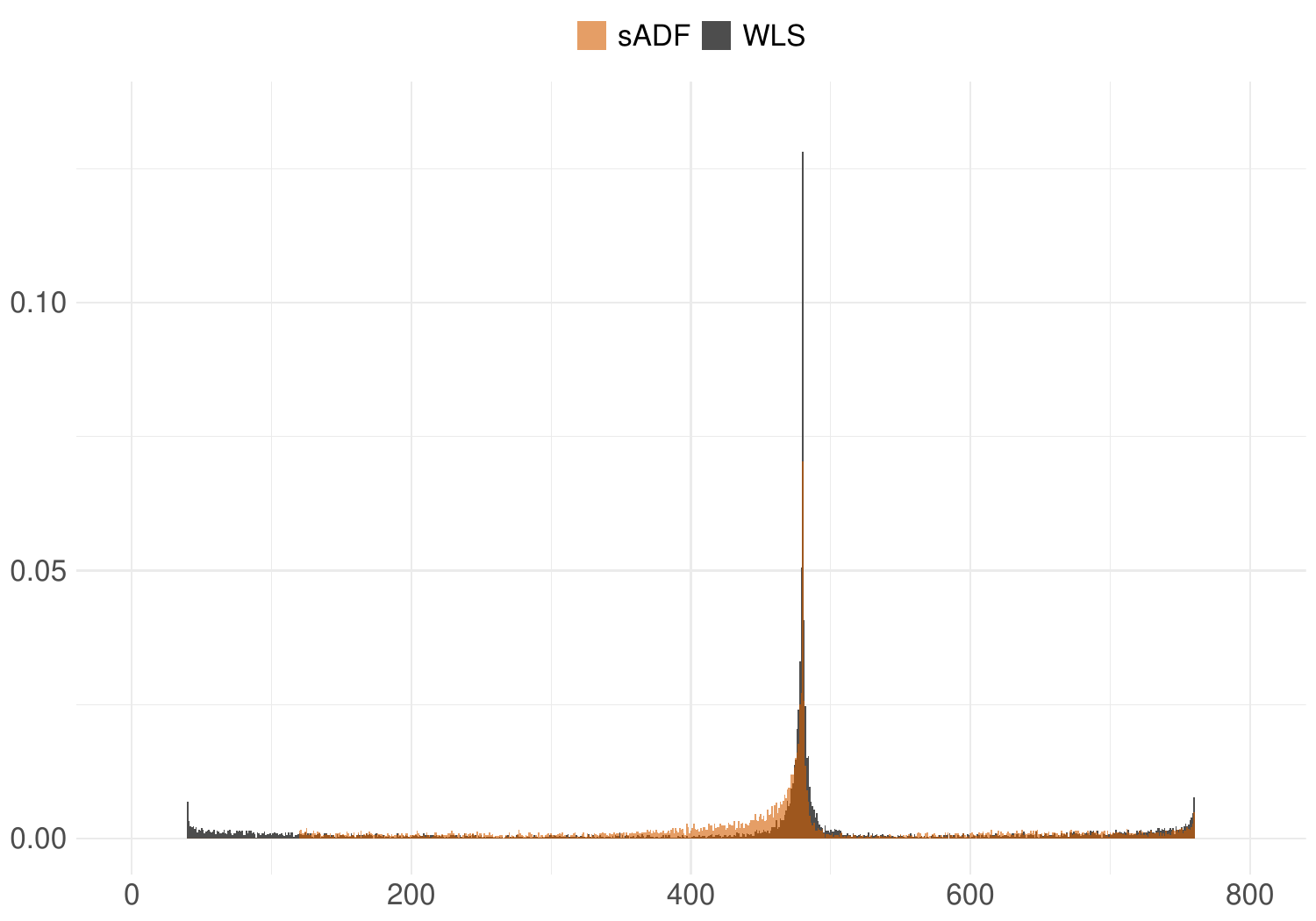}\label{fig3:4:4}}\\
\subfigure[$T=800$, $c_a=6$, $c_b=6$, $s_0/s_1=1$]{\includegraphics[width=0.45\linewidth]{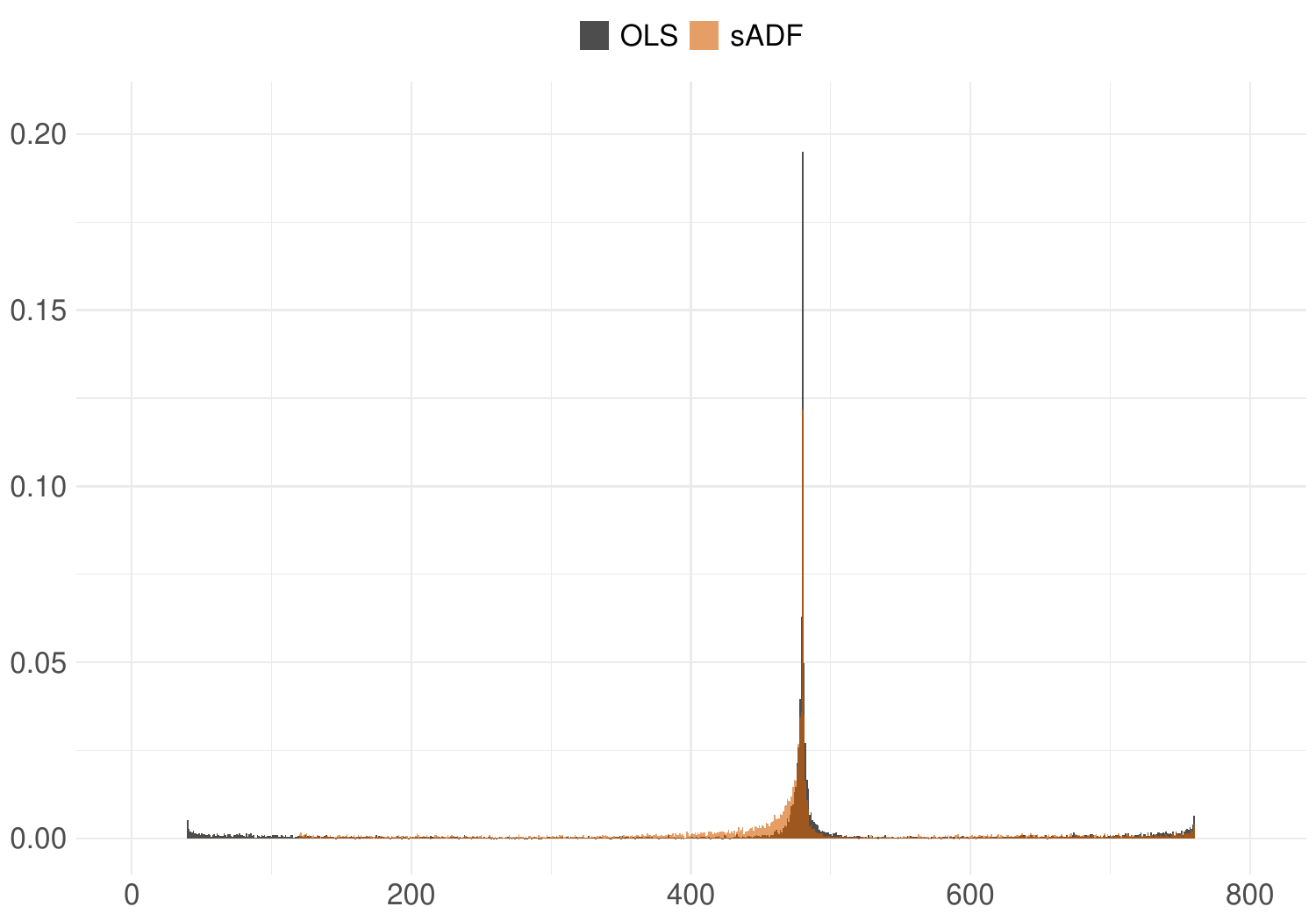}\label{fig3:4:5}}
\subfigure[$T=800$, $c_a=6$, $c_b=6$, $s_0/s_1=1$]{\includegraphics[width=0.45\linewidth]{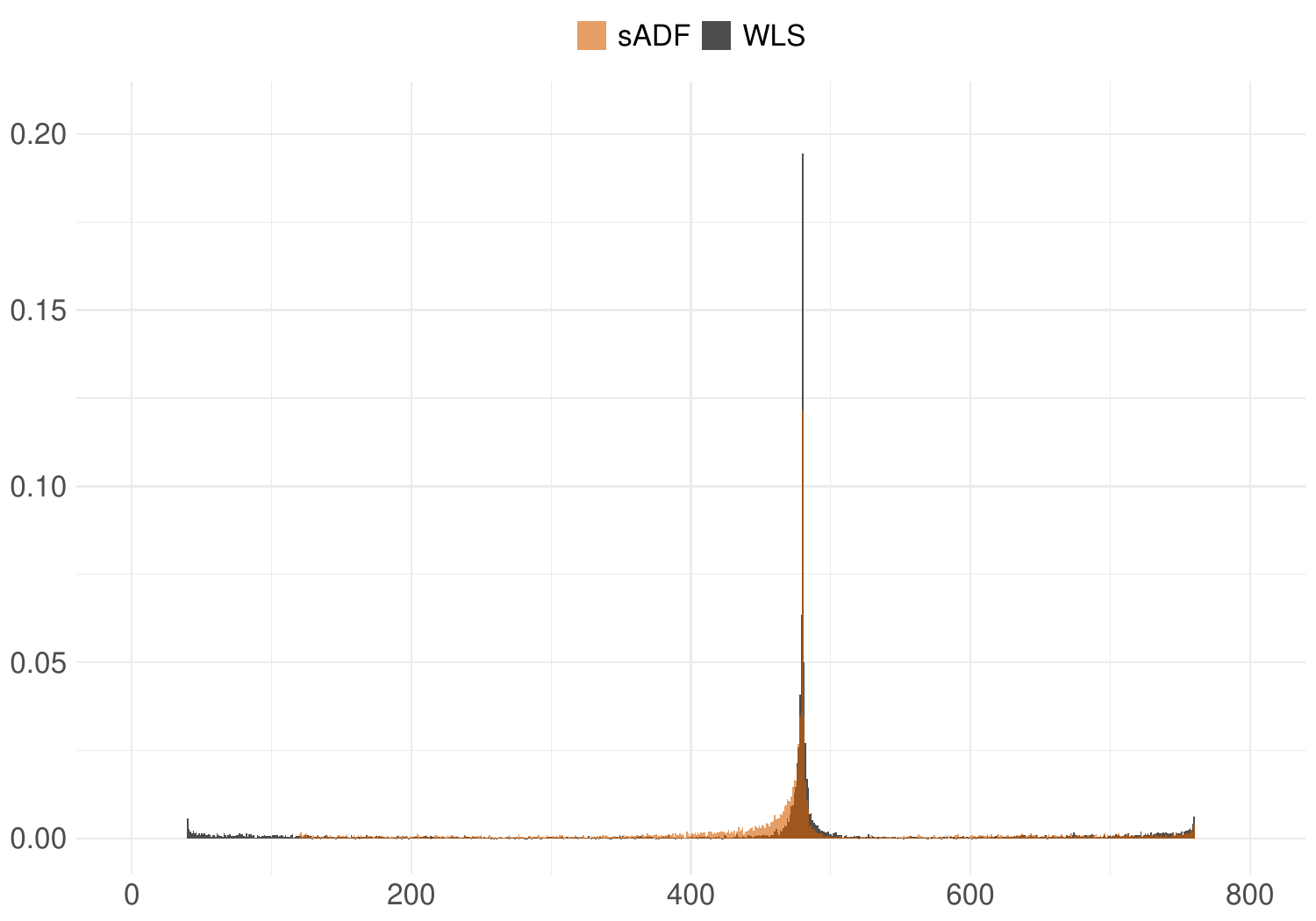}\label{fig3:4:6}}\\
\end{center}%
\caption{Histograms of $\hat{k}_c$ 
for $(\tau_e,\tau_c,\tau_r)=(0.4,0.6,0.7)$,  $\tau=0.2$, $s_0/s_1=1$, $T=800$}
\label{fig34}
\end{figure}

\newpage

\begin{figure}[h!]%
\begin{center}%
\subfigure[$T=400$, $c_a=4$, $c_b=6$, $s_0/s_1=5$]{\includegraphics[width=0.45\linewidth]{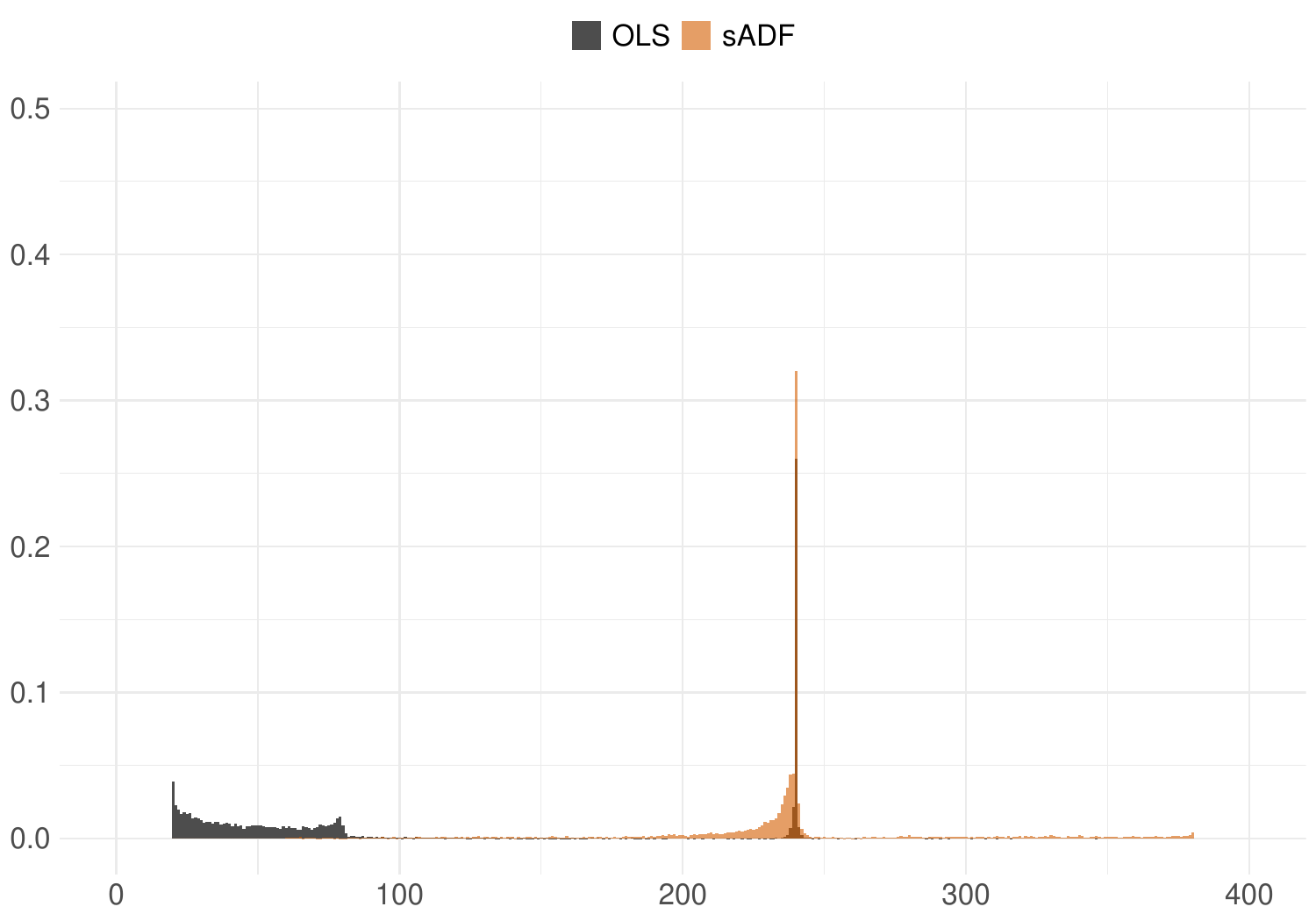}\label{fig3:5:1:app}}
\subfigure[$T=400$, $c_a=4$, $c_b=6$, $s_0/s_1=5$]{\includegraphics[width=0.45\linewidth]{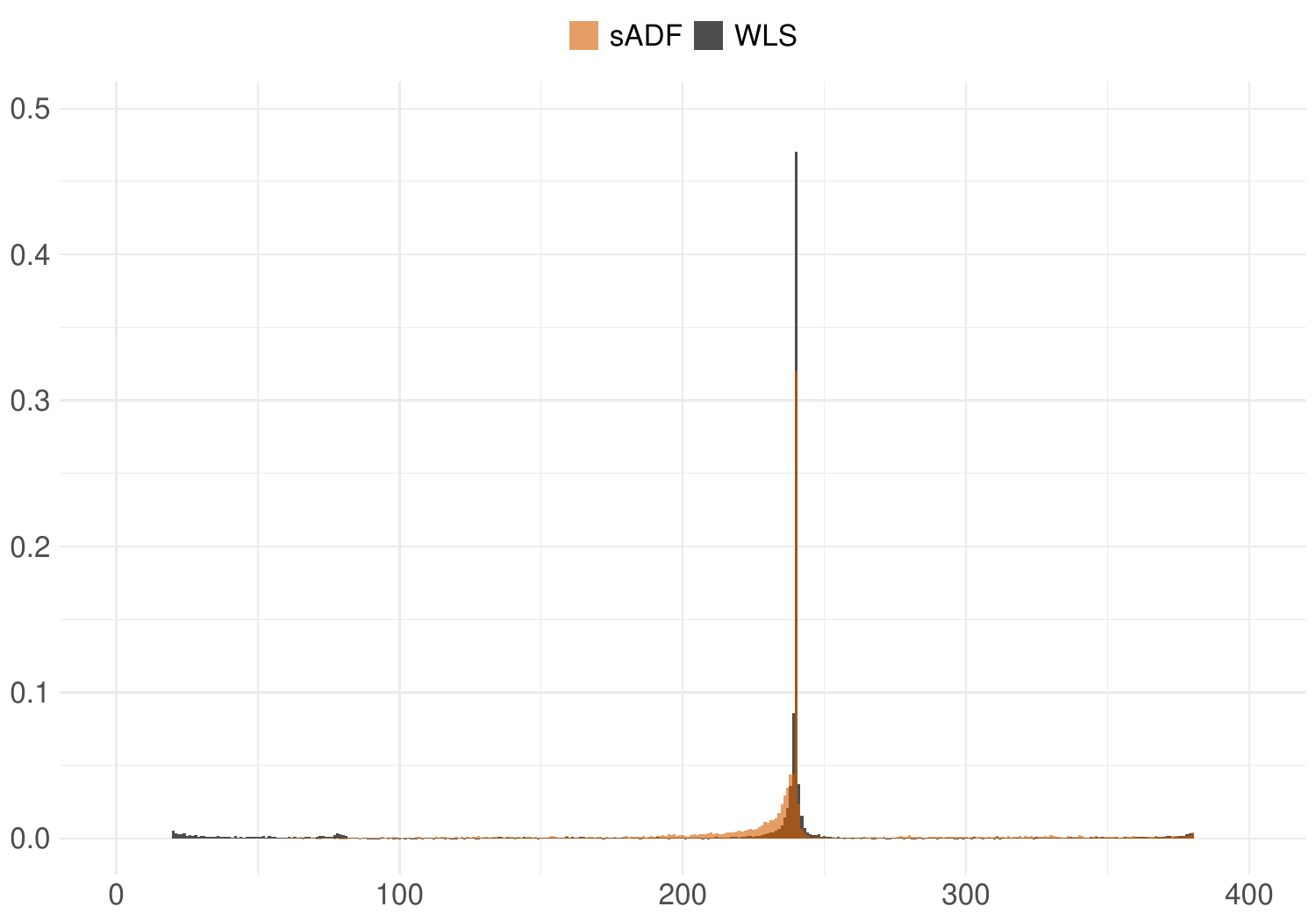}\label{fig3:5:2:app}}\\
\subfigure[$T=400$, $c_a=5$, $c_b=6$, $s_0/s_1=5$]{\includegraphics[width=0.45\linewidth]{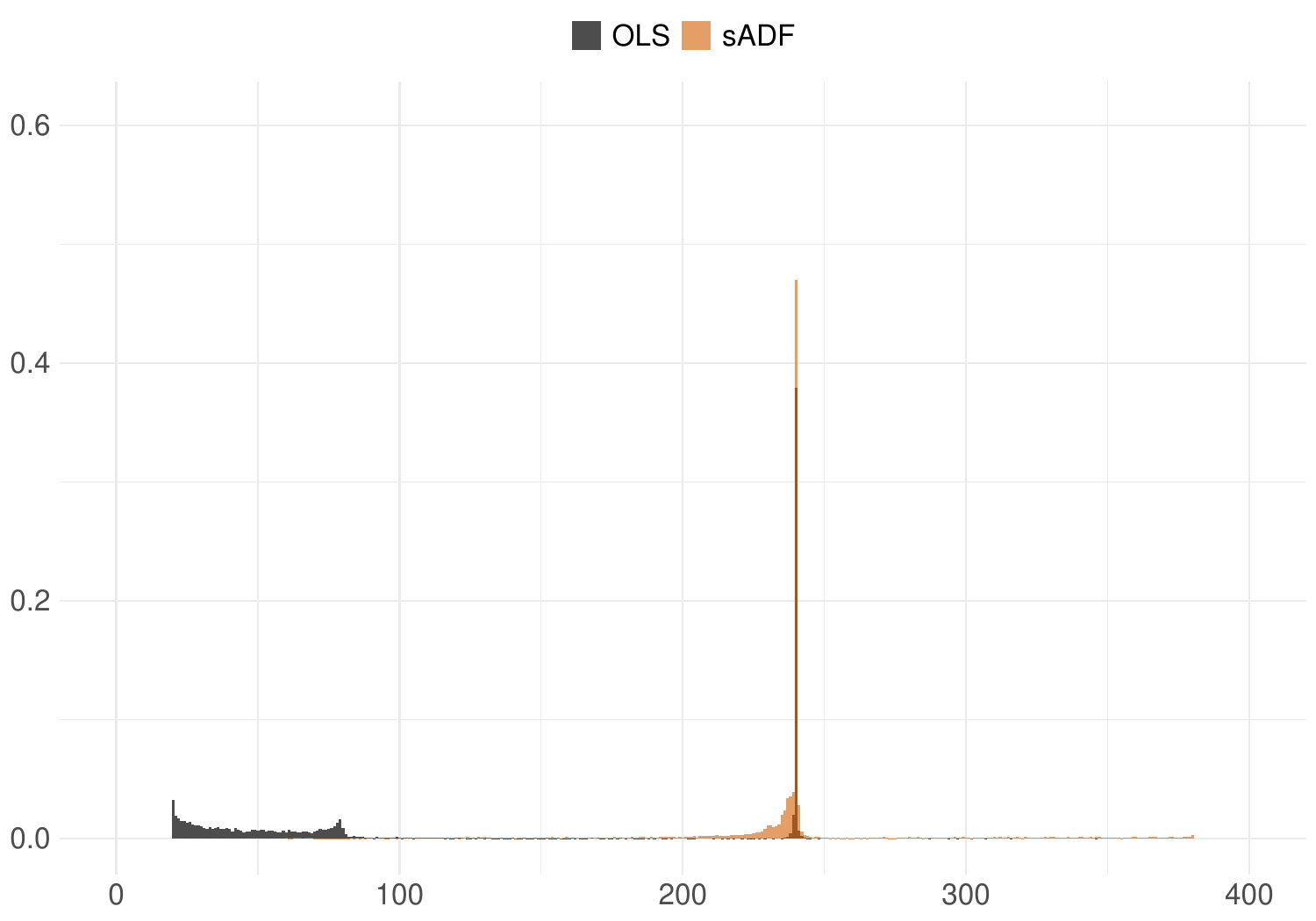}\label{fig3:5:3:app}}
\subfigure[$T=400$, $c_a=5$, $c_b=6$, $s_0/s_1=5$]{\includegraphics[width=0.45\linewidth]{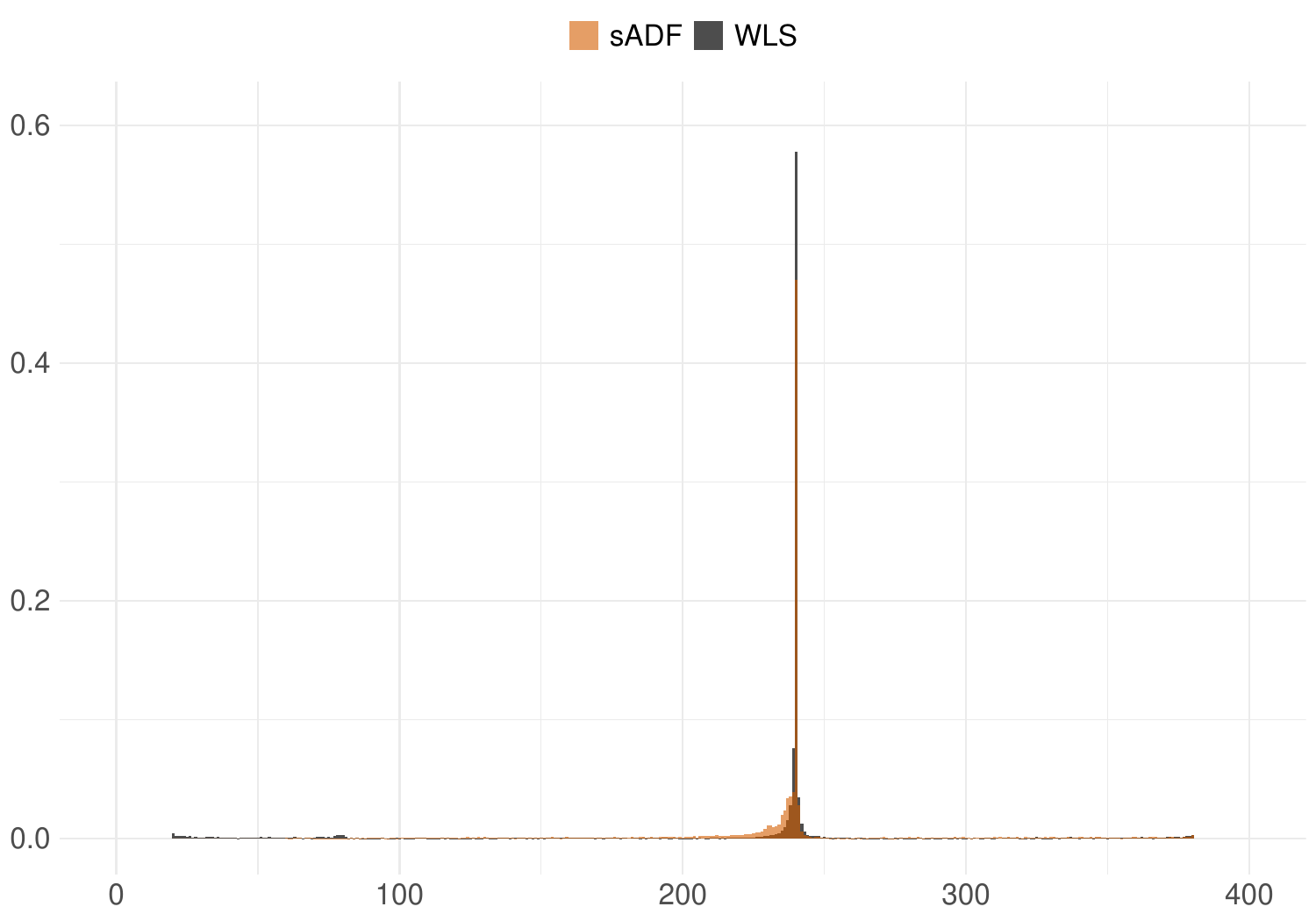}\label{fig3:5:4:app}}\\
\subfigure[$T=400$, $c_a=6$, $c_b=6$, $s_0/s_1=5$]{\includegraphics[width=0.45\linewidth]{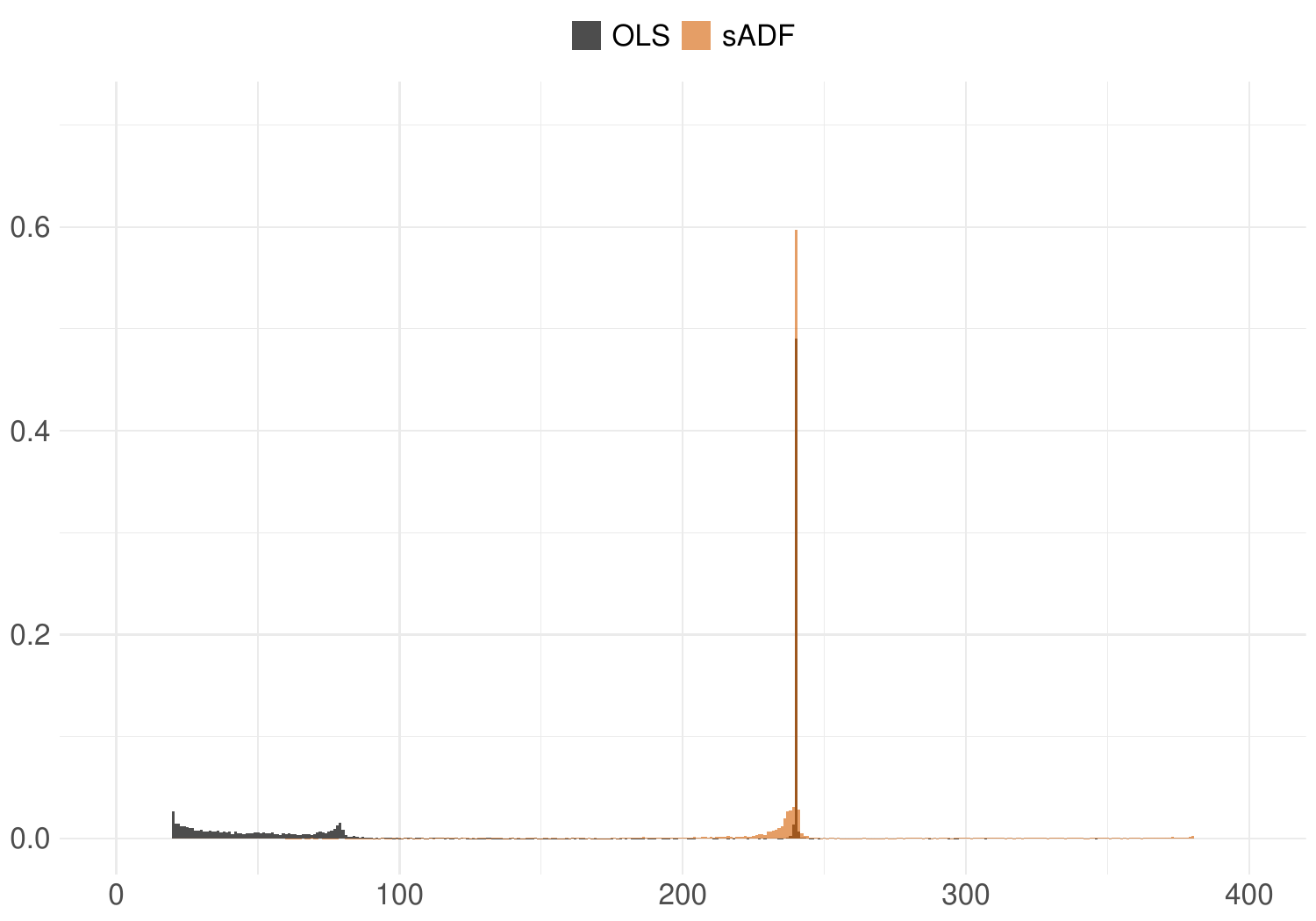}\label{fig3:5:5:app}}
\subfigure[$T=400$, $c_a=6$, $c_b=6$, $s_0/s_1=5$]{\includegraphics[width=0.45\linewidth]{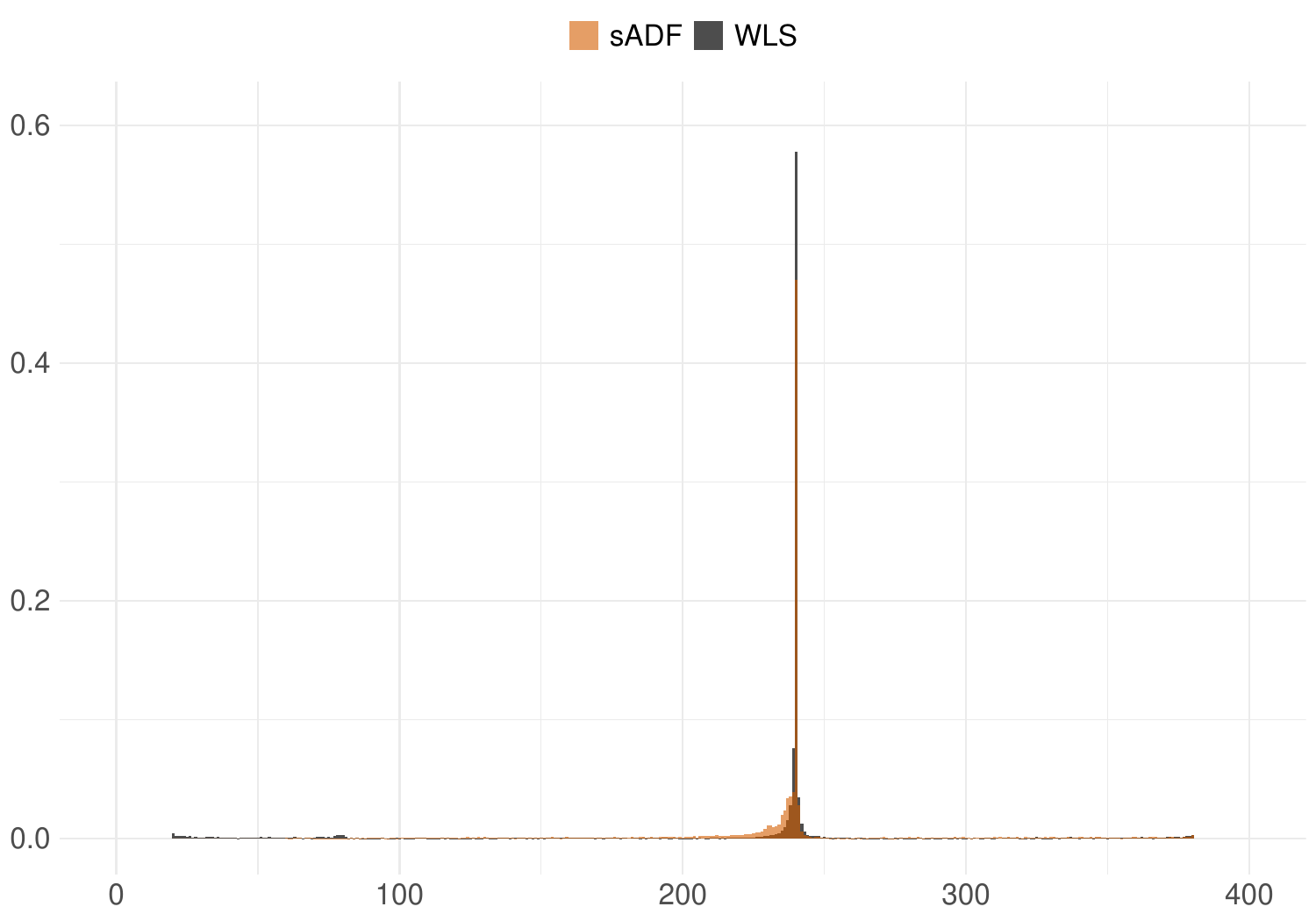}\label{fig3:5:6:app}}\\
\end{center}%
\caption{Histograms of $\hat{k}_c$ 
for $(\tau_e,\tau_c,\tau_r)=(0.4,0.6,0.7)$,  $\tau=0.2$, $s_0/s_1=5$, $T=400$}
\label{fig35:app}
\end{figure}

\newpage

\begin{figure}[h!]%
\begin{center}%
\subfigure[$T=800$, $c_a=4$, $c_b=6$, $s_0/s_1=5$]{\includegraphics[width=0.45\linewidth]{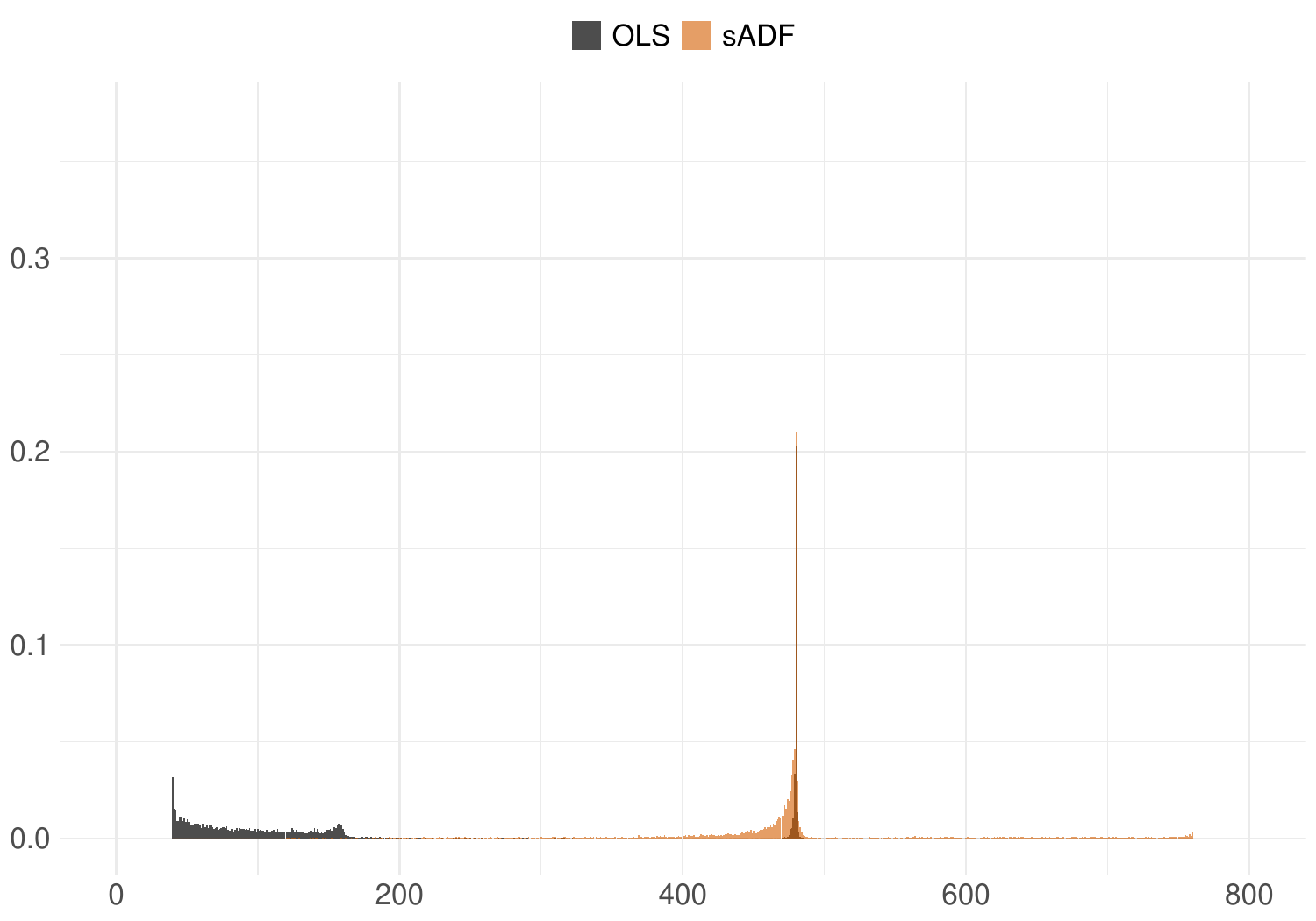}\label{fig3:6:1}}
\subfigure[$T=800$, $c_a=4$, $c_b=6$, $s_0/s_1=5$]{\includegraphics[width=0.45\linewidth]{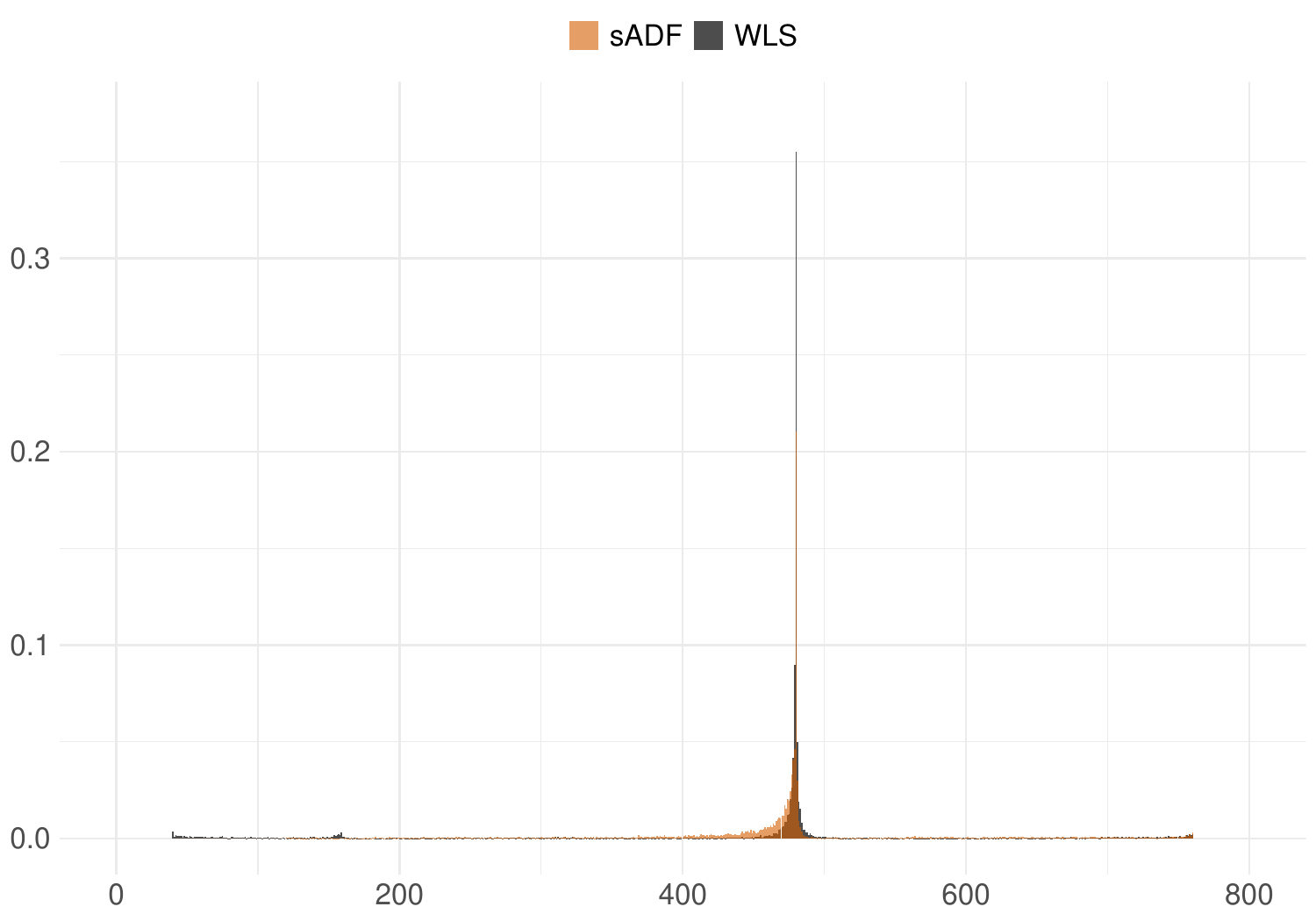}\label{fig3:6:2}}\\
\subfigure[$T=800$, $c_a=5$, $c_b=6$, $s_0/s_1=5$]{\includegraphics[width=0.45\linewidth]{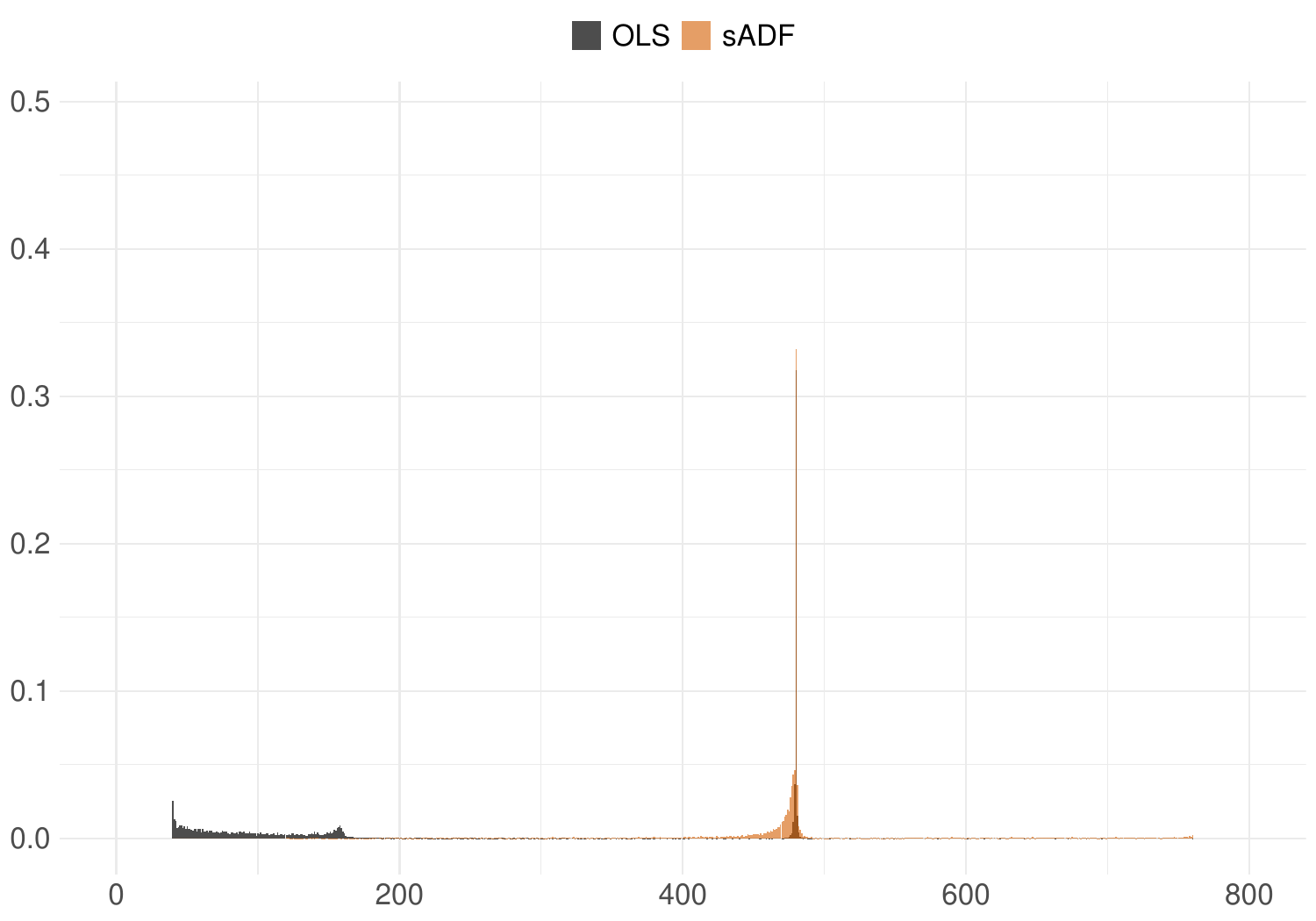}\label{fig3:6:3}}
\subfigure[$T=800$, $c_a=5$, $c_b=6$, $s_0/s_1=5$]{\includegraphics[width=0.45\linewidth]{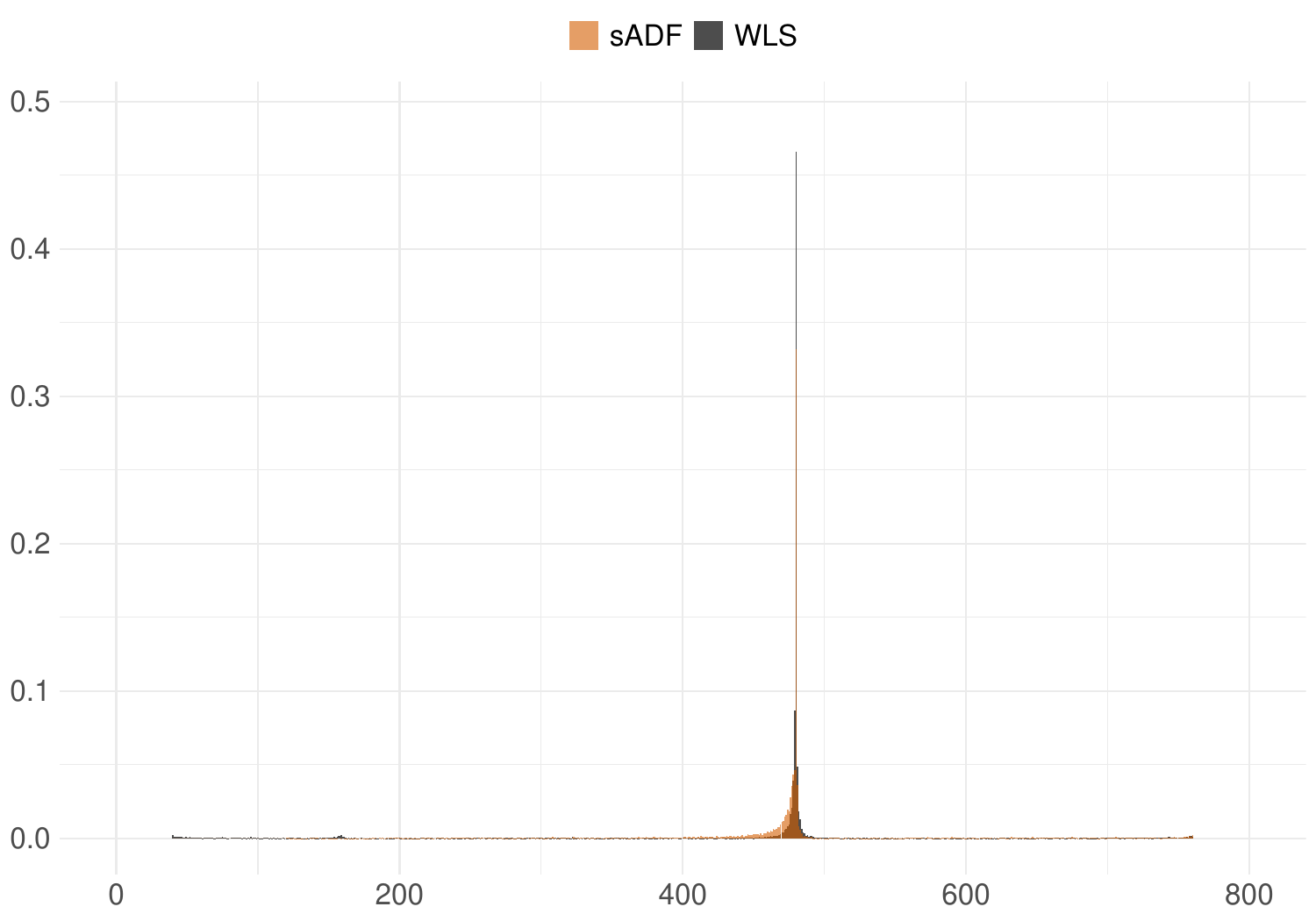}\label{fig3:6:4}}\\
\subfigure[$T=800$, $c_a=6$, $c_b=6$, $s_0/s_1=5$]{\includegraphics[width=0.45\linewidth]{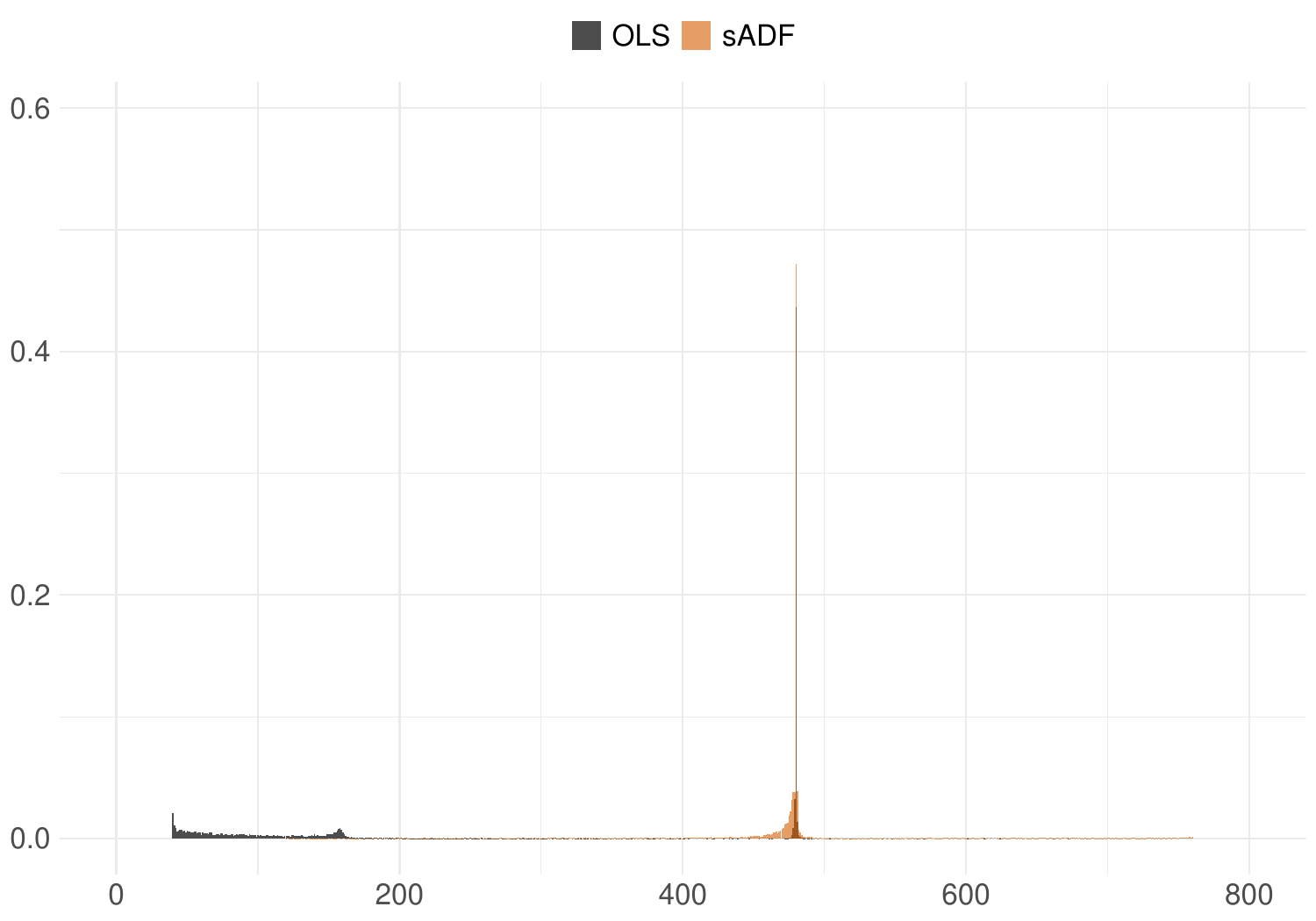}\label{fig3:6:5}}
\subfigure[$T=800$, $c_a=6$, $c_b=6$, $s_0/s_1=5$]{\includegraphics[width=0.45\linewidth]{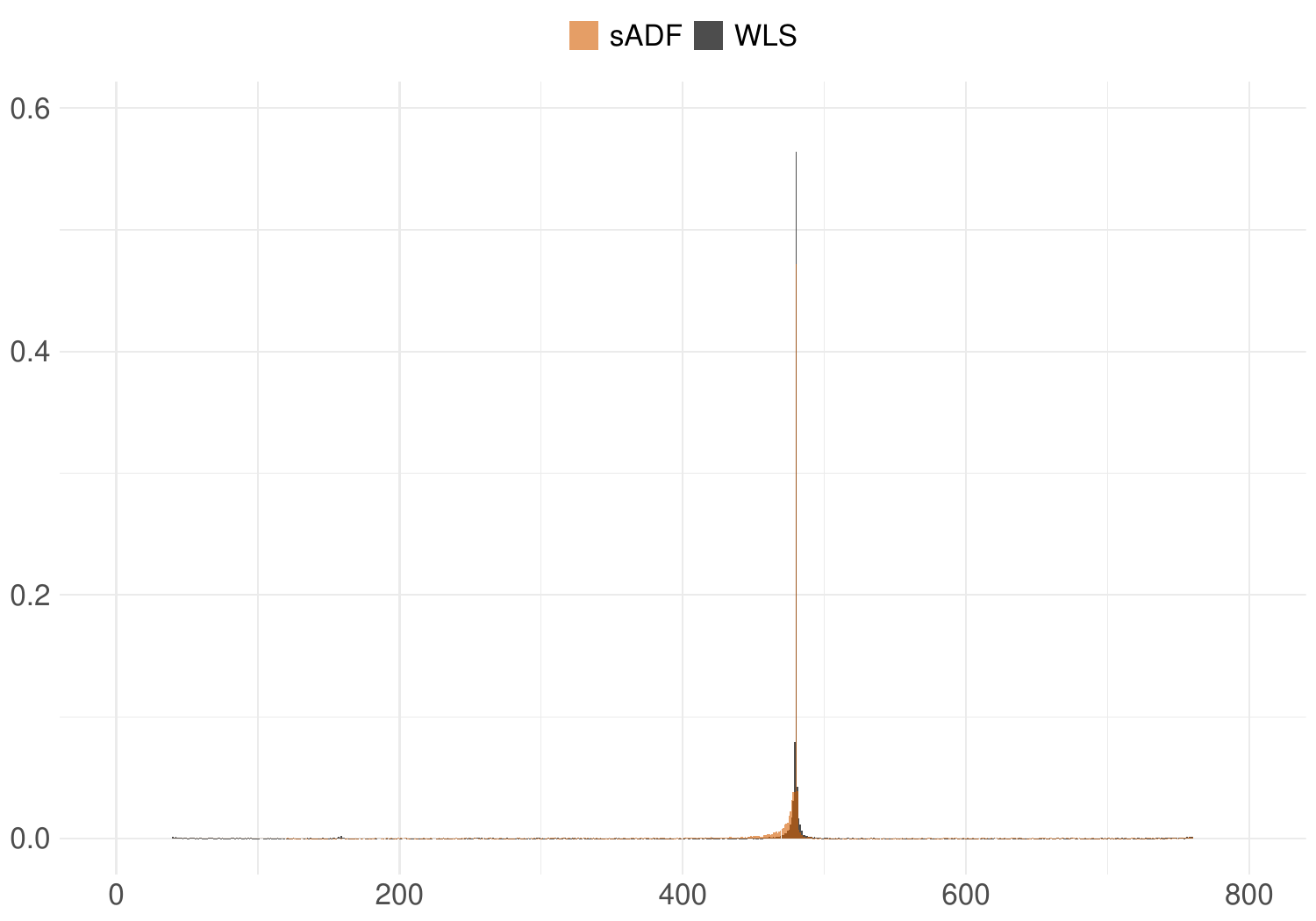}\label{fig3:6:6}}\\
\end{center}%
\caption{Histograms of $\hat{k}_c$ 
for $(\tau_e,\tau_c,\tau_r)=(0.4,0.6,0.7)$,  $\tau=0.2$, $s_0/s_1=5$, $T=800$}
\label{fig36}
\end{figure}

\newpage

\section{$\tau=0.2$, $\hat{k}_e$}
\setcounter{figure}{0}

\begin{figure}[h!]%
\begin{center}%
\subfigure[$T=400$, $c_a=4$, $c_b=6$, $s_0/s_1=1/5$]{\includegraphics[width=0.45\linewidth]{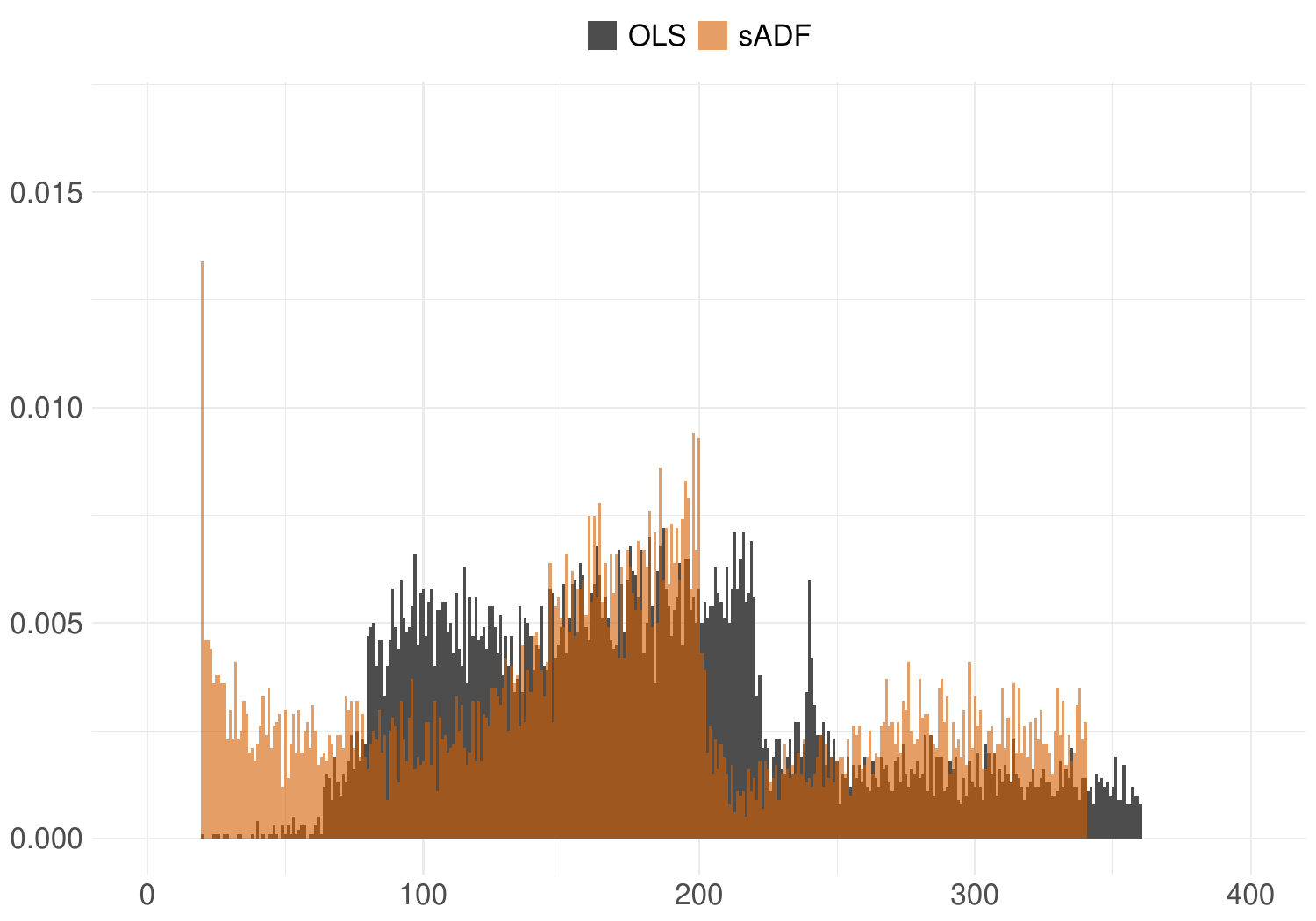}\label{fig4:1:1}}
\subfigure[$T=400$, $c_a=4$, $c_b=6$, $s_0/s_1=1/5$]{\includegraphics[width=0.45\linewidth]{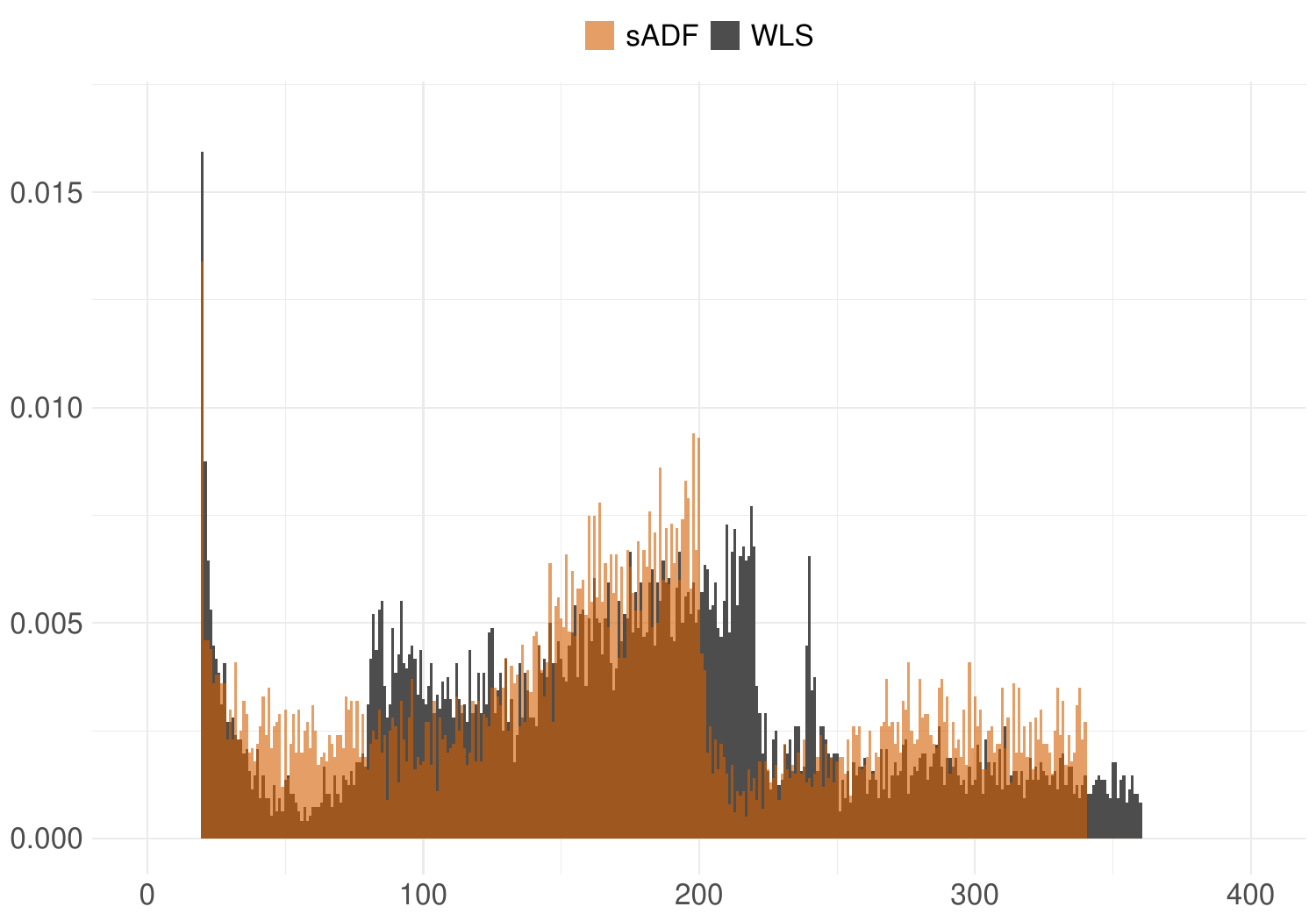}\label{fig4:1:2}}\\
\subfigure[$T=400$, $c_a=5$, $c_b=6$, $s_0/s_1=1/5$]{\includegraphics[width=0.45\linewidth]{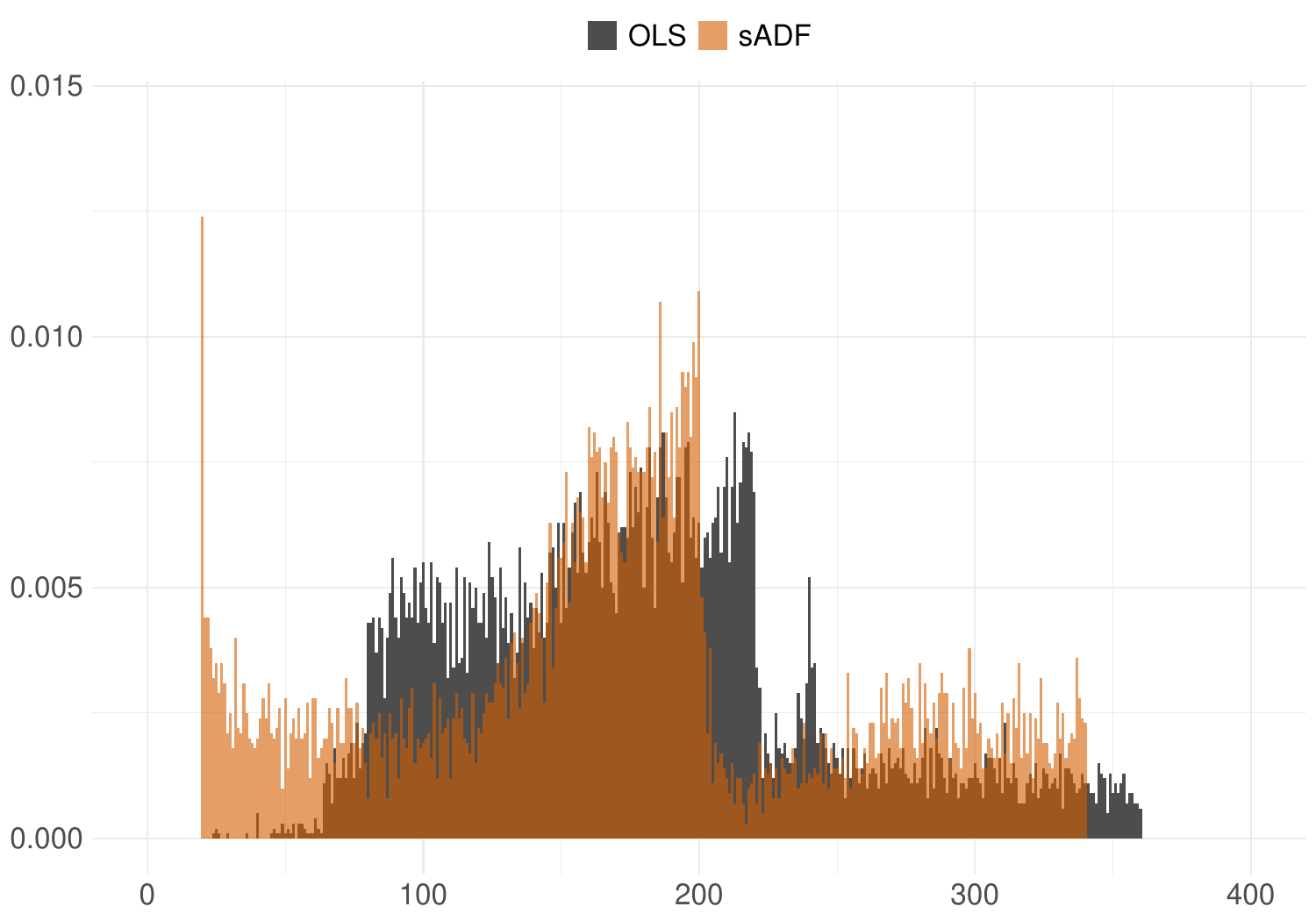}\label{fig4:1:3}}
\subfigure[$T=400$, $c_a=5$, $c_b=6$, $s_0/s_1=1/5$]{\includegraphics[width=0.45\linewidth]{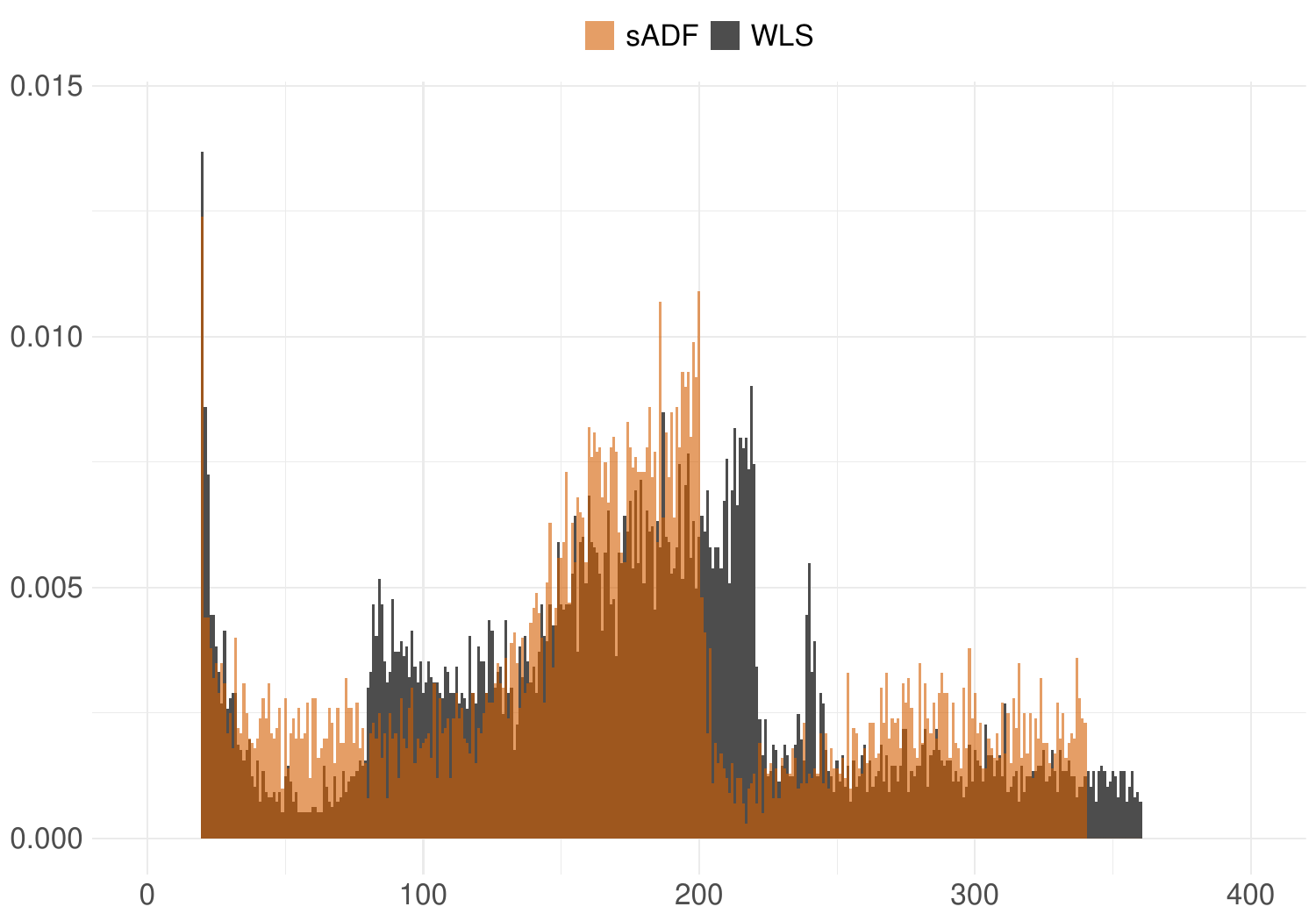}\label{fig4:1:4}}\\
\subfigure[$T=400$, $c_a=6$, $c_b=6$, $s_0/s_1=1/5$]{\includegraphics[width=0.45\linewidth]{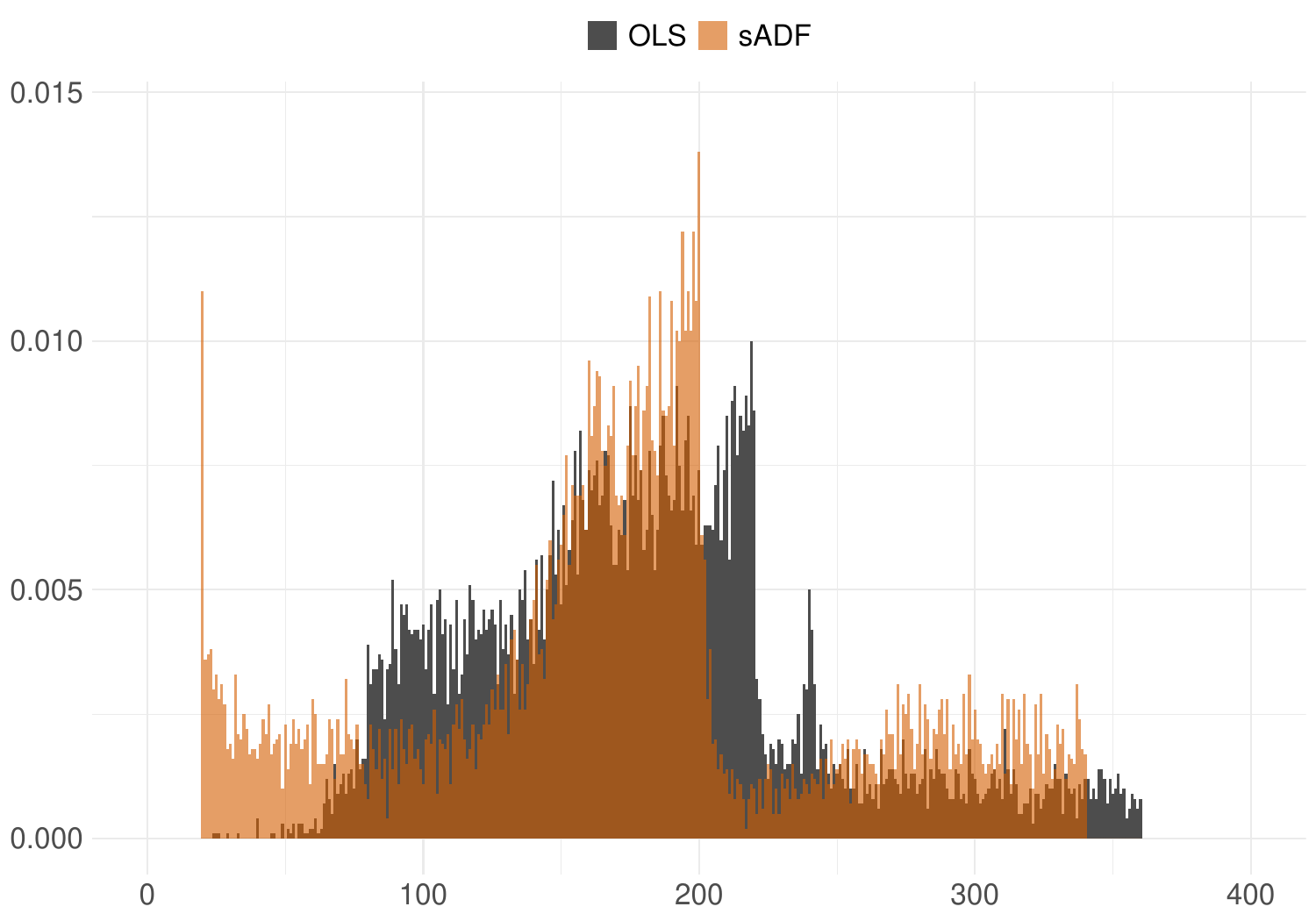}\label{fig4:1:5}}
\subfigure[$T=400$, $c_a=6$, $c_b=6$, $s_0/s_1=1/5$]{\includegraphics[width=0.45\linewidth]{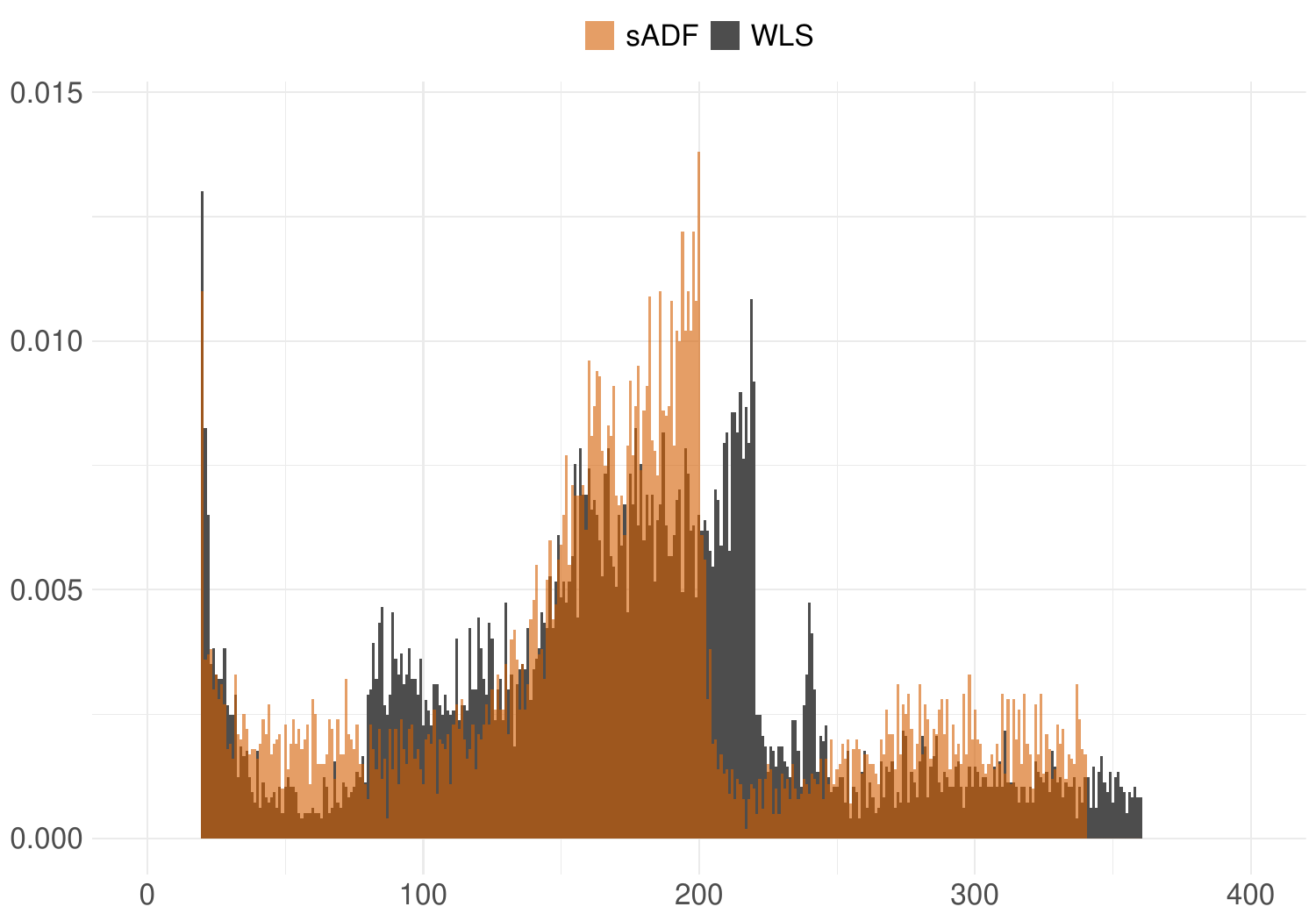}\label{fig4:1:6}}\\
\end{center}%
\caption{Histograms of $\hat{k}_e$ 
for $(\tau_e,\tau_c,\tau_r)=(0.4,0.6,0.7)$,  $\tau=0.2$, $s_0/s_1=1/5$, $T=400$}
\label{fig41}
\end{figure}

\newpage

\begin{figure}[h!]%
\begin{center}%
\subfigure[$T=800$, $c_a=4$, $c_b=6$, $s_0/s_1=1/5$]{\includegraphics[width=0.45\linewidth]{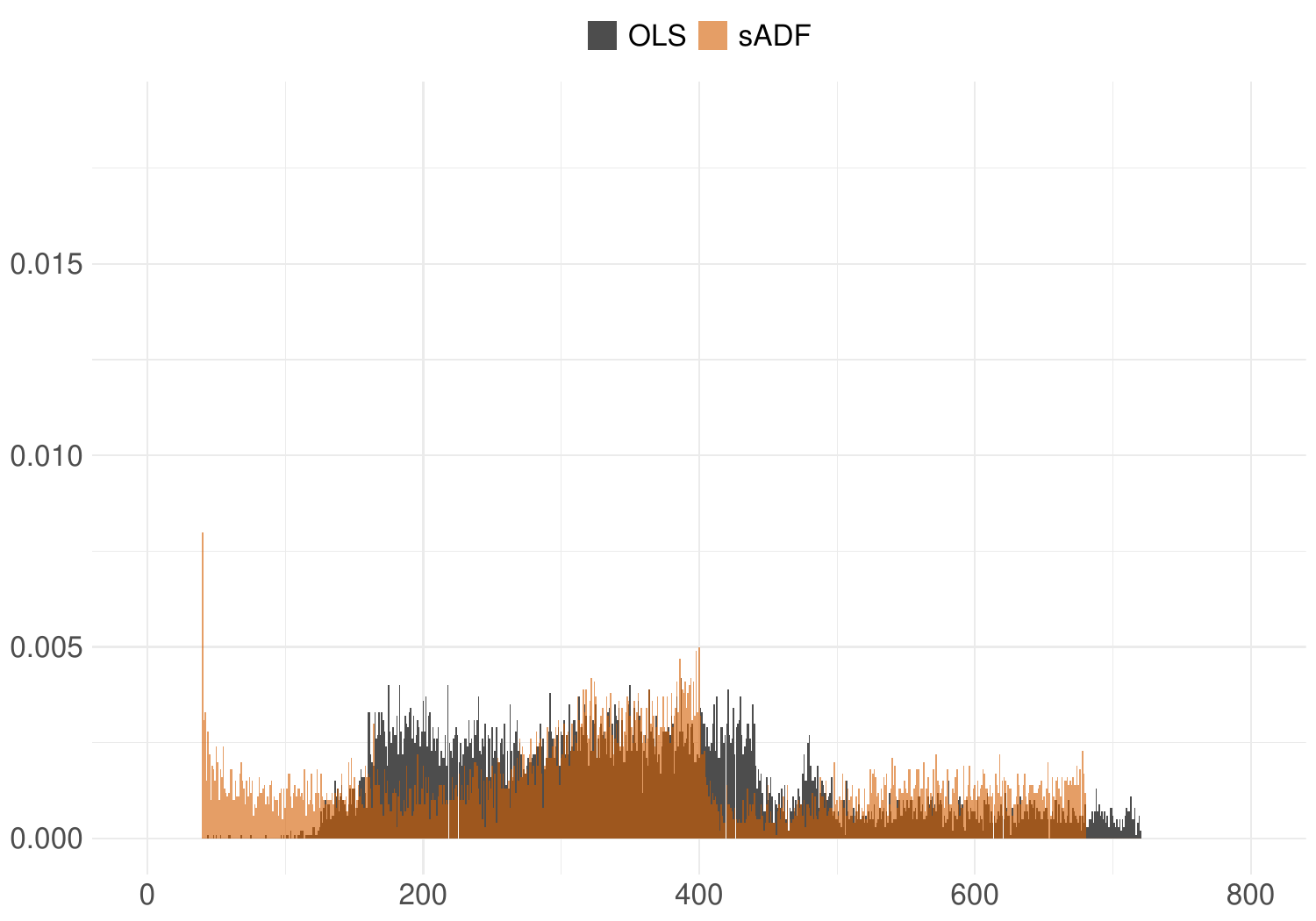}\label{fig4:2:1}}
\subfigure[$T=800$, $c_a=4$, $c_b=6$, $s_0/s_1=1/5$]{\includegraphics[width=0.45\linewidth]{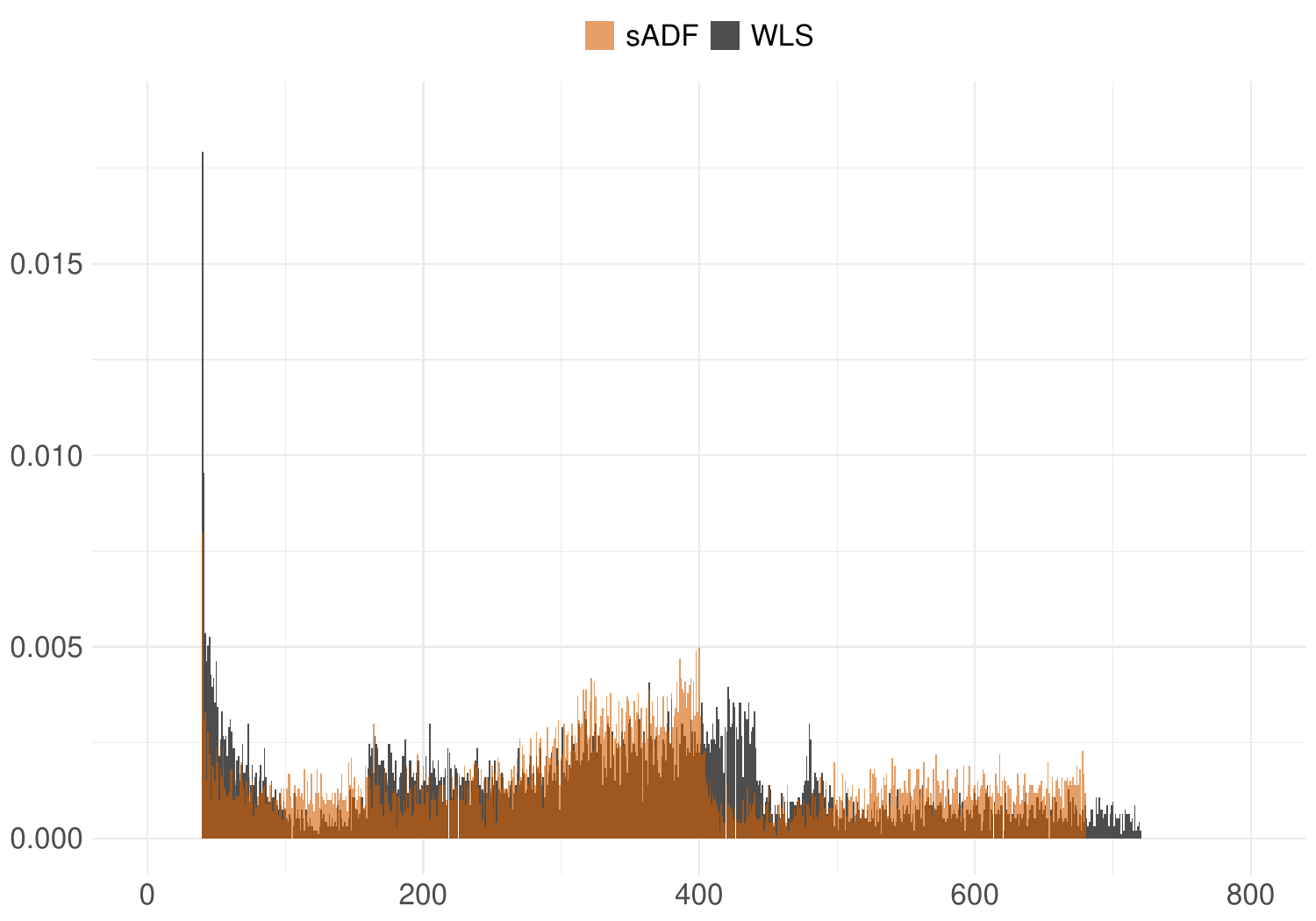}\label{fig4:2:2}}\\
\subfigure[$T=800$, $c_a=5$, $c_b=6$, $s_0/s_1=1/5$]{\includegraphics[width=0.45\linewidth]{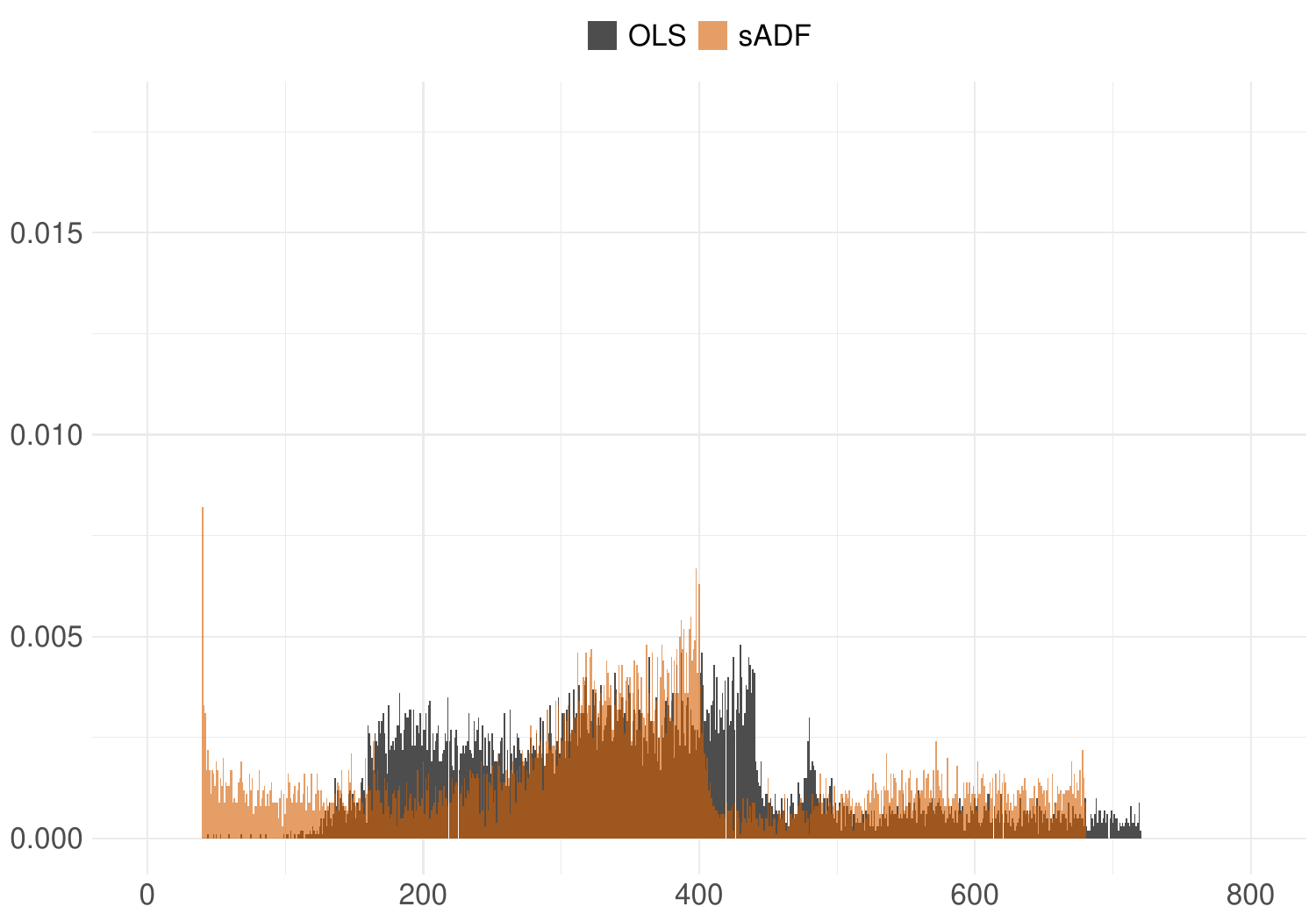}\label{fig4:2:3}}
\subfigure[$T=800$, $c_a=5$, $c_b=6$, $s_0/s_1=1/5$]{\includegraphics[width=0.45\linewidth]{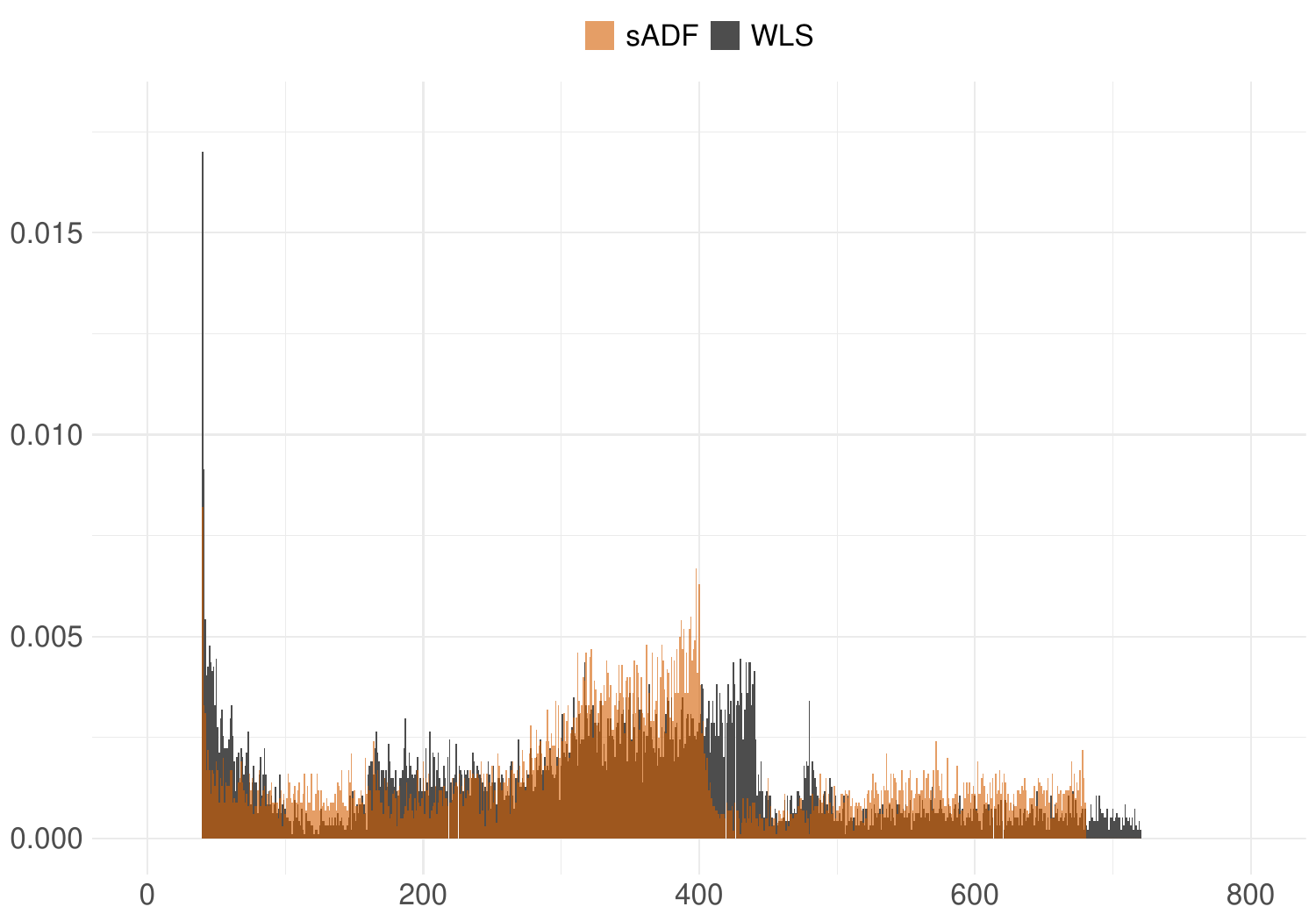}\label{fig4:2:4}}\\
\subfigure[$T=800$, $c_a=6$, $c_b=6$, $s_0/s_1=1/5$]{\includegraphics[width=0.45\linewidth]{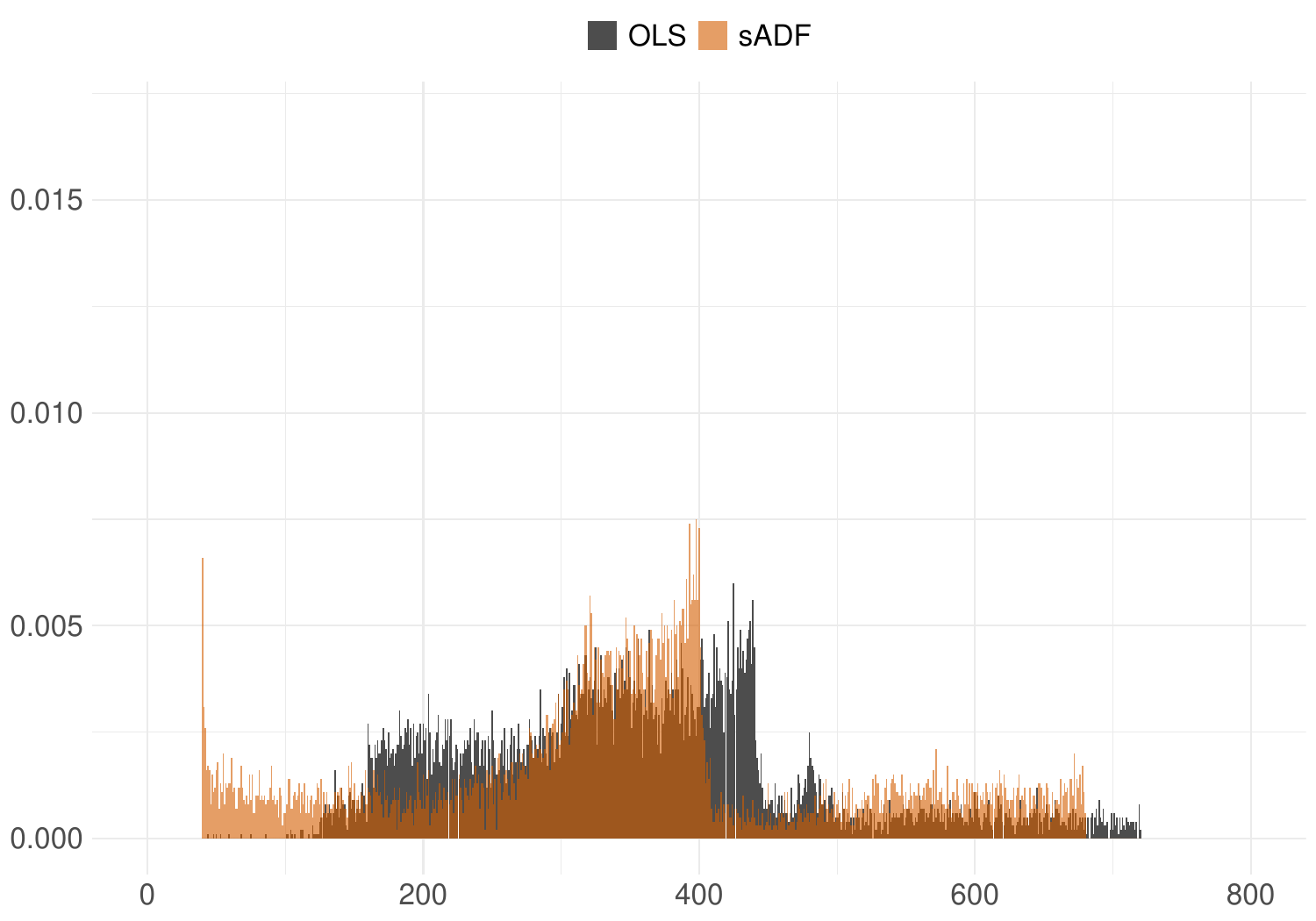}\label{fig4:2:5}}
\subfigure[$T=800$, $c_a=6$, $c_b=6$, $s_0/s_1=1/5$]{\includegraphics[width=0.45\linewidth]{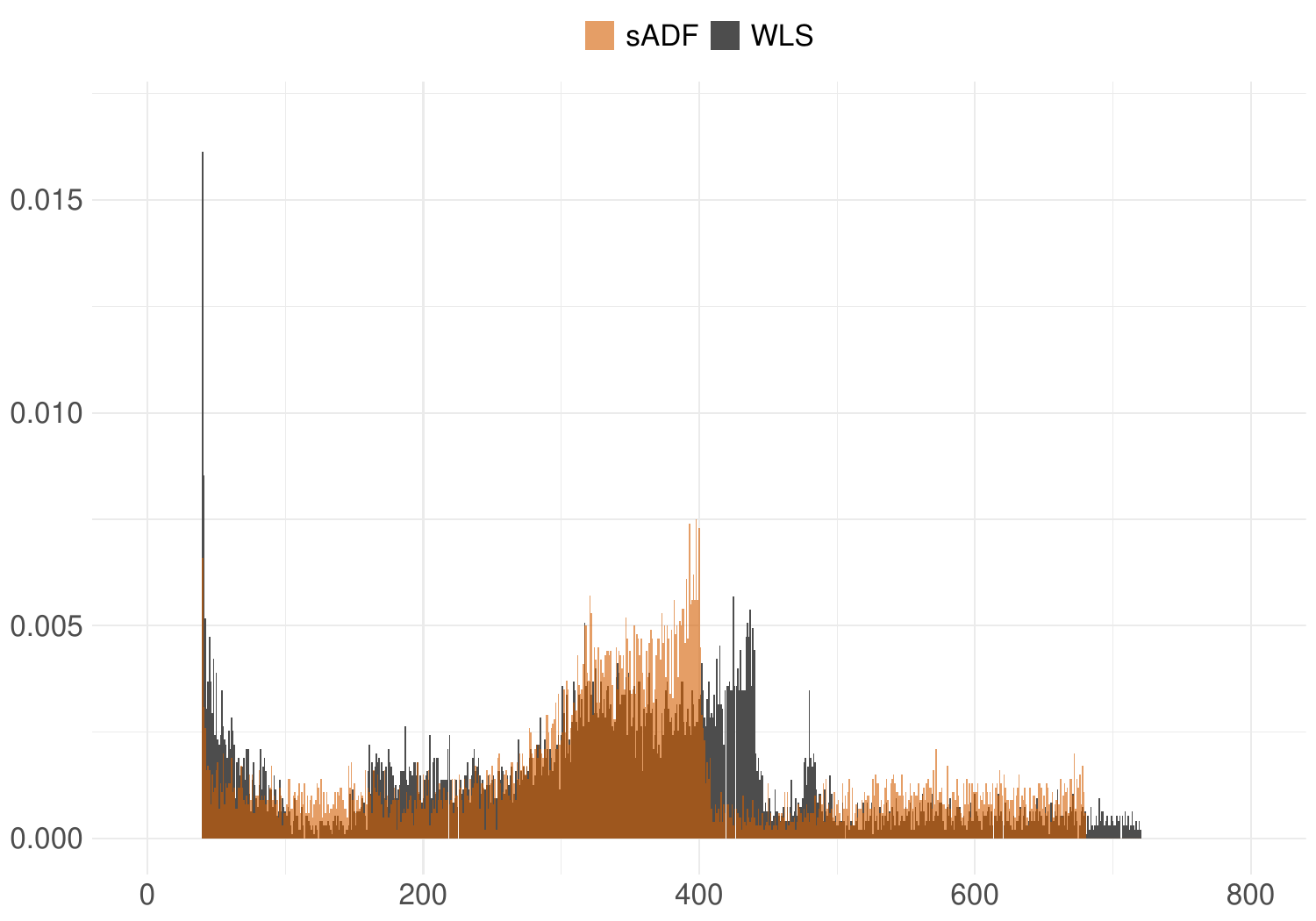}\label{fig4:2:6}}\\
\end{center}%
\caption{Histograms of $\hat{k}_e$ 
for $(\tau_e,\tau_c,\tau_r)=(0.4,0.6,0.7)$,  $\tau=0.2$, $s_0/s_1=1/5$, $T=800$}
\label{fig42}
\end{figure}

\newpage

\begin{figure}[h!]%
\begin{center}%
\subfigure[$T=400$, $c_a=4$, $c_b=6$, $s_0/s_1=1$]{\includegraphics[width=0.45\linewidth]{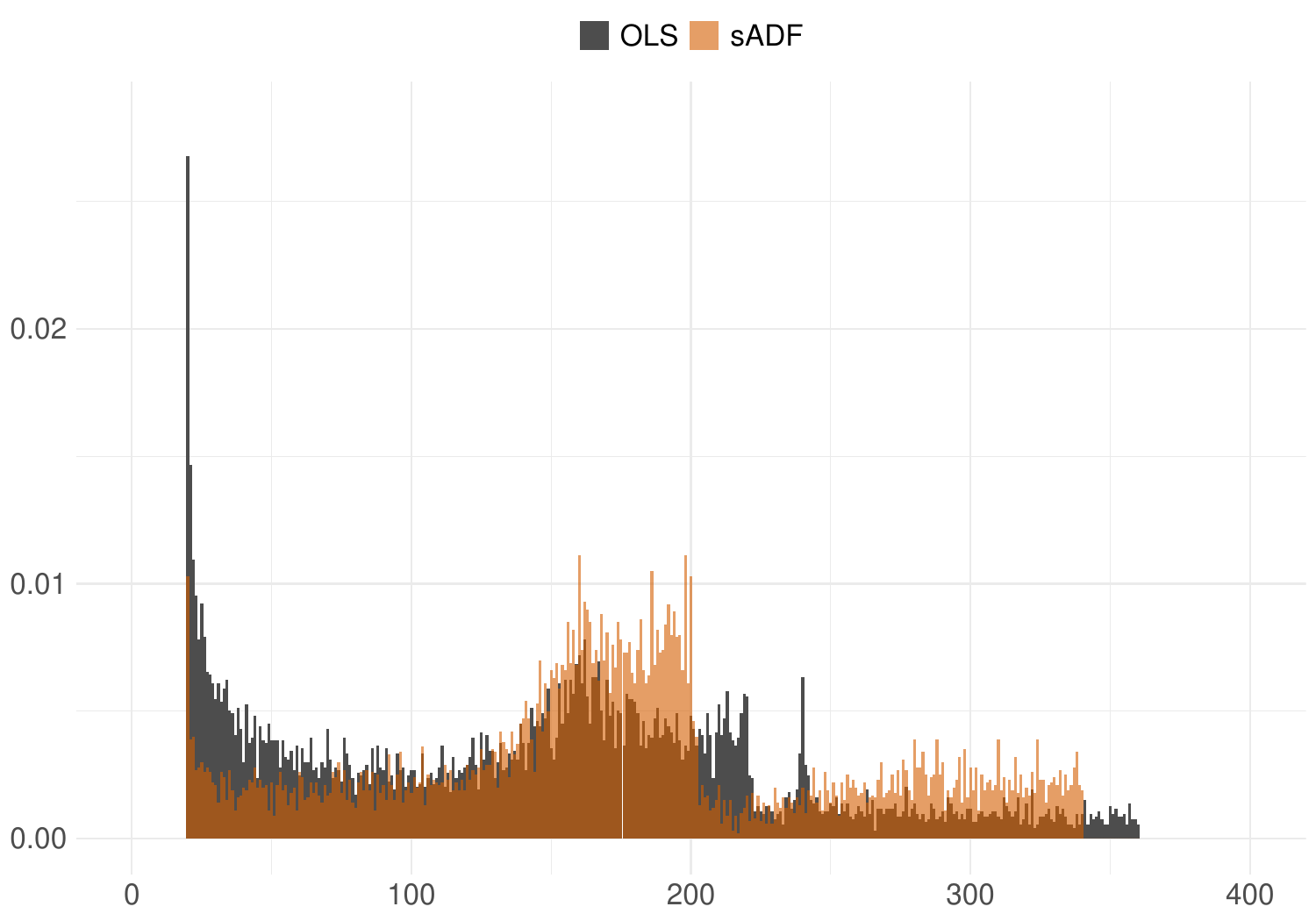}\label{fig4:3:1}}
\subfigure[$T=400$, $c_a=4$, $c_b=6$, $s_0/s_1=1$]{\includegraphics[width=0.45\linewidth]{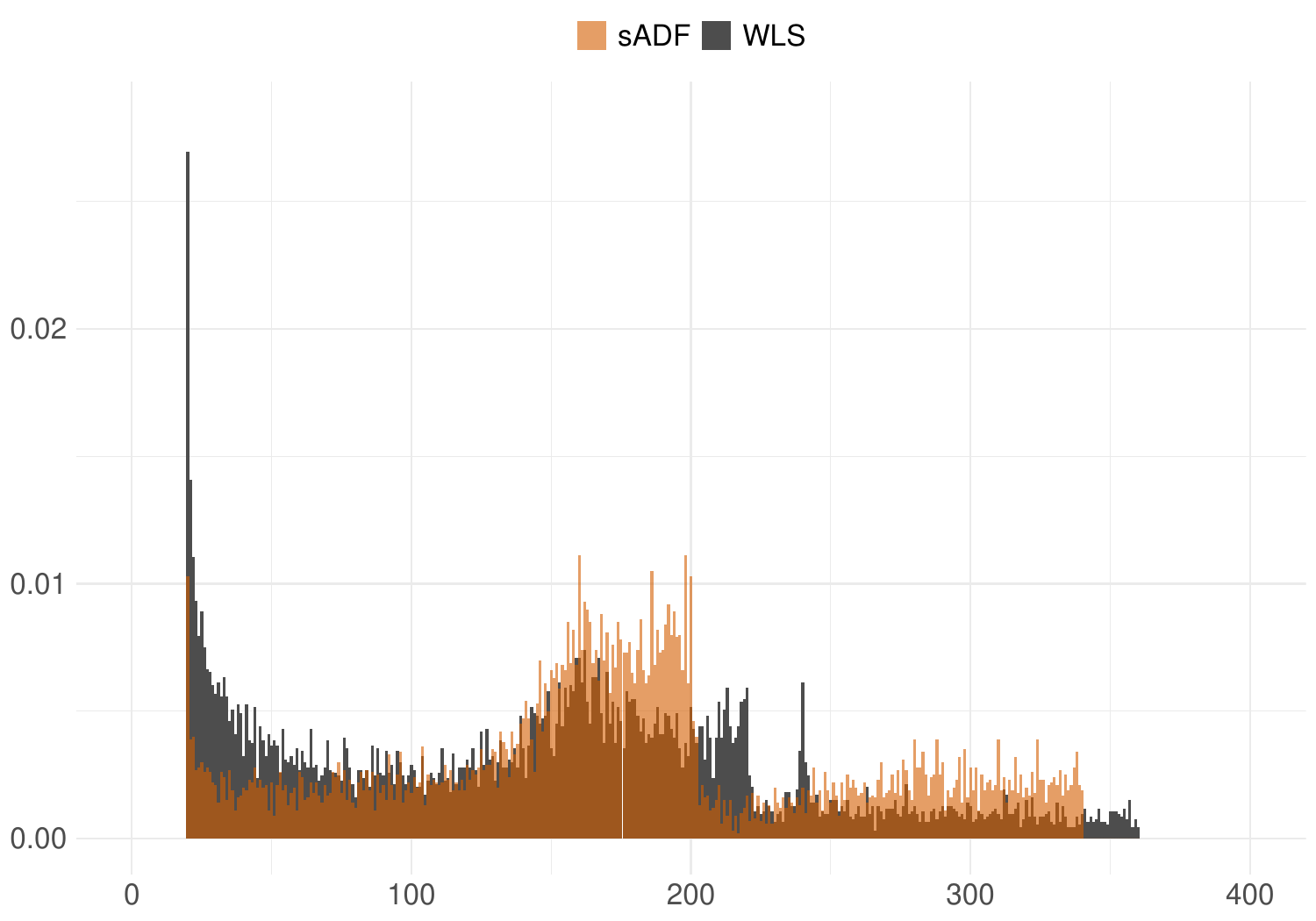}\label{fig4:3:2}}\\
\subfigure[$T=400$, $c_a=5$, $c_b=6$, $s_0/s_1=1$]{\includegraphics[width=0.45\linewidth]{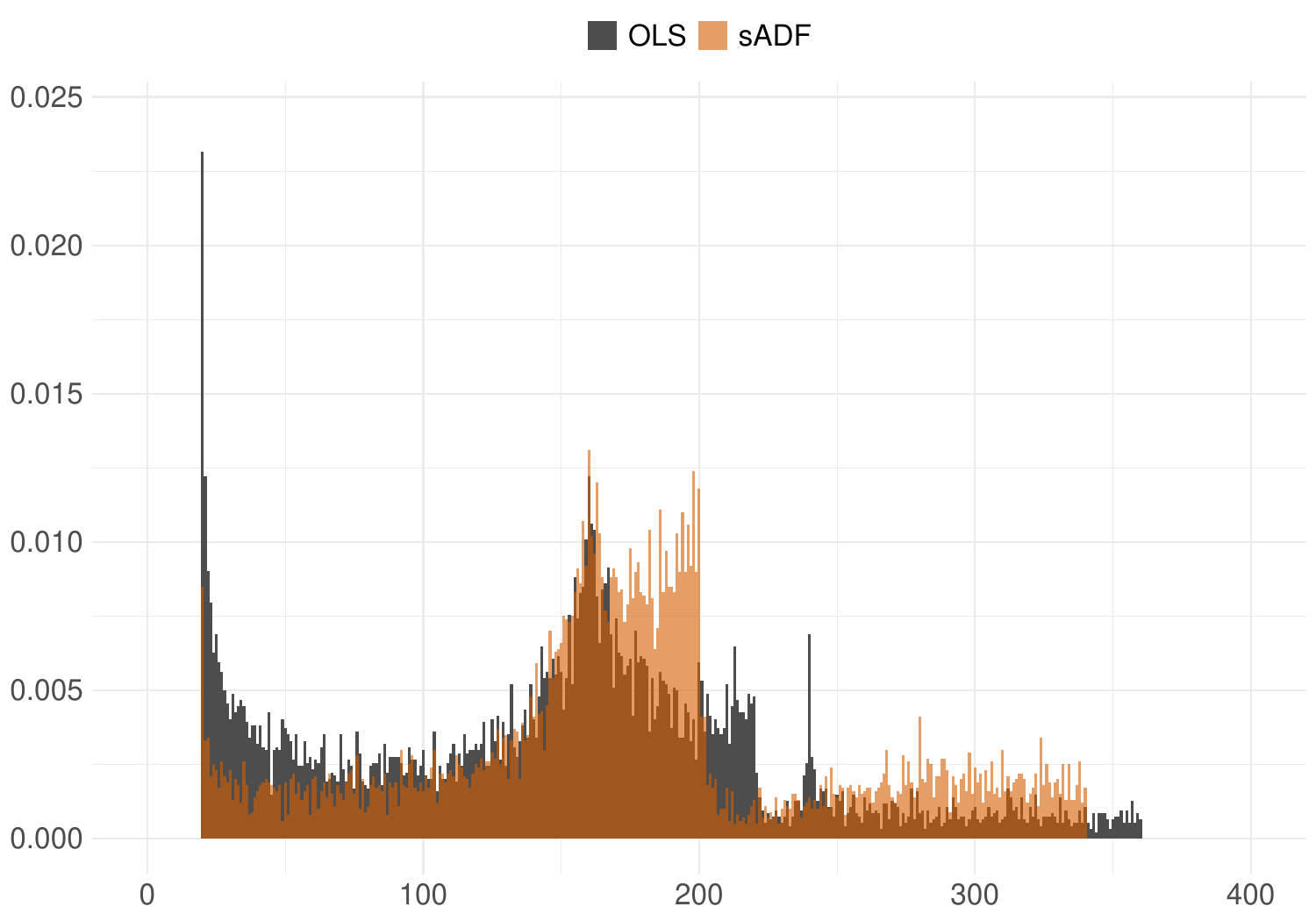}\label{fig4:3:3}}
\subfigure[$T=400$, $c_a=5$, $c_b=6$, $s_0/s_1=1$]{\includegraphics[width=0.45\linewidth]{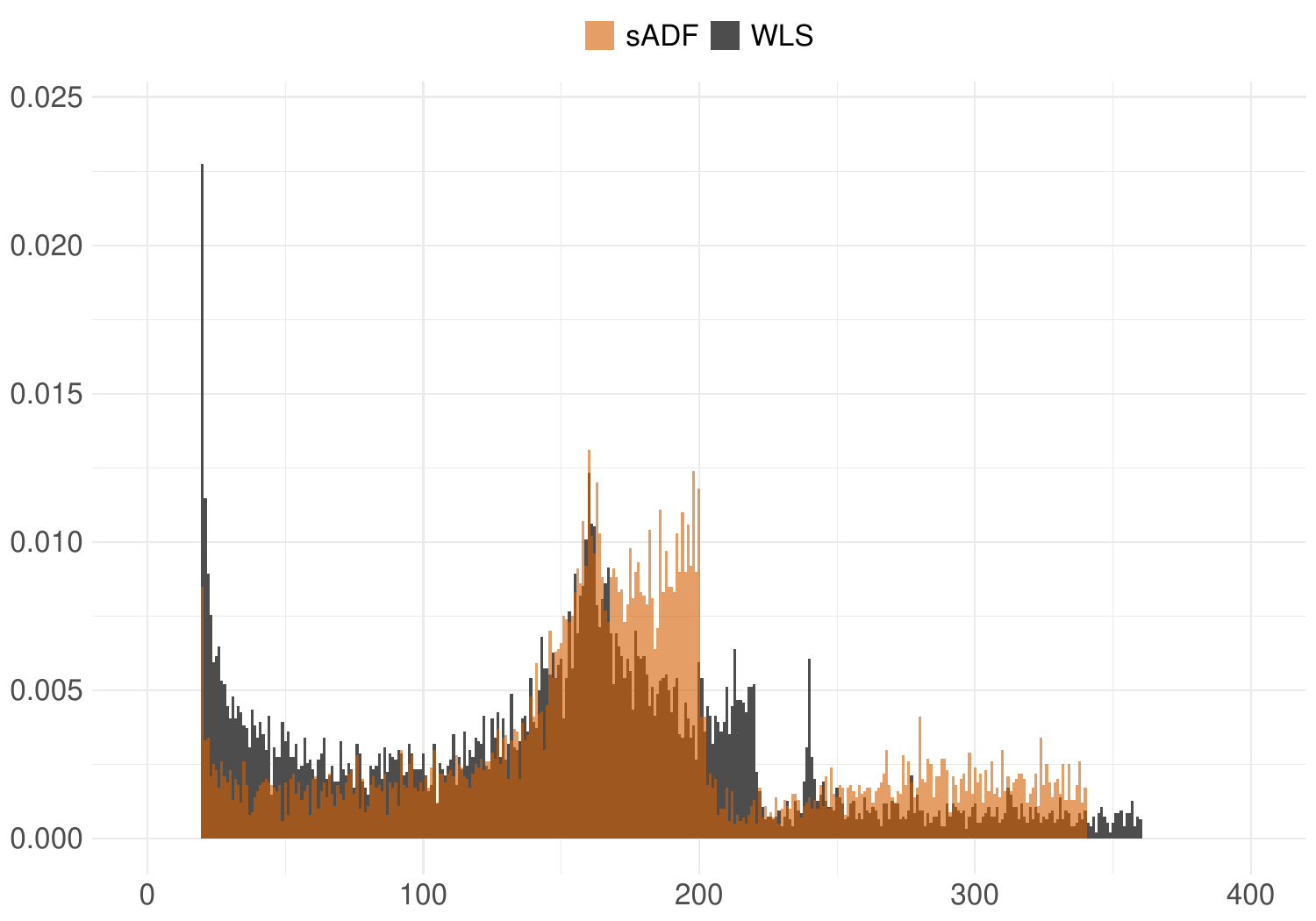}\label{fig4:3:4}}\\
\subfigure[$T=400$, $c_a=6$, $c_b=6$, $s_0/s_1=1$]{\includegraphics[width=0.45\linewidth]{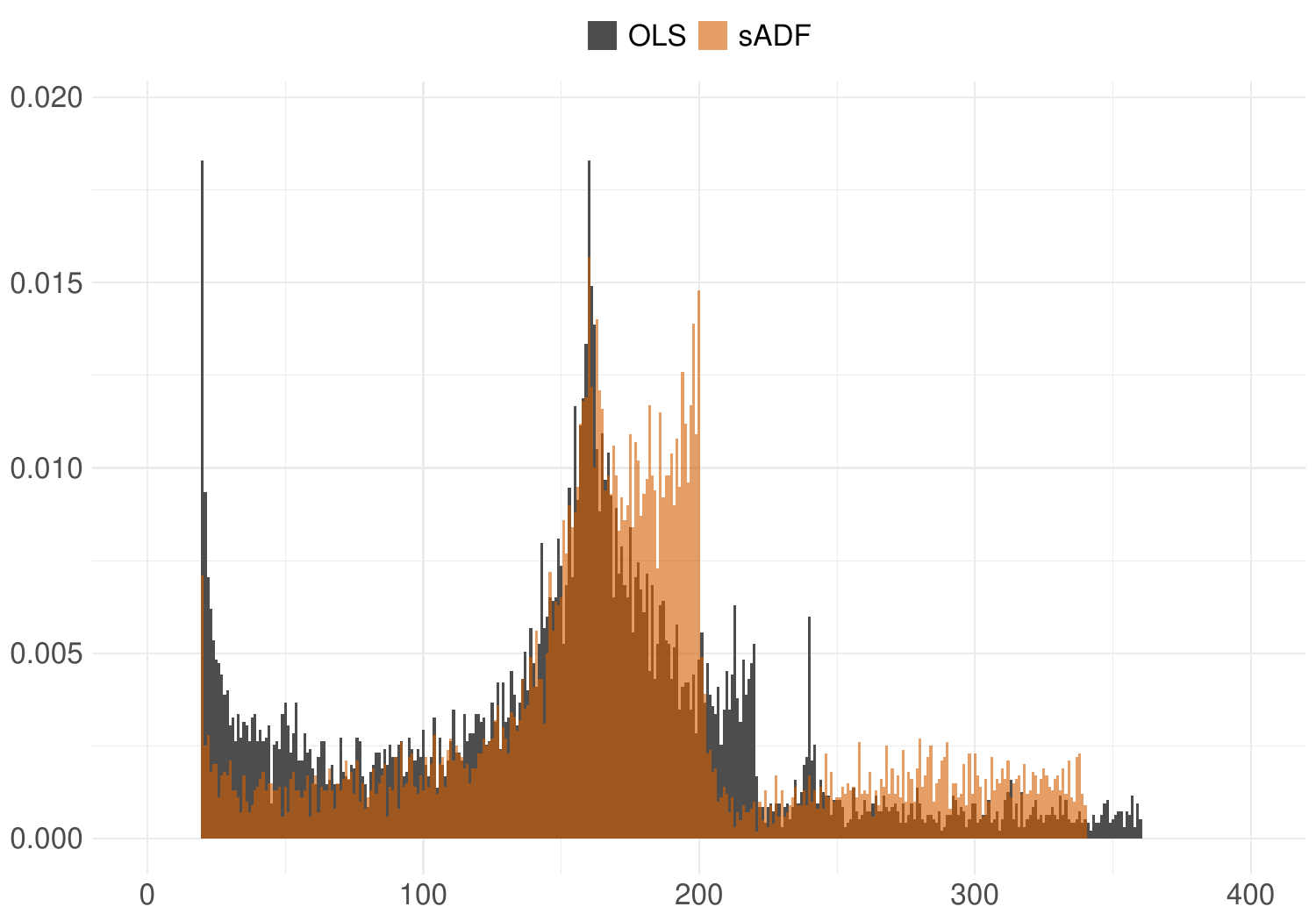}\label{fig4:3:5}}
\subfigure[$T=400$, $c_a=6$, $c_b=6$, $s_0/s_1=1$]{\includegraphics[width=0.45\linewidth]{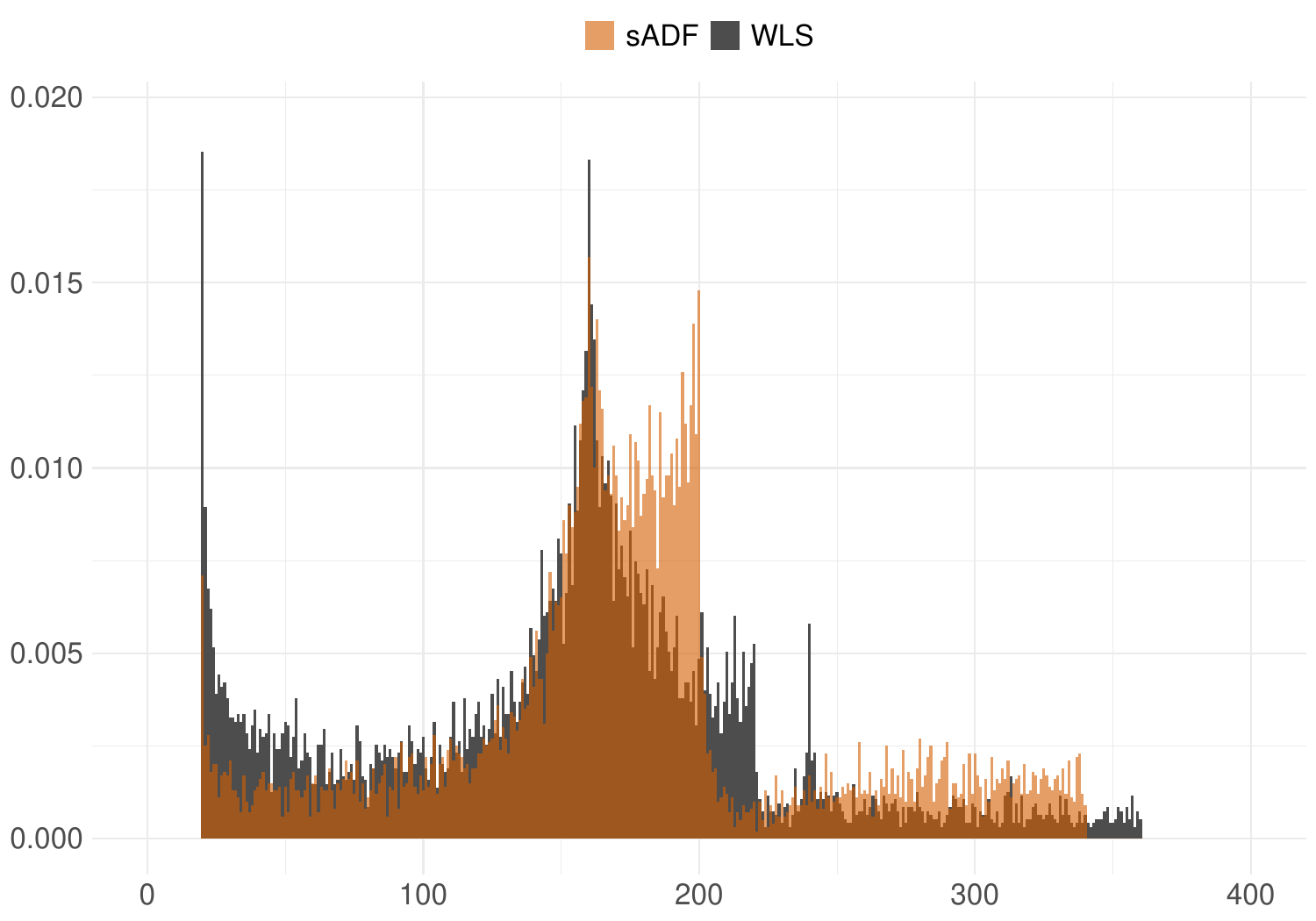}\label{fig4:3:6}}\\
\end{center}%
\caption{Histograms of $\hat{k}_e$ 
for $(\tau_e,\tau_c,\tau_r)=(0.4,0.6,0.7)$,  $\tau=0.2$, $s_0/s_1=1$, $T=400$}
\label{fig43}
\end{figure}

\newpage

\begin{figure}[h!]%
\begin{center}%
\subfigure[$T=800$, $c_a=4$, $c_b=6$, $s_0/s_1=1$]{\includegraphics[width=0.45\linewidth]{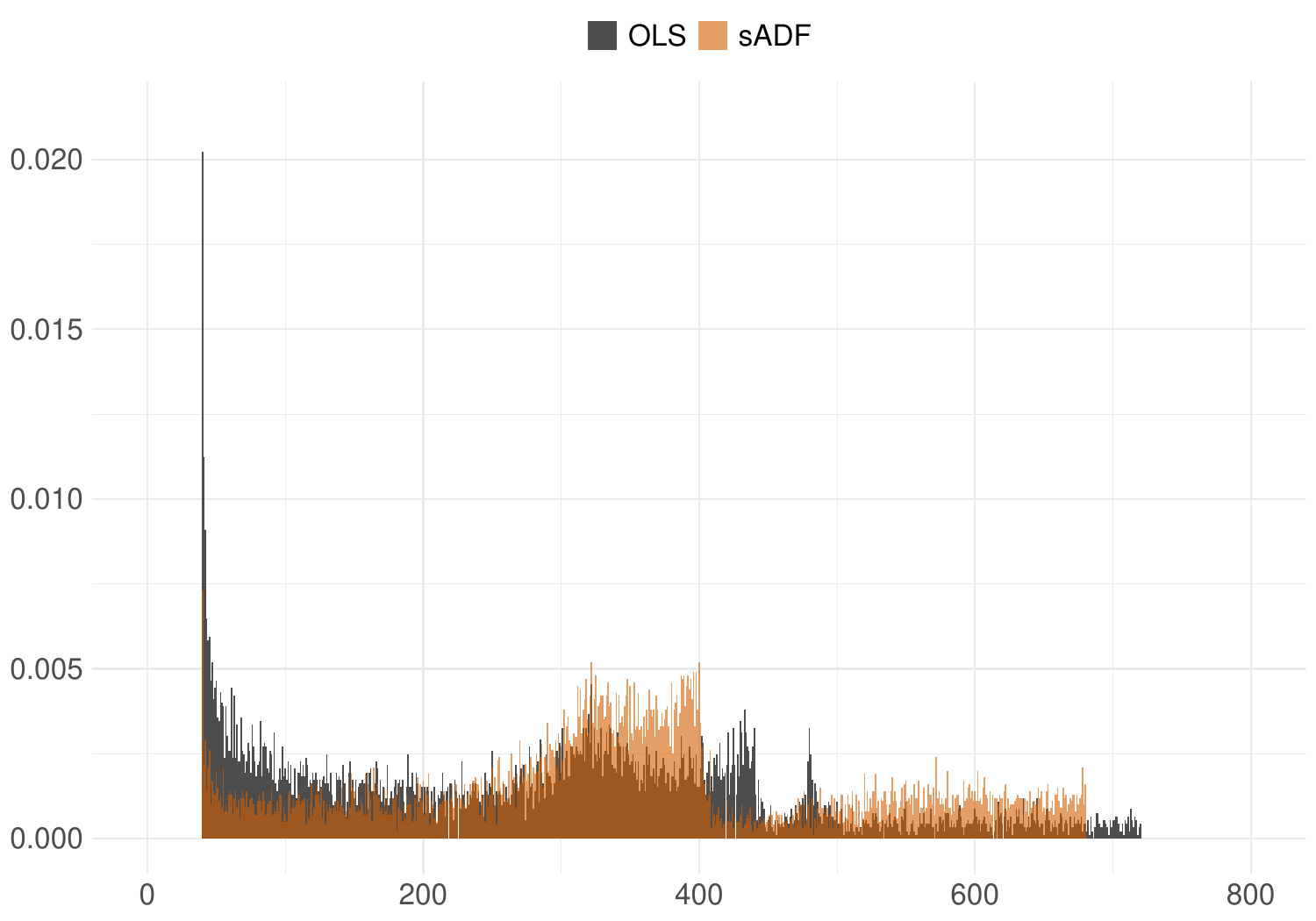}\label{fig4:4:1}}
\subfigure[$T=800$, $c_a=4$, $c_b=6$, $s_0/s_1=1$]{\includegraphics[width=0.45\linewidth]{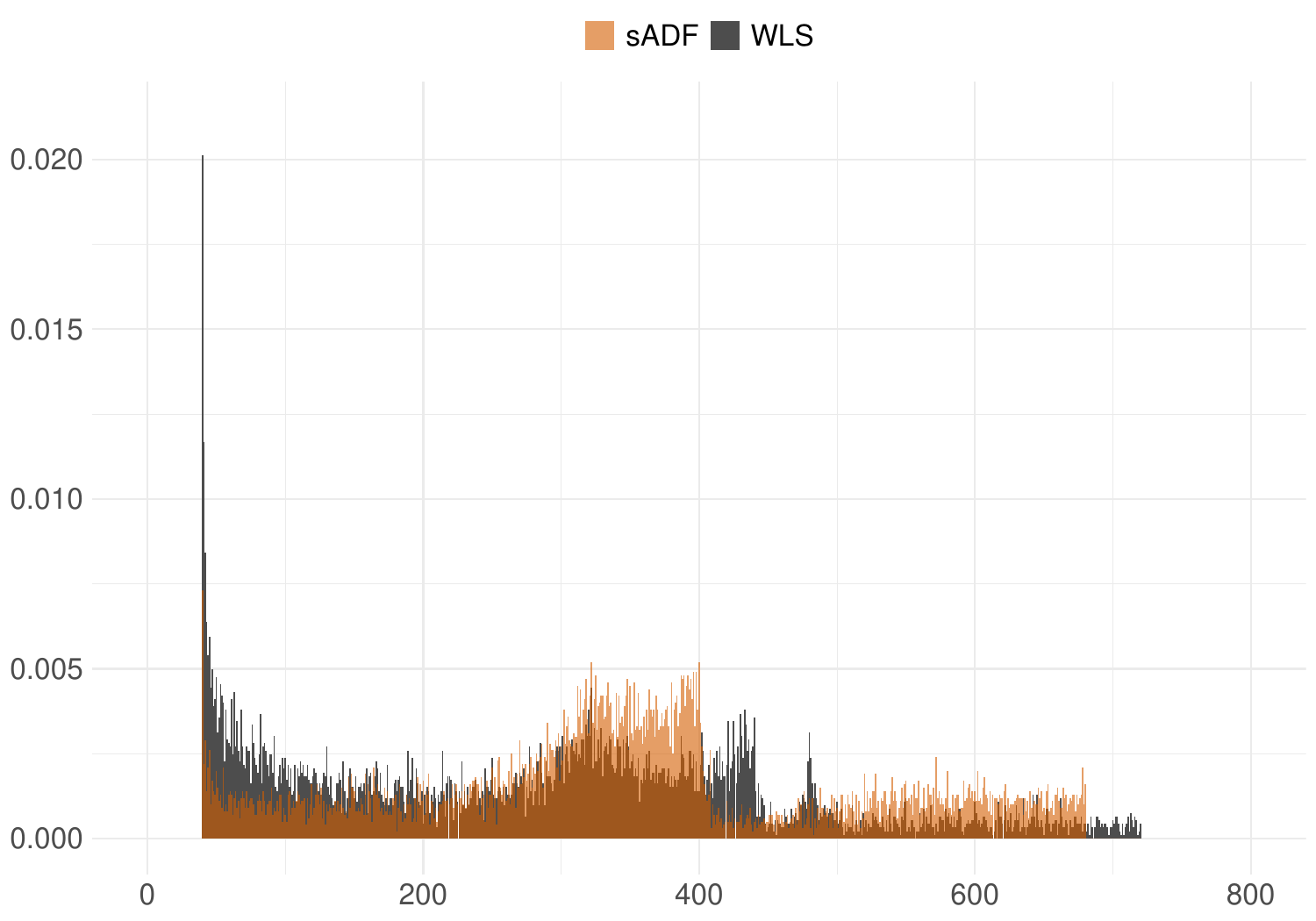}\label{fig4:4:2}}\\
\subfigure[$T=800$, $c_a=5$, $c_b=6$, $s_0/s_1=1$]{\includegraphics[width=0.45\linewidth]{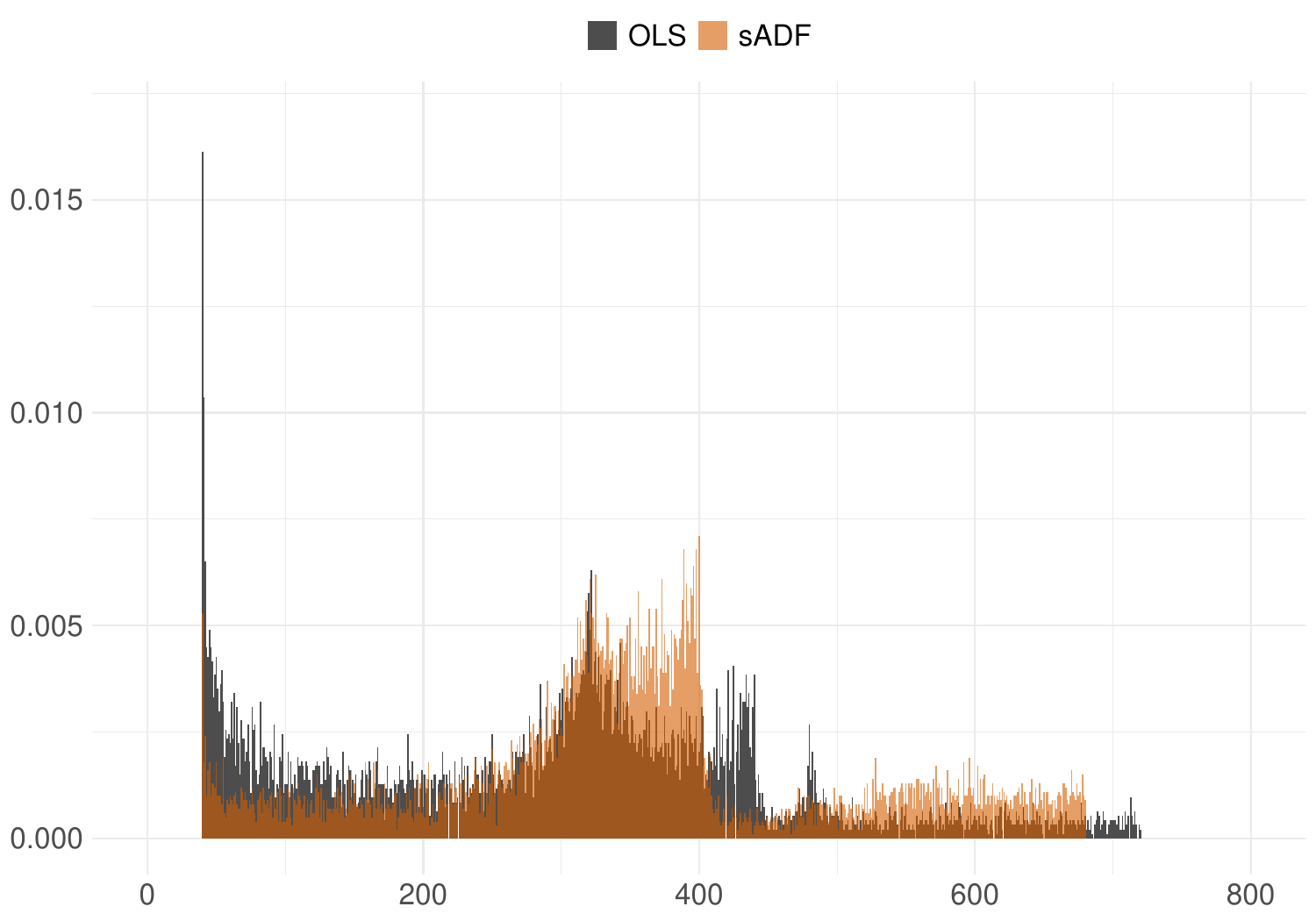}\label{fig4:4:3}}
\subfigure[$T=800$, $c_a=5$, $c_b=6$, $s_0/s_1=1$]{\includegraphics[width=0.45\linewidth]{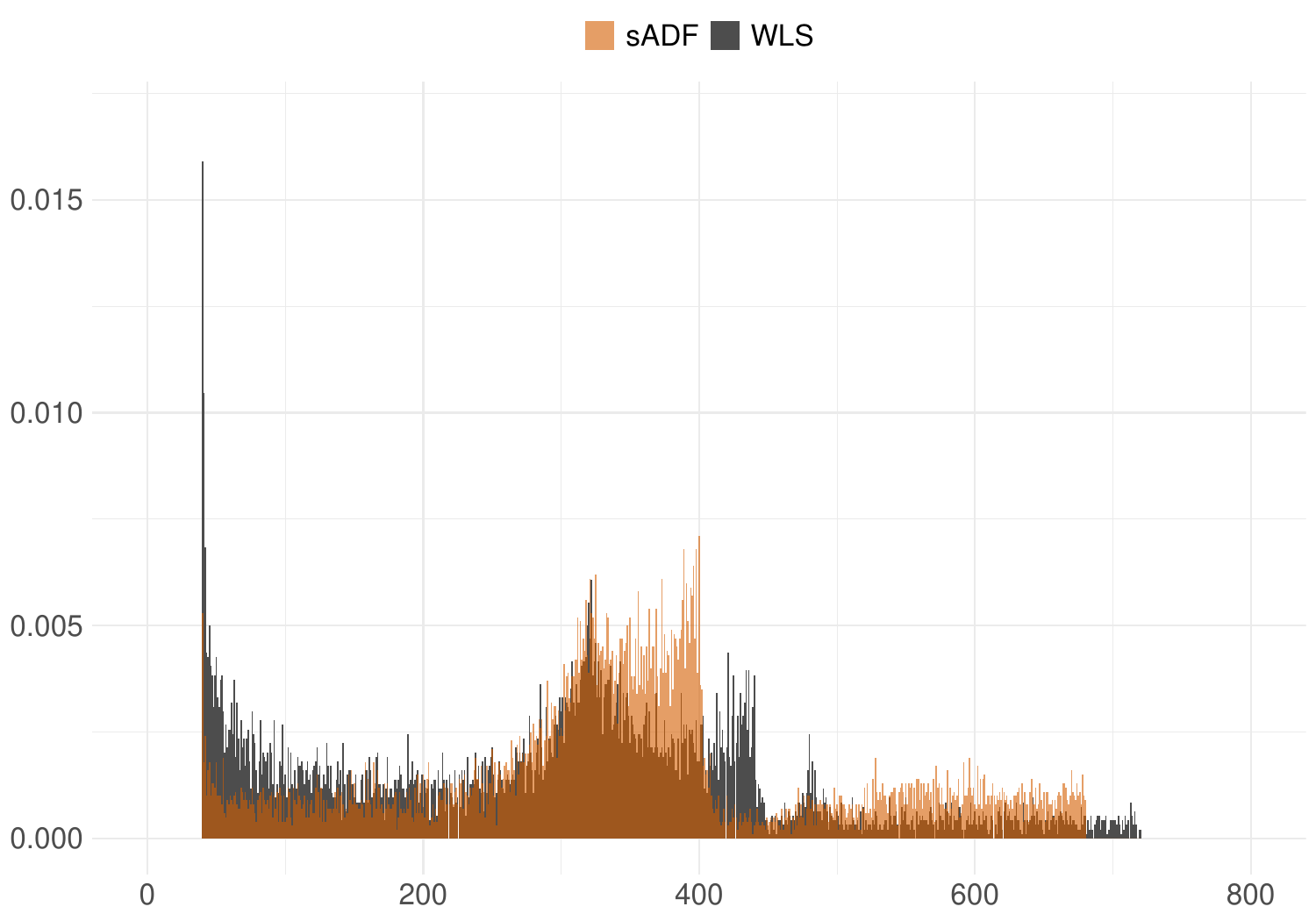}\label{fig4:4:4}}\\
\subfigure[$T=800$, $c_a=6$, $c_b=6$, $s_0/s_1=1$]{\includegraphics[width=0.45\linewidth]{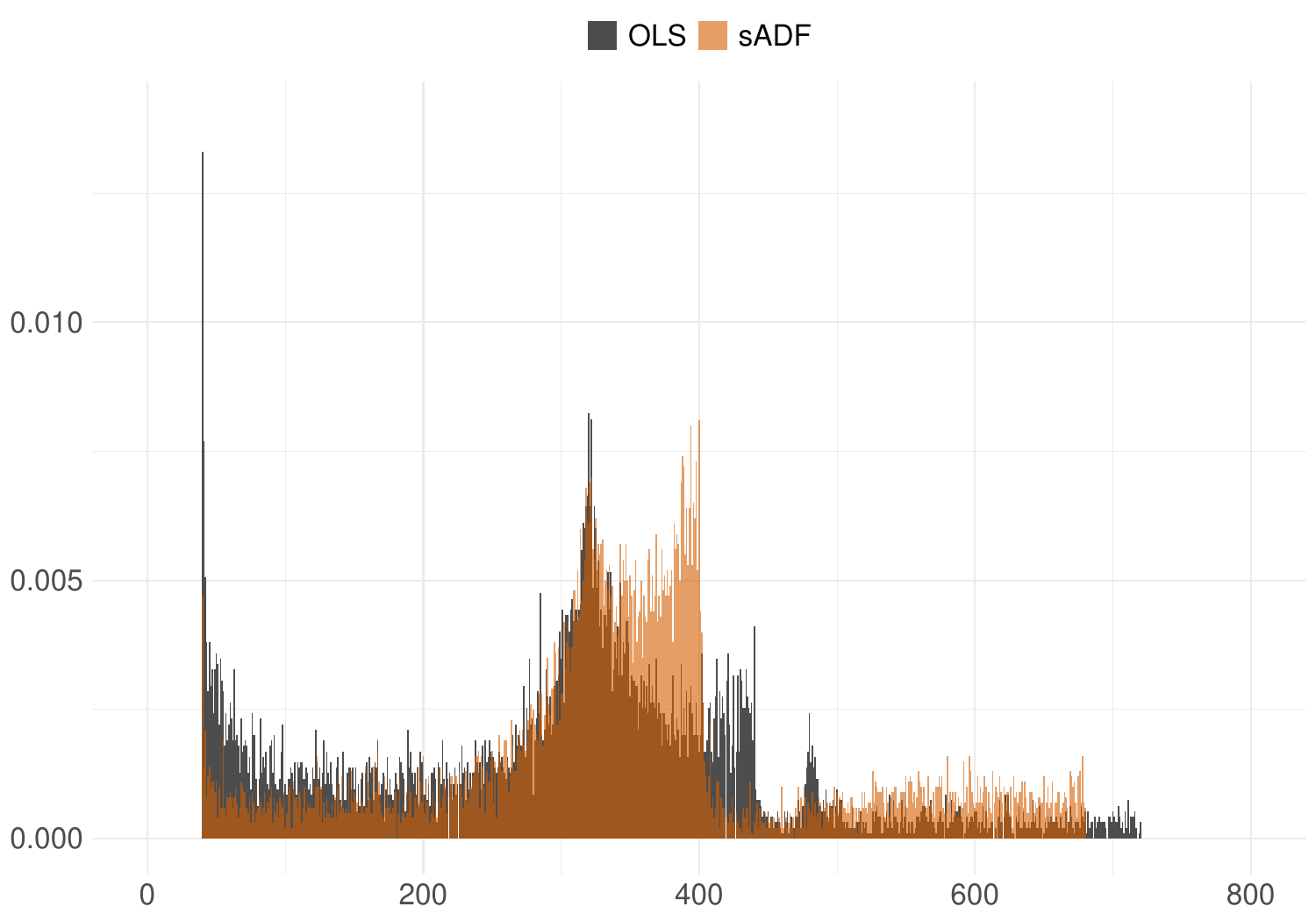}\label{fig4:4:5}}
\subfigure[$T=800$, $c_a=6$, $c_b=6$, $s_0/s_1=1$]{\includegraphics[width=0.45\linewidth]{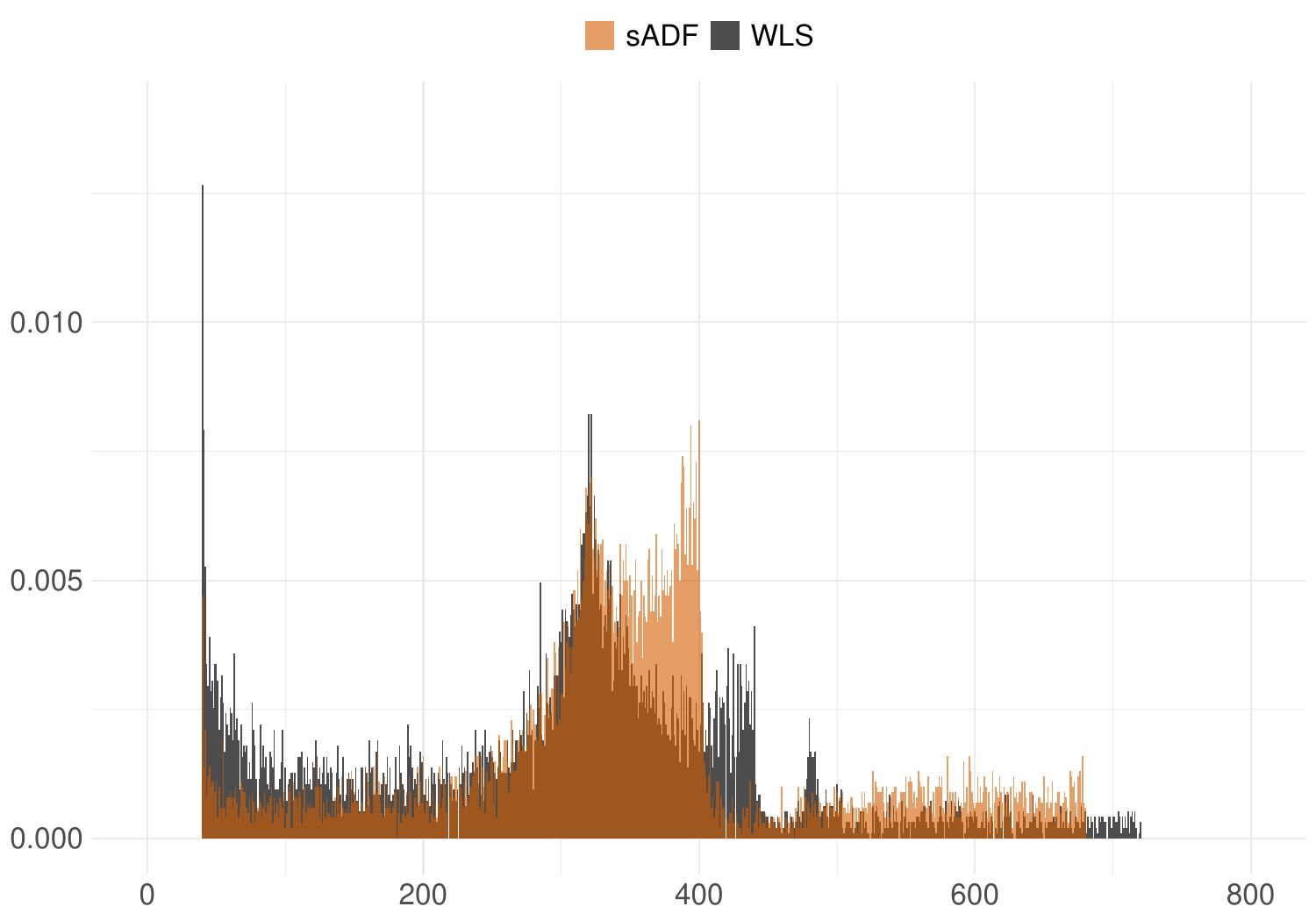}\label{fig4:4:6}}\\
\end{center}%
\caption{Histograms of $\hat{k}_e$ 
for $(\tau_e,\tau_c,\tau_r)=(0.4,0.6,0.7)$,  $\tau=0.2$, $s_0/s_1=1$, $T=800$}
\label{fig44}
\end{figure}

\newpage

\begin{figure}[h!]%
\begin{center}%
\subfigure[$T=400$, $c_a=4$, $c_b=6$, $s_0/s_1=5$]{\includegraphics[width=0.45\linewidth]{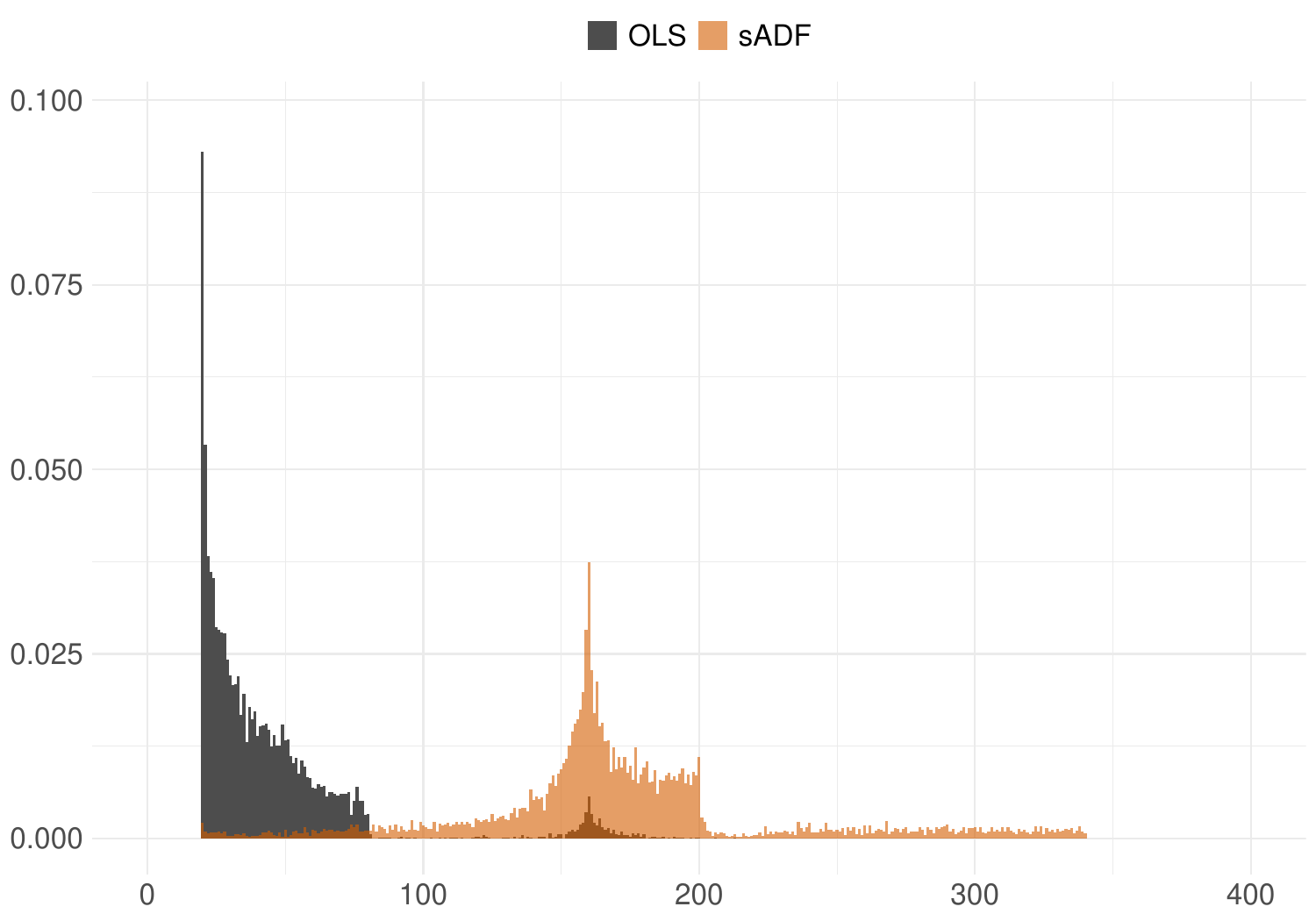}\label{fig4:5:1:app}}
\subfigure[$T=400$, $c_a=4$, $c_b=6$, $s_0/s_1=5$]{\includegraphics[width=0.45\linewidth]{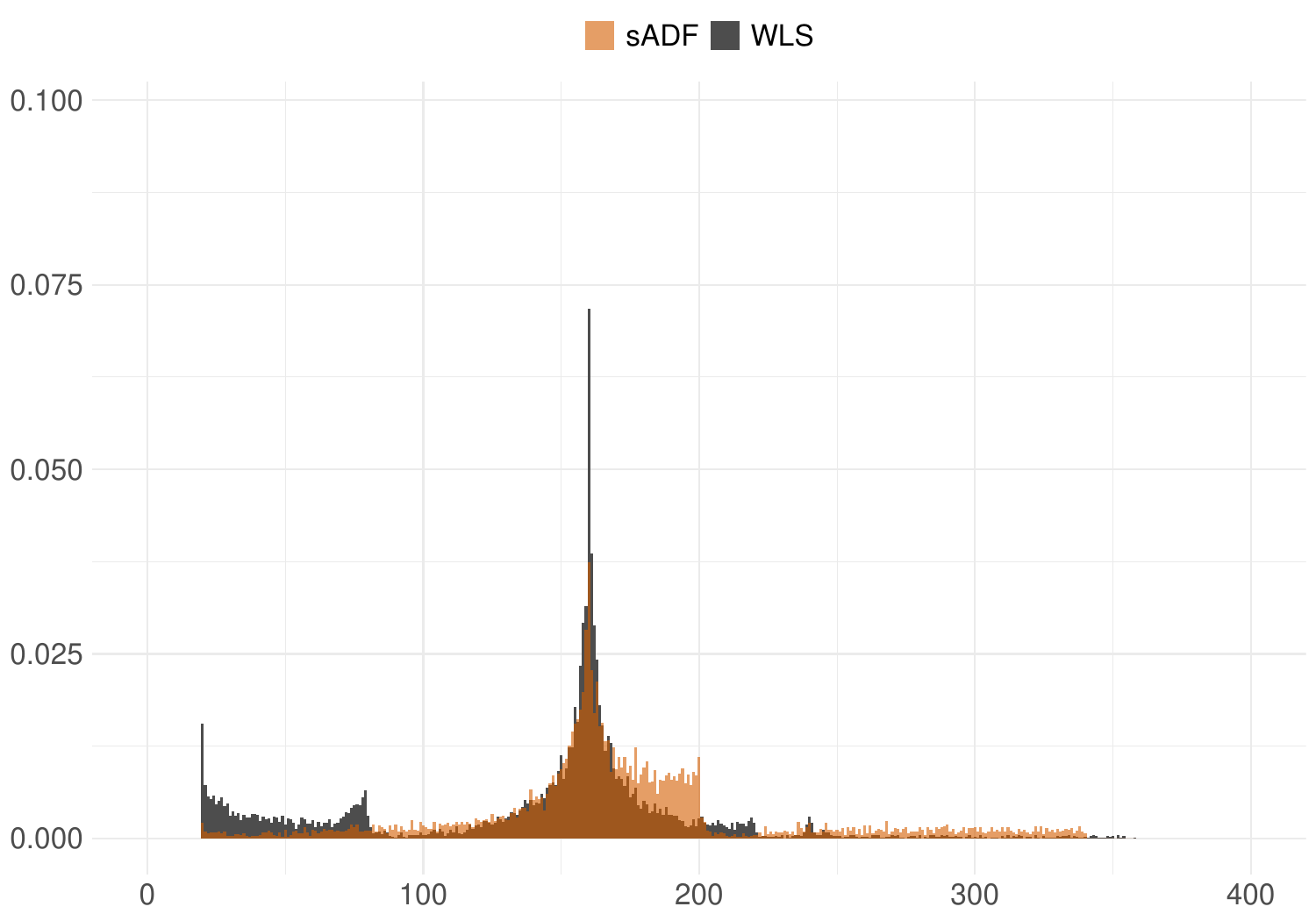}\label{fig4:5:2:app}}\\
\subfigure[$T=400$, $c_a=5$, $c_b=6$, $s_0/s_1=5$]{\includegraphics[width=0.45\linewidth]{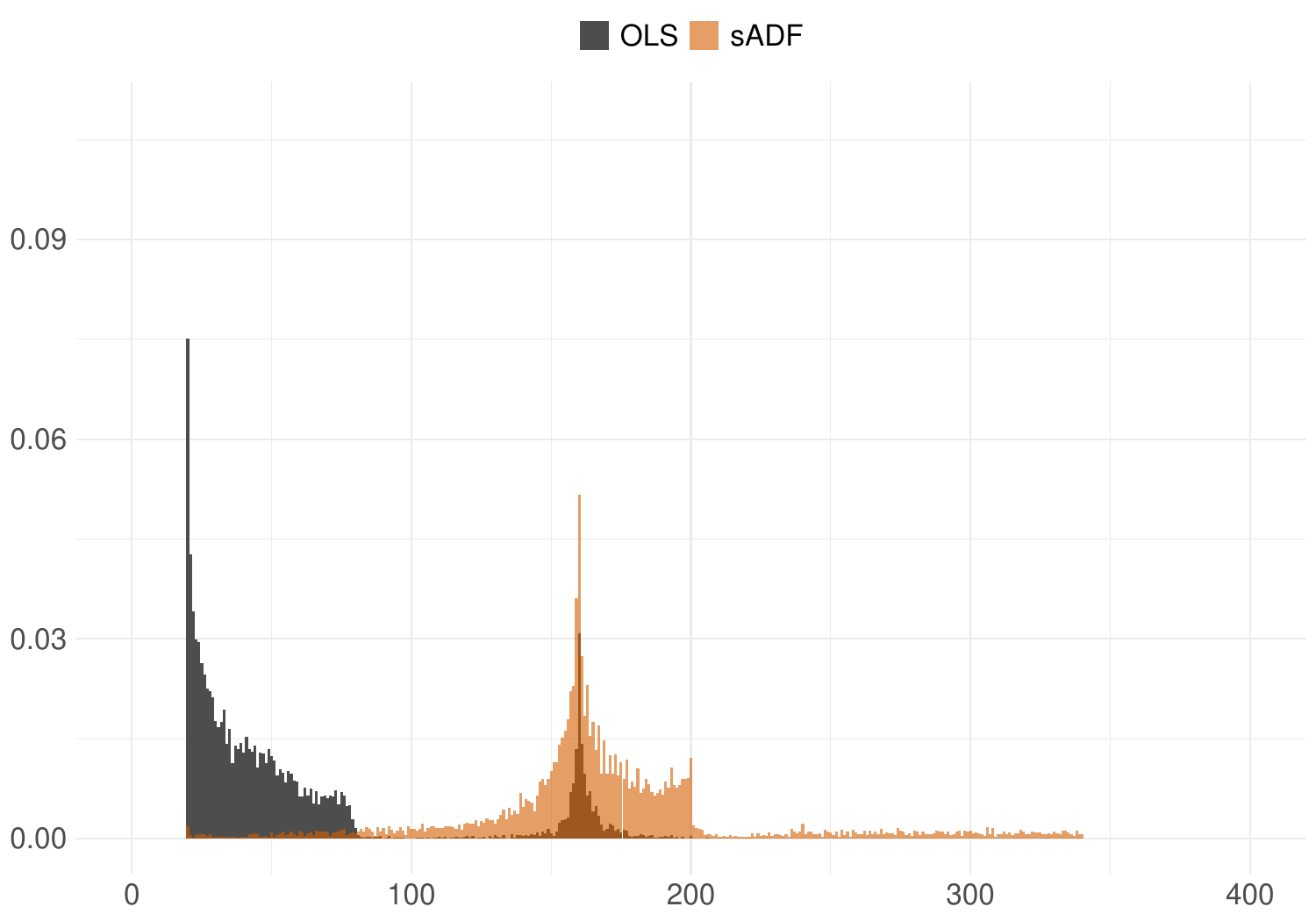}\label{fig4:5:3:app}}
\subfigure[$T=400$, $c_a=5$, $c_b=6$, $s_0/s_1=5$]{\includegraphics[width=0.45\linewidth]{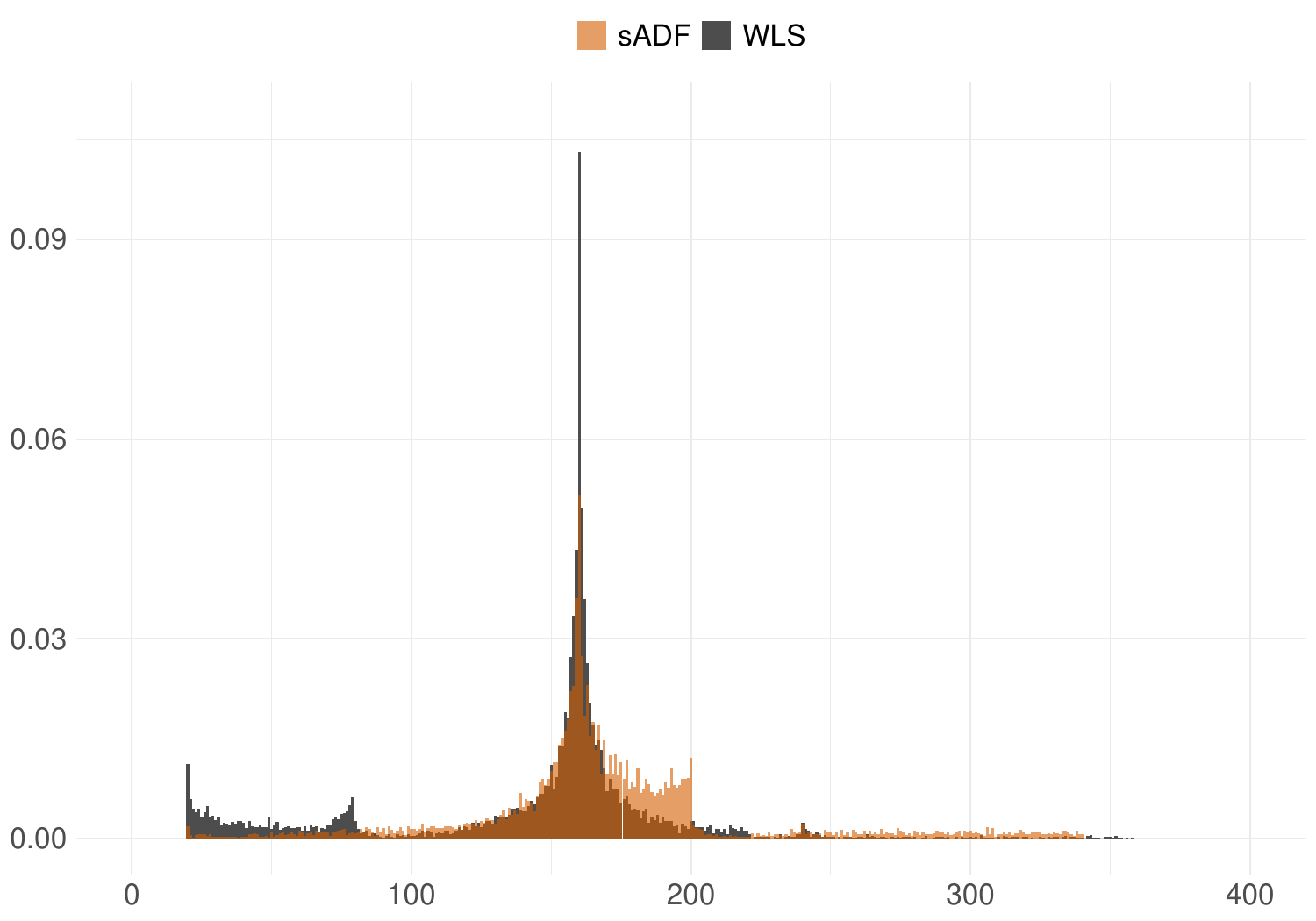}\label{fig4:5:4:app}}\\
\subfigure[$T=400$, $c_a=6$, $c_b=6$, $s_0/s_1=5$]{\includegraphics[width=0.45\linewidth]{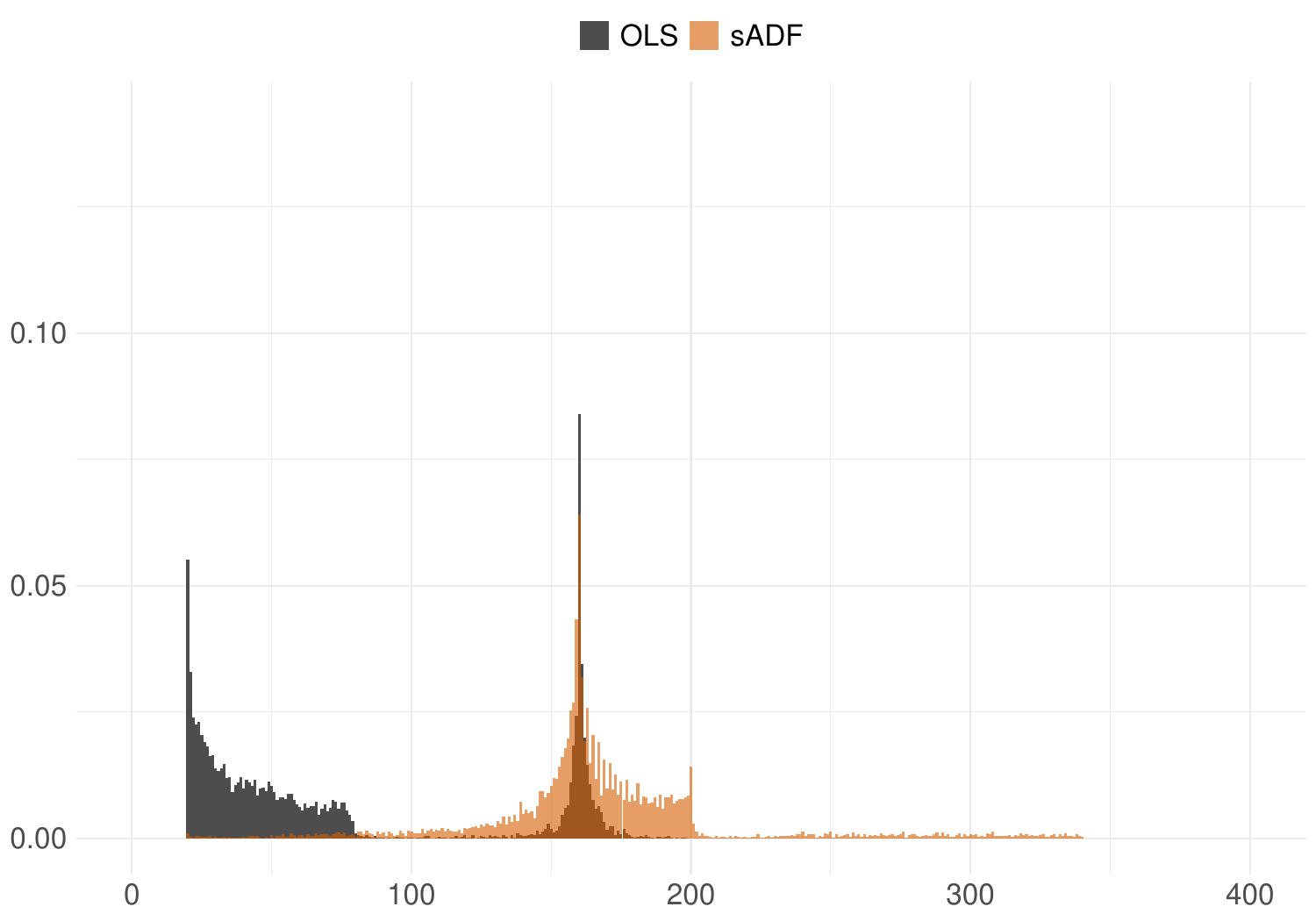}\label{fig4:5:5:app}}
\subfigure[$T=400$, $c_a=6$, $c_b=6$, $s_0/s_1=5$]{\includegraphics[width=0.45\linewidth]{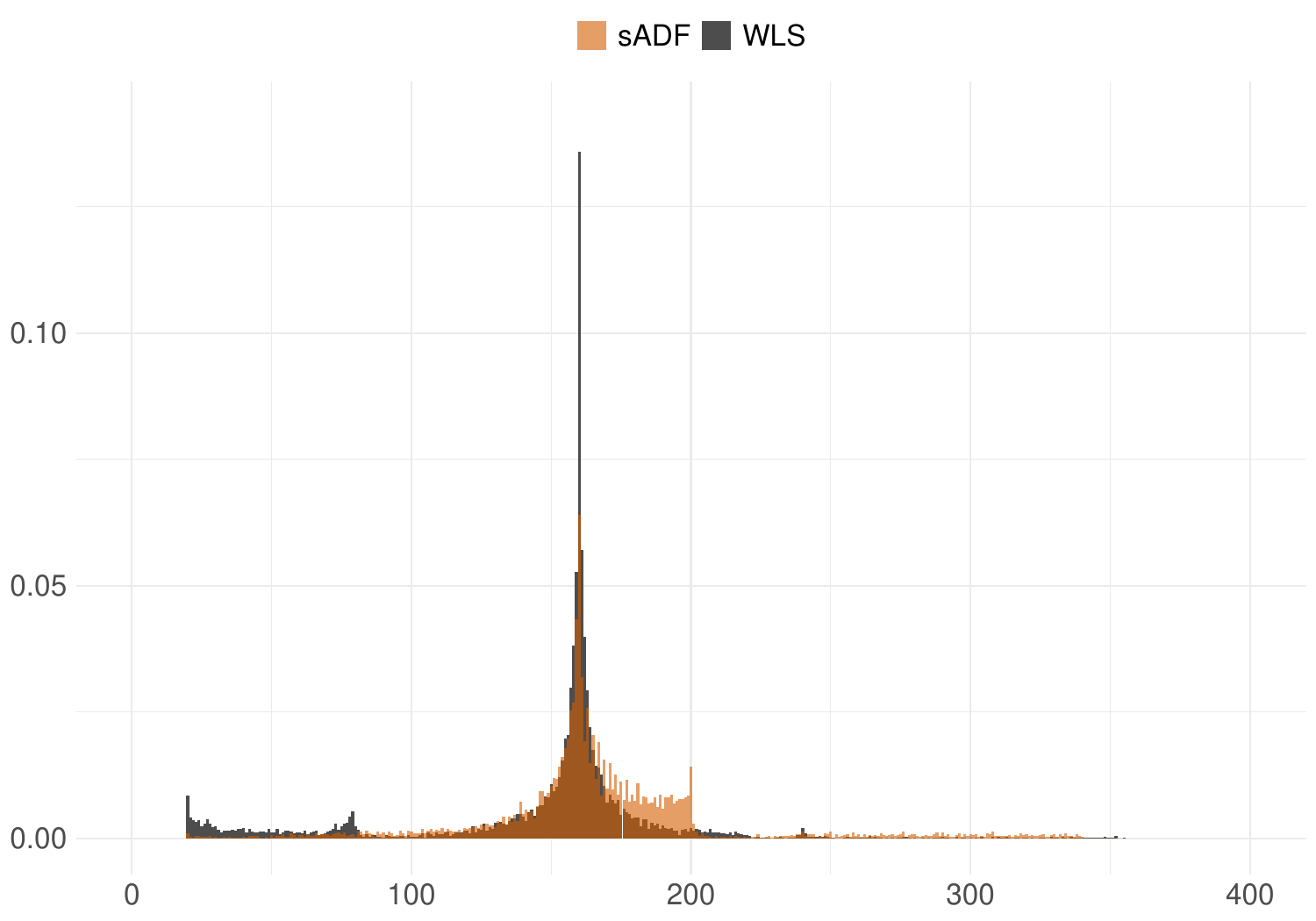}\label{fig4:5:6:app}}\\
\end{center}%
\caption{Histograms of $\hat{k}_e$ 
for $(\tau_e,\tau_c,\tau_r)=(0.4,0.6,0.7)$,  $\tau=0.2$, $s_0/s_1=5$, $T=400$}
\label{fig45:app}
\end{figure}

\newpage

\begin{figure}[h!]%
\begin{center}%
\subfigure[$T=800$, $c_a=4$, $c_b=6$, $s_0/s_1=5$]{\includegraphics[width=0.45\linewidth]{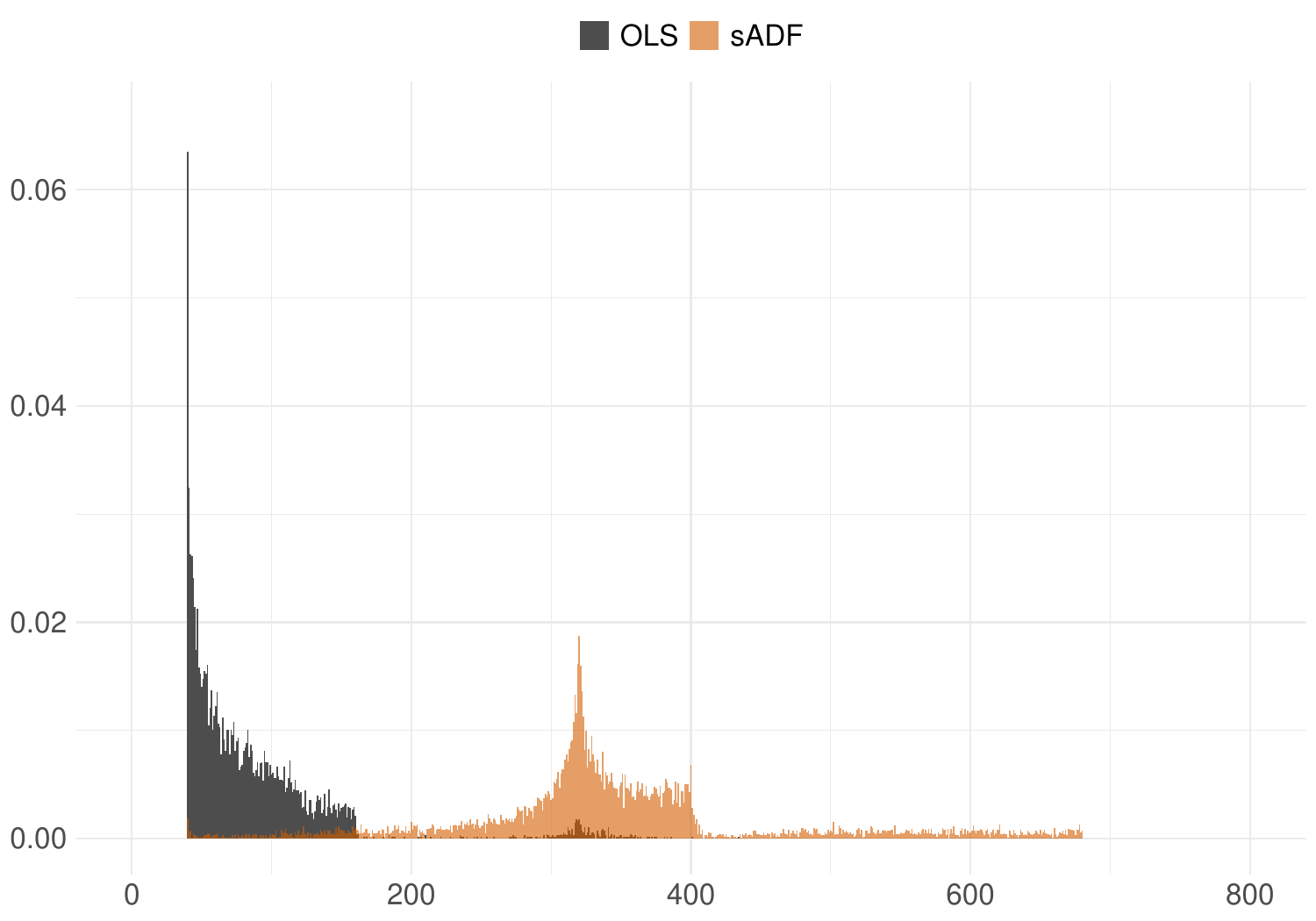}\label{fig4:6:1}}
\subfigure[$T=800$, $c_a=4$, $c_b=6$, $s_0/s_1=5$]{\includegraphics[width=0.45\linewidth]{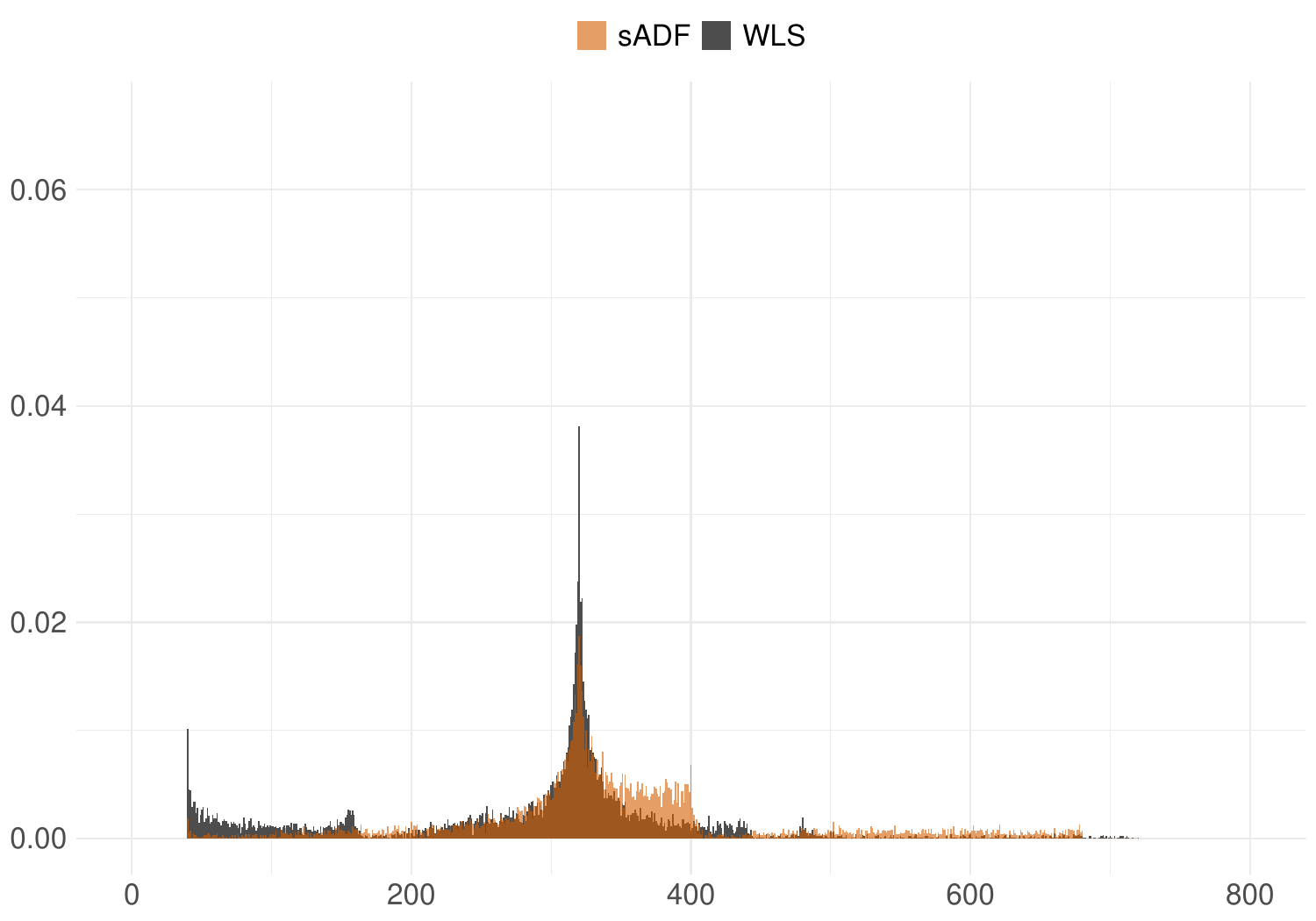}\label{fig4:6:2}}\\
\subfigure[$T=800$, $c_a=5$, $c_b=6$, $s_0/s_1=5$]{\includegraphics[width=0.45\linewidth]{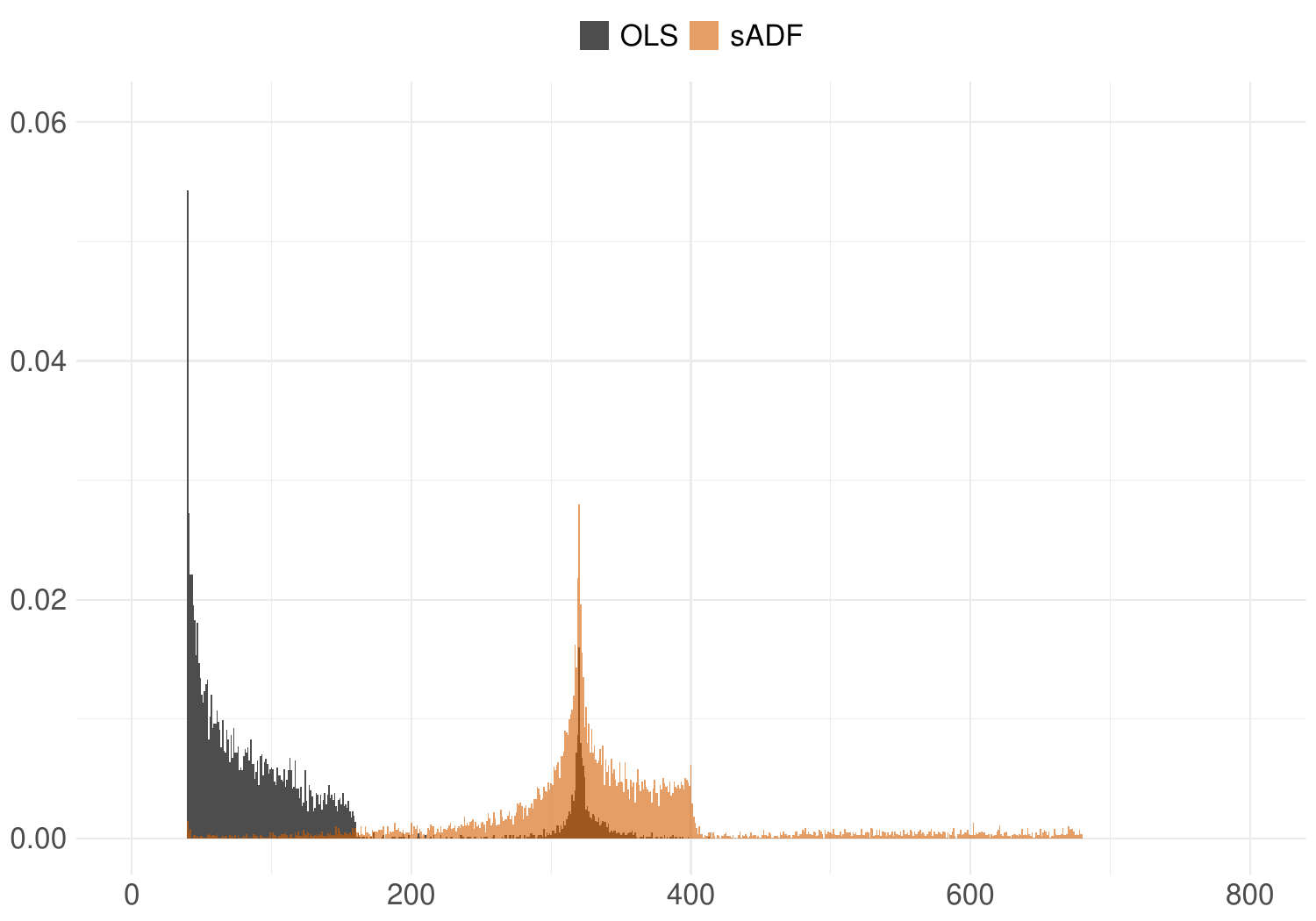}\label{fig4:6:3}}
\subfigure[$T=800$, $c_a=5$, $c_b=6$, $s_0/s_1=5$]{\includegraphics[width=0.45\linewidth]{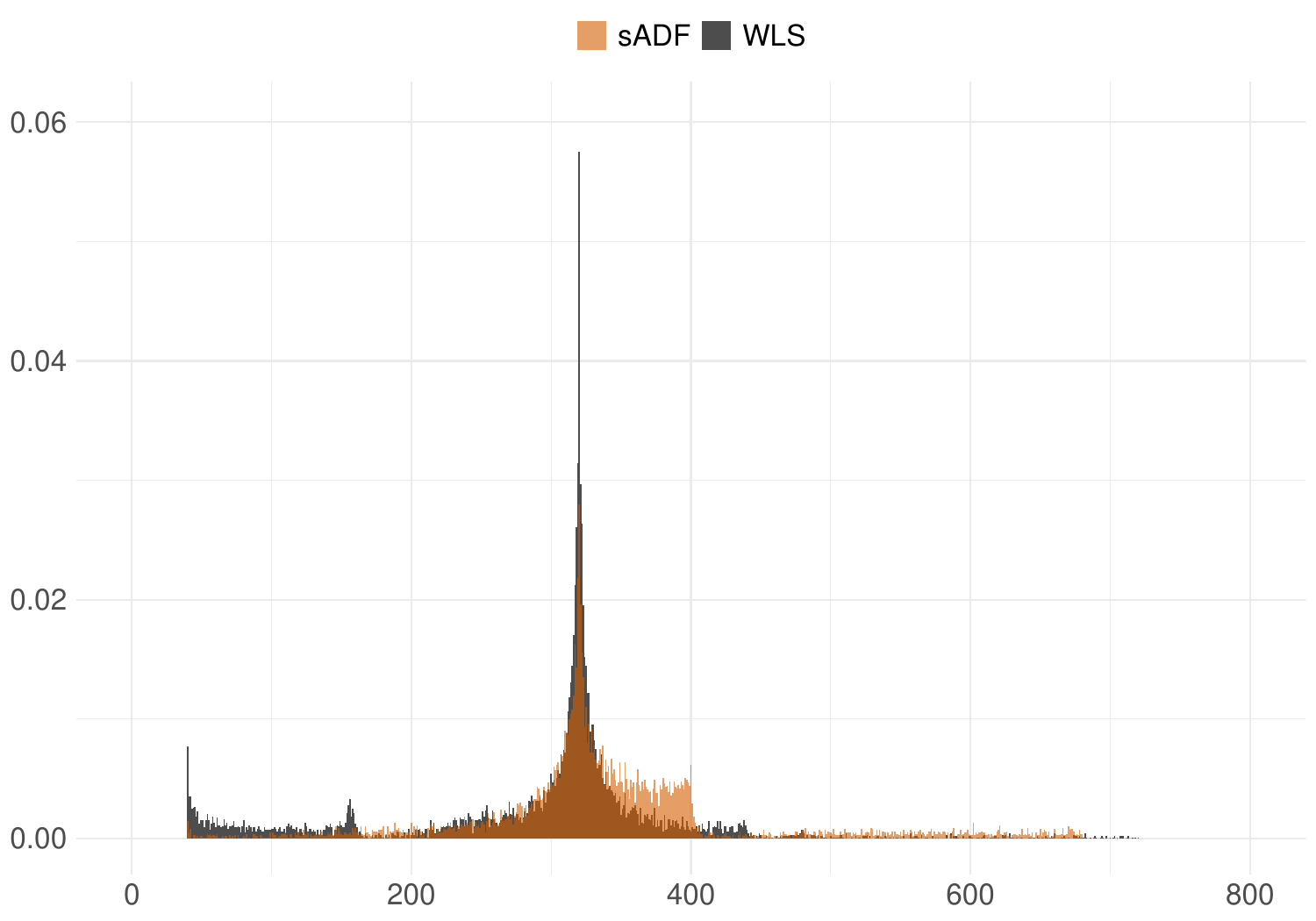}\label{fig4:6:4}}\\
\subfigure[$T=800$, $c_a=6$, $c_b=6$, $s_0/s_1=5$]{\includegraphics[width=0.45\linewidth]{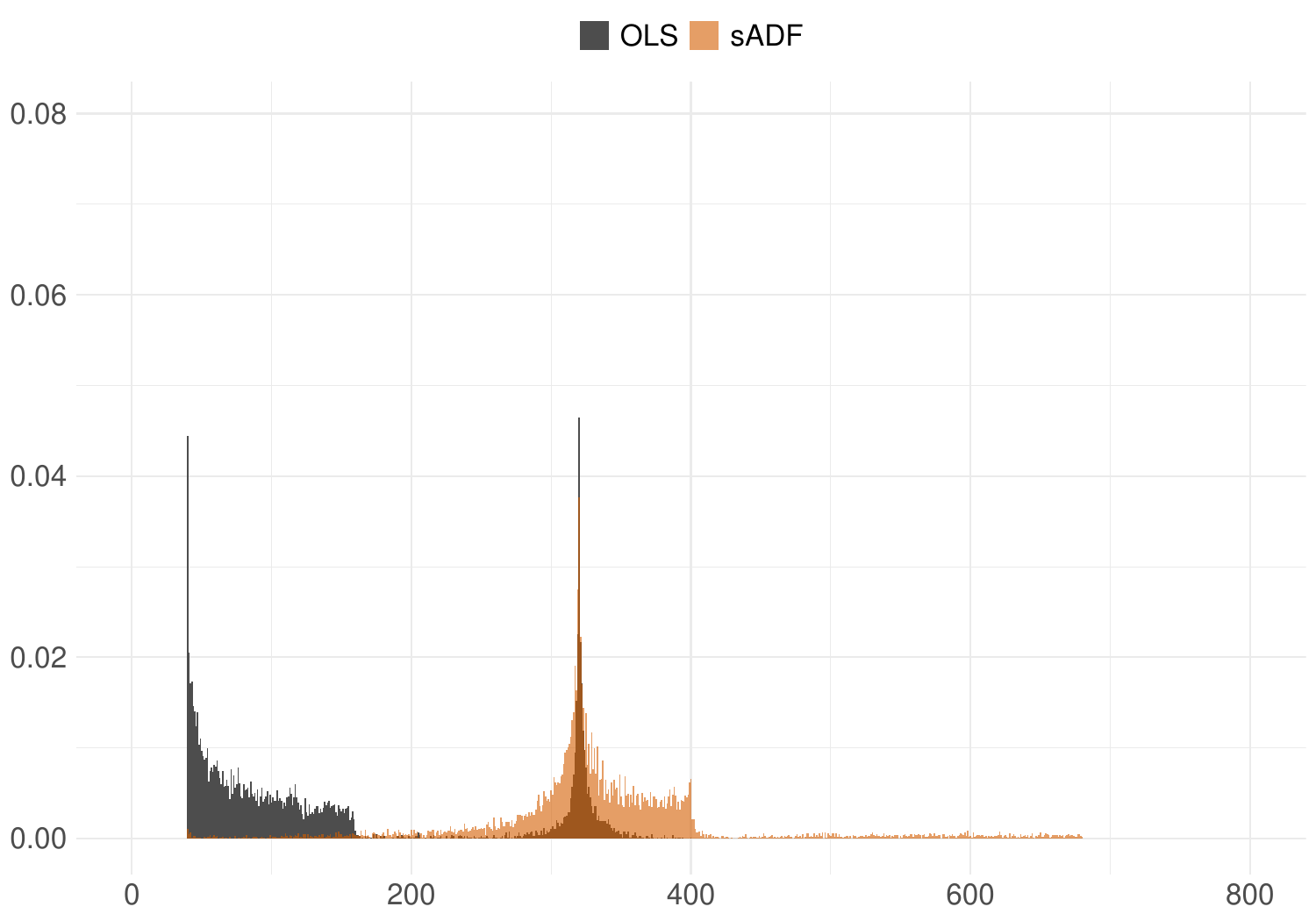}\label{fig4:6:5}}
\subfigure[$T=800$, $c_a=6$, $c_b=6$, $s_0/s_1=5$]{\includegraphics[width=0.45\linewidth]{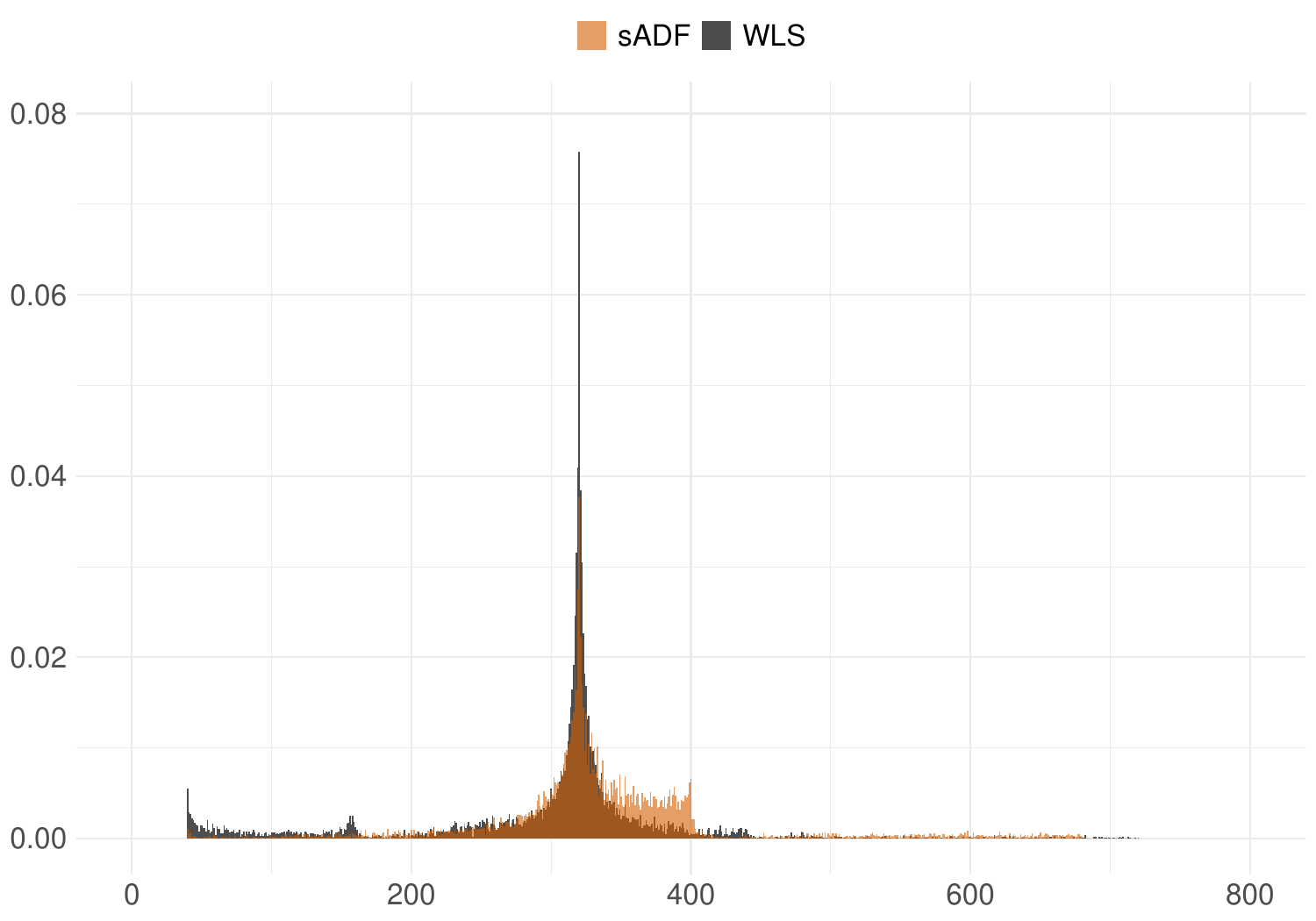}\label{fig4:6:6}}\\
\end{center}%
\caption{Histograms of $\hat{k}_e$ 
for $(\tau_e,\tau_c,\tau_r)=(0.4,0.6,0.7)$,  $\tau=0.2$, $s_0/s_1=5$, $T=800$}
\label{fig46}
\end{figure}

\newpage

\section{$\tau=0.2$, $\hat{k}_r$}
\setcounter{figure}{0}

\begin{figure}[h!]%
\begin{center}%
\subfigure[$T=400$, $c_a=4$, $c_b=6$, $s_0/s_1=1/5$]{\includegraphics[width=0.45\linewidth]{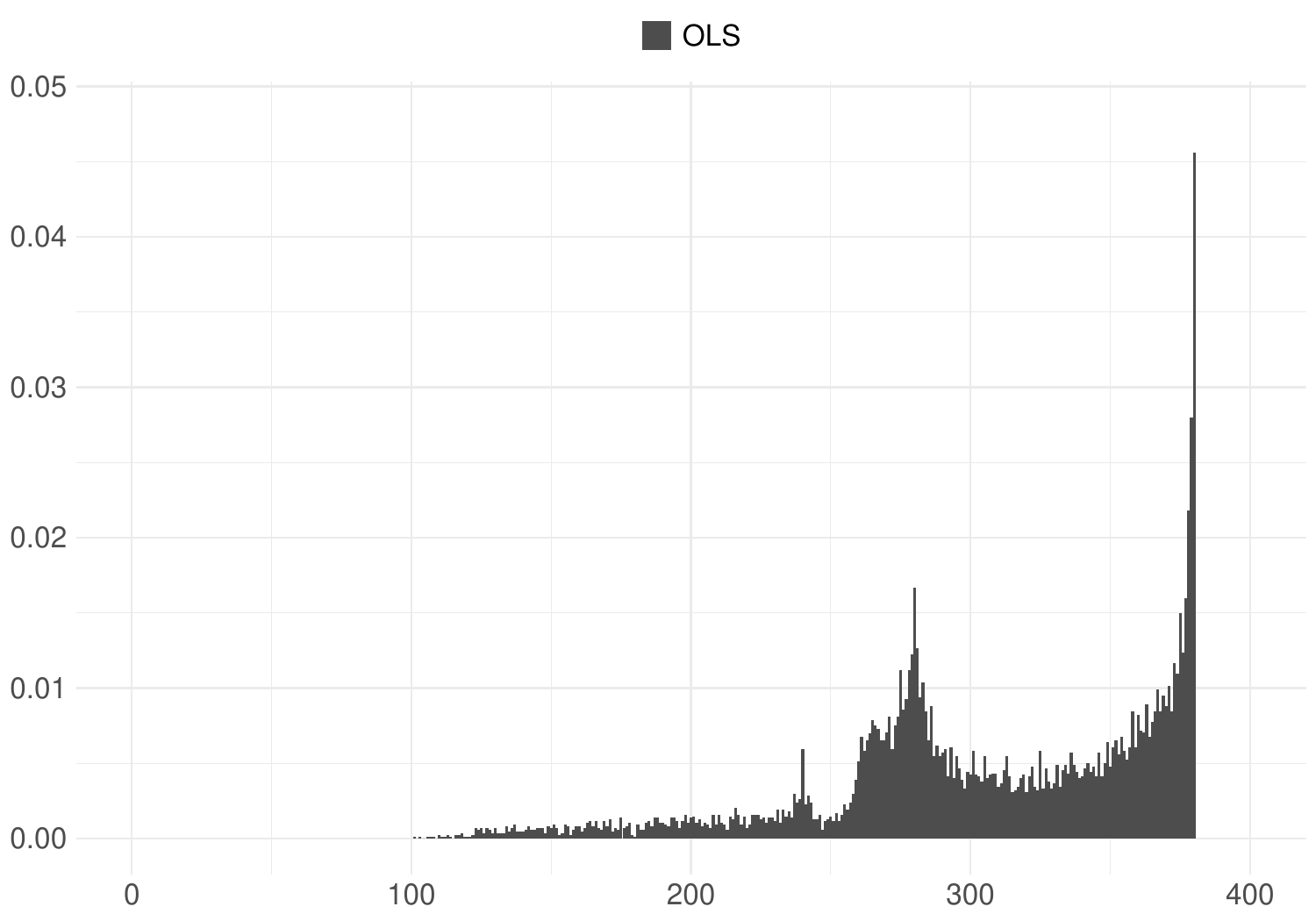}\label{fig2:1r:1}}
\subfigure[$T=400$, $c_a=4$, $c_b=6$, $s_0/s_1=1/5$]{\includegraphics[width=0.45\linewidth]{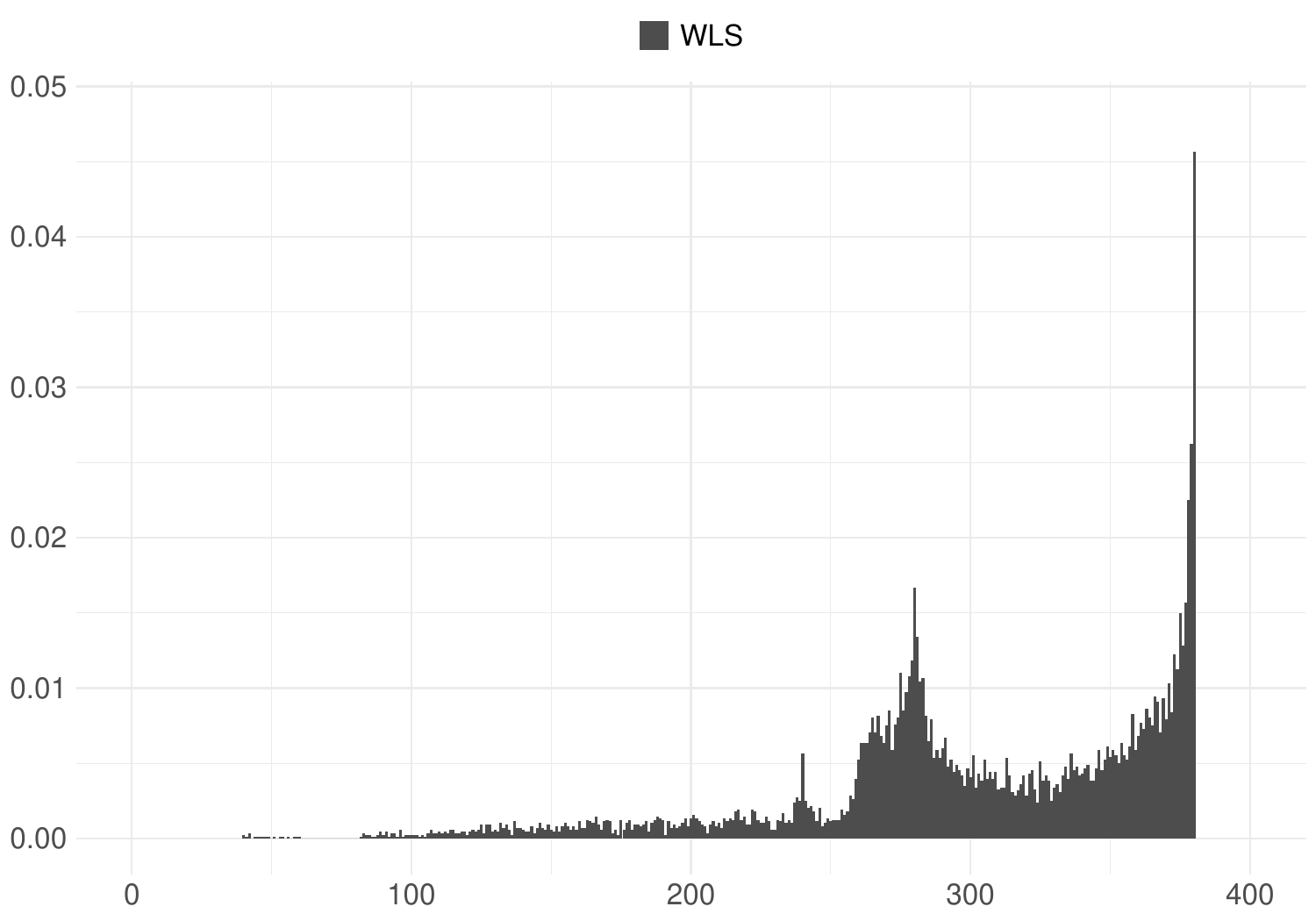}\label{fig2:1r:2}}\\
\subfigure[$T=400$, $c_a=5$, $c_b=6$, $s_0/s_1=1/5$]{\includegraphics[width=0.45\linewidth]{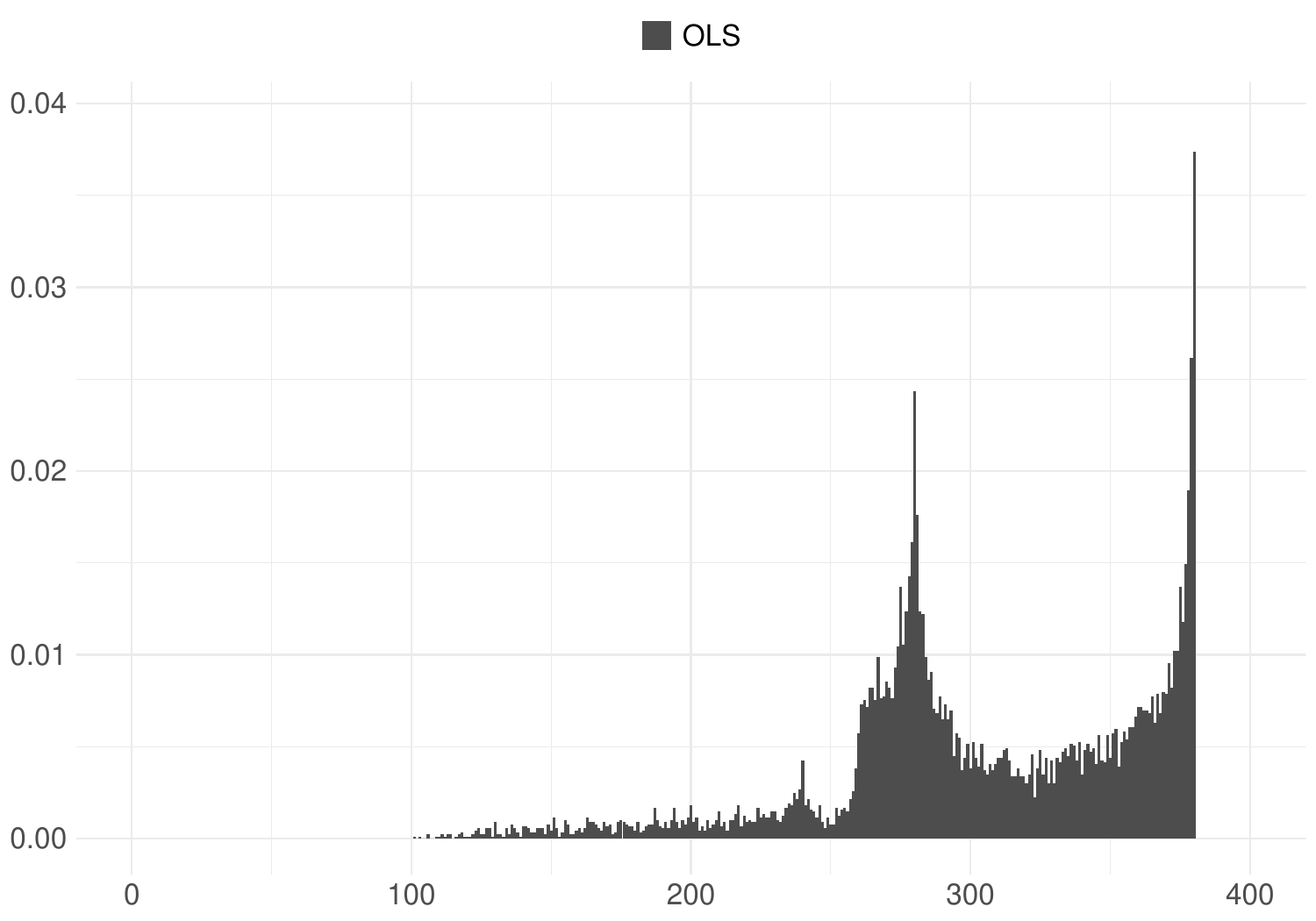}\label{fig2:1r:3}}
\subfigure[$T=400$, $c_a=5$, $c_b=6$, $s_0/s_1=1/5$]{\includegraphics[width=0.45\linewidth]{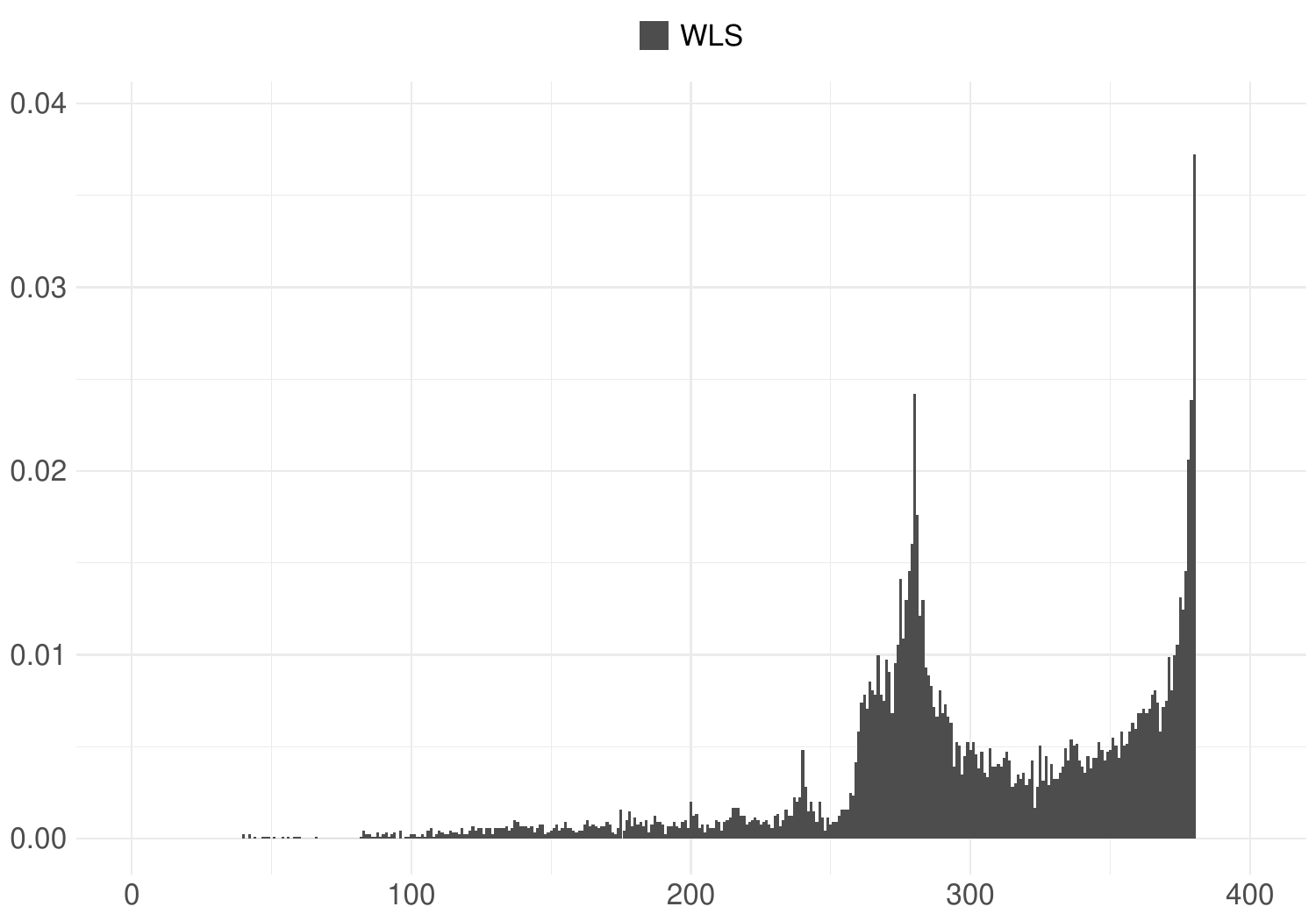}\label{fig2:1r:4}}\\
\subfigure[$T=400$, $c_a=6$, $c_b=6$, $s_0/s_1=1/5$]{\includegraphics[width=0.45\linewidth]{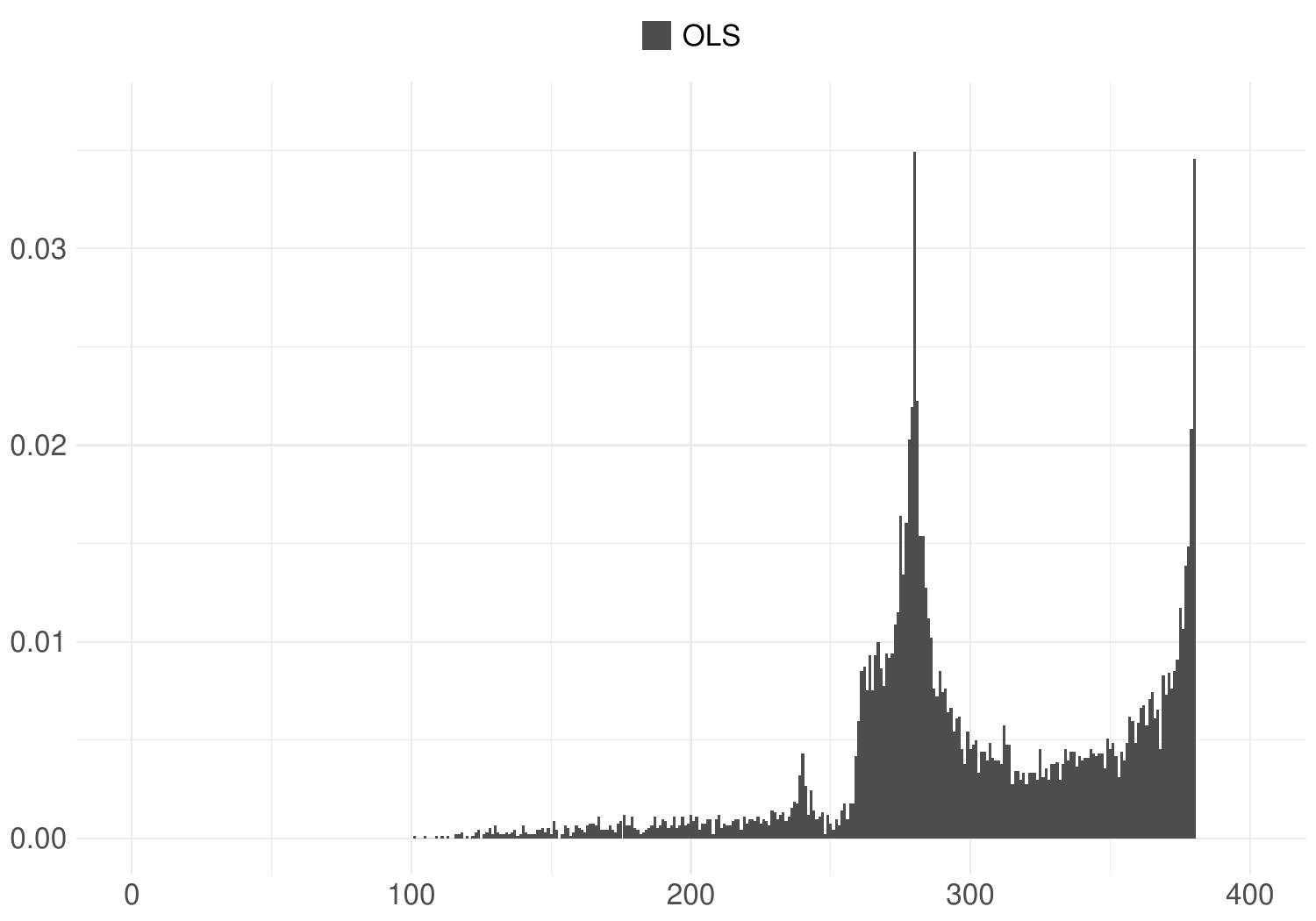}\label{fig2:1r:5}}
\subfigure[$T=400$, $c_a=6$, $c_b=6$, $s_0/s_1=1/5$]{\includegraphics[width=0.45\linewidth]{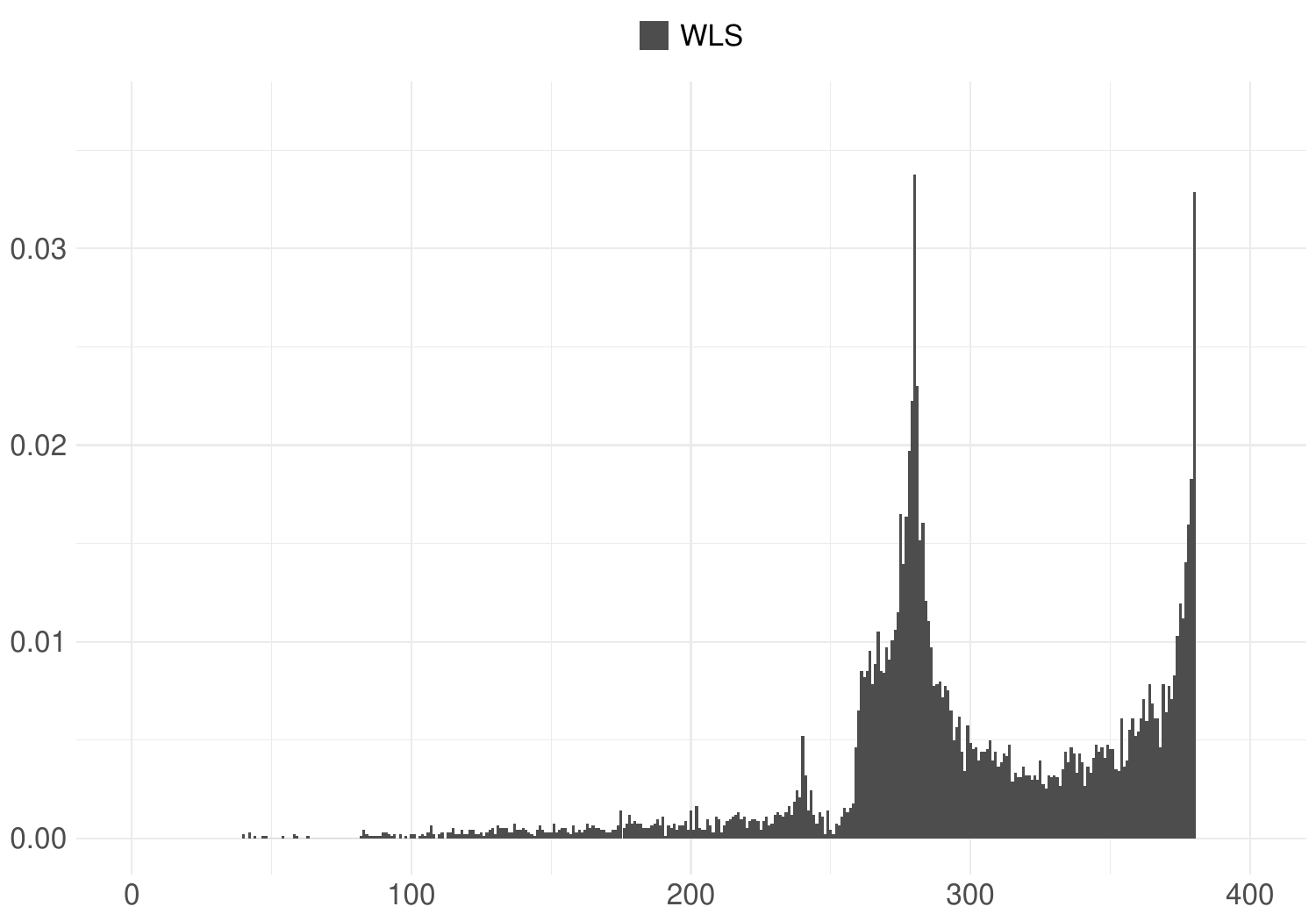}\label{fig2:1r:6}}\\
\end{center}%
\caption{Histograms of $\hat{k}_r$ 
for $(\tau_e,\tau_c,\tau_r)=(0.4,0.6,0.7)$,  $\tau=0.2$, $s_0/s_1=1/5$, $T=400$}
\label{fig21_r}
\end{figure}

\newpage

\begin{figure}[h!]%
\begin{center}%
\subfigure[$T=800$, $c_a=4$, $c_b=6$, $s_0/s_1=1/5$]{\includegraphics[width=0.45\linewidth]{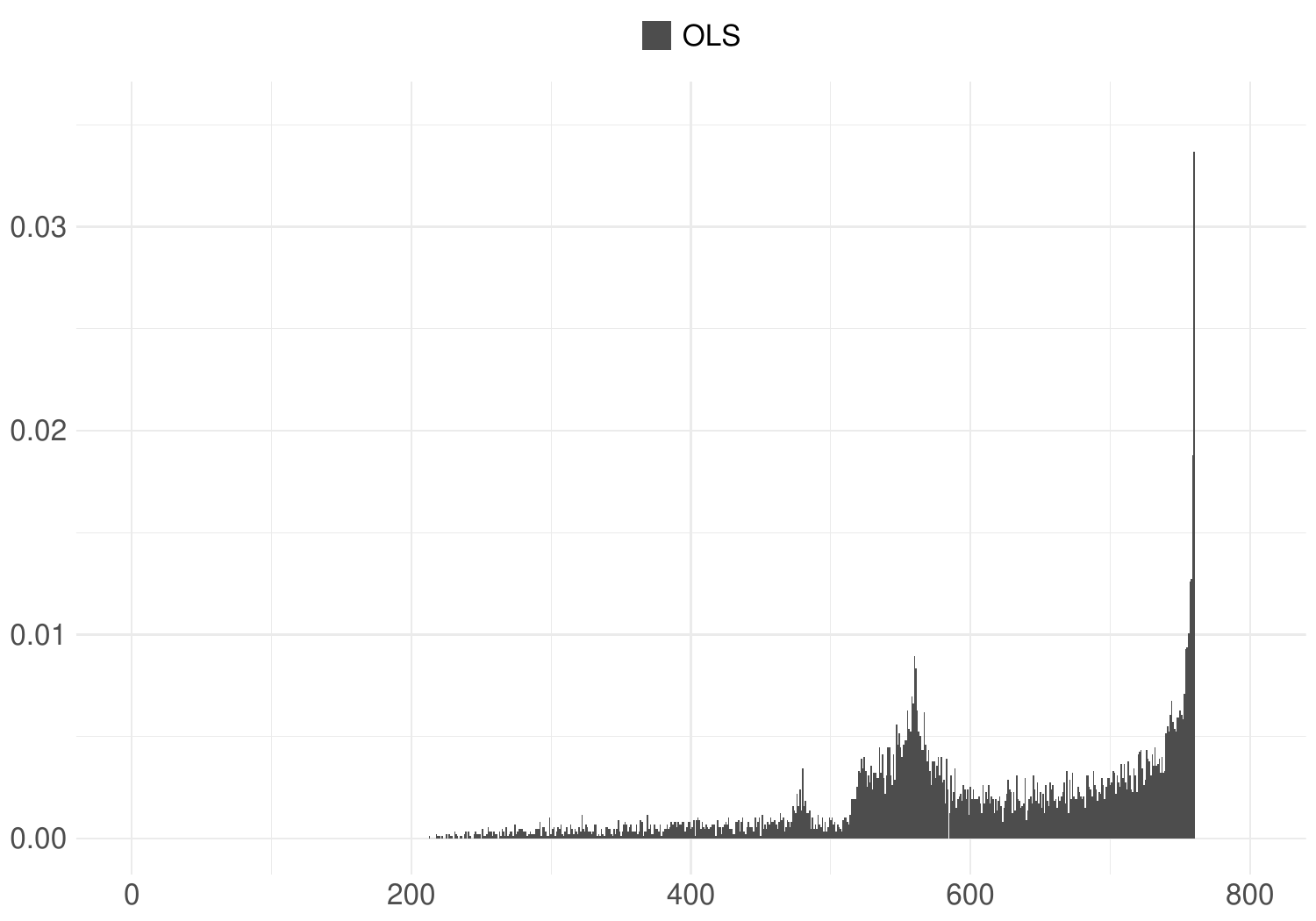}\label{fig2:2r:1}}
\subfigure[$T=800$, $c_a=4$, $c_b=6$, $s_0/s_1=1/5$]{\includegraphics[width=0.45\linewidth]{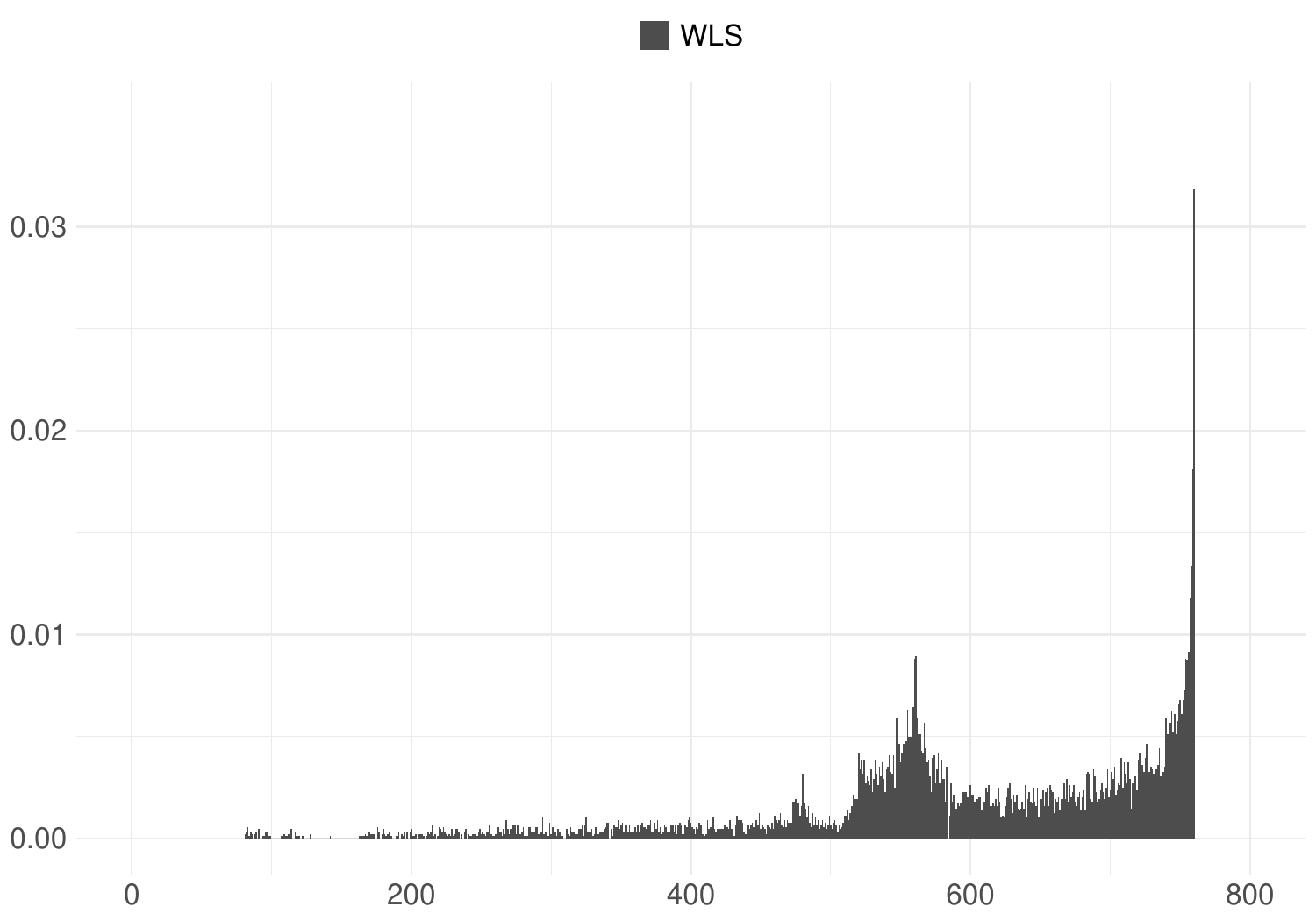}\label{fig2:2r:2}}\\
\subfigure[$T=800$, $c_a=5$, $c_b=6$, $s_0/s_1=1/5$]{\includegraphics[width=0.45\linewidth]{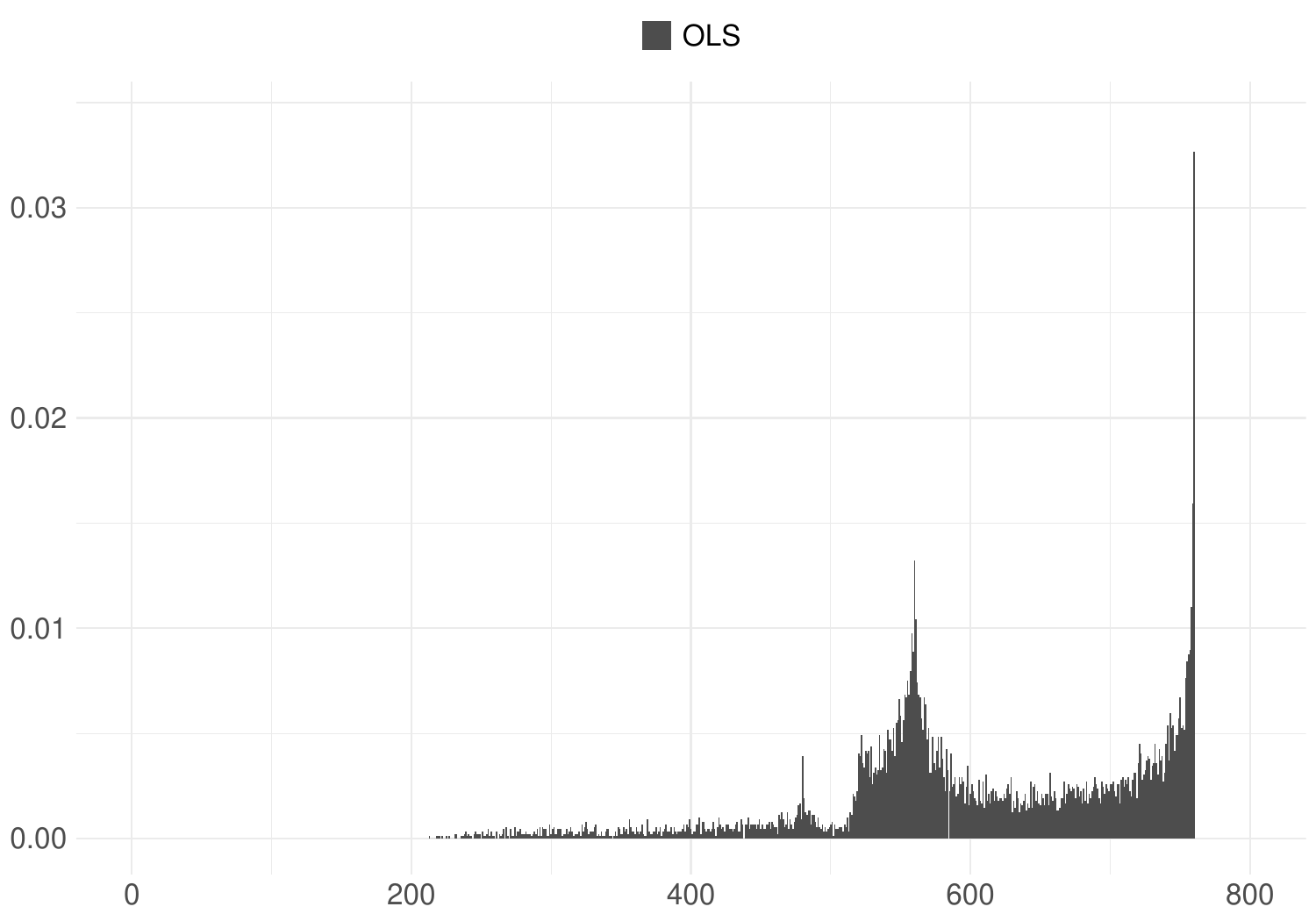}\label{fig2:2r:3}}
\subfigure[$T=800$, $c_a=5$, $c_b=6$, $s_0/s_1=1/5$]{\includegraphics[width=0.45\linewidth]{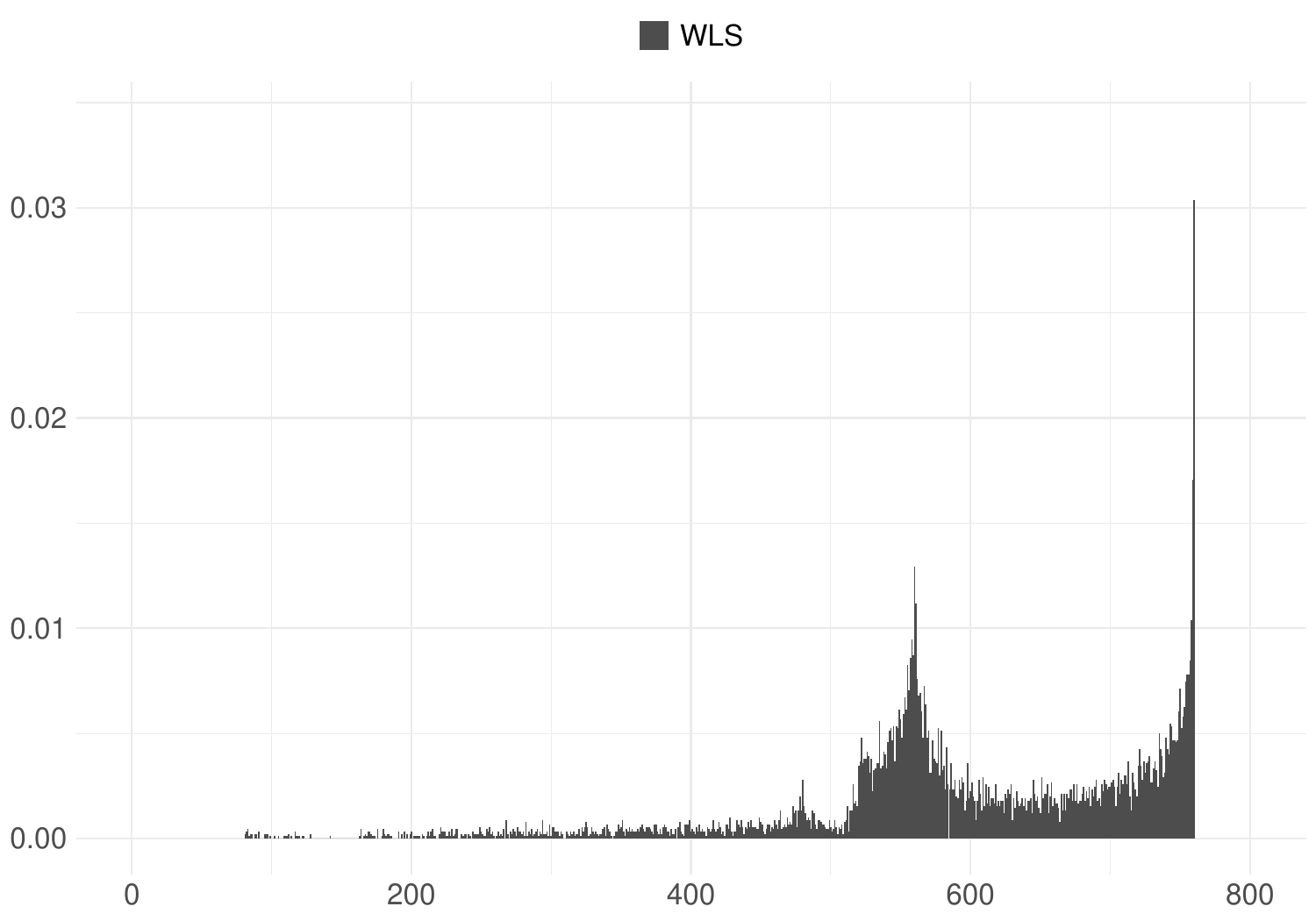}\label{fig2:2r:4}}\\
\subfigure[$T=800$, $c_a=6$, $c_b=6$, $s_0/s_1=1/5$]{\includegraphics[width=0.45\linewidth]{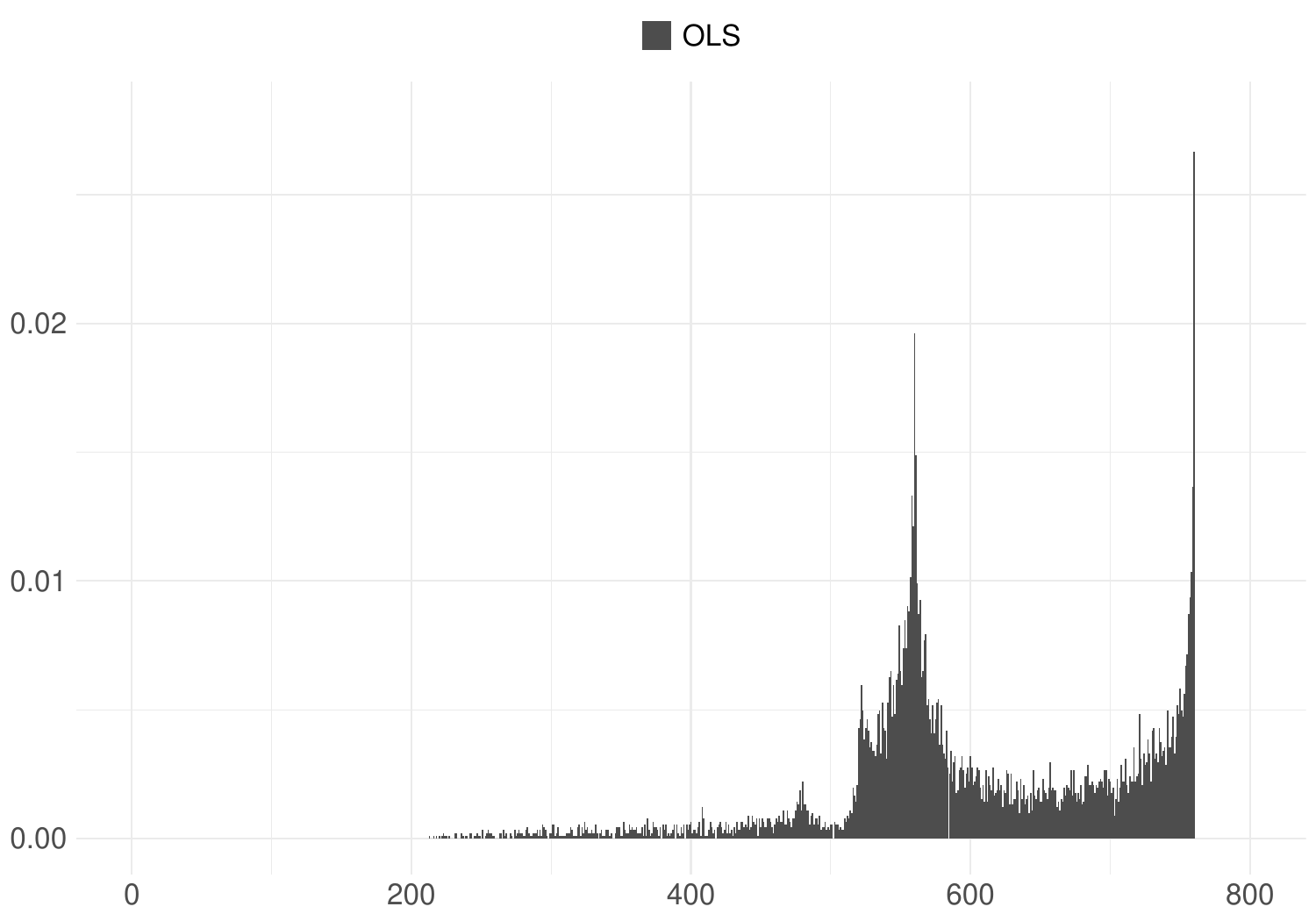}\label{fig2:2r:5}}
\subfigure[$T=800$, $c_a=6$, $c_b=6$, $s_0/s_1=1/5$]{\includegraphics[width=0.45\linewidth]{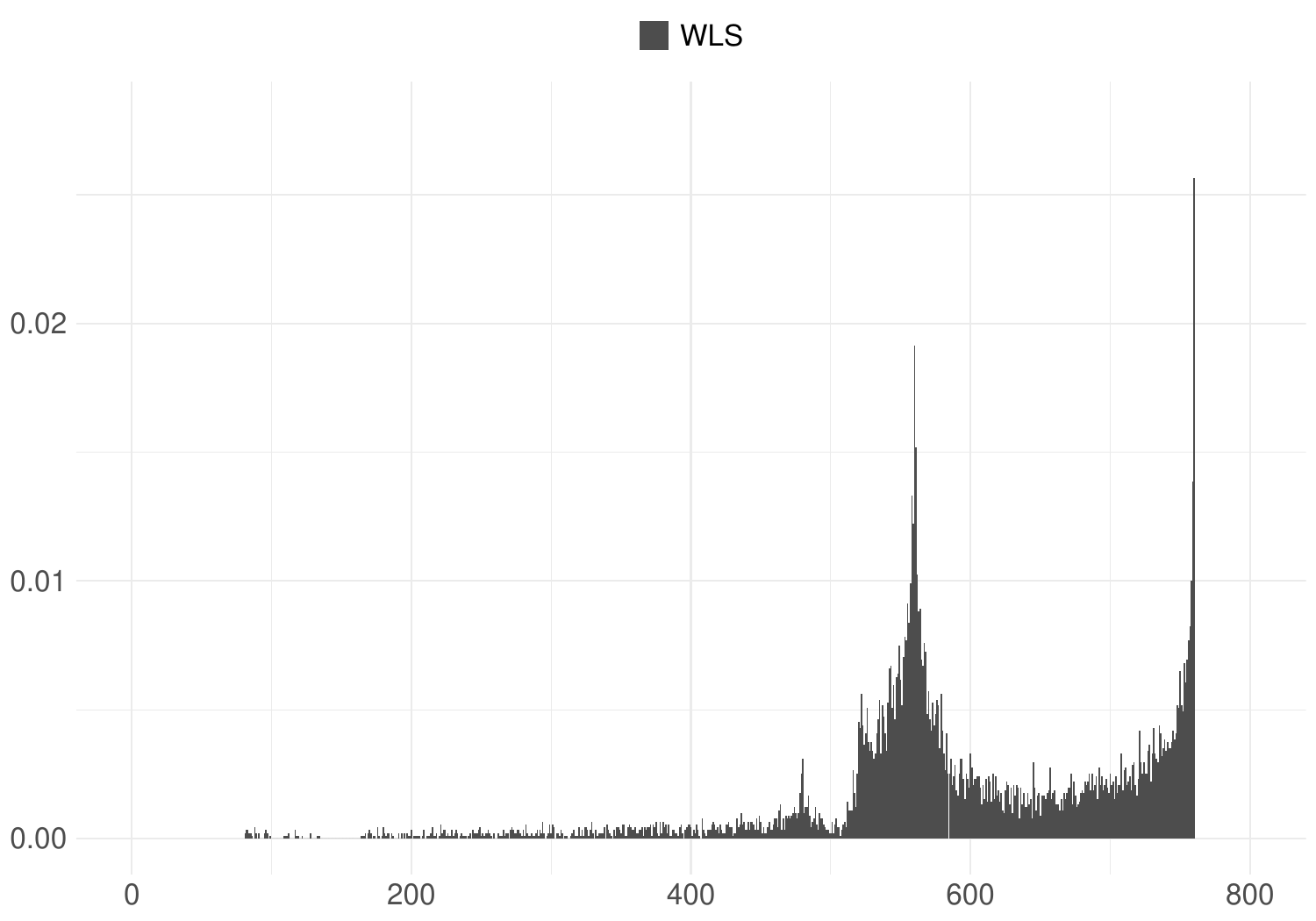}\label{fig2:2r:6}}\\
\end{center}%
\caption{Histograms of $\hat{k}_r$ 
for $(\tau_e,\tau_c,\tau_r)=(0.4,0.6,0.7)$,  $\tau=0.2$, $s_0/s_1=1/5$, $T=800$}
\label{fig22_r}
\end{figure}

\newpage

\begin{figure}[h!]%
\begin{center}%
\subfigure[$T=400$, $c_a=4$, $c_b=6$, $s_0/s_1=1$]{\includegraphics[width=0.45\linewidth]{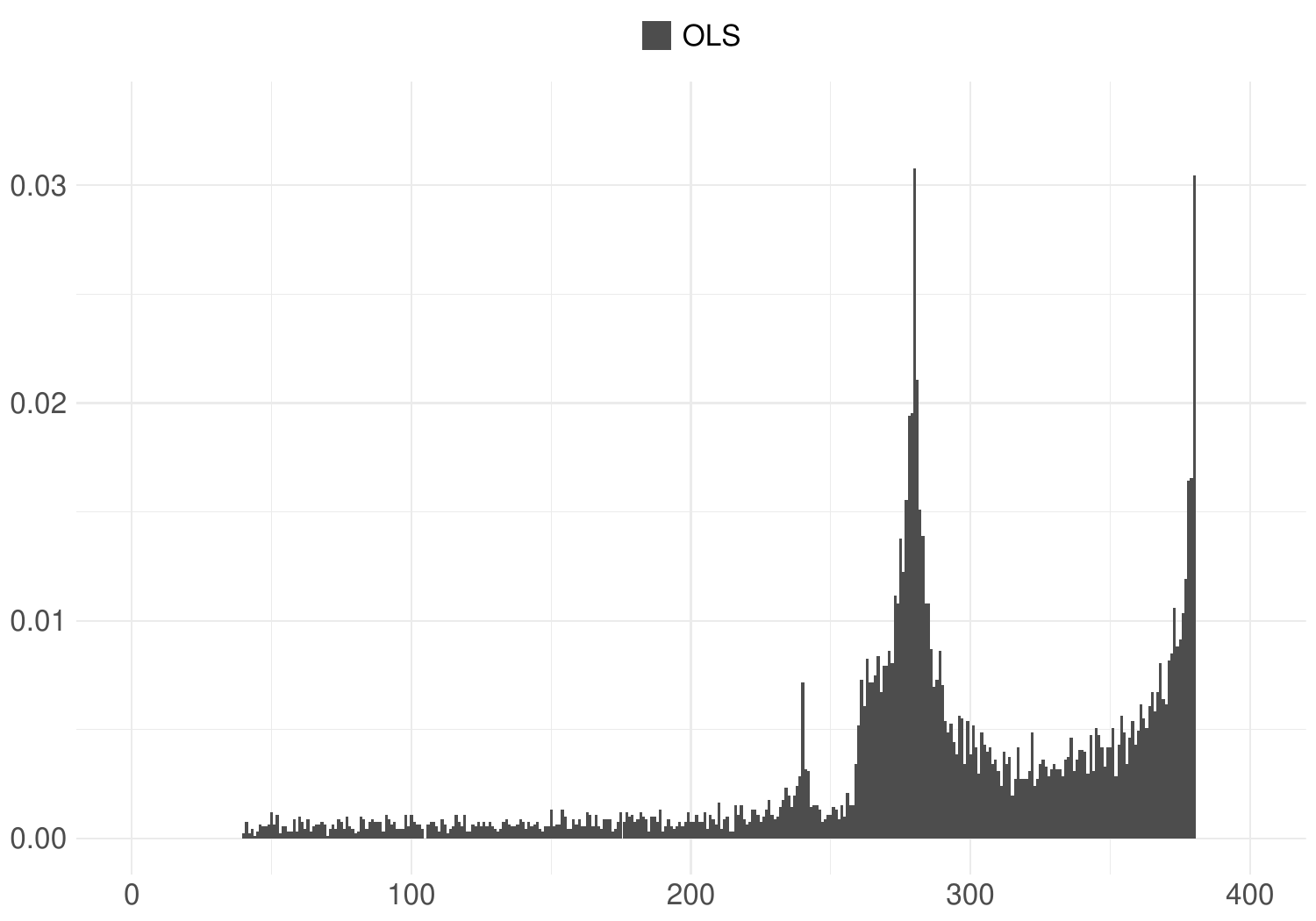}\label{fig2:3r:1}}
\subfigure[$T=400$, $c_a=4$, $c_b=6$, $s_0/s_1=1$]{\includegraphics[width=0.45\linewidth]{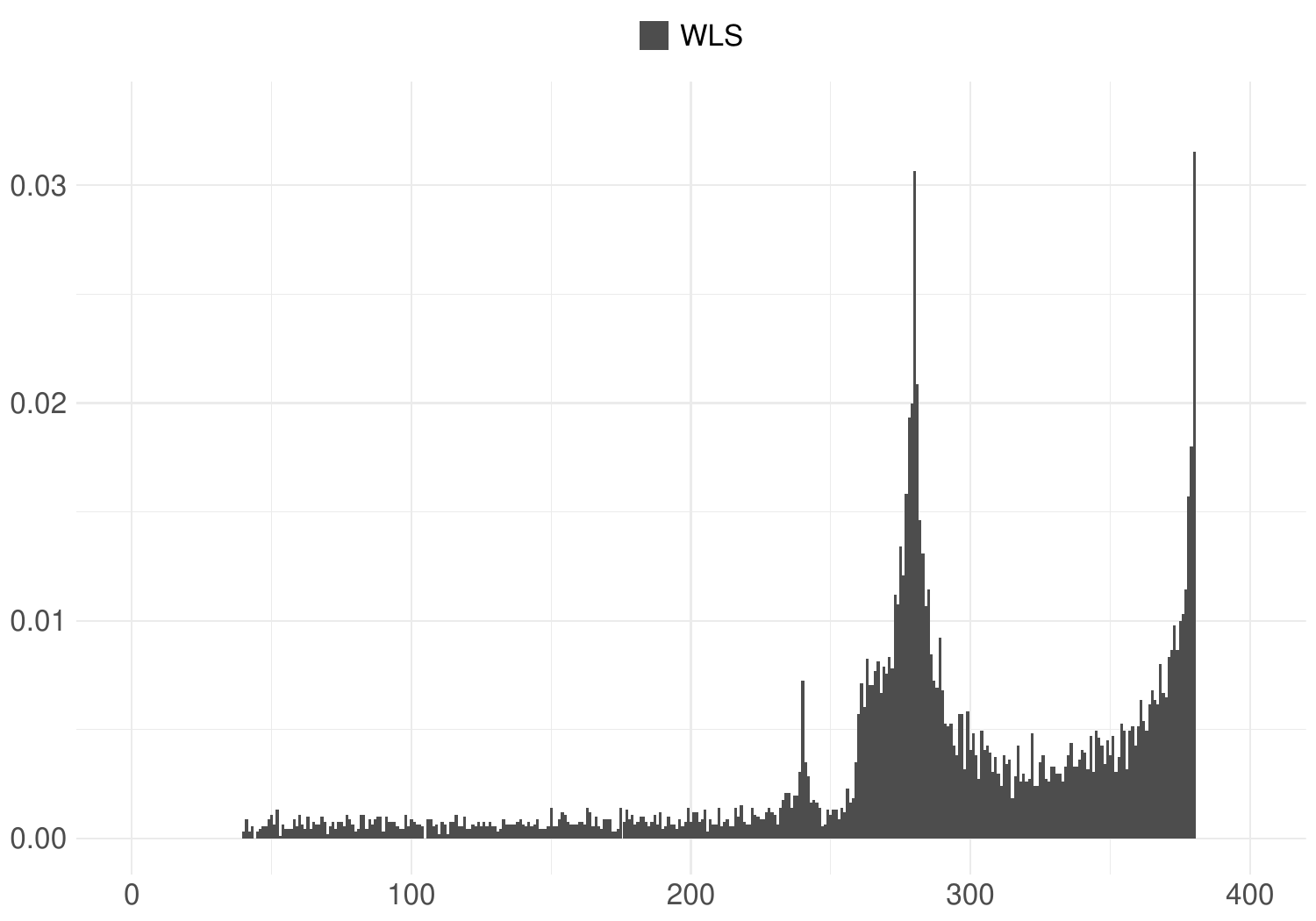}\label{fig2:3r:2}}\\
\subfigure[$T=400$, $c_a=5$, $c_b=6$, $s_0/s_1=1$]{\includegraphics[width=0.45\linewidth]{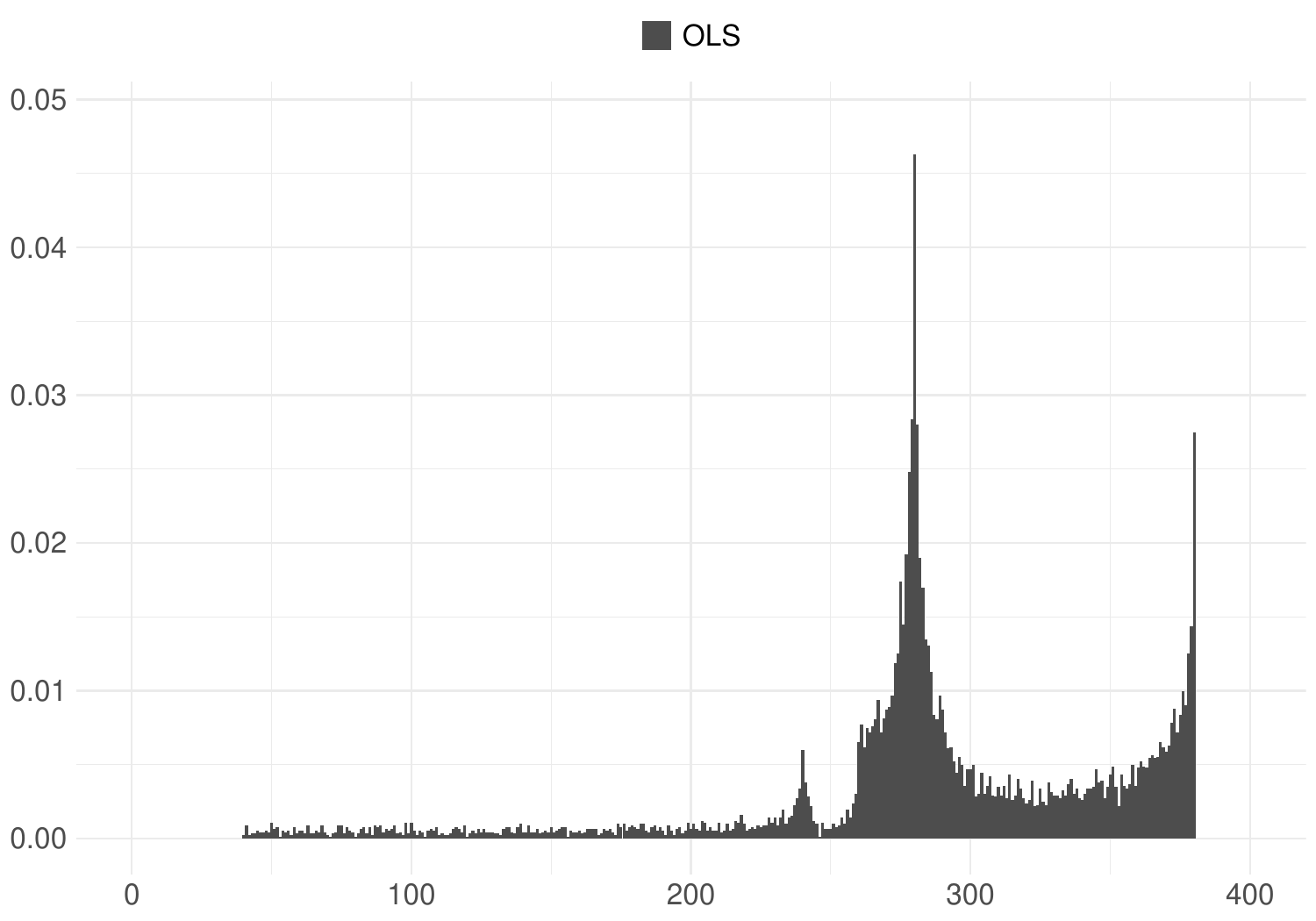}\label{fig2:3r:3}}
\subfigure[$T=400$, $c_a=5$, $c_b=6$, $s_0/s_1=1$]{\includegraphics[width=0.45\linewidth]{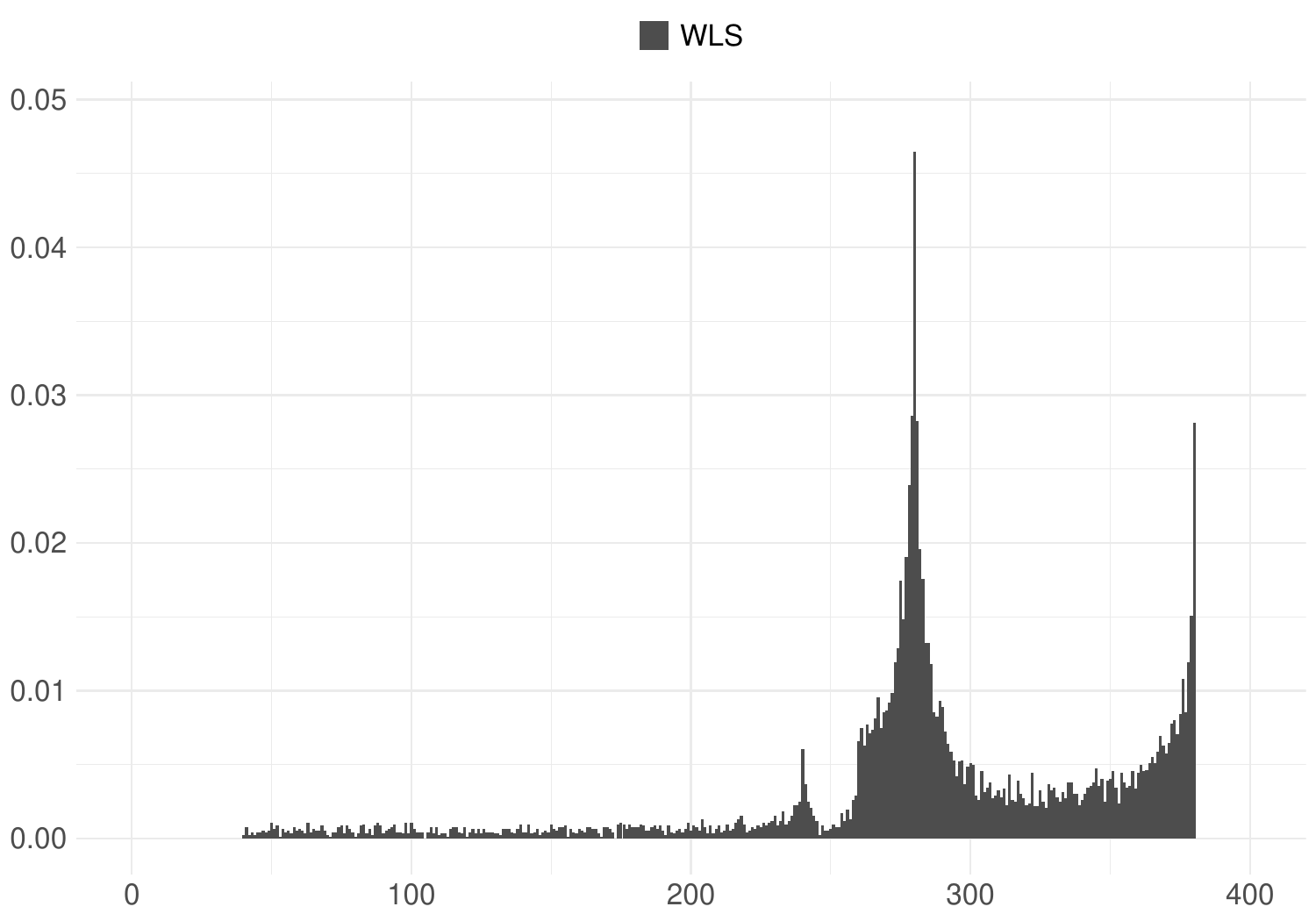}\label{fig2:3r:4}}\\
\subfigure[$T=400$, $c_a=6$, $c_b=6$, $s_0/s_1=1$]{\includegraphics[width=0.45\linewidth]{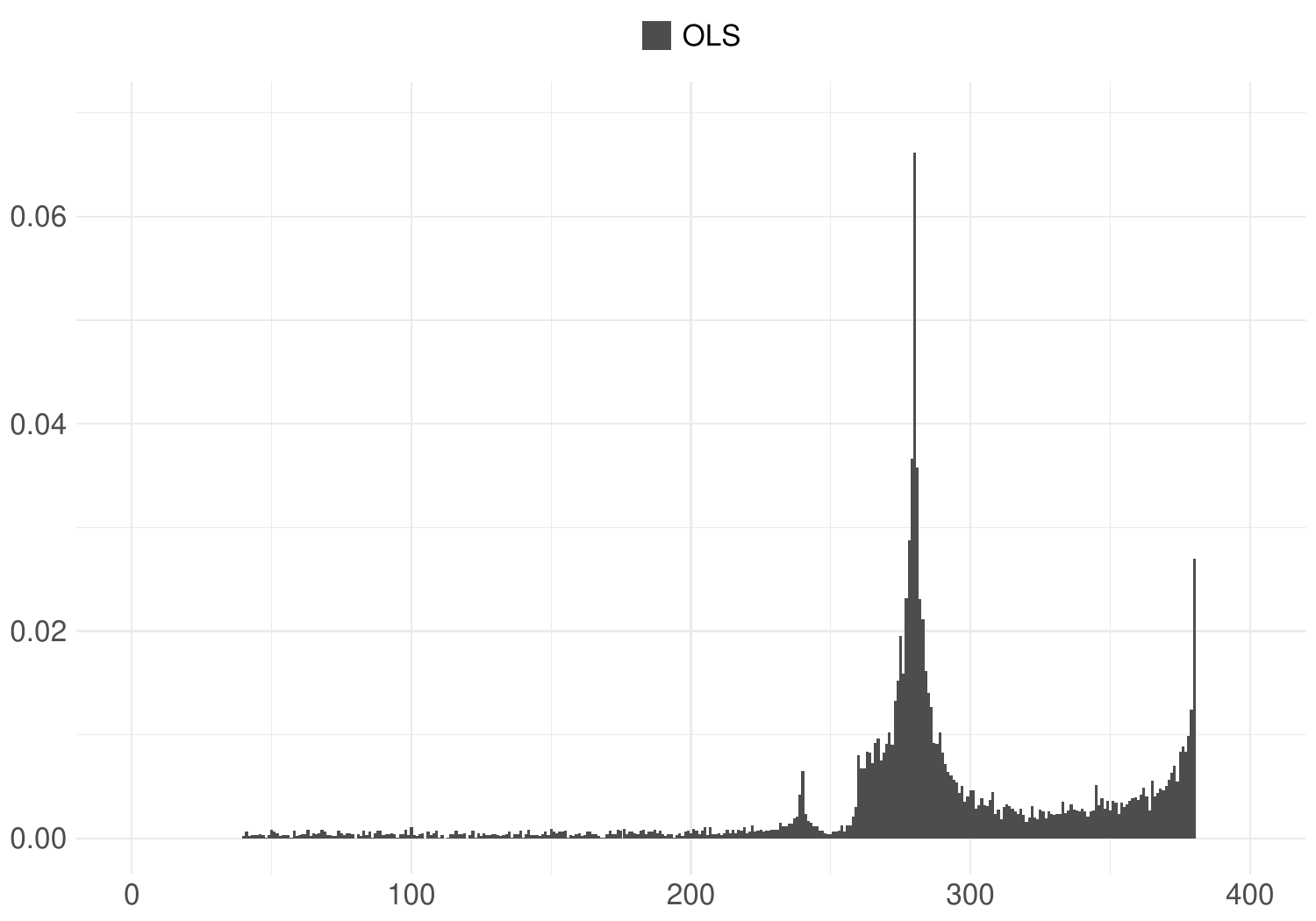}\label{fig2:3r:5}}
\subfigure[$T=400$, $c_a=6$, $c_b=6$, $s_0/s_1=1$]{\includegraphics[width=0.45\linewidth]{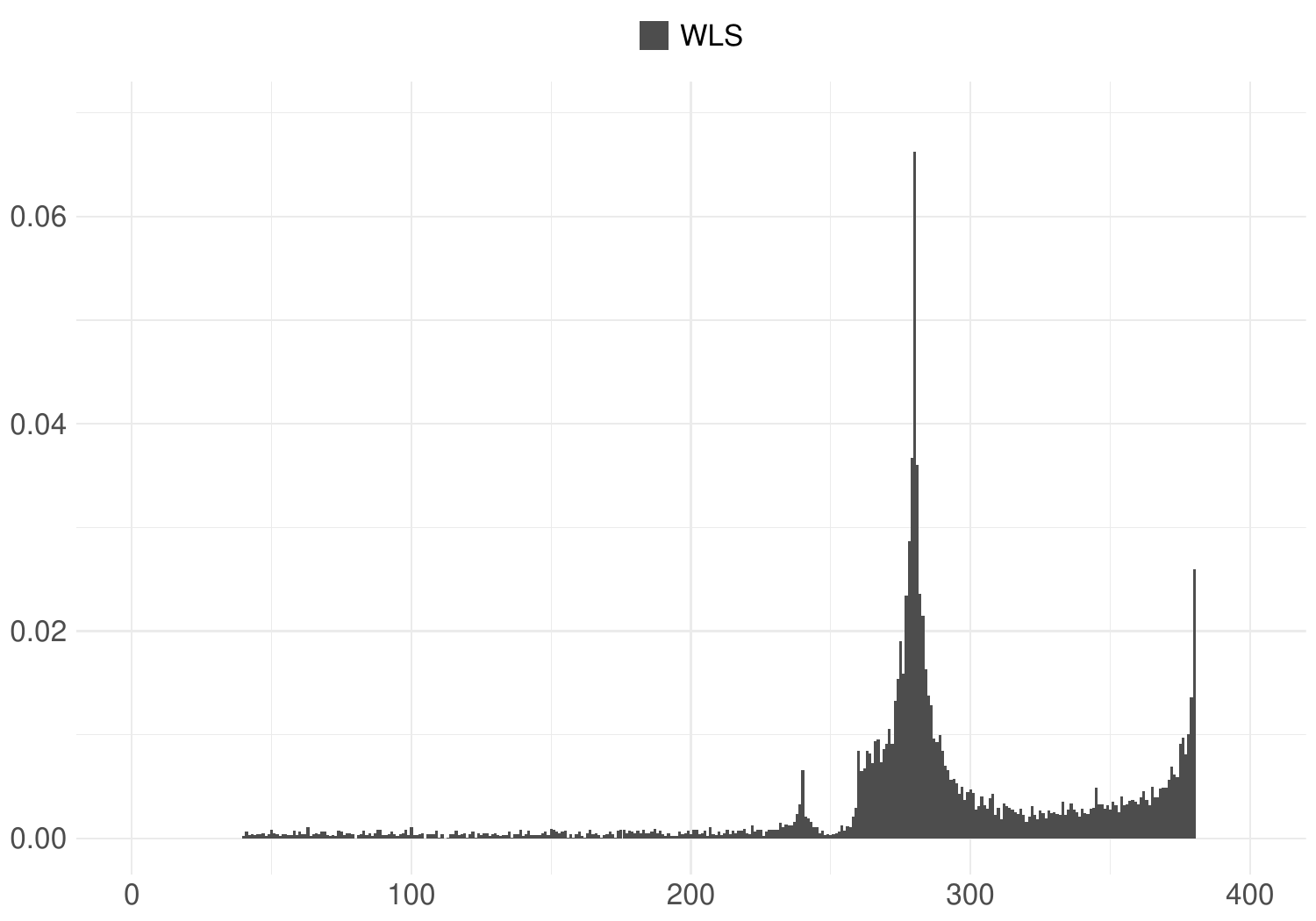}\label{fig2:3r:6}}\\
\end{center}%
\caption{Histograms of $\hat{k}_r$ 
for $(\tau_e,\tau_c,\tau_r)=(0.4,0.6,0.7)$,  $\tau=0.2$, $s_0/s_1=1$, $T=400$}
\label{fig23_r}
\end{figure}

\newpage

\begin{figure}[h!]%
\begin{center}%
\subfigure[$T=800$, $c_a=4$, $c_b=6$, $s_0/s_1=1$]{\includegraphics[width=0.45\linewidth]{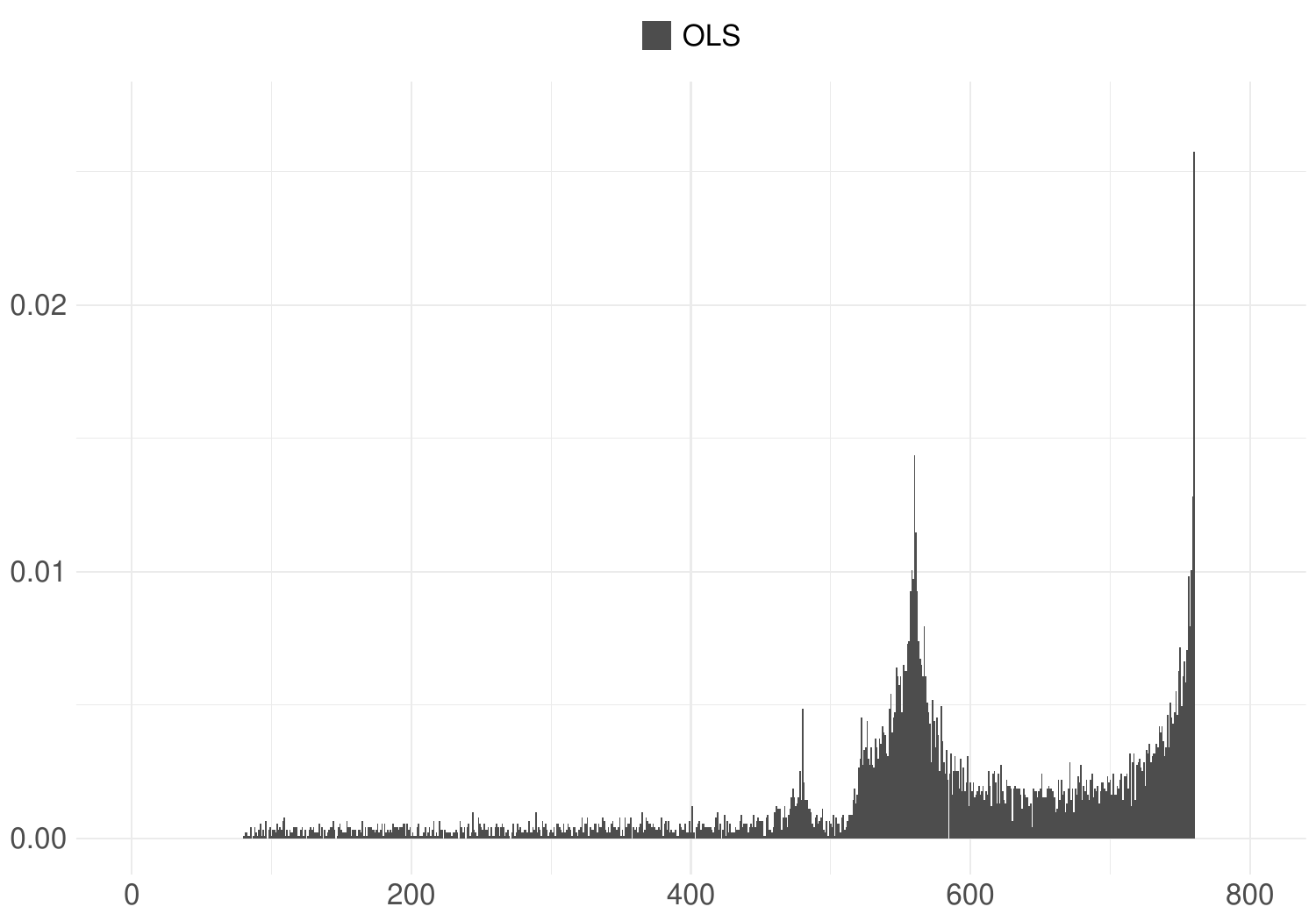}\label{fig2:4r:1}}
\subfigure[$T=800$, $c_a=4$, $c_b=6$, $s_0/s_1=1$]{\includegraphics[width=0.45\linewidth]{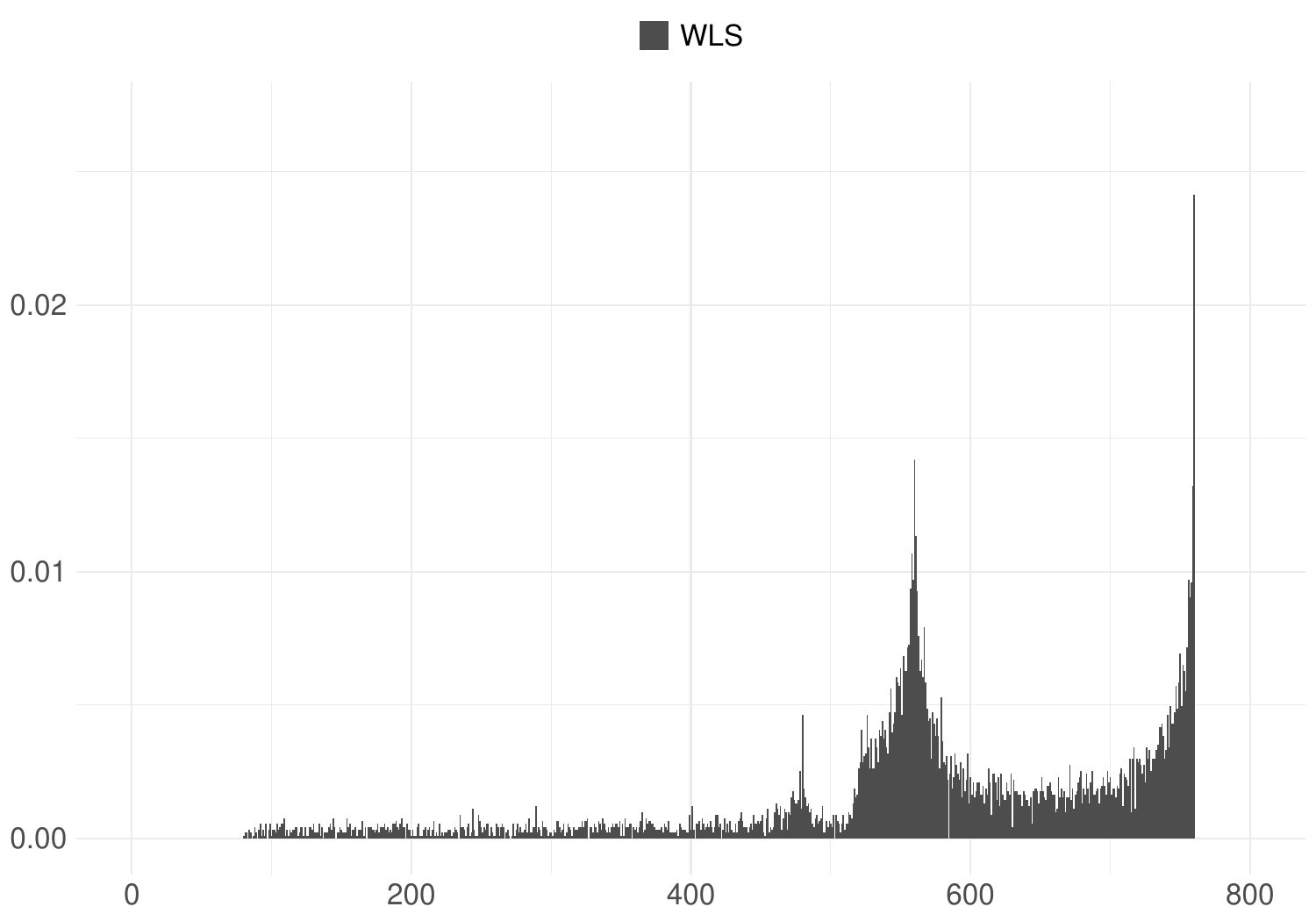}\label{fig2:4r:2}}\\
\subfigure[$T=800$, $c_a=5$, $c_b=6$, $s_0/s_1=1$]{\includegraphics[width=0.45\linewidth]{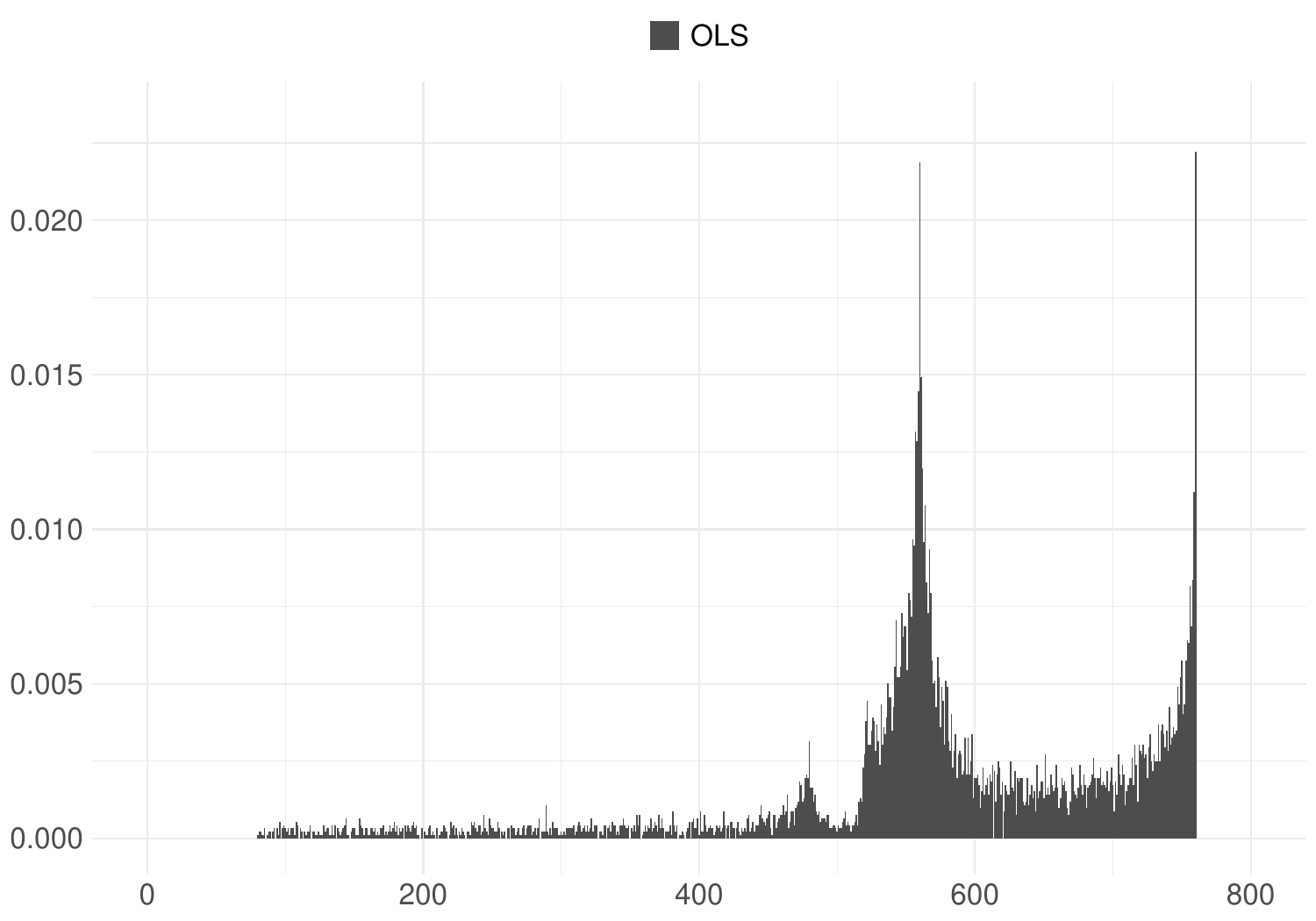}\label{fig2:4r:3}}
\subfigure[$T=800$, $c_a=5$, $c_b=6$, $s_0/s_1=1$]{\includegraphics[width=0.45\linewidth]{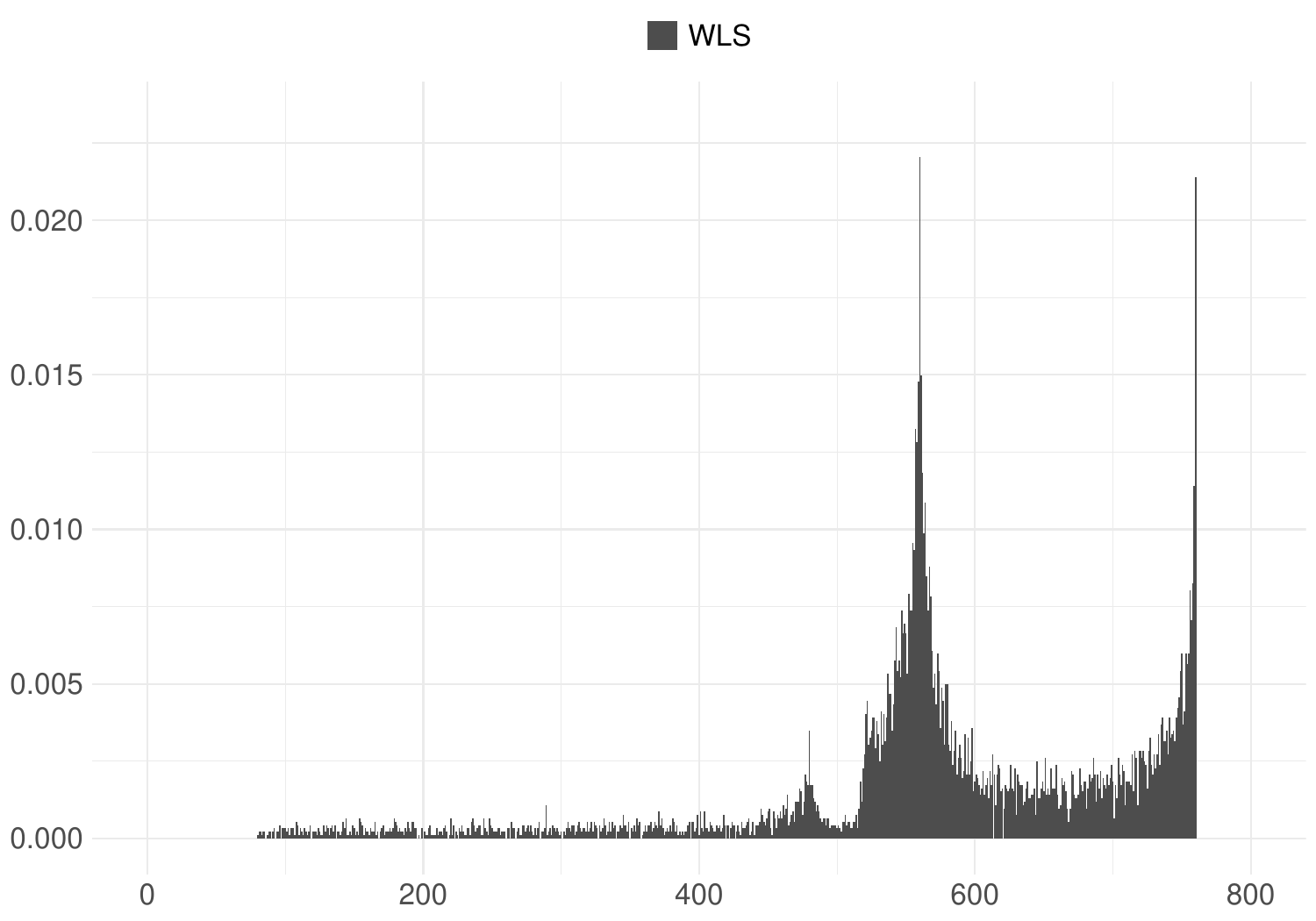}\label{fig2:4r:4}}\\
\subfigure[$T=800$, $c_a=6$, $c_b=6$, $s_0/s_1=1$]{\includegraphics[width=0.45\linewidth]{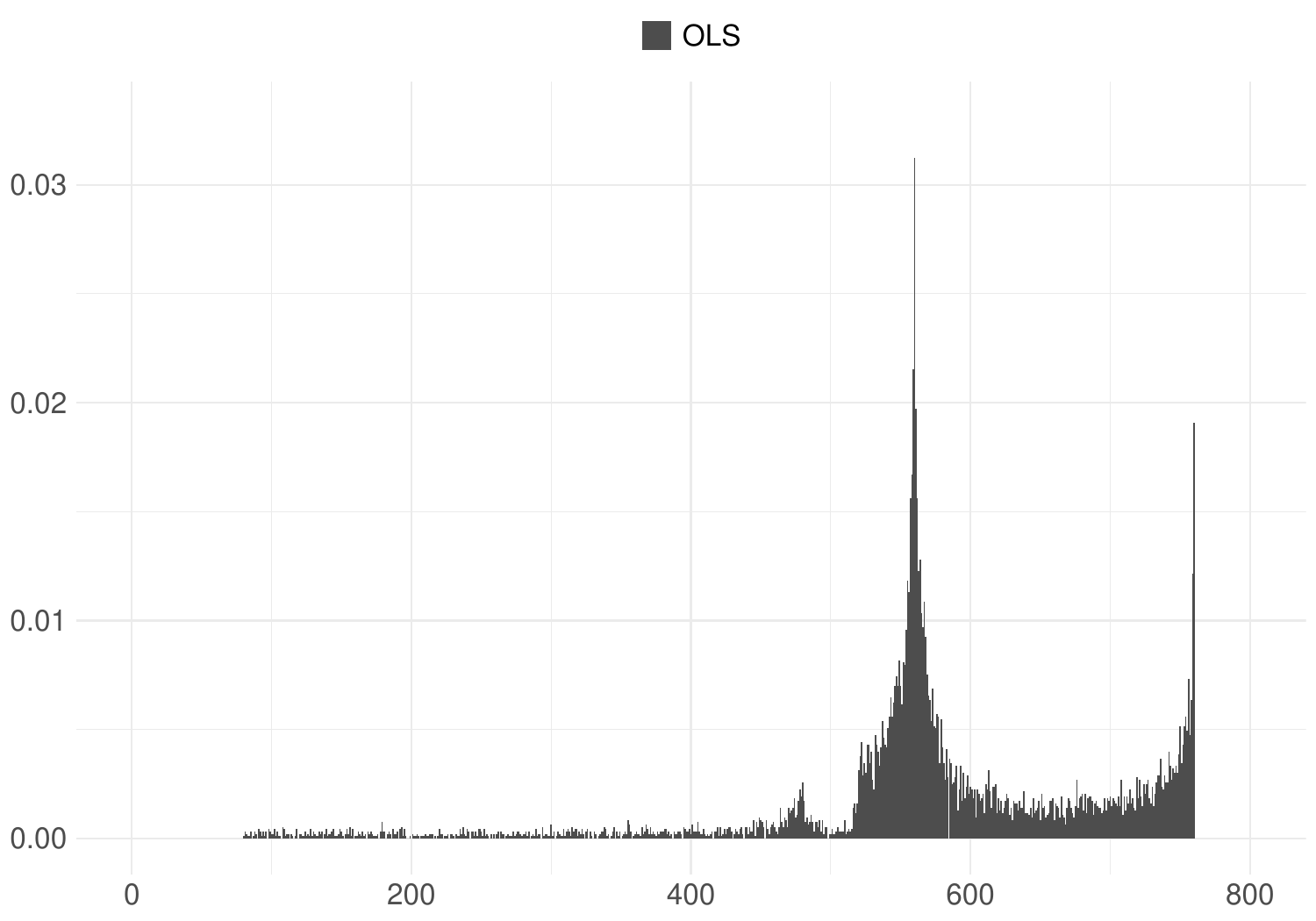}\label{fig2:4r:5}}
\subfigure[$T=800$, $c_a=6$, $c_b=6$, $s_0/s_1=1$]{\includegraphics[width=0.45\linewidth]{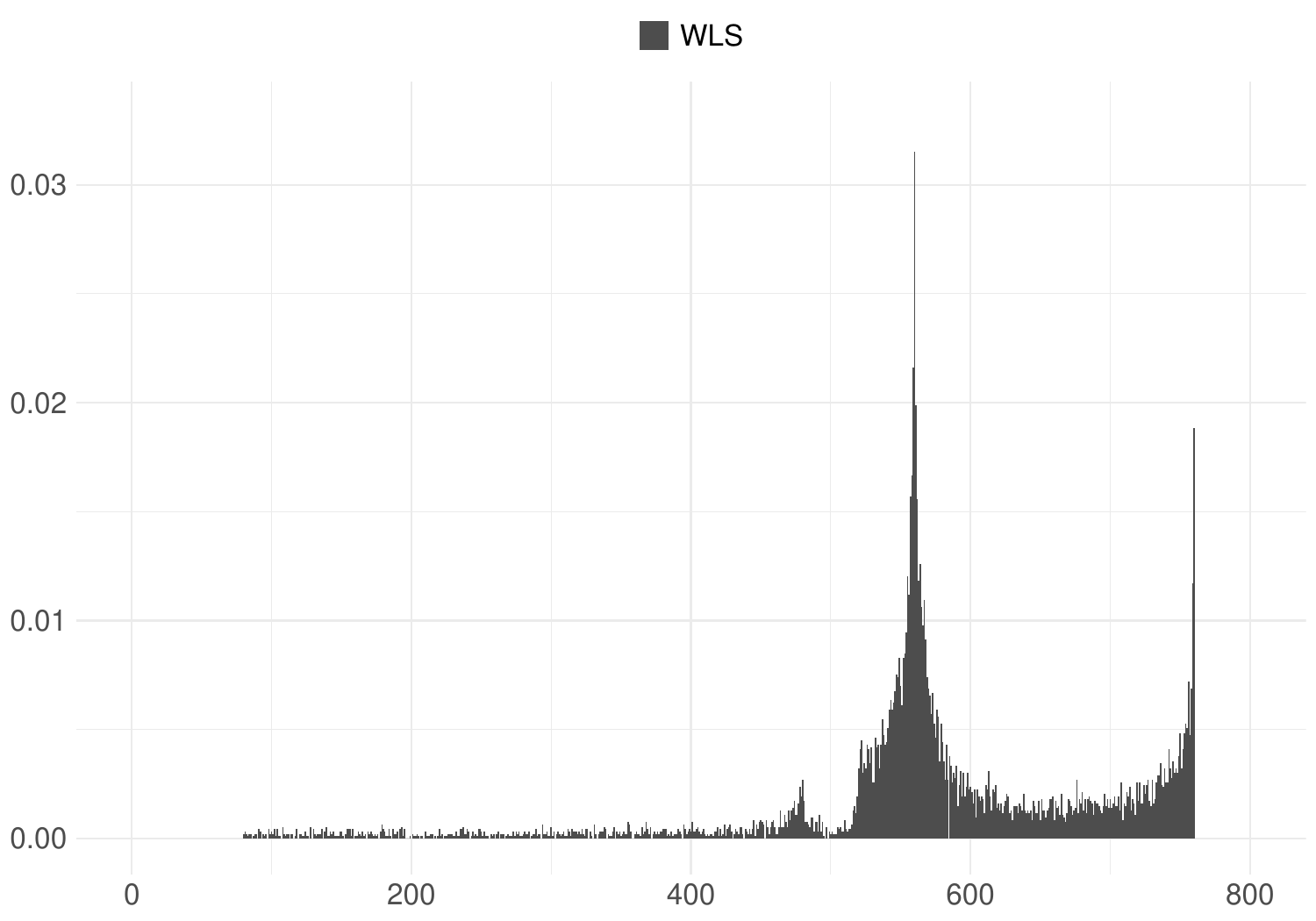}\label{fig2:4r:6}}\\
\end{center}%
\caption{Histograms of $\hat{k}_r$ 
for $(\tau_e,\tau_c,\tau_r)=(0.4,0.6,0.7)$,  $\tau=0.2$, $s_0/s_1=1$, $T=800$}
\label{fig24_r}
\end{figure}

\newpage

\begin{figure}[h!]%
\begin{center}%
\subfigure[$T=400$, $c_a=4$, $c_b=6$, $s_0/s_1=5$]{\includegraphics[width=0.45\linewidth]{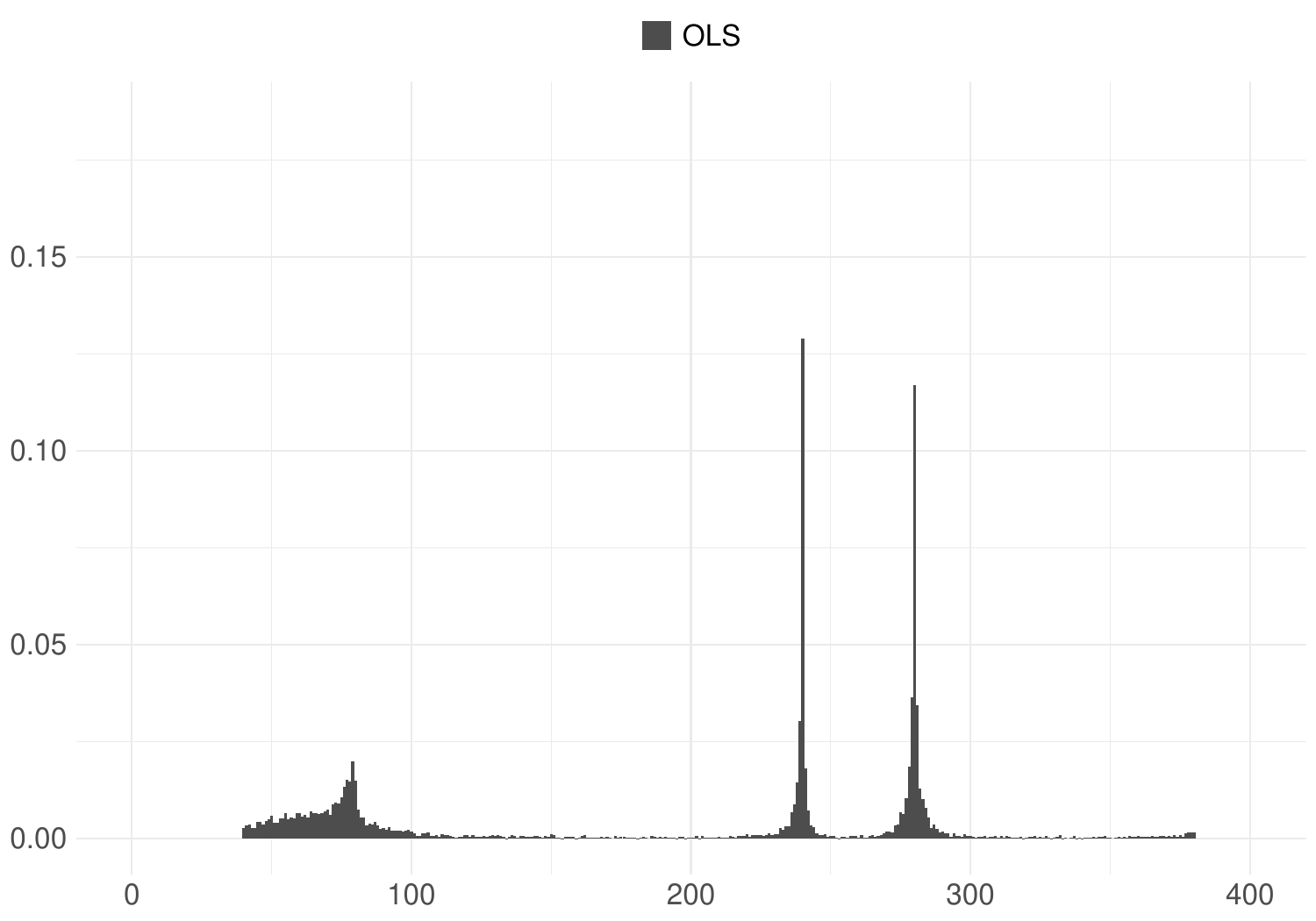}\label{fig2:5r:1}}
\subfigure[$T=400$, $c_a=4$, $c_b=6$, $s_0/s_1=5$]{\includegraphics[width=0.45\linewidth]{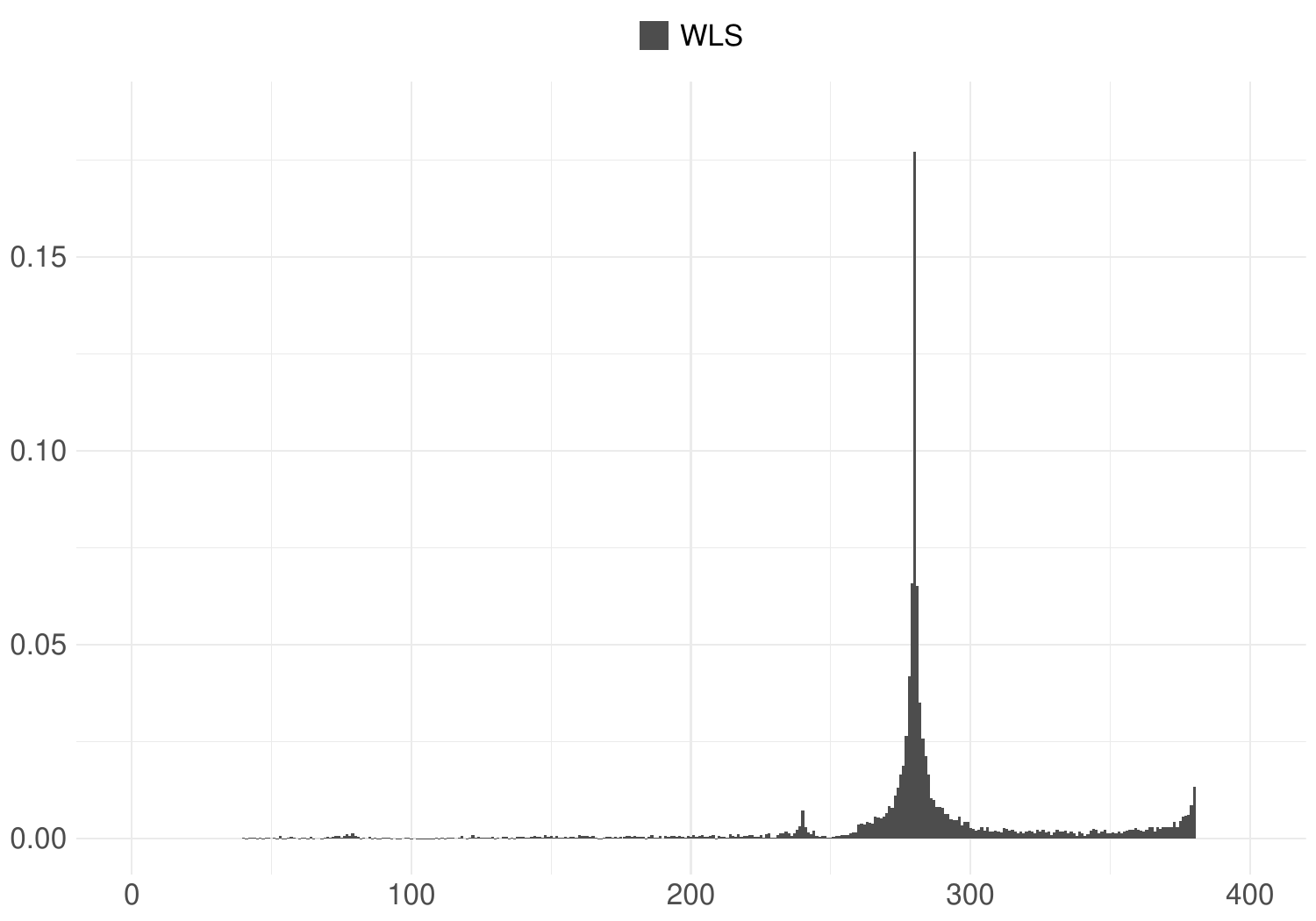}\label{fig2:5r:2}}\\
\subfigure[$T=400$, $c_a=5$, $c_b=6$, $s_0/s_1=5$]{\includegraphics[width=0.45\linewidth]{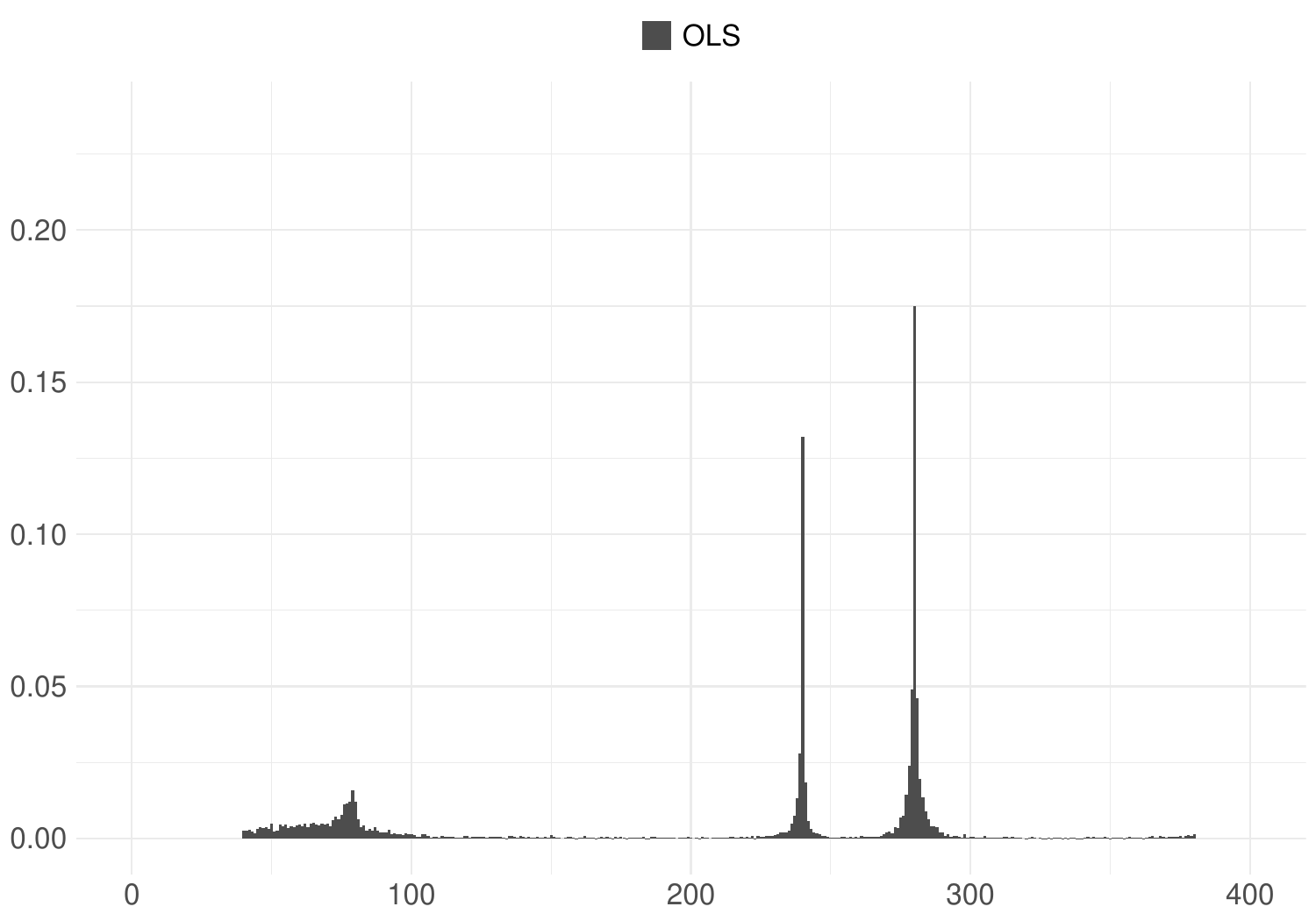}\label{fig2:5r:3}}
\subfigure[$T=400$, $c_a=5$, $c_b=6$, $s_0/s_1=5$]{\includegraphics[width=0.45\linewidth]{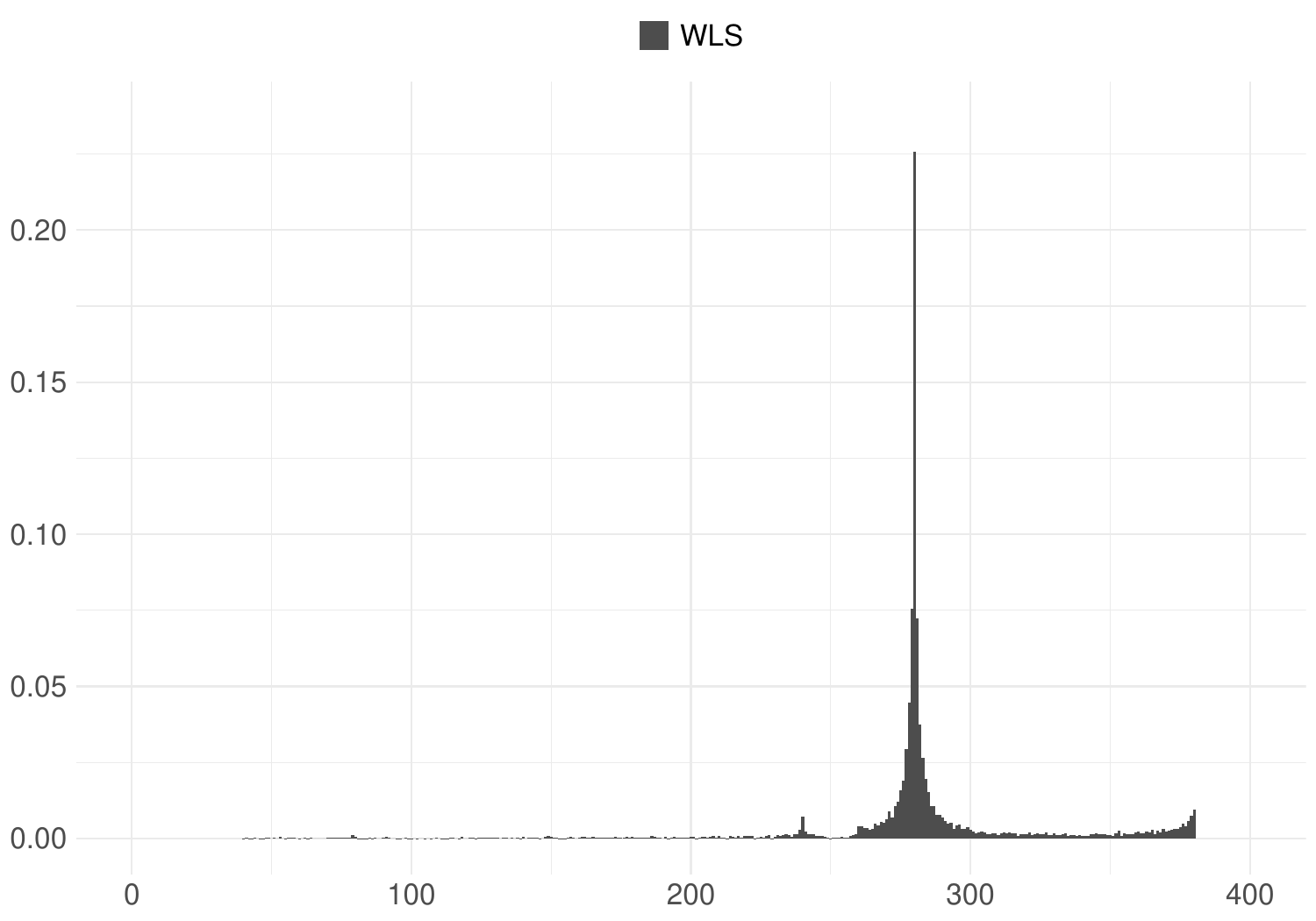}\label{fig2:5r:4}}\\
\subfigure[$T=400$, $c_a=6$, $c_b=6$, $s_0/s_1=5$]{\includegraphics[width=0.45\linewidth]{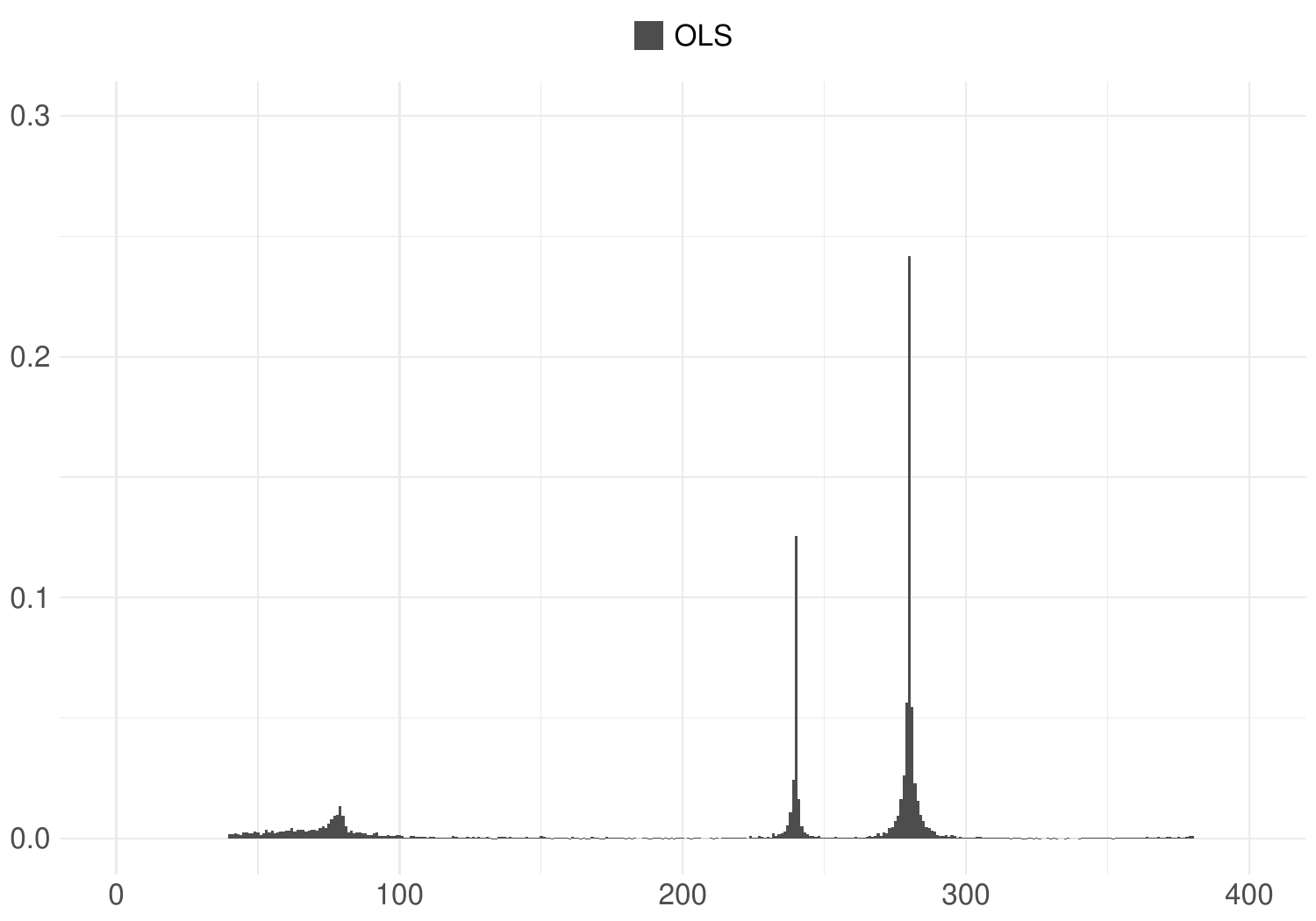}\label{fig2:5r:5}}
\subfigure[$T=400$, $c_a=6$, $c_b=6$, $s_0/s_1=5$]{\includegraphics[width=0.45\linewidth]{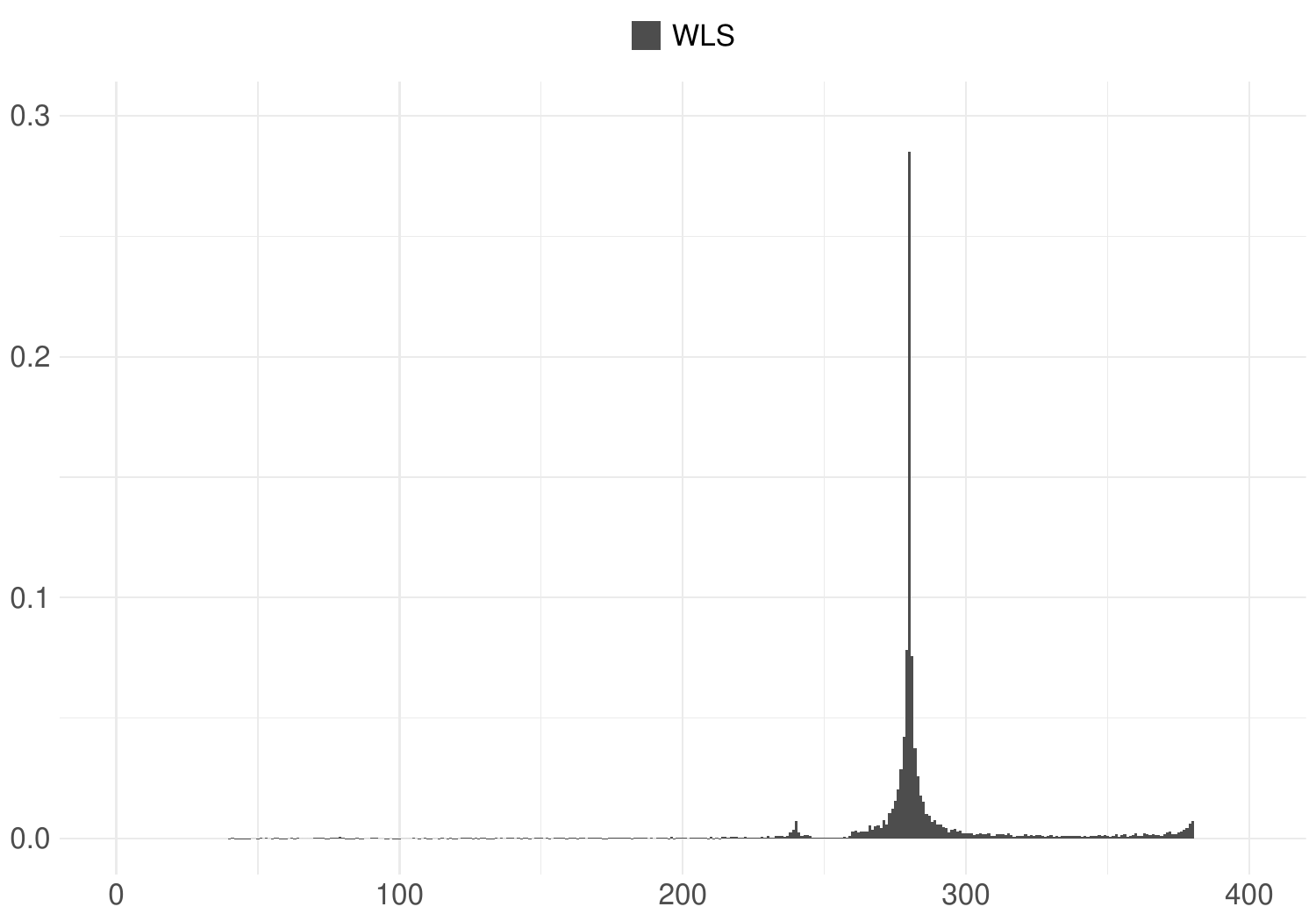}\label{fig2:5r:6}}\\
\end{center}%
\caption{Histograms of $\hat{k}_r$ 
for $(\tau_e,\tau_c,\tau_r)=(0.4,0.6,0.7)$,  $\tau=0.2$, $s_0/s_1=5$, $T=400$}
\label{fig25_r}
\end{figure}

\newpage

\begin{figure}[h!]%
\begin{center}%
\subfigure[$T=800$, $c_a=4$, $c_b=6$, $s_0/s_1=5$]{\includegraphics[width=0.45\linewidth]{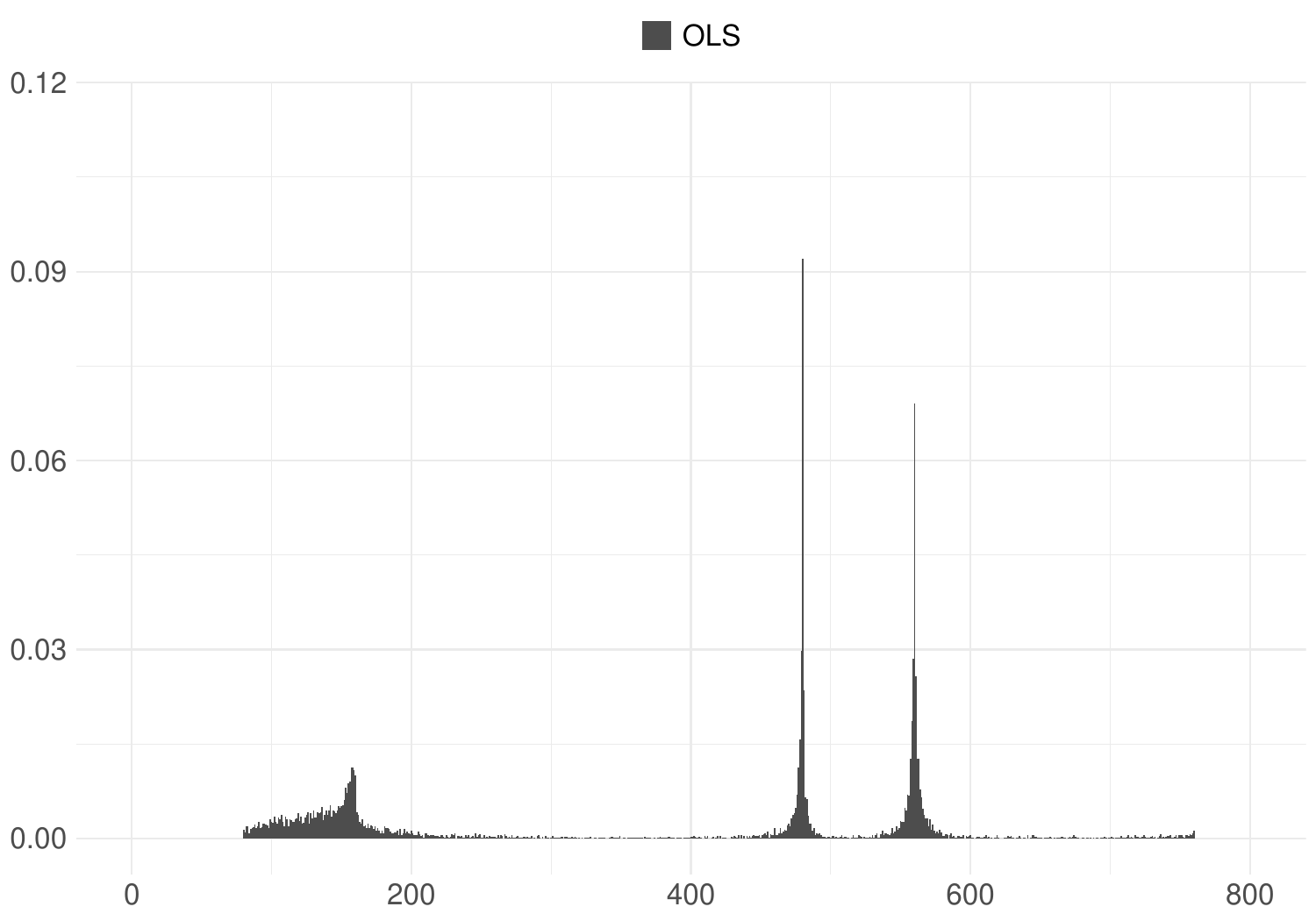}\label{fig2:6r:1}}
\subfigure[$T=800$, $c_a=4$, $c_b=6$, $s_0/s_1=5$]{\includegraphics[width=0.45\linewidth]{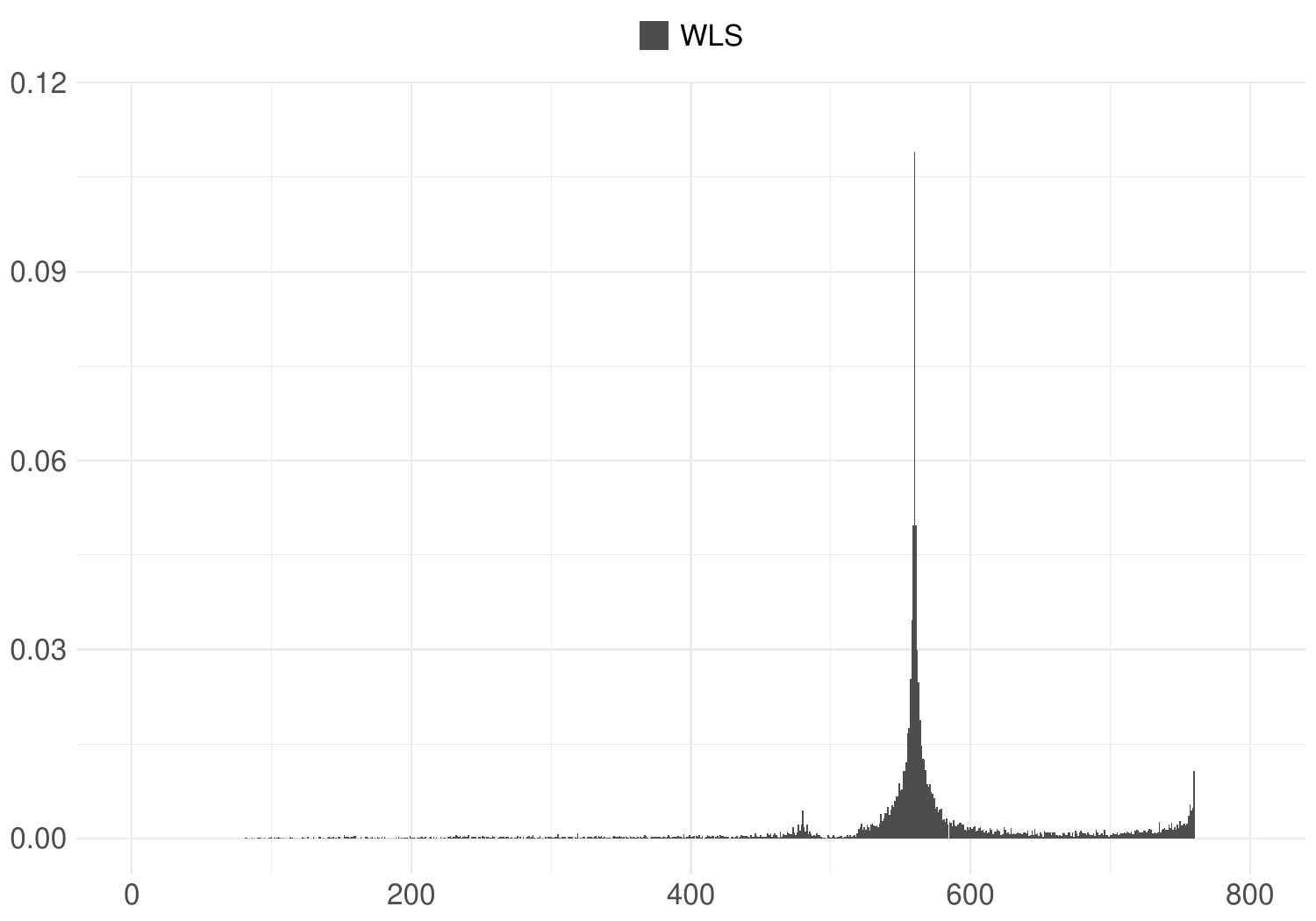}\label{fig2:6r:2}}\\
\subfigure[$T=800$, $c_a=5$, $c_b=6$, $s_0/s_1=5$]{\includegraphics[width=0.45\linewidth]{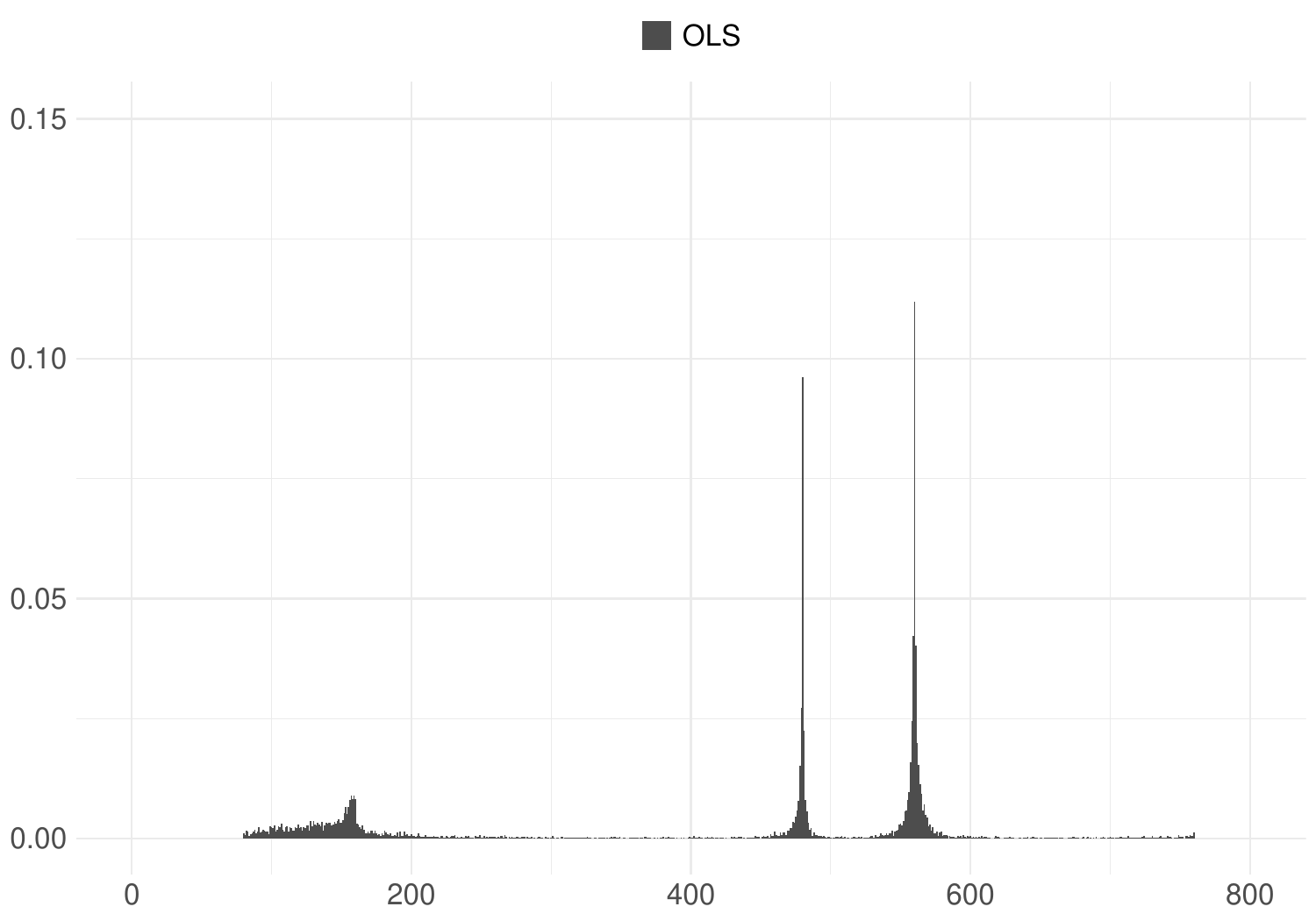}\label{fig2:6r:3}}
\subfigure[$T=800$, $c_a=5$, $c_b=6$, $s_0/s_1=5$]{\includegraphics[width=0.45\linewidth]{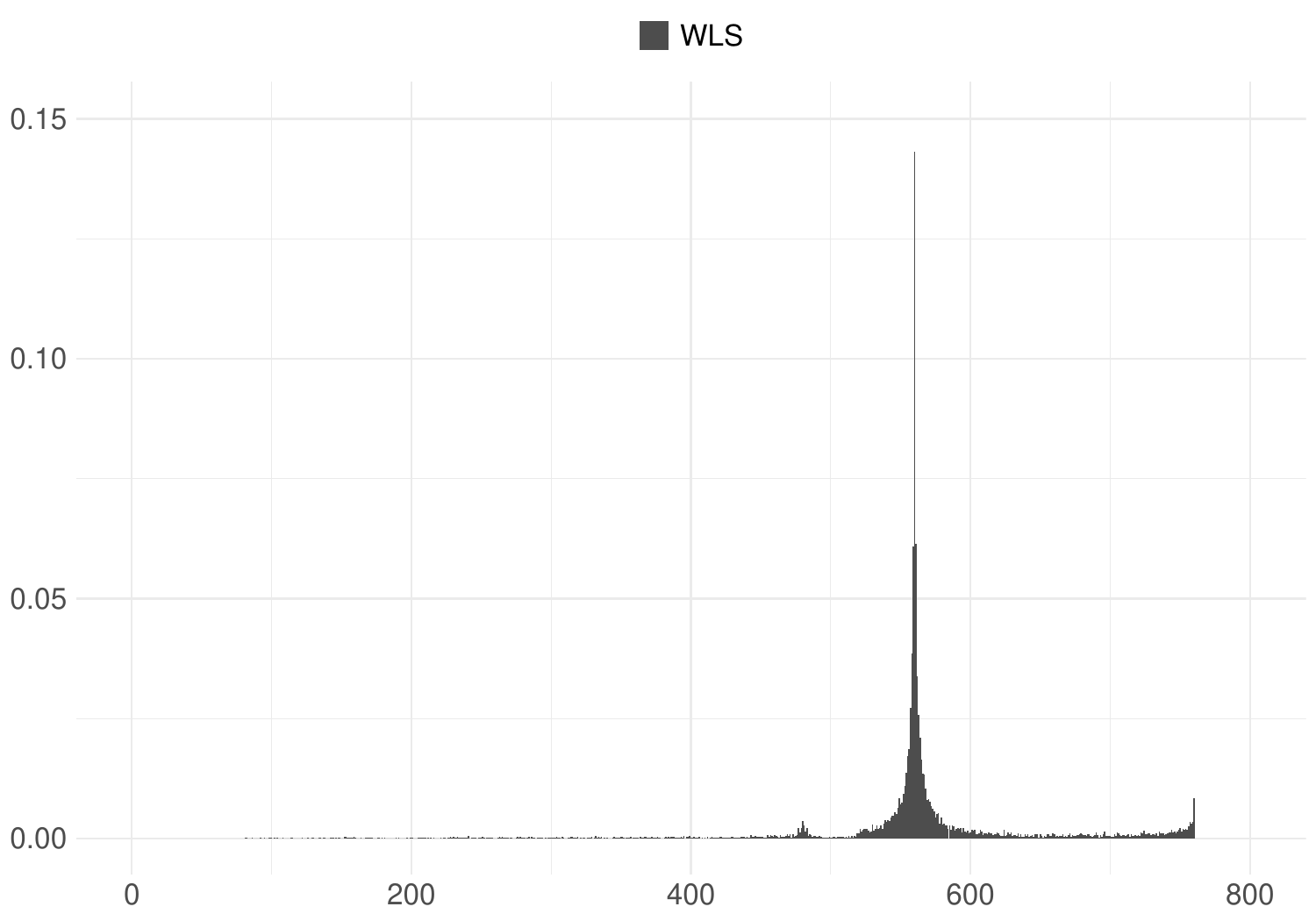}\label{fig2:6r:4}}\\
\subfigure[$T=800$, $c_a=6$, $c_b=6$, $s_0/s_1=5$]{\includegraphics[width=0.45\linewidth]{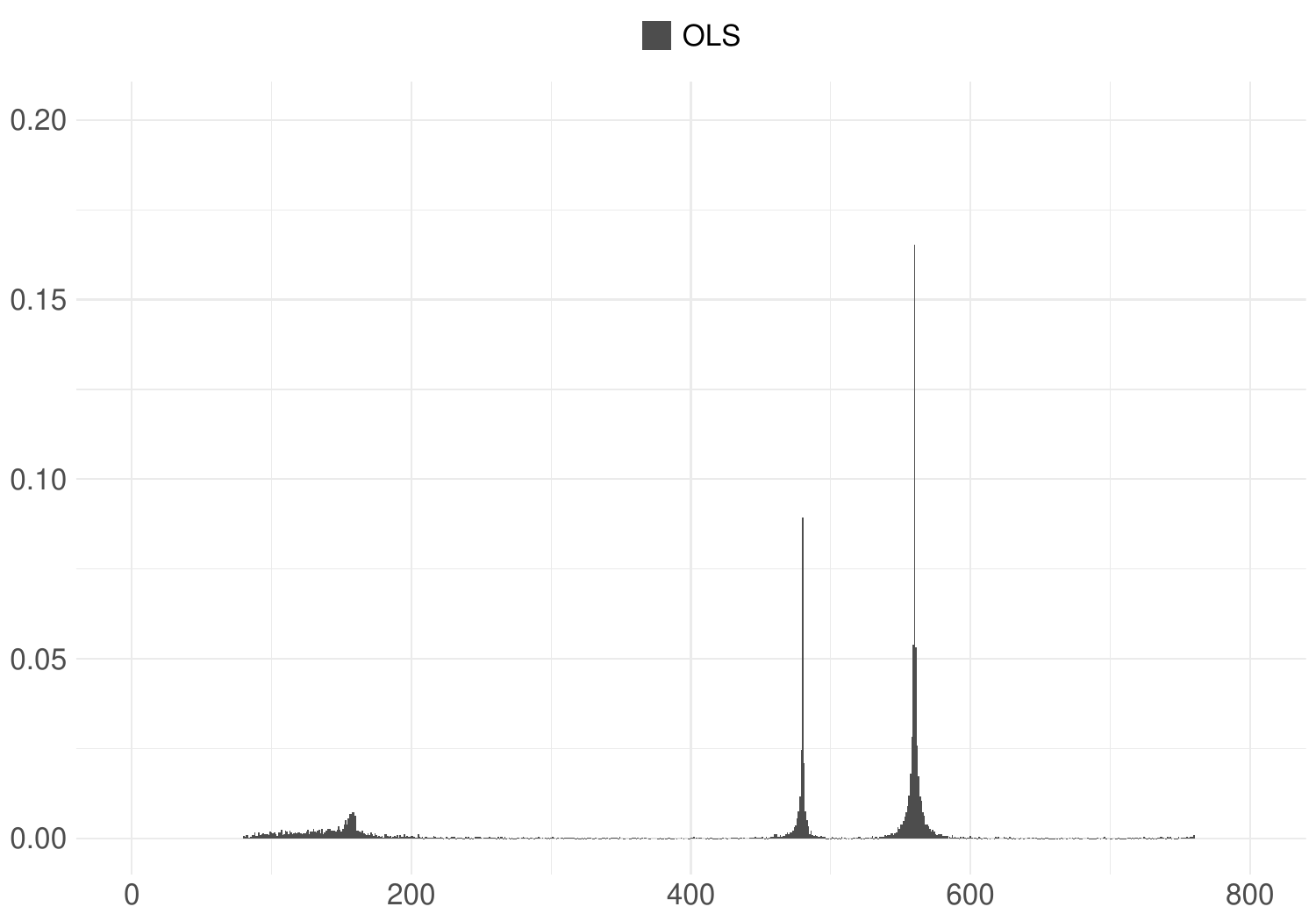}\label{fig2:6r:5}}
\subfigure[$T=800$, $c_a=6$, $c_b=6$, $s_0/s_1=5$]{\includegraphics[width=0.45\linewidth]{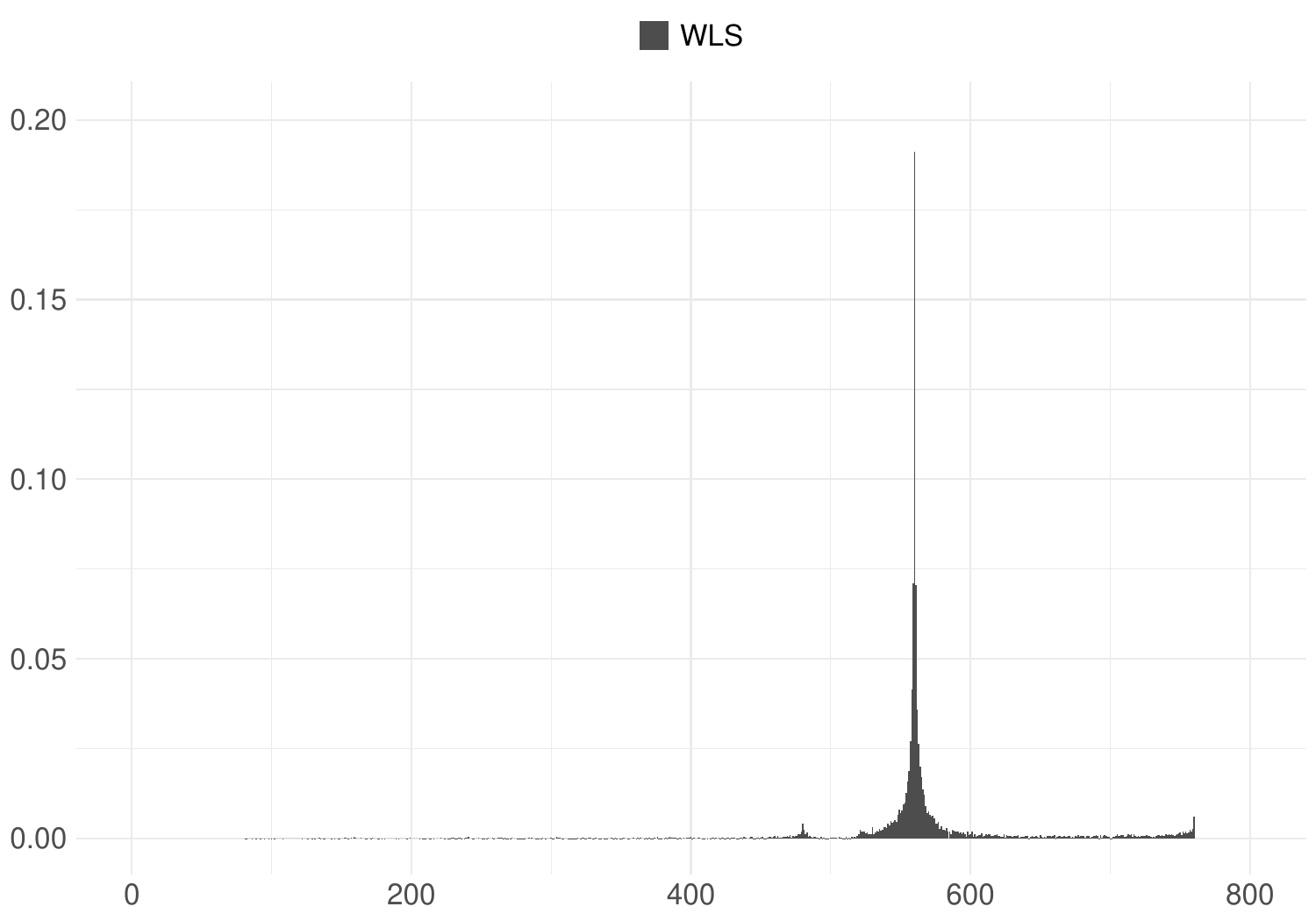}\label{fig2:6r:6}}\\
\end{center}%
\caption{Histograms of $\hat{k}_r$ 
for $(\tau_e,\tau_c,\tau_r)=(0.4,0.6,0.7)$,  $\tau=0.2$, $s_0/s_1=5$, $T=800$}
\label{fig26_r}
\end{figure}

\newpage
\renewcommand{\thesection}{\Alph{section}.}
\section{Accuracy tables}\label{sec:acc_app}
{\color{black} The tables report $P_5$, $P_{10}$ and the RMSE of the estimators of $k_e$, $k_c$ and $k_r$ for all the combinations of the volatility design, $s_0/s_1$, $T$ and $c_a$ consideblack in the paper (the homoskedastic case $s_0/s_1=1$ is reported once). The tables with the bias and MAE are available from the authors.}

{\color{black}
\scriptsize\setlength{\tabcolsep}{3pt}
\begin{longtable}{lrr ccc ccc ccc}
\caption{{\color{black}Accuracy of $\hat k_e$: $P_5$, $P_{10}$ and RMSE (10,000 replications; $c_b=6$). RMSE is computed over the replications in which the date is estimated.}}\label{tab:acc_k_e}\\
\toprule
 & & & \multicolumn{3}{c}{$P_5$} & \multicolumn{3}{c}{$P_{10}$} & \multicolumn{3}{c}{RMSE}\\
\cmidrule(lr){4-6}\cmidrule(lr){7-9}\cmidrule(lr){10-12}
Design & $T$ & $c_a$ & OLS & WLS & sADF & OLS & WLS & sADF & OLS & WLS & sADF\\
\midrule
\endfirsthead
\toprule
Design & $T$ & $c_a$ & OLS & WLS & sADF & OLS & WLS & sADF & OLS & WLS & sADF\\
\midrule
\endhead
\bottomrule
\endfoot
homoskedastic & 400 & 4 & 0.063 & 0.062 & 0.089 & 0.109 & 0.109 & 0.159 & 87.2 & 87.1 & 81.1 \\
 & 400 & 5 & 0.094 & 0.093 & 0.110 & 0.155 & 0.155 & 0.188 & 80.2 & 79.9 & 75.0 \\
 & 400 & 6 & 0.128 & 0.126 & 0.129 & 0.205 & 0.202 & 0.218 & 73.2 & 73.1 & 68.6 \\
 & 800 & 4 & 0.031 & 0.031 & 0.044 & 0.054 & 0.054 & 0.082 & 177.5 & 177.3 & 163.6 \\
 & 800 & 5 & 0.047 & 0.047 & 0.055 & 0.079 & 0.079 & 0.099 & 164.6 & 164.7 & 151.4 \\
 & 800 & 6 & 0.067 & 0.067 & 0.065 & 0.109 & 0.110 & 0.118 & 151.0 & 150.2 & 138.6 \\
\midrule
shift, $\tau=0.8$, $s_0/s_1=1/5$ & 400 & 4 & 0.010 & 0.064 & 0.089 & 0.017 & 0.110 & 0.159 & 155.4 & 87.4 & 81.1 \\
 & 400 & 5 & 0.023 & 0.095 & 0.110 & 0.037 & 0.157 & 0.188 & 149.9 & 80.6 & 75.0 \\
 & 400 & 6 & 0.045 & 0.131 & 0.129 & 0.068 & 0.208 & 0.218 & 142.8 & 73.5 & 68.6 \\
 & 800 & 4 & 0.004 & 0.030 & 0.044 & 0.007 & 0.052 & 0.082 & 318.6 & 175.8 & 163.6 \\
 & 800 & 5 & 0.011 & 0.047 & 0.055 & 0.018 & 0.078 & 0.099 & 309.3 & 163.3 & 151.4 \\
 & 800 & 6 & 0.021 & 0.066 & 0.065 & 0.033 & 0.109 & 0.118 & 296.3 & 148.1 & 138.6 \\
\midrule
shift, $\tau=0.8$, $s_0/s_1=5$ & 400 & 4 & 0.050 & 0.052 & 0.084 & 0.089 & 0.089 & 0.148 & 85.0 & 86.5 & 83.1 \\
 & 400 & 5 & 0.069 & 0.069 & 0.098 & 0.120 & 0.118 & 0.175 & 79.4 & 80.8 & 77.4 \\
 & 400 & 6 & 0.099 & 0.096 & 0.119 & 0.162 & 0.159 & 0.205 & 73.3 & 74.8 & 71.7 \\
 & 800 & 4 & 0.028 & 0.027 & 0.043 & 0.051 & 0.048 & 0.078 & 171.0 & 178.6 & 166.0 \\
 & 800 & 5 & 0.040 & 0.039 & 0.054 & 0.067 & 0.066 & 0.097 & 160.8 & 167.4 & 153.5 \\
 & 800 & 6 & 0.058 & 0.056 & 0.066 & 0.095 & 0.092 & 0.115 & 147.4 & 153.9 & 140.7 \\
\midrule
shift, $\tau=0.2$, $s_0/s_1=1/5$ & 400 & 4 & 0.062 & 0.051 & 0.067 & 0.113 & 0.093 & 0.125 & 72.7 & 83.2 & 86.5 \\
 & 400 & 5 & 0.066 & 0.059 & 0.077 & 0.121 & 0.106 & 0.144 & 68.8 & 80.0 & 82.3 \\
 & 400 & 6 & 0.077 & 0.070 & 0.087 & 0.139 & 0.125 & 0.160 & 65.6 & 76.1 & 77.9 \\
 & 800 & 4 & 0.032 & 0.027 & 0.037 & 0.059 & 0.048 & 0.067 & 143.5 & 174.6 & 173.6 \\
 & 800 & 5 & 0.036 & 0.031 & 0.042 & 0.065 & 0.055 & 0.074 & 137.0 & 167.8 & 165.2 \\
 & 800 & 6 & 0.041 & 0.037 & 0.048 & 0.073 & 0.066 & 0.084 & 129.1 & 159.2 & 156.1 \\
\midrule
shift, $\tau=0.2$, $s_0/s_1=5$ & 400 & 4 & 0.019 & 0.302 & 0.227 & 0.025 & 0.407 & 0.341 & 120.0 & 61.5 & 57.0 \\
 & 400 & 5 & 0.081 & 0.382 & 0.269 & 0.096 & 0.494 & 0.396 & 111.8 & 54.3 & 50.2 \\
 & 400 & 6 & 0.191 & 0.451 & 0.309 & 0.219 & 0.565 & 0.439 & 99.8 & 47.7 & 45.1 \\
 & 800 & 4 & 0.008 & 0.201 & 0.139 & 0.012 & 0.289 & 0.217 & 241.1 & 120.8 & 114.2 \\
 & 800 & 5 & 0.052 & 0.266 & 0.172 & 0.067 & 0.367 & 0.260 & 225.9 & 105.9 & 101.4 \\
 & 800 & 6 & 0.138 & 0.329 & 0.205 & 0.167 & 0.437 & 0.301 & 202.4 & 94.4 & 89.5 \\
\midrule
double shift, $s_0/s_1=1/5$ & 400 & 4 & 0.126 & 0.022 & 0.050 & 0.224 & 0.038 & 0.091 & 37.2 & 100.4 & 86.2 \\
 & 400 & 5 & 0.112 & 0.025 & 0.055 & 0.203 & 0.042 & 0.102 & 37.4 & 97.2 & 84.1 \\
 & 400 & 6 & 0.100 & 0.028 & 0.061 & 0.183 & 0.048 & 0.111 & 38.0 & 93.7 & 81.3 \\
 & 800 & 4 & 0.067 & 0.007 & 0.025 & 0.119 & 0.012 & 0.046 & 73.9 & 198.5 & 174.2 \\
 & 800 & 5 & 0.060 & 0.006 & 0.028 & 0.106 & 0.012 & 0.052 & 74.2 & 193.3 & 169.2 \\
 & 800 & 6 & 0.053 & 0.005 & 0.031 & 0.094 & 0.011 & 0.055 & 75.2 & 189.6 & 164.0 \\
\midrule
double shift, $s_0/s_1=5$ & 400 & 4 & 0.030 & 0.117 & 0.278 & 0.051 & 0.172 & 0.405 & 104.6 & 88.4 & 51.4 \\
 & 400 & 5 & 0.080 & 0.198 & 0.324 & 0.118 & 0.277 & 0.459 & 95.9 & 76.8 & 45.8 \\
 & 400 & 6 & 0.152 & 0.277 & 0.361 & 0.213 & 0.376 & 0.504 & 85.1 & 66.2 & 40.9 \\
 & 800 & 4 & 0.017 & 0.068 & 0.175 & 0.028 & 0.105 & 0.274 & 211.0 & 171.0 & 100.8 \\
 & 800 & 5 & 0.050 & 0.123 & 0.220 & 0.077 & 0.180 & 0.327 & 193.2 & 148.5 & 88.1 \\
 & 800 & 6 & 0.098 & 0.185 & 0.258 & 0.143 & 0.262 & 0.373 & 171.7 & 127.4 & 78.3 \\
\midrule
logistic, $s_0/s_1=1/5$ & 400 & 4 & 0.008 & 0.014 & 0.064 & 0.013 & 0.026 & 0.116 & 109.8 & 107.1 & 90.2 \\
 & 400 & 5 & 0.011 & 0.022 & 0.077 & 0.018 & 0.036 & 0.137 & 106.3 & 104.6 & 87.4 \\
 & 400 & 6 & 0.017 & 0.034 & 0.092 & 0.026 & 0.055 & 0.163 & 102.8 & 101.4 & 84.2 \\
 & 800 & 4 & 0.004 & 0.010 & 0.033 & 0.005 & 0.018 & 0.061 & 222.0 & 214.0 & 180.6 \\
 & 800 & 5 & 0.004 & 0.015 & 0.039 & 0.006 & 0.025 & 0.070 & 216.9 & 209.3 & 175.0 \\
 & 800 & 6 & 0.006 & 0.023 & 0.047 & 0.010 & 0.037 & 0.084 & 209.4 & 203.8 & 169.6 \\
\midrule
logistic, $s_0/s_1=5$ & 400 & 4 & 0.042 & 0.055 & 0.075 & 0.074 & 0.097 & 0.134 & 96.0 & 85.6 & 63.1 \\
 & 400 & 5 & 0.077 & 0.090 & 0.093 & 0.130 & 0.156 & 0.163 & 87.2 & 75.9 & 56.6 \\
 & 400 & 6 & 0.113 & 0.124 & 0.111 & 0.187 & 0.208 & 0.191 & 77.8 & 66.5 & 51.7 \\
 & 800 & 4 & 0.024 & 0.029 & 0.038 & 0.042 & 0.052 & 0.069 & 193.0 & 168.3 & 122.4 \\
 & 800 & 5 & 0.042 & 0.050 & 0.046 & 0.072 & 0.085 & 0.080 & 175.8 & 149.0 & 110.2 \\
 & 800 & 6 & 0.067 & 0.068 & 0.055 & 0.111 & 0.115 & 0.095 & 156.4 & 130.8 & 101.1 \\
\midrule
linear trend, $s_0/s_1=1/5$ & 400 & 4 & 0.038 & 0.045 & 0.066 & 0.069 & 0.081 & 0.122 & 109.0 & 95.2 & 88.0 \\
 & 400 & 5 & 0.046 & 0.057 & 0.079 & 0.084 & 0.099 & 0.142 & 104.0 & 91.0 & 84.1 \\
 & 400 & 6 & 0.059 & 0.073 & 0.090 & 0.104 & 0.124 & 0.160 & 98.3 & 86.2 & 79.9 \\
 & 800 & 4 & 0.019 & 0.023 & 0.035 & 0.034 & 0.043 & 0.064 & 221.2 & 188.2 & 175.7 \\
 & 800 & 5 & 0.022 & 0.028 & 0.040 & 0.042 & 0.052 & 0.074 & 210.5 & 180.7 & 168.3 \\
 & 800 & 6 & 0.029 & 0.036 & 0.048 & 0.054 & 0.065 & 0.084 & 198.7 & 172.2 & 159.3 \\
\midrule
linear trend, $s_0/s_1=5$ & 400 & 4 & 0.034 & 0.061 & 0.098 & 0.060 & 0.107 & 0.177 & 100.0 & 85.8 & 76.8 \\
 & 400 & 5 & 0.066 & 0.095 & 0.123 & 0.110 & 0.155 & 0.209 & 92.0 & 78.1 & 70.3 \\
 & 400 & 6 & 0.106 & 0.131 & 0.147 & 0.165 & 0.208 & 0.244 & 83.1 & 70.4 & 63.7 \\
 & 800 & 4 & 0.020 & 0.035 & 0.053 & 0.035 & 0.060 & 0.097 & 200.2 & 171.8 & 152.3 \\
 & 800 & 5 & 0.038 & 0.055 & 0.066 & 0.062 & 0.091 & 0.117 & 185.3 & 155.3 & 138.1 \\
 & 800 & 6 & 0.061 & 0.079 & 0.083 & 0.098 & 0.127 & 0.141 & 167.2 & 138.6 & 125.0 \\
\midrule
\end{longtable}
}

{\color{black}
\scriptsize\setlength{\tabcolsep}{3pt}
\begin{longtable}{lrr ccc ccc ccc}
\caption{{\color{black}Accuracy of $\hat k_c$: $P_5$, $P_{10}$ and RMSE (10,000 replications; $c_b=6$). RMSE is computed over the replications in which the date is estimated.}}\label{tab:acc_k_c}\\
\toprule
 & & & \multicolumn{3}{c}{$P_5$} & \multicolumn{3}{c}{$P_{10}$} & \multicolumn{3}{c}{RMSE}\\
\cmidrule(lr){4-6}\cmidrule(lr){7-9}\cmidrule(lr){10-12}
Design & $T$ & $c_a$ & OLS & WLS & sADF & OLS & WLS & sADF & OLS & WLS & sADF\\
\midrule
\endfirsthead
\toprule
Design & $T$ & $c_a$ & OLS & WLS & sADF & OLS & WLS & sADF & OLS & WLS & sADF\\
\midrule
\endhead
\bottomrule
\endfoot
homoskedastic & 400 & 4 & 0.376 & 0.374 & 0.225 & 0.437 & 0.436 & 0.297 & 96.7 & 97.1 & 78.0 \\
 & 400 & 5 & 0.476 & 0.476 & 0.323 & 0.529 & 0.529 & 0.399 & 89.1 & 89.3 & 71.2 \\
 & 400 & 6 & 0.568 & 0.568 & 0.440 & 0.609 & 0.610 & 0.513 & 81.8 & 81.9 & 63.9 \\
 & 800 & 4 & 0.271 & 0.270 & 0.141 & 0.344 & 0.345 & 0.198 & 199.5 & 199.4 & 158.4 \\
 & 800 & 5 & 0.376 & 0.377 & 0.231 & 0.449 & 0.450 & 0.298 & 183.0 & 182.7 & 143.8 \\
 & 800 & 6 & 0.474 & 0.476 & 0.330 & 0.543 & 0.544 & 0.408 & 167.9 & 167.5 & 129.3 \\
\midrule
shift, $\tau=0.8$, $s_0/s_1=1/5$ & 400 & 4 & 0.063 & 0.398 & 0.225 & 0.069 & 0.455 & 0.297 & 115.6 & 95.6 & 78.0 \\
 & 400 & 5 & 0.107 & 0.492 & 0.323 & 0.112 & 0.544 & 0.399 & 112.6 & 87.9 & 71.2 \\
 & 400 & 6 & 0.174 & 0.576 & 0.440 & 0.177 & 0.619 & 0.513 & 108.3 & 80.7 & 63.9 \\
 & 800 & 4 & 0.041 & 0.300 & 0.141 & 0.049 & 0.373 & 0.198 & 234.1 & 193.9 & 158.4 \\
 & 800 & 5 & 0.075 & 0.395 & 0.231 & 0.083 & 0.468 & 0.298 & 229.6 & 179.7 & 143.8 \\
 & 800 & 6 & 0.128 & 0.495 & 0.330 & 0.139 & 0.559 & 0.408 & 222.5 & 164.3 & 129.3 \\
\midrule
shift, $\tau=0.8$, $s_0/s_1=5$ & 400 & 4 & 0.355 & 0.301 & 0.202 & 0.421 & 0.364 & 0.271 & 106.3 & 111.3 & 80.4 \\
 & 400 & 5 & 0.454 & 0.403 & 0.299 & 0.514 & 0.465 & 0.371 & 98.1 & 103.1 & 73.7 \\
 & 400 & 6 & 0.550 & 0.505 & 0.402 & 0.600 & 0.556 & 0.476 & 89.6 & 93.8 & 67.2 \\
 & 800 & 4 & 0.268 & 0.208 & 0.134 & 0.348 & 0.274 & 0.187 & 214.3 & 229.7 & 160.9 \\
 & 800 & 5 & 0.365 & 0.307 & 0.214 & 0.449 & 0.383 & 0.280 & 197.8 & 210.5 & 146.3 \\
 & 800 & 6 & 0.467 & 0.414 & 0.315 & 0.544 & 0.488 & 0.392 & 181.5 & 193.7 & 131.4 \\
\midrule
shift, $\tau=0.2$, $s_0/s_1=1/5$ & 400 & 4 & 0.272 & 0.274 & 0.159 & 0.343 & 0.343 & 0.218 & 76.3 & 87.9 & 85.7 \\
 & 400 & 5 & 0.364 & 0.363 & 0.225 & 0.434 & 0.430 & 0.292 & 70.8 & 82.4 & 80.7 \\
 & 400 & 6 & 0.462 & 0.459 & 0.319 & 0.525 & 0.518 & 0.392 & 65.6 & 76.8 & 75.0 \\
 & 800 & 4 & 0.193 & 0.194 & 0.101 & 0.263 & 0.260 & 0.146 & 155.2 & 191.3 & 170.2 \\
 & 800 & 5 & 0.277 & 0.277 & 0.158 & 0.358 & 0.357 & 0.215 & 143.4 & 177.9 & 160.3 \\
 & 800 & 6 & 0.370 & 0.368 & 0.236 & 0.454 & 0.450 & 0.301 & 132.1 & 165.7 & 149.4 \\
\midrule
shift, $\tau=0.2$, $s_0/s_1=5$ & 400 & 4 & 0.304 & 0.704 & 0.534 & 0.305 & 0.738 & 0.605 & 160.1 & 69.7 & 50.7 \\
 & 400 & 5 & 0.414 & 0.773 & 0.663 & 0.415 & 0.797 & 0.721 & 145.7 & 61.1 & 43.7 \\
 & 400 & 6 & 0.516 & 0.822 & 0.752 & 0.516 & 0.839 & 0.797 & 131.2 & 54.6 & 38.4 \\
 & 800 & 4 & 0.277 & 0.643 & 0.425 & 0.280 & 0.700 & 0.506 & 326.4 & 136.5 & 101.7 \\
 & 800 & 5 & 0.395 & 0.729 & 0.559 & 0.398 & 0.770 & 0.638 & 296.5 & 118.9 & 89.1 \\
 & 800 & 6 & 0.500 & 0.790 & 0.677 & 0.501 & 0.821 & 0.736 & 267.2 & 104.1 & 77.7 \\
\midrule
double shift, $s_0/s_1=1/5$ & 400 & 4 & 0.237 & 0.107 & 0.093 & 0.334 & 0.134 & 0.130 & 39.4 & 114.4 & 81.2 \\
 & 400 & 5 & 0.292 & 0.145 & 0.120 & 0.393 & 0.173 & 0.162 & 37.0 & 110.2 & 78.5 \\
 & 400 & 6 & 0.357 & 0.197 & 0.159 & 0.459 & 0.226 & 0.207 & 35.1 & 104.9 & 74.9 \\
 & 800 & 4 & 0.159 & 0.034 & 0.054 & 0.228 & 0.046 & 0.077 & 77.8 & 217.7 & 162.9 \\
 & 800 & 5 & 0.204 & 0.045 & 0.071 & 0.283 & 0.059 & 0.099 & 72.8 & 208.3 & 157.1 \\
 & 800 & 6 & 0.261 & 0.059 & 0.099 & 0.346 & 0.073 & 0.133 & 68.4 & 197.3 & 150.5 \\
\midrule
double shift, $s_0/s_1=5$ & 400 & 4 & 0.350 & 0.486 & 0.627 & 0.384 & 0.536 & 0.698 & 111.8 & 90.3 & 45.7 \\
 & 400 & 5 & 0.447 & 0.581 & 0.734 & 0.479 & 0.623 & 0.787 & 101.4 & 79.7 & 39.6 \\
 & 400 & 6 & 0.542 & 0.664 & 0.805 & 0.568 & 0.700 & 0.844 & 91.1 & 70.0 & 34.5 \\
 & 800 & 4 & 0.298 & 0.436 & 0.516 & 0.339 & 0.506 & 0.608 & 228.7 & 172.2 & 90.1 \\
 & 800 & 5 & 0.401 & 0.541 & 0.650 & 0.444 & 0.606 & 0.723 & 207.5 & 150.0 & 76.5 \\
 & 800 & 6 & 0.496 & 0.634 & 0.749 & 0.536 & 0.688 & 0.801 & 187.0 & 130.8 & 66.7 \\
\midrule
logistic, $s_0/s_1=1/5$ & 400 & 4 & 0.134 & 0.093 & 0.078 & 0.200 & 0.136 & 0.121 & 87.4 & 122.6 & 89.7 \\
 & 400 & 5 & 0.167 & 0.126 & 0.098 & 0.239 & 0.172 & 0.144 & 84.5 & 119.1 & 86.7 \\
 & 400 & 6 & 0.206 & 0.166 & 0.120 & 0.286 & 0.217 & 0.172 & 81.1 & 115.2 & 83.0 \\
 & 800 & 4 & 0.074 & 0.049 & 0.042 & 0.121 & 0.077 & 0.071 & 176.7 & 253.5 & 179.6 \\
 & 800 & 5 & 0.097 & 0.068 & 0.053 & 0.151 & 0.103 & 0.085 & 171.9 & 247.2 & 173.7 \\
 & 800 & 6 & 0.125 & 0.094 & 0.066 & 0.189 & 0.136 & 0.104 & 165.7 & 241.0 & 167.5 \\
\midrule
logistic, $s_0/s_1=5$ & 400 & 4 & 0.391 & 0.556 & 0.636 & 0.392 & 0.560 & 0.677 & 128.4 & 84.7 & 54.5 \\
 & 400 & 5 & 0.491 & 0.662 & 0.737 & 0.491 & 0.665 & 0.763 & 116.7 & 75.4 & 47.3 \\
 & 400 & 6 & 0.579 & 0.741 & 0.803 & 0.579 & 0.744 & 0.824 & 104.8 & 66.8 & 41.8 \\
 & 800 & 4 & 0.380 & 0.583 & 0.564 & 0.381 & 0.599 & 0.626 & 261.9 & 151.3 & 105.3 \\
 & 800 & 5 & 0.481 & 0.688 & 0.685 & 0.482 & 0.699 & 0.732 & 238.5 & 136.4 & 91.8 \\
 & 800 & 6 & 0.569 & 0.764 & 0.770 & 0.569 & 0.771 & 0.803 & 216.2 & 121.4 & 80.6 \\
\midrule
linear trend, $s_0/s_1=1/5$ & 400 & 4 & 0.186 & 0.208 & 0.134 & 0.243 & 0.268 & 0.192 & 92.0 & 99.7 & 86.8 \\
 & 400 & 5 & 0.260 & 0.284 & 0.189 & 0.319 & 0.345 & 0.256 & 87.2 & 94.4 & 82.0 \\
 & 400 & 6 & 0.354 & 0.374 & 0.265 & 0.411 & 0.433 & 0.339 & 81.7 & 88.3 & 76.9 \\
 & 800 & 4 & 0.129 & 0.143 & 0.080 & 0.180 & 0.198 & 0.121 & 185.2 & 208.3 & 173.5 \\
 & 800 & 5 & 0.188 & 0.204 & 0.122 & 0.251 & 0.272 & 0.172 & 175.7 & 196.8 & 164.9 \\
 & 800 & 6 & 0.266 & 0.285 & 0.179 & 0.337 & 0.358 & 0.241 & 164.3 & 183.6 & 154.5 \\
\midrule
linear trend, $s_0/s_1=5$ & 400 & 4 & 0.368 & 0.438 & 0.307 & 0.396 & 0.491 & 0.376 & 132.8 & 104.1 & 72.5 \\
 & 400 & 5 & 0.479 & 0.546 & 0.427 & 0.500 & 0.589 & 0.498 & 121.0 & 93.9 & 65.1 \\
 & 400 & 6 & 0.570 & 0.642 & 0.548 & 0.584 & 0.673 & 0.610 & 110.5 & 84.4 & 57.6 \\
 & 800 & 4 & 0.312 & 0.359 & 0.226 & 0.359 & 0.434 & 0.290 & 269.9 & 204.5 & 144.1 \\
 & 800 & 5 & 0.420 & 0.476 & 0.339 & 0.462 & 0.541 & 0.415 & 248.7 & 183.1 & 127.4 \\
 & 800 & 6 & 0.524 & 0.584 & 0.455 & 0.559 & 0.640 & 0.534 & 225.4 & 164.4 & 113.3 \\
\midrule
\end{longtable}
}

{\color{black}
\scriptsize\setlength{\tabcolsep}{3pt}
\begin{longtable}{lrr cc cc cc}
\caption{{\color{black}Accuracy of $\hat k_r$: $P_5$, $P_{10}$ and RMSE (10,000 replications; $c_b=6$). RMSE is computed over the replications in which the date is estimated.}}\label{tab:acc_k_r}\\
\toprule
 & & & \multicolumn{2}{c}{$P_5$} & \multicolumn{2}{c}{$P_{10}$} & \multicolumn{2}{c}{RMSE}\\
\cmidrule(lr){4-5}\cmidrule(lr){6-7}\cmidrule(lr){8-9}
Design & $T$ & $c_a$ & OLS & WLS & OLS & WLS & OLS & WLS\\
\midrule
\endfirsthead
\toprule
Design & $T$ & $c_a$ & OLS & WLS & OLS & WLS & OLS & WLS\\
\midrule
\endhead
\bottomrule
\endfoot
homoskedastic & 400 & 4 & 0.166 & 0.166 & 0.243 & 0.242 & 73.7 & 74.9 \\
 & 400 & 5 & 0.222 & 0.224 & 0.312 & 0.315 & 67.7 & 68.5 \\
 & 400 & 6 & 0.280 & 0.282 & 0.379 & 0.382 & 62.4 & 63.2 \\
 & 800 & 4 & 0.090 & 0.090 & 0.144 & 0.144 & 150.3 & 151.3 \\
 & 800 & 5 & 0.126 & 0.127 & 0.191 & 0.191 & 137.0 & 137.7 \\
 & 800 & 6 & 0.166 & 0.167 & 0.243 & 0.244 & 125.8 & 125.3 \\
\midrule
shift, $\tau=0.8$, $s_0/s_1=1/5$ & 400 & 4 & 0.005 & 0.115 & 0.006 & 0.168 & 89.3 & 78.9 \\
 & 400 & 5 & 0.011 & 0.153 & 0.013 & 0.216 & 87.9 & 74.5 \\
 & 400 & 6 & 0.026 & 0.198 & 0.030 & 0.270 & 85.4 & 69.3 \\
 & 800 & 4 & 0.002 & 0.087 & 0.002 & 0.140 & 179.7 & 154.8 \\
 & 800 & 5 & 0.006 & 0.117 & 0.008 & 0.181 & 177.5 & 144.7 \\
 & 800 & 6 & 0.014 & 0.155 & 0.018 & 0.229 & 173.0 & 133.4 \\
\midrule
shift, $\tau=0.8$, $s_0/s_1=5$ & 400 & 4 & 0.185 & 0.149 & 0.289 & 0.223 & 74.7 & 81.4 \\
 & 400 & 5 & 0.231 & 0.193 & 0.345 & 0.278 & 68.3 & 74.1 \\
 & 400 & 6 & 0.283 & 0.247 & 0.404 & 0.347 & 61.7 & 67.3 \\
 & 800 & 4 & 0.105 & 0.082 & 0.175 & 0.130 & 152.7 & 167.5 \\
 & 800 & 5 & 0.138 & 0.116 & 0.217 & 0.179 & 139.8 & 153.4 \\
 & 800 & 6 & 0.177 & 0.153 & 0.268 & 0.228 & 126.6 & 138.0 \\
\midrule
shift, $\tau=0.2$, $s_0/s_1=1/5$ & 400 & 4 & 0.102 & 0.104 & 0.162 & 0.164 & 65.7 & 69.1 \\
 & 400 & 5 & 0.136 & 0.137 & 0.208 & 0.211 & 61.8 & 64.7 \\
 & 400 & 6 & 0.180 & 0.181 & 0.263 & 0.263 & 57.6 & 60.0 \\
 & 800 & 4 & 0.060 & 0.060 & 0.100 & 0.098 & 131.9 & 143.6 \\
 & 800 & 5 & 0.081 & 0.084 & 0.133 & 0.135 & 123.5 & 133.1 \\
 & 800 & 6 & 0.111 & 0.111 & 0.173 & 0.171 & 115.4 & 124.7 \\
\midrule
shift, $\tau=0.2$, $s_0/s_1=5$ & 400 & 4 & 0.267 & 0.494 & 0.291 & 0.583 & 130.8 & 49.4 \\
 & 400 & 5 & 0.371 & 0.567 & 0.399 & 0.654 & 114.6 & 43.2 \\
 & 400 & 6 & 0.468 & 0.630 & 0.500 & 0.710 & 100.4 & 38.4 \\
 & 800 & 4 & 0.208 & 0.378 & 0.242 & 0.477 & 269.3 & 96.2 \\
 & 800 & 5 & 0.308 & 0.452 & 0.354 & 0.552 & 235.8 & 83.7 \\
 & 800 & 6 & 0.405 & 0.524 & 0.456 & 0.619 & 206.7 & 73.0 \\
\midrule
double shift, $s_0/s_1=1/5$ & 400 & 4 & 0.267 & 0.156 & 0.330 & 0.187 & 43.4 & 81.8 \\
 & 400 & 5 & 0.323 & 0.191 & 0.385 & 0.224 & 41.2 & 78.7 \\
 & 400 & 6 & 0.381 & 0.238 & 0.444 & 0.272 & 38.9 & 75.1 \\
 & 800 & 4 & 0.193 & 0.066 & 0.255 & 0.092 & 87.3 & 161.3 \\
 & 800 & 5 & 0.240 & 0.076 & 0.308 & 0.102 & 82.9 & 157.3 \\
 & 800 & 6 & 0.297 & 0.082 & 0.370 & 0.107 & 78.5 & 151.8 \\
\midrule
double shift, $s_0/s_1=5$ & 400 & 4 & 0.118 & 0.144 & 0.180 & 0.225 & 82.4 & 69.1 \\
 & 400 & 5 & 0.165 & 0.188 & 0.242 & 0.283 & 74.0 & 61.7 \\
 & 400 & 6 & 0.219 & 0.238 & 0.309 & 0.345 & 66.6 & 55.3 \\
 & 800 & 4 & 0.071 & 0.088 & 0.113 & 0.141 & 167.5 & 133.2 \\
 & 800 & 5 & 0.102 & 0.117 & 0.159 & 0.185 & 150.8 & 119.0 \\
 & 800 & 6 & 0.139 & 0.152 & 0.211 & 0.234 & 135.1 & 108.1 \\
\midrule
logistic, $s_0/s_1=1/5$ & 400 & 4 & 0.036 & 0.033 & 0.068 & 0.060 & 71.6 & 95.3 \\
 & 400 & 5 & 0.042 & 0.036 & 0.077 & 0.067 & 70.3 & 92.8 \\
 & 400 & 6 & 0.051 & 0.045 & 0.091 & 0.081 & 68.7 & 90.0 \\
 & 800 & 4 & 0.018 & 0.015 & 0.035 & 0.028 & 143.3 & 198.7 \\
 & 800 & 5 & 0.021 & 0.017 & 0.039 & 0.032 & 141.5 & 193.8 \\
 & 800 & 6 & 0.026 & 0.022 & 0.048 & 0.039 & 138.8 & 189.0 \\
\midrule
logistic, $s_0/s_1=5$ & 400 & 4 & 0.370 & 0.498 & 0.387 & 0.542 & 104.1 & 55.9 \\
 & 400 & 5 & 0.468 & 0.600 & 0.486 & 0.641 & 93.2 & 48.8 \\
 & 400 & 6 & 0.560 & 0.682 & 0.574 & 0.718 & 82.4 & 43.1 \\
 & 800 & 4 & 0.333 & 0.448 & 0.362 & 0.513 & 214.4 & 102.4 \\
 & 800 & 5 & 0.428 & 0.544 & 0.461 & 0.613 & 191.5 & 87.9 \\
 & 800 & 6 & 0.518 & 0.621 & 0.549 & 0.687 & 170.1 & 76.2 \\
\midrule
linear trend, $s_0/s_1=1/5$ & 400 & 4 & 0.047 & 0.061 & 0.078 & 0.102 & 73.4 & 78.1 \\
 & 400 & 5 & 0.066 & 0.085 & 0.106 & 0.137 & 70.7 & 74.3 \\
 & 400 & 6 & 0.091 & 0.114 & 0.141 & 0.177 & 67.5 & 70.1 \\
 & 800 & 4 & 0.024 & 0.032 & 0.042 & 0.057 & 147.6 & 162.0 \\
 & 800 & 5 & 0.035 & 0.044 & 0.060 & 0.076 & 142.4 & 153.9 \\
 & 800 & 6 & 0.052 & 0.063 & 0.084 & 0.104 & 135.9 & 146.0 \\
\midrule
linear trend, $s_0/s_1=5$ & 400 & 4 & 0.248 & 0.287 & 0.321 & 0.373 & 103.6 & 73.4 \\
 & 400 & 5 & 0.326 & 0.360 & 0.407 & 0.454 & 93.3 & 66.5 \\
 & 400 & 6 & 0.401 & 0.435 & 0.485 & 0.531 & 84.3 & 59.0 \\
 & 800 & 4 & 0.169 & 0.194 & 0.236 & 0.272 & 211.0 & 142.9 \\
 & 800 & 5 & 0.235 & 0.259 & 0.310 & 0.345 & 191.7 & 126.9 \\
 & 800 & 6 & 0.306 & 0.323 & 0.388 & 0.417 & 170.0 & 112.7 \\
\midrule
\end{longtable}
}

\end{document}